\documentclass[11pt]{cernrepdbg}
\usepackage{graphicx}
\usepackage{epsfig}
\usepackage{color}
\usepackage{bbm}
\usepackage{cite}
\usepackage{amssymb}
\usepackage{here}
\usepackage{xspace}
\usepackage{epsf}
\usepackage{feynmp}
\begin{document}
\bibliographystyle{lesHouches}

%%%%FROM csection_houch.tex
\def\GGHPSsigtot  {\mbox{$\sigma_{\rm tot}^{\rm pp}$}}
\def\GGHPSrs{\mbox{$\sqrt{s}$}}
\newcommand{\GGHPScomment}[1]{}

%%%%FROM nnlo.tex
\def\NNLObeq{\begin{equation}}
\def\NNLOeeq{\end{equation}}
\def\NNLObeeq{\begin{eqnarray}}
\def\NNLOeeeq{\end{eqnarray}}
\def\NNLOaand{\!\!\!\!\!\!\!\!&&}
\def\NNLOd{{\rm d}}

\newcommand\NNLOla         {\langle}
\newcommand\NNLOra         {\rangle}
\newcommand\NNLObom[1]     {{\mbox{\boldmath $#1$}}}
\newcommand\NNLORef[1]     {Ref.\,\cite{#1}}
\newcommand\NNLORefs[1]    {Refs.\,\cite{#1}}
\newcommand\NNLOeqn[1]     {Eq.\,(\ref{#1})}
\newcommand\NNLOeqns[2]    {Eqs.\,(\ref{#1}) and~(\ref{#2})}
\newcommand\NNLOeqnss[2]   {Eqs.\,(\ref{#1})--(\ref{#2})}
\newcommand\NNLOnn         {\nonumber}
\newcommand\NNLOas         {\ensuremath{\alpha_{\mathrm{s}}}}
\newcommand{\NNLObT}       {\NNLObom{T}}
\newcommand{\NNLOeps}      {\varepsilon}        
\newcommand{\NNLOrd}       {{\mathrm{d}}}
\newcommand\NNLOtsig[1]    {\sigma^{\mathrm{#1}}}
\newcommand\NNLOdsig[1]    {\NNLOrd\sigma^{{\rm #1}}}
\newcommand\NNLOdsiga[2]   {\NNLOrd\sigma^{{\rm #1,A}_{\scriptscriptstyle #2}}}
\newcommand\NNLOM[2]       {\ensuremath{|{\cal{M}}_{#1}^{#2}|^2}}
\newcommand\NNLObra[3]     {\NNLOla {\cal M}_{#1}^{#2}#3|}
\newcommand\NNLOket[3]     {|{\cal M}_{#1}^{#2}#3\NNLOra}
\newcommand{\NNLObA}[1]    {\NNLObom{\mathrm A}_{#1}}
\newcommand{\NNLOhP}       {\hat{P}}
\newcommand{\NNLOcS}       {{\cal S}}
\newcommand{\NNLOcC}[1]    {{\cal C}_{#1}}
\newcommand{\NNLOcCS}[1]   {{\cal C}\kern-2pt{\cal S}_{#1}}
\newcommand{\NNLOPS}[2]    {\NNLOrd\phi^{(#1)}{#2}}
\newcommand{\NNLOti}[1]    {\tilde{#1}}
\newcommand{\NNLOwti}[1]   {\widetilde{#1}}
\newcommand\NNLOtz[1]      {\tilde z_{#1}}
\newcommand\NNLOtzz[2]     {\tilde z_{#1,#2}}
\newcommand\NNLOkT[1]      {k_{\perp,#1}}
\newcommand\NNLOkTt[1]     {\tilde{k}_{\perp,#1}}

%%%%FROM reweight.tex
\newcommand\SFsss{\scriptscriptstyle}
\newcommand\SFhalf{\frac{1}{2}}

%%%%FROM DYcomparison.tex
\newcommand{\DYPV}{\mathrm{V}}
\newcommand{\DYPW}{\mathrm{W}}
\newcommand{\DYPZ}{\mathrm{Z}}
\newcommand{\DYPH}{\mathrm{H}}
\newcommand{\DYPe}{\mathrm{e}}
\newcommand{\DYPp}{\mathrm{p}}
\newcommand{\DYPd}{\mathrm{d}}
\newcommand{\DYPu}{\mathrm{u}}
\newcommand{\DYPs}{\mathrm{s}}
\newcommand{\DYPc}{\mathrm{c}}
\newcommand{\DYPb}{\mathrm{b}}
\newcommand{\DYPt}{\mathrm{t}}
\newcommand{\DYPq}{\mathrm{q}}
\newcommand{\DYPep}{\mathrm{{e^+}}}
\newcommand{\DYPem}{\mathrm{{e^-}}}
\newcommand{\DYPWp}{\mathrm{{W^+}}}
\newcommand{\DYPWm}{\mathrm{{W^-}}}
% various shorthands 
\newcommand{\DYsw}{{s_\rw}}
\newcommand{\DYcw}{{c_\rw}}
\newcommand{\DYQq}{{Q_\Pq}}
\newcommand{\DYGF}{{G_\mu}}
%physical units
\newcommand{\DYTeV}{\unskip\,\mathrm{TeV}}
\newcommand{\DYGeV}{\unskip\,\mathrm{GeV}}
\newcommand{\DYMeV}{\unskip\,\mathrm{MeV}}
\newcommand{\DYpba}{\unskip\,\mathrm{pb}}
\newcommand{\DYfb}{\unskip\,\mathrm{fb}}

%%%%FROM gg_bbh.tex
%%%%%%%%%%%%%%%% two floating figures, side by side %%%%%%%%%%%%%%%
\newenvironment{BBH2figures}[1]{\begin{figure}[#1] 
  \begin{center}
    \begin{tabular}{p{.47\textwidth}p{.47\textwidth}} }
 {  \end{tabular}
  \end{center} 
 \end{figure}
}
%%%%%%%%%%%%%%%%%%%%%%%%%%%%%%%%%%%%%%%%%%%%%%%%%%%%%%%%%%%%%%%%%%%

%%%%FROM ttbarNormalization.tex
\newcommand{\DGZra}{\rightarrow}

%%%%FROM LH05_proceedings.tex
\newcommand{\ABBBWWL}{\mbox{${\mathrm W_L}{\mathrm W_L}$}\xspace}
\newcommand{\ABBBVVL}{\mbox{${\mathrm V_L}{\mathrm V_L}$}\xspace}
\newcommand{\ABBBWW}{\mbox{${\mathrm W}{\mathrm W}$}\xspace}
\newcommand{\ABBBZZ}{\mbox{${\mathrm Z}{\mathrm Z}$}\xspace}
\newcommand{\ABBBVV}{\mbox{${\mathrm V}{\mathrm V}$}\xspace}
\newcommand{\ABBBZW}{\mbox{${\mathrm Z}{\mathrm W}$}\xspace}
\newcommand{\ABBBVW}{\mbox{${\mathrm V}{\mathrm W}$}\xspace}
\newcommand{\ABBBW}{\mbox{${\mathrm W}$}\xspace}
\newcommand{\ABBBV}{\mbox{${\mathrm V}$}\xspace}
\newcommand{\ABBBVL}{\mbox{${\mathrm V_L}$}\xspace}
\newcommand{\ABBBZ}{\mbox{${\mathrm Z}$}\xspace}
\newcommand{\ABBBpt}{\mbox{${\mathrm p_T}$}\xspace}
\newcommand{\ABBBGeV}{\mbox{${\mathrm GeV}$}\xspace}
\newcommand{\ABBBTeV}{\mbox{${\mathrm TeV}$}\xspace}
\newcommand{\ABBBtoptop}{\mbox{$\mathrm {t\bar t}$}\xspace}

\newcommand{\ABBBtbn}[1]{Tab.~\ref{#1}}
\newcommand{\ABBBtbns}[2]{Tabs.~\ref{#1}--\ref{#2}}
\newcommand{\ABBBtbnsc}[2]{Tabs.~\ref{#1},~\ref{#2}}
\newcommand{\ABBBfig}[1]{Fig.~\ref{#1}}
\newcommand{\ABBBfigs}[2]{Figs.~\ref{#1}--\ref{#2}}
\newcommand{\ABBBfigsc}[2]{Figs.~\ref{#1},~\ref{#2}}

\newcommand{\ABBBPhase}{{\tt PHASE}\xspace}

%%%%FROM lh.tex
\newcommand{\CDmsbar}{$\overline{\mbox{MS}}$ }

%%%%FROM sx.tex
\newcommand{\sxleqsim}{\,\raisebox{-0.6ex}{$\buildrel < \over \sim$}\,}
\newcommand{\sxgeqsim}{\,\raisebox{-0.6ex}{$\buildrel > \over \sim$}\,}

%%%%FROM lhproc.tex
\newcommand{\KPSpT}{p_{\mathrm{T}}}
\newcommand{\KPSthw}{{\theta_\mathrm{w}}}
\newcommand{\KPSsw}{{s_{\mathrm{\scriptscriptstyle W}}}}
\newcommand{\KPScw}{{c_{\mathrm{\scriptscriptstyle W}}}}
\newcommand{\KPScew}{C^{\mathrm{ew}}}
\newcommand{\KPSDeltamsbar}{\bar \Delta_{\mathrm{UV}}}
\newcommand{\KPSlogar}[2]{\mathrm{L}^{#1}_{#2}}
\newcommand{\KPSnewA}[2]{A^{#1,(#2)}}

%%%%FROM colrec.tex
\newcommand{\sktsc}[1]{\textsc{#1}}
\newcommand{\skttt}[1]{\texttt{#1}}
\newcommand{\skmrm}[1]{\mathrm{#1}}
\newcommand{\skGeV}{\mathrm{GeV}}
\newcommand{\skMeV}{\mathrm{MeV}}
\newcommand{\skfm}{\mathrm{fm}}
\newcommand{\ske}{\mathrm{e}}
\newcommand{\skg}{\mathrm{g}}
\newcommand{\skq}{\mathrm{q}}
\newcommand{\skqbar}{\bar{\mathrm{q}}}
\newcommand{\skb}{\mathrm{b}}
\newcommand{\skp}{\mathrm{p}}
\newcommand{\skbbar}{\bar{\mathrm{b}}}
\newcommand{\skpbar}{\bar{\mathrm{p}}}
\newcommand{\skZ}{\mathrm{Z}}
\newcommand{\skW}{\mathrm{W}}
\newcommand{\skt}{\mathrm{t}}
\newcommand{\sktbar}{\bar{\mathrm{t}}}

\title{\centering{THE QCD, EW, AND HIGGS WORKING GROUP:\\
          \textbf{Summary Report}}}

  \author{\underline{Convenors}:
%  all conveners, in alphabetical order
C.~Buttar$^1$,
S.~Dittmaier$^2$,
V.~Drollinger$^3$,
S.~Frixione$^4$,
A.~Nikitenko$^5$,
S.~Willenbrock$^6$
\\
\underline{Contributing authors}:
%  all the others, in alphabetical order
S.~Abdullin$^7$,
E.~Accomando$^{8,9}$,
D.~Acosta$^{10}$,
A.~Arbuzov $^{70}$
% ...
\\\mbox{} }
\institute{\centering{\small
  $^1$ Department of Physics and Astronomy
  University of Glasgow, G12 8QQ, Glasgow, UK\\
  $^2$ Max-Planck-Institut f\"ur Physik (Werner-Heisenberg-Institut),
  D-80805 M\"unchen, Germany\\
  $^3$ Dipartimento di Fisica "Galileo Galilei", Universit\`a di Padova,
  Padova, Italy\\
  $^4$ INFN, Sezione di Genova, Genoa, Italy\\
  $^5$ Imperial College, London, UK\\
  $^6$ Department of Physics, University of Illinois at Urbana-Champaign,
  IL, USA\\
  $^7$ Fermi National Accelerator Laboratory, Batavia, IL-60510, USA\\
  $^8$ INFN, Sezione di Torino, Turin, Italy\\
  $^9$ Dipartimento di Fisica Teorica, Universit\`a di Torino, Turin, Italy\\
  $^{10}$ University of Florida, Gainesville, FL, USA\\
  $^{11}$ School of Physics, University of Edinburgh,
  Edinburgh EH9 3JZ, Scotland, UK\\
  $^{12}$ Joint Institute for Nuclear Research, Joliot-Curie 6
  Dubna, Moscow region, Russia, 141980\\
  $^{13}$ Michigan State University, East Lansing, MI, USA\\
  $^{14}$ Dipartimento di Fisica Sperimentale, Universit\`a di Torino,
  Turin, Italy\\
  $^{15}$ Scobeltsyn Institute of Nuclear Physics of Lomonosov Moscow State
  University, Moscow, Russia, 119992\\
  $^{16}$ CERN, Geneva, Switzerland\\
  $^{17}$ INFN, Sezione di Firenze, Florence, Italy\\
  $^{18}$ Dipartimento di Fisica, Universit\`a di Firenze,
  Sesto Fiorentino, Florence, Italy\\
  $^{19}$ Penn State University, University Park, PA, USA\\
  $^{20}$ Particle Physics, Oxford University,
  Oxford, OX1 3RH, UK\\
  $^{21}$ Dipartimento di Fisica,
  Universit\`a degli Studi di Roma `La Sapienza', Rome, Italy\\
  $^{22}$ Institute for Particle Physics, ETH Z\"urich, Switzerland\\
  $^{23}$ Paul Scherrer Institut,
  CH-5232 Villigen PSI, Switzerland\\
  $^{24}$ H. H. Wills Physics Laboratory, University of Bristol,
  United Kingdom\\
  $^{25}$ Institut de Physique Nucl\'eaire de Lyon/Universit\'e Claude
  Bernard Lyon I, Villeurbanne, France\\
  $^{26}$ Institut f\"ur Theoretische Physik, Universit\"at Z\"urich,
  CH-8057 Z\"urich, Switzerland\\
  $^{27}$ Centre for High Energy Physics,
  Indian Institute of Science, Bangalore, 560012, India\\
  $^{28}$ Departamento de Fisica Teorica y del Cosmos,
  Universidad de Granada, Spain\\
  $^{29}$ Institute For High Energy Physics, Protvino, Russia\\
  $^{30}$ Institut f\"ur Theoretische Teilchenphysik,
  Universit\"at Karlsruhe, D-76128 Karlsruhe, Germany\\
  $^{31}$ Institute of Experimental Physics, Warsaw University, Poland\\
  $^{32}$ Brookhaven National Laboratory, Upton, NY, USA\\
  $^{33}$ Institut f\"ur Theoretische Physik E, EWTH Aachen, Germany\\
  $^{34}$ Helsinki Institute of Physics, 00014 University of
  Helsinki, Finland\\
  $^{35}$ Physics Department, Mount Allison University, Sackville NB,
  E4L 1E6 Canada\\
  $^{36}$ Institut de Physique Th\'eorique, Universit\'e Catholique de
  Louvain, Belgium\\
  $^{37}$ University of Wisconsin - Madison,
  Department of Physics,
  Madison, WI, USA\\
  $^{38}$ Department of Physics, Ferrara University, Ferrara, Italy\\
  $^{39}$ School of Physics and Astronomy, University of Southampton,
  Highfield, Southampton SO17 1BJ, UK\\
  $^{40}$ Department of Physics, Okayama University, Okayama, 700-8530, Japan\\
  $^{41}$ INFN, Sezione di Milano, Milan, Italy\\
  $^{42}$ Southern Methodist University, Dallas, TX, USA\\
  $^{43}$ INFN, Laboratori Nazionali di Frascati, Frascati, Italy\\
  $^{44}$ INFN, Sezione di Pavia, Pavia, Italy\\
  $^{45}$ Dipartimento di Fisica Nucleare e Teorica, Universit\'a di Pavia,
  Pavia, Italy\\
  $^{46}$ Bergische Universit\"at Wuppertal,
  Fachgruppe Physik, 42097 Wuppertal, Germany\\
  $^{47}$ Institute for Theoretical Physics, University of Dresden, Germany\\
  $^{48}$ Institute of Nuclear Research of
  the Hungarian Academy of Sciences, H-4001 Debrecen, Hungary\\
  $^{49}$ C.~N.~Yang Institute for Theoretical Physics, SUNY
  Stony Brook, USA\\
  $^{50}$ Physics Department, University of Perugia, Perugia, Italy\\
  $^{51}$ Justus-Liebig Universit\"at, Giessen, Germany\\
  $^{52}$ Department of Physics, Tohoku University, Sendai, 980-8578, Japan\\
  $^{53}$ INFN, Sezione di Ferrara, Ferrara, Italy\\
  $^{54}$ INFN, Sezione di Perugia, Perugia, Italy\\
  $^{55}$ University of Debrecen, H-4001 Debrecen, Hungary\\
  $^{56}$ Inter-University Institute for High Energies (IIHE),
  Brussels, Belgium\\
  $^{57}$ Institut f\"ur Experimentelle Kernphysik, Universit\"at
  Karlsruhe, D-76128 Karlsruhe, Germany\\
  $^{58}$ LAPTH CNRS, B.P.110, Annecy-le-Vieux F-74941, France\\
  $^{59}$ KEK, Oho 1-1, Tsukuba, Ibaraki 305-0801, Japan\\
  $^{60}$ Kogakuin University, Nishi-Shinjuku 1-24, Shinjuku, Tokyo
  163-8677, Japan\\
  $^{61}$ Physikalisches Institut, Universit\"at Freiburg,
  D-79104 Freiburg, Germany\\
  $^{62}$ INFN, Sezione di Pisa, Pisa, Italy\\
  $^{63}$ Universit\'e Catholique de Louvain, Belgium\\
  $^{64}$ Institut f\"ur Theoretische Physik und Astrophysik,
  Universit\"at W\"urzburg, D-97074 W\"urzburg, Germany\\
  $^{65}$ Department of Physics, University of Durham,
  Durham DH1 3LE, UK\\
$^{66}$ Dept. of Physics, State University of New York, Buffalo,
NY 14260, USA\\
$^{67}$ II Institute for  Theoretical Physics,
   University of Hamburg, Germany\\
$^{68}$  Institut f\"ur Physik, Albert-Ludwigs Universit\"at Freiburg, \\
         Hermann-Herder-Str. 3, 79104, Freiburg, Germany \\
$^{69}$ INR, Moscow 117312, Russia\\
$^{70}$ Bogoliubov Laboratory of Theoretical Physics, JINR,\
Dubna, \ 141980 \ \  Russia \\
$^{71}$ Dzhelepov Laboratory of Nuclear Physics, JINR,\
Dubna, \ 141980 \ \  Russia \\
$^{72}$ Dipartimento di Fisica, Universita degli Studi di
Milano, Italy\\
}}
 \maketitle

 \begin{center}
   \textit{Report of the Working Group on Quantum Chromodynamics,
     Electroweak, and Higgs Physics for the Workshop ``Physics at TeV
     Colliders'', Les Houches, France, 2--20 May, 2005.  }
\end{center}
\newpage

\setcounter{tocdepth}{1}
\tableofcontents
\setcounter{footnote}{0}

%%%%%%%%%%%%%%%%%%%%%%%%%%%%%%%%%%%%%%%%%%%%%%%%%%%%%%%%%%%%%%%%%%%%%%%%%%%%%
\section[Foreword]{FOREWORD}
The primary goal of the LHC will be that of finding evidence or hints
of physics whose signals have not been detected yet by collider
experiments. This includes any physics beyond that so successfully
described by the Standard Model, but also that relevant to the only
sector of the Standard Model which has not been probed directly so
far, namely the Higgs sector. The signatures of new physics are vastly
diverse, but in the majority of the cases they imply chain decays of
massive particles, which in turn will appear in detectors as many-jet
events. Although a good understanding of the continuum many-jet QCD
production will be needed in order to disentangle such signals from
the background, we may consider this situation as a favourable one,
since the discovery of new physics will be relatively quick and
independent of theoretical assumptions (a much more difficult problem
will then be that of understanding which kind of underlying theory
is responsible for the signals detected). An even easier case will be
that of a very heavy neutral vector boson, whose dilepton decay
should be basically background-free. On the other hand, the detection
of a Standard Model Higgs will be a pretty complicated affair, since
the signal is overwhelmed by huge QCD backgrounds, whose good control
is therefore mandatory in order to claim a discovery.

In all cases, the reliability of the outcomes of LHC experiments
will depend on their capability of reproducing, and improving, what
we know about the Standard Model and QCD, through the studies of a
few benchmark processes, the ``standard candles'', such as $W$,
$Z$, and $t\bar{t}$ production.

The aim of the SM and Higgs Working Group in Les Houches has
therefore been twofold. On one hand, we performed a variety of
experimental and theoretical studies on standard candles, treating
them either as proper signals of known physics, or as backgrounds
to unknown physics; we also addressed issues relevant to those
non-perturbative or semi-perturbative ingredients, such as Parton
Density Functions and Underlying Events, whose understanding will
be crucial for a proper simulation of the actual events taking
place in the detectors. On the other hand, several channels for
the production of the Higgs, or involving the Higgs, have been
considered in some details.

This report is organized into four main parts. The first one deals with
Standard Model physics, except the Higgs. A variety of arguments are
treated here, from full simulation of processes constituting a background
to Higgs production, to studies of uncertainties due to PDFs and to
extrapolations of models for underlying events, from small-$x$ issues
to electroweak corrections which may play a role in vector boson
physics. The second part of the report treats Higgs physics from
the point of view of the signal. In the third part, reviews are
presented on the current status of multi-leg, next-to-leading
order and of next-to-next-to-leading order QCD computations.
Finally, the fourth part deals with the use of Monte Carlos
for simulation of LHC physics.

%%%%%%%%%%%%%%%%%%%%%%%%%%%%%PART%%%%%%%%%%%%%%%%%%%%%%%%%%%%%
\part[STANDARD MODEL BENCHMARKS AND BACKGROUNDS]
{STANDARD MODEL BENCHMARKS AND BACKGROUNDS}
%%%%%%%%%%%%%%%%%%%%%%%%%%%%%PART%%%%%%%%%%%%%%%%%%%%%%%%%%%%%

%%%%%%%%%%%%%%%%%%%%%%%%%%%%%%%%%%%%%%%%%%%%%%%%%%%%%%%%%%%%%%%%%%%%%%%%%%%%%
\section[Model predictions for $\sigma^{\rm tot}$ at the LHC]
{MODEL PREDICTIONS FOR $\sigma^{\rm tot}$ AT THE LHC~\protect
\footnote{Contributed by: R.M.~Godbole, A.~Grau, R.~Hegde,
G.~Pancheri, Y.~Srivastava}}
\subsection{Introduction}
Energy dependence of total hadronic cross-sections has been the focus
of intense theoretical interest as a sensitive probe of strong
interactions long before the establishment of QCD as ``the'' theory
of hadrons. Even now, notwithstanding creditable successes of
perturbative and lattice QCD, alas a first principle description
of total/elastic and inelastic hadronic cross-section is unavailable.
More pragmatically, for a correct projection of the expected
underlying activity at LHC, a reliable prediction of total
non-diffractive cross-section is essential to ensure the extraction
of new physics from the LHC data. Surely we will have to depend -at
the initial stages of LHC- upon predictions based on our current
understanding of these matters. Only much later it may become
feasible to use the LHC data itself towards this goal. Hence,
a critical evaluation of the range of theoretical predictions
is absolutely essential.

\noindent
The hadronic cross-section data exhibit, and require explanation of, 
three basic features: 
\begin{description}
\item (i) the normalization of the cross section, 
\item (ii)
an initial decrease and 
\item (iii) a  subsequent rise with energy
\end{description}
Various theoretical models exist which are motivated by our theoretical
understanding of the strong interactions. The parameters in these models, in
most cases, are fitted to explain the observed low-energy data and the model
predictions are then extrapolated to give the \GGHPSsigtot\ at the LHC
energies.  There are different classes of models. The highly successful
Donnachie-Landschoff parameterisation~\cite{Donnachie:1992ny} of the form
\begin{equation}
\sigma_{tot}(s)=Xs^\epsilon+Ys^{-\eta},
\label{DL}
\end{equation}
has been used for a very long time. Here the two terms are understood as 
arising 
from the Regge and the Pomeron trajectories, the $\epsilon$ being approximately
close to zero and $\eta$ being close to $0.5$. These values seem to be 
consistent with a large, but not all, body of the hadronic cross-sections.  
In this note we will  first present phenomenological arguments for the 
approximate values of these parameters which seem to be required to describe 
the data satisfactorily.  As a matter of fact, there also exist in the 
literature discussions of the `hard' pomeron~\cite{DLnew} motivated by the 
discrepancies in the rate of energy rise observed by E710~\cite{Amos:1991bp}, 
E811~\cite{Avila:1998ej} and the CDF~\cite{Abe:1993xy}.  In addition, a 
variety of models exist wherein the observed energy dependence of the
cross-section, along with few very general requirements of factorisation,
unitarity and/or ideas of Finite Energy Sum Rules (FESR), is used to determine 
the values of model parameters~\cite{Block:2005ka, Block:2005pt, Igi:2005jm,
Avila:2002tk,Cudell:2002xe,Cudell:2002sy}. The so-obtained parameterisations
are then extended to make predictions at the LHC energies.  There also exist 
QCD-motivated models based on the  mini-jet 
formalism~\cite{Cline:1973kv,Pancheri:1985sr,Gaisser:1984pg}, 
wherein the energy rise of the total cross-sections is driven by the 
increasing number of the low-$x$ gluon-gluon collisions. These models need to 
be embdedded in an eikonal formalism~\cite{Durand:1987yv} 
to soften the violent energy  rise of the mini-jet cross-sections. Even 
after eikonalisation the  predicted energy rise is harder than the gentle 
one observed experimentally~\cite{Gaisser:1984pg,Pancheri:1986qg}.
A QCD-based model where the rise is further tamed by the  phenomenon of 
increasing emission of soft gluons by the valence quarks in the colliding 
hadrons, with increasing energy~\cite{Grau:1999em,Godbole:2004kx}, 
offers a consistent description of \GGHPSsigtot\ .  Thus we have a variety of 
model predictions for \GGHPSsigtot\ at the LHC.  In this note we compare these 
predictions with each other in order to obtain an estimate of the 
``theoretical'' uncertainty on them.

\subsection{Phenomenological models}
The two terms of eq.~(\ref{DL})~\cite{Donnachie:1992ny} reflect the
well known duality between resonance and Regge pole exchange on the one hand
and background and Pomeron exchange on the other, established in the late
60's through FESR \cite{Igi:1967}. This correspondence  meant that, while at 
low energy the cross-section could be written as due to a  background term
 and a sum of resonances,  at higher energy it could be written as 
a sum of Regge trajectory exchanges and  a Pomeron exchange.  

\GGHPScomment
{It is well to ask (i) where the  ``two component'' structure 
of eq.~(\ref{DL}) comes from and (ii) why the difference in the two powers 
(in $s$) is approximately a half.}
Our present knowledge of QCD and its 
employment for a description of hadronic phenomena can be used to provide 
some insight into the ``two component'' structure of the eq.~(\ref{DL}).
This begins with considerations about the bound state nature of hadrons 
which necessarily transcends perturbative QCD. For hadrons made of light 
quarks ($q$) and gluons ($g$), the two terms 
arise from $q\bar{q}$ and $gg$ excitations. For these, the energy is given 
by a sum of three terms: (i) the rotational energy, (ii) the Coulomb 
energy and (iii) the ``confining'' energy. If we accept the Wilson area 
conjecture in QCD, (iii) reduces to the linear potential\cite{Srivastava:2000fb,
Landshoff:2001pp}. Then the hadronic rest mass for a state of angular momentum
J can be obtained by minimising  the expression for the energy of two 
massless particles ($q \bar q$ or $gg$) separated by a distance $r$.
\GGHPScomment{
Explicitly, in the CM frame of two massless particles,
either a $q\bar{q}$ or a $gg$ pair separated by a relative 
distance $r$ with relative angular momentum $J$, the energy is given by
\begin{equation} \label{string1}
E_i(J, r) = {{2J}\over{r}} - {{C_i \bar{\alpha}}\over{r}} + C_i \tau r,
\end{equation}
where $i\ =\ 1$ refers to $q\bar{q}$,  $i\ =\ 2$ refers to $gg$, $\tau$
is the ``string tension'' and the Casimir's are $C_1\ =\ C_F\ =\ 4/3$, $C_2\
=\ C_G\ =\ 3$. $\bar{\alpha}$ is the QCD coupling constant evaluated at 
some average value of $r$ and whose precise  value
will disappear in the ratio to be considered. The hadronic
rest mass for a state of angular momentum $J$ is then computed through
minimising the above energy
\begin{equation} \label{string2}
M_i(J) = Min_r[ {{2J}\over{r}} - {{C_i \bar{\alpha}}\over{r}} + C_i \tau r ],
\end{equation}
This then is given by,
\begin{equation} \label{string3}
M_i(J) = 2 \sqrt{(C_i \tau)[2J - C_i \bar{\alpha}]},
\end{equation}
}
This can then be used to obtain the two sets of linear Regge 
trajectories $\alpha_i(s)$ 
\begin{equation} \label{string4}
\alpha_i(s) = {{C_i \bar{\alpha}}\over{2}} + ({{1}\over{8 C_i \tau}}) s
= \alpha_i(0) + \alpha_i' s.
\end{equation} 
where $i\ =\ 1$ refers to $q\bar{q}$,  $i\ =\ 2$ refers to $gg$, $\tau$
is the ``string tension'' and the Casimir's are $C_1\ =\ C_F\ =\ 4/3$, $C_2\
=\ C_G\ =\ 3$. $\bar{\alpha}$ is the QCD coupling constant evaluated at
some average value of $r$. 
Note that $\alpha_i$ are  {\it not} the coupling constants.
\GGHPScomment
{Thus, the ratio of the intercepts is given by
\begin{equation}\label{string5}
{{\alpha_{gg}(0)}\over{\alpha_{q\bar{q}}(0)}} = C_G/C_F = {{9}\over{4}}. 
\end{equation}
}
Employing our present understanding that resonances are $q{\bar q}$ bound 
states while the background, dual to the Pomeron,
is provided by gluon-gluon exchanges\cite{Landshoff:2001pp}, the above 
equation can be rewritten  as 
\begin{equation} \label{string5qhd}
{{\alpha_{P}(0)}\over{\alpha_R(0)}} = C_G/C_F = {{9}\over{4}}. 
\end{equation}
If we restrict our attention to the leading Regge trajectory, namely
the degenerate $\rho-\omega-\phi$ trajectory, then
 $\alpha_R(0)=\eta\ \approx\ 0.48-0.5$, and we obtain
for $\epsilon\ \approx\ 0.08-0.12$, a rather  satisfactory value. 
The same argument for the slopes gives 
\begin{equation}\label{string6}
{{\alpha_{gg}'}\over{\alpha_{q\bar{q}}'}} = C_F/C_G = {{4}\over{9}}.
\end{equation}
so that if we take for the Regge slope $\alpha_R'\ \approx\ 0.88-0.90$, we get
for $\alpha_P'\ \approx\ 0.39-0.40$,  in fair
agreement with lattice estimates\cite{lattice:2005}.

We now have good reasons for a break up of the amplitude into 
two components. To proceed further, it is necessary to realize that 
precisely because massless hadrons do not exist, eq.~(\ref{DL}) violates 
the Froissart bound and thus must be unitarized. To begin this task, 
let us first rewrite eq.~(\ref{DL}) by putting in the  ``correct'' dimensions
\begin{equation} \label{DL1} 
\bar{\sigma}_{tot}(s)= \sigma_1 (s/\bar{s})^\epsilon+ \sigma_2
(\bar{s}/s)^{1/2},
\end{equation}
where we have imposed the nominal value $\eta\ =\ 1/2$.  It is possible further,
to  obtain~\cite{Godbole:2004kx} rough estimates for the size of the 
parameters in eq.~(\ref{DL1}).  A minimum occurs
in $\bar{\sigma}_{tot}(s)$ at $s\ =\ \bar{s}$, for $\sigma_2\ =\ 2\epsilon
\sigma_1$.  If we make this choice, then eq.~(\ref{DL1}) becomes
\GGHPScomment
{has one less parameter 
and it reduces to
\begin{equation} \label{DL2} 
\bar{\sigma}_{tot}(s)= \sigma_1 [(s/\bar{s})^\epsilon+ 2 \epsilon
(\bar{s}/s)^{1/2}].
\end{equation}
We can isolate the rising part of the cross-section by rewriting the above
as}
\begin{equation}\label{DL3}
\bar{\sigma}_{tot}(s)= \sigma_1 [ 1 + 2\epsilon(\bar{s}/s)^{1/2}] 
+ \sigma_1 [(s/\bar{s})^\epsilon - 1].
\end{equation}
eq.~(\ref{DL3}) separates cleanly the cross-section into two parts: the first
part is a ``soft'' piece which shows a saturation
to a constant value (but which contains no rise) and the second a ``hard'' 
piece which has all the rise. 
\GGHPScomment
{Morover, $\bar{s}$ naturally provides the 
scale beyond which the cross-sections would begin to rise. Thus, our
``Born''  term assumes the generic form
\begin{equation} \label{DL4}
\sigma_{tot}^B(s)= \sigma_{soft}(s) + \vartheta (s - \bar{s})
\sigma_{hard}(s).
\end{equation}
with $\sigma_{soft}$ containing a constant ( the ``old'' Pomeron
with $\alpha_P(0)\ =\ 1$) plus a (Regge) term decreasing
as $1/\sqrt{s}$ and with an estimate for their relative magnitudes
($\sigma_2/\sigma_1\ \sim\ 2\epsilon$).}
 In the eikonalised mini-jet model used 
by us~\cite{Godbole:2004kx} the rising 
part of the cross-section $\sigma_{hard}$ is provided by jets which are 
calculable in perturbative QCD, obviating (at least in principle) the 
need of an arbitrary parameter $\epsilon$. An estimate of $\sigma_1$ can also
be obtained~\cite{Godbole:2004kx} and is $\sim 40$ mb.

As said earlier, the DL parameterisation~\cite{Donnachie:1992ny} is a 
fit to the existing data of the form given by eq.~(\ref{DL}), with
$\epsilon =0.0808, \eta = 0.4525$. This fit has been extended to include a 
'hard' pomeron~\cite{DLnew} due to the discrepancey between different data 
sets. The BH model~\cite{Block:2005ka} gives a fit to the data using duality
constraints. The BH fit for $\sigma^{\pm} = \sigma^{\bar p p}/ \sigma^{p p}$
as a function of beam energy $\nu$, is given as,
$$
\sigma^{\pm} = c_0 +c_1 \ln(\nu/m) + c_2 \ln^2 (\nu /m) 
+ \beta_{P'} (\nu/m)^{\mu-1} \pm \delta (\nu/m)^{\alpha -1},
$$
where  $\mu = 0.5, \alpha = 0.415$ and all the other parameters in mb are
$c_0 = 37.32, c_1 = -1.440 \pm 0.07, c_2 = 0.2817 \pm 0.0064, \beta_{P'} = 
37.10, \delta = -28.56$.
The fit obtained by Igi et al\cite{Igi:2005jm} using the finite energy sum 
rules (FESR) gives LHC predictions very similar to those given by the BH fit.
Avila et al. give a fit~\cite{Avila:2002tk} using analyticity arguments
whereas Cudell et al~\cite{Cudell:2002xe} give predictions at the LHC energies 
by extrapolating fits obtained to the current data using again constraints
from unitarity, analyticity of the S-matrix, factorisation,
coupled with a requirement that the cross-section asymptotically goes to
a constant plus a $\ln s$ or $\ln^2 s$ term, in the framework of the 
COMPETE program.

In the mini-jet models the energy rise of \GGHPSsigtot\ is driven by the 
increase with energy of the $\sigma_{jet}^{ab}$ given by
\begin{equation}
\label{sigjet}
\sigma^{ab}_{\rm jet} (s) = \int_{p_{tmin}}^{\GGHPSrs/2} d p_t
\int_{4 p_t^2/s}^1 d x_1 \int_{4 p_t^2/(x_1 s)}^1 d x_2 \sum_{i,j,k,l}
f_{i|a}(x_1) f_{j|b}(x_2) \frac { d \hat{\sigma}_{ij \rightarrow kl}(\hat{s})}
{d p_t},
\end{equation}
where subscripts $a$ and $b$ denote particles ($\gamma, \ p, \dots$),
$i, \ j, \ k, \ l$ are partons and $x_1,x_2$ the fractions of the
parent particle momentum carried by the parton. $\hat{s} = x_1 x_2 s$  and
$\hat{ \sigma}$ are hard partonic scattering cross--sections. As said before,
the rise with energy of this cross-section is too steep, and hence it
has to be imbedded in an eikonal formulation given by,
\begin{equation}
\sigma^{ab}_{tot}=2\int d^2{\vec b}[1-e^{-\Im m\chi(b,s)}]
\label{stot}
\end{equation}
where $ 2\Im m~\chi(b,s)=n(b,s)$ is the average number of multiple collisions 
which are Poisson distributed. As outlined in eq.~(\ref{DL3}) this quantity
too has contributions coming from soft and hard physics and can be 
written as 
\begin{equation}
n(b,s)=n_{soft}+n_{hard}\simeq A_{soft}(b)\sigma_{soft}(s)+
A_{jet}(b)\sigma_{jet}(s).
\label{nsplit}
\end{equation}
In the second step the number $n(b,s)$  has been  assumed  to be factorizable
into an overlap function $A(b)$ and the cross-section $\sigma$. The assumption
of factorisation as well as the split up between the two contributions, hard 
and soft, are only approximate. The extent to which this softens the energy
rise, depends on the $b$ dependence of $n(b,s)$, i.e., that of $A(b)$ in the
factorised case. The normal assumption of using the same form of $A(b)$ for 
both the hard and the soft part, given  by the Fourier transform of the 
electromagnetic Form Factor (FF), still gives too steep a rise even 
in this Eikonalised Mini-jet Model (EMM)~\cite{Durand:1987yv}. In our model
this rise is tamed by including the effect on the transverse  momentum 
distribution of the partons in the proton,  of the soft gluon emission from 
the valence quarks in the proton~\cite{Grau:1999em}; the effect increases with 
increasing energy.  The non-perturbative soft part of the eikonal includes 
only limited low-energy gluon emission and leads to the initial decrease in 
the proton-proton cross-section. On the other hand, the rapid rise in the 
hard, perturbative jet part of the eikonal is tamed into the experimentally 
observed mild increase by soft gluon radiation whose maximum energy ($q_{max}$)
rises slowly with energy. Thus the overlap functions $A(b)$  are 
no longer a function of $b$ alone. We denote the corresponding overlap 
function by $A_{BN} (b,q_{max})$
\GGHPScomment{
\GGHPSsigtot\ is given by 
\begin{equation}
2\Im m~\chi(b,s)=n(b,s;q_{max},p_{tmin})=n_{soft}+
n_{jet}=A_{soft}(b)\sigma_{soft}+
A_{BN}(b,q_{max})\sigma_{jet}.
\end{equation}}
~\cite{Grau:1999em} determined by  $q_{max}$, which  depends on
the energy and the kinematics of the subprocess. What we use is an average 
value over all the momentum fractions of the parent partons. We need to further
make a model for the 'soft' part which is determined by the nonperturbative 
dynamics. It is this part of the eikonal that contributes to the \GGHPSsigtot\
at high energies, the turn around from the decreasing Regge behaviour to the 
softly rising behaviour around $\sqrt{s} \simeq 15$ GeV, where the  hard part 
contribution is miniscule.  
\GGHPScomment{We have taken $n_{soft}$  to be  factorised into
a constant soft cross-section $\sigma_{soft}$ and taken 
$A_{soft} = A{BN}(b,q_{max})$.}
We have further postulated  that the $q_{max}$ 
is the same for the hard and soft processes at low energy, parting company 
around $10$ GeV where the hard processes start becoming important. A good fit 
to the data requires that $q_{max}$ at low energies to be a very slowly 
increasing function of energy, with a value around $0.20$ MeV at $\sqrt{s} = 5$
GeV rising to about $0.24$ MeV, $\sqrt{s} \geq 10$ GeV, the upper value
of this soft scale being completely consistent with our picture of the proton.
Further, we need to fix one more parameter for nonperturbative 
region, the $\sigma_{soft}$. For the $pp$ case it is a constant $\sigma_0$
which will fix the normalization of \GGHPSsigtot\ , whereas for the $p\bar p$ 
case the duality arguments suggest that there be an additional 
$\sqrt{s}$-dependent piece $\simeq 1/\sqrt{s}$. Thus neglecting the real 
part of the eikonal, $n(b,s)$ in our model is given by
\GGHPScomment{
\begin{equation} 
\sigma_{tot}=2\int d^2{\vec b}[1-e^{-\Im m~\chi (b,s)}]
\end{equation}
with}
\begin{equation}
n(b,s)  = A_{BN}(b,q_{max}^{soft}) \sigma_{soft}^{pp,{\bar p}}+
A_{BN}(b,q_{max}^{jet})
 \sigma_{jet}(s;p_{tmin}),
\end{equation}
where 
\begin{equation}
\sigma_{soft}^{pp}=\sigma_0, \ \ \ \ \ \ \ \ \ \ \
\ 
\sigma_{soft}^{p{\bar p}}=\sigma_0
(1+{{2}\over{\sqrt{s}}}) 
\end{equation}

Thus the parameters of the model are $p_{tmin}$ and $\sigma_0$. In addition,
the evaluation of $A_{BN}$ involves the $\alpha_s$ in the infrared region, 
for which we use a phenomenological form inspired by the Richardson Potential
\cite{Grau:1999em}. This involves a parameter $p$ which for the
Richardson Potential takes value $1$.
Values of $p_{tmin}, \sigma_0$ and $p$  which give a good fit to the data 
with the GRV parameterisation of the proton densities~\cite{Gluck:1991ng} 
are  $1.15$ GeV,  $48$ mb  and $3/4$   respectively, as presented in 
Ref.~\cite{Godbole:2004kx}. These values are  consistent with the
expectations of the general argument~\cite{Godbole:2004kx}.  We expect these
best fit values to change somewhat with the choice of parton density functions 
(PDF).  Since we are ultimately interested in the predictions of the model
at TeV energies, we need PDF parameterisations which cover a $Q^2$ range 
between $2$ and $10^4$~GeV$^2$, as well as are valid up to rather small 
values of $x~(\sim 10^{-5})$. Further, since our calculation here is only
LO, for consistency we have to use LO densities. We have repeated the exercise
then for a range of PDF's~\cite{Gluck:1994uf,Gluck:1998xa,Martin:2002dr} 
meeting these requirements. 
%For each PDF, it is onset of the rise that 
%fixes the $p_{tmin}$, $\sigma_0$ controlling the normalisation and 
%$p$ determining the slope of the rising part of the cross-section.
We find that it is
possible to get a satisfactory description of all the current data, 
for all the choices of PDF's considered. The corresponding range of values of 
$p_{tmin}, \sigma_0$ and $p$ are given in Table 1. 
The predictions of this modified EMM model span a range which are presented 
and discussed in the next section.
\begin{table}
\begin{footnotesize}
\begin{center}
\caption{
Values of $p_{tmin}$ and $\sigma_0$ corresponding to the different 
parton densities in the proton, for  which the EMM (as described in 
Ref.~\protect\cite{Godbole:2004kx}) gives a satisfactory description of 
\GGHPSsigtot\ .
}
  \vspace*{1mm}
\begin{tabular}{|c|c|c|c|}
\hline
&&&\\
PDF&$p_{tmin}$ (GeV) & $\sigma_0$ (mb)&p\\
\hline
&&&\\
GRV \cite{Gluck:1991ng}&1.15&48&0.75\\
\hline
&&&\\
GRV94lo\cite{Gluck:1994uf} &1.10&46&0.72\\
&1.10&51&0.78\\
\hline
&&&\\
GRV98lo\cite{Gluck:1998xa} &1.10&45&0.70\\
&1.10&50&0.77\\
\hline
&&&\\
MRST\cite{Martin:2002dr} &1.25&47.5&0.74\\
&1.25&44&0.66\\
\hline
\end{tabular}
\end{center}
\end{footnotesize}
\end{table}
\subsection{Model predictions for \GGHPSsigtot\ at the LHC}
\begin{figure}
  \begin{center}
    \includegraphics[scale=0.65]{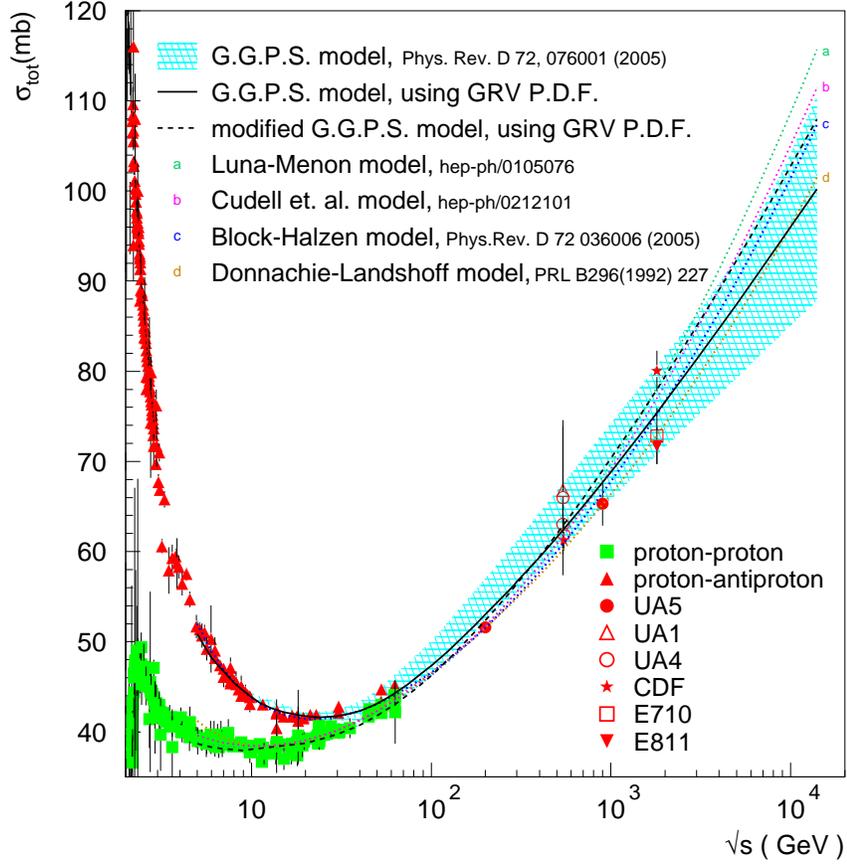}
\caption{\label{fig:csection}
Predictions for \protect\GGHPSsigtot\ in various models.
The shaded area gives
the range of results in the Eikonalised mini-jet model with soft gluon
resummation~\protect\cite{Godbole:2004kx} (the G.G.P.S. model) the solid
line giving the prediction obtained using the GRV parton
densities~\protect\cite{Gluck:1991ng} in the model. The long-dashed dotted 
curve ($d$), indicates the predictions for the DL 
fit~\protect\cite{Donnachie:1992ny}.  The dotted (BH) curve ($c$) and the 
uppermost dashed curve ($a$), are the results of two
analytical models incoporating constraints from unitarity and analyticity,
from~\protect\cite{Block:2005ka} and ~\protect\cite{Avila:2002tk},
respectively. The prediction obtained by Igi et al, using FESR follows
very closely that given by the BH curve. The short dash dotted curve
($b$), is the result of a fit by the COMPETE
collaboration~\protect\cite{Cudell:2002xe}.}
\end{center}
\end{figure}

Figure~\ref{fig:csection}  
summarises the predictions of the different models described in
the previous section.  The shaded area gives
the range of predictions in the Eikonalised mini-jet model with soft gluon 
resummation~\protect\cite{Godbole:2004kx} (the G.G.P.S. model), the different 
PDF's used giving the range as described in the earlier section.
The solid line gives prediction obtained using the GRV parton 
densities~\protect\cite{Gluck:1991ng} in the model. The long-dashed 
dotted curve ($d$), indicates the predictions of the DL 
fit~\protect\cite{Donnachie:1992ny}.  The dotted (BH) curve ($c$) and 
the uppermost dashed curve ($a$), are the results of two analytical models 
incorporating constraints from unitarity and analyticity, 
from~\cite{Block:2005ka} and~\cite{Avila:2002tk}, respectively.
The prediction obtained by Igi et al using FESR follows 
very closely that given by the BH curve. Furthermore, the short-dash 
dotted curve ($b$) is the result of a fit by the COMPETE 
collaboration~\protect\cite{Cudell:2002xe}.
The parameterisation for the DL curve and BH curve is already given in the 
last section.
It is gratifying to see that the range of results of our QCD motivated 
minijet models for the LHC span the other predictions based on models using 
unitarity, factorisation, analyticity fitting the current data. Thus the 
predictions seem consistent with each other. 

In case of the EMM model results we have parameterised them 
with a $\ln^2 s$ fit. We found that in most cases this gave a better 
representation of our results than a fit of the Regge-Pomeron type 
of the form of eq.~(\ref{DL}). 
\begin{table}
\label{param}
\begin{footnotesize}
\begin{center}
\caption{
Values of  $a_0,a_1,a_2,a_3$ and $b$ 
parton densities in the proton, for  which the EMM (as described in 
Ref.~\cite{Godbole:2004kx}) gives a satisfactory description of \GGHPSsigtot\ .
}
  \vspace*{1mm}
\begin{tabular}{|c|c|c|c|c|c|}
\hline
&&&&&\\
&$a_0$ (mb)& $a_1$ (mb)&b & $a_2$ (mb)& $a_3$ (mb)\\
\hline
&&&&&\\
Top edge&23.61&54.62&-0.52&1.15&.17\\
\hline
&&&&&\\
Center&-139.80&193.89&-0.11&13.98&-.14\\
\hline
&&&&&\\
Lower edge&-68.73&125.80&-0.16&11.05&-.16\\
\hline
\end{tabular}
\end{center}
\end{footnotesize}
\end{table}
The top edge of the EMM model prediction is 
obtained for the MRST parameterisation whereas the lower edge for the GRV98lo.
We give  fits to our results for $\sigma^{pp}$ of the form,
\begin{equation}
\label{emmfits}
\GGHPSsigtot = a_0 + a_1 s^{b}  + a_2 \ln (s) + a_3 \ln^2 (s).
\end{equation}

\subsection{Conclusions}
We thus see that the range of the results for the \GGHPSsigtot\ from our
QCD motivated EMM model~\cite{Godbole:2004kx} spans the range of predictions
made using the current data and general arguments of unitarity and/or 
factorization. Furthermore, we give $\ln^2 (s)$ parameterisation of the model 
results for \GGHPSsigtot\  which may be used in evaluating the  range of the 
predictions for the underlying event at the LHC.

%%%%%%%%%%%%%%%%%%%%%%%%%%%%%%%%%%%%%%%%%%%%%%%%%%%%%%%%%%%%%%%%%%%%%%%%%%%%%
\section[Tuning models for minimum bias and the underlying event]
{TUNING MODELS FOR MINIMUM BIAS AND THE UNDERLYING EVENT~\protect
\footnote{Contributed by: A.~Moraes, C.~Buttar}}
\subsection{Introduction}

PYTHIA version 6.3 introduces major changes related to the description
of minimum bias interactions and the underlying
event (UE) \cite{Sjostrand:2003wg, PYTHIA-web}. There is a new, more
sophisticated scenario for multiple interactions, new p$_{T}$-ordered
initial- and final-state showers (ISR and FSR), and a new treatment of
beam remnants \cite{Sjostrand:2003wg, PYTHIA-web}.

PYTHIA6.2 has been shown to describe both, minimum bias and
underlying event data reasonably well when appropriately tuned
\cite{Sjostrand:1987su,Moraes:2005,CDF-tuneA}.
A tuning for PYTHIA6.3 as successful as the ATLAS \cite{Moraes:2005}
and CDF \cite{CDF-tuneA} tunings for 6.2, has yet to be
proposed. However, sets of tuned parameters for PYTHIA6.3 which
generate minimum bias and underlying event distributions with
reasonably good agreement with the data are presented in this report.

JIMMY \cite{JIMMY} is a library of routines which should be
linked to the HERWIG Monte Carlo (MC) event generator \cite{Corcella:2000bw}
and is designed to generate multiple parton scattering events in
hadron-hadron events.
JIMMY implements ideas of the eikonal model which are discussed in
more detail in Ref. \cite{Butterworth:1996zw}. The multiparton interaction
is calculated using the cross-section for the hard subprocess, the
conventional parton densities and the area overlap function, $A(b)$
\cite{JIMMY}. JIMMY, however, is limited to the description of the
underlying event and should not be used to predict minimum bias events
\cite{JIMMY}.
                                                                                
In this report, in addition to the tunings for PYTHIA6.323 to both,
minimum bias \cite{Breakstone:1983ns, Alner:1987wb, Abe:1989td, Alexopoulos:1998bi,
  Matinyan:1998ja} and the underlying event \cite{Affolder:2001yp, Acosta:2004wq},
we also propose a tuning for JIMMY4.1 to the underlying event. 

\subsection{Minimum bias events}

Table~\ref{tab:PYTHIA-tunings} displays the relevant PYTHIA6.3
parameters tuned to the minimum bias data \cite{Breakstone:1983ns,
  Alner:1987wb, Abe:1989td, Alexopoulos:1998bi, Matinyan:1998ja}. It shows the ATLAS
tuning \cite{Moraes:2005} used in PYTHIA6.2 in recent ATLAS data
challenges \cite{Rome2005, Rome-generators}, and the new proposed
PYTHIA6.3 tuning which is labelled as \textit{Min-bias}. The
PYTHIA6.323 tuning for the underlying event is also shown in Table~\ref{tab:PYTHIA-tunings}.
The parameters in \textit{Min-bias} were specifically obtained for
PYTHIA6.323 with CTEQ6L as the selected PDF set.
For the purpose of comparison, the corresponding default values in
PYTHIA6.323 \cite{PYTHIA-web} are also shown in the table. 
\begin{table}[h!]
\centering
\caption{PYTHIA6.323 default, Min-bias, UE and PYTHIA6.2 - ATLAS
  parameters.}
  \vspace*{1mm}
\begin{tabular}{|c|c|c|c|c|}
\hline
\multicolumn{1}{|c}{\scriptsize \textbf{Default}} &
\multicolumn{1}{|c}{\scriptsize \textbf{ATLAS}} &
\multicolumn{1}{|c}{\scriptsize \textbf{Min-bias} } &
\multicolumn{1}{|c}{\scriptsize \textbf{UE} } &
\multicolumn{1}{|c|}{\scriptsize \textbf{Comments}}\\  

\multicolumn{1}{|c}{\scriptsize \textbf{(PYTHIA6.323)} \cite{PYTHIA-web}} &
\multicolumn{1}{|c}{\scriptsize \textbf{(PYTHIA6.214)}
  \cite{Moraes:2005}} & 
\multicolumn{1}{|c}{\scriptsize \textbf{(PYTHIA6.323)}} &
\multicolumn{1}{|c}{\scriptsize \textbf{(PYTHIA6.323)}} &
\multicolumn{1}{|c|}{}\\  

\hline

\scriptsize MSTP(51)=7 & \scriptsize MSTP(51)=7 & \scriptsize
MSTP(51)=10042 & \scriptsize MSTP(51)=10042 & \scriptsize PDF set \\
\scriptsize CTEQ5L & \scriptsize CTEQ5L & \scriptsize MSTP(52)=2 &
\scriptsize MSTP(52)=2 &  \\
 & & \scriptsize CTEQ6L \scriptsize{(from LHAPDF)} & \scriptsize
CTEQ6L \scriptsize{(from LHAPDF)} &  \\[0.2cm]

\scriptsize MSTP(68)=3 & \scriptsize MSTP(68)=1 & \scriptsize
 MSTP(68)=1 & \scriptsize MSTP(68)=1  & \scriptsize max. virtuality
 scale\\
 & & & & \scriptsize and ME matching for ISR  \\[0.2cm]

\scriptsize MSTP(70)=1 & - & \scriptsize MSTP(70)=2 & \scriptsize
MSTP(70)=2 & \scriptsize regul. scheme for ISR \\[0.2cm]

\scriptsize MSTP(82)=3 & \scriptsize MSTP(82)=4 & \scriptsize
 MSTP(82)=4 & \scriptsize MSTP(82)=4 & \scriptsize complex scenario +
 double\\[-0.05cm] 
 & & & & \scriptsize Gaussian matter distribution\\[0.2cm]

- & \scriptsize PARP(67)=1 & - & - & \scriptsize parameter regulating
 \\[-0.05cm] 
 & & & & \scriptsize ISR \\[0.2cm]

\scriptsize PARP(82)=2.0 & \scriptsize PARP(82)=1.8 & \scriptsize
PARP(82)=2.3 & \scriptsize PARP(82)=2.6 & \scriptsize
p$_{t_{\rm{min}}}$ \scriptsize parameter \\[0.2cm] 

\scriptsize PARP(84)=0.4 & \scriptsize PARP(84)=0.5 & \scriptsize
 PARP(84)=0.5 & \scriptsize PARP(84)=0.3 & \scriptsize hadronic core
 radius \\[-0.05cm] 
 & & & & \scriptsize (only for MSTP(82)=4) \\[0.2cm]

\scriptsize PARP(89)=1.8 & \scriptsize PARP(89)=1.0 & \scriptsize
 PARP(89)=1.8 & \scriptsize PARP(89)=1.8 & \scriptsize energy scale
 (TeV) used to \\[-0.05cm]
 & & & & \scriptsize calculate p$_{t_{\rm{min}}}$ \\[0.2cm]

\scriptsize PARP(90)=0.25 & \scriptsize PARP(90)=0.16 & \scriptsize
 PARP(90)=0.20 & \scriptsize PARP(90)=0.24 & \scriptsize power of the
 p$_{t_{\rm{min}}}$  \\[-0.05cm]
 & & & & \scriptsize energy dependence  \\[0.1cm]

\hline
\end{tabular}
\label{tab:PYTHIA-tunings}
\end{table}         

\subsubsection{Predictions vs. minimum bias data}

Throughout this report, minimum bias events will be defined as
non-single diffractive inelastic (NSD) interactions, following the
experimental definition used in
\cite{Breakstone:1983ns,Alner:1987wb,Abe:1989td,Alexopoulos:1998bi,Matinyan:1998ja}.
In the PYTHIA language, this means that subprocesses 94 and 95 are
switched on (MSUB(94)=1 and MSUB(95)=1). 
The MC distributions have also been adapted to reproduce the particle
selection requirements applied to the data by
setting $\pi^{0}, K_{s}$ and $\Lambda^{0}$ as stable particles. 

\begin{figure*}
\begin{tabular}{cc}
\resizebox{0.45\textwidth}{!}{\includegraphics{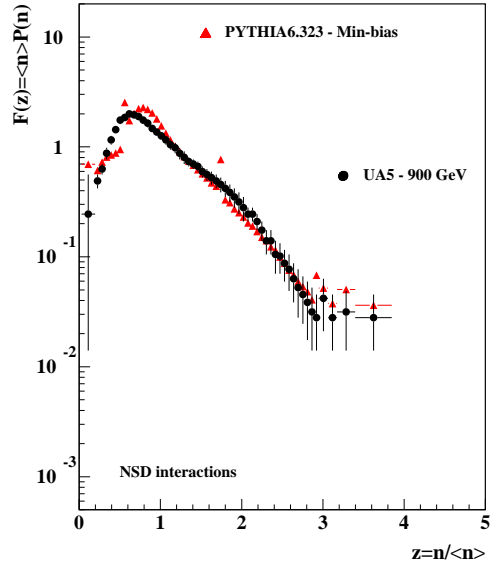}}
&
\resizebox{0.45\textwidth}{!}{\includegraphics{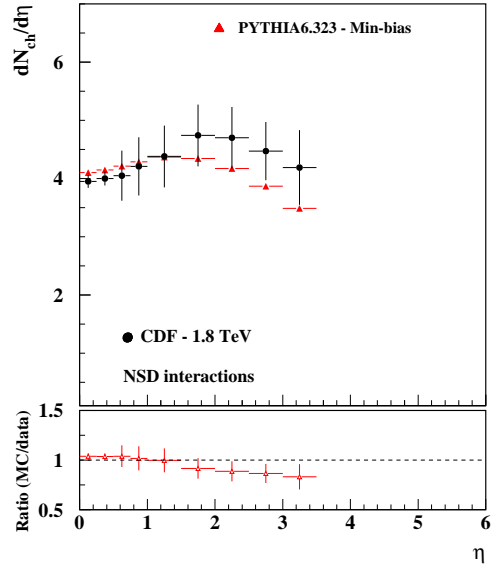}}
\\
(a) & (b)
\\
\resizebox{0.45\textwidth}{!}{\includegraphics{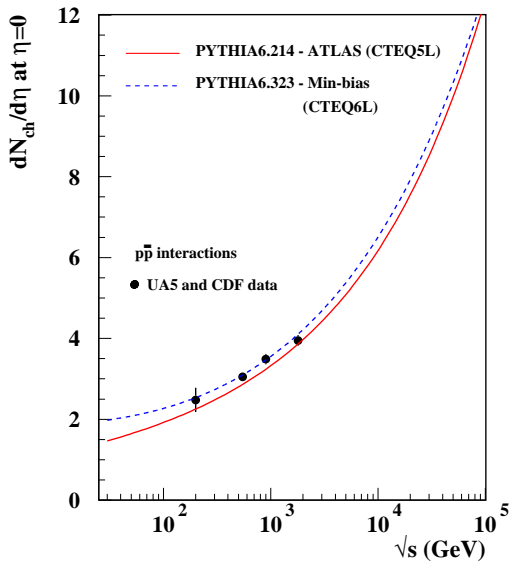}}
&
\resizebox{0.5\textwidth}{!}{\includegraphics{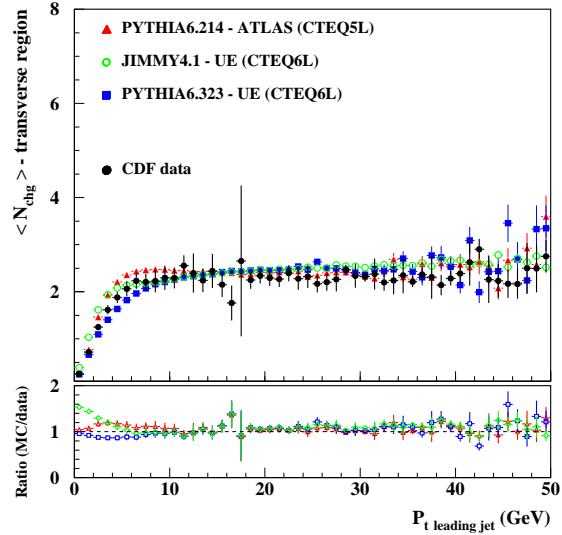}}
\\
(c) & (d)

\end{tabular}
\caption{(a) Charged multiplicity distributions for NSD
  p$\overline{\rm{p}}$ collisions at $\sqrt{\rm{s}}$ = 900
  GeV; (b) dN$_{ch}$/d$\eta$ for NSD p$\overline{\rm{p}}$
  collisions at 1.8 TeV; (c) dN$_{ch}$/d$\eta$ at $\eta =0$ for a wide
  range of $\sqrt{\rm{s}}$ shown for PYTHIA6.214 - Rome and
  PYTHIA6.323 - Min-bias and (d) average charged particles
  multiplicity in the UE compared to CDF data.}
\label{fig:comparison1}
\end{figure*}

Figure~\ref{fig:comparison1} shows
distributions generated by PYTHIA6.323 - Min-bias compared to
minimum bias data. In Fig.~\ref{fig:comparison1}(a) the generated
charged particle multiplicity distribution (KNO variables) is compared to
data measured at $\sqrt{\rm{s}}$ = 900 GeV. Figure~\ref{fig:comparison1}(b)
 compares the Min-bias tuning prediction to
the charged particle density distribution, dN$_{ch}$/d$\eta$, at
$\sqrt{\rm{s}}$ = 1.8 TeV. In Fig.~\ref{fig:comparison1}(c)
dN$_{ch}$/d$\eta$ at $\eta =0$ for a wide range of $\sqrt{\rm{s}}$
is shown for PYTHIA6.214 - ATLAS and PYTHIA6.323 - Min-bias. There is
a reasonably good agreement between distributions generated with the
PYTHIA6.323 - Min-bias tuning and the data. 

At the qualitative level, the agreement between data and the
PYTHIA6.323 - Min-bias tuning is very similar to the agreement seen
between the previous ATLAS tuning (PYTHIA6.2 - ATLAS) and the
minimum bias data (see Ref. \cite{Moraes:2005}).

\subsubsection{LHC predictions for minimum bias events}

\begin{figure}
\begin{tabular}{cc}
\resizebox{0.45\textwidth}{!}{\includegraphics{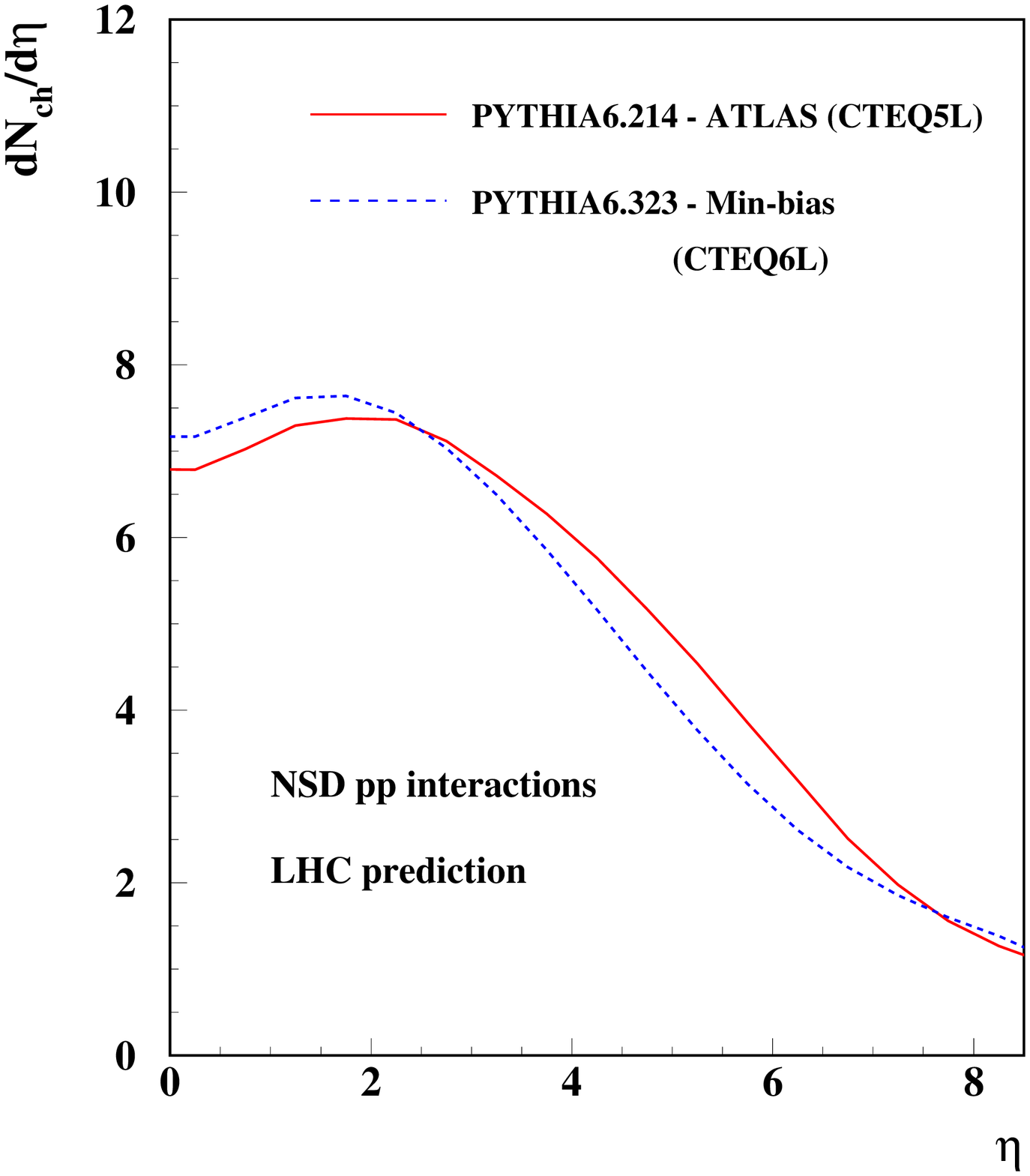}} 
&
\resizebox{0.45\textwidth}{!}{\includegraphics{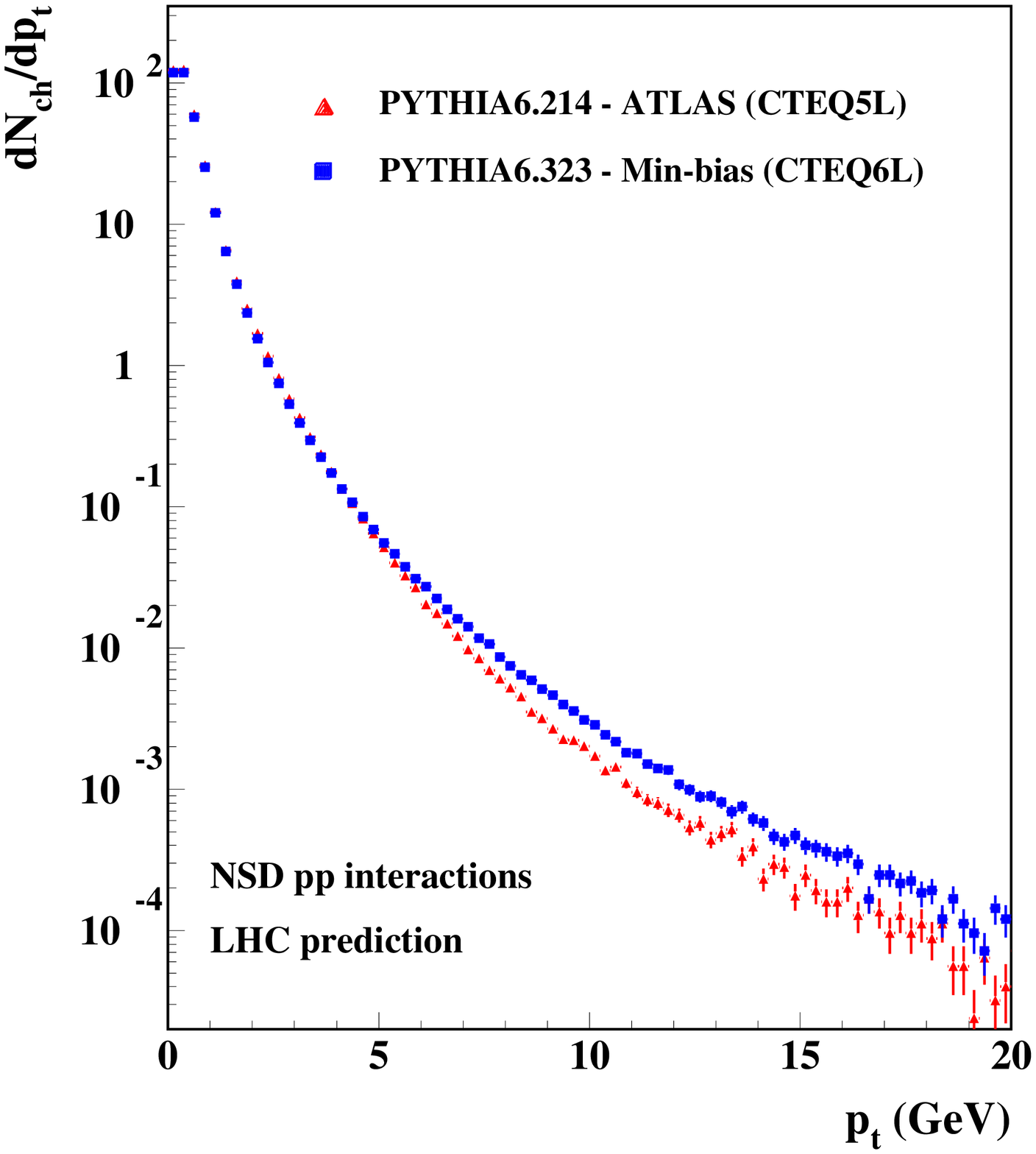}} 
\\
(a) & (b)

\end{tabular}
\caption{(a) Charged particle density distributions,
  dN$_{ch}$/d$\eta$, for NSD pp collisions at $\sqrt{\rm{s}}$ = 14
  TeV. Predictions generated by PYTHIA6.214 - ATLAS and PYTHIA6.323 -
  Min-bias. (b) Charged particle p$_{t}$ spectrum for NSD pp
  collisions at $\sqrt{\rm{s}}$ = 14 TeV.}
\label{fig:minbias-lhc}
\end{figure}

Figure~\ref{fig:minbias-lhc}(a) shows charged particle density
distributions in pseudorapidity for minimum bias pp collisions at
$\sqrt{\rm{s}}$ = 14 TeV generated by PYTHIA6.214 - ATLAS and
PYTHIA6.323 - Min-bias. The charged particle density generated by
PYTHIA6.214 - ATLAS and PYTHIA6.323 - Min-bias at $\eta = 0$ is 6.8
and 7.1, respectively. Note that the dN$_{ch}$/d$\eta$ shape is
slightly different in the two predictions, especially in the range
$2.5 < \eta < 6.5$.

Compared to the charged particle density dN$_{ch}$/d$\eta$ measured by
CDF at 1.8 TeV (Fig.~\ref{fig:comparison1}(b)), both models indicate
a plateau rise of $\sim 70\%$ at the LHC in the central
region.

The average charged particle multiplicity in LHC minimum bias collisions,
$<n_{ch}>$, is 91.04 and 88.72 charged particles as predicted by
PYTHIA6.214 - ATLAS and PYTHIA6.323 - Min-bias, respectively. Even
though PYTHIA6.323 - Min-bias predicts a higher central plateau,
the integrated multiplicity is smaller than that predicted by
PYTHIA6.214 - ATLAS because the former also generates a slightly
narrower dN$_{ch}$/d$\eta$ spectrum compared to the latter. 

The p$_{t}$ spectrum of charged particles
produced in LHC minimum bias events is displayed in Fig.~\ref{fig:minbias-lhc}(b).
 Once again, PYTHIA6.323 - Min-bias is
compared to PYTHIA6.214 - ATLAS. The ``soft'' part of the spectrum
(p$_{t} < 5$ GeV) is very similar as predicted by the two models,
however PYTHIA6.323 - Min-bias generates a harder high-p$_{t}$ tail
than PYTHIA6.214 - ATLAS. 

\subsection{Underlying event} 

The PYTHIA6.323 tuning for the underlying event, labelled \textit{UE},
is also shown in Table~\ref{tab:PYTHIA-tunings}. As for the
\textit{Min-bias} tuning, the parameters in PYTHIA6.323 - UE were
specifically obtained for CTEQ6L as the selected PDF set. Note that
there are differences between the UE and Min-bias tunings. These can
be seen in the p$_{t_{\rm{min}}}$ parameter PARP(82) and
in the choice for the hadronic core radius (PARP(84)).

\subsubsection{JIMMY4.1 tuning}

JIMMY4.1 linked to HERWIG6.507 has been tuned to describe the UE as
measured by CDF \cite{Affolder:2001yp, Acosta:2004wq} and the resulting set of
parameters, labelled UE, is shown in Table~\ref{tab:JIMMY-tunings}. 
As for PYTHIA6.323, the tuned settings were
obtained for CTEQ6L. The default parameters are also included in Table~\ref{tab:JIMMY-tunings}
 for comparison. 
\begin{table}[h!]
\begin{center}
\caption{JIMMY4.1 default and \textit{UE} parameters for the underlying
  event.}  
  \vspace*{1mm}
\begin{tabular}{|c|c|c|}
\hline
\multicolumn{1}{|c}{\scriptsize \textbf{Default}} &
\multicolumn{1}{|c}{\scriptsize \textbf{UE}} &
\multicolumn{1}{|c|}{\scriptsize \textbf{Comments}}\\  

\multicolumn{1}{|c}{} &
\multicolumn{1}{|c}{} &
\multicolumn{1}{|c|}{}\\  

\hline

\scriptsize JMUEO=1 & \scriptsize JMUEO=1 & \scriptsize multiparton
  interaction \\[-0.05cm] 
  & & \scriptsize model\\[0.2cm]

\scriptsize PTMIN=10.0 & \scriptsize PTMIN=10.0 & \scriptsize minimum
  p$_{T}$ in \\[-0.05cm] 
  & & \scriptsize hadronic jet production \\[0.2cm]

\scriptsize PTJIM=3.0 & \scriptsize PTJIM=$2.8 \times \left( \frac{\sqrt{s}}{1.8
  ~\rm{TeV}} \right)^{0.274} $ & \scriptsize minimum p$_{T}$ of
  secondary \\[-0.05cm] 
  & & \scriptsize scatters when JMUEO=1 or 2 \\[0.2cm]

\scriptsize JMRAD(73)=0.71 & \scriptsize JMRAD(73)=1.8 & \scriptsize
  inverse proton \\[-0.05cm]  
  & & \scriptsize radius squared \\[0.2cm]

\scriptsize PRSOF=1.0 & \scriptsize PRSOF=0.0 & \scriptsize
  probability of a soft \\[-0.05cm] 
  & & \scriptsize underlying event \\[0.2cm]

\hline
\end{tabular}
\label{tab:JIMMY-tunings}
\end{center}
\end{table}         
JMRAD(91) should also be changed to the same value used for JMRAD(73)
when antiprotons are used in the simulation (e.g. Tevatron events).

Notice that an energy dependent term has been introduced in PTJIM for
the UE tuning. This leads to a value of PTJIM=2.1 for
p$\overline{\rm{p}}$ collisions at $\sqrt{\rm{s}}$ = 630 GeV and
PTJIM=4.9 for the LHC centre-of-mass energy in pp collisions. 

\subsubsection{Predictions vs. UE data}

Based on CDF measurements \cite{Affolder:2001yp}, the UE is defined as the
angular region in $\phi$ which is transverse to the leading charged
particle jet.  

Figure~\ref{fig:comparison1}(d) shows PYTHIA6.323 - UE
(Table~\ref{tab:PYTHIA-tunings}) and JIMMY4.1 - UE (Table~\ref{tab:JIMMY-tunings}) 
predictions for the UE compared to CDF data
\cite{Affolder:2001yp} for the average charged particle ($p_{t} >0.5~$GeV
and $\left| \eta \right| <1$) multiplicity in the underlying
event. A distribution, generated with the
ATLAS tuning for PYTHIA6.2 and used in recent ATLAS data challenges is
also included in the plot for comparison. There is a reasonably good
agreement between the proposed tunings and the data. The distribution
shapes are slightly different in the region of P$_{t_{\rm{ljet}}}
\lesssim 15$ GeV. PYTHIA6.323 - UE underestimates the data while
JIMMY4.1 - UE overestimates it at low P$_{t_{\rm{ljet}}}$. 

\begin{figure}[h!]
\begin{center}
\begin{tabular}{cc}
\resizebox{0.525\textwidth}{!}{\includegraphics{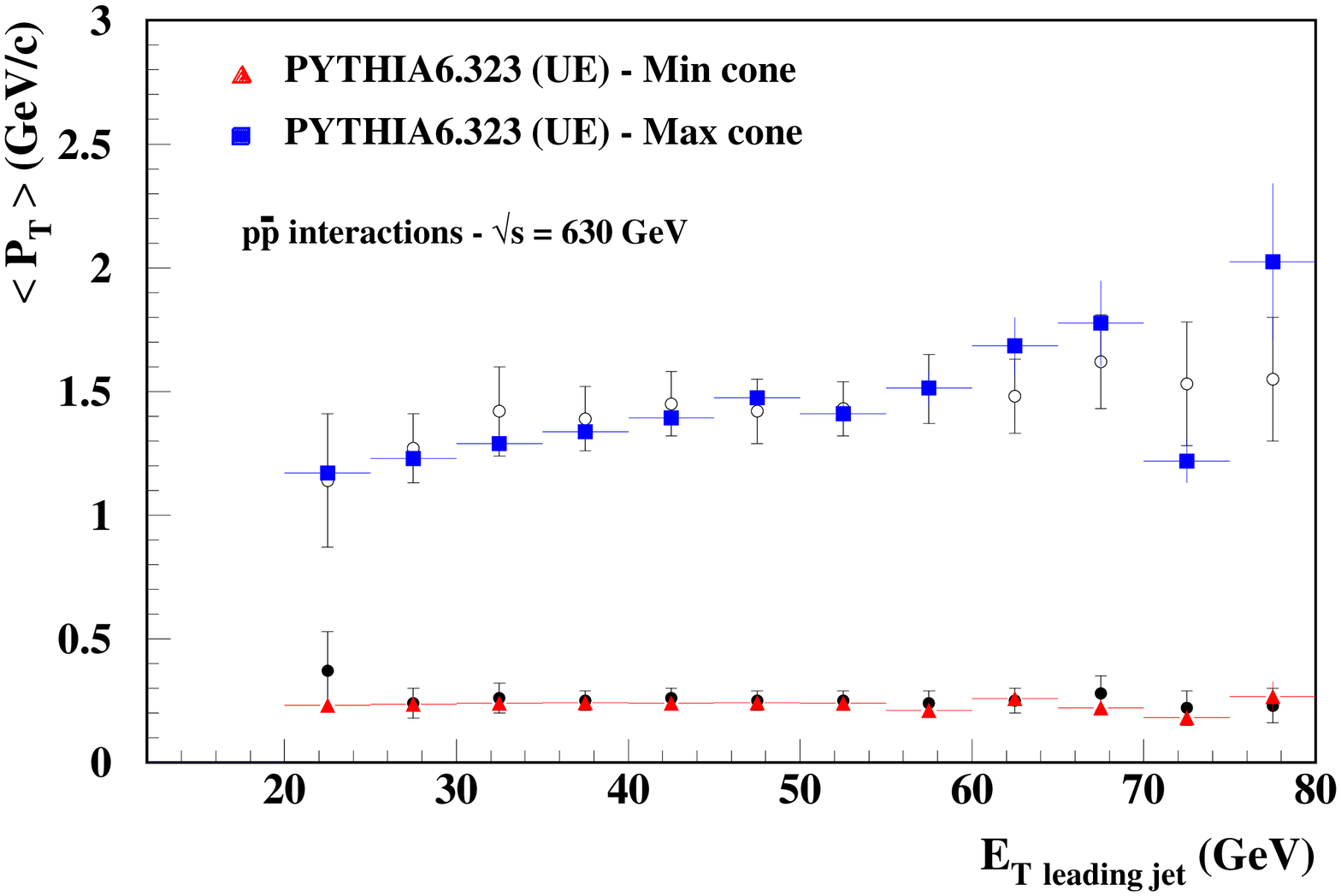}} 
&
\resizebox{0.525\textwidth}{!}{\includegraphics{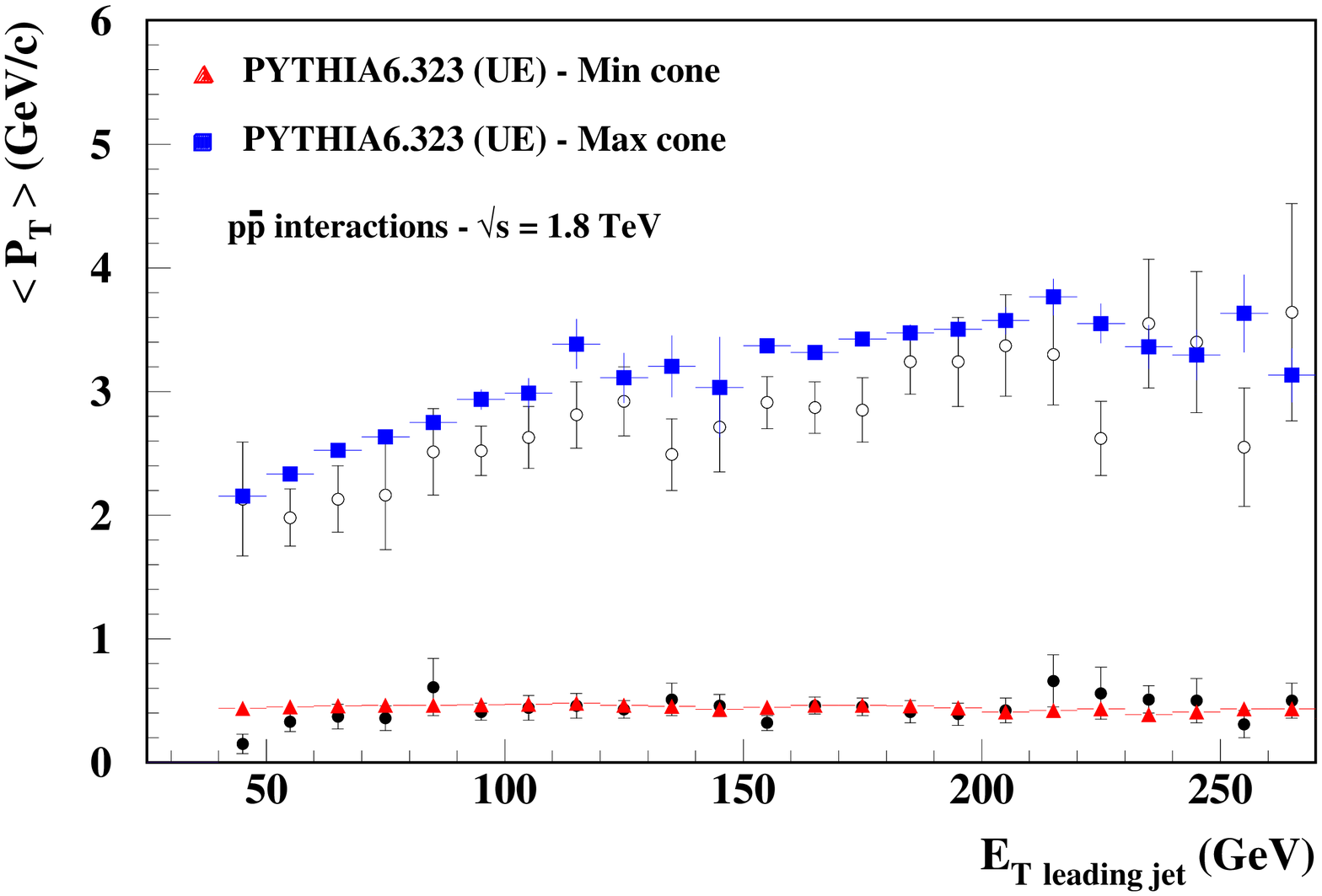}} 
\\
(a) & (b)

\end{tabular}
\caption{PYTHIA6.323 - UE predictions for the UE
  compared to the  $<p_{t}>$ in MAX and MIN
  cones for (a) p$\overline{\rm{p}}$ collisions at $\sqrt{\rm{s}}$
  = 630 GeV  and (b) 1.8 TeV.} 
\label{fig:maxmin-pythia}
%\end{figure}

%\begin{figure}
\begin{tabular}{cc}
\resizebox{0.525\textwidth}{!}{\includegraphics{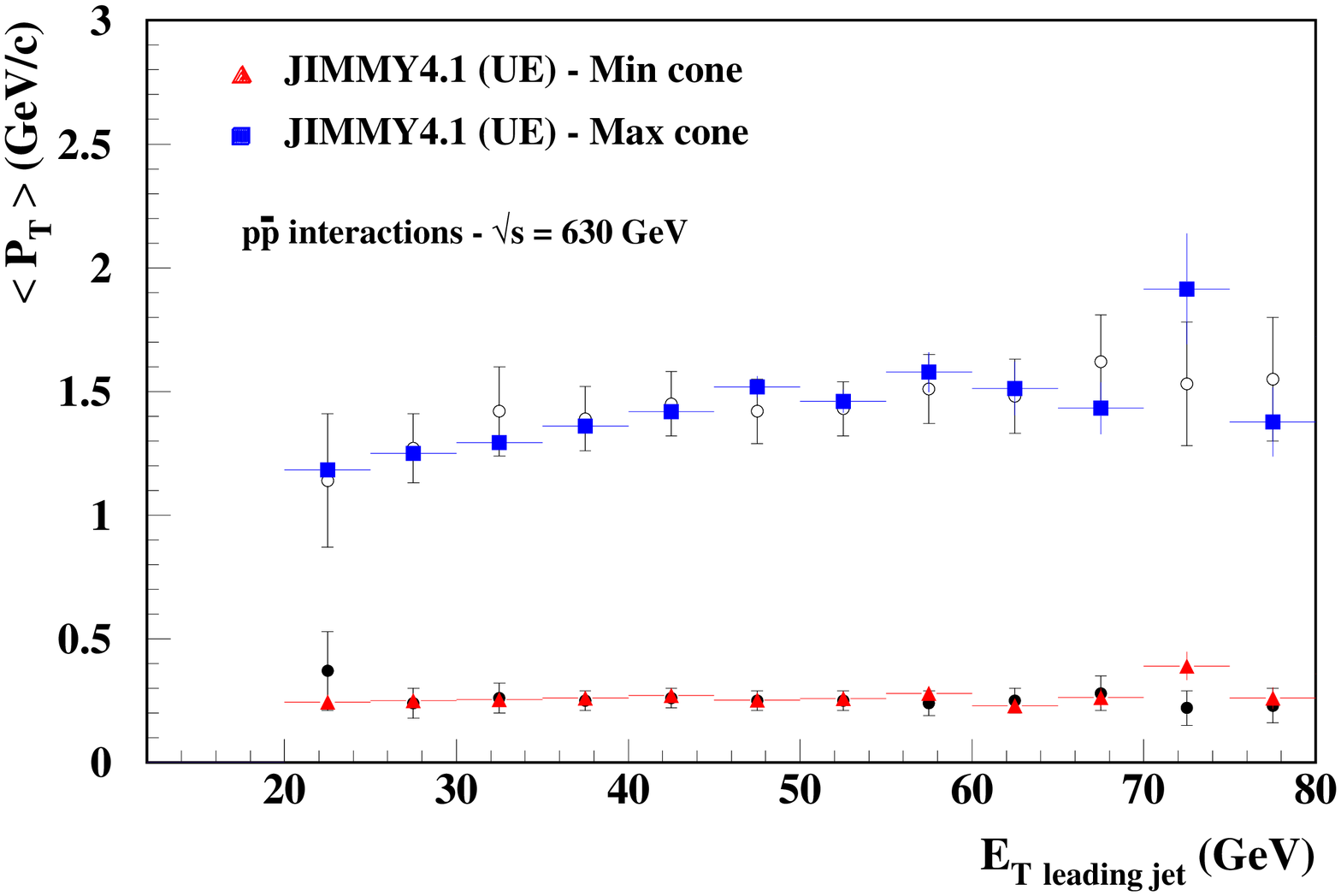}} 
&
\resizebox{0.525\textwidth}{!}{\includegraphics{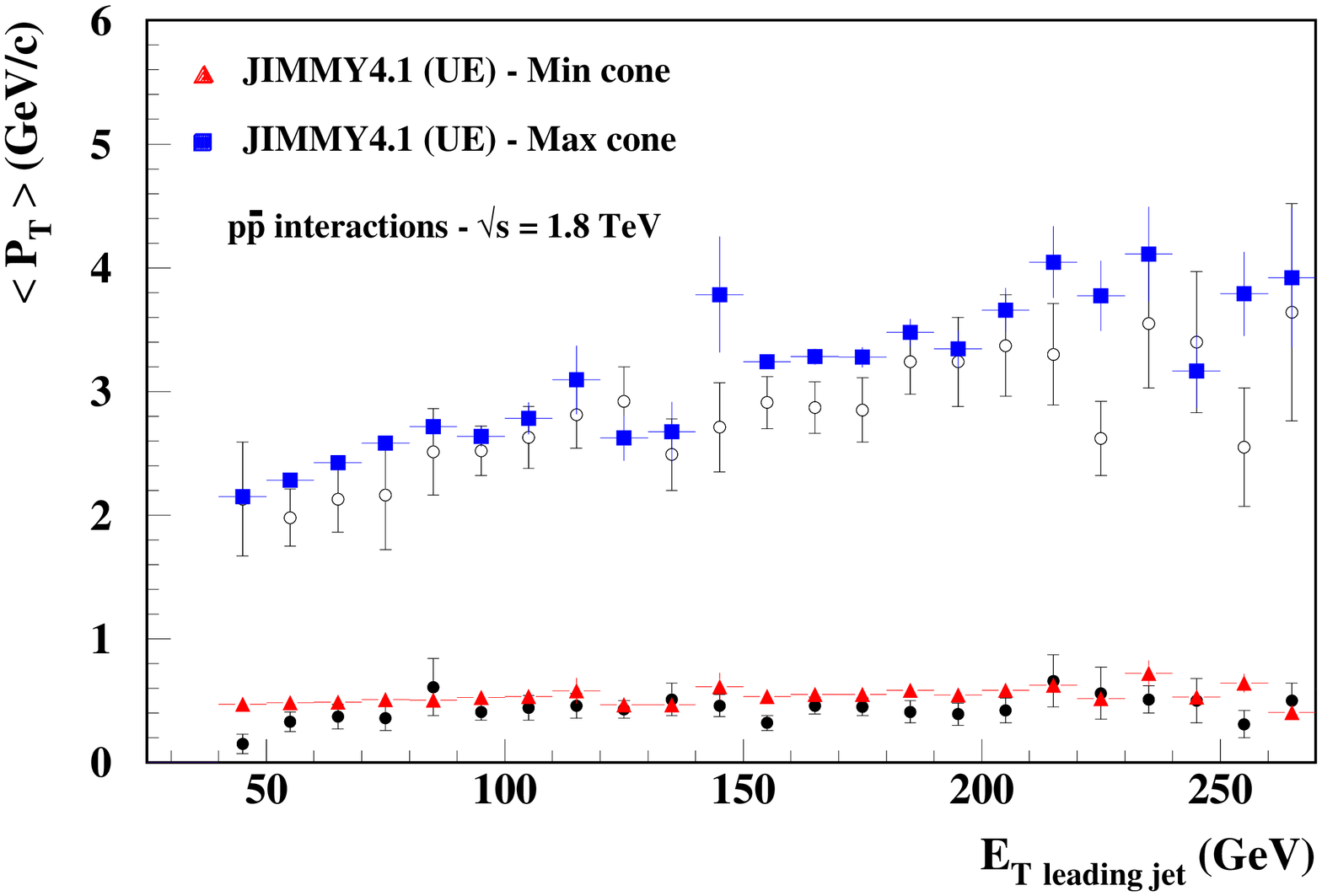}} 
\\
(a) & (b)

\end{tabular}
\caption{JIMMY4.1 - UE predictions for the UE
  compared to the $<p_{t}>$ in MAX and MIN
  cones for (a) p$\overline{\rm{p}}$ collisions at $\sqrt{\rm{s}}$
  = 630 GeV  and (b) 1.8 TeV.} 
\label{fig:maxmin-jmy}

\end{center}
\end{figure}

Another measurement of the UE event is made by defining two cones in
$\eta - \phi$ space, at the same pseudorapidity $\eta$ as the leading
E$_{T}$ jet (calorimeter jet) and $\pm 90^{\circ}$ in the azimuthal
direction, $\phi$ \cite{Acosta:2004wq}. The total charged track momentum
inside each of the two cones is then measured and the higher of the
two values used to define the ``MAX'' cone, with the remaining cone
being labelled ``MIN'' cone. 
Figure~\ref{fig:maxmin-pythia} shows PYTHIA6.323 - UE
predictions for the UE compared to CDF data \cite{Acosta:2004wq} for the
$<p_{t}>$ of charged particles in the MAX and MIN cones for
p$\overline{\rm{p}}$ collisions at (a) $\sqrt{\rm{s}}$ = 630 GeV
and (b) 1.8 TeV. JIMMY4.1 - UE predictions are compared to the data
in Fig.~\ref{fig:maxmin-jmy}. 
Both tunings describe the data with good agreement, however, this only
has become possible by tuning the energy depence terms which regulate
the minimum p$_{t}$ cut-off parameters in both generators (PARP(82),
(89) and (90) for PYTHIA6.3 and PTJIM for JIMMY4.1). 

\subsubsection{LHC predictions for the UE}

\begin{figure}[h!]
\begin{center}
\scalebox{0.4}[0.375]{\includegraphics{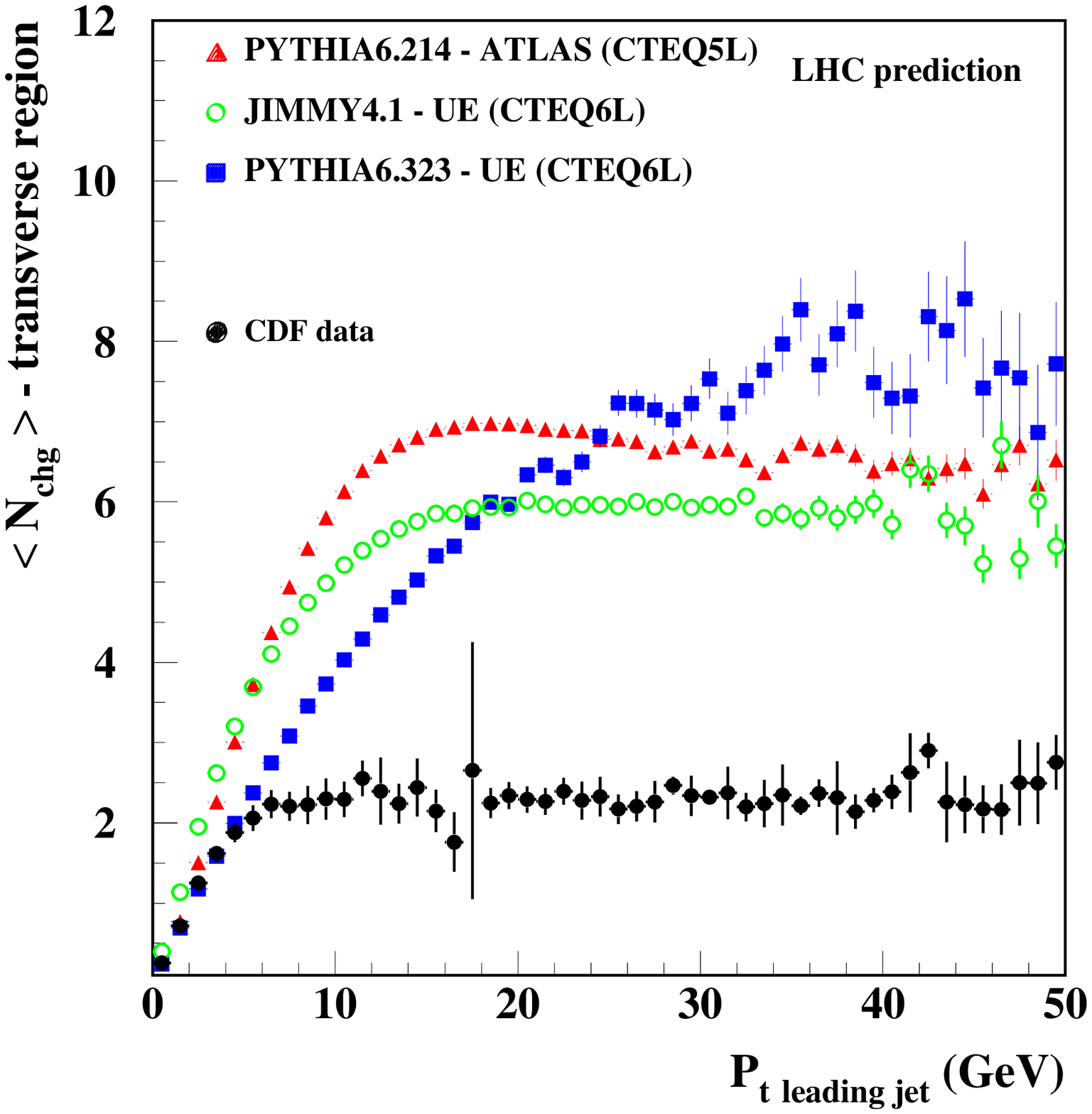}} 
\caption{PYTHIA6.323 - UE, JIMMY4.1 - UE and PYTHIA6.214 - ATLAS
  predictions for the average multiplicity in the UE for LHC pp
  collisions.}    
\label{fig:ue-lhc}
\end{center}
\end{figure}

Figure~\ref{fig:ue-lhc} shows PYTHIA6.323 - UE
(Table~\ref{tab:PYTHIA-tunings}), JIMMY4.1 - UE (Table~\ref{tab:JIMMY-tunings})
 and PYTHIA6.2 - ATLAS predictions for the
average multiplicity in the UE for LHC pp collisions. The CDF data
(p$\overline{\rm{p}}$ collisions at $\sqrt{\rm{s}}$ = 1.8 TeV)
for the average multiplicity in the UE is also included in
Fig.~\ref{fig:ue-lhc}. 

A close inspection of predictions for the UE given in
Fig.~\ref{fig:ue-lhc}, shows that the average charged particle
multiplicity in the UE for leading jets with P$_{t_{\rm{ljet}}} >
20$ GeV reaches a plateau at $\sim 6$ charged particles according to
JIMMY4.1 - UE, $\sim 6.5$ for PYTHIA6.214 - ATLAS and $\sim 7.5$
according to PYTHIA6.323 - UE. Expressed as particle densities per
unit $\eta - \phi$, where the UE  phase-space is given by $\Delta \eta
\Delta \phi = 4 \pi / 3$ \cite{CDF-tuneA, Affolder:2001yp}, these
multiplicities correspond to 1.43, 1.56 and 1.79 charged particles per
unit $\eta - \phi$ (p$_{t} >0.5~$GeV), as predicted by JIMMY4.1 - UE,
PYTHIA6.214 - ATLAS and PYTHIA6.323 - UE, respectively. 

The distribution shapes also show significant differences
between the model predictions. The shape of the multiplicity
distribution, generated by PYTHIA6.323 - UE, is considerably different
from the other two models in the region of P$_{t_{\rm{ljet}}}
\lesssim 25$ GeV.

\subsection{Conclusions}

In this report we have proposed minimum bias and underlying event
tunings for PYTHIA6.323 and JIMMY4.1 (see Tables~\ref{tab:PYTHIA-tunings}
 and \ref{tab:JIMMY-tunings}). 

The minimum bias tuning for PYTHIA6.323 (\textit{Min-bias} - Table~\ref{tab:PYTHIA-tunings}) 
has been shown to describe the minimum bias
data at different colliding energies (figs. \ref{fig:comparison1}(a) -
(c)).

LHC predictions from PYTHIA6.323 - Min-bias and PYTHIA6.214 - ATLAS do
not show any severe differences. There are some noticeable 
differences though. The shape of dN$_{ch}$/d$\eta$ is narrower in the
distribution generated by PYTHIA6.323 - Min-bias compared to that
from PYTHIA6.214 - ATLAS. Another difference is seen for
dN$_{ch}$/dp$_{t}$. PYTHIA6.323 - Min-bias generates a p$_{t}$
spectrum with a harder tail compared to the prediction from
PYTHIA6.214 - ATLAS.

As for the minimum bias tuning for PYTHIA6.323, the underlying event
tunings for PYTHIA6.323 and JIMMY4.1 (Table~\ref{tab:PYTHIA-tunings})
have also been shown to describe most of the UE data made available by
the CDF Collaboration \cite{Affolder:2001yp, Acosta:2004wq}.
However, comparisons to data also indicate that these models need
improvements, especially regarding their capability to correctly
describe the ratio $<p_{t}>$/$<N_{chg}>$ in the UE.

Comparing PYTHIA6.323 - UE, JIMMY4.1 - UE and PYTHIA6.214 - ATLAS
at the LHC, one can clearly notice differences in the shapes of the
distributions predicted by PYTHIA6.323 - UE and the other two models,
as shown in Fig.~\ref{fig:ue-lhc}. 

Tuning the JIMMY parameter PTJIM to include an energy dependent factor
made it possible to describe the MAX-MIN $<p_{t}>$ distributions at
different energies. 

As a final point, we would like to mention that this is an
``\textit{ongoing}'' study. At the moment these are the best
parameters we have found to describe the data, but as the models are
better understood, the tunings could be improved in the near future.

%%%%%%%%%%%%%%%%%%%%%%%%%%%%%%%%%%%%%%%%%%%%%%%%%%%%%%%%%%%%%%%%%%%%%%%%%%%%%
\section[Small $x$]
{SMALL $x$~\protect\footnote{Contributed by: R.D.~Ball, M.~Cooper-Sarkar,
V.~Del~Duca}}
Almost every event recorded at the LHC will involve collisions of partons,
mostly gluons, carrying a relatively small proportion of the longitudinal 
momenta of the colliding beams. Even benchmark cross sections such as 
$W$ and $Z$ production are largely made up of contributions from partons 
carrying rather small values of $x$. However, parton distribution functions
and parton evolution are relatively poorly understood when $x$ is small, due 
to the small $x$ logarithms which render the usual (fixed order) 
perturbation expansion unreliable. This is a serious problem since to 
make a theoretical prediction for an LHC process we must first obtain 
reliable parton distribution functions (typically by analysis of data 
from HERA), and then evolve these partons to scales appropriate for the LHC.

Here we will consider three separate aspects of this problem. Firstly, we
consider the sensitivity of the $W$ and $Z$ cross sections, and in 
particular their rapidity distributions, to small $x$ parton distributions.
We also consider how from an experimental perspective these cross sections 
may eventually be used to improve our knowledge of parton distribution 
functions. Secondly, we will consider the current theoretical status of 
small $x$ resummation, using collinear resummation of the BFKL kernel at 
LO and NLO, and the prospects for accurate calculations by the 
time we have LHC data. Finally, we consider how we might search for 
footprints of BFKL dynamics in LHC data at large rapidities.

\subsection{Low-$x$ physics and $W$ and $Z$ production at the LHC~\protect
\footnote{Author: A.M.~Cooper-Sarkar}}
\label{sec:amcs}

\subsubsection{Introduction}
\label{subsec:lowx;amcs_intro}

\begin{figure}[tbp]
%\vspace{-2.0cm} 
%\vspace*{5pt}
\centerline{
\epsfig{figure=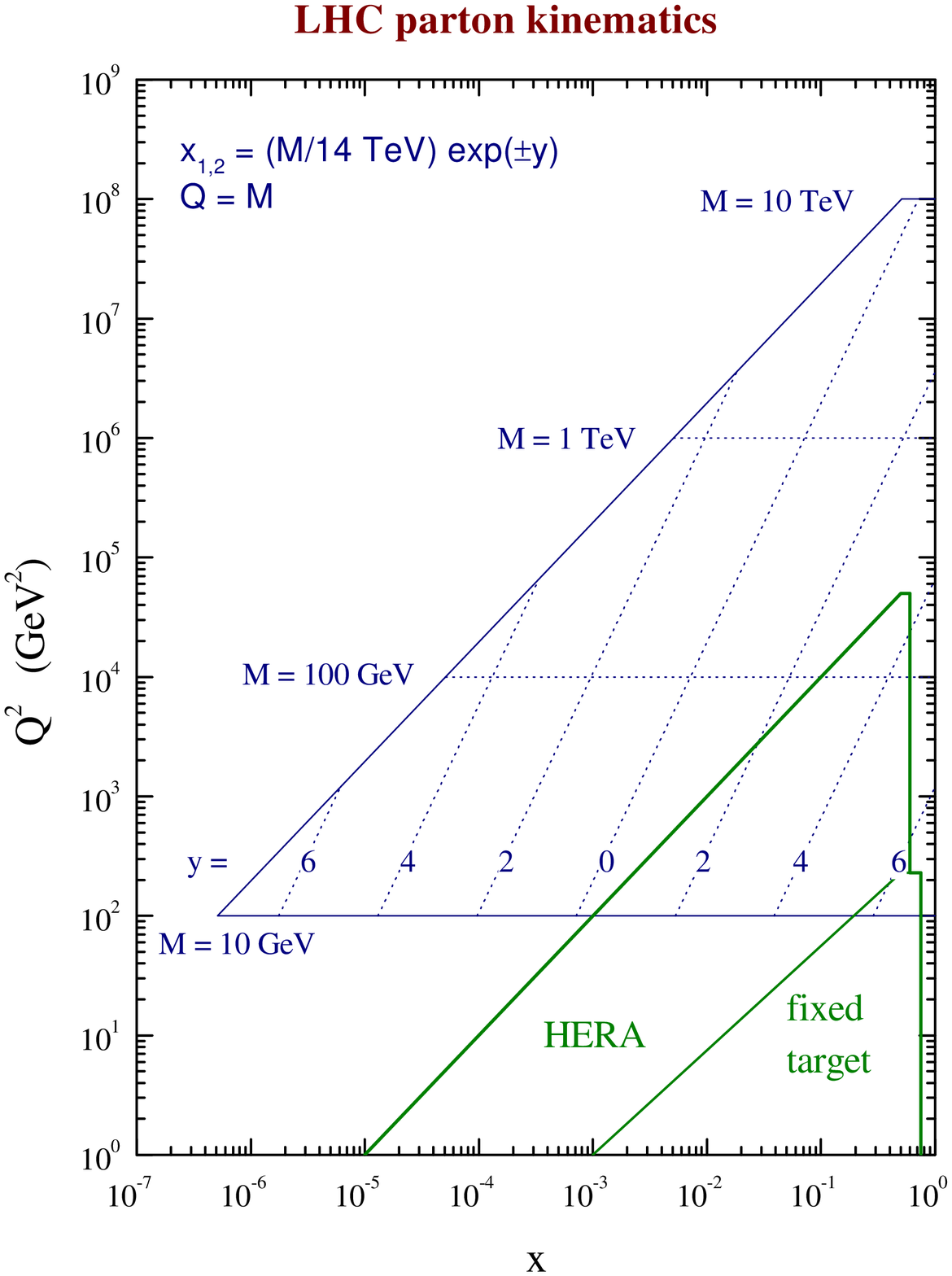,width=0.5\textwidth}
\epsfig{figure=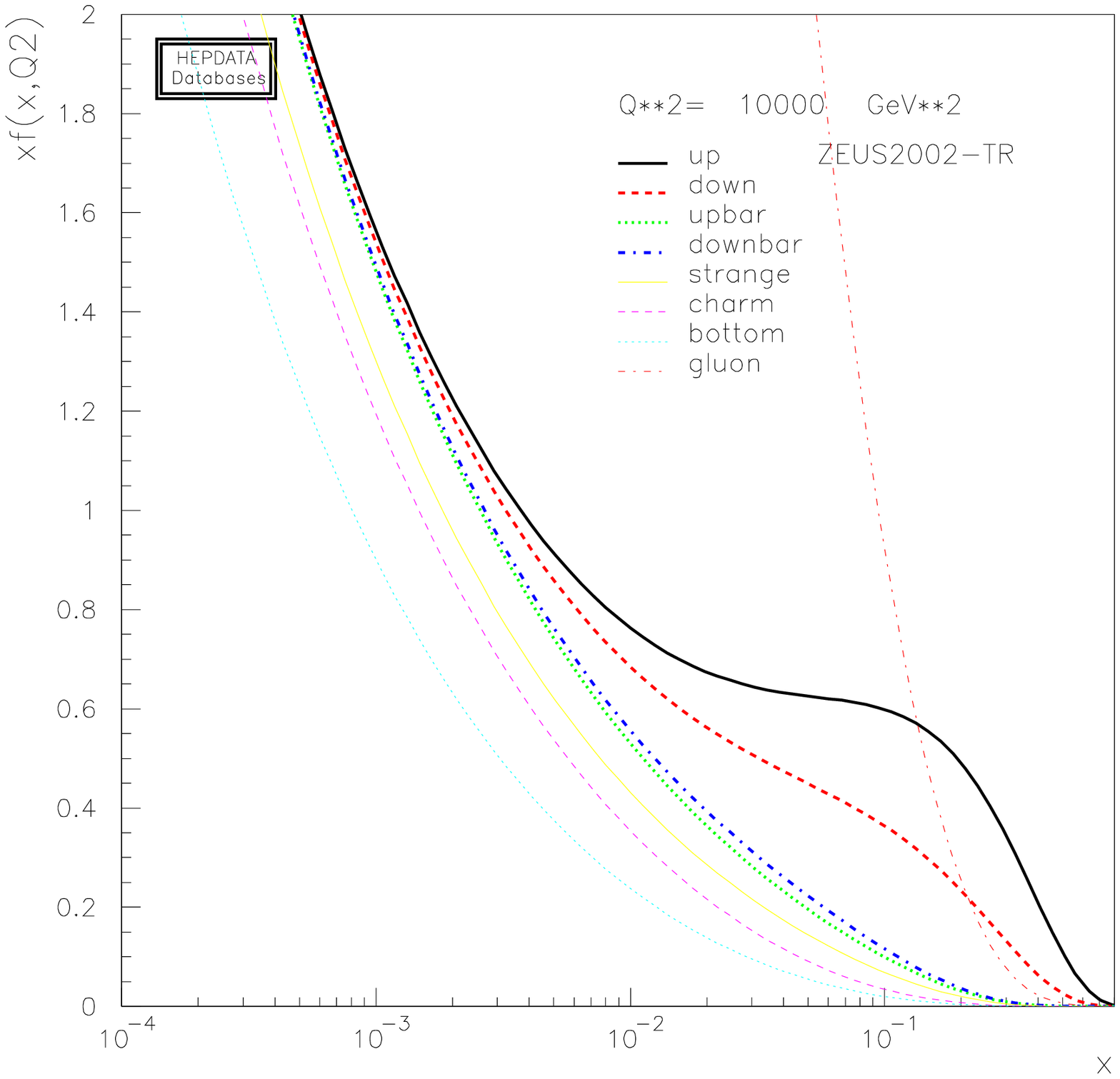,width=0.5\textwidth}}
\caption {Left plot: The LHC kinematic plane (thanks to James Stirling).
Right plot: PDF distributions at $Q^2 = 10,000$~GeV$^2$.}
\label{fig:kin/pdfs}
\end{figure}

The kinematic plane for the LHC is shown in Fig.~\ref{fig:kin/pdfs}, which 
translates the kinematics for producing a state of mass $M$ and rapidity $y$ 
into the deep inelastic scattering variables, $Q^2$, the scale of the 
hard sub-process, and the Bjorken $x$ 
values of the participating partons. The scale of the process is given by
$Q^2 = M^2$ and the Bjorken $x$ values by, $x_1 = (M/ \surd{s}) exp(y)$, and, 
$x_2 = (M/ \surd{s}) exp(-y)$, where $y$ is the parton rapidity,  
$y = \frac{1}{2} \ln {\frac{(E+pl)}{(E-pl)}}$. Thus, at central rapidity, 
these $x$ 
values are equal, but as one moves away from central rapidity, one 
parton moves to higher $x$ and one to lower $x$, as illustrated by 
the lines of constant $y$ 
on the plot.
The first physics to be studied at the LHC will be at relatively 
low scales, where the large cross sections ensure that even low 
luminosity running will yield copious numbers of events. Thus the LHC will 
begin by studying standard model (SM) physics, calibrating our knowledge of 
the detectors on these well known processes. Study of 
Fig.~\ref{fig:kin/pdfs} makes it clear that the cross sections for these 
processes are only well known if 
the parton distribution functions (PDFs) of the proton are 
well known at small-$x$. This assumes that the theoretical 
formalism of NLO QCD, as embodied in the DGLAP
equations, is valid at small-$x$, since this is the formalism used for 
determining PDFs. In the present contribution we address the
question of how PDF uncertainties at low $x$ affect the SM processes 
of $W$ and $Z$ production at the LHC.

The major source of information on low-$x$ physics in the last decade has been
the HERA data. One of the most striking results of HERA was observation
of an unexpected rise
of the $F_2$ structure function at low-$x$.  
The interpretation of the rise in $F_2$, 
in the DGLAP formalism, attributes it to a strong rise in the 
gluon distribution function at low-$x$, since the gluon drives the sea 
distributions by $g \to q \bar{q}$ splitting. 
In fact, the DGLAP equations predict that, at high 
$Q^2 (\sxgeqsim 100~$GeV$^2$), 
a steep rise of the gluon and the sea at low-$x$ will evolve from flat input
shapes at a low $Q^2(\sim 4~$GeV$^2$). Nevertheless, the rise was unexpected, 
firstly, because most theoreticians expected any such tendency
 to be tamed either by screening effects, or by gluon recombination 
at high gluon density. Secondly, because the rise was 
already present for low $Q^2(\sim 1-2~$GeV$^2$) - even lower than the 
conventional starting scale for QCD evolution. Hence
the observation of the rise led to excitement in a somewhat orthogonal section
 of the theoretical community, where a steep rise at low $Q^2$ had been 
predicted in the BFKL formalism, which resums diagrams involving
$\ln (1/x)$. Such resummations are not part of the 
conventional DGLAP $\ln (Q^2)$ summations.  

However, even though the observation of a rise of $F_2$ at low $x$ and 
low $Q^2$ defied conventional prejudice, it can be accommodated within the
conventional DGLAP formalism provided sufficiently flexible input shapes are 
used at a low enough input scale (now taken to be $Q^2 \sim 1~$GeV$^2$). In fact
it turns out that whereas the input sea distribution is still rising at 
low-$x$, the input gluon distribution has turned over to become valence-like, 
and is even allowed to become negative in some parameterizations. 
%Fig.~\ref{fig:NCglusea} illustrates this behaviour in the sea and gluon PDFs,
%together with the data used to extract them.
%\begin{figure}[tbp] 
%*\vspace*{5pt}
%\vspace{-1.0cm}
%\centerline{
%\epsfig{figure=s_ball/NClowq2.eps,width=0.45\textwidth}
%\epsfig{figure=s_ball/glusea.eps,width=0.4\textwidth}}
%\caption {Left side: ZEUS data on $F_2$ showing the rise at low $x$, for 
%various $Q^2$. Right side: the gluon and sea PDFs extracted from these data in
%the ZEUSJETS PDF fit, for various $Q^2$, illustrating the turnover of the gluon
%PDF at low $Q^2$ }
%\label{fig:NCglusea}
%\end{figure}
This counter intuitive behaviour has 
led many QCD theorists to believe that the conventional formalism is in need 
of extension~\cite{Devenish:2004pb}. In Sec.~\ref{sec:rdb} we describe 
recent 
work in this area. The present 
section is concerned with how well the PDFs are known at low-$x$, within 
the conventional framework, and how this affects the predictions for $W$ and 
$Z$ production at the LHC. These processes have been suggested as 
`standard-candles' for the measurement of luminosity because their 
cross sections are `well known'. In the present contribution we investigate to 
what extent this is really true - and what might be done about it.
  
\subsubsection{$W$ and $Z$ Production at the LHC} 
\label{subsec:lowx;amcs_wzpred}

At leading order (LO), $W$ and $Z$ production occur by the process, 
$q \bar{q} \rightarrow W/Z$.
Consulting Fig.~\ref{fig:kin/pdfs}, we see that at central rapidity, 
the participating partons have small momentum fractions, $x \sim 0.005$, and
over the measurable rapidity range, $|y| < 2.4$, 
$x$ values remain in the range, $5.10^{-4} < x < 0.05$. 
Thus, in contrast to the situation at the TeVatron, 
the scattering is happening dominantly between sea quarks and anti-quarks. 
Furthermore, the high scale of the process $Q^2 = M^2 \sim 10,000~$GeV$^2$ 
ensures that the gluon is the dominant 
parton as also illustrated in Fig.~\ref{fig:kin/pdfs}, where the PDFs for all 
parton flavours 
are shown for $Q^2 = \sim 10,000~$GeV$^2$. Hence the sea quarks have 
mostly been generated by the flavour blind $g \to q \bar{q}$ splitting 
process. Thus the precision of our knowledge of $W$ and $Z$ cross sections at 
the LHC is crucially dependent on the uncertainty on 
the momentum distribution of the gluon at low-$x$. This is where the HERA data
come in. 
%\begin{figure}
%\begin{center}
%\includegraphics[width=0.5\textwidth]{lhc-kinplane.eps}
% \caption{The LHC kinematic plane (thanks to James Stirling).
%}
%\label{search}
%\end{center}
%\end{figure}

\begin{figure}[tbp] 
%\vspace*{5pt}
\vspace{-1.0cm}
\centerline{
\epsfig{figure=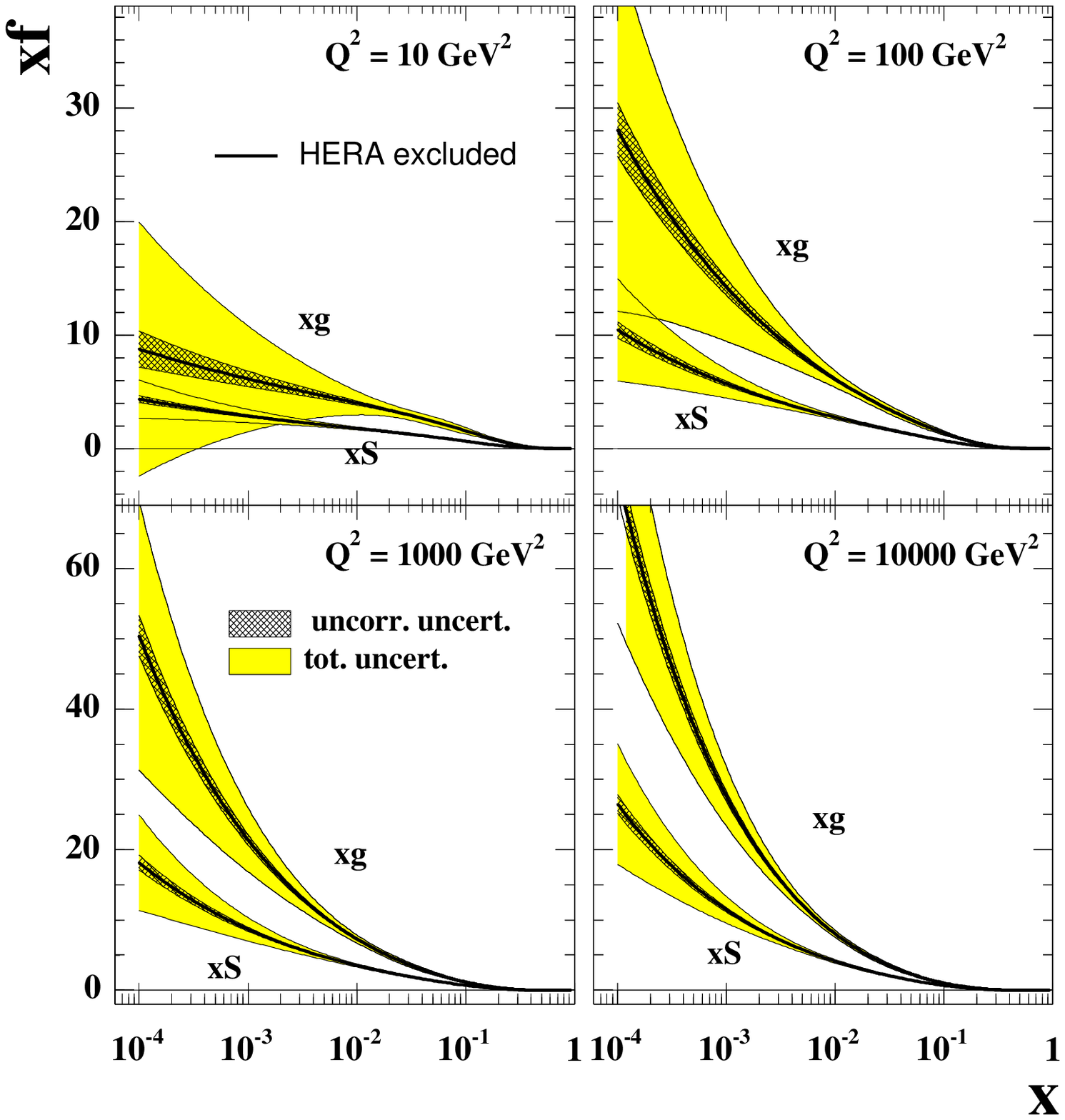,width=0.5\textwidth}
\epsfig{figure=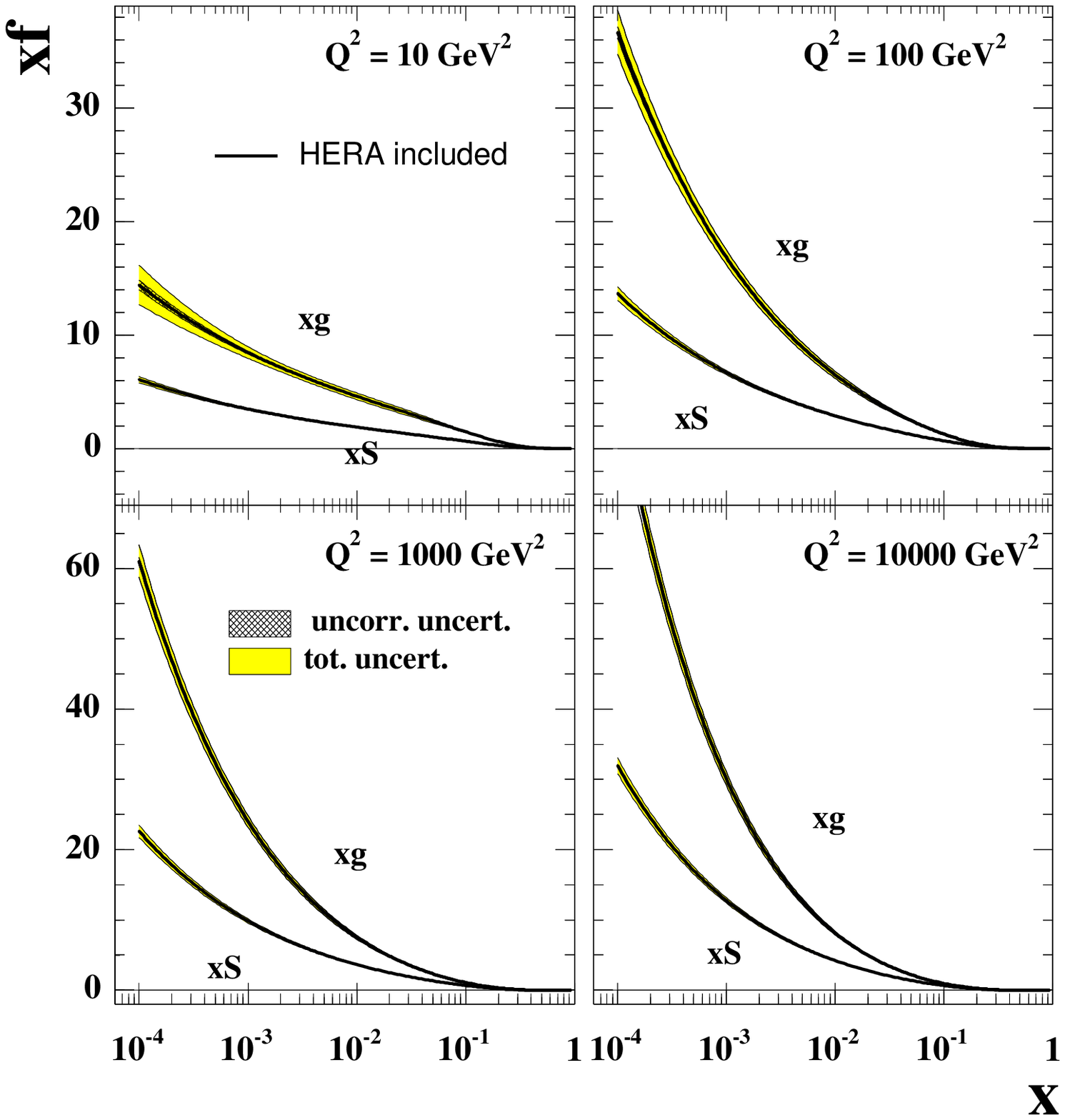,width=0.5\textwidth}}
\caption {Sea ($xS$) and gluon ($xg$) PDFs, as a function of $x$, for various 
$Q^2$ values; left plot: from the ZEUS-S global PDF analysis not including
 HERA data; right plot: from the ZEUS-S global PDF analysis including HERA data.}
\label{fig:pre/postPDFs}
\end{figure}

Figure~\ref{fig:pre/postPDFs} shows the sea and gluon PDFs (and their 
uncertainties) extracted from an
NLO QCD PDF fit analysis to world data on deep inelastic 
scattering, before and after HERA data are included. 
The latter fit is the ZEUS-S global fit~\cite{Chekanov:2002pv}, whereas the former is 
a fit using the same fitting analysis but leaving out the ZEUS data. The full 
PDF uncertainties for both fits are calculated from the eigenvector PDF sets 
of the ZEUS-S analysis using LHAPDF~\cite{Whalley:2005nh}. The 
improvement in the 
level of uncertainty is striking.
 
\begin{figure}[tbp] 
%\vspace*{5pt}
\vspace{-1.0cm}
\centerline{
\epsfig{figure=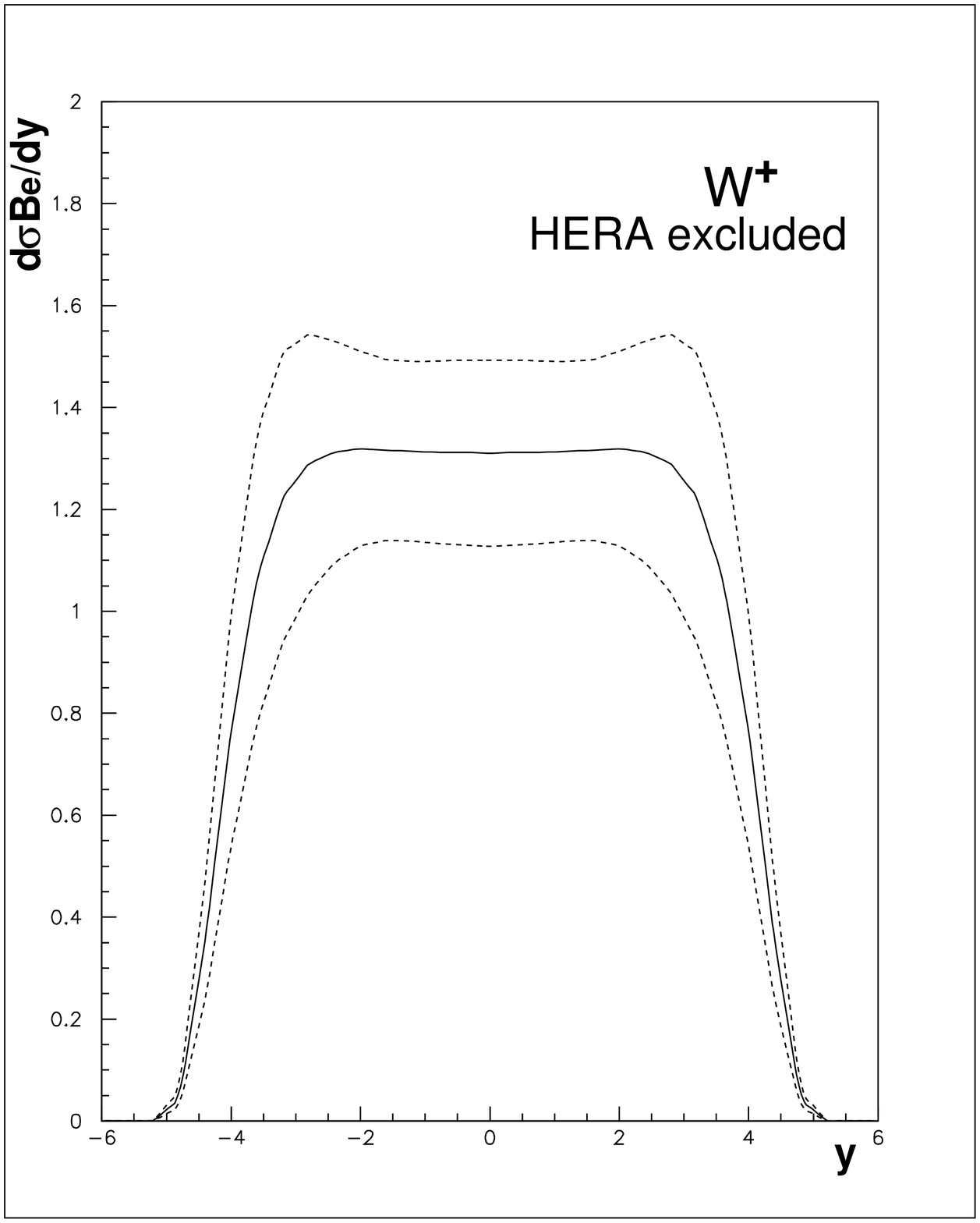,width=0.3\textwidth,height=4.5cm}
\epsfig{figure=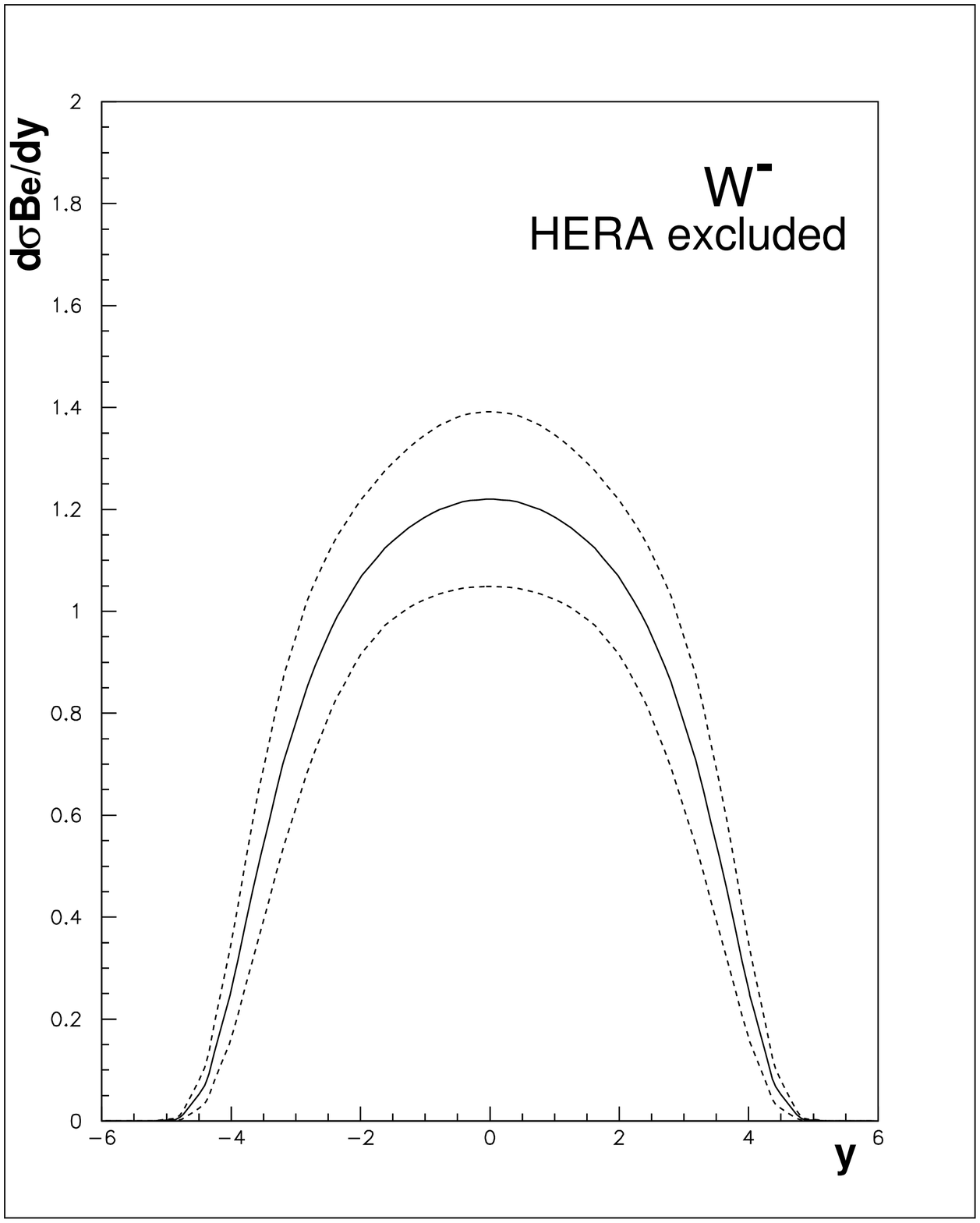,width=0.3\textwidth,height=4.5cm}
\epsfig{figure=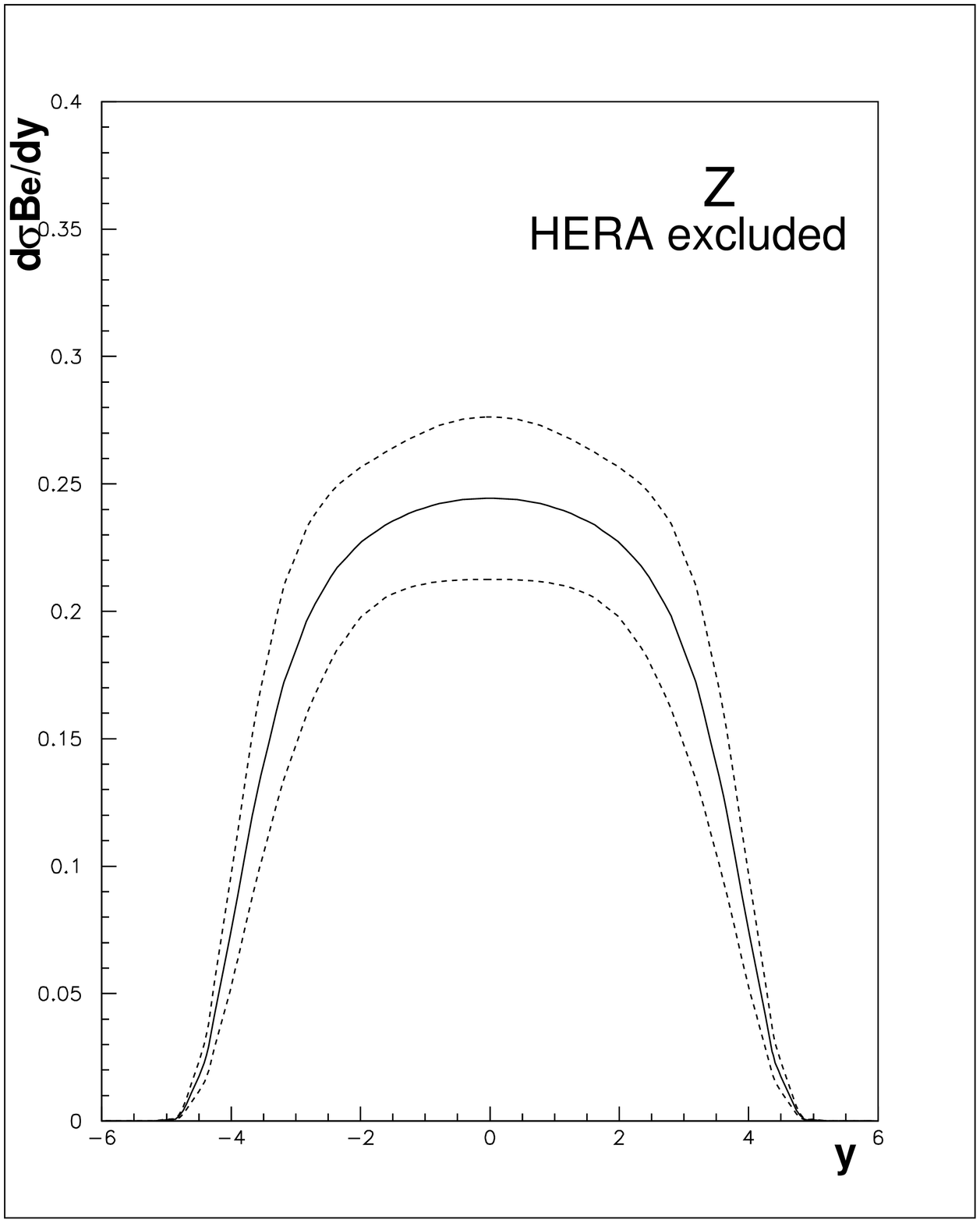,width=0.3\textwidth,height=4.5cm} 
}
\centerline{
\epsfig{figure=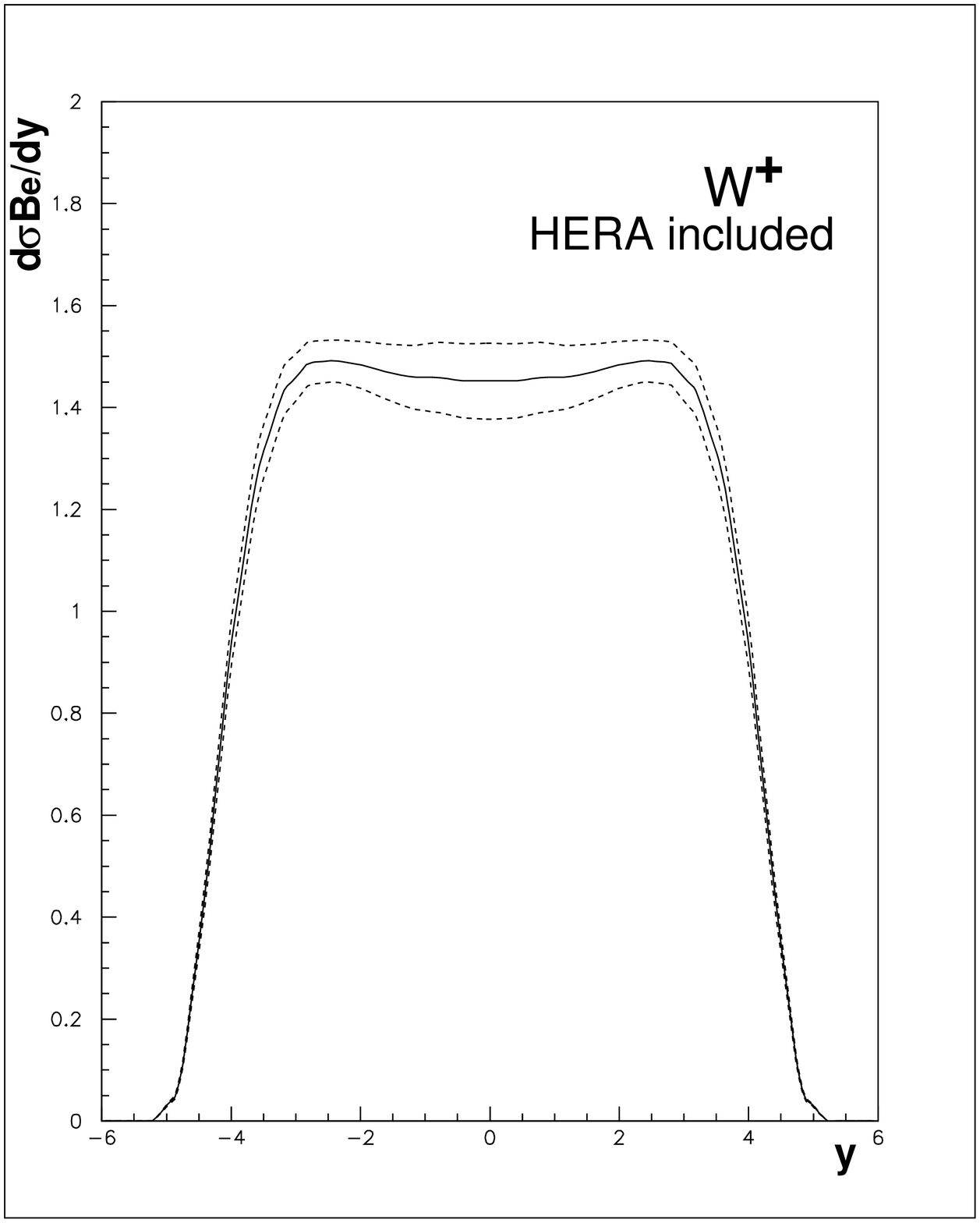,width=0.3\textwidth,height=4.5cm}
\epsfig{figure=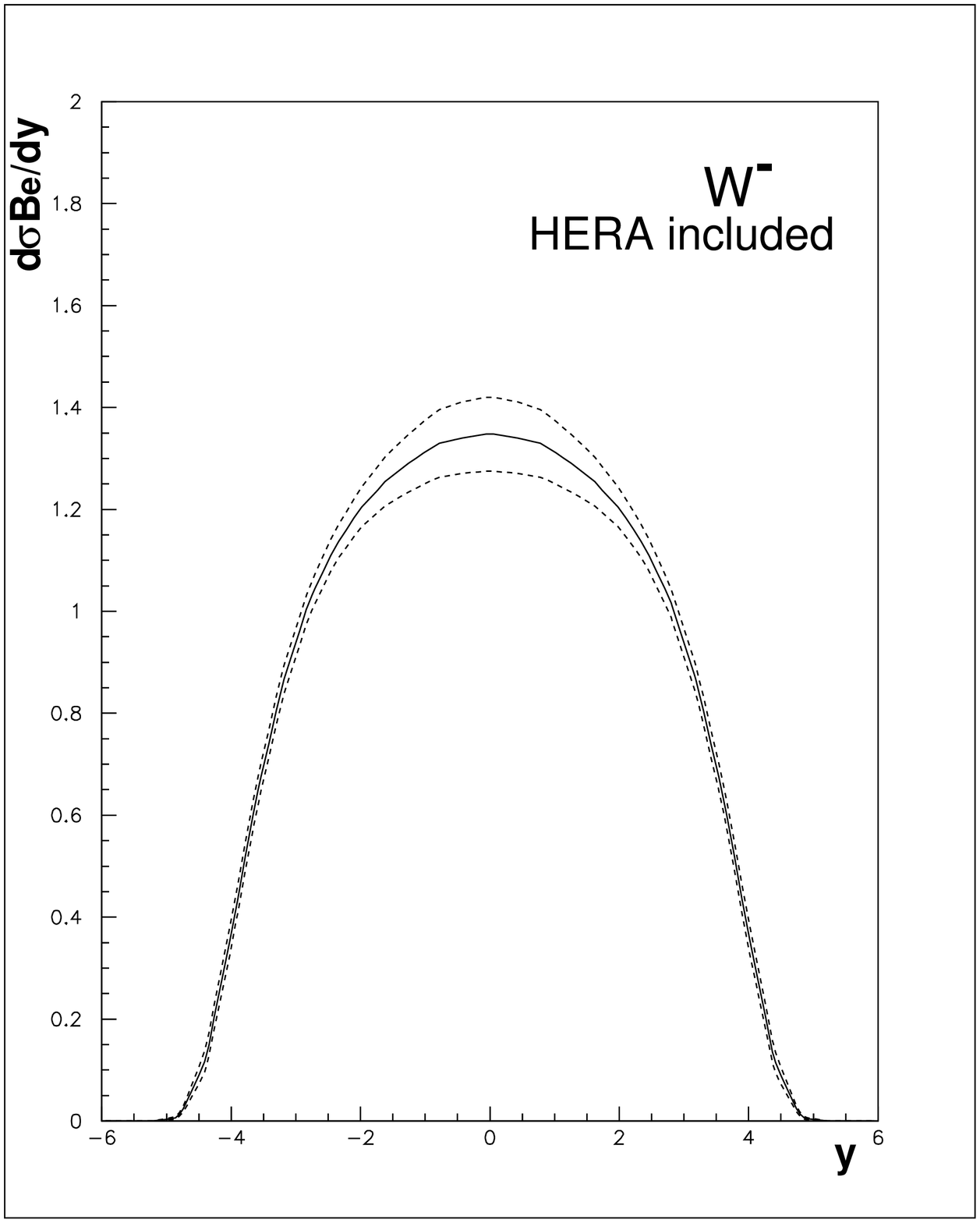,width=0.3\textwidth,height=4.5cm}
\epsfig{figure=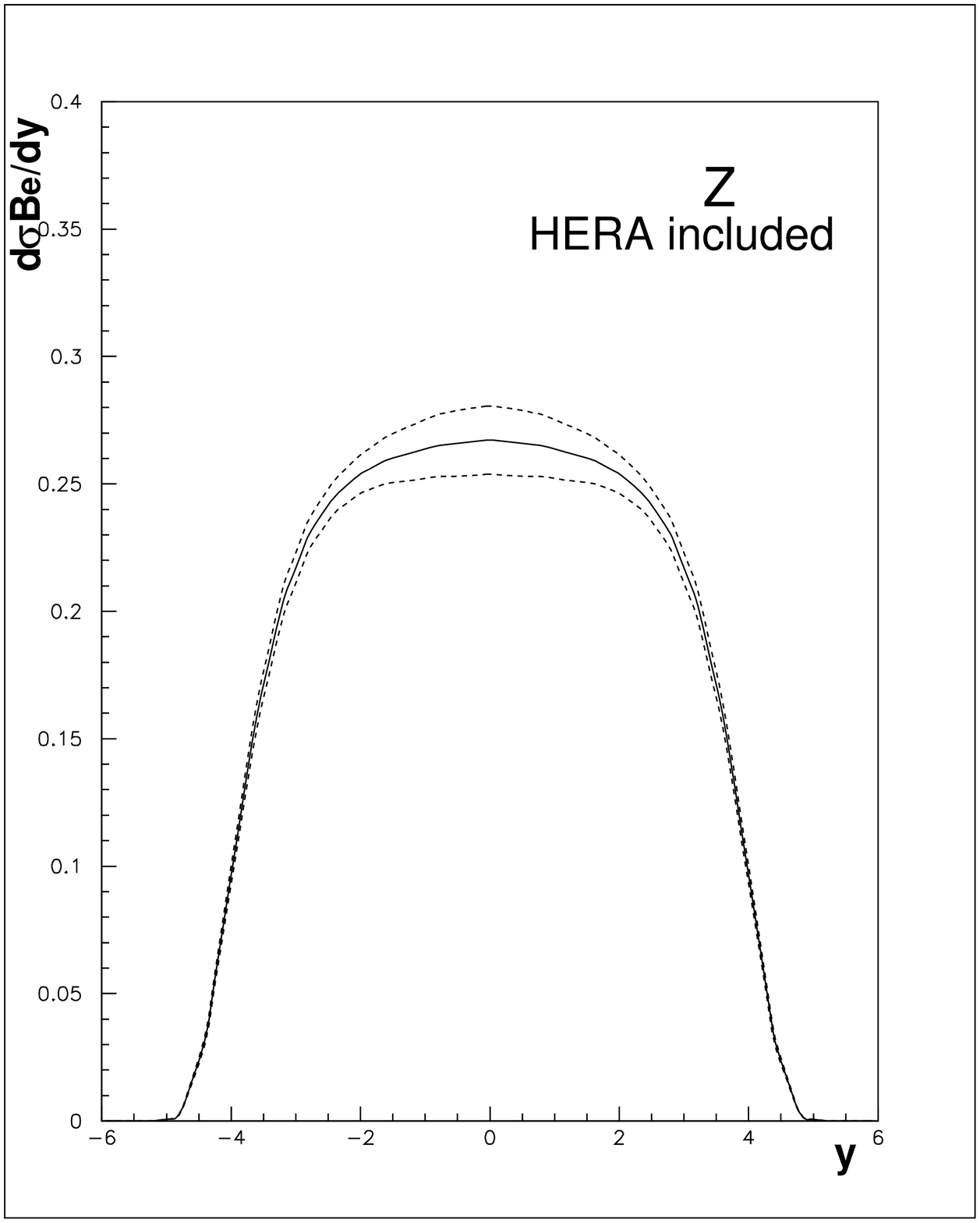,width=0.3\textwidth,height=4.5cm}
}
\caption {LHC $W^+,W^-,Z$ rapidity distributions and their PDF uncertainties: top row: from the ZEUS-S 
global PDF analysis
not including HERA data; left plot $W^+$; middle plot $W^-$; right plot $Z$; bottom row: from the ZEUS-S
global PDF analysis including HERA data; left plot: $W^+$; middle plot: $W^-$; right plot: $Z$.}
\label{fig:WZrapFTZS13}
\end{figure}

Figure~\ref{fig:WZrapFTZS13} illustrates how this improved knowledge of the gluon
and sea distributions has improved our knowledge of $W$ and $Z$ 
cross sections.
It shows $W$ and $Z$ rapidity spectra predicted using the PDFs extracted from
the global PDF fit which does not include the HERA data, compared to 
those extracted from the similar global PDF fit which does include HERA data. 
The corresponding predictions for the $W$ and $Z$ cross sections, 
decaying to the lepton decay mode, are summarized in Table~\ref{tab:datsum}.
\begin{table}[t]
\caption{LHC $W$ and $Z$ cross sections for decay via the lepton mode, for various PDFs}
  \vspace*{-1mm}
\centering{\small
\begin{tabular}{llllcccc}\\
%\vspace{-1.0cm}
 \hline
PDF Set  & $\sigma(W^+).B(W^+ \rightarrow l^+\nu_l)$ & $\sigma(W^-).B(W^- \rightarrow l^-\bar{\nu}_l)$ & 
$\sigma(Z).B(Z \rightarrow l^+ l^-)$\\
 \hline
 ZEUS-S no HERA  & $10.63 \pm 1.73 $~nb & $7.80 \pm 1.18 $~nb & $1.69 \pm 0.23$~nb \\
 ZEUS-S  & $12.07 \pm 0.41 $~nb & $8.76 \pm 0.30 $~nb & $1.89 \pm 0.06$~nb\\
 CTEQ6.1 & $11.66 \pm 0.56 $~nb & $8.58 \pm 0.43 $~nb & $1.92 \pm 0.08$~nb\\
 MRST01 & $11.72 \pm 0.23 $~nb & $8.72 \pm 0.16 $~nb & $1.96 \pm 0.03$~nb\\
 \hline\\
\end{tabular}}
\label{tab:datsum}
\end{table}
The uncertainties in the predictions for these cross sections have decreased 
from $\sim 16\%$ pre-HERA to $\sim 3.5\%$ post-HERA. There could clearly have 
been no talk of using these processes as standard candle processes, 
without the HERA data.

\begin{figure}[tbp] 
%\vspace*{5pt}
%\vspace{-1.0cm}
\centerline{
\epsfig{figure=s_ball/my_wprap_lha_zs13.eps,width=0.3\textwidth,height=4.5cm}
\epsfig{figure=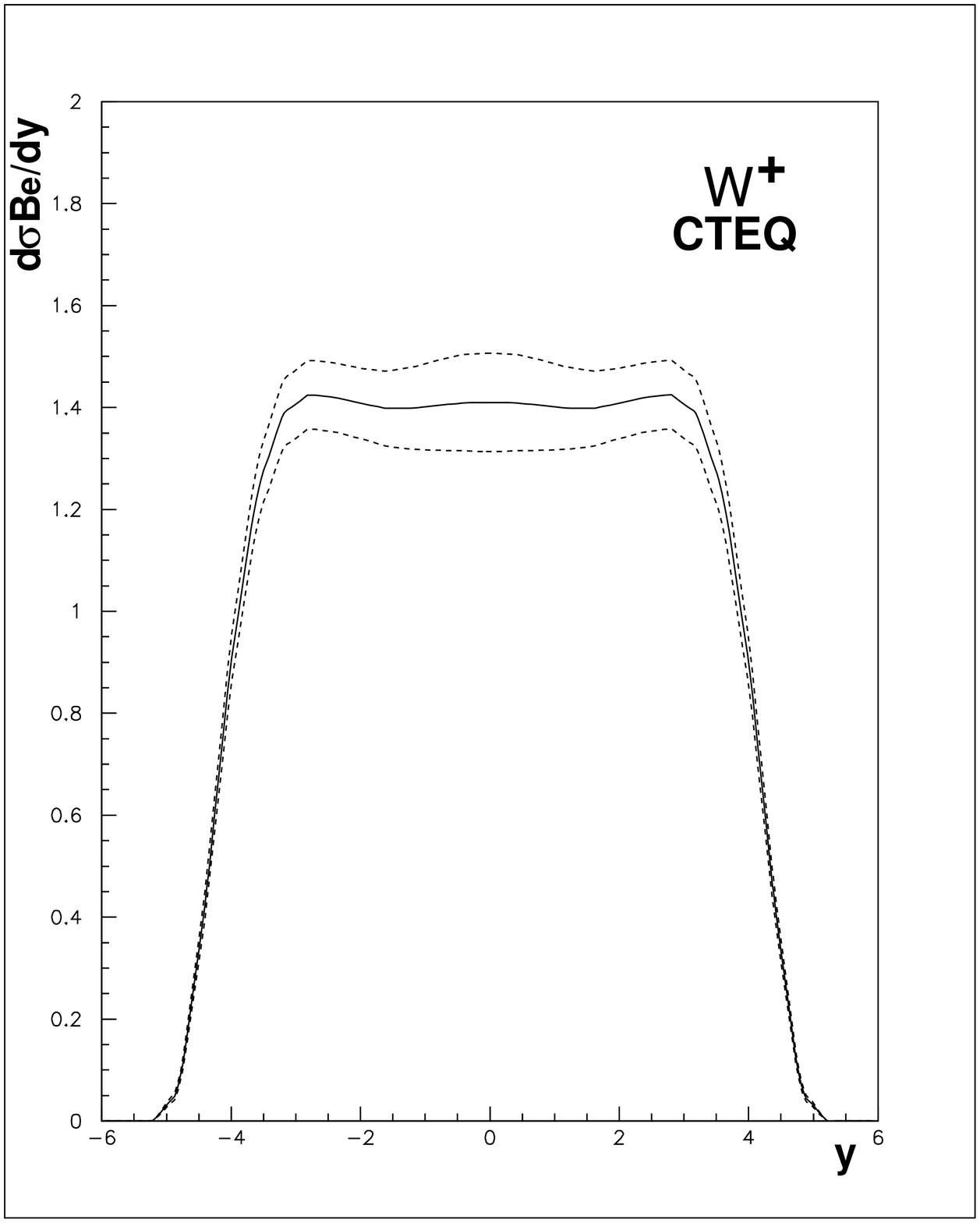,width=0.3\textwidth,height=4.5cm}
\epsfig{figure=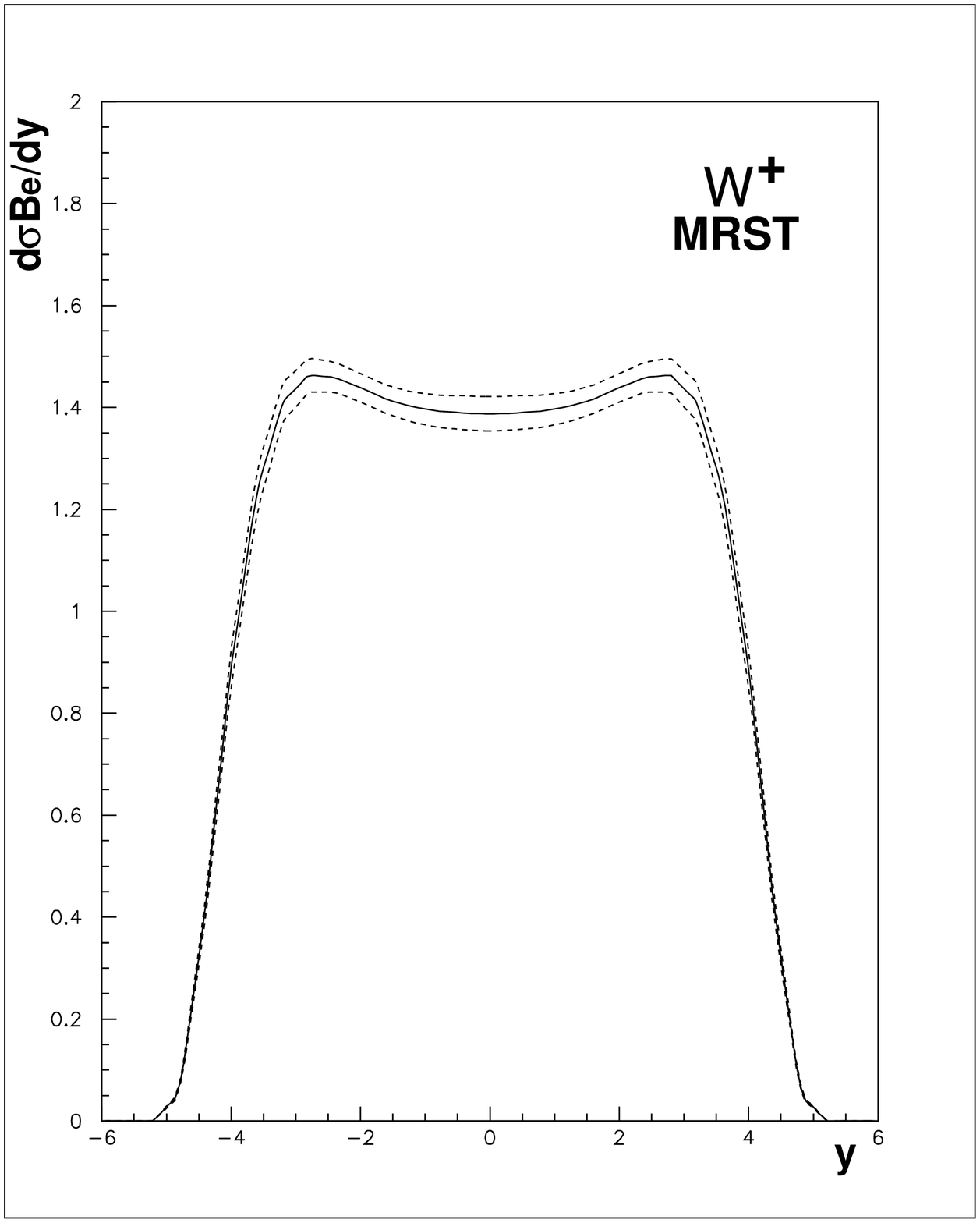,width=0.3\textwidth,height=4.5cm}
}
\caption {LHC $W^+$ rapidity distributions and their PDF uncertainties;
 left plot: ZEUS-S PDFs; middle plot: CTEQ6.1 PDFs;
right plot: MRST01 PDFs.}
\label{fig:mrstcteq}
\end{figure}

The post-HERA level of precision, illustrated in Fig.~\ref{fig:WZrapFTZS13}, 
is taken for granted in modern analyses. However, when 
considering the PDF uncertainties on the Standard Model (SM) predictions it 
is necessary not 
only to consider the uncertainties of one particular PDF analysis, but also to 
compare PDF analyses. Figure~\ref{fig:mrstcteq} compares the predictions for 
$W^+$ production for the ZEUS-S PDFs with those of 
the CTEQ6.1\cite{Pumplin:2002vw} PDFs and the MRST01\cite{Martin:2001es} 
PDFs\footnote{MRST01 PDFs are used because the 
full error analysis is available for this PDF set.}. 
The corresponding $W^+$ cross sections for decay to the leptonic mode are 
given in Table~\ref{tab:datsum}.
Comparing the uncertainty at central rapidity, rather 
than the total cross section, we see that the uncertainty estimates are 
somewhat larger: $\sim 6\%$ for ZEUS-S; $\sim 8\%$ 
for CTEQ6.1M and $\sim 3\%$ for MRST01. 
The difference in the central value between 
ZEUS-S and CTEQ6.1 is $\sim 4\%$. Thus the spread in the predictions of the 
different PDF sets is 
comparable to the uncertainty estimated by the individual analyses. 
Since the measurable rapidity range is restricted to central rapidity, it is 
more prudent to use these uncertainty estimates when considering if $W,Z$
cross sections can be used as luminosity monitors. Comparing the results from
the three PDF extractions it seems reasonable to use the generous estimate of the
CTEQ6.1 analysis, $8\%$, as an estimate of how well the luminosity could be
measured, at the present level of uncertainty. We subject this estimate to 
some further reality checks below and in 
Sec.~\ref{subsec:lowx;amcs_reality} and 
we discuss the possibility of improving this estimate with early LHC data in
Sec.~\ref{subsec:lowx;amcs_improve}

\begin{figure}[tbp] 
%\vspace*{5pt}
%\vspace{-1.0cm}
\centerline{
\epsfig{figure=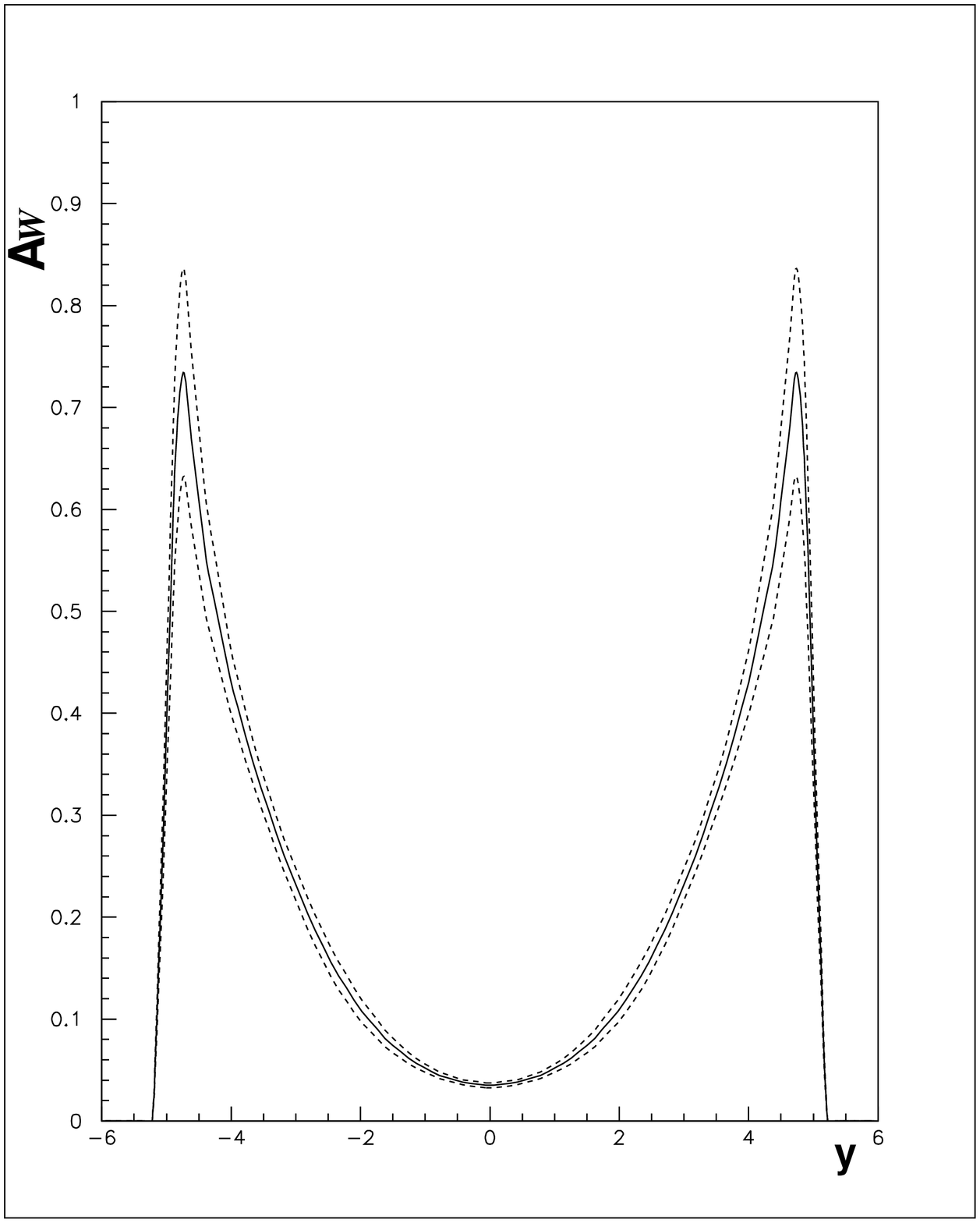,width=0.3\textwidth,height=4.5cm}
\epsfig{figure=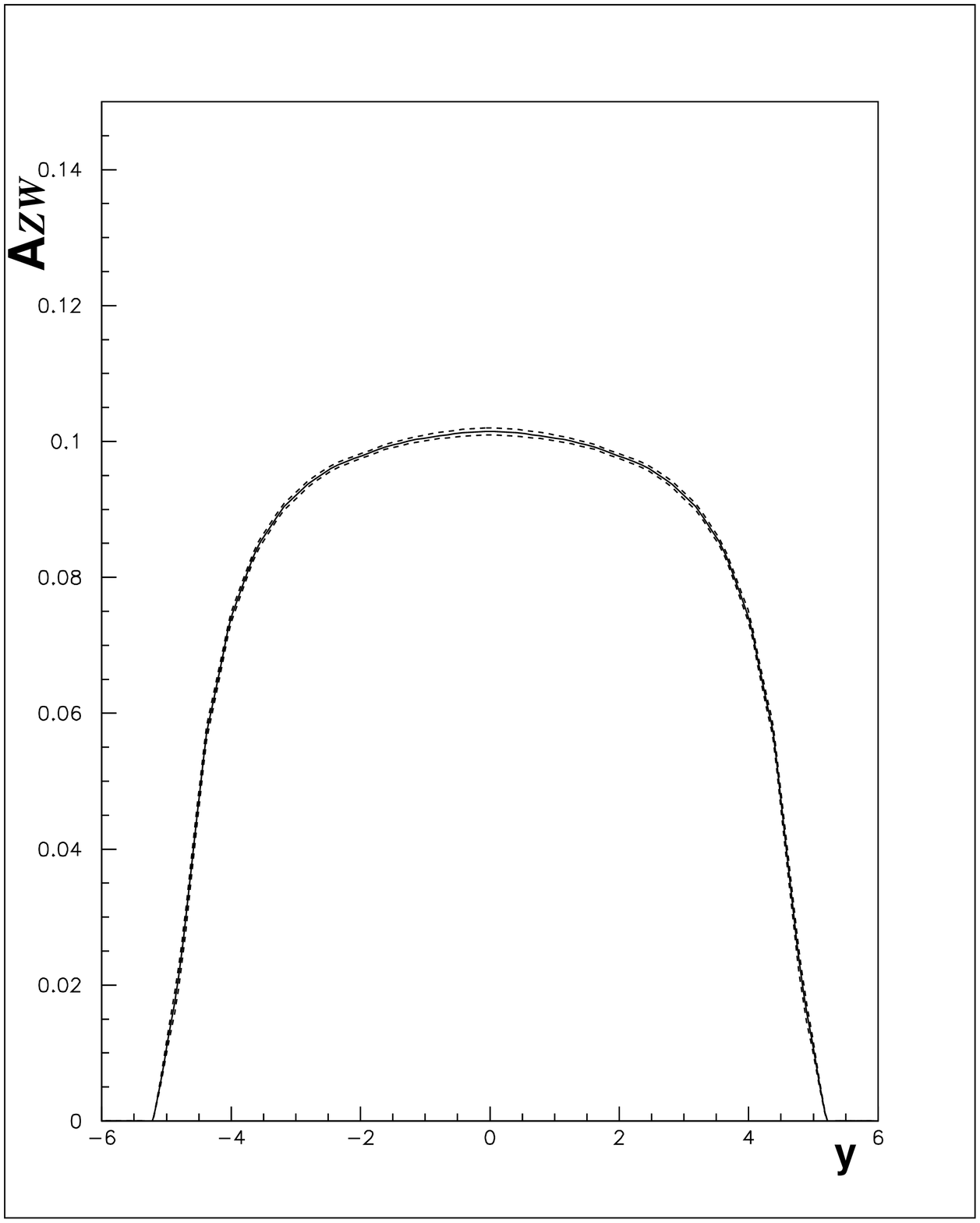,width=0.3\textwidth,height=4.5cm}
\epsfig{figure=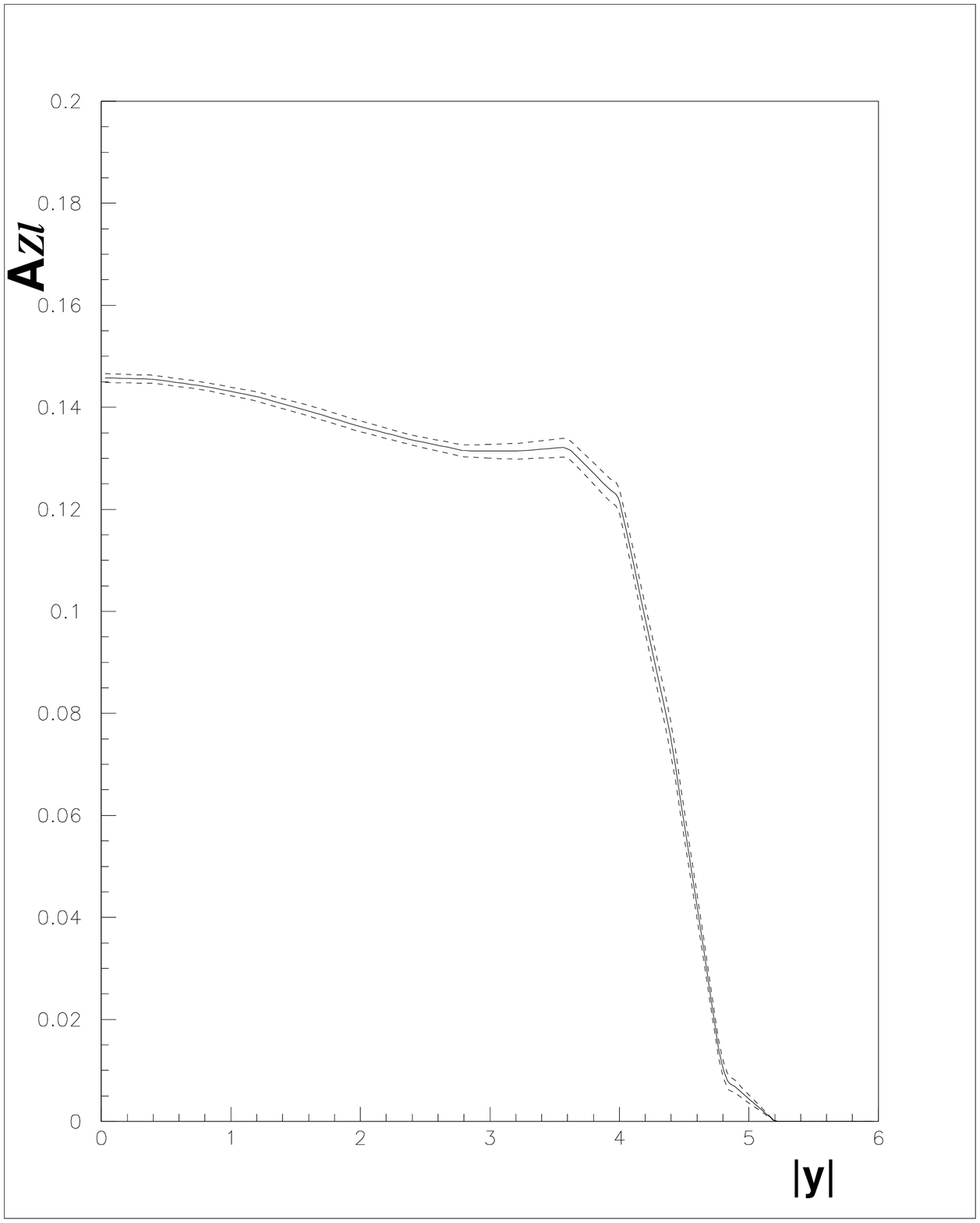,width=0.3\textwidth,height=4.5cm}
}
\centerline{
\epsfig{figure=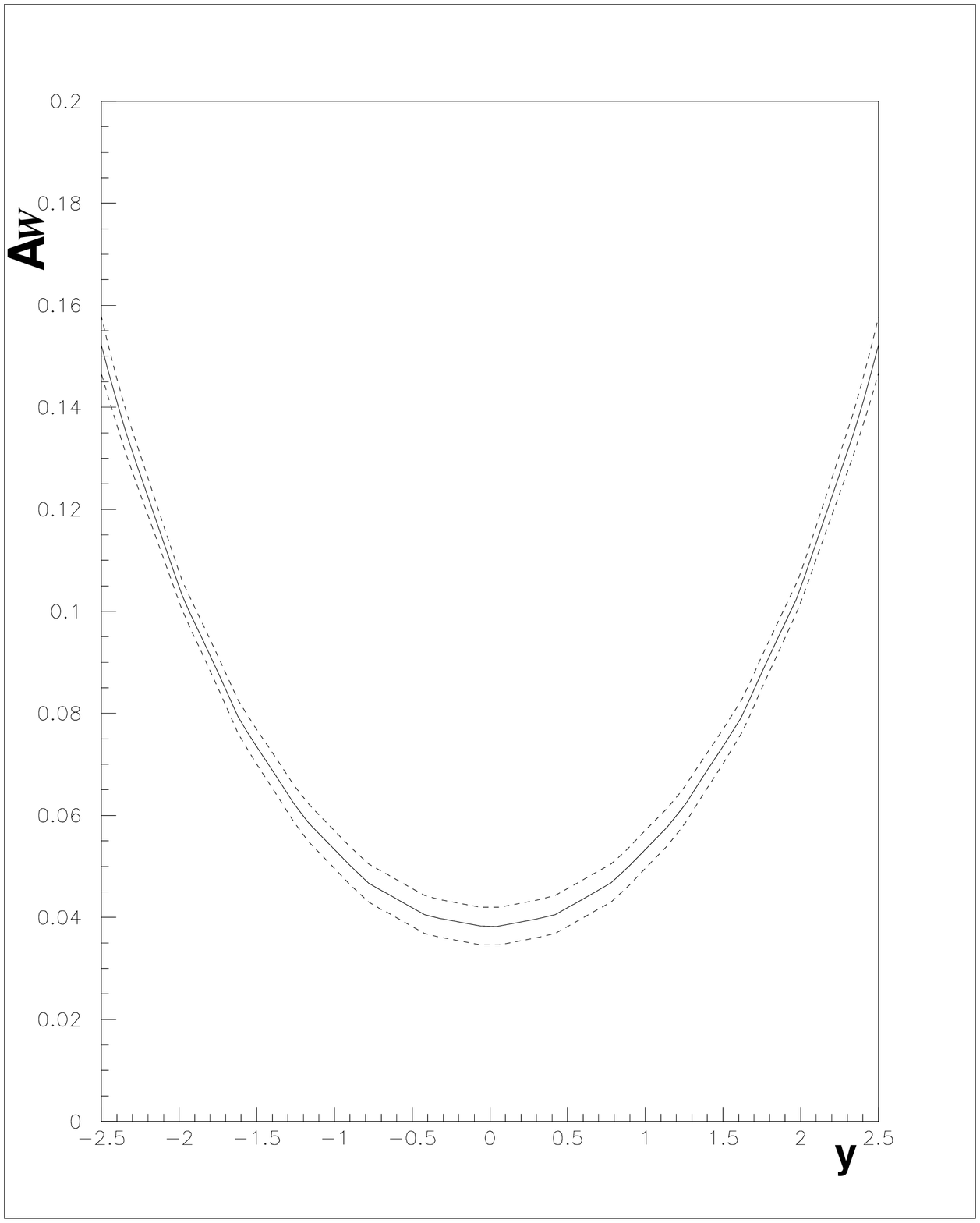,width=0.3\textwidth,height=4.5cm}
\epsfig{figure=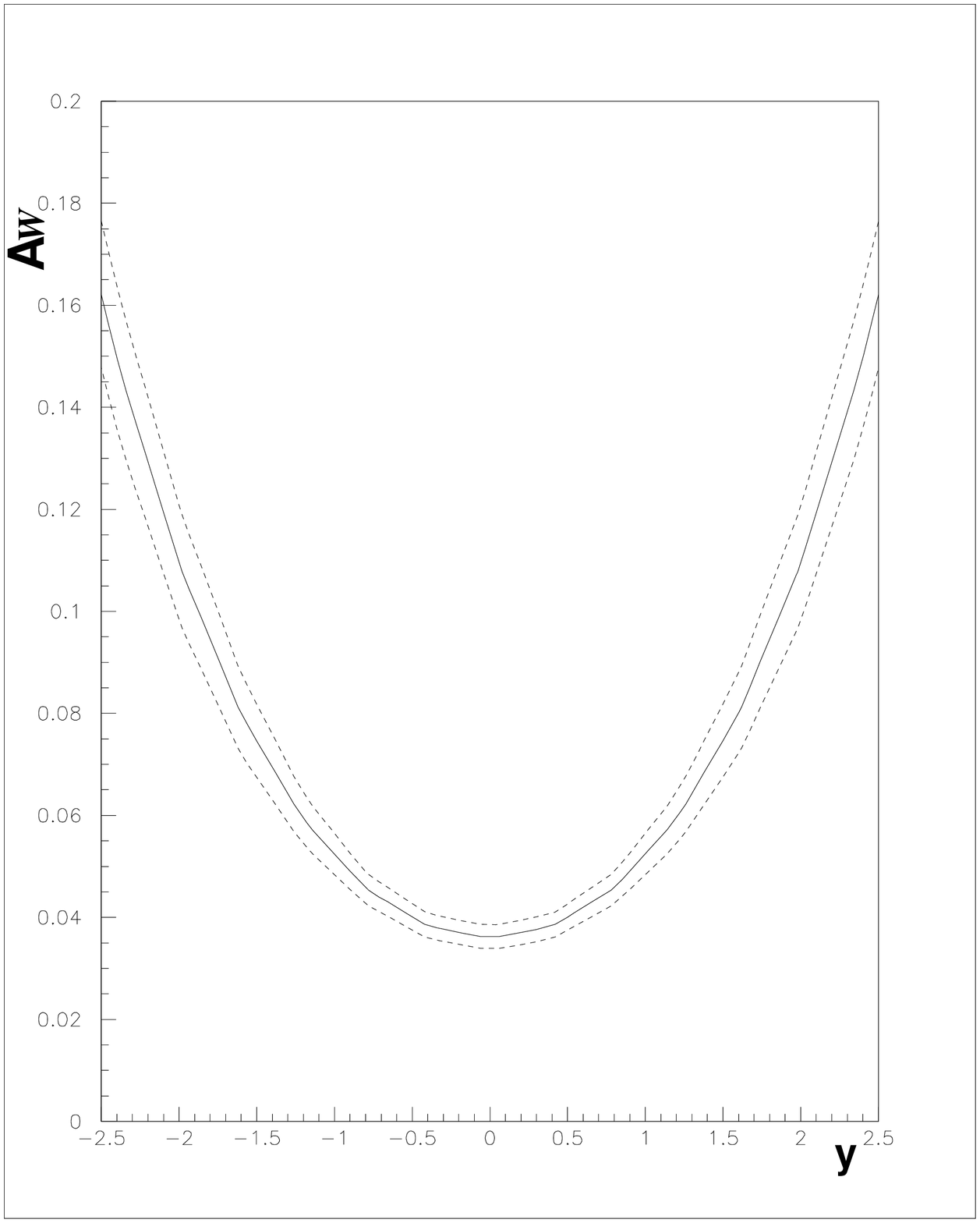,width=0.3\textwidth,height=4.5cm}
\epsfig{figure=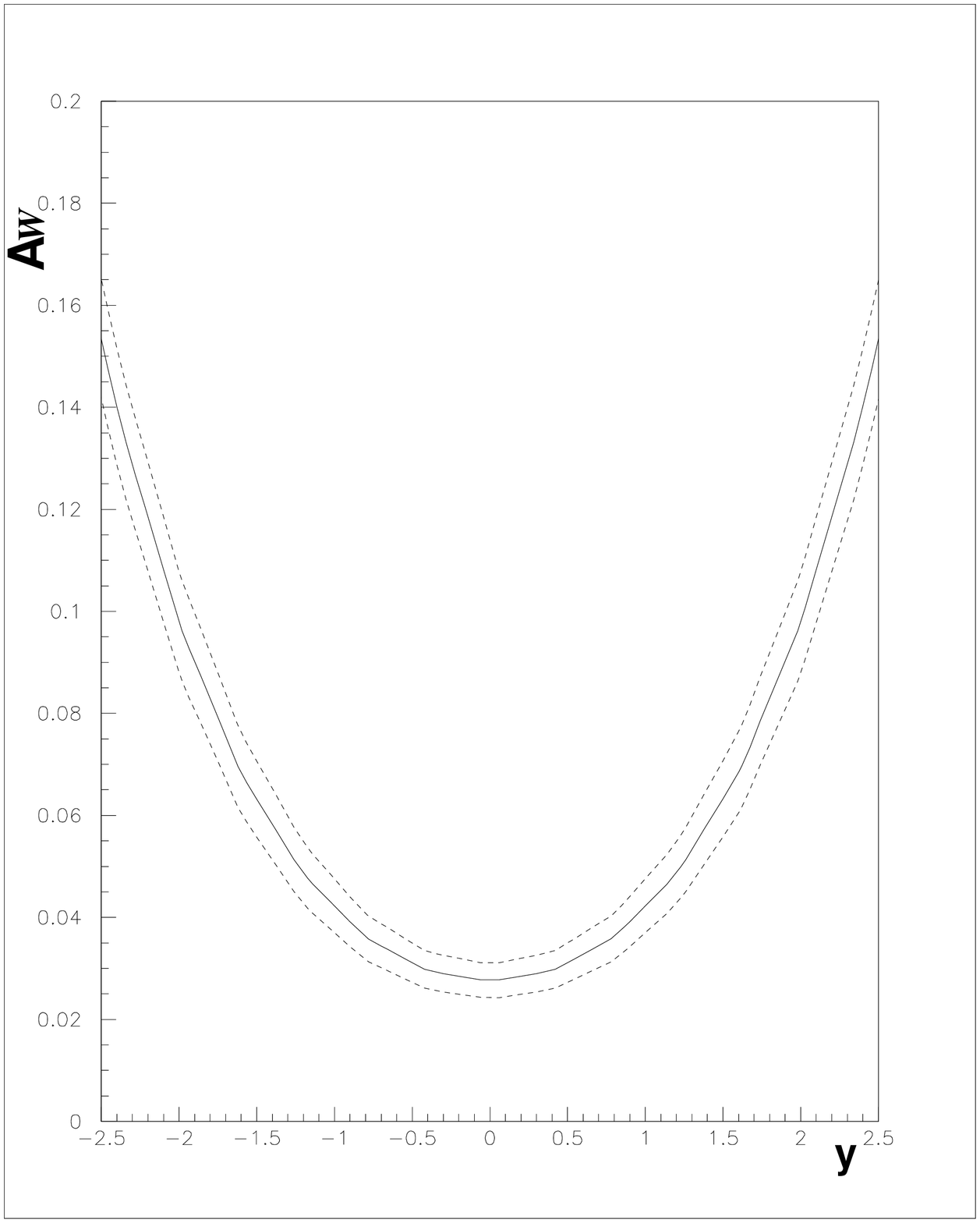,width=0.3\textwidth,height=4.5cm} 
}
\caption {
Top row: predictions from the CTEQ6.1 PDFs: left plot: the $W$ asymmetry 
$A_W$; middle plot: the ratio $A_{ZW}$; right plot: the ratio $A_{Zl}$.  
Bottom row: the $W$ asymmetry $A_W$ 
within the measurable rapidity range, as predicted using different PDF 
analyses: left plot: ZEUS-S; middle plot: CTEQ6.1; right plot: MRST01.}
\label{fig:awzw}
\end{figure}

Since the PDF uncertainty feeding into the $W^+, W^-$ and $Z$ production is 
mostly coming from the
gluon PDF for all three processes, there is a correlation in their 
uncertainties, which can be 
removed by taking ratios. The upper half of 
Fig.~\ref{fig:awzw} shows the $W$ asymmetry 
\[A_W = (W^+ - W^-)/(W^+ + W^-).\] for CTEQ6.1 PDFs. 
The PDF uncertainties on the asymmetry at central rapidity are about
$~5\%$, smaller than those on the $W$ spectra themselves, and
a PDF eigenvector decomposition indicates that 
sensitivity to $u$ and $d$ quark flavour distributions is now evident.
Even this residual flavour 
sensitivity can be removed by taking the ratio \[A_{ZW} = Z/(W^+ +W^-)\] 
as also shown in Fig.~\ref{fig:awzw}. 
This quantity is almost independent of PDF uncertainties, which are now as 
small as $~0.5\%$, within the CTEQ6.1 PDF analysis. 

However, as before, 
it is necessary to compare these quantities between different PDF 
analyses. The variation in the predictions for the ratio $A_{ZW}$ between PDF 
analyses (MRST01, ZEUS-S and Alekhin02 PDFs have been compared to CTEQ6.1) is 
outside the PDF uncertainty estimates of the different analyses, but it is 
still only $\sim 5\%$. Hence this ratio could be a used as an SM
benchmark measurement. The ratio $A_W$ shows
a much more striking difference between MRST01 PDFs and the others. 
This is illustrated in the lower half of 
Fig.~\ref{fig:awzw} for the ZEUS-S, CTEQ6.1 and MRST01 PDFs, 
in the measurable rapidity range. There is a difference of $\sim 25\%$ in the 
predictions.  The origin of this difference between MRST and other PDFs is in the valence spectra. At leading order, the dominant contribution to $A_W$ is
\begin{equation}
 A_W = \frac{u\bar{d} - d \bar{u}}{u\bar{d} + d \bar{u}}.
\end{equation}
At central rapidity, $x \sim 0.005$, for both partons and consequently $\bar{u} \approx \bar{d}$ \footnote{Even if some fairly wild assumptions as to the 
shape of $\bar{d}- \bar{u}$ are made for low $Q^2$, 
the absolute size of 
$\bar{q}$ evolves with $Q^2$ to become very large at $Q^2 = M_W^2$, whereas 
the difference does not evolve, and becomes relatively small.}. Thus
\begin{equation}
 A_W = \frac{u - d}{u + d} = \frac{u_v - d_v}{u_v + d_v +2 \bar{q}}
\end{equation}
and $A_W$ depends on the difference of the valence quarks. 
The quantity 
$u_v- d_v$, is different for MRST and CTEQ, and this difference is outside 
the PDF uncertainty estimates of either
analysis. However, these uncertainty estimates are themselves 
unreliable for valence spectra 
at $x \sim 0.005$, since there is no data on valence quantities at such small 
$x$. The LHC can provide the first such measurement.    

\begin{figure}[tbp] 
%\vspace*{5pt}
\vspace{-1.0cm}
\centerline{
\epsfig{figure=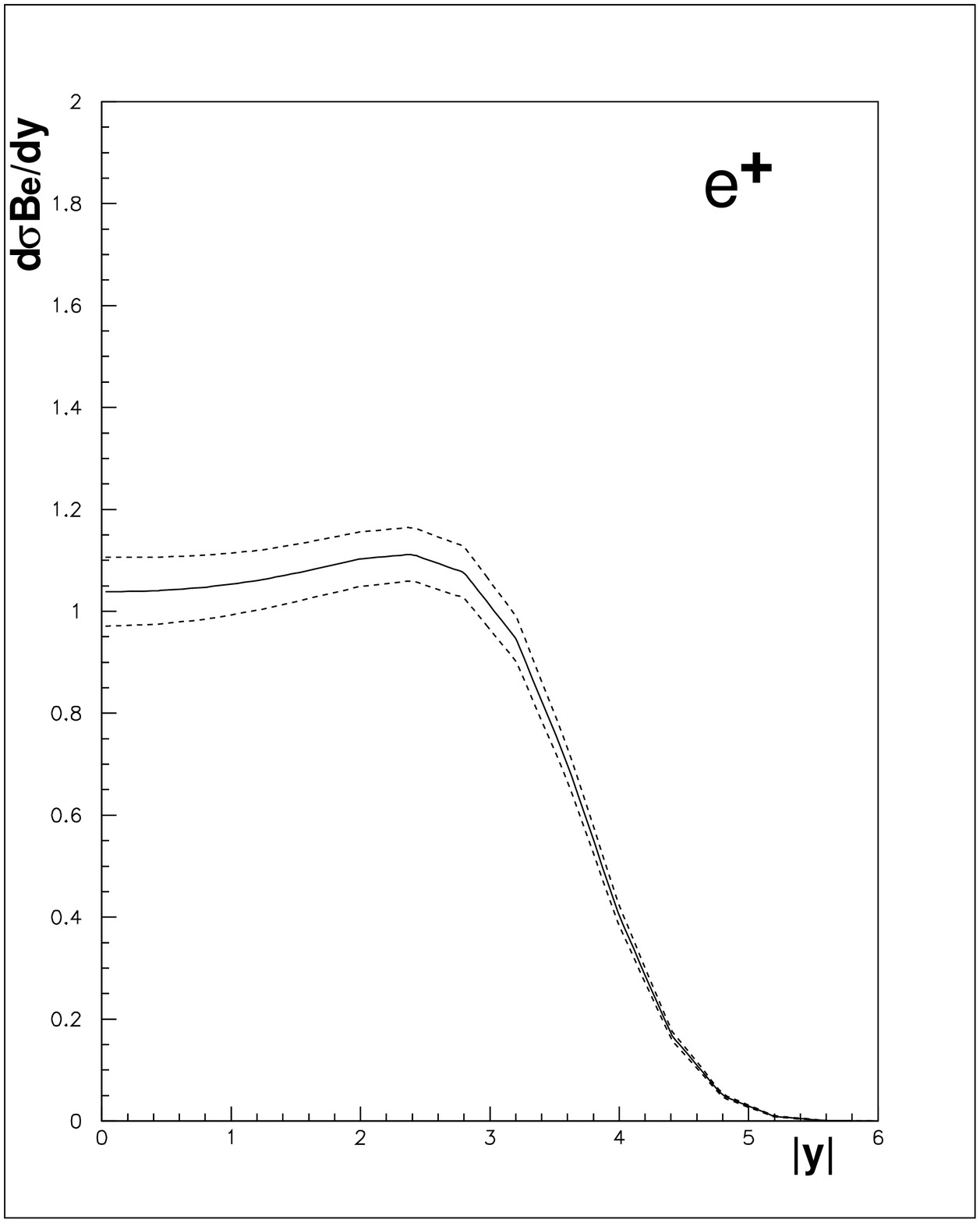,width=0.3\textwidth,height=4.5cm}
\epsfig{figure=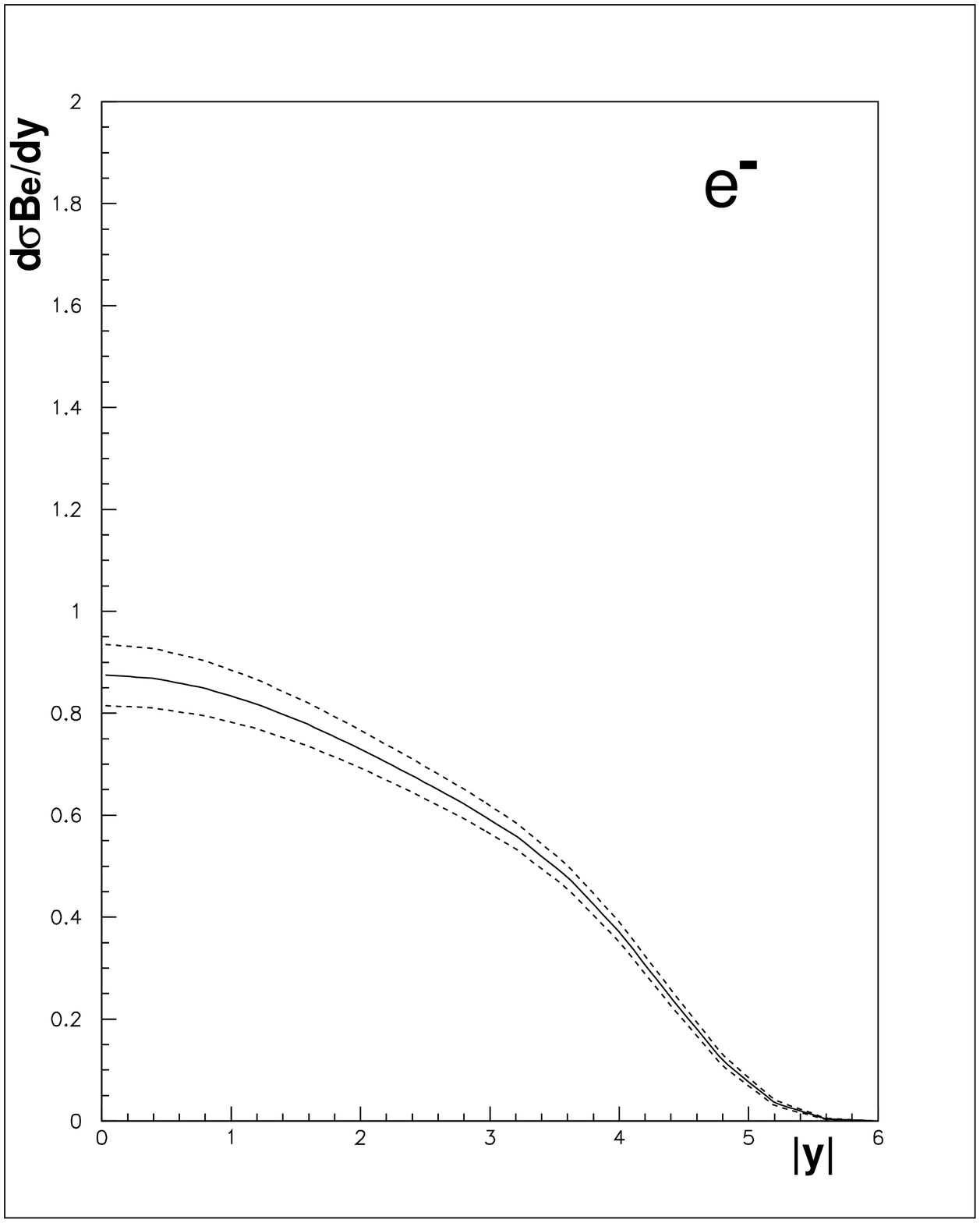,width=0.3\textwidth,height=4.5cm}
\epsfig{figure=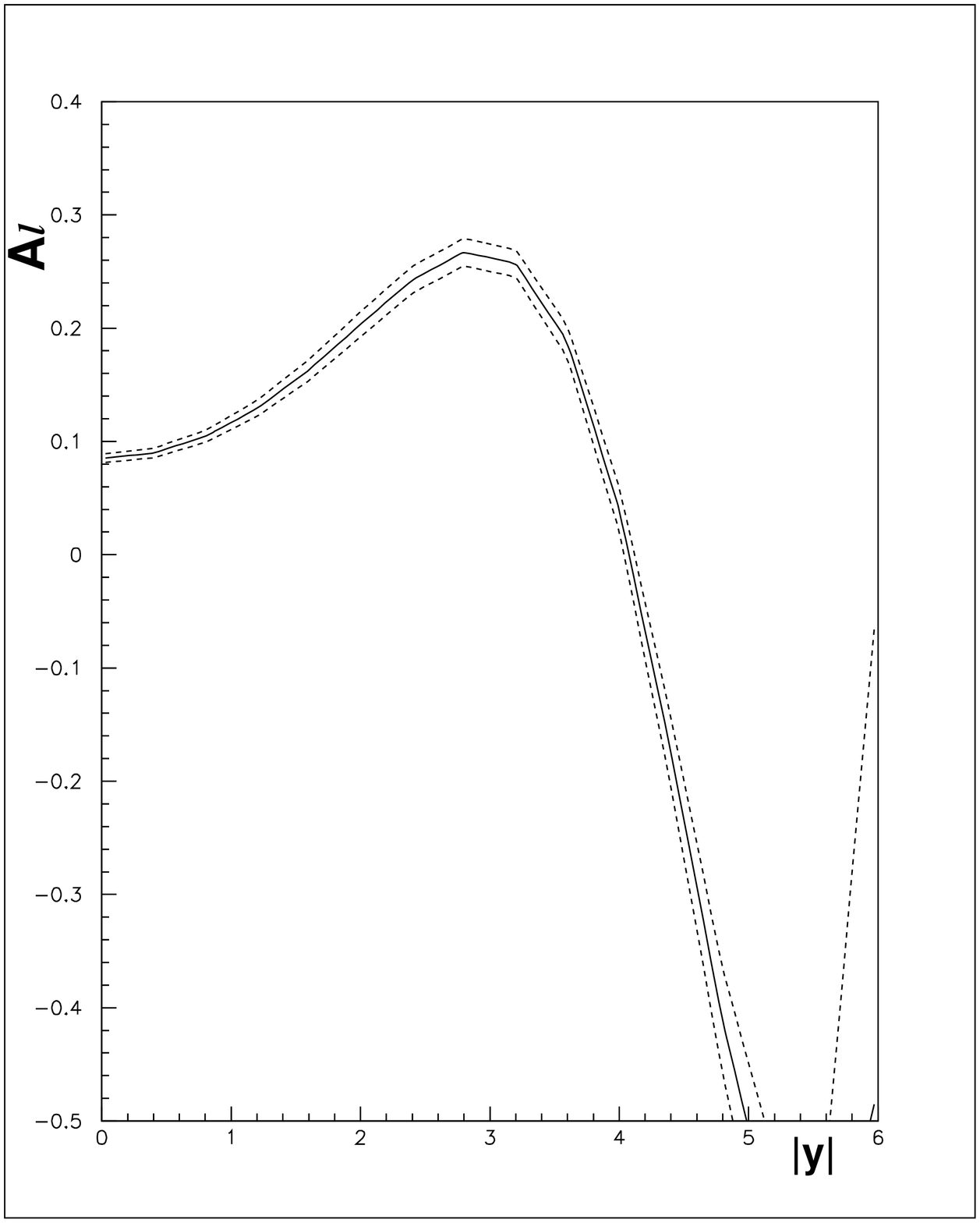,width=0.3\textwidth,height=4.5cm}
}
\centerline{
\epsfig{figure=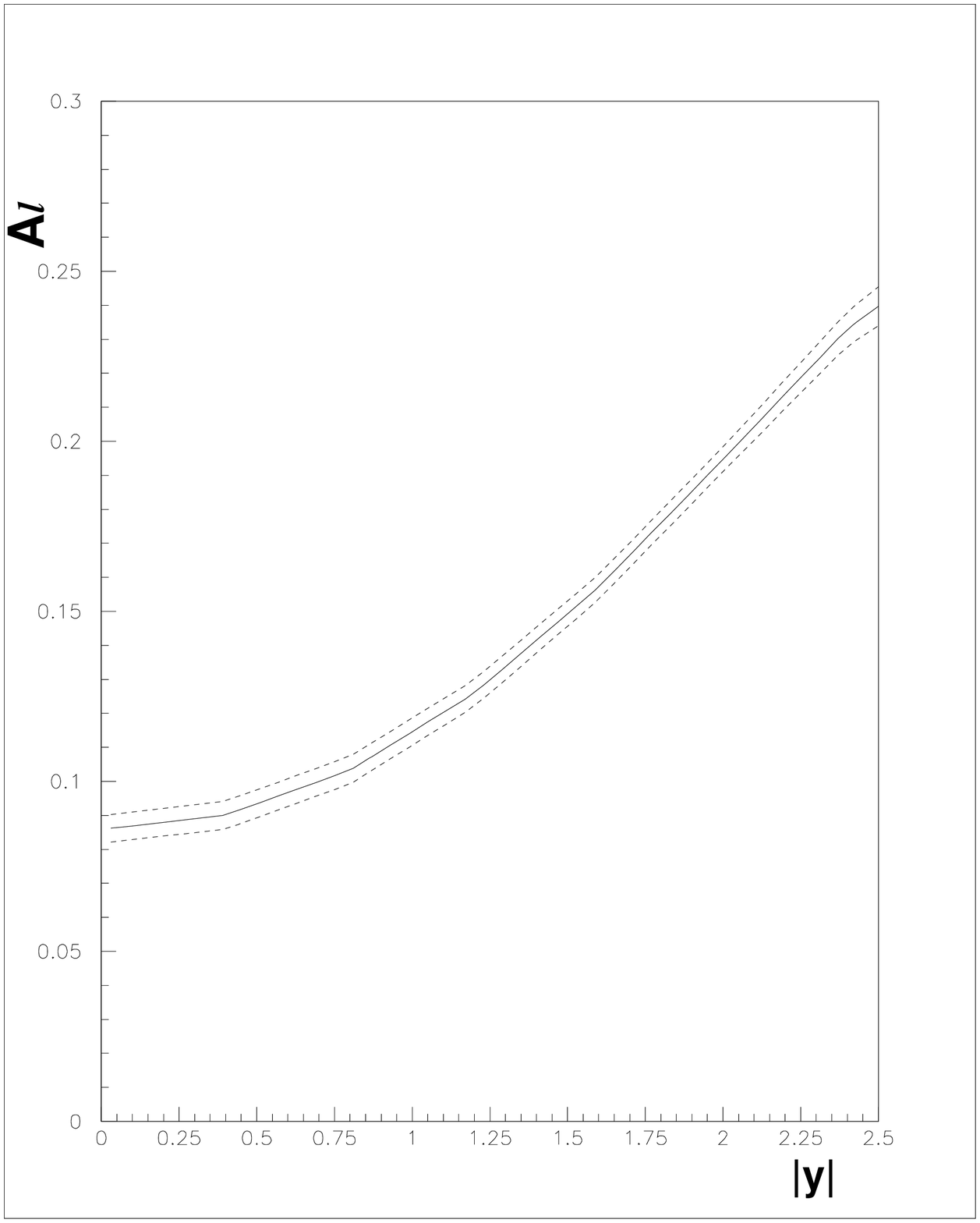,width=0.3\textwidth,height=4.5cm}
\epsfig{figure=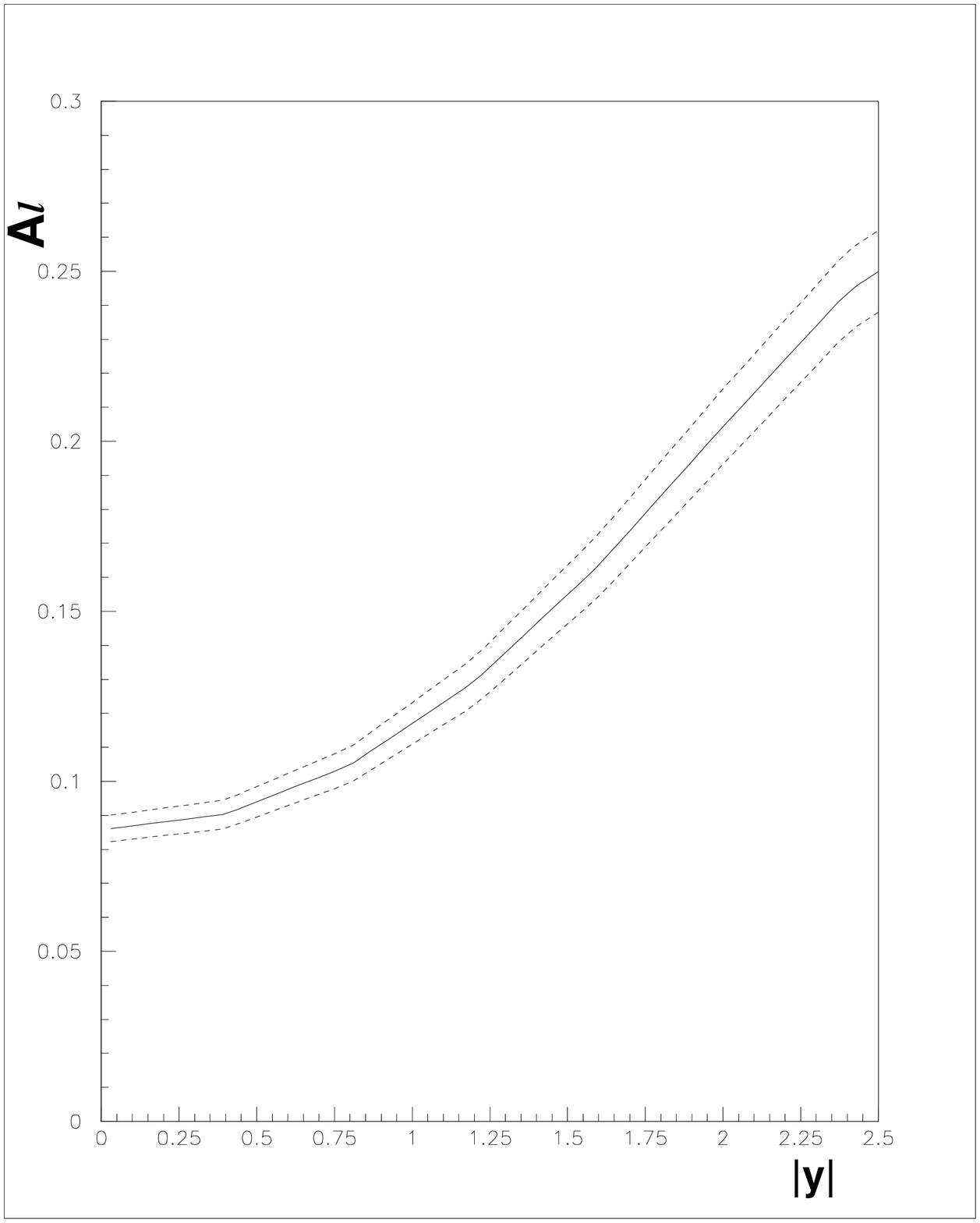,width=0.3\textwidth,height=4.5cm}
\epsfig{figure=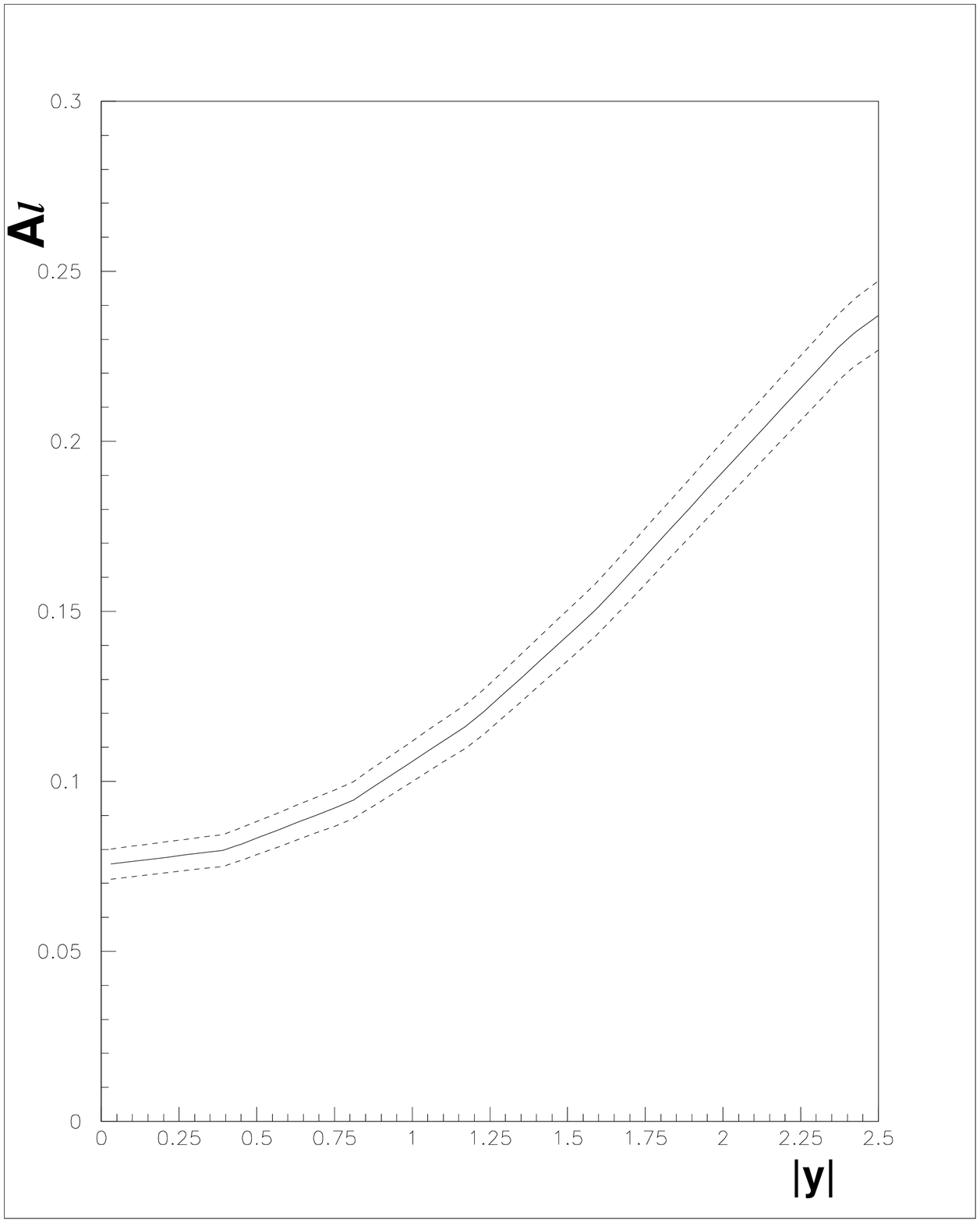,width=0.3\textwidth,height=4.5cm}
}
\caption {Top row: lepton spectra from the CTEQ6.1 PDFs; left plot: 
decay $e^+$ rapidity spectrum; middle plot: decay $e^-$ rapidity spectrum; 
right plot: lepton asymmetry $A_l$. Bottom row: the lepton asymmetry 
$A_l$ from different PDF analyses: left plot: ZEUS-S; 
middle plot: CTEQ6.1; right plot: MRST01.}
\label{fig:leptons}
\end{figure}

In order to assess, if LHC measurements will actually be discriminating, we must first 
account for the fact that
$W$ bosons decay and are most easily detected from their leptonic final states. 
Thus we actually measure the decay lepton 
rapidity spectra rather than the $W$ rapidity spectra. The upper half of  
Fig.~\ref{fig:leptons} shows these rapidity spectra for positive and 
negative leptons from $W^+$ and $W^-$ decay together with
 the lepton asymmetry, 
\[A_l = (l^+ - l^-)/(l^+ + l^-)\] for the CTEQ6.1 PDFs. 
A cut of, $p_{t} > 25~$GeV, 
has been applied on the decay lepton, since it will not be possible to 
identify leptons with small $p_{t}$. 
A particular lepton rapidity can be fed from a range 
of $W$ rapidities so that the contributions of partons at different $x$ 
values are smeared out 
in the lepton spectra. Nevertheless, the broad features 
of the $W$ spectra and the 
sensitivity to the gluon parameters are reflected in the lepton spectra,
resulting in a similar estimate ($\sim 8\%$) of PDF uncertainty  
at central rapidity for the CTEQ6.1 PDFs. 
The lepton asymmetry shows the change of sign at large $y$ which is 
characteristic of the $V-A$ 
structure of the lepton decay. The cancellation of the 
uncertainties due to the gluon PDF is not so 
perfect in the lepton asymmetry as in the $W$ asymmetry. Even so, in the 
measurable rapidity range, the PDF uncertainty in the asymmetry is smaller 
than in the lepton spectra, being $\sim 5\%$, for the 
CTEQ6.1 PDFs. The $Z$ to $W$ ratio $A_{ZW}$ has 
also been recalculated as a $Z$ to leptons ratio, \[A_{Zl} = Z/(l^+ +l^-)\] 
illustrated in Fig.~\ref{fig:awzw}. Just as for $A_{ZW}$, the overall 
uncertainty in $A_{Zl}$ is very small ($\sim 0.5\%$) for CTEQ6.1 PDFs.

It is again necessary to consider the difference between different 
PDF analyses for the predictions of the lepton spectra, $A_{Zl}$ and $A_l$. 
For the lepton spectra, the spread in the predictions of the 
different PDF analyses of MRST01, CTEQ6.1 and ZEUS-S is 
comparable to the uncertainty estimated by the individual analyses, just as 
for the $W$ spectra, and this is shown later in Fig.~\ref{fig:gendet}.  
Just as for $A_{ZW}$, there
are greater differences in the predictions for $A_{Zl}$ between PDF analyses 
than within any PDF analysis, but these differences remain within $\sim 5\%$
preserving this quantity as an SM benchmark measurement. Thus our 
previous estimate of the usefulness of these processes as luminosity monitors
and SM benchmarks
survives the reality check of the fact that we will measure the leptons, not
the $W$ bosons.

The significant differences which we noticed between the 
predictions of the different PDF analyses for $A_W$, remain in the predictions
for $A_l$. The lower half of Fig.~\ref{fig:leptons} 
compares these predictions for the ZEUS-S PDFs with those of 
the CTEQ6.1 PDFs and the MRST01 PDFs, in the measurable rapidity range.
The discrepancy of $\sim 25\%$ which was found in $A_W$ has been somewhat 
diluted to $\sim 15\%$ in $A_l$, but this should still be large enough for LHC 
measurements to discriminate, and hence to give information on the low-$x$
valence distributions. 

\subsubsection{How well can we actually measure $W$ spectra at the LHC?}
\label{subsec:lowx;amcs_reality}

The remainder of this contribution will be concerned with the question: 
how accurately can we  
measure the lepton rapidity spectra and can we use the early LHC data 
to improve on the current level of uncertainty?

We have simulated one million signal, $W \rightarrow e \nu_e$, events for 
each of the PDF sets CTEQ6.1, MRST2001 and ZEUS-S using HERWIG (6.505). 
For each of these PDF sets the eigenvector error PDF sets have been simulated 
by PDF re-weighting and k-factors have been applied to approximate an NLO generation. A study has been made of the validity of both PDF re-weighting and 
k-factor re-weighting and this is reported in ref.~\cite{Tricoli:2005nx}. The 
conclusion is that PDF re-weighting is valid for re-weighting the rapidity 
spectra when the PDF sets are broadly similar, as they are within any one PDF 
analysis. The k-factor re-weighting to simulate NLO is also valid for the 
rapidity spectra for which it was designed.

\begin{figure}[tbp] 
%\vspace*{5pt}
\vspace{-1.0cm}
\centerline{
\epsfig{figure=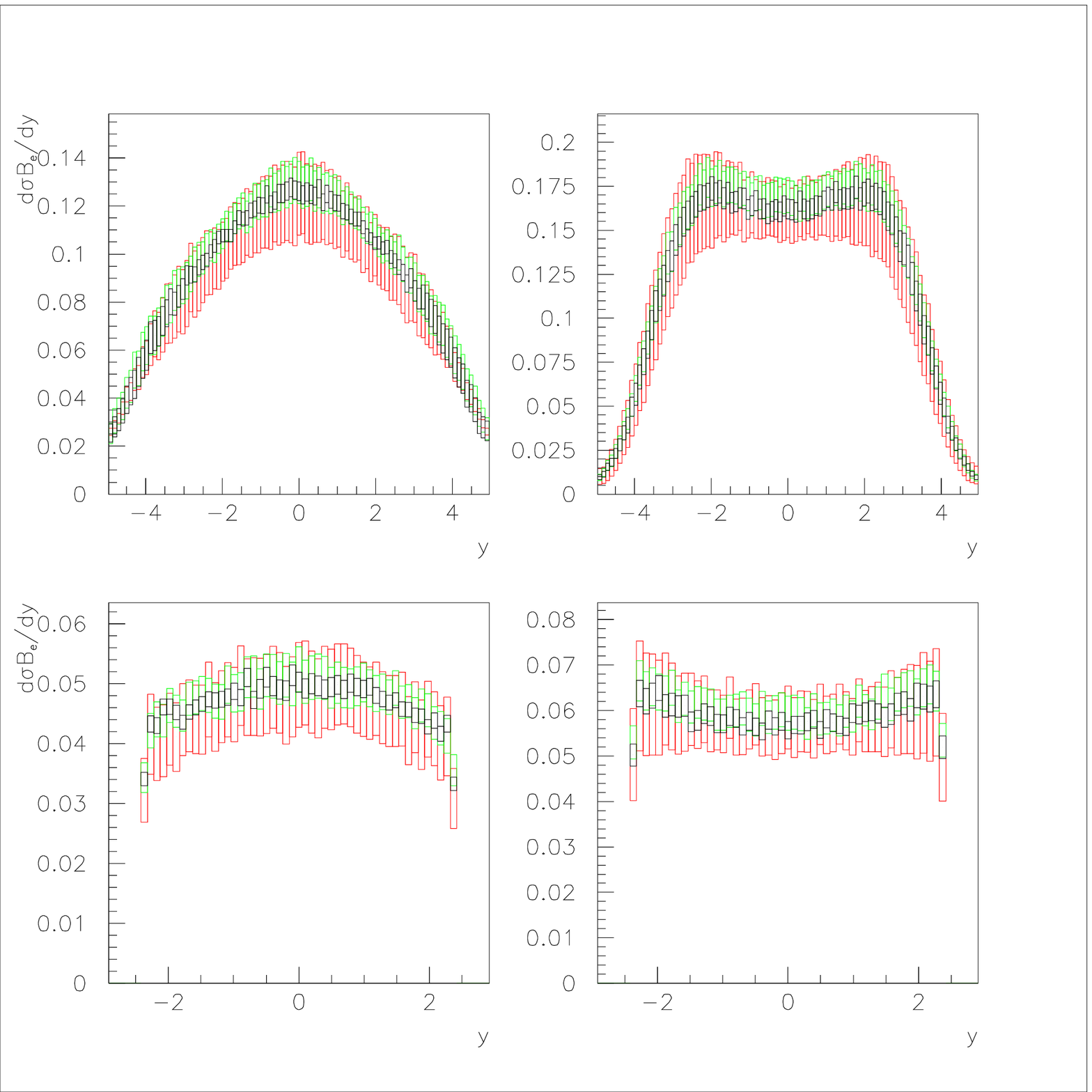,width=0.66\textwidth,height=9cm }
\epsfig{figure=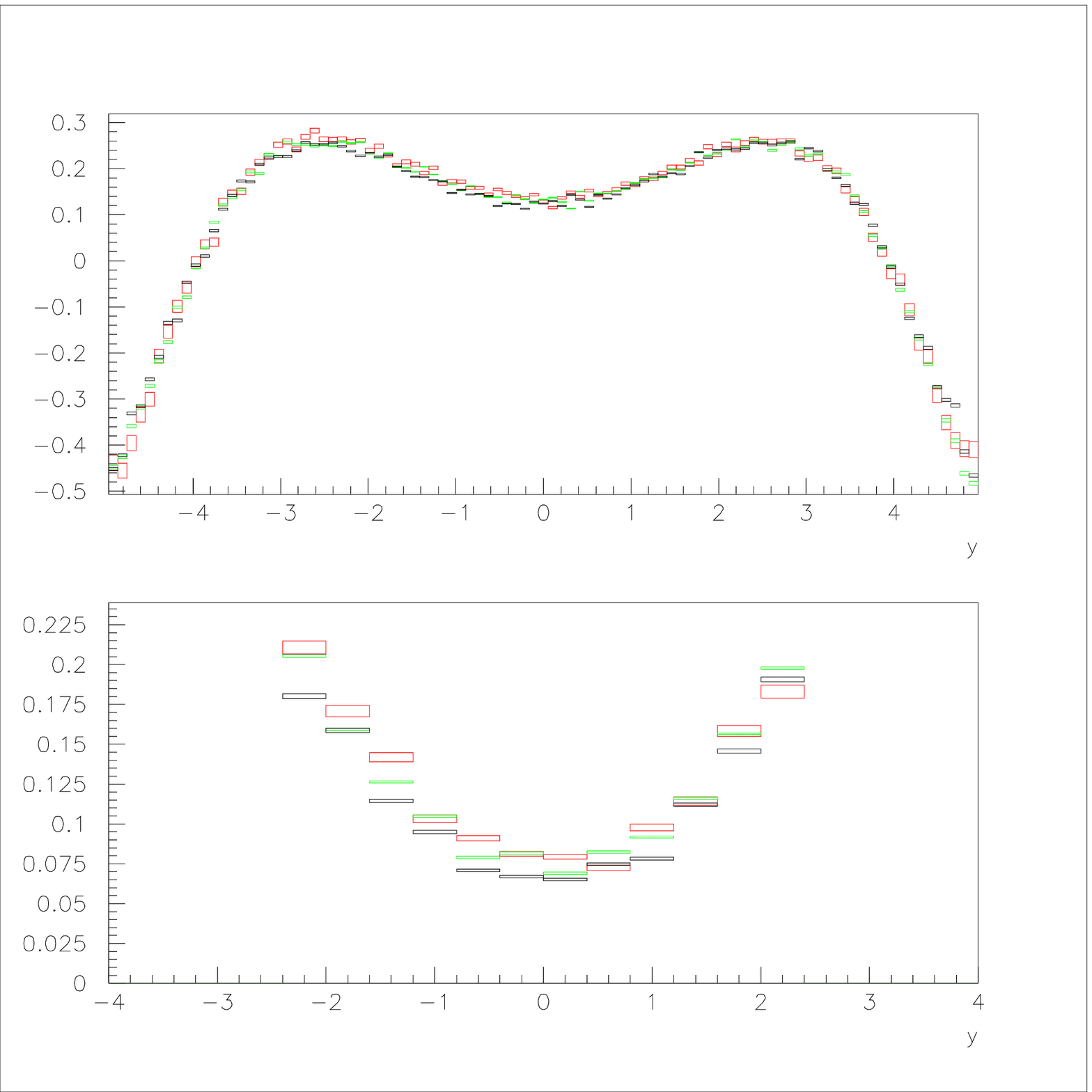,width=0.33\textwidth,height=9cm}
}
\caption {Top row: $e^-$, $e^+$ and $A_e$ rapidity spectra for the lepton from the $W$ decay, 
generated using HERWIG + k factors and CTE6.1 (red),
ZEUS-S (green) and MRST2001 (black) PDF sets with full uncertainties. Bottom row: the same spectra after passing 
through the ATLFAST detector simulation and selection cuts.}
\label{fig:gendet}
\end{figure}

The top part of Fig.~\ref{fig:gendet}, shows the $e^{\pm}$ and $A_l$ 
spectra at the generator level, for all of the PDF sets superimposed. As mentioned before, it is 
clear that the lepton spectra as predicted by the different PDF analyses are 
compatible, within the PDF uncertainties of the analyses.
The events are then passed through ATLFAST, the fast simulation of the ATLAS 
detector. This applies loose kinematic cuts: 
$|\eta| < 2.5$, $p_{te} > 5\ GeV$, and electron isolation criteria. 
It also smears the 4-momenta of the 
leptons to mimic momentum dependent detector resolution. 
We then apply further cuts, designed to eliminate the background preferentially. These criteria are:
\begin{itemize}
\item pseudorapidity, $|\eta| <2.4$, to avoid bias at the edge of the measurable rapidity range
\item  $p_{te} > 25\ GeV$, high $p_t$ is necessary for efficient electron identification
\item  missing $E_t > 25$~GeV, the $\nu_e$ in a signal event will have a correspondingly large missing $E_t$
\item  no reconstructed jets in the event with $p_t > 30\ GeV$, to discriminate against QCD background 
\item  recoil of the $W$ boson in the transverse plane $p_t^{recoil} < 20\ GeV$, to discriminate against QCD background
\end{itemize}
These cuts ensure that background from the processes: 
$W \rightarrow \tau \nu_\tau$; $Z \rightarrow \tau^+ \tau^-$; and $Z \rightarrow e^+ e^-$, is negligible ($\sxleqsim 1\%$)~\cite{Tricoli:2005nx}. Furthermore, 
a study of charge misidentification has established that the lepton asymmetry 
will need only very small corrections ($\sxleqsim 0.5\%$), 
within the measurable rapidity range~\cite{Tricoli:2005nx}.

The lower half of
Fig.~\ref{fig:gendet}, shows the $e^{\pm}$ and $A_l$ spectra at the detector level after application 
of these cuts for all of the PDF sets superimposed. 
The level of precision of each PDF set, seen in the analytic calculations of Fig.~\ref{fig:mrstcteq}, appears 
somewhat degraded  at detector level, so that a net level of PDF 
uncertainty in the lepton spectra of $\sim 10\%$ is expected at central 
rapidity. Thus the usefulness of these processes as a luminosity monitor
is somewhat compromised if a measurement to better than $10\%$ is required.

The anticipated cancellation of PDF 
uncertainties in the asymmetry spectrum is observed, within each PDF set, 
such that the uncertainties predicted by each PDF set are $\sim 5\%$, but the 
spread between the MRST and CTEQ/ZEUS-S PDF sets is as large as $\sim 15\%$.
Thus measurements, which are accurate to about $\sim 5\%$, could provide useful
information on the valence distributions at low $x$.

\subsubsection{Using LHC data to improve precision of PDFs}
\label{subsec:lowx;amcs_improve}

We now consider the possibility of improving on the current level of PDF 
uncertainty by using LHC data itself.
 The high cross sections for $W$ production at 
the LHC ensure that it will be the experimental systematic errors, rather than the statistical errors, which 
are determining. Our experience with the detector simulation leads us to believe that a systematic precision of $\sim 5\%$ could be achievable. 
We have optimistically imposed a random  $4\%$
scatter on our samples of one million $W$ events, 
generated using different PDFs, in order to 
investigate if measurements at this level of precision will improve PDF 
uncertainties at central rapidity significantly, if they 
are input to a global PDF fit. 

\begin{figure}[tbp] 
%\vspace*{5pt}
\vspace{-1.5cm}
\centerline{
\epsfig{figure=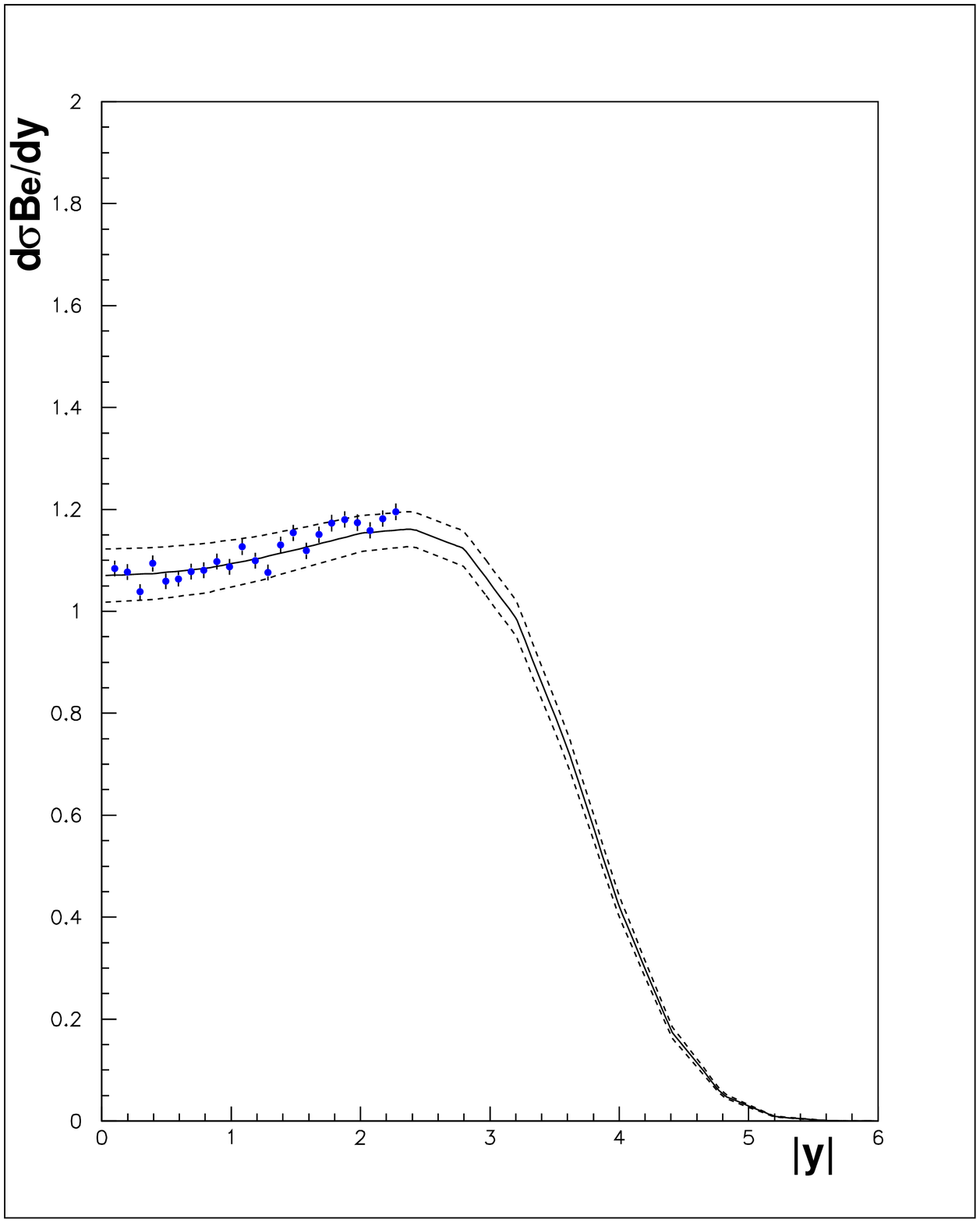,width=0.3\textwidth,height=4cm}
\epsfig{figure=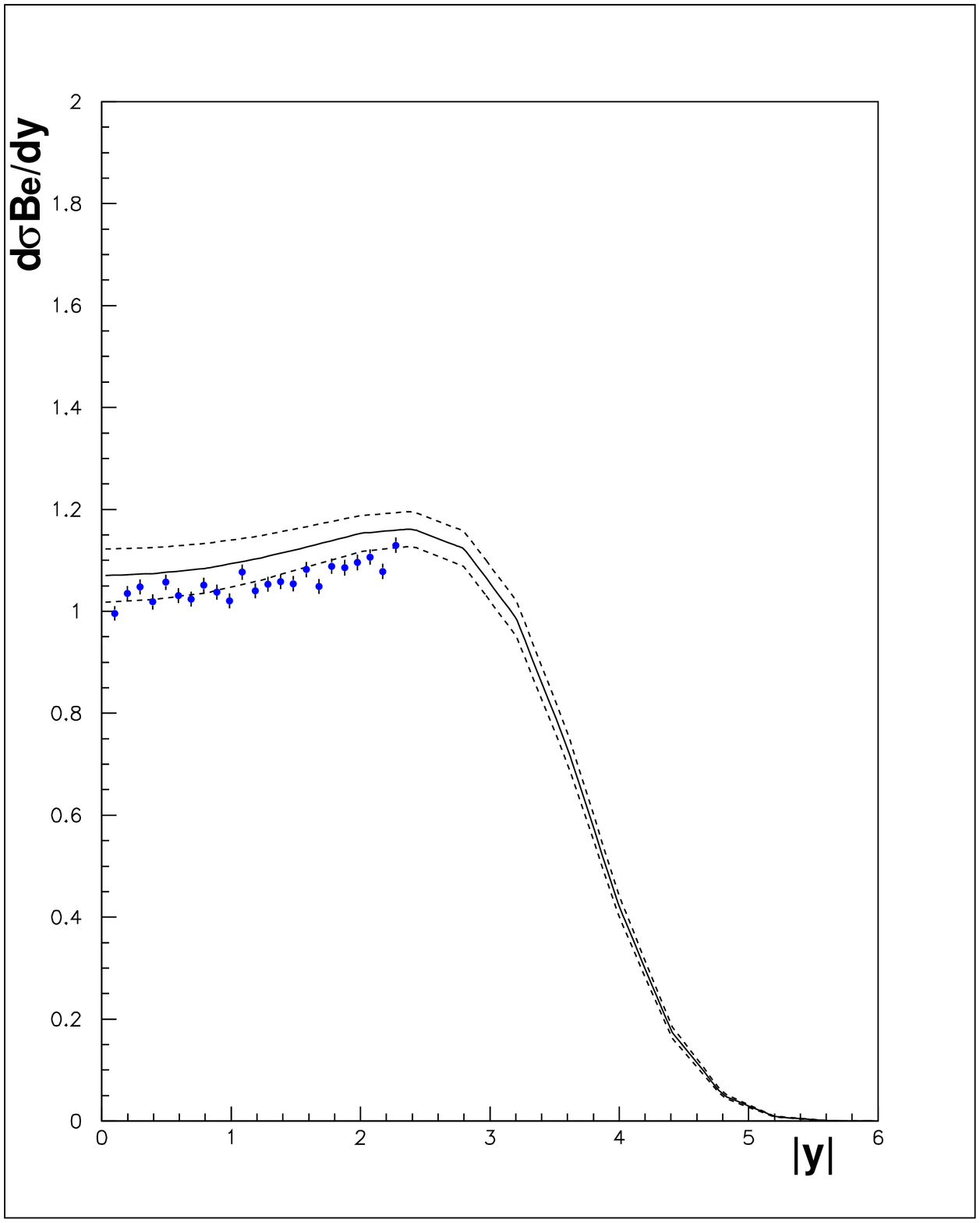,width=0.3\textwidth,height=4cm}
\epsfig{figure=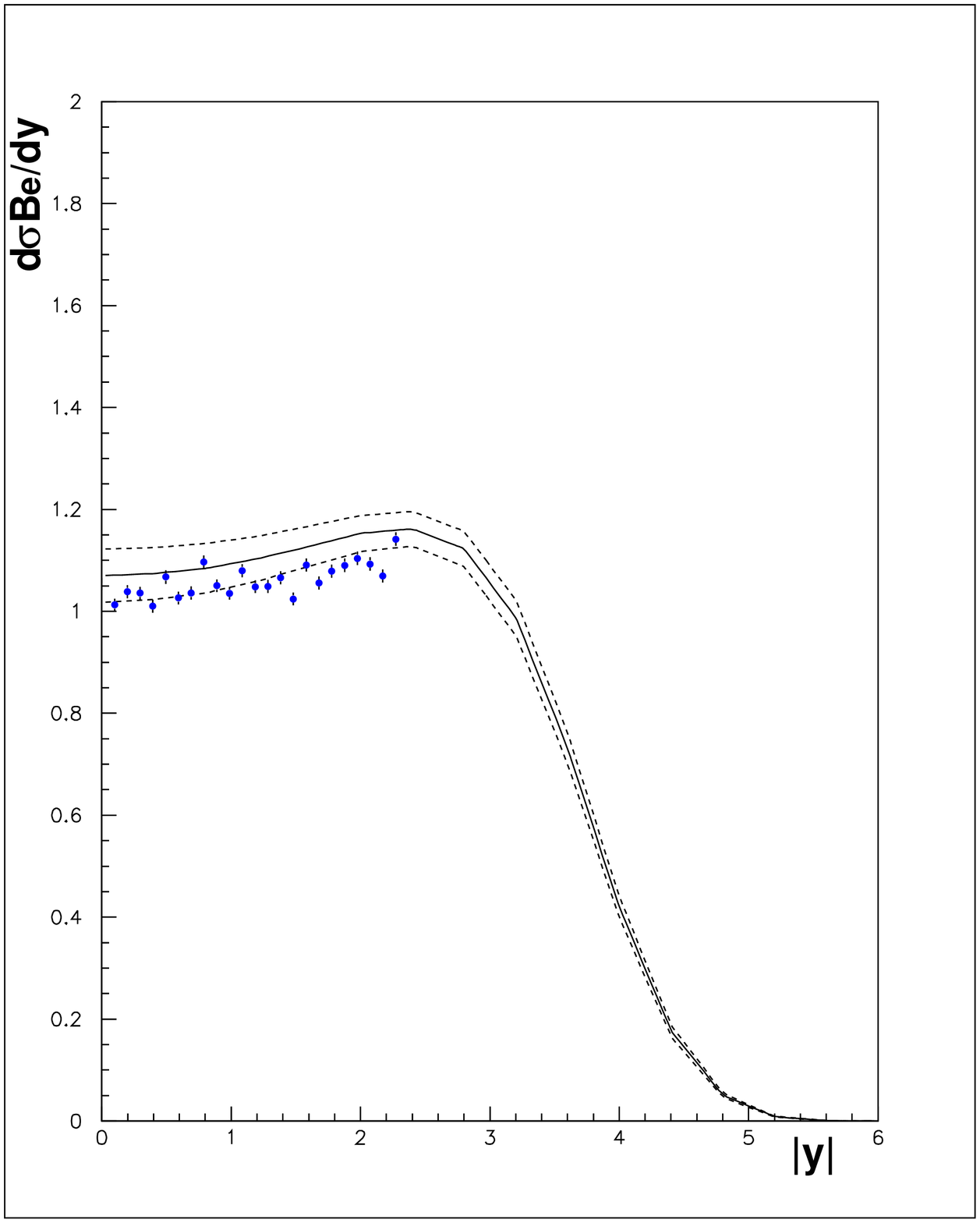,width=0.3\textwidth,height=4cm}
}
\centerline{
\epsfig{figure=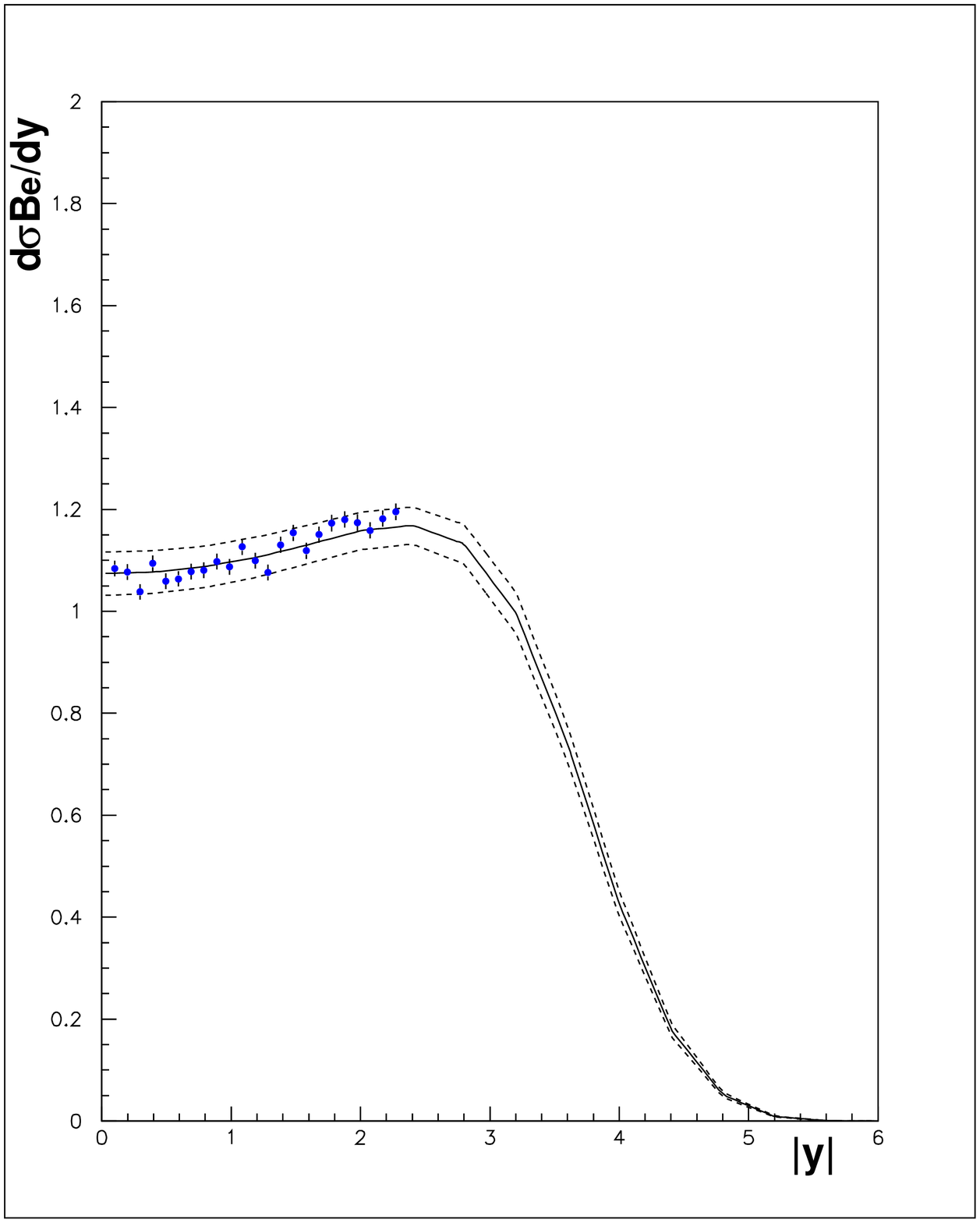,width=0.3\textwidth,height=4cm}
\epsfig{figure=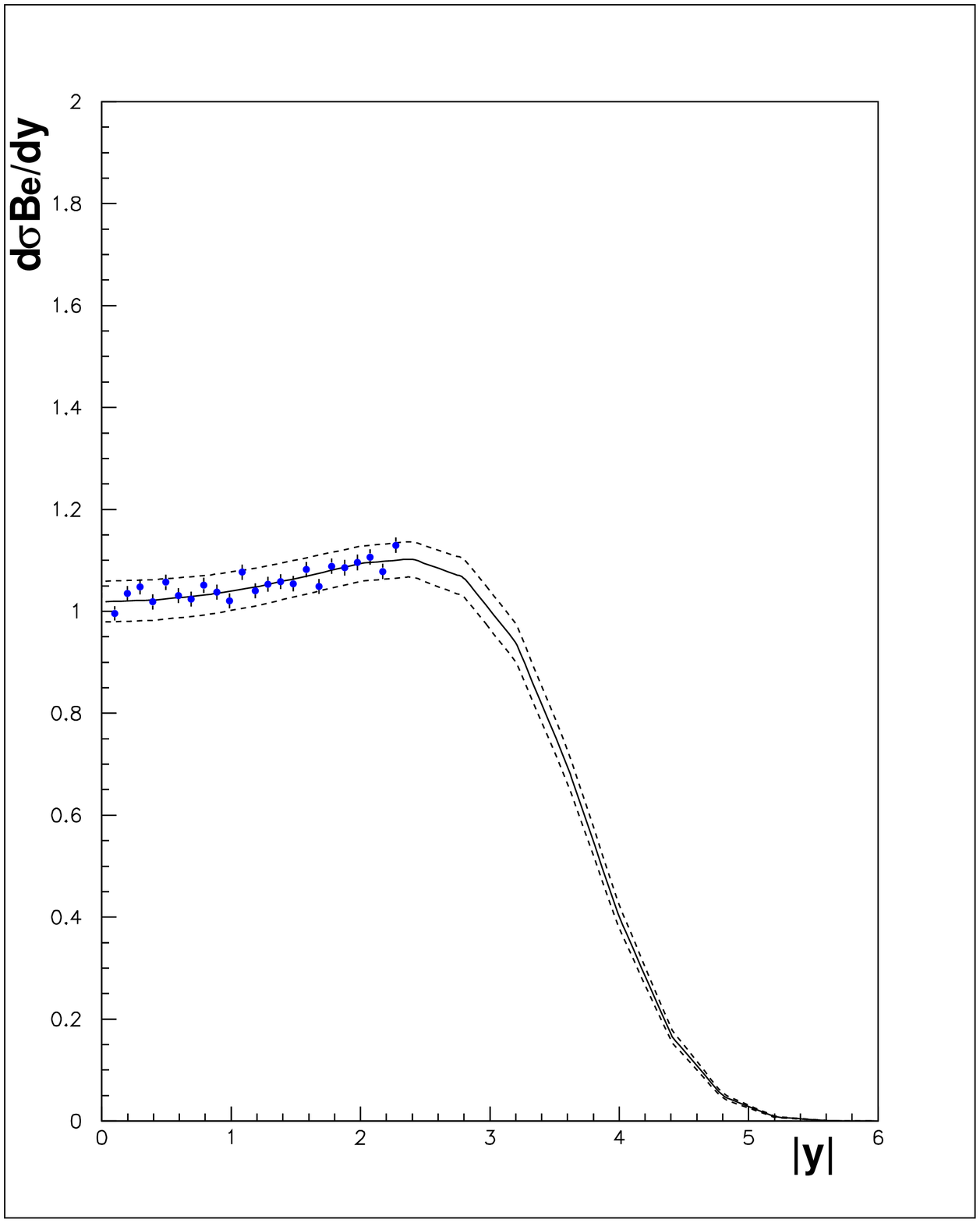,width=0.3\textwidth,height=4cm}
\epsfig{figure=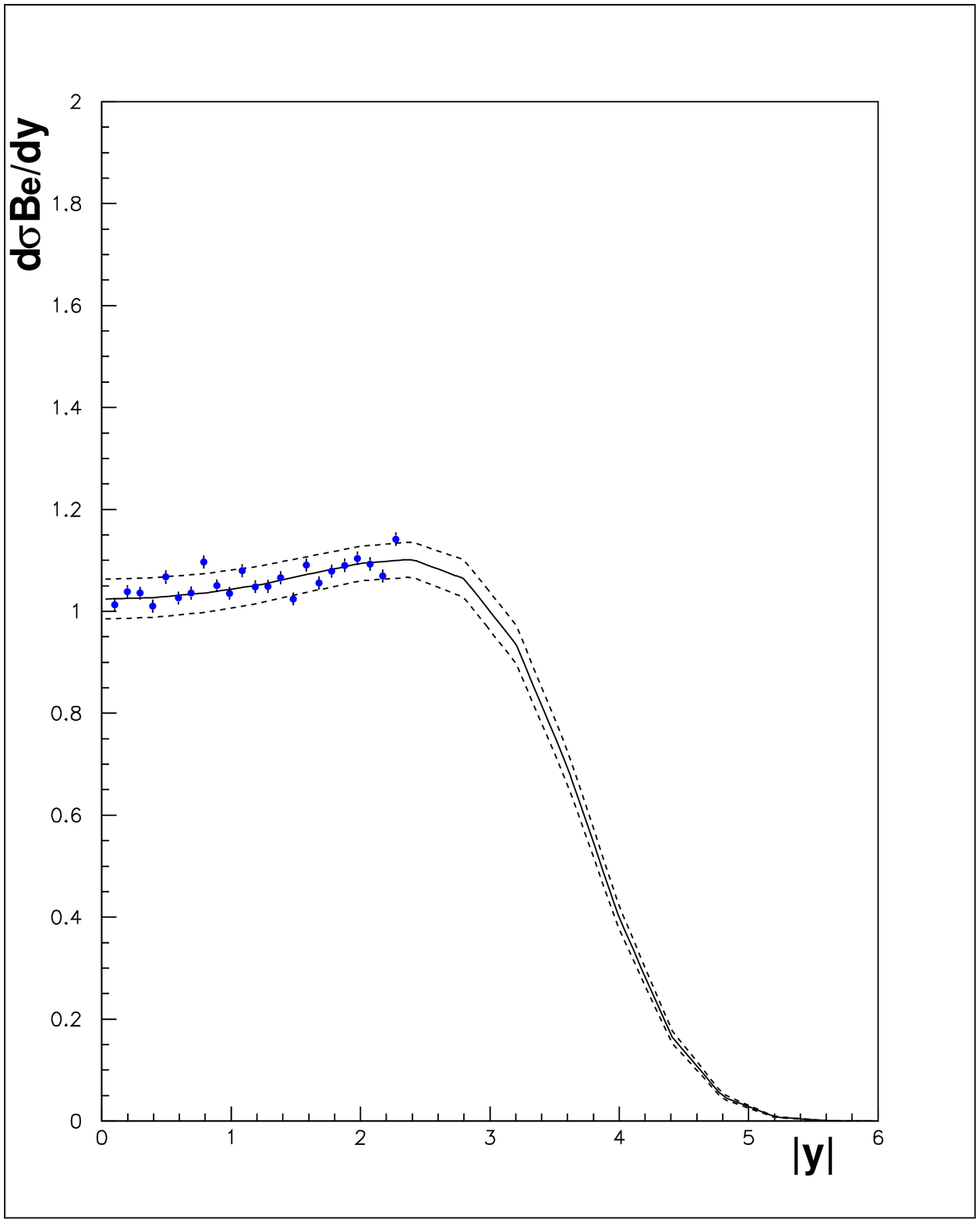,width=0.3\textwidth,height=4cm}
}
\caption {Top row: $e^+$ rapidity spectra generated from: left plot: ZEUS-S 
PDFS; middle plot: CTEQ6.1 PDFs; right plot: CTEQ6.1 PDFs which have been 
passed through the ATLFAST detector simulation and corrected back to generator
level using ZEUS-S PDFs; compared to the analytic prediction
using ZEUS-S PDFs. Bottom row: the same lepton rapidity spectra as above 
compared to the analytic 
prediction {\it after} including these lepton pseudo-data in the ZEUS-S PDF fit.}
\label{fig:zeusfit}
\end{figure}

The upper left hand plot of Fig.~\ref{fig:zeusfit} shows the $e^+$ rapidity 
spectrum for events generated from the ZEUS-S PDFs compared to the analytic 
predictions for the same
 ZEUS-S PDFs. The lower left hand plot illustrates the result if 
these events are then 
included in the ZEUS-S PDF fit (together with the $e^-$ spectra which are 
not illustrated). The size of the PDF uncertainties, at $y=0$, 
decreases from $6\%$ to $4.5\%$.  
The largest improvement is in the PDF parameter $\lambda_g$, controlling the 
low-x gluon at the input scale, $Q^2_0$: $xg(x) \sim x^{\lambda_g}$ at 
low-$x$, $\lambda_g = -0.199 \pm 0.046$, 
before the input of the LHC pseudo-data, compared to, 
$\lambda_g = -0.196 \pm 0.029$, after input. 
Note that whereas the relative normalizations of the $e^+$ and $e^-$ spectra 
are set by the PDFs, 
the absolute normalization of the data is free in the fit, so that 
no assumptions are made on our ability to measure luminosity.
Secondly, we repeat this procedure for events generated using the CTEQ6.1 PDFs.
This is illustrated in the middle section of Fig.~\ref{fig:zeusfit}. 
Before they are input to the fit, the cross section for these events is on the 
lower edge of the uncertainty band of the ZEUS-S predictions 
(upper middle plot). If these events are then input to the fit the 
central value shifts and the uncertainty decreases (lower middle plot). 
The value of the parameter $\lambda_g$ becomes, 
$\lambda_g = -0.189 \pm 0.029$, after input of these pseudo-data.
Finally, to simulate the situation which really faces experimentalists, 
we generate events with CTEQ6.1, 
and pass them through the ATLFAST detector simulation and cuts. We then 
correct back from detector level 
to generator level using a different PDF set - in this cases the ZEUS-S PDFs - 
since in practice we will not know 
the true PDFs. The upper right hand plot of Fig.~\ref{fig:zeusfit} shows 
that the resulting corrected data look 
pleasingly like CTEQ6.1, but they are more smeared. When these data are input 
to the PDF fit the central values shift and 
errors decrease (lower right plot) just as for the perfect CTEQ6.1 
pseudo data. The value of $\lambda_g$ becomes,
 $\lambda = -0.181 \pm 0.030$, after input of these pseudo data. 
Thus we see that the bias introduced by the 
correction procedure from detector to generator level is small compared to 
the PDF uncertainty, and that measurements at the $\sim 4\%$ level should be 
able to improve the level of uncertainty of the PDF predictions.

\subsubsection{Conclusions}
\label{subsec:lowx;amcs_conc}

We have investigated the PDF uncertainty on the predictions for $W$ and $Z$ production at the LHC, using the electron decay channel for the $W$ bosons and taking 
into account realistic expectations for the measurement accuracy and the cuts on data which will be needed to 
identify signal events from background processes. We conclude that, at the present level of PDF uncertainty, the 
decay lepton spectra can be used as a luminosity monitor but it is only good 
to $\sim 10\%$. However, 
we have also investigated the measurement accuracy 
necessary for early measurements of these decay lepton spectra to be useful in further constraining the 
PDFs. A systematic measurement error of $\sim 4\%$ could provide 
useful extra constraints.

The ratio of $Z$ to $W^+ + W^-$ production (measured via the lepton spectra) can provide an SM measurement
which is relatively insensitive to PDF uncertainties. 
By contrast, a measurement of the lepton asymmetry can provide the 
first measurements of the valence difference $u_v - d_v$ at small $x$.

We now return to the caveat made in the introduction: 
the current study has been performed using 
standard PDF sets which are extracted using NLO QCD in the 
DGLAP formalism. The extension to NNLO is 
straightforward, giving small corrections $\sim 1\%$. PDF analyses at NNLO 
including full accounting of the PDF 
uncertainties are not extensively available yet, so this small correction 
has not been pursued here. However, there may be much larger  
uncertainties in the theoretical calculations because the kinematic region 
involves  low-$x$.  
The MRST group recently produced a PDF set, MRST03, which does not include 
any data for $x < 5\times 10^{-3}$, in order to avoid 
the inappropriate use of the DGLAP formalism at small-$x$. 
Thus the  MRST03 PDF set should only be used for $x > 5\times 10^{-3}$. 
What is needed is an alternative theoretical formalism
for smaller $x$, as explained in Sec.~\ref{sec:rdb}. It is clear 
that the use of this formalism would bring greater changes than the small 
corrections involved in going to NNLO. There may even be a need for  
more radical extensions of the theory at low-$x$ due to high density effects.

\begin{figure}[tbp] 
%\vspace*{5pt}
%\vspace{-1.0cm}
\centerline{
\epsfig{figure=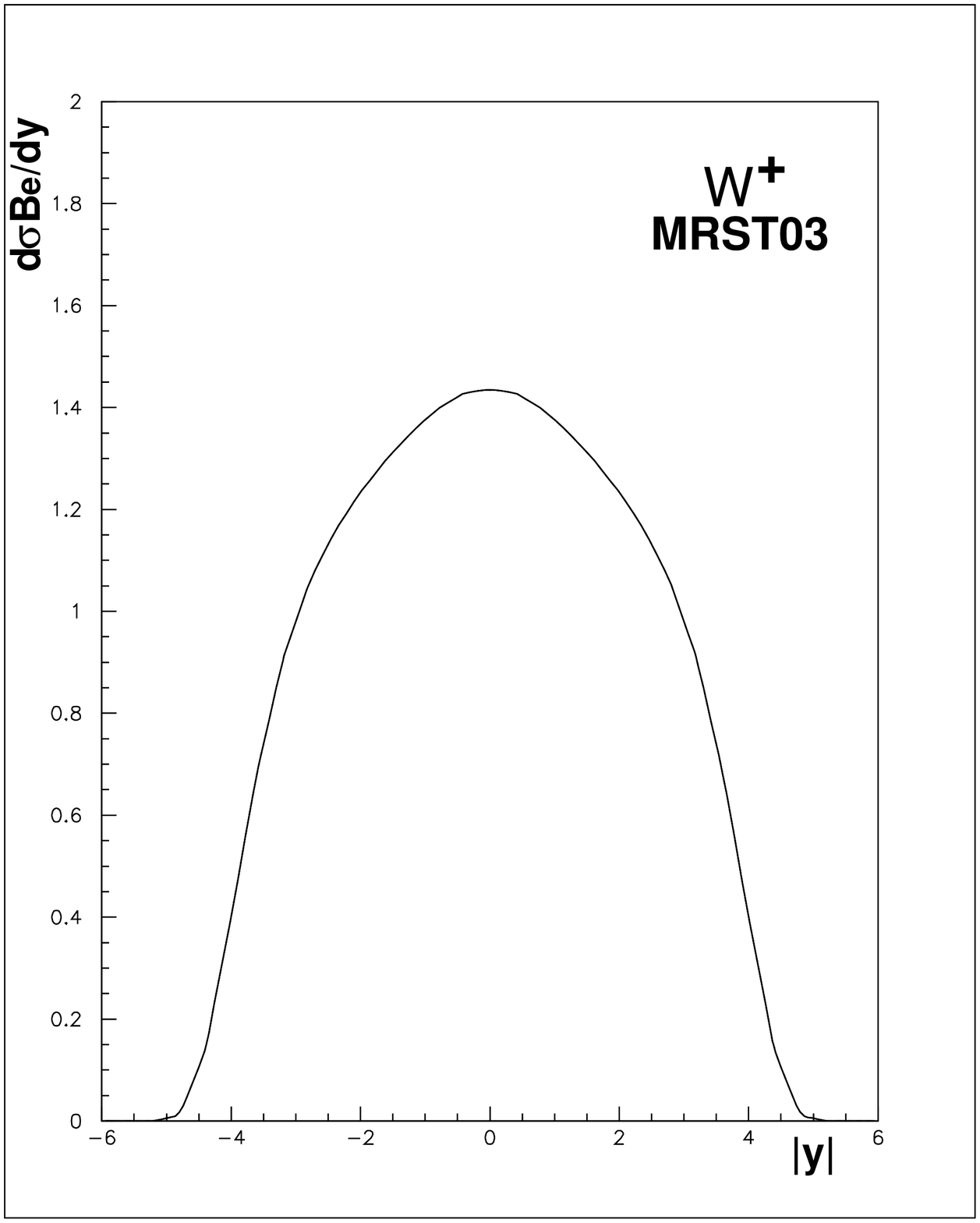,width=0.3\textwidth,height=4cm}
\epsfig{figure=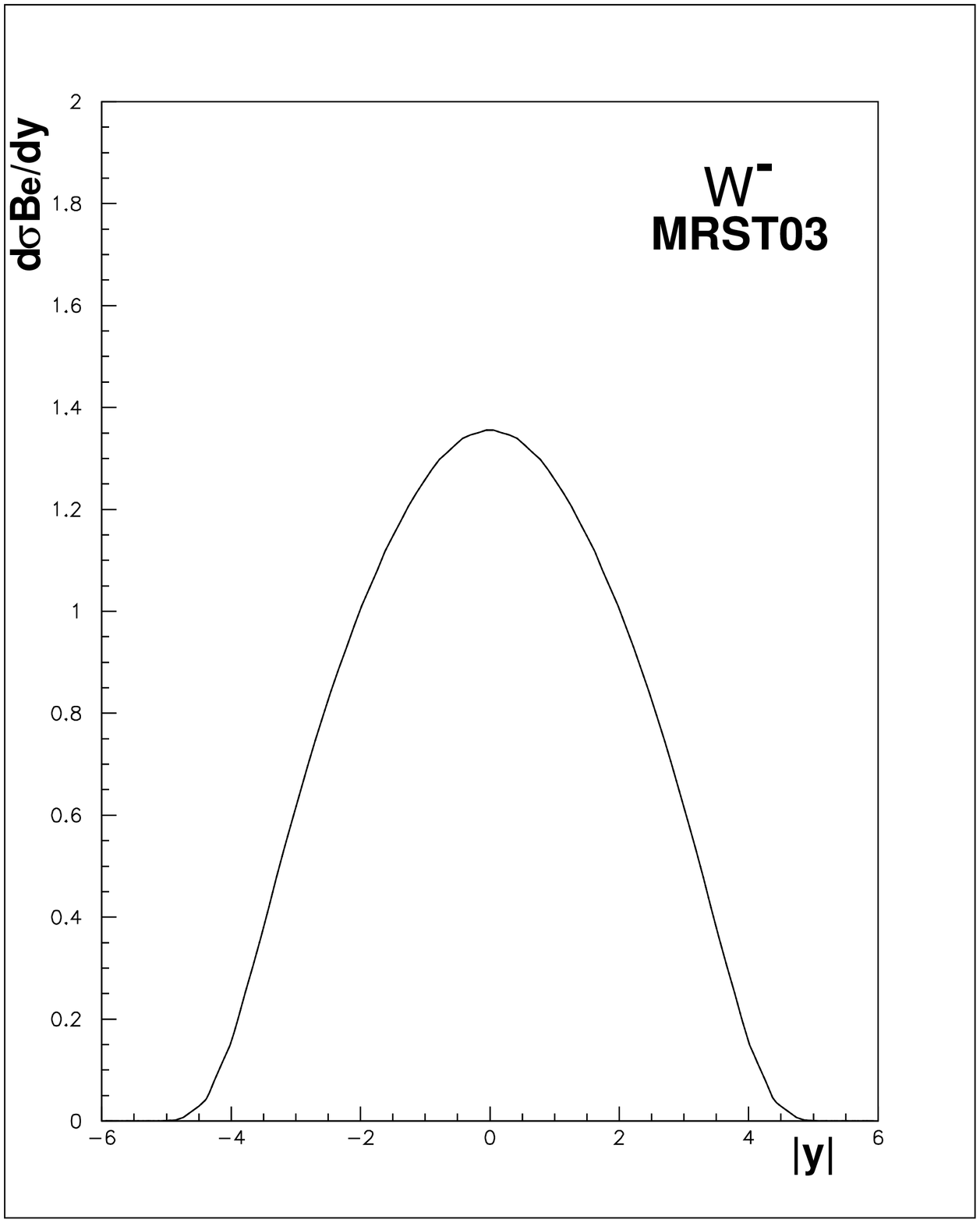,width=0.3\textwidth,height=4cm}
\epsfig{figure=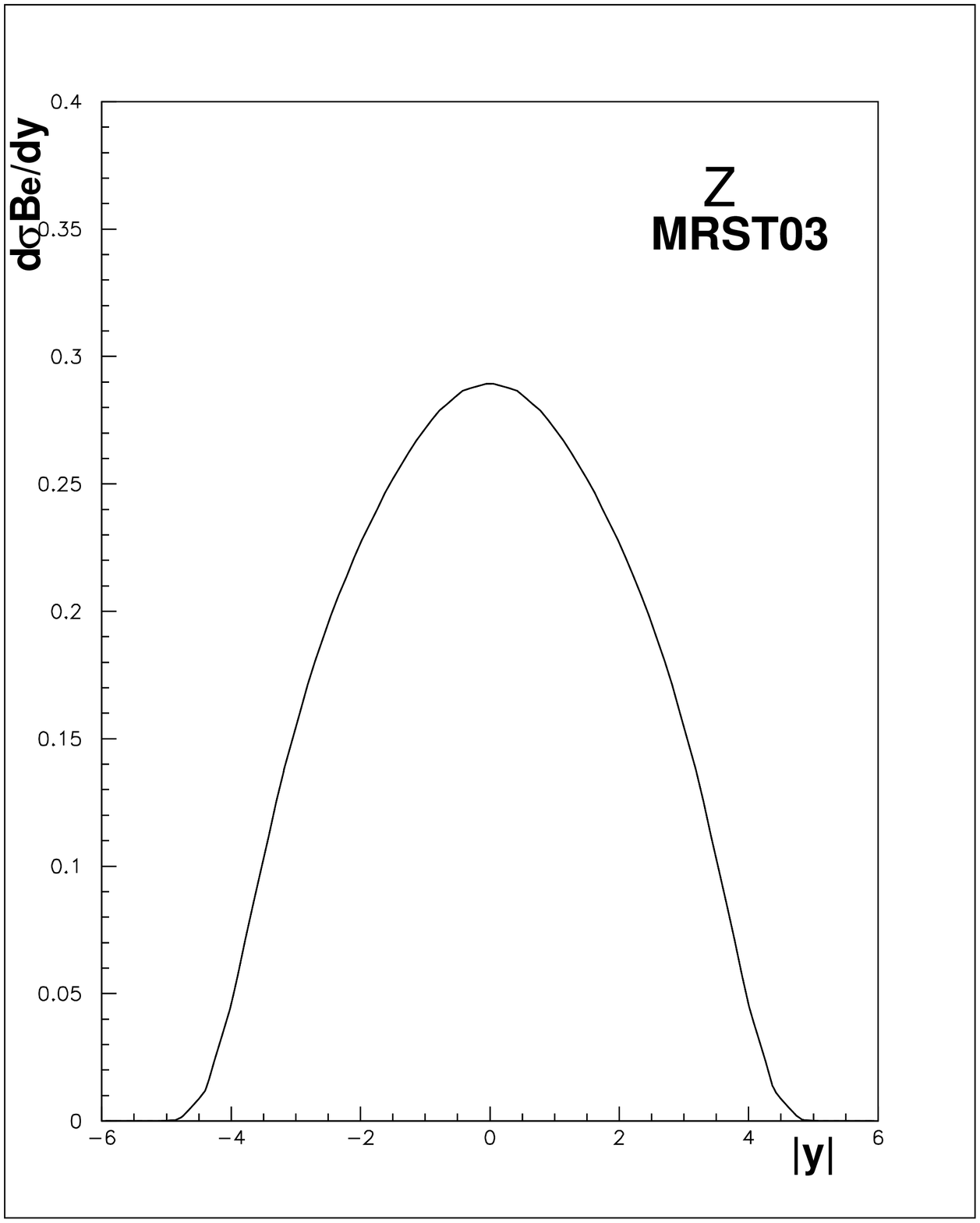,width=0.3\textwidth,height=4cm} 
}
\caption {$W^+$, $W^-$ and $Z$ rapidity distributions for the MRST03 PDFs at the LHC: left plot: $W^+$; middle plot: $W^-$; 
right plot: $Z$.}
\label{fig:mrst03pred}
\end{figure}

The MRST03 PDF set may be used as a toy PDF set, 
to illustrate the effect of using 
very different PDF sets on our predictions. 
A comparison of Fig.~\ref{fig:mrst03pred} with 
Fig.~\ref{fig:WZrapFTZS13} or Fig.~\ref{fig:mrstcteq} shows how different 
the analytic predictions are from the conventional ones, and thus illustrates 
where we might  expect to see differences due 
to the need for an alternative formalism at small-$x$. Whereas these results 
may seem far fetched, we should remind ourselves that moving into a different 
kinematic regime can provide surprises - as it did with the HERA data itself!  

\subsection{Resummed Perturbative Evolution at High Energy~\protect
\footnote{Author: R.D.~Ball}}
\label{sec:rdb}

\subsubsection{Introduction}
\label{subsec:rdbintro}

Logarithmic enhancement of higher order perturbative result
may take place when more than one large scale ratio is present. In DIS 
and DY this happens in the two opposite limits when the centre-of-mass 
energy of the partonic collision is
close to the threshold for the production of the 
final state, or much higher than the characteristic scale of the
process. These correspond respectively to the small-$x$ and large-$x$
kinematical regions, where $0\le x\le 1$ is defined in terms of
the invariant mass of the non-leptonic final
state  $\frac{ (1 - x) Q^2}{x}$. The corresponding
perturbative contributions are respectively enhanced by powers of $
\ln\frac{1}{x}$ and $\ln (1-x)$, or, equivalently, in the space
of Mellin moments, by powers of $\frac{1}{N}$ and $\ln N$, where
$N\to0 $ moments dominate as $x\to 0$ while $N\to\infty$ moments
dominate as $x\to 1$. Here we will be concerned only with the small-$x$ 
(and thus small $N$) region.

In the DGLAP evolution equation one resums collinear logarithms first,
resulting at LO and NLO in a good description of many data sets, and in 
particular the HERA $F_2$ data at perturbative $Q^2$ and values of 
$x$ as low as $10^{-4}$ \cite{Ball:1994du,Ball:1994kc,Ball:1995qd}.
However in the singlet (gluonic) channel the fixed order splitting 
functions contain small-$x$ logarithms also, which at NNLO~\cite{Vogt:2004mw}
begin to destabilise the perturbative expansion, so that 
a further resummation is needed if the evolution is to be reliable at 
small $x$. Small-$x$ logarithms may be resummed by using the 
BFKL equation~\cite{Kuraev:1976ge,Kuraev:1977fs,Balitsky:1978ic}.
However the fixed order kernels of this equation, currently 
known to NLO \cite{Fadin:1998py} are perturbatively unstable for 
all realistic values of $\alpha_s$, and are thus by themselves 
useless for reliable calculations \cite{Ball:1998be,Ross:1998xw,Ball:1999sh}. 
This is because they contain collinear logarithms, which must be resummed even 
at LO if reliable predictions are to be obtained \cite{Salam:1998tj}.
This collinear resummation of the BFKL kernel restores longitudinal 
momentum conservation \cite{Altarelli:1999vw}, and leads ultimately 
to a stable expansion.

Two approaches to the simultaneous resummation of these two classes of
logs have recently reached the stage where their phenomenological
application can be envisaged. 
In the duality (ABF) approach~\cite{Altarelli:1999vw,
Altarelli:2000mh,Altarelli:2001ji,Altarelli:2003hk,Altarelli:2003kt,
Altarelli:2004dq,Altarelli:2005ni} 
one concentrates on the problem of obtaining an improved anomalous
dimension (splitting function) 
for DIS which reduces to the ordinary perturbative result at large 
$N$ (large $x$),
while including resummed BFKL corrections at small $N$ (small $x$),
determined through the BFKL kernel. The CCSS
approach~\cite{Ciafaloni:1999yw,Ciafaloni:2003ek,Ciafaloni:2003rd,
Ciafaloni:2005cg}
is built up within the BFKL 
framework, by improving the whole hierarchy of subleading kernels in
the collinear region consistently with the 
renormalization group. The BFKL equation is then solved and a perturbative 
splitting function extracted numerically.

Here we will briefly review the theoretical underpinnings of the duality
approach, and then compare phenomenological results
obtained in both approaches. 

\subsubsection{High Energy Duality}
\label{subsec:rdbduality}

In  the ABF approach one
constructs  an improved anomalous
dimension (splitting function) for DIS which reduces to the ordinary
perturbative result at large $N$ (large $x$) given by: 
\begin{equation}
\gamma(N,\alpha_s)=\alpha_s \gamma_0(N)~+~\alpha_s^2
\gamma_1(N)~+~\alpha_s^3
\gamma_2(N)~~\dots .
\label{gammadef}
\end{equation}
while including resummed BFKL corrections at small $N$ (small $x$)
which are determined  by the aforementioned BFKL
kernel $\chi(M,\alpha_s)$:
\begin{equation}
\chi(M,\alpha_s)=\alpha_s \chi_0(M)~+~\alpha_s^2 \chi_1(M)~+~\dots ,
\label{chidef}
\end{equation} 
which is the Mellin transform of the angular-averaged 
kernel $K$ with respect to $t=\ln \frac{k^2}{k_0^2}$.
The main theoretical tool which enables this construction is 
the duality relation between the kernels $\chi$ and $\gamma$
\begin{equation}
\chi(\gamma(N,\alpha_s),\alpha_s)=N,
\label{dualdef}
\end{equation}
which is a consequence of the fact that the solutions of the BFKL and DGLAP
equations coincide at 
leading twist~\cite{Ball:1997vf,Ball:1999sh,Altarelli:1999vw}.
Further improvements are obtained exploiting
the symmetry
under gluon interchange of the BFKL gluon-gluon kernel and through the
inclusion of running coupling effects. 

By using duality, one can
construct a more balanced expansion for both $\gamma$ and $\chi$,
the "double leading" (DL) expansion, where the
information from $\chi$ is used to include in $\gamma$ all powers of
$\alpha_s/N$ and, conversely, $\gamma$ is used to improve $\chi$ by all
powers of $\alpha_s/M$. A great advantage of the DL expansion is that
it resums the collinear poles of $\chi$ at $M=0$, enabling the
imposition of the physical requirement of momentum conservation
$\gamma(1,\alpha_s)=0$, whence, by duality: 
\begin{equation}
\chi(0,\alpha_s)=1.
\label{mom}
\end{equation}
This procedure eliminates in a model-independent way the alternating
sign poles $+1/M, -1/M^2,.....$ that appear in $\chi_0$,
$\chi_1$,\dots. These poles make the perturbative expansion of $\chi$
unreliable even in the central region of $M$: e.g., 
$\alpha_s \chi_0$ has a minimum at $M=1/2$, while, at realistic values
of $\alpha_s$,  $\alpha_s \chi_0+\alpha_s^2 \chi_1$ has a maximum. 

At
this stage, while the poles at $M=0$ are eliminated, those at $M=1$
remain, so that the DL expansion is still not finite near
$M=1$. The resummation of the $M=1$ poles can be accomplished by
exploiting the collinear-anticollinear symmetry,
as in the CCSS approach \cite{Ciafaloni:1999yw,Ciafaloni:2003ek,
Ciafaloni:2003rd}. In Mellin space, this symmetry
implies that at the fixed-coupling level the kernel $\chi$
for evolution in  $\ln \frac{s}{k k_0}$ must satisfy
$\chi(M)=\chi(1-M)$ order by order in perturbation theory. 
This symmetry is however broken by the DIS choice
of variables $\ln\frac{1}{ x}=\ln\frac {s}{Q^2}$ and by the running of the
coupling. In the fixed-coupling limit 
the kernel $\chi_{\rm DIS}$, dual to the DIS anomalous
dimension, is related to the symmetric one
$\chi_{\sigma}$ through  the implicit equation~\cite{Fadin:1998py} 
\begin{equation}
\chi_{\rm DIS}\left(M+\frac{1}{2}\chi_{\rm \sigma}(M)\right)=\chi_{\sigma}(M).
\label{symm}
\end{equation} 
Hence, the $M=1$ poles can be resummed  by performing the double-leading
resummation of $M=0$ poles of $\chi_{\rm DIS}$, determining the
associated $\chi_\sigma$ through eq. (\ref{symm}), then symmetrizing
it, and finally going back to DIS
variables by using eq. (\ref{symm}) again in reverse.
Using the momentum conservation  eq. (\ref{mom}) and eq. (\ref{symm}), 
it is easy to show that $\chi_\sigma(M)$ is an entire
function of M, with 
$\chi_\sigma(-1/2)=\chi_\sigma(3/2)=1$, 
and thus necessarily has a minimum at $M=1/2$. Through this 
procedure one obtains order
by order from the DL expansion a symmetrized DL
kernel $\chi_{\rm DIS}$, and its corresponding dual anomalous dimension 
$\gamma$. The
kernel $\chi_{\rm DIS}$ has to all orders a minimum and satisfies a
momentum conservation constraint $\chi_{\rm DIS}(0)=\chi_{\rm DIS}(2)=1$.

The final ingredient of the ABF approach is a treatment of the running
coupling corrections to the resummed terms. Indeed, their inclusion
in the resummed anomalous dimension greatly softens
the asymptotic behavior near $x=0$. Hence, the 
dramatic rise of structure functions at small $x$, 
which characterized resummations
based on leading--order BFKL evolution, and is
ruled out phenomenologically, is replaced by a much milder rise. This
requires a running coupling generalization of the duality
equation~(\ref{dualdef}), which is possible noting that in $M$ space the
running coupling $\alpha_s(t)$ becomes a differential operator,  
since $t \rightarrow d/dM$. Hence, the BFKL evolution equation for
double moments $G(N,M)$, which is an algebraic equation at fixed
coupling, becomes a differential equation in $M$ for running
coupling. In the ABF approach, one solves this differential equation
analitically when the kernel is replaced by its quadratic
approximation near the minimum. The solution is expressed in terms of an
Airy  function if the kernel is linear in $\alpha_s$ \cite{Altarelli:2001ji}, 
for example in
the case of $\alpha_s \chi_0$, or of a Bateman function in the more
general case of a non linear dependence on $\alpha_s$ 
\cite{Altarelli:2005ni} as is the case
for the DL kernels. The final result for the improved anomalous
dimension is given in terms of the DL expansion plus the ``Airy'' or
``Bateman'' anomalous dimension, with the terms already included in the DL
expansion subtracted away.  

For example, at leading DL order, i.e. only using $\gamma_0(N)$ and
$\chi_0(M)$, the improved anomalous dimension is
\begin{eqnarray}
\gamma_I^{NL}(\alpha_s, N) &=& 
\big[\alpha_s\gamma_0(N)+ \alpha_s^2 \gamma_1(N) 
+\gamma_s(\frac{\alpha_s}{N}) -\frac{n_c\alpha_s}{\pi N}\big]\nonumber \\*
&&\qquad +\gamma_A(c_0,\alpha_s,N)-\frac{1}{2} +
\sqrt{\frac{2}{\kappa_0\alpha_s}[N-\alpha_s
c_0]} .\label{gamimp}
\end{eqnarray}
The terms within square brackets
give the LO DL approximation, i.e. they contain the
fixed--coupling information from
$\gamma_0$ and (through $\gamma_s$) from $\chi_0$. The 
``Airy'' anomalous dimension
$\gamma_A(c_0,\alpha_s,N)$ contains the running coupling resummation,
i.e. it is the exact solution of the 
running coupling BFKL equation which corresponds to a quadratic 
approximation to $\chi_0$ near $M=1/2$.
The last two terms
subtract the contributions to $\gamma_A(c_0,\alpha_s,N)$
which are already included in
$\gamma_s$ and $\gamma_0$. In the limit $\alpha_s
\rightarrow 0$ with $N$ fixed, $\gamma_I(\alpha_s,N)$ reduces to
$\alpha_s\gamma_0(N)+O(\alpha_s^2)$. For  $\alpha_s
\rightarrow 0$ with $\alpha_s/N$ fixed, $\gamma_I(\alpha_s,N)$  reduces to
$\gamma_s(\frac{\alpha_s}{N})+O(\alpha_s^2/N)$, i.e. the leading term of the
small-$x$ expansion. Thus the Airy term is subleading in both
limits. However, if $N\to 0$ at fixed $\alpha_s$, the Airy term
replaces the leading singularity of the DL anomalous dimension,
which is a square root branch cut, with a simple pole, located on the
real axis at rather smaller $N$, thereby softening the small-$x$ behaviour.
The quadratic approximation is sufficient to
give the correct asymptotic behaviour up to terms which
are of subleading order in comparison to those included in the DL
expression in eq.~(\ref{gamimp}). 

The running coupling resummation 
procedure can be applied to a symmetrized kernel, which possesses
a minimum to all orders, and then extended to next-to-leading
order~\cite{Altarelli:2004dq,Altarelli:2005ni}. This entails
various technical complications, specifically related to the nonlinear
dependence of the symmetrized kernel on $\alpha_s$, to the need to
include interference between running coupling effects and 
the small $x$ resummation, and to the consistent treatment
of next-to-leading log $Q^2$ terms, in particular those related to the
running of the coupling. 

\subsubsection{Results}
\label{subsec:rdbres}

Even though the basic underlying physical principles of the ABF and
CCSS approaches are close, there are technical differences in the
construction of the symmetrized DL 
kernel, in the derivation from it of an anomalous dimension and associated
splitting function, and in the inclusion of running coupling effects.
Therefore, we will compare results for the resummed
fixed-coupling $\chi$ kernel (BFKL characteristic function), then the
corresponding fixed-coupling splitting functions, and finally the
running coupling splitting functions which provide the final result
in both approaches. In order to assess the phenomenological impact on
parton evolution we will finally compare the convolution of the
splitting function with a ``typical'' gluon distrubution.

\begin{figure}[htbp]
  \includegraphics[width=0.75\textwidth]{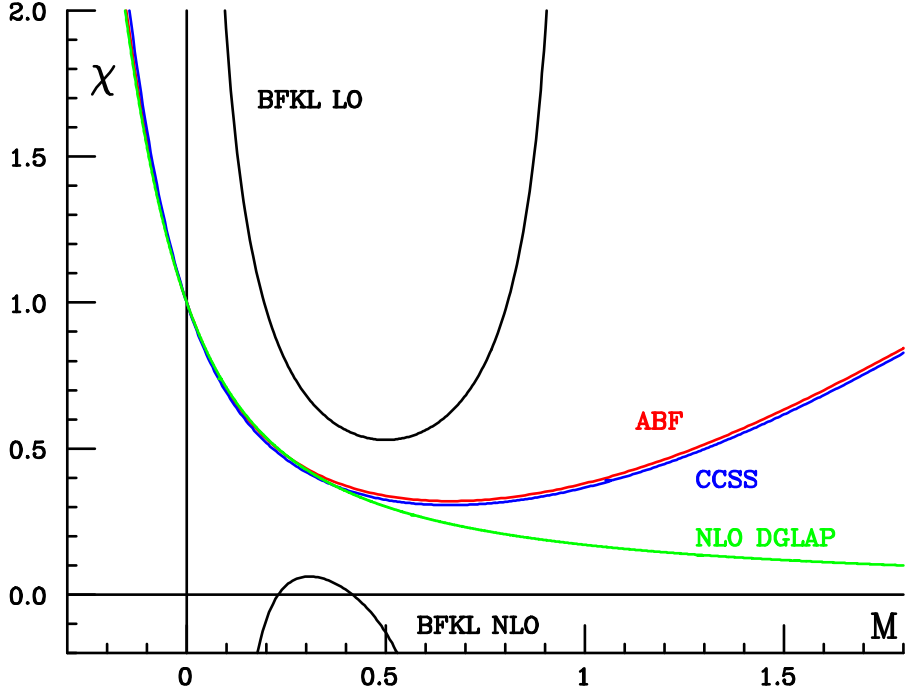}\centering
  \caption{The kernel $\chi$ (BFKL characteristic function) for fixed coupling
    ($\beta_0=0$) $\alpha_s=0.2$ and
    ${n_{\! f}}=0$. The BFKL curves are the LO and NLO truncations of
    eq.~(\ref{chidef}), the DGLAP curve is the dual
    eq.~(\ref{dualdef}) of the NLO anomalous dimension
    eq.~(\ref{gammadef}), while the ABF and CCSS curves are
    respectively the solution $\chi_{\rm DIS}$ of eq.~(\ref{symm}) 
    and the solution for $\omega$ of a similar eqn. in the CCSS approach.}
  \label{fig:chi}
\end{figure}

In fig.~\ref{fig:chi} we show the
solution $\chi_{\rm DIS}$ of eq.~(\ref{symm}) for the symmetrized NLO DL 
kernel. The pure L$x$ and NL$x$ (BFKL) and next-to-leading $\ln Q^2$
(DGLAP) are also shown. All curves are determined  with frozen
coupling ($\beta_0=0$), and with
$n_f=0$, in order to avoid complications related to the diagonalization of
the DGLAP anomalous dimension matrix and to the choice of scheme for the
quark parton distribution.  Also shown is the corresponding resummed kernel 
in the RGI CCSS approach. The resummed ABF and CCSS results are very
close, the main difference being due to the fact that 
at small $M$ the ABF result coincides by construction with NLO 
DGLAP, whereas for very small $M$ (i.e. large $x$) the CCSS result 
reduces to LO DGLAP. Because of the underlying symmetry this 
small difference is also seen in the anticollinear region $M\sim 1$, 
though the two curves coincide by construction at the momentum 
conservation points $M=0$
and $M=2$. In comparison to DGLAP, the resummed kernels have a minimum,
related to the fact that both collinear and anticollinear logs are
resummed.
In comparison to BFKL, which has a minimum at LO but
not NLO, the resummed kernels always have a perturbatively stable  minimum,
characterized by a lower intercept than leading--order BFKL: specifically,
when $\alpha_s=0.2$, $\lambda\sim 0.25$  instead of $\lambda\sim0.5$. This
corresponds to a softer small-$x$ rise of the associated splitting
function. 

\begin{figure}[htbp]
\includegraphics[width=0.75\textwidth]{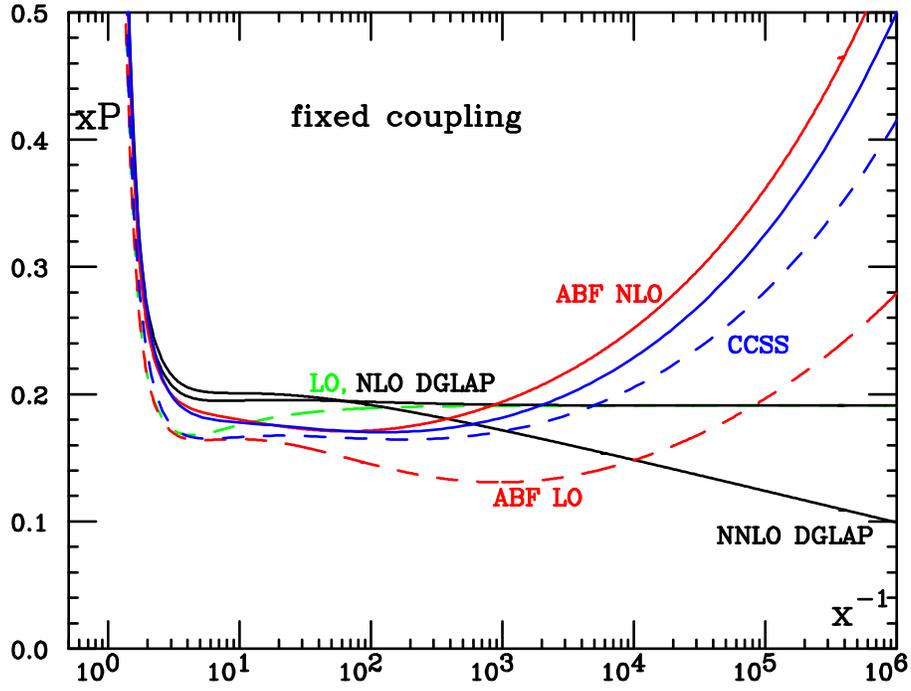}\centering
  \caption{The fixed coupling
    ($\beta_0=0$) $xP_{gg}(x)$ splitting
 function, evaluated with $\alpha_s=0.2$  and $n_f=0$. 
The dashed curves are LO for DGAP, symmetrized LO DL for ABF, and 
NL$x$+LO for CCSS, while the solid curves are NLO and NNLO for DGLAP, 
symmetrized NLO DL for ABF and NL$x$+NLO for CCSS. }
  \label{fig:Pggb0}
\end{figure}

The fixed--coupling resummed splitting functions up to NLO are shown in 
figure~\ref{fig:Pggb0}, along with the unresummed DGLAP splitting
functions up to NNLO. For ${n_{\! f}}=0$ the NLO DGLAP splitting 
function has the
property that it vanishes at small $x$ --- this makes it relatively
straightforward to combine not just LO DGLAP but also NLO DGLAP with
the NLLx resummation. 
In the ABF approach the splitting function is the 
inverse Mellin transform of the anomalous dimension obtained using
duality eq.~(\ref{dualdef}) from the symmetrized DL $\chi$
kernel. Hence, the LO and NLO resummed result respectively reproduce all
information contained in the LO and NLO $\chi$ and $\gamma$  kernel
with the additional constraint of collinear-anticollinear symmetry.
Both the ABF LO and NLO results are shown in figure~\ref{fig:Pggb0}.
The CCSS NL$x$+LO and NL$x$+NLO curves are also shown for comparison.

In comparison to unresummed results, the resummed splitting functions
display the characteristic rise at small $x$ of fixed-coupling
leading-order BFKL
resummation, though the small-$x$ rise is rather milder ($\sim x^{-0.25}$
instead of $\sim x^{-0.5}$ for $\alpha_s=0.2$).  
At large $x$ there is good agreement between the resummed results and
the corresponding LO (dashed) or NLO (solid) DGLAP curves. At small $x$ 
the difference
between the ABF LO and CCSS NL$x$+LO  (dashed) curves is mostly due to
the different way the symmetrization is implemented, as both
approaches contain the same dominant small-$x$ terms. This difference
is reduced when one compares the CCSS NL$x$+nLO with ABF NLO, and it
might be taken as an estimate of the intrinsic ambiguity of the
fixed--coupling resummation procedure.
At intermediate $x$  the NLO resummed splitting functions
is of a similar order of magnitude as the NLO DGLAP result even down to quite
small $x$, but with a rather different shape, characterized  by a
dip at $x\sim 10^{-3}$. 
The unstable small-$x$ drop of the  NNLO DGLAP result is 
consequence of the unresummed $\frac{\alpha_s^3}{N^2}$ 
double pole in the NNLO anomalous dimension.

\begin{figure}[htbp]
 \includegraphics[width=0.75\textwidth]{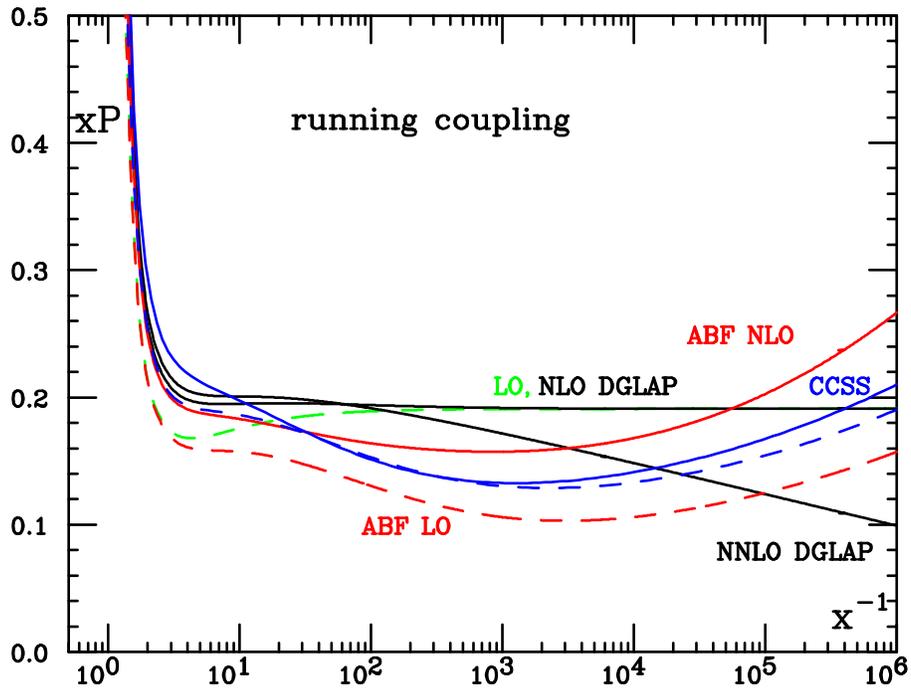}\centering
  \caption{The running coupling
     $xP_{gg}(x)$ splitting
 function, evaluated with  $\alpha_s=0.2$ and $n_f=0$. The various
     curves correspond to 
     the same cases as in figure~\ref{fig:Pggb0}. }
  \label{fig:Pgg}
\end{figure}

The running-coupling resummed splitting functions are displayed in
figure~\ref{fig:Pgg}. Note that the unresummed curves are the same as
in the fixed coupling case since their dependence on $\alpha_s$ is
just through a prefactor of $\alpha_s^k$, whereas in the resummed case
there is an interplay between the running of the coupling and the
structure of the small-$x$ logs. All the resummed curves display a
considerable softening of the small-$x$ behaviour in comparison to their
fixed-coupling counterparts, due to the softening of the leading 
small-$x$ singularity in the running-coupling 
case~\cite{Ciafaloni:1999yw,Altarelli:2001ji}.
As a consequence, the various resummed results are closer to each
other than in the fixed-coupling case, and also closer to the
unresummed LO and NLO DGLAP results. The resummed perturbative
expansion appears to be stable, subject to moderate theoretical
ambiguity, and qualitatively close to NLO DGLAP.

\begin{figure}[htbp]
 \includegraphics[width=0.5\textwidth]{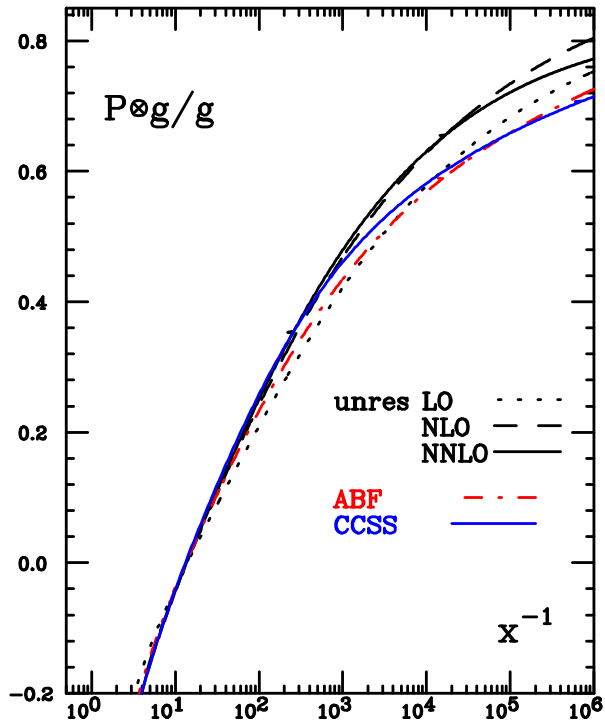}\centering
  \caption{Convolution of resummed and fixed-order $P_{gg}$ splitting
    functions with a toy gluon distribution,
    eq.~(\ref{eq:toy_gluon_for_pgg_test}),  normalised to
    the gluon distribution itself, with $\alpha_s=0.2$ and  ${n_{\!
    f}} = 0$. The resummed ABF and CCSS curves are obtained using
    respectively the ABF NLO and 
    CCSS NL$x$+NLO splitting function shown in fig.~\ref{fig:Pgg}.
}
  \label{fig:conv}
\end{figure}

Finally, to appreciate the impact of resummation it is useful to investigate
not only the properties of the splitting function, but also its
convolution with a physically reasonable gluon distribution. We take
the following toy gluon
\begin{equation}
  \label{eq:toy_gluon_for_pgg_test}
  xg(x) = x^{-0.18} (1-x)^5\,,
\end{equation}
and show in fig.~\ref{fig:conv} the result of its convolution with
various splitting functions of fig.~\ref{fig:Pgg}.
The differences between resummed and unresummed results,
and between the ABF and CCSS  resummations are as expected partly washed out by
the convolution, even though the difference between the unresummed LO
and NLO DGLAP results is clearly visible. In particular,
differences between the fixed-order and resummed
convolution start to become significant only for $x$ below $10^{-2}-10^{-3}$,
even though resummation effects started to be visible in the splitting
functions at somewhat larger $x$. However it should be clear from this figure
why the structure function data from HERA were so well described by LO and NLO 
GLAP evolution \cite{Ball:1994du,Ball:1994kc,Ball:1995qd}: significant 
deviations from GLAP will only be seen for smaller $x$ and larger $Q^2$ 
than is accessible at HERA, but may well be important for many important 
processes at the LHC.

\subsection{Hunting BFKL at Hadron Colliders~\protect
\footnote{Author: V.~Del~Duca}}
\label{sec:vdd}

\subsubsection{Introduction}
\label{subsec:vddintro}

With its unprecedented kinematic range, the LHC offers an unique opportunity
to explore strong-interaction processes characterised by two large and 
disparate scales. In inclusive di-jet production, for instance, 
jets of transverse energy
$E_T = 50$~GeV may span a rapidity interval of up to about 10 units of
rapidity. Processes with two large and disparate scales may contain large
logarithms of the ratio of those scales. The interest in such processes
arises from the Balitsky-Fadin-Kuraev-Lipatov (BFKL) 
equation~\cite{Kuraev:1976ge,Kuraev:1977fs,Balitsky:1978ic},
which performs an all-order resummation in $\alpha_S$ of the leading
logarithms (LL) of $\ln(\hat s/|\hat t|)$, with
$\hat s$ the squared parton centre-of-mass energy, $\hat t$ a typical
momentum transfer, and $\hat s \gg |\hat t|$. The generality of the resummation
is based on the fact that gluon exchange in the $t$ channel dominates
in any scattering process with $\hat s \gg |\hat t|$. 
The BFKL formalism
then re-sums the multiple gluon radiation out of the gluon exchanged in
the $t$ channel. The LL terms are obtained in the approximation of a strong
rapidity ordering of the emitted gluons. The resummation yields an integral
equation which describes the evolution of the gluon propagator in the $t$
channel, and whose kernel is formed by the emission of a gluon along the
ladder and by the LL contribution to a gluon-loop exchange in the ladder. 
By putting together the emission of two close-in-rapidity 
gluons~\cite{Fadin:1989kf,DelDuca:1995ki} and a $q\bar{q}$ 
pair~\cite{Fadin:1996nw,DelDuca:1996me,Fadin:1997hr,Catani:1990xk,Catani:1990eg}
along the ladder, the one-loop 
corrections~\cite{Fadin:1993wh,Fadin:1994fj,Fadin:1996yv,DelDuca:1998cx} to 
the emission of a gluon along the ladder, and the next-to-leading-logarithmic 
(NLL)~\cite{Fadin:1995xg,Fadin:1996tb,Fadin:1995km,Blumlein:1998ib,DelDuca:2001gu}
contribution to a gluon two-loop exchange in the ladder, also the NLL
corrections~\cite{Fadin:1998py,Ciafaloni:1998gs,Camici:1997ij} 
to the BFKL equation have been computed. 

\subsubsection{Jets at large rapidity intervals}
\label{subsec:vddjets}

During the last two decades, a large body of work has been dedicated to 
predict and detect
footprints of emission of BFKL gluon radiation in strong-interaction 
processes, like di-jet production at large rapidity
intervals~\cite{Abachi:1996et,Abbott:1999ai,Mueller:1986ey,DelDuca:1993mn,
Stirling:1994zs,Andersen:2001kt},
$W$-boson production in association with jets~\cite{Andersen:2001ja},
heavy-quark production at hadron 
colliders~\cite{Catani:1990xk,Catani:1990eg,Ellis:1990hw,Andersen:2004nm}; 
forward-jet production~\cite{Aid:1995we,Breitweg:1998ed,Adloff:1998fa,
Aktas:2005up,Mueller:1990er,Tang:1991am,Bartels:1991tf,Askew:1992tw,
Bartels:1996gr,Mirkes:1996jd,Bartels:1996wx}, 
forward-pion production~\cite{Adloff:1998fa,Adloff:1999zx,Aktas:2004rb} 
and trasverse-energy flow~\cite{Aid:1995we,Aktas:2004rb,Adloff:1999ws} in DIS;
$\gamma^*\gamma^*$ collisions in double-tag events, 
$e^+\, e^- \to e^+\, e^- +$ 
hadrons~\cite{Acciarri:1998ix,Abbiendi:2001tv,Achard:2001kr,Heister:2003jb,
Bartels:1996ke,Brodsky:1997sd,Bartels:2000sk,Cacciari:2000cb,DelDuca:2002qt}.
All that the processes above have in common is a large logarithm:
in di-jet production at large rapidity intervals, for instance, the large 
logarithm is the rapidity interval between the jets,
$\Delta y\simeq\ln(\hat s/E_{T1}E_{T2})$, with
$E_{T1}$ and $E_{T2}$ the transverse energies of the two tagged jets;
in forward-jet production in DIS the large logarithm is $\ln(x/x_{bj})$, 
where $x_{bj}$ is the Bjorken scaling variable and $x$ the momentum fraction 
of the parton entering the hard scattering; in $\gamma^*\gamma^*$ collisions
in $e^+\, e^- \to e^+\, e^- +$ hadrons, the large logarithm is 
$Y= \ln(y_1y_2S/\sqrt{Q_1^2Q_2^2})$, with $S$ the $e^+\, e^-$ centre-of-mass 
energy, and $y_i$ and $Q_i^2$ the light-cone momentum fraction and the 
virtuality of the virtual photon $i$, with $i=1, 2$.

From a theoretical point of view,
di-jet production at large rapidity intervals is the simplest process
to which to apply the BFKL resummation, because at leading order in
$\alpha_S$ it is just parton-parton scattering, which at large rapidity
intervals is dominated by gluon exchange in the $t$ channel.
In fact, the $t$-channel gluon dominance is also used as a diagnostic tool for 
discriminating between different dynamical models for parton scattering.
In the measurement of di-jet angular distributions,
models which feature gluon exchange in the $t$ channel, like QCD, 
predict the characteristic $\sin^{-4}(\theta^\star/2)$
di-jet angular distribution~\cite{Ellis:1992qq,Abe:1992sj}, 
while models featuring contact-term interactions, which do not have gluon 
exchange in the $t$ channel,
predict a flattening of the di-jet angular distribution
at large $\hat s/|\hat t|$~\cite{Abe:1996mj,Abbott:1997nf}.
The phenomenological analysis of di-jet production at large rapidity intervals,
though, is not so simple as its theoretical construct: since
$\hat s = x_ax_b S$, with $S$ the hadron centre-of-mass energy and $x_a, x_b$
the momentum fractions of the partons entering the hard scattering,
once the jet transverse energies are fixed, there are two ways of increasing
$\Delta y\simeq\ln(x_ax_b s/E_{T1}E_{T2})$: by increasing the $x$'s in a 
fixed-energy collider; 
or viceversa, by fixing the $x$'s and letting $S$ grow, in a
ramping-run collider experiment. The former set-up, the only feasible at
a collider run at fixed energy, like the Tevatron or the LHC,
has been proven to be unpractical, 
since in the di-jet production rate $d\sigma/d\Delta y$ as a function of 
$\Delta y$ it is difficult to disentangle the BFKL-driven rise of 
the parton cross section from the steep fall-off of the parton 
densities~\cite{DelDuca:1993mn}. The latter set-up,
even though the first to be proposed~\cite{Mueller:1986ey}, has been 
analysed only in the late 90's~\cite{Abbott:1999ai}, because 
it required a collider running at different centre-of-mass energies, and
it has become feasible only when the Tevatron has undertaken for a few days
a run at $\sqrt{S} = 630$~GeV, in addition to the usual $\sqrt{S} = 1800$~GeV
of Run I. However, a careful analysis at fixed $x$'s~\cite{Andersen:2001kt}
has shown that in the kinematic reach of Tevatron di-jet production at 
large rapidity intervals is far from the BFKL asymptotic regime.
Other processes of BFKL interest at hadron colliders are 
$W$-boson production in association with jets and heavy-quark production.
The former is simpler to analyse experimentally than di-jet production,
which is hindered by the complexity of triggering on jets in the forward 
calorimetry. In fact, if the $W$ boson decays leptonically, it is more 
advantageous to trigger on the $W$ decay products~\cite{Andersen:2001ja}. 
Heavy-quark production, although potentially interesting, is hampered by
the fact that gluon exchange in the $t$ channel is a higher-order effect:
it occurs only at two orders of $\alpha_S$ higher than the leading order,
and the logarithms $\ln(\hat s/|\hat t|)$ are not large enough, within
the kinematic reach of Tevatron and LHC, to offset that initial 
handicap~\cite{Andersen:2004nm}.

Forward-jet production in DIS consists of tagging a jet far in
rapidity from the current-fragmentation region. In such a way, it is
guaranteed that the momentum fraction $x$ of the parton entering 
the hard scattering is much larger than the Bjorken-scaling variable $x_{bj}$,
and a large logarithm $\ln(x/x_{bj})$ arises. In such a case, gluon exchange
in the $t$ channel occurs at NLO (the leading order is the creation of
a quark pair or of a quark and a gluon, because the parton-model process
of a virtual photon knocking a quark off is constrained by $x=x_{bj}$, 
and thus it is
forbidden by the forward-jet requirement). Although the NLO analysis falls
short of describing the data~\cite{Aid:1995we,Breitweg:1998ed,Adloff:1998fa},
the LL BFKL prediction overshoots the data, calling for a NLL BFKL analysis
which so far has not yet been performed. Recently, the improved statistics 
have allowed for an analysis of three-jet production in DIS, with one 
forward jet out of the three jets~\cite{Aktas:2005up}.
This process offers the advantage of having gluon exchange
in the $t$ channel already at leading order, thus the NLO analysis guarantees
a better control over the theoretical uncertainties. In this case, the data
are in good agreement with the NLO prediction for three-jet 
production~\cite{Nagy:2001xb}.

$\gamma^*\gamma^*$ collisions in double-tag events, $e^+\, e^- \to e^+\, e^- +$ 
hadrons have been analysed by the LEP collaborations~\cite{Acciarri:1998ix,
Abbiendi:2001tv,Achard:2001kr,Heister:2003jb}. There is good agreement with 
the NLO analysis~\cite{Cacciari:2000cb}, except for the higher values of $Y$.
This is understandable, because gluon exchange in the $t$ channel
occurs only at NNLO. However, although a complete NNLO calculation is
beyond the reach of the present technology, the contributions which feature
gluon exchange in the $t$ channel can be included exactly in the theoretical
prediction. Doing so~\cite{DelDuca:2002qt} improves the agreement between
the data and the theory at the higher values of $Y$.

In conclusion, it is difficult to find strong, compelling evidence of
the BFKL resummation in the comparison between the data and the
theoretical analysis. That may be in part due to the fact that within
the kinematic reach of the present colliders, the asymptotic region
where the BFKL resummation is supposed to be applicable does not seem to
have been reached yet. But also to the fact that the LL BFKL predictions 
are plagued by large theoretical uncertainties, while the NLL resummation,
for which the analytic solution~\cite{Fadin:1998py} and numerical Monte Carlo 
studies~\cite{Schmidt:1996fg,Orr:1997im,Orr:1998hc,Andersen:2003an,
Andersen:2003wy} exist, is not in a form yet that can be readily applied
to the phenomenological analyses outlined above.

%%%%%%%%%%%%%%%%%%%%%%%%%%%%%%%%%%%%%%%%%%%%%%%%%%%%%%%%%%%%%%%%%%%%%%%%%%%%%
\section[Parton-parton luminosity functions for the LHC]{PARTON-PARTON
LUMINOSITY FUNCTIONS FOR THE LHC~\protect
\footnote{Contributed by: A.~Belyaev, J.~Huston, J.~Pumplin}}
\subsection{Introduction}

\begin{figure}[htb]
\begin{center}
\vskip -0.5cm
\includegraphics[width=0.5\textwidth]{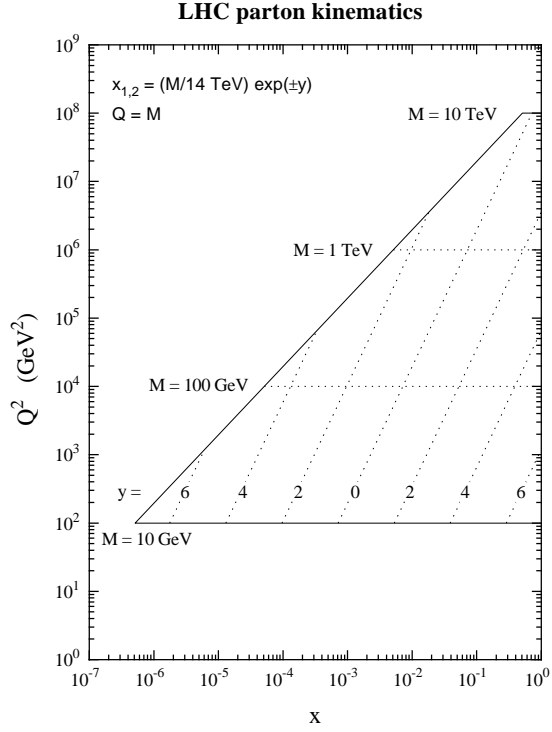}
\vskip -0.7cm
\caption{Parton kinematics for the LHC.\label{lhcgrid}}
\end{center}
\end{figure}

The number of events anticipated at the LHC for a process with a cross
section $\sigma$ can be calculated by multiplying the cross section times
the beam-beam luminosity. There are a number of programs available to
calculate cross sections for processes of interest at leading-order,
next-to-leading order and next-to-next-to-leading order, and in some cases
with parton showering and hadronization effects included~\cite{Dobbs:2004qw}. But it is
sometimes also useful to be able to make quick order-of-magnitude estimates
for the sizes of cross sections. For hard interactions, the collision is not
between the protons per se but between the partons in the two protons, 
carrying  fractions $x_1$ and $x_2$ of their parent proton's momentum. A plot showing the
parton kinematics at the LHC is shown in Fig.~\ref{lhcgrid}, indicating the relationship
between the two parton $x$ values and the mass $M=\sqrt{\hat{s}}$ 
and rapidity $y=\frac{1}{2}\ln(x_1/x_2)$ of the produced system
Thus, for example, a final state with a mass $M=100\,\mathrm{GeV}$ and a 
rapidity $y=4$ is produced by two partons with $x$ values of approximately 
$0.00015$ and $0.35$. 

Because the interacting partons carry only a fraction of the parent proton's
momentum, it is useful to define the differential parton-parton luminosity
$dL_{ij}/d\hat{s}\,dy$ and its integral $dL_{ij}/d\hat{s}$:
\begin{equation}
\frac{d L_{ij}}{d\hat{s}\,dy} = 
\frac{1}{s} \, \frac{1}{1+\delta_{ij}} \, 
[f_i(x_1,\mu) f_j(x_2,\mu) + (1\leftrightarrow 2)] \; .
\label{lum:eq1}
\end{equation}
The prefactor with the Kronecker delta avoids double-counting in case the
partons are identical.  The generic parton-model formula 
\begin{equation}
\sigma = \sum_{i,j} \int_0^1 dx_1 \, dx_2 \, 
f_i(x_1,\mu) \, f_j(x_2,\mu) \, \hat{\sigma}_{ij}
\label{lum:eq2}
\end{equation}
can then be written as 
\begin{equation}
\sigma = \sum_{i,j} \int \left(\frac{d\hat{s}}{\hat{s}} \, dy\right) 
\, \left(\frac{d L_{ij}}{d\hat{s}\,dy}\right) \, 
\left(\hat{s} \,\hat{\sigma}_{ij} \right) \; .
\label{lum:eq3}
\end{equation}
(This result is easily derived by defining $\tau = x_1 \, x_2 = \hat{s}/s$ 
and observing that the Jacobian 
$\frac{\partial(\tau,y)}{\partial(x_1,x_2)} = 1$.)

\begin{figure}[htb]
\includegraphics[width=0.5\textwidth,height=0.5\textwidth]{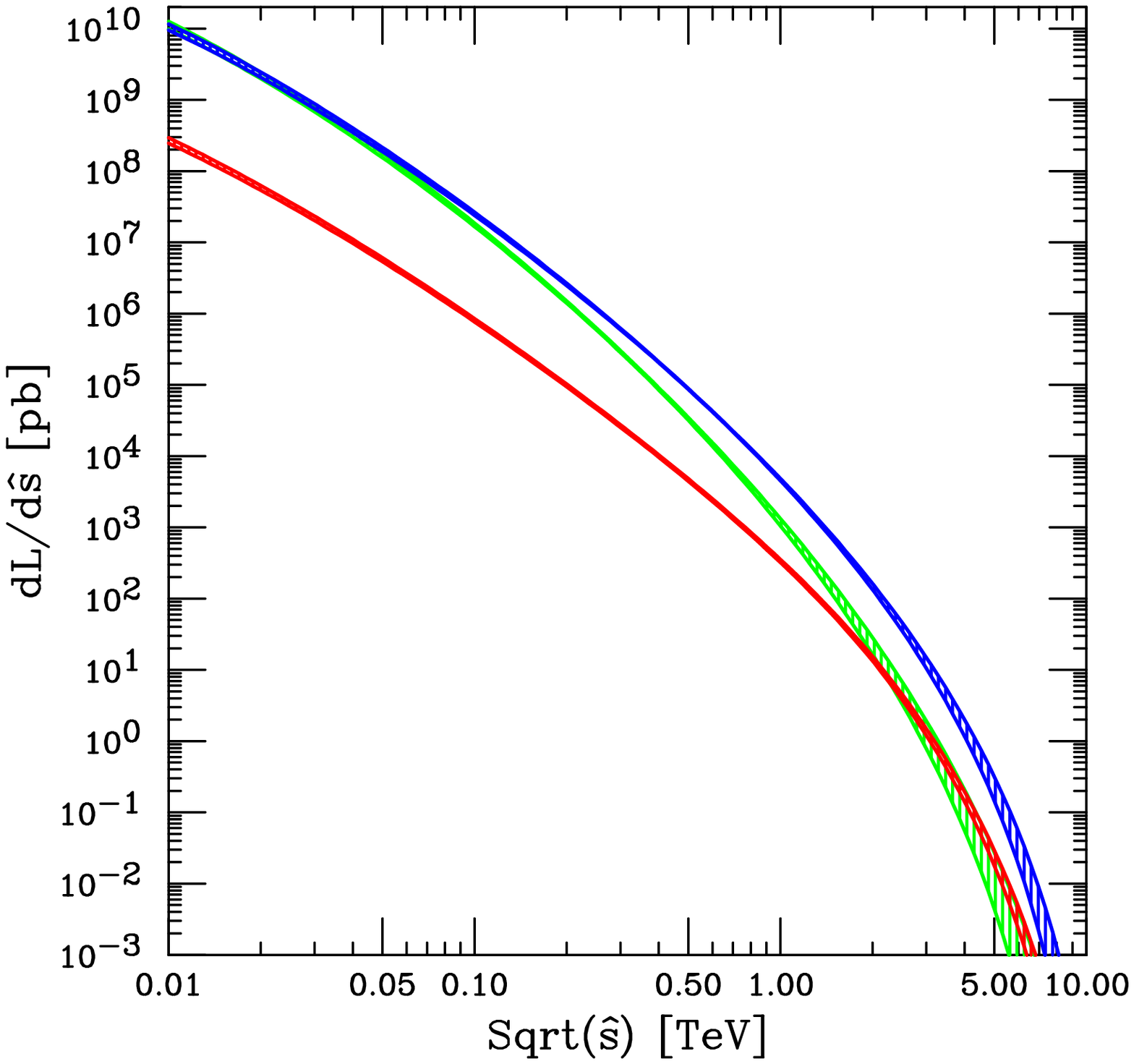}
\includegraphics[trim=0 25 0 0,width=0.53\textwidth]{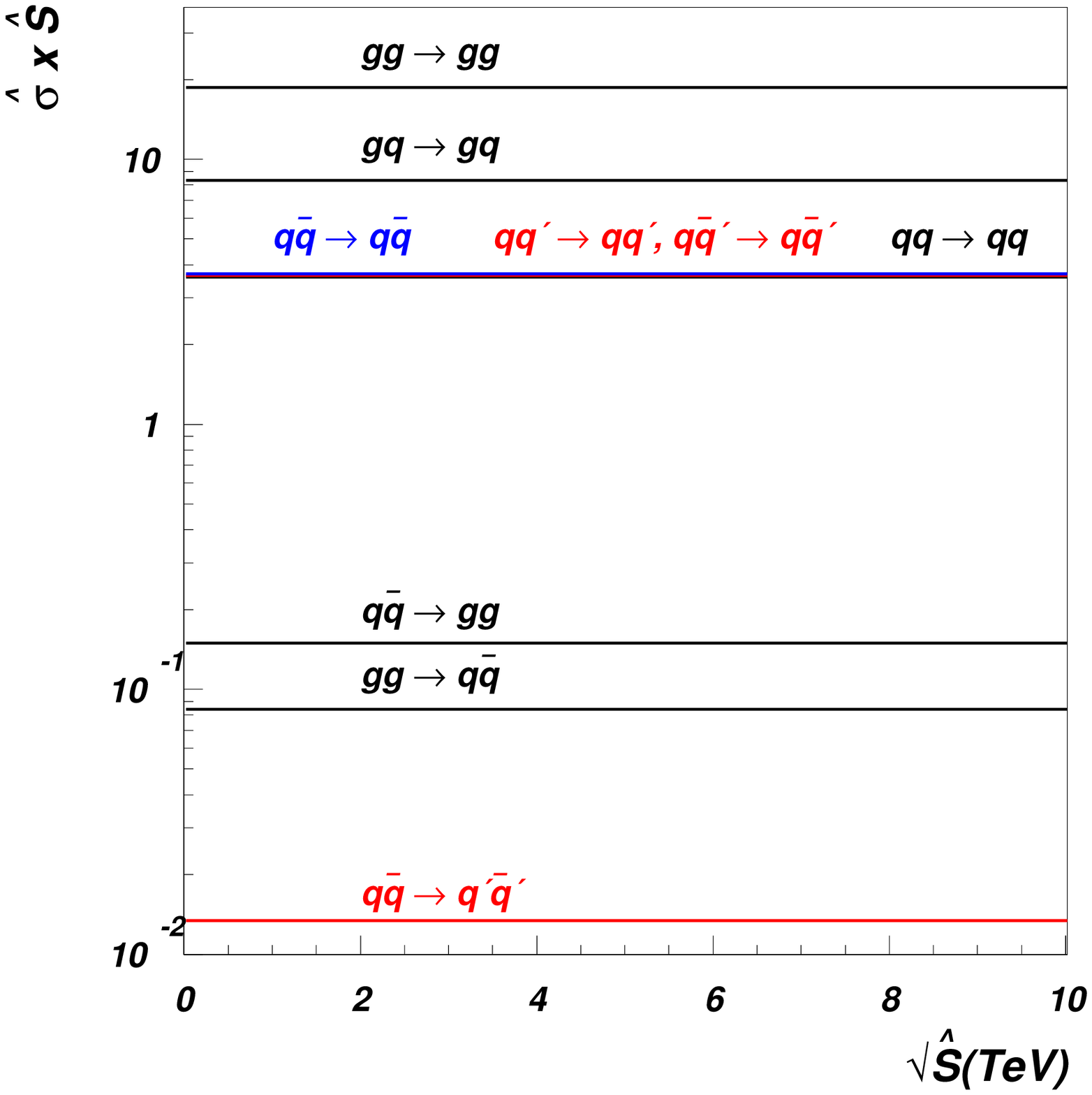}
\caption{ Left: 
luminosity$\left[\frac{1}{\hat{s}}\frac{dL_{ij}}{d\tau}\right]$ in pb
integrated over $y$. 
Green=$gg$, 
Blue=$g (d + u + s + c + b) 
 + g (\bar{d} + \bar{u} + \bar{s} + \bar{c} + \bar{b})
 + (d + u + s + c + b)g 
 + (\bar{d} + \bar{u} + \bar{s} + \bar{c} + \bar{b})g$,
Red=$d\bar{d} + u\bar{u} + s\bar{s} + c\bar{c} + b\bar{b}
 + \bar{d}d + \bar{u}u + \bar{s}s + \bar{c}c + \bar{b}b$.
Right: parton level cross sections
 \label{figlum4}$\left[\hat{s}\hat{\sigma}_{ij}\right]$
for various processes\label{sigma}}
\end{figure}

Equation~\ref{lum:eq3} can be used to estimate the production rate for a 
hard scattering process at the LHC as follows. 
Figure~\ref{figlum4}(left) shows a plot of the luminosity function integrated 
over rapidity,
$dL_{ij}/d\hat{s} = \int (dL_{ij}/d\hat{s}\,dy) \, dy$, at the LHC
value $\sqrt{s} = 14 \, \mathrm{TeV}$ for various parton flavor combinations,
calculated using the CTEQ6.1 parton distribution 
functions~\cite{Stump:2003yu}.  The widths of the curves indicate an estimate 
for the PDF uncertainties.  We assume $\mu = \sqrt{\hat{s}}$ for the scale.
(Similar plots made with earlier PDFs are shown in Ellis, Stirling, 
Webber~\cite{Ellis:1991qj}.) 
On the other hand, Fig.~\ref{sigma}(right)
presents the second product, 
$\left[\hat{s}\hat{\sigma}_{ij}\right]$,
for various $2 \rightarrow 2$ partonic processes.
The parton level cross sections are given for a
parton $p_T> 0.1\times\sqrt{\hat{s}}$ cut and for fixed $\alpha_s=0.118$.
We have used the CalcHEP package~\cite{Pukhov:2004ca} to estimate
these cross sections.
As expected, the $gg$ luminosity is large at low $\sqrt{\hat{s}}$ but falls rapidly
with respect to the other parton luminosities. The $gq$ luminosity is large over
the entire kinematic region plotted.

One can use  Equation~\ref{lum:eq3} in the form
\begin{equation}
\sigma=\frac{\Delta\hat{s}}{\hat{s}}
\left(\frac{d L_{ij}}{d \hat{s}}\right)
\left(\hat{s} \,\hat{\sigma}_{ij} \right).
\end{equation}
and Fig.~\ref{sigma}
to estimate  the production cross sections for  QCD jets
for a given $\Delta \hat{s}$ interval.
For example, for the gluon  pair production rate
for $\hat{s}$=1~TeV  and $\Delta\hat{s}=0.01\hat{s}$,
we have  $\frac{d L_{gg}}{d \hat{s}}\simeq 10^3$~pb and 
$\hat{s} \,\hat{\sigma}_{gg}\simeq 20$
leading to $\sigma\simeq 200$~pb 
(for the $p_T^g> 0.1\times\sqrt{\hat{s}}$ cut we have assumed above).
Note that for a given small $\Delta\hat{s}/\hat{s}$ interval,
the corresponding  invariant mass 
$\Delta\sqrt{\hat{s}}/\sqrt{\hat{s}}$ interval, is 
$\Delta\sqrt{\hat{s}}/\sqrt{\hat{s}}\simeq \frac{1}{2}\Delta\hat{s}/\hat{s}$.
One should also mention  that all hard cross sections presented in Fig.\ref{sigma}
are proportional  to $\alpha_s^2$ and have been calculated for
$\alpha_s=0.118$, so production rates can be easily rescaled 
for a certain  $\alpha_s$ at a given scale.

\begin{figure}
\begin{center}
\includegraphics[scale=1.0]{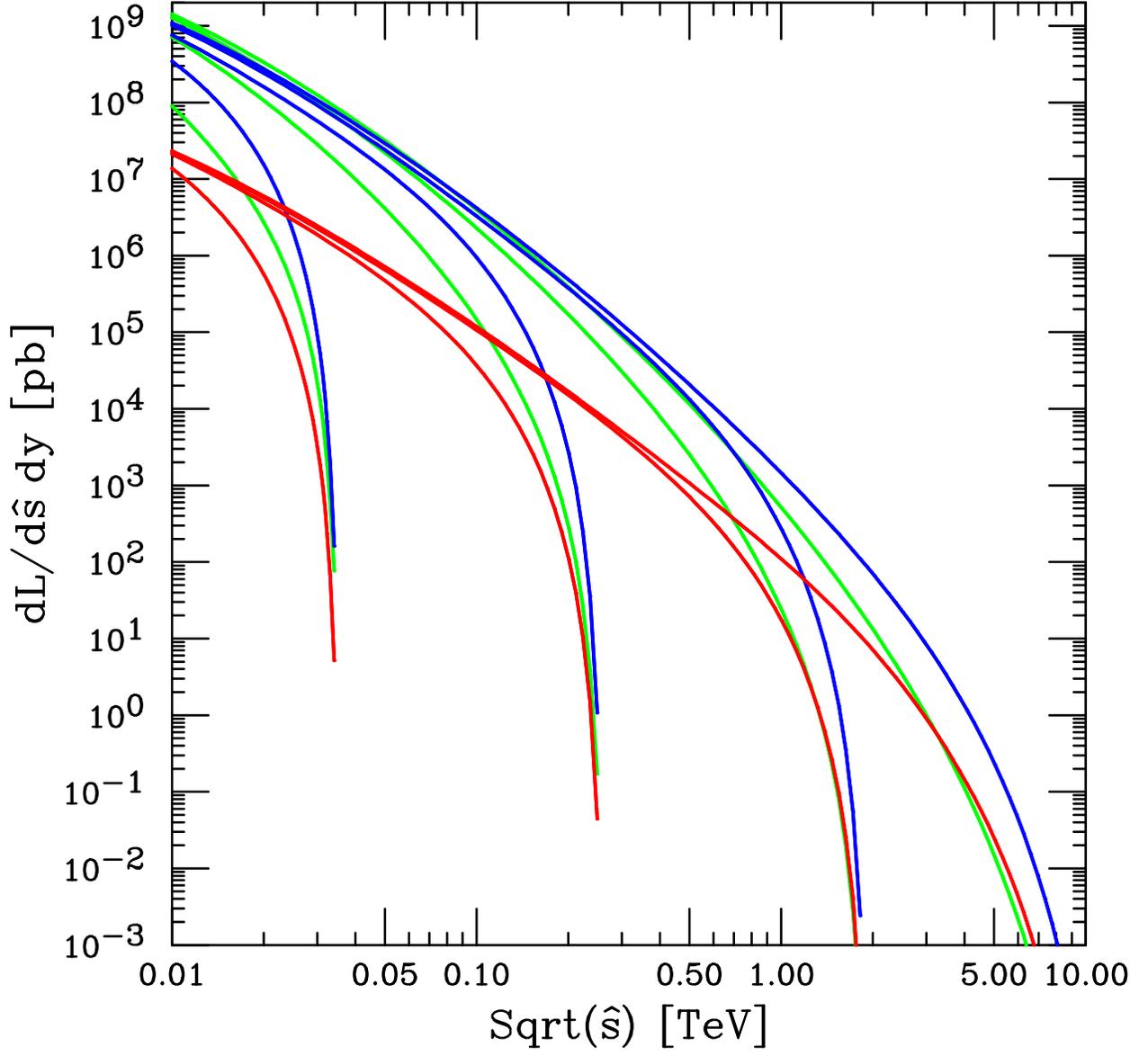}
\end{center}
\caption{ dLuminosity/dy at $y=0, 2, 4, 6$.
Green=$gg$, 
Blue=$g (d + u + s + c + b) 
  + g (\bar{d} + \bar{u} + \bar{s} + \bar{c} + \bar{b})
  + (d + u + s + c + b)g 
  + (\bar{d} + \bar{u} + \bar{s} + \bar{c} + \bar{b})g$,
Red=$d\bar{d} + u\bar{u} + s\bar{s} + c\bar{c} + b\bar{b}
  + \bar{d}d + \bar{u}u + \bar{s}s + \bar{c}c + \bar{b}b$.
  \label{figlum5}}
\end{figure}

One can further specify the parton-parton luminosity for a specific 
rapidity
$y$ and $\hat{s}$, $dL_{ij}/d\hat{s}\, dy$. 
If one is interested in a specific partonic initial state, then the resulting
differential luminosity can be displayed in families of curves as shown in
Fig.~\ref{figlum5}, where the differential parton-parton luminosity at the LHC is
shown as a function of the subprocess center-of-mass energy $\sqrt{\hat{s}}$ at
various values of rapidity for the produced system for several different
combinations of initial state partons. One can read from the curves the
parton-parton luminosity for a specific value of mass fraction and 
rapidity. (It is also easy to use the Durham PDF plotter to generate the pdf
curve for any desired flavor and kinematic configuration~\cite{durham}.) 

\begin{figure}
\begin{center}
\includegraphics[scale=0.6]{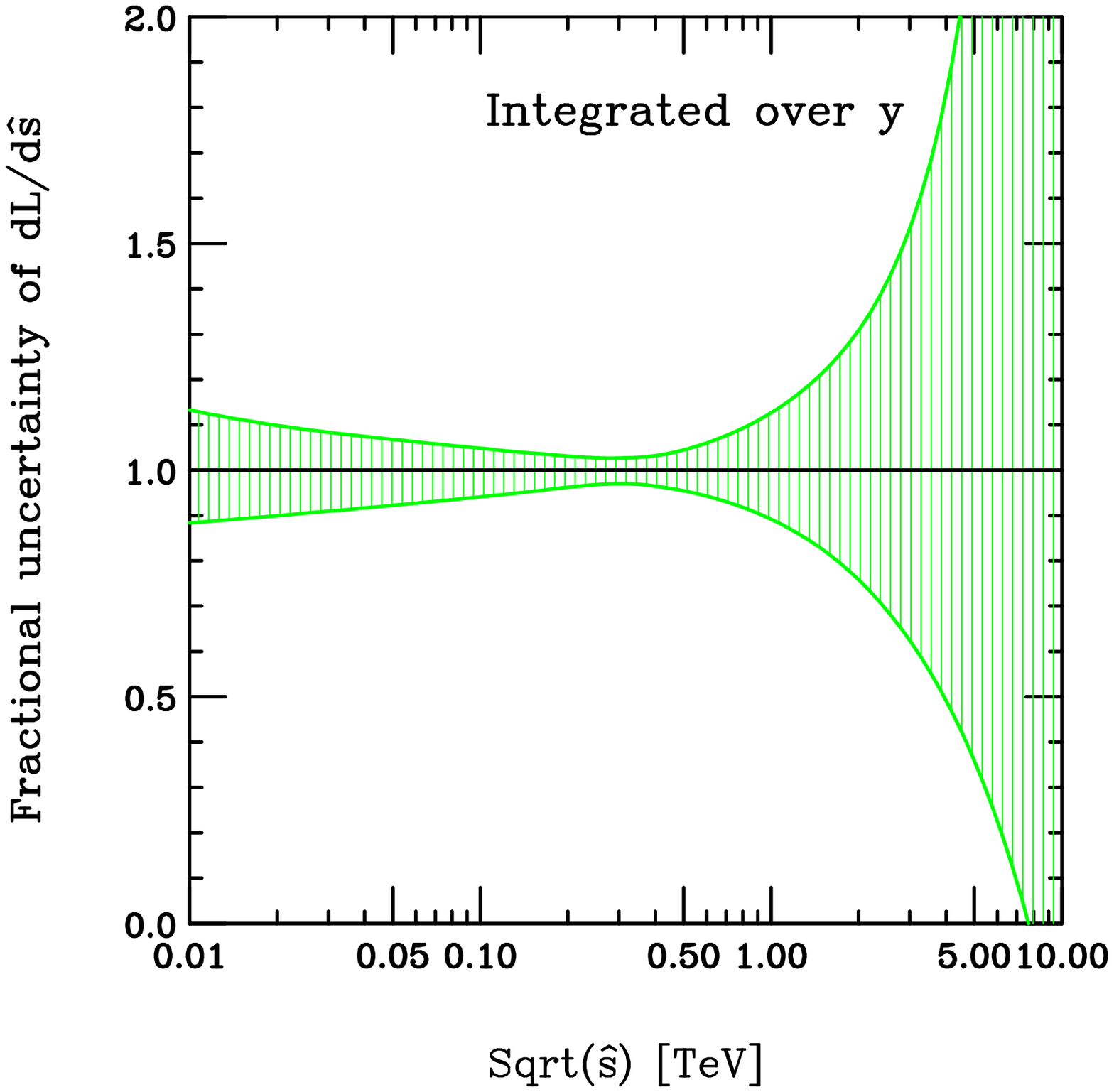}
\end{center}
\caption{ Fractional uncertainty of $gg$ luminosity integrated over $y$. \label{figlum6}}
\end{figure}

It is also of great interest to understand the uncertainty for the parton-parton luminosity
for specific kinematic configurations. Some representative parton-parton luminosity
uncertainties are shown in Figs.~\ref{figlum6}-\ref{figlum8}. The PDF uncertainties were generated from the CTEQ6.1
Hessian error analysis using the standard $\Delta \chi^2$  100 criterion. Except for
kinematic  regions where one or both partons is a gluon at high $x$, the pdf uncertainties
are of  the order of 5-10\%. Even tighter constraints will be possible once the LHC
Standard Model  data is included in the global pdf fits. Again, the uncertainties for
individual PDF's can also be calculated online using the Durham pdf plotter. 

\begin{figure}
\begin{center}
\includegraphics[scale=0.6]{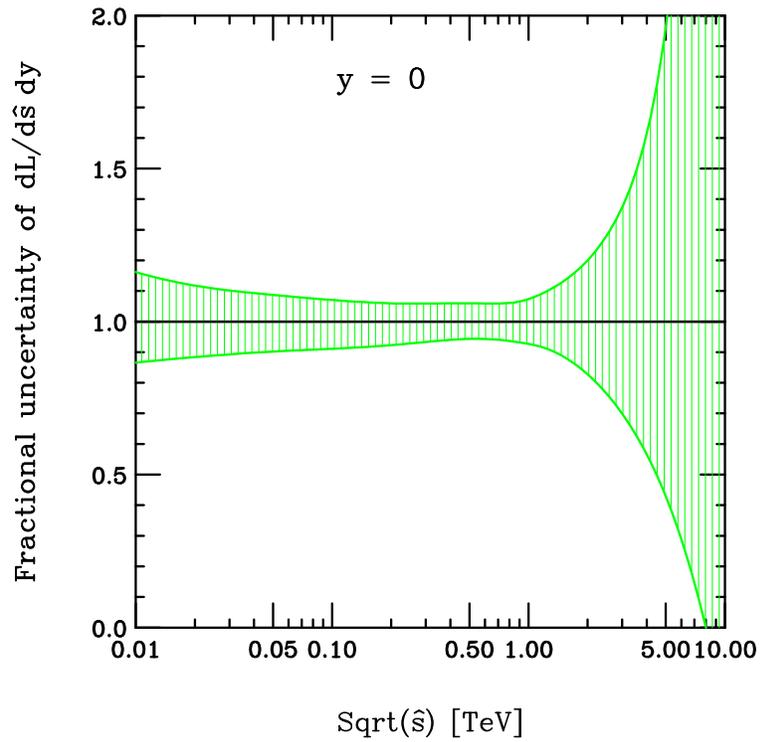}
\end{center}
\caption{ Fractional uncertainty of $gg$ luminosity at $y=0$. }
\end{figure}

\begin{figure}
\begin{center}
\includegraphics[scale=0.6]{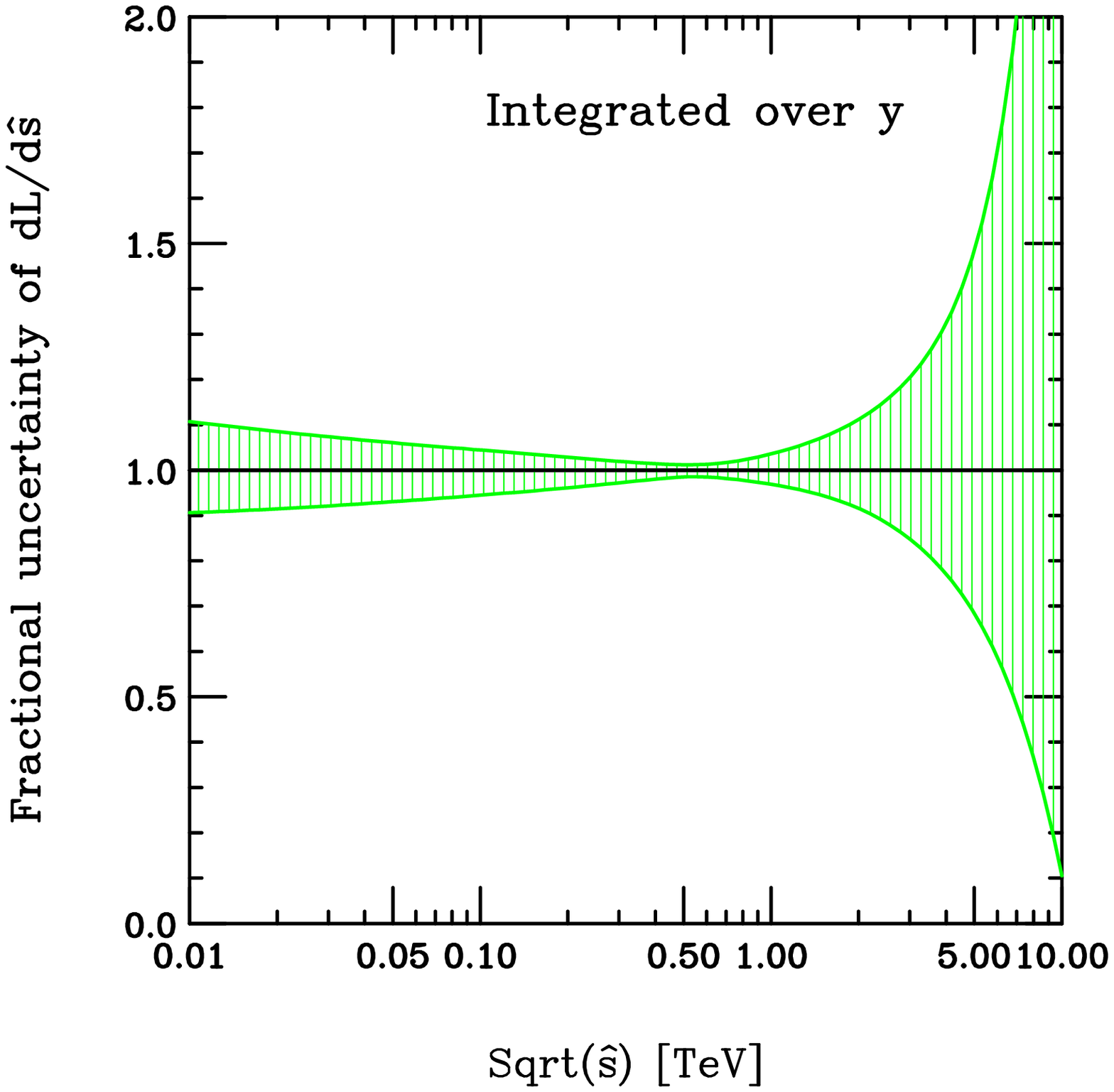}
\end{center}
\caption{ 
Fractional uncertainty for Luminosity integrated over $y$
for 
$g (d + u + s + c + b) 
  + g (\bar{d} + \bar{u} + \bar{s} + \bar{c} + \bar{b})
  + (d + u + s + c + b)g 
  + (\bar{d} + \bar{u} + \bar{s} + \bar{c} + \bar{b})g$,
}
\end{figure}

\begin{figure}
\begin{center}
\includegraphics[scale=0.6]{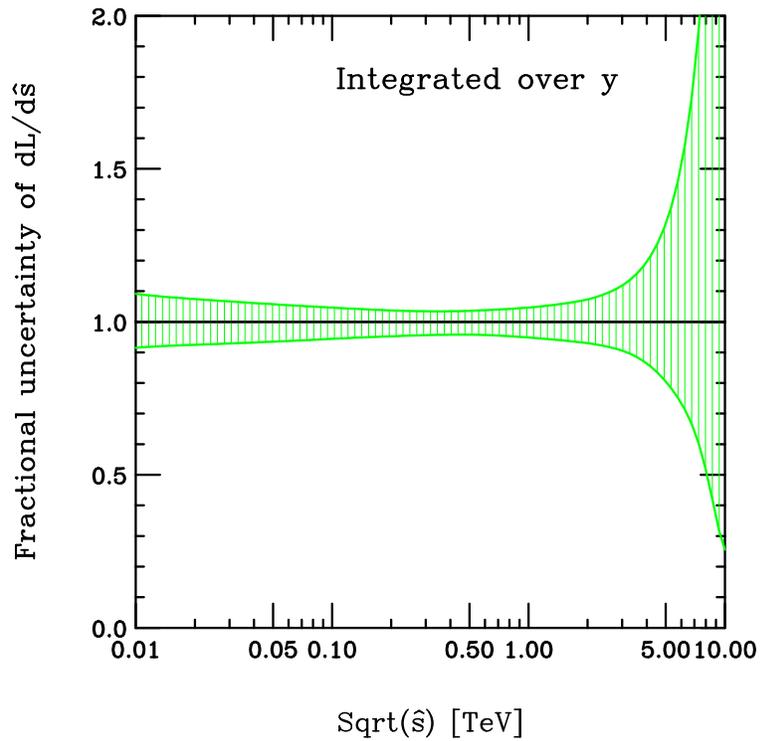}
\end{center}
\caption{ 
Fractional uncertainty for Luminosity integrated over $y$
for 
$d\bar{d} + u\bar{u} + s\bar{s} + c\bar{c} + b\bar{b}
  + \bar{d}d + \bar{u}u + \bar{s}s + \bar{c}c + \bar{b}b$.
\label{figlum8}}
\end{figure}

\subsection{Conclusions}

Some representative parton-parton luminosity and luminosity uncertainty plots
have been presented. A more complete set will be maintained at the Standard
Model benchmark website started at Les Houches 2005:
\verb|www.pa.msu.edu/~huston/Les_Houches_2005/Les_Houches_SM.html| and will also be included in a
review article to be published in the near future.

\subsection*{Acknowledgements}
We would like to thank J.~Campbell, W.J.~Stirling and W.K.~Tung for useful discussions and comments.

%%%%%%%%%%%%%%%%%%%%%%%%%%%%%%%%%%%%%%%%%%%%%%%%%%%%%%%%%%%%%%%%%%%%%%%%%%%%%
\section[A simple model for large-$x$ resummed parton distributions]
{A SIMPLE MODEL FOR LARGE-$x$ RESUMMED PARTON DISTRIBUTIONS~\protect
\footnote{Contributed by: G.~Corcella, L.~Magnea}}
The achievement of precision goals of the LHC and other high-energy
colliders crucially depends on the knowledge of parton distribution
functions (PDF's). One of the phase space regions in which parton
densities are less constrained is the large-$x$ region, where limited
data exist and NLO fits do not work well due to the presence of large
higher-order and power corrections. The range of applicability of
perturbative methods can be extended in this region by applying 
soft-gluon resummation techniques, which are available for many hard
processes. Here we will be concerned with estimating the effects of
resummation on PDF's. To this end, we will present a
simple analysis of Deep Inelastic Scattering (DIS) structure function
data, which will allow us to extract next-to-leading order (NLO) and
resummed parton distribution functions at large $x$.

DIS structure functions $F_i(x,Q^2)$ are given by the convolution of
coefficient functions and PDF's. NLO coefficient functions are known
to contain terms that become large and dominant at large $x$,
originating from soft and collinear gluon emission.  These
contributions need to be resummed to all orders to improve the
validity of the perturbative prediction.  Large-$x$ resummation for
the DIS coefficient function was performed in
\cite{Sterman:1986aj,Catani:1989ne} in the massless approximation, and
in \cite{Laenen:1998kp,Corcella:2003ib} with the inclusion of
quark-mass effects, which are relevant when the focus is on heavy
quark production.

Soft resummation is naturally performed in Mellin moment space, where
large-$x$ terms correspond, at ${\cal O}(\alpha_s)$, to single
($\alpha_s \ln N$) and double ($\alpha_s \ln^2 N$) logarithms of the
variable $N$.  Such logarithms exponentiate in a Sudakov form factor:
in the next-to-leading logarithmic (NLL) approximation, terms ${\cal
O} (\alpha_s^n \ln^{n+1} N )$ (LL) and ${\cal O} (\alpha_s^n \ln^n N)$
(NLL) are resummed in the Sudakov exponent.  Using large-$x$ resummed
coefficient functions, we can extract resummed PDF's from DIS
structure function data, and compare them with a NLO fit.  We shall
consider recent charged-current (CC) data from neutrino-iron
scattering, collected by the NuTeV collaboration \cite{Tzanov:2005kr},
and neutral-current (NC) data from the NMC \cite {Arneodo:1996qe} and
BCDMS \cite{Benvenuti:1989rh,Benvenuti:1989fm} collaborations.

\begin{figure}[ht!]
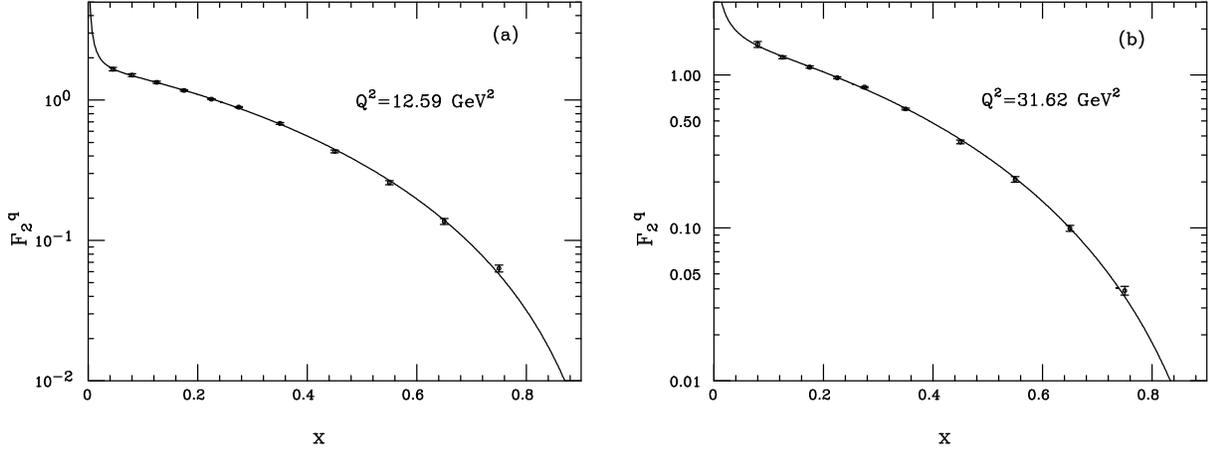

\centerline{\resizebox{0.48\textwidth}{!}{\includegraphics{s_corcella2/f2_1259.ps}}%
\hfill%
\resizebox{0.48\textwidth}{!}{\includegraphics{s_corcella2/f2_3162.ps}}}
\caption{NuTeV data and best-fit curves for the 
structure function $F_2^q$ at $Q^2=12.59$~GeV$^2$ (a) 
and 31.62 GeV$^2$ (b).}
\label{fits}
\end{figure}

\begin{figure}[ht!]
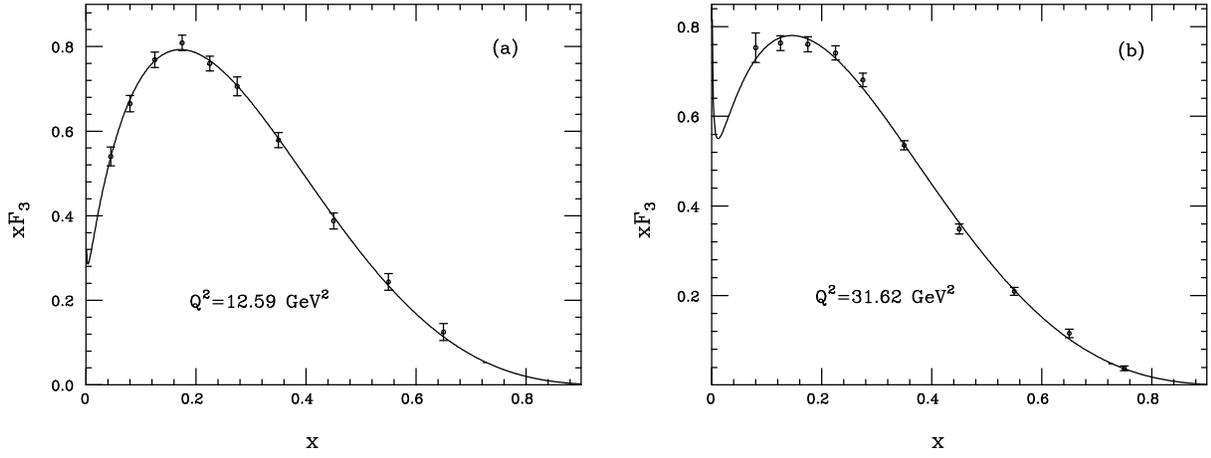

\centerline{\resizebox{0.48\textwidth}{!}{\includegraphics{s_corcella2/f3_1259.ps}}%
\hfill%
\resizebox{0.48\textwidth}{!}{\includegraphics{s_corcella2/f3_3162.ps}}}
\caption{As in Fig.~\ref{fits}, but for the structure function $F_3$.} 
\label{fits1}
\end{figure}

We have used NuTeV data on $F_2(x)$ and $x F_3(x)$ at the test values
of $Q^2 = 31.62$~GeV$^2$ and 12.59~GeV$^2$. The structure function
$F_2$ contains a gluon-initiated contribution $F_2^g$, which is not
soft-enhanced and is very small at large $x$: we extracted 
$F_2^g$ from a global fit, e.g. CTEQ6M \cite{Pumplin:2002vw}, and
limited our fit to the quark-initiated term $F_2^q$. We chose a
parametrization of the form $F_2^q (x) = F_2 (x) - F_2^g (x) = A x^{-
\alpha} (1 - x)^\beta (1 + b x)$; $ x F_3(x) = C x^{-\rho} ( 1 - x
)^\sigma ( 1 + k x )$. The best-fit parameters and the $\chi^2$ per
degree of freedom from the fit are quoted in
\cite{Corcella:2005us}. In Figs.~\ref{fits} and \ref{fits1}, we
present the NuTeV data on $F_2(x)$ and $x F_3(x)$ at $Q^2 =
12.59$~GeV$^2$, along with the best-fit curves.
\begin{figure}[ht!]
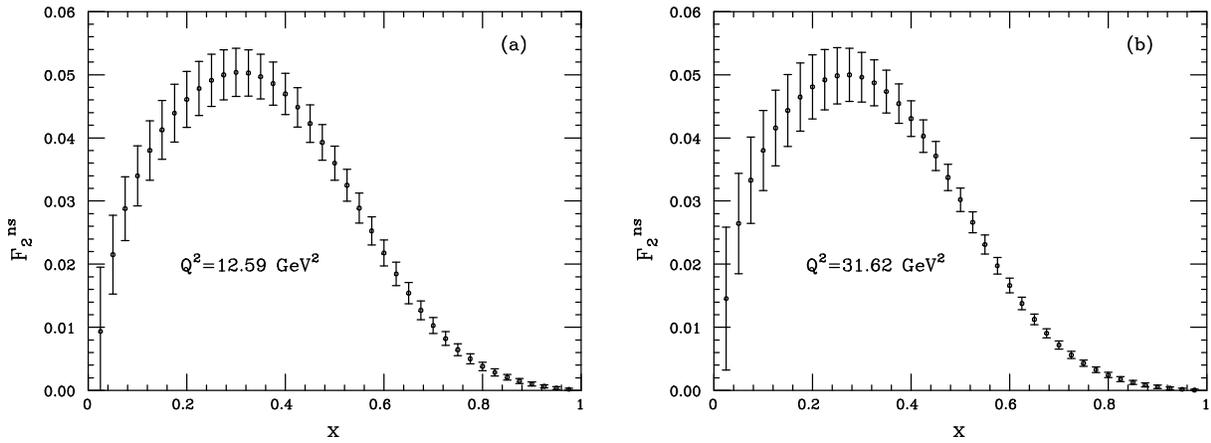

\centerline{\resizebox{0.48\textwidth}{!}{\includegraphics{s_corcella2/nn1259.ps}}%
\hfill%
\resizebox{0.48\textwidth}{!}{\includegraphics{s_corcella2/nn3162.ps}}}
\caption{A sampling of the neural parametrization of NMC and BCDMS
data for $F_2^{\rm ns} (x, Q^2)$ at $Q^2 = 12.59 ~{\rm GeV}^2$ (a) and
at $Q^2 = 31.62~{\rm GeV}^2$ (b), from the NNPDF
Collaboration~\cite{DelDebbio:2004qj}.}
\label{nn}
\end{figure}
In order to extract individual quark distributions, we need to
consider also NC data. We use NMC and BCDMS results, and
employ the parametrization of the nonsinglet structure function
$F_2^{\mathrm{ns}} = F_2^p - F_2^D$ provided by the NNPDF
collaboration \cite{DelDebbio:2004qj,Forte:2002fg}.  The
parametrization~\cite{DelDebbio:2004qj,Forte:2002fg} is based on
neural networks trained on Monte-Carlo copies of the data set, which
include all information on errors and correlations: this gives an
unbiased representation of the probability distribution in the space
of structure functions.  Fig.~\ref{nn} shows $F_2^{\rm ns} (x, Q^2)$,
computed with the neural parametrization at our chosen values of
$Q^2$, for $x = n/40$, $n = 1, \ldots, 39$.  The central values are
given by the averages of the results obtained with the one thousand
neural networks of the NNPDF collaboration, and the error bars are the
corresponding standard deviations. The errors are quite large, because
$F_2^{\rm ns} (x, Q^2)$ is the difference between proton and deuteron
structure functions, which implies a loss of precision.

Writing $F_2$, $x F_3$ and $F_2^\mathrm{ns}$ in terms of their parton
content, we can extract NLO and NLL-resummed quark distributions,
according to whether we use NLO or NLL coefficient functions. In order
to solve for individual quark distributions, we assume isospin
symmetry of the sea, i.e. $s = \bar s$ and $\bar u = \bar d$, we
neglect the charm density, and impose a relation of the form $\bar s =
\kappa \, \bar u$. We obtain a system of three equations, explicitly
presented and solved in \cite{Corcella:2005us} in terms of $u$, $d$
and $s$. We begin by working in $N$-space, where the resummation has a
simpler form and quark distributions are just the ratio of the
appropriate structure function and coefficient function. We will then
invert the results to $x$-space using a simple parametrization, $q(x)
= D x^{-\gamma}(1 - x)^\delta$.

\begin{figure}[ht!]
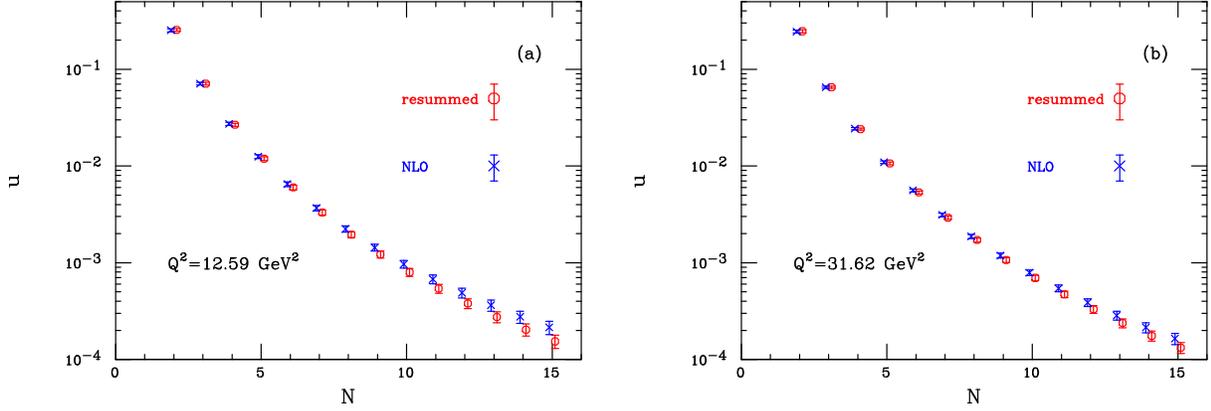

\centerline{\resizebox{0.48\textwidth}{!}{\includegraphics{s_corcella2/un2.ps}}%
\hfill%
\resizebox{0.48\textwidth}{!}{\includegraphics{s_corcella2/un3.ps}}}
\caption{NLO and resummed moments of the
up quark distribution at $Q^2 = 12.59$~GeV$^2$ (a) and 31.62 GeV$^2$.}
\label{up}
\end{figure}

\begin{figure}[ht!]
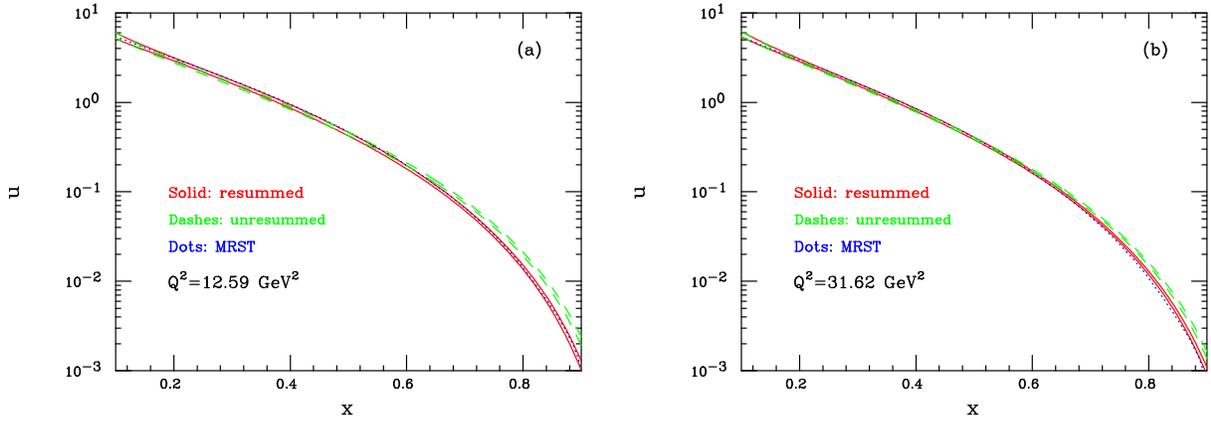

\centerline{\resizebox{0.48\textwidth}{!}{\includegraphics{s_corcella2/uxa.ps}}%
\hfill%
\resizebox{0.48\textwidth}{!}{\includegraphics{s_corcella2/uxb.ps}}}
\caption{NLO and NLL up-quark density in $x$-space at $Q^2=$~12.59
GeV$^2$ (a) and 31.62 GeV$^2$ (b).  Plotted are the edges of
statistical bands at one-standard-deviation confidence level.  For the
sake of comparison, the MRST2001 result is also shown.}
\label{up1}
\end{figure}

Figs.~\ref{up}--\ref{up1} show the effect of the resummation on the
up-quark distribution at $Q^2 = 12.59$ and $31.62$~GeV$^2$, in $N$-
and $x$-space, for $\kappa = 1/2$.  As for the result in $x$-space,
the best-fit values of $D$, $\gamma$ and $\delta$, along with the
$\chi^2/\mathrm{dof}$, are quoted in Table~\ref{tabb}.

The impact of the resummation is noticeable at large $N$ and $x$: the
coefficient function and its moments in that region are enhanced by
the resummation, and therefore quark densities are suppressed when
extracted from a given set of structure function data.  The effect is
larger at $Q^2 = 12.59$~GeV$^2$, as expected from the running of the
strong coupling. In Fig.~\ref{up1} we also present the up-quark
density according to the MRST2001 set \cite{Martin:2002dr},
in the NLO approximation. Given
the various approximations which we made in our analysis, the limited
data set and the emphasis on large-$x$ data, we do not expect that our
results should agree with the MRST2001 global fit. We note however
that at low $x$ the MRST2001 up-quark distribution is within the error
range of the densities extracted form our fit. At large $x$ it looks
closer to our NLL-resummed PDF rather than to the NLO one. In fact, as
observed in \cite{Corcella:2005ig}, the MRST2001 set was fitted to
CCFR structure function data \cite{Yang:2000ju}, which are lower than
NuTeV at large $x$. It is therefore reasonable that the NLO MRST2001
PDF's be lower than the NLO ones which we extracted from a fit to
NuTeV, and therefore closer to our resummed PDF's.  The discrepancy
between NuTeV and CCFR at large $x$ is now described as understood
\cite{Tzanov:2005kr}.  In principle, also $d$ and $s$ densities are
affected by the resummation; we found, however, that the errors on
these PDF's within our fit are too large to display sensitivity to
soft resummation.
\begin{table}[ht!]
\caption{Best-fit values and errors for the up-quark $x$-space parametrization,
at the chosen values of $Q^2$.}
  \vspace*{1mm}
\centering{
\begin{tabular}{llllccccc}\\
\hline
$\hspace{3mm} Q^2$ & PDF & $ \hspace{1cm} D$ & $ \hspace{1cm} \gamma$ & 
$\hspace{2mm} \delta$ \\ 
\hline
12.59 & NLO & $3.025 \pm 0.534$ & $0.418 \pm 0.101$ & $3.162 \pm 0.116$ \\
      & RES & $4.647 \pm 0.881$ & $0.247 \pm 0.109$ & $3.614 \pm 0.128$ \\
31.62 & NLO & $2.865 \pm 0.420$ & $0.463 \pm 0.086$ & $3.301 \pm 0.098$ \\
      & RES & $3.794 \pm 0.583$ & $0.351 \pm 0.090$ & $3.598 \pm 0.104$ \\
\hline \\
\end{tabular}}
\label{tabb}
\end{table}
\begin{figure}
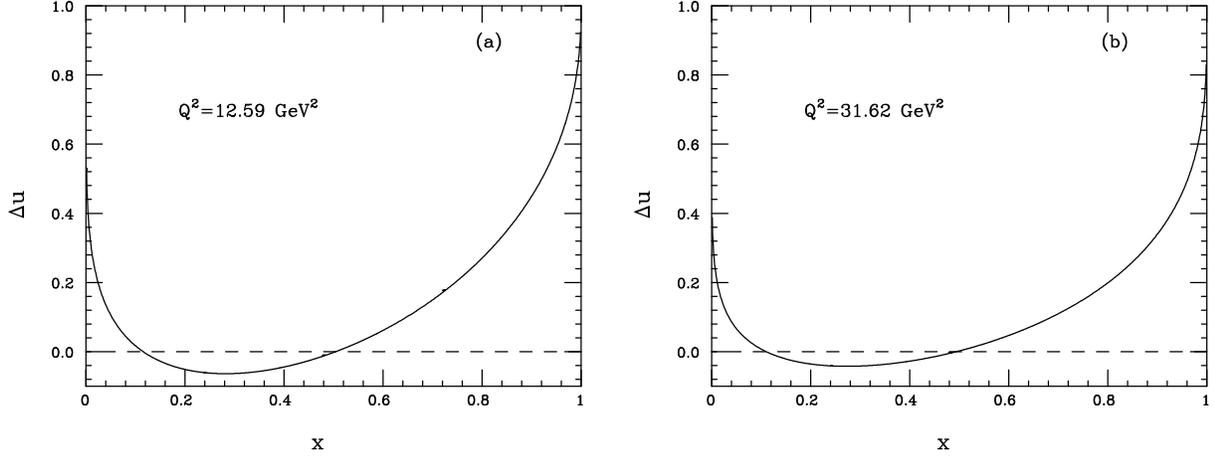

\centerline{\resizebox{0.48\textwidth}{!}{\includegraphics{s_corcella2/deltau1.ps}}%
\hfill%
\resizebox{0.48\textwidth}{!}{\includegraphics{s_corcella2/deltau.ps}}}
\caption{Relative effect of large-$x$ resummation on the
up-quark density at $Q^2 = 12.59$~GeV$^2$ (a) and 31.62 GeV$^2$.}
\label{delta}
\end{figure}
\par Fig.~\ref{delta} shows the impact of the resummation on the ratio
$\Delta u(x) = \left(u_{\rm NLO} (x) - u_{\rm res} (x) \right)/u_{\rm
NLO} (x)$, at both values of $Q^2$, for the central values of the
best-fit parameters, as quoted in Table~\ref{tabb}.  We observe that
the suppression of the resummed up quark distribution with respect to
the NLO one is about $5\%$ at $x \simeq 0.58$, $10\%$ at $x \simeq
0.65$ and $20\%$ at $x \simeq 0.75$ for $Q^2 = 12.59$ GeV$^2$, while
for $Q^2 = 31.62$ GeV$^2$ the same suppression factors are reached at
$x \simeq 0.61$, $x \simeq 0.69$ and $x \simeq 0.8$, respectively.

\begin{figure}[ht!]
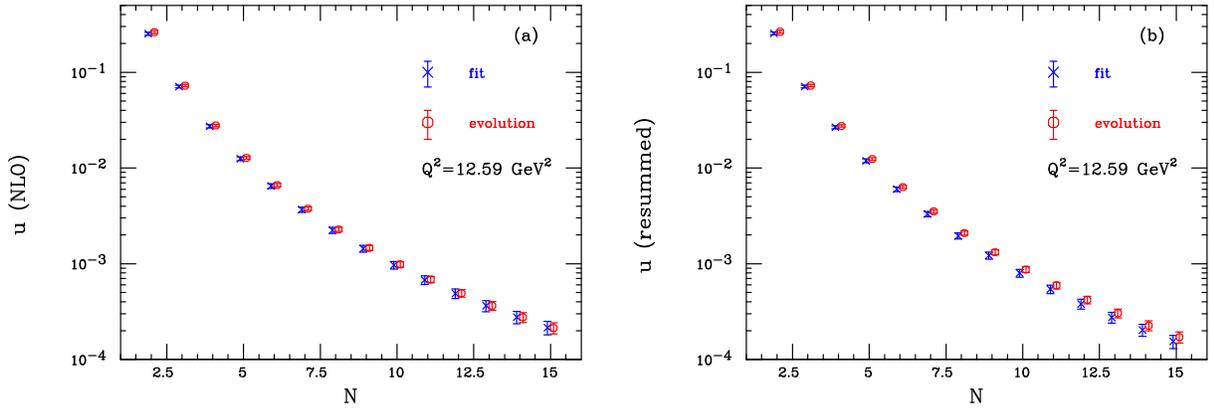

\centerline{\resizebox{0.48\textwidth}{!}{\includegraphics{s_corcella2/evnlo1.ps}}%
\hfill%
\resizebox{0.48\textwidth}{!}{\includegraphics{s_corcella2/evres2.ps}}}
\caption{Comparison of fitted moments of the up-quark
distributions at $Q^2=12.59$~GeV$^2$, with moments obtained via NLO evolution 
from $Q^2 = 31.62$~GeV$^2$.} 
\label{figev}
\end{figure}

We note that our results on fixed-order and resummed quark
distributions at the two values of $Q^2$ are consistent with NLO
perturbative evolution.  This is shown in Fig.~\ref{figev}: NLO
and NLL-resummed moments obtained from a fit of the data at 12.59
GeV$^2$ are consistent with the ones obtained via NLO evolution from
the values fitted at 31.62 GeV$^2$, just within one standard
deviation. It should be observed, however, that the evolution of
resummed moments is less consistent than the NLO one, which might be
due to effects of power corrections, which are entangled to the
resummation and have not been accounted for in our work.
\par In conclusion, we have fitted fixed-target large-$x$ DIS
structure function data, and extracted NLO and NLL-resummed parton
densities.  We found an impact of the resummation on valence quark
densities, which are suppressed by about $10-20 \%$ at $x > 0.5$ and
moderate $Q^2$. Our results show that higher-order perturbative
effects should not be neglected in parton fits at large $x$. A NNLO
analysis \cite{Martin:2002dr} is bound to include some of the effects
discussed here and is a step in right direction. Whenever resummed
hard partonic cross sections are employed, however, it would be
desirable to have at hand resummed PDF's as well: our results in fact
show that in some cases the Sudakov enhancement of the partonic cross
section which is typical of soft resummation could be partly
compensated by a suppression of large-$x$ partons when the same
physical effects are consistently included in their determination.

%%%%%%%%%%%%%%%%%%%%%%%%%%%%%%%%%%%%%%%%%%%%%%%%%%%%%%%%%%%%%%%%%%%%%%%%%%%%%
\section[Bottom-quark fragmentation: from $e^+e^-$ data to top and Higgs
decays]
{BOTTOM-QUARK FRAGMENTATION: FROM $e^+e^-$ DATA TO TOP AND HIGGS
DECAYS~\protect\footnote{Contributed by: G.~Corcella, V.~Drollinger}}
\subsection{Introduction}
We investigate $B$-hadron production in $e^+e^-$ annihilation
($e^+e^-\to b\bar b$), top decay ($t\to bW$) and the Standard-Model-Higgs 
decay $H\to b\bar b$, which is relevant at the LHC for $m_H<135$~GeV, and 
to Higgs production in association with vector bosons \cite{Drollinger:2002uj}
or $t\bar t$ pairs \cite{Richter-Was:1999sa,Drollinger:2001ym}.
 
We describe $b$-quark production using resummed 
calculations, based on the formalism of perturbative fragmentation functions
\cite{Mele:1990cw}, and the HERWIG \cite{Corcella:2000bw}
and PYTHIA \cite{Sjostrand:2001yu,Sjostrand:2003wg} 
Monte Carlo event generators.
We use $e^+e^-$ data on the $B$-hadron spectrum collected by the
SLD \cite{Abe:1999ki}, ALEPH \cite{Heister:2001jg} 
and OPAL \cite{Abbiendi:2002vt}
experiments to fit the cluster and string 
hadronization models, implemented by HERWIG 
and PYTHIA, and
the Kartvelishvili non-perturbative
fragmentation function \cite{Kartvelishvili:1977pi}, 
used in the framework of the resummed calculation.
We shall employ the fitted models to predict the $B$-energy 
distribution in top and Higgs decays.
Furthermore, we shall use data from DELPHI \cite{delphi}
in Mellin moment space to extract
the moments of the non-perturbative fragmentation function and predict the
$N$-space $B$-spectrum in $t\to bW$ and $H\to b\bar b$.

\subsection{Bottom-quark production and multi-parton radiation}

We shall first consider $b$-quark production at the next-to-leading order 
(NLO) in $Z$, $H$ and top  decays:
\begin{equation}
P(Q) \to b(p_b) \bar b(p_{\bar b}) \left( g(p_g)\right),
\end{equation}
with $P=Z$ or $H$, and
\begin{equation}
t(Q) \to b(p_b)  W(p_W) \left( g(p_g)\right).
\end{equation}
We shall neglect powers of $(m_b^2/Q^2)^p$ and consider the $b$-quark
energy fraction 
\begin{equation}
x_b={1\over{1-w}}{{2p_b\cdot Q}\over{Q^2}},
\label{xb}
\end{equation}
with $w=0$ in Higgs and $Z$ decays and $w=m_W^2/m_t^2$ in $t\to bW$.
The approach of perturbative fragmentation functions \cite{Mele:1990cw}
expresses the $x_b$ 
distribution as the convolution of a massless coefficient
function and a perturbative fragmentation function $D(m_b,\mu_F)$,
associated with the transition of a light parton into a heavy quark:
\begin{equation}
{1\over {\Gamma_0}} {{d\Gamma_b}\over{dx_b}} (x_b,Q,m_b) =
\int_{x_b}^1{{{dz}\over z}\left[{1\over{\Gamma_0}}
{{d\hat\Gamma_b}\over {dz}}(z,Q,\mu,\mu_F)
\right]^{\overline{\mathrm{MS}}}
D_b^{\overline{\mathrm{MS}}}\left({x_b\over z},\mu_F,m_b \right)}.
\label{pff}
\end{equation}
In Eq.~(\ref{pff}), $d\hat\Gamma_b /dz$ is the differential width for the 
production of a massless $b$,
after subtraction of the collinear singularity in the \CDmsbar factorization
scheme (\CDmsbar coefficient function), $\mu$ and $\mu_F$ are the
renormalization and factorization scales.
The NLO coefficient functions have been computed in 
\cite{Mele:1990cw,Corcella:2001hz,Corcella:2004xv} for
$e^+e^-$ collisions, top and Higgs decays, respectively.

The perturbative fragmentation function follows the 
Dokshitzer--Gribov--Lipatov--Altarelli--Parisi (DGLAP) evolution equations
and its initial condition at a scale $\mu_{0F}$ is process-independent
\cite{Mele:1990cw,Cacciari:2001cw}.
Solving the DGLAP equations for an evolution from 
$\mu_{0F}$ to $\mu_F$, with an NLO kernel, allows one to resum leading (LL)
$\alpha_S^n\ln^n(\mu_F^2/\mu_{0F}^2)$ and next-to-leading (NLL) 
$\alpha_S^n \ln^{n-1}(\mu_F^2/\mu_{0F}^2)$ logarithms (collinear
resummation).
Setting $\mu_{0F}\simeq m_b$ and $\mu_F\simeq Q$, one resums the large
mass logarithm
$\ln(Q^2/m_b^2)$, which appears in the massive spectrum \cite{Mele:1990cw}.

Moreover, the coefficient functions and the initial condition of the 
perturbative fragmentation present terms which become large for
$x_b\to 1$, which corresponds to soft-gluon radiation.
NLL soft resummation in the initial condition of the perturbative fragmentation
function is process-independent and
has been carried out in \cite{Cacciari:2001cw}.
In \cite{Cacciari:2001cw,Cacciari:2002re,Corcella:2004xv}, NLL soft terms in
 the coefficient functions
of $Z\to b\bar b$, $t\to bW$ and $H\to b\bar b$ processes
have been resummed.
In terms of the Mellin variable $N$, such calculations resum
LLs ($\alpha_S^n\ln^{n+1}N$) and NLLs ($\alpha_S^n\ln^nN$) in the Sudakov
exponent. 

As far as Monte Carlo event generators are concerned, HERWIG and PYTHIA 
implement LO processes, such as
$Z(H)\to b\bar b$ and $t\to bW$, and the subsequent parton radiation is
treated in the collinear approximation. As discussed, e.g.,
in \cite{Catani:1990rr},
this is equivalent to a LL resummation, 
with the inclusion of some NLL terms as well. 

In order to allow hard and large-angle radiation, parton showers are
provided with matrix-element corrections. PYTHIA uses the collinear 
approximation to populate the full phase space and the tree-level exact
matrix element corrects the first emission 
\cite{Miu:1998ju,Norrbin:2000uu}. PYTHIA 6.220, which we shall
use hereafter, contains matrix-element corrections to all the considered 
processes.
Unlike PYTHIA, 
the standard HERWIG algorithm completely suppresses the radiation in the
so-called `dead zone', corresponding to hard and/or large-angle radiation.
The exact matrix element populates the dead zone (hard correction) and
corrects the shower every time an emission is the `hardest so far'
(soft correction) \cite{Seymour:1994df}.
HERWIG 6.506, our default version, includes the corrections to
$e^+e^-$ annihilation \cite{Seymour:1991xa} and top decay 
\cite{Corcella:1998rs}. More recently, the corrections to 
$H\to b\bar b$ processes have been implemented \cite{Corcella:2005dk} , and we
shall account for them in the following.

\subsection{\boldmath{$B$}-hadron spectrum in \boldmath{$x_B$}-space}

In order to describe hadron production, both resummed calculations and 
Monte Carlo parton showers need to be supplemented by hadronization models,
which contain parameters which need to be tuned to experimental data.
In particular, PYTHIA and HERWIG implement the string and the cluster model,
respectively. As far as the resummed computation is concerned, we shall
convolute the $b$ spectrum with the Kartvelishvili non-perturbative
fragmentation function:
\begin{equation}
D^{\mathrm{np}}(x;\gamma)=(1+\gamma)(2+\gamma) (1-x) x^\gamma,
\label{kk}
\end{equation}
and fit the parameter $\gamma$ to data.

We shall consider data from SLD, ALEPH and OPAL on the $B$-hadron energy
fraction $x_B$, which is the hadron-level counterpart of Eq.~(\ref{xb}).
As in \cite{Corcella:2005dk}, when using the resummation,
we consider the data in the range
$0.18\leq x_B\leq 0.94$, to avoid the regions $x_B\to 0$ and $x_B\to 1$, where
the calculation is unreliable. In fact, the predicted 
parton- and hadron-level spectra become negative at very small and very 
large $x$, owing to the presence of unresummed terms 
and to non-perturbative corrections, relevant especially at large $x$. 
In the considered range, we obtain:
$\gamma=17.178\pm 0.303$, with $\chi^2/\mathrm{dof}=46.2/53$.

\begin{figure}[t!]
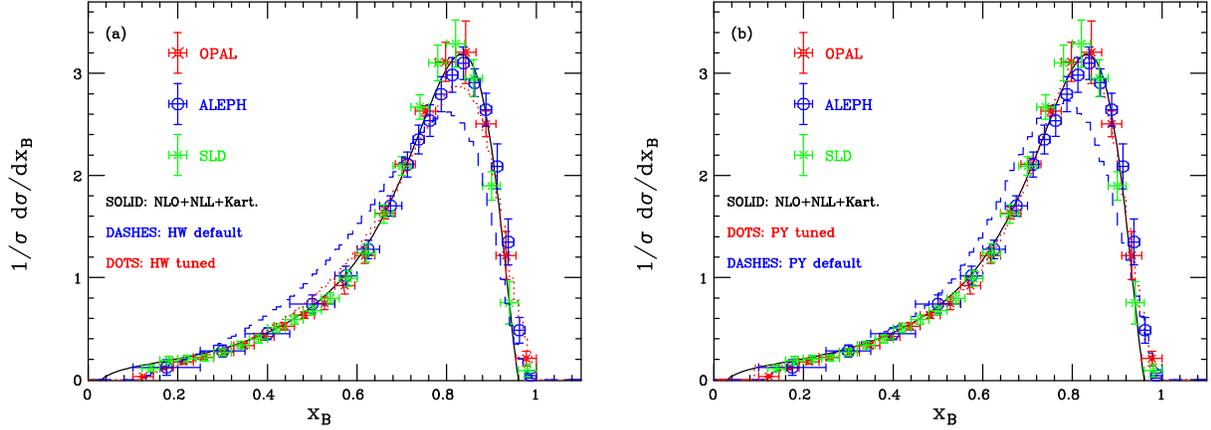

\centerline{\resizebox{0.48\textwidth}{!}{\includegraphics{s_corcella/hee1.ps}}%
\hfill%
\resizebox{0.48\textwidth}{!}{\includegraphics{s_corcella/pyee1.ps}}}
\caption{Comparison of OPAL, ALEPH and SLD data 
with HERWIG (a), PYTHIA (b) and the NLO+NLL calculation.}
\label{ee}
\end{figure}

As for HERWIG and PYTHIA, the default parametrization of the hadronization
models would not be able to fit such data, as one would get
$\chi^2/\mathrm{dof}=739.4/61$ for HERWIG and $\chi^2/\mathrm{dof}=467.9/61$ 
for PYTHIA.
In \cite{Corcella:2005dk}, 
the cluster and string models were tuned to the data and the results are 
reported in Table~\ref{para}.
The new $\chi^2$ are  $\chi^2/\mathrm{dof}$ = 45.7/61 
for PYTHIA and $\chi^2/\mathrm{dof}$ = 222.4/61 for HERWIG:
although HERWIG is still not able to fit well the data, the comparison
is greatly improved after the tuning. Major improvements in the
description of $b$-fragmentation are nonetheless present in the C++ version
HERWIG++ (whose use is beyond the goals of our analysis), which is able to 
describe well the data, after fitting only the shower cutoff
\cite{Gieseke:2003hm}.
More details about the
fits are discussed in \cite{Corcella:2005dk}, where it
is also pointed out that our tuning works well also for
the new model implemented in PYTHIA 6.3 \cite{Sjostrand:2003wg}.
Using options and parameters as they are defined in the new scenario (model 1),
we get $\chi^2/\mathrm{dof}=45.7/61$.
\begin{table}[t]
\caption{\label{para} Parameters of HERWIG and PYTHIA
hadronization models tuned to
$e^+e^-$ data, along with the $\chi^2/\mathrm{dof}$.}
  \vspace*{1mm}
\centering
\begin{tabular}{|c|c|}\hline
HERWIG & PYTHIA \\
\hline\hline
CLSMR(1) = 0.4  &                 \\
 \hline
CLSMR(2) = 0.3  & PARJ(41) = 0.85 \\
\hline
DECWT = 0.7     & PARJ(42) = 1.03 \\
\hline
CLPOW = 2.1     & PARJ(46) = 0.85 \\
\hline
PSPLT(2) = 0.33 &                \\
\hline
\hline
$\chi^2/\mathrm{dof}$ = 222.4/61 & $\chi^2/\mathrm{dof}$ = 45.7/61 \\
\hline
\end{tabular}
\end{table}
In Fig.~\ref{ee} we present the $x_B$ data, and the spectra given by the
resummed calculation, convoluted with the Kartvelishvili model, by HERWIG
and by PYTHIA. Default HERWIG and PYTHIA are quite far from the data; after the
tuning, PYTHIA reproduces the data quite well, while HERWIG yields a broader
distribution.

\begin{figure}[ht!]
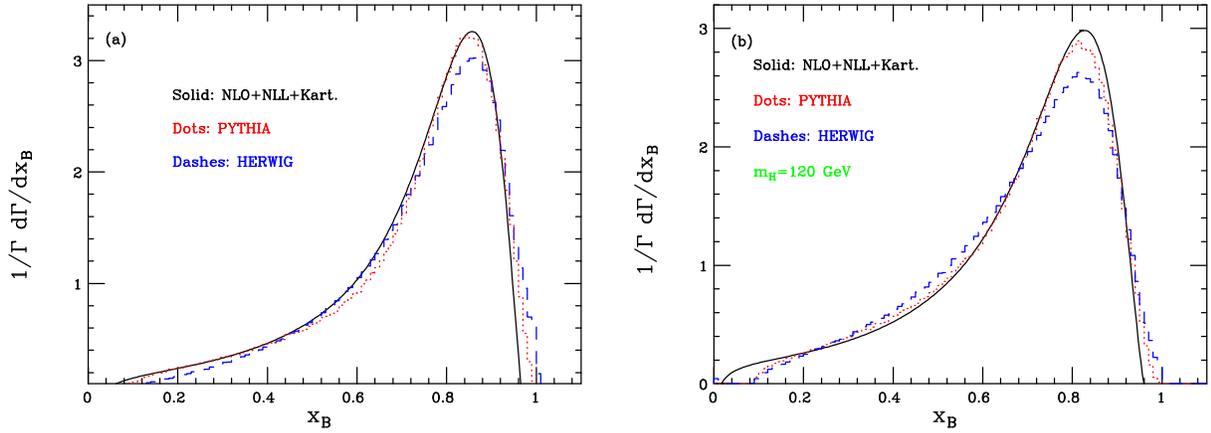

\centerline{\resizebox{0.48\textwidth}{!}{\includegraphics{s_corcella/hptt1.ps}}%
\hfill%
\resizebox{0.48\textwidth}{!}{\includegraphics{s_corcella/hph1.ps}}}
\caption{$B$ spectrum in top (a) and Higgs (b) decays, using
HERWIG, PYTHIA and NLO+NLL calculations.}
\label{tophig}
\end{figure}

Using the fitted parameters, we can predict the $B$ spectrum in top and
Higgs decays: this is shown in Fig.~\ref{tophig}, for $m_t=175$~GeV and 
$m_H=120$~GeV.
In top decay, PYTHIA reproduces the 
peak of the resummed calculation rather well, while it is below the NLL
prediction at $x_B<0.7$ and
above at $x_B>0.9$. HERWIG is below the resummed spectrum in most the
$x_B$ range, and above only at large $x_B$.
As for $H\to b\bar b$ processes,
PYTHIA fares rather well with respect to the
NLL calculation and, although  small discrepancies are visible, the
overall agreement looks acceptable. HERWIG yields instead a broader 
spectrum, which is higher than the NLO+NLL one at intermediate and 
large $x_B$, and lower around the peak.
\subsection{Results in moment space}
We now turn to the results on $B$-hadron production in Mellin moment 
space, where the moments $\Gamma_N$ of the differential width read:
\begin{equation}
\Gamma_N  =\int_0^1 {dz  \ z^{N-1}
{1\over{\Gamma}}{{d\Gamma}\over{dz}}(z) }.
\label{moment}
\end{equation}
In Ref.~\cite{delphi}, the DELPHI collaboration presented the moments
for $B$ production in $e^+e^-$ annihilation.
From the point of view of resummed calculations, 
working in moment space is better since,
in $N$-space, convolutions become ordinary products, and the thus 
relation between parton- and hadron-level cross sections becomes
$\sigma_N^B=\sigma_N^bD_N^{\mathrm{np}}$, where 
$D_N^{\mathrm{np}}$ the $N$-space counterpart of the 
non-perturbative fragmentation function.
Therefore, there is no need to assume any functional form for the 
non-perturbative fragmentation function in $x$-space.
Moreover, resummed calculations 
are well defined in $N$-space, and do not exhibit the problems of the 
$x_B$ spectra, which become negative at small or large $x_B$.

In Table~\ref{tab} we quote the data from DELPHI, the moments yielded
by HERWIG, PYTHIA and the NLO+NLL calculation
in $Z$, $t$ and $H$ decays.
The moments given by HERWIG and PYTHIA in $e^+e^-$ annihilation
are consistent, within the error
ranges, with the ones measured by DELPHI. 
Although problems are present when fitting the $x_B$ data
from LEP and SLD, it is remarkable that HERWIG is
compatible with the DELPHI moments within one standard deviation.

The results for top and Higgs decays
exhibit similar features to the $x_B$ spectra.
In top decay, PYTHIA is very close to the NLL calculation which uses
$D_N^{\mathrm{np}}$ extracted from the DELPHI data, while
HERWIG, whose predictions are shifted toward larger $x_B$,
yields larger moments. For $H\to b\bar b$, PYTHIA and
HERWIG  give moments which are compatible within $1\%$.

\begin{table}[t]
\caption{\label{tab}\small  Moments
$\sigma^B_N$ from
DELPHI~\protect\cite{delphi}, and moments
in $e^+e^-$ annihilation, Higgs ($H$) and top ($t$) decay, 
using NLL resummed calculations, HERWIG (HW) and PYTHIA (PY).}
\centering
\vspace*{1mm}
\begin{tabular}{| c | c c c c |}
\hline
& $\langle x\rangle$ & $\langle x^2\rangle$ & $\langle x^3\rangle$
& $\langle x^4\rangle$ \\
\hline
\hline
$e^+e^-$ data $\sigma_N^B$&0.7153$\pm$0.0052 &0.5401$\pm$0.0064 &
0.4236$\pm$0.0065 &0.3406$\pm$0.0064  \\
\hline
\hline
$e^+e^-$ NLL $\sigma_N^b$   & 0.7801 & 0.6436 & 0.5479 & 0.4755  \\
\hline
$D^{\mathrm{np}}_N$ & 0.9169 & 0.8392 & 0.7731 & 0.7163 \\
\hline
\hline
$e^+e^-$ HW $\sigma_N^B$   & 0.7113 & 0.5354 & 0.4181 & 0.3353  \\
\hline
$e^+e^-$ PY $\sigma_N^B$   & 0.7162 & 0.5412 & 0.4237 & 0.3400  \\
\hline
\hline
$H$-dec.NLL $\Gamma^b_N$ & 0.7580 & 0.6166 & 0.5197 & 0.4477  \\
\hline
$H$-dec $\Gamma^B_N$
& 0.6950 & 0.5175 & 0.4018 & 0.3207 \\
\hline
$H$-dec. HW  $\Gamma^B_N$ & 0.6842 & 0.5036 & 0.3877 & 0.3076 \\
\hline
$H$-dec. PY $\Gamma^B_N$ & 0.6876 & 0.5080 & 0.3913 & 0.3099 \\
\hline
\hline
$t$-dec. NLL $\Gamma^b_N$ & 0.7883 & 0.6615 & 0.5735 & 0.5071 \\
\hline
$t$-dec. NLL $\Gamma^B_N$ & 0.7228 & 0.5551 & 0.4434 & 0.3632 \\
\hline
$t$-dec. HW $\Gamma^B_N$ & 0.7325 & 0.5703 & 0.4606 & 0.3814 \\
\hline
$t$-dec. PY $\Gamma^B_N$ & 0.7225 & 0.5588 & 0.4486 & 0.3688 \\
\hline
\end{tabular}
\end{table}
\subsection{Conclusions}
In summary, we have investigated $b$-quark fragmentation in $e^+e^-$
annihilation, top and Higgs $H\to b\bar b$ decays. We have described 
$b$ production using resummed calculations, based on the
perturbative fragmentation approach, 
HERWIG and PYTHIA. We have fitted a few hadronization models to
LEP and SLD data on $B$-hadron production and performed predictions
on the $B$ spectrum in top and Higgs decays.
Tuning the HERWIG and PYTHIA hadronization models played a crucial role,
as the default parametrizations
would not be able to reproduce the $e^+e^-$ data. 
Moreover, we have analysed data from the DELPHI Collaboration in moment space,
extracted the non-perturbative fragmentation function in $N$-space, 
and compared the moments given by resummed calculations, HERWIG and PYTHIA.

%%%%%%%%%%%%%%%%%%%%%%%%%%%%%%%%%%%%%%%%%%%%%%%%%%%%%%%%%%%%%%%%%%%%%%%%%%%%%
\section[Study of jet clustering algorithms at the LHC]
{STUDY OF JET CLUSTERING ALGORITHMS AT THE LHC~\protect
\footnote{Contributed by: D.~Benedetti, S.~Cucciarelli, J.~D'Hondt,
A.~Giammanco, J.~Heyninck, A.~Schmidt, C.~Weiser}}
\subsection{Introduction}
A wide spectrum of new physics topologies, as well as
known processes like top quark production, will have quarks in the final
state of proton-proton collisions at the LHC. When reconstructing the quark's kinematics, jet
reconstruction is of major importance, which is a complex task and not necessarily robust. Ambiguities
in the jet definition not only arise from the theoretical point of view
if higher order corrections are taken into account, but also
experimentally, due to the magnetic field, the calorimeter response and
the different configuration possibilities of jet clustering algorithms.

This study concentrates on the algorithmic task of
clustering the input objects (e.g simulated particles or 
calorimeter cells) into jets. This is performed
from an analysis perspective, which
means that  the jet clustering is considered to be optimal if the
reconstruction efficiency of the complete kinematics of the primary quark event topology is maximized. This
reconstruction efficiency will be defined in terms of
quality criteria or quality markers, relative to the performance of a
typical analysis like the reconstruction of the mass of a resonance
decaying into quarks. The distance between the generated primary partons $i$ and
the reconstructed jets $j$, and therefore the error of the jets, should be minimized in both
energy and momentum (angular) space, for example
$\epsilon_{\theta}=\theta_{j}^{jet}-\theta_{i}^{quark}$.

Physics effects like pile-up, underlying event and radiation enlarge this mean error. The scope of this study is to
find the most efficient configuration for jet finding algorithms in the presence of these
effects, in order to maximize the fraction of events for which all quarks have smaller errors than some predefined criteria. Hence, events 
suffering from a large amount of hard gluon radiation will be
rejected.

The resulting efficiency does not
only depend on the configuration of the jet finding but also on the 
event topology. This will be  investigated  for processes with two, four, six
and eight primary quarks in the final state, covering a significant spectrum of
physics processes at the LHC.

To disentangle detector effects
from pure algorithmic and physics effects, the study is performed with simulated particle
information as input to the jet finding algorithms. The
comparison with a realistic detector is beyond the general
scope of this contribution. It will 
be described in dedicated papers for the specific experiments.

\subsection{Jet Clustering Algorithms \label{algos}} 
The following jet reconstruction algorithms are considered in this
study: the {\it Iterative Cone} algorithm~(IC), the inclusive {$k_{T}$}
algorithm~($k_{T}$) and the {\it MidPoint~Cone} algorithm~(MC).
%While the algorithms cone like work mainly in a $(\eta,\phi)$-metric
%the KT works in a more complex phase-space using both energetic and
%space angle information allowing this algorithm to be collinear and
%infrared safe. 
A description of these algorithms and the definition of their parameters can be found in Ref. \cite{Blazey:2000qt}.  For all
algorithms the energy recombination scheme and the $\eta,\phi$-metric
is used. The main parameters that are varied for the different
algorithms are: the cone radius for the {\it Iterative Cone} algorithm;
the R-parameter that reflects a radius-like role for the
{$k_{T}$}~algorithm; the cone radius and the shared energy fraction
threshold for merging for the {\it MidPoint Cone} algorithm.

For all algorithms generated and stable final state particles are used
as input. Muons and neutrinos are excluded, and the effects of the magnetic field are not taken into account. All particles are assumed to emerge from the primary vertex, where the clustering is performed.

%\item MidPointConeOverlapThreshold: $E_T$-fraction that decides
%whether a pair of proto-jets should be merged or splitted: 50\% 
%  \item MidPointConeConeAreaFraction: The cone radius used for the
%  merging and splitting proto-jets. $R_{merge-split} = 2 \times
%  R_{cone}$ 

\subsection{Event Generation}
Proton collisions at 14 TeV have been generated at a luminosity of
2 $\times$ 10\textrm{$^{33}$} cm\textrm{$^{-2}$}s\textrm{$^{-1}$}. 
Final states like fully leptonic and semileptonic $t\bar{t}$, and semileptonic and
hadronic $t\bar{t}H\rightarrow t\bar{t} b \bar{b}$ events, have been selected to represent topologies with two, four,
six and eight primary quarks.
The $t\bar{t}$ events were generated using PYTHIA version 6.2 and the
$t\bar{t}H$ events were generated with compHEP version 41.10, interfaced to
PYTHIA version 6.215.
For the leptonic decay of the W boson, only electrons and muons
are considered.

\subsection{Event Selection \label{sec:EventSelection}}
%The event selection for our clustering study is done requiring a
%number of jets greater then the number of the final state parton.  
A realistic event selection is applied. The reconstructed jets are required
to have a transverse energy larger than 20~GeV, and to be within the tracker
acceptance for a proper $b$-tagging performance ($|\eta|< 2.4$ for the CMS experiment).   
%This energy and space cuts are given to bring the analysis close to
%the real top mass reconstruction where a cut on calibrated energy
%greater then 20 GeV hould be applied as the $ |\eta| $ space
%restriction to apply the b-tagging. 
Isolated signal leptons from the W-decay are removed from the jet
finding input. Only if the number of jets passing these criteria is
larger than or equal to the number of primary partons the event is
considered for the analysis. 

An iterative procedure is used to match the reconstructed jets to the generated primary partons based on
the $\Delta R$ distance in the ($\eta$,$\phi$) plane.  For each possible jet-quark couple the
$\Delta R$-value is calculated, and the smallest value is considered as
a correct jet-quark matching and is removed from the list for the next iteration. When more jets have a
minimal $\Delta R$-value with the same quark, the couple with the
lowest $\Delta R$-value is taken. This procedure is iterated until
all jets have their respective quark match.

\subsection{Description of the Quality Markers \label{QMs}}
In order to obtain an efficient reconstruction of the kinematics of the primary partons, the selected jets should match both in energy and direction the primary partons. Variables called quality markers are defined to quantify the goodness of the event reconstruction from that perspective. Although physics effects of pile-up, gluon radiation and underlying event will degrade the overall event reconstruction efficiency, their magnitude is equal for all considered jet definitions. Hence, the relative comparison between jet definitions is meaningful.

%Two important requirements for a properly reconstructed jet are the
%direction of the jet, which should match the direction of the
%generated parton, and the energy which should differ from the
%parton energy only due to hadronization effects and neutrinos. As a
%consequence, events suffering from hard gluon radiation will be
%considered as badly reconstructed events, because these events will be
%difficult to handle anyway, e.g. with respect to a resonance
%reconstruction. These demands lead to the definition of the following
%quality markers:

\subsubsection{Event Selection Efficiency ``$\epsilon_{s}$'':\label{subsec:seleff}}
This efficiency is defined as the fraction of events that pass
the event selection.  
When the selection is applied on quark level, the efficiency is equal to 80\% for the two quarks final state, 62\% for the four 
quarks final state, 61\% for the six quarks final state and  52\% for the eight quarks final state.

\subsubsection{Angular Distance between Jet and Parton ``Frac $\alpha_{jp}^{max}$'':\label{subsec:alpha}}
\begin{figure}[!htp] 
  \centering
  \begin{minipage}[t]{0.245\linewidth}
    \centering
    \includegraphics[width=0.99\textwidth]{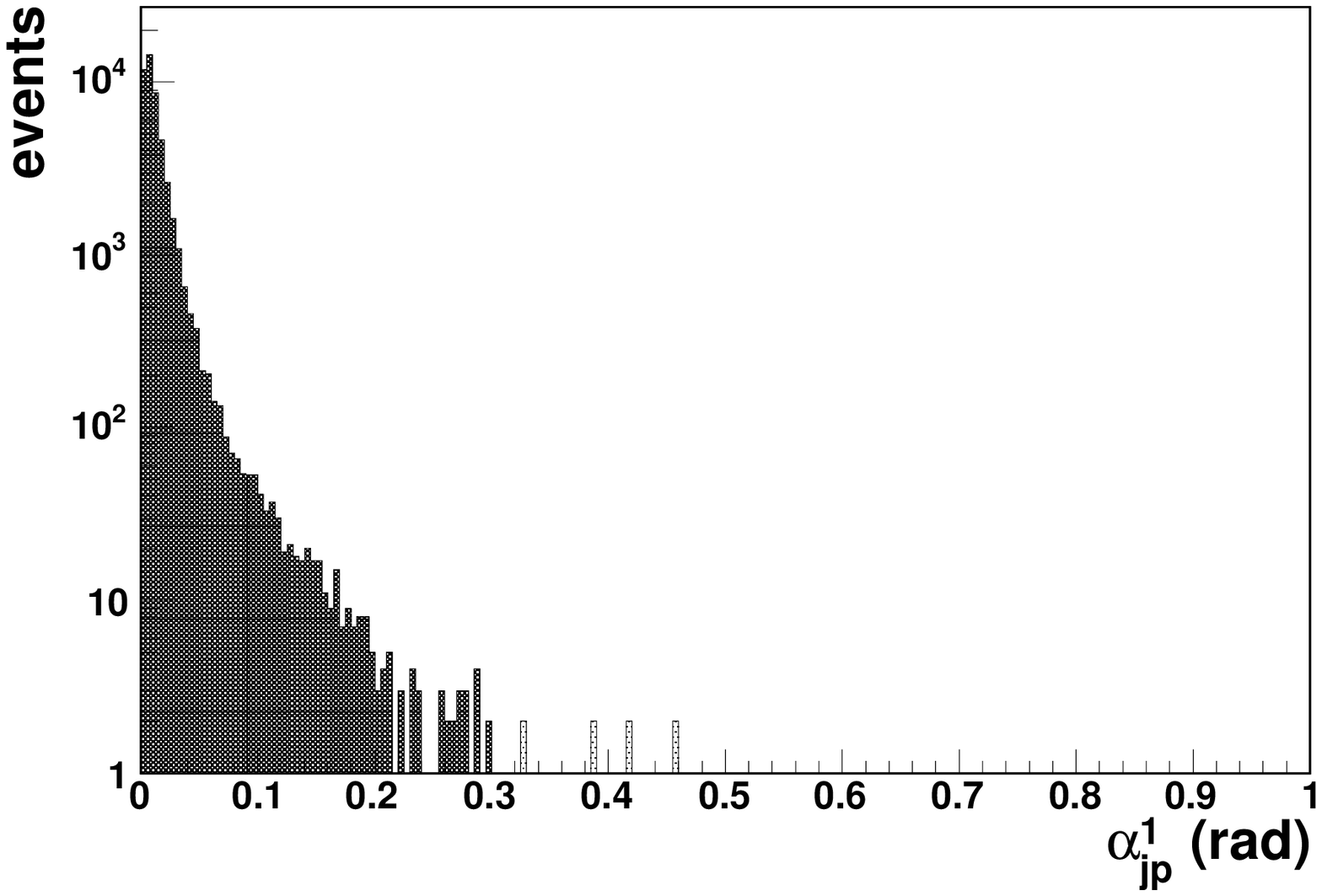} 
  \end{minipage}
  \begin{minipage}[t]{0.245\linewidth}
    \centering
    \includegraphics[width=0.99\textwidth]{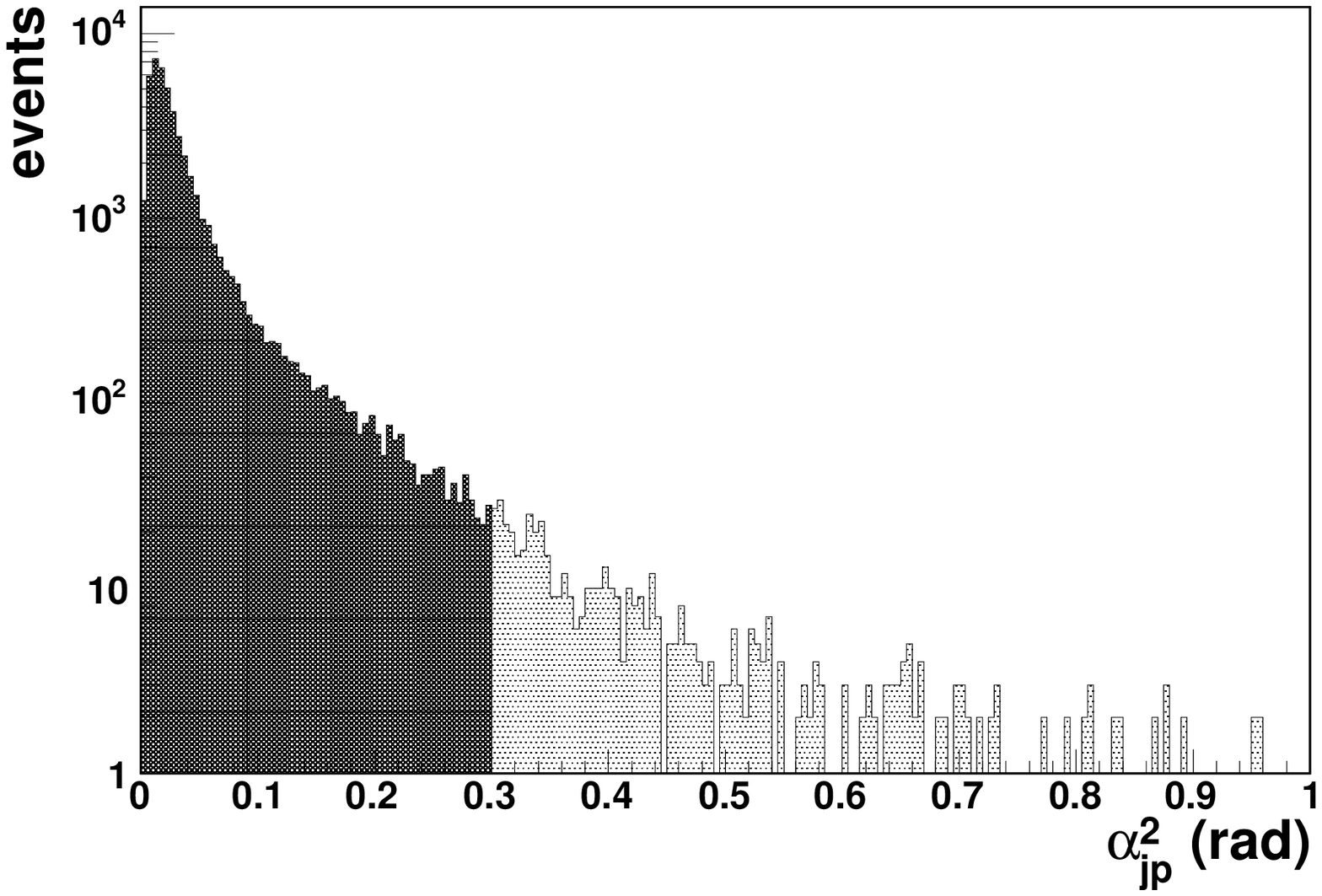} 
  \end{minipage}    
  \begin{minipage}[t]{0.245\linewidth}
    \centering
    \includegraphics[width=0.99\textwidth]{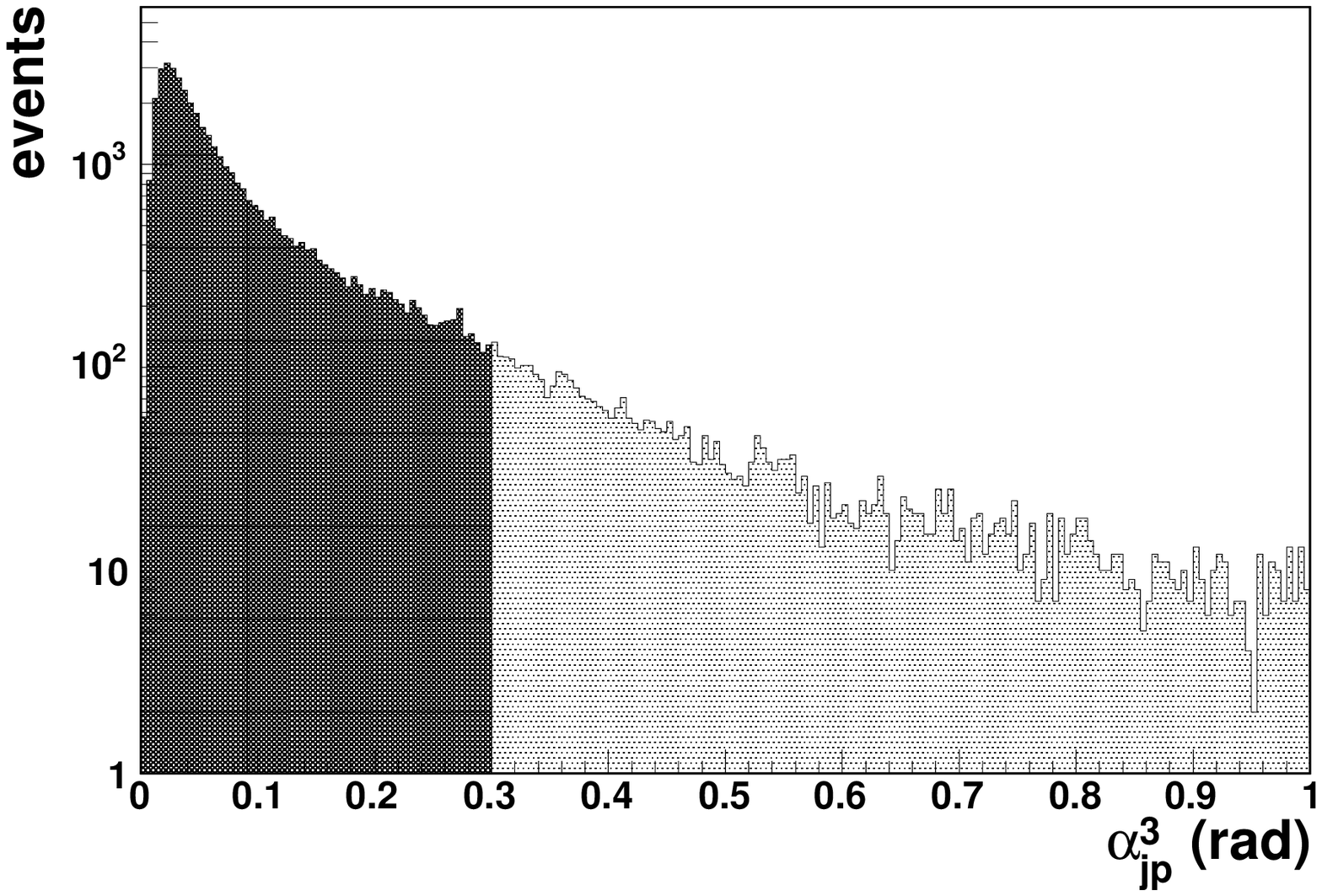} 
  \end{minipage}
  \begin{minipage}[t]{0.245\linewidth}
    \centering
    \includegraphics[width=0.99\textwidth]{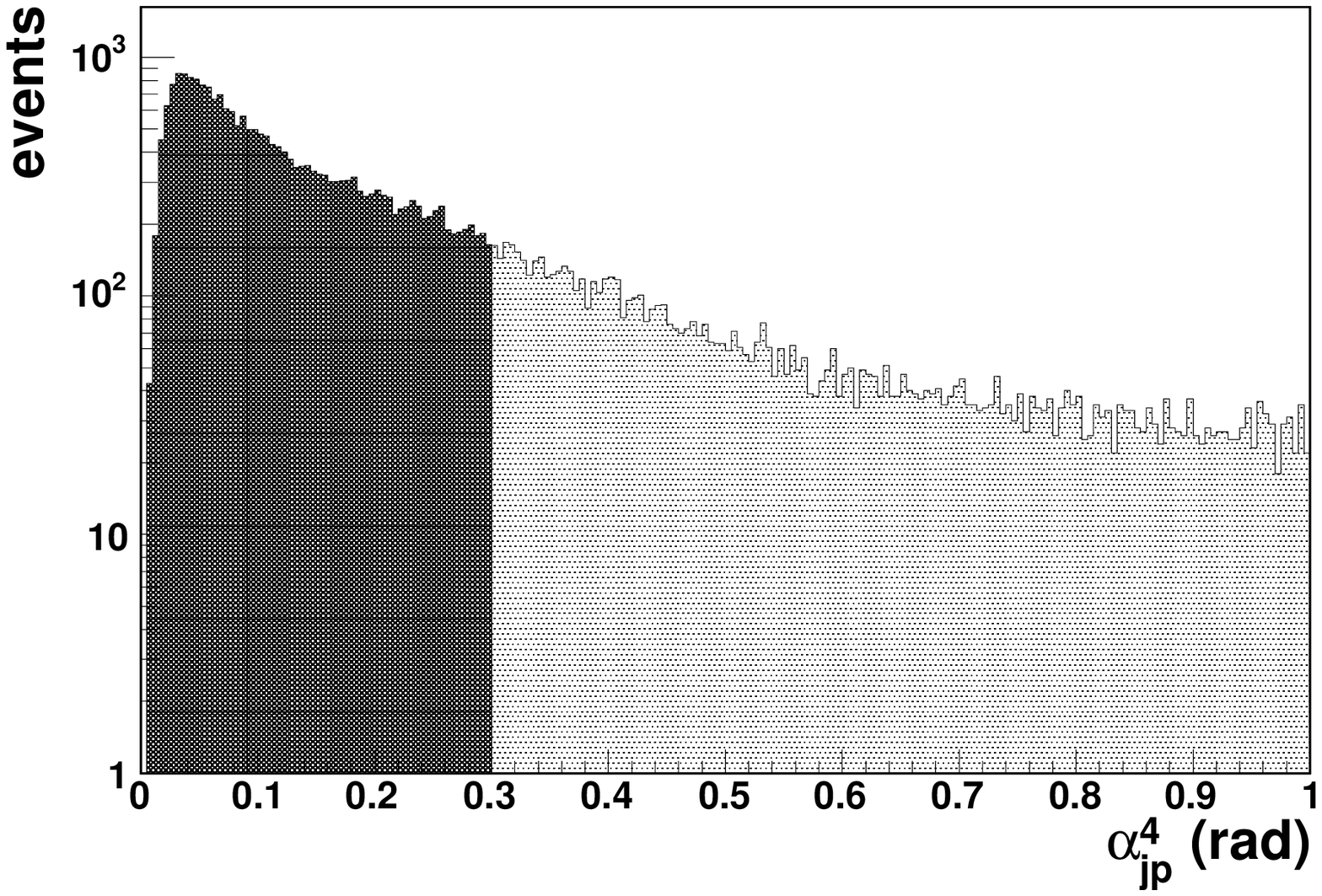} 
  \end{minipage}    
  \caption{Distributions of $\alpha_{jp}^{i}$  in increasing order
          for the IC algorithm with a cone radius of 0.4 in the case
          of a final state with four quarks. The 0.3 rad criteria as
          discussed in the text is indicated.} 
  \label{maxalpha}
 \end{figure}

A jet is considered to be well reconstructed, if the $\Delta R$ distance
between its direction and its best matched quark direction,
$\alpha_{jp}$, is sufficiently small. For each event, this results in a list of increasing $\alpha_{jp}^{i}$-values, $\{\alpha_{jp}^{1},...,\alpha_{jp}^{n}=\alpha_{jp}^{max}\}$, where $n$ is the amount of primary quarks in the considered event topology.
 Hence, $\alpha_{jp}^{max}$ is defined as the maximum $\alpha_{jp}^{i}$-value
of all $i$ jet-quark pairs in the event. The $\alpha_{jp}^{i}$ distributions for a four quarks final state are shown in Fig.\ref{maxalpha}. 
%\begin{figure}[hbtp!] 
%  \begin{center}
%        \includegraphics[width=0.4\textwidth]{s_schmidt/MaxAlphaIC041} 
%        \caption{Distributions of $\alpha_{jp}$  in increasing order
%          for the {\it Iterative Cone} algorithm in the case of the
%          four jets final state. } 
%        \label{maxalpha}
%  \end{center} 
%\end{figure}
The last one of these plots represents the  $\alpha_{jp}^{max}$
variable. 
To quantify the angular reconstruction performance of  a particular jet definition, a quality marker is defined as the
fraction of events with a $\alpha_{jp}^{max}$ value lower than 0.3 and denoted as ``Frac~$\alpha_{jp}^{max}$''. The choice of
the worst jet is motivated by the reasoning that the directions of all primary quarks in the event are required to be well determined.

\subsubsection{Energy Difference ``Frac $\beta_{jp}^{max}$'':\label{efrac}}
\begin{figure}[!htp] 
  \begin{minipage}[t]{0.5\linewidth}
    \centering
    \includegraphics[width=0.6\textwidth]{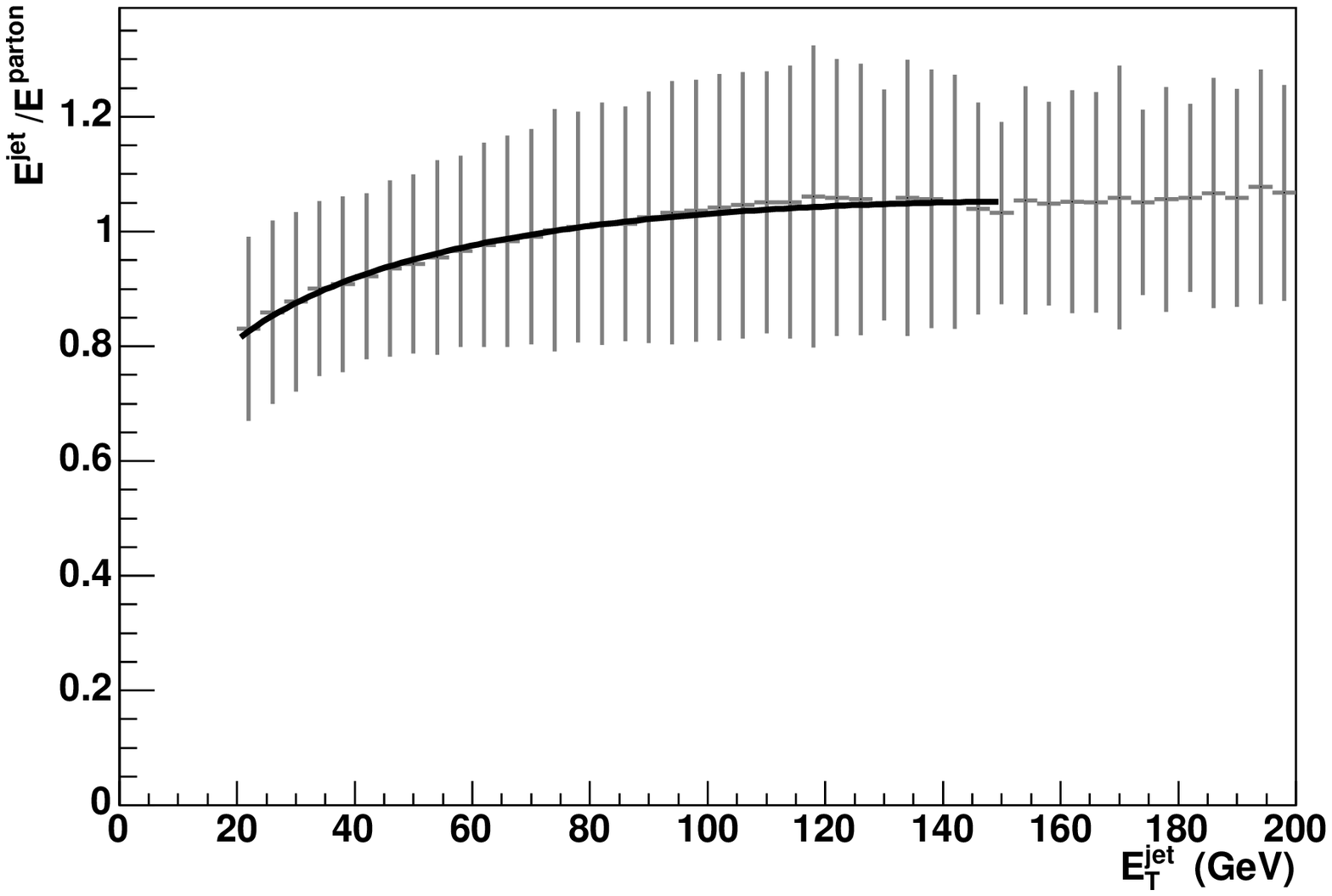} 
%\setcaptionwidth{0.9\textwidth}
    \caption{Example of a $\frac{E^{jet}}{E^{parton}}$
          vs. $E_{T}^{jet}$ curve for the IC algorithm with a cone
          radius of 0.4, applied on a final state with four primary quarks. The vertical bars illustrate the resolution.}
    \label{calibration}
  \end{minipage}
  \begin{minipage}[t]{0.5\linewidth}
    \centering
    \includegraphics[width=0.6\textwidth]{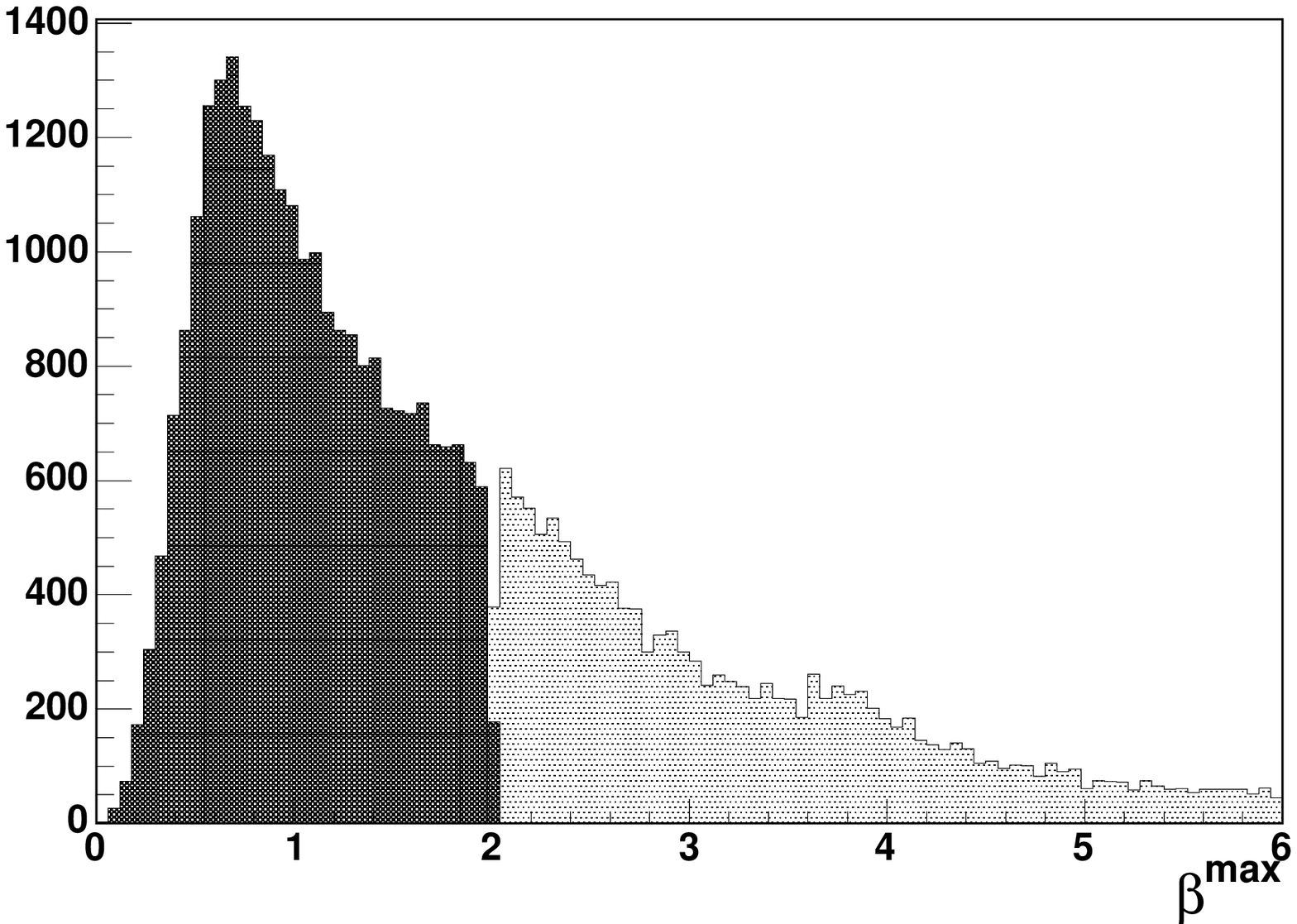} 
%\setcaptionwidth{0.9\textwidth}
    \caption{Distribution of $\beta_{jp}^{max}$  for the
          IC algorithm with a cone radius of 0.4, applied on a final
          state with four primary quarks.}
    \label{maxbeta}
  \end{minipage}    
 \end{figure}

The reconstructed energy of the primary quarks  is usually biased and
has a broad resolution.  Figure~\ref{calibration} shows the average
fraction of the quark energy that is reconstructed for a specific  
algorithm configuration as a function of the reconstructed transverse 
jet energy. Such a calibration curve can be interpreted as an
estimator for the expected reconstructed energy. For this plot only well matched ($\alpha_{jp}<$0.3),  
non-overlapping jets were taken into account. For the iterative cone algorithm, a jet is considered
to be non-overlapping, if its $\Delta R$ distance to any other jet is
larger than twice the value of the cone radius parameter of the algorithm.
It is the aim of jet calibration studies to determine these average corrections to be applied on the reconstructed jet energies. Therefore the remaining component is the energy resolution.

%The effect is shown of the non gaussian
%behaviour of the distributions in the considered  
%$E_T^{parton}$-bins. With respect to the curve with the gaussian bin
%fits, for each jet configuration and each jet the number of  
%standard 
%deviations the jet's
%$E^{rec}/E^{gen}$-value is away from the expected value in the curve
%($\beta_{jp}$). 
The $\beta_{jp}^{i}$ values are defined for each primary quark $i$ as the distance from
the expected energy fraction (deduced from the fitted function in Fig.~\ref{calibration}) in units of standard deviations.
For each selected event, the primary quark with the highest
$\beta_{jp}^{i}$ value, called $\beta_{jp}^{max}$ is considered to be
the one with the worst reconstruction performance from the energy point of view. An example for the $\beta_{jp}^{max}$ distribution is shown in
Fig.~\ref{maxbeta}.  An energy related quality marker is defined as the
fraction of events with a $\beta_{jp}^{max}$ lower than 2 standard deviations, and denoted as ``Frac $\beta_{jp}^{max}$''.

\subsubsection{Combined Variable ``Frac($\alpha_{jp}^{max}$+$\beta_{jp}^{max}$)'':\label{subsec:combined}} 
\begin{figure}[!htp] 
  \begin{minipage}[c]{0.5\linewidth}
    \centering
    \includegraphics[width=0.7\textwidth]{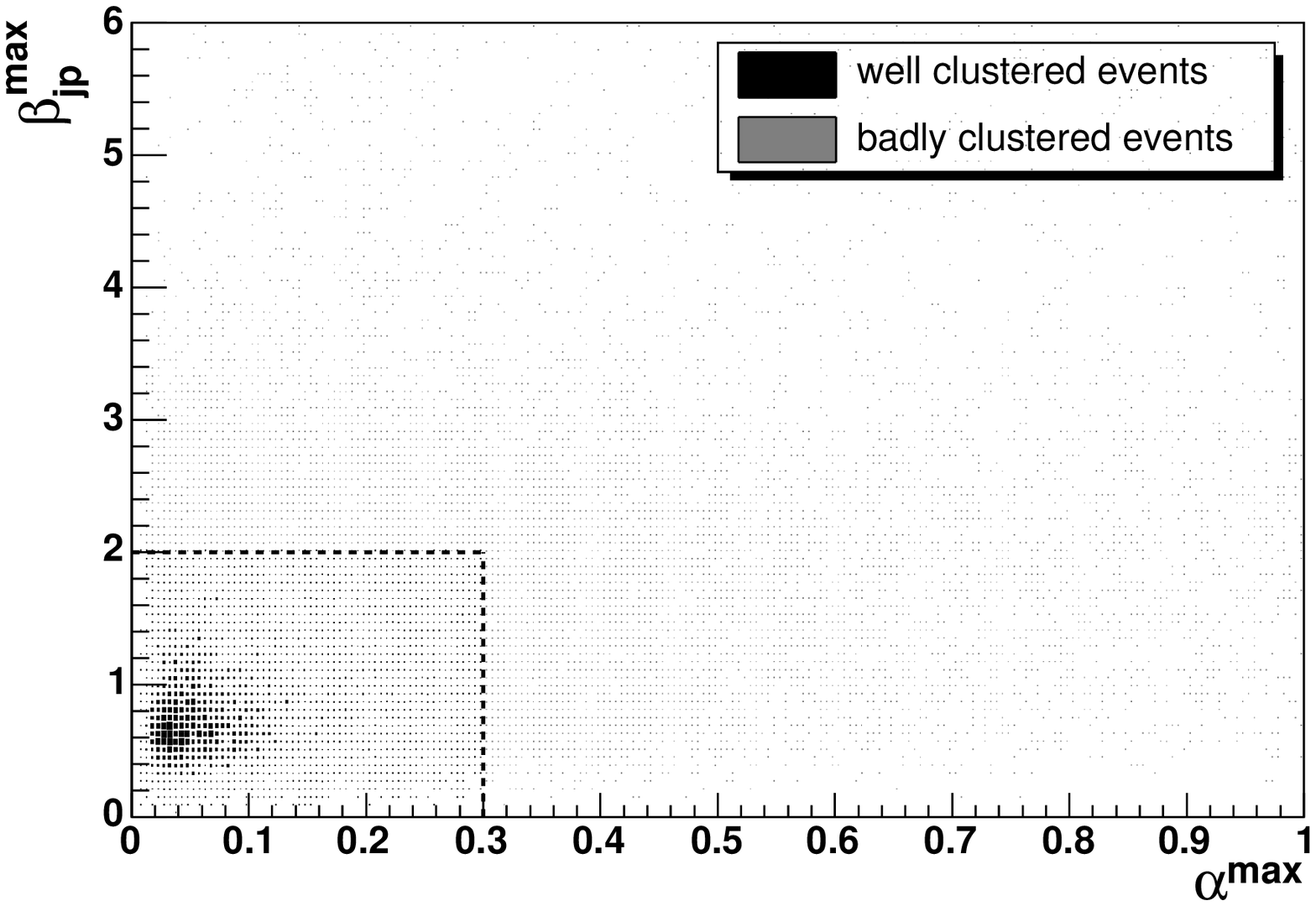} 
%\setcaptionwidth{0.9\textwidth}
    \caption{Box plot of $\beta_{jp}^{max}$
        vs. $\alpha_{jp}^{max}$ for the IC algorithm with a cone
        radius of 0.4, applied on a final state with four primary quarks.}
    \label{exFracIn}
  \end{minipage}
  \begin{minipage}[c]{0.5\linewidth}
    \centering
    \includegraphics[width=0.7\textwidth]{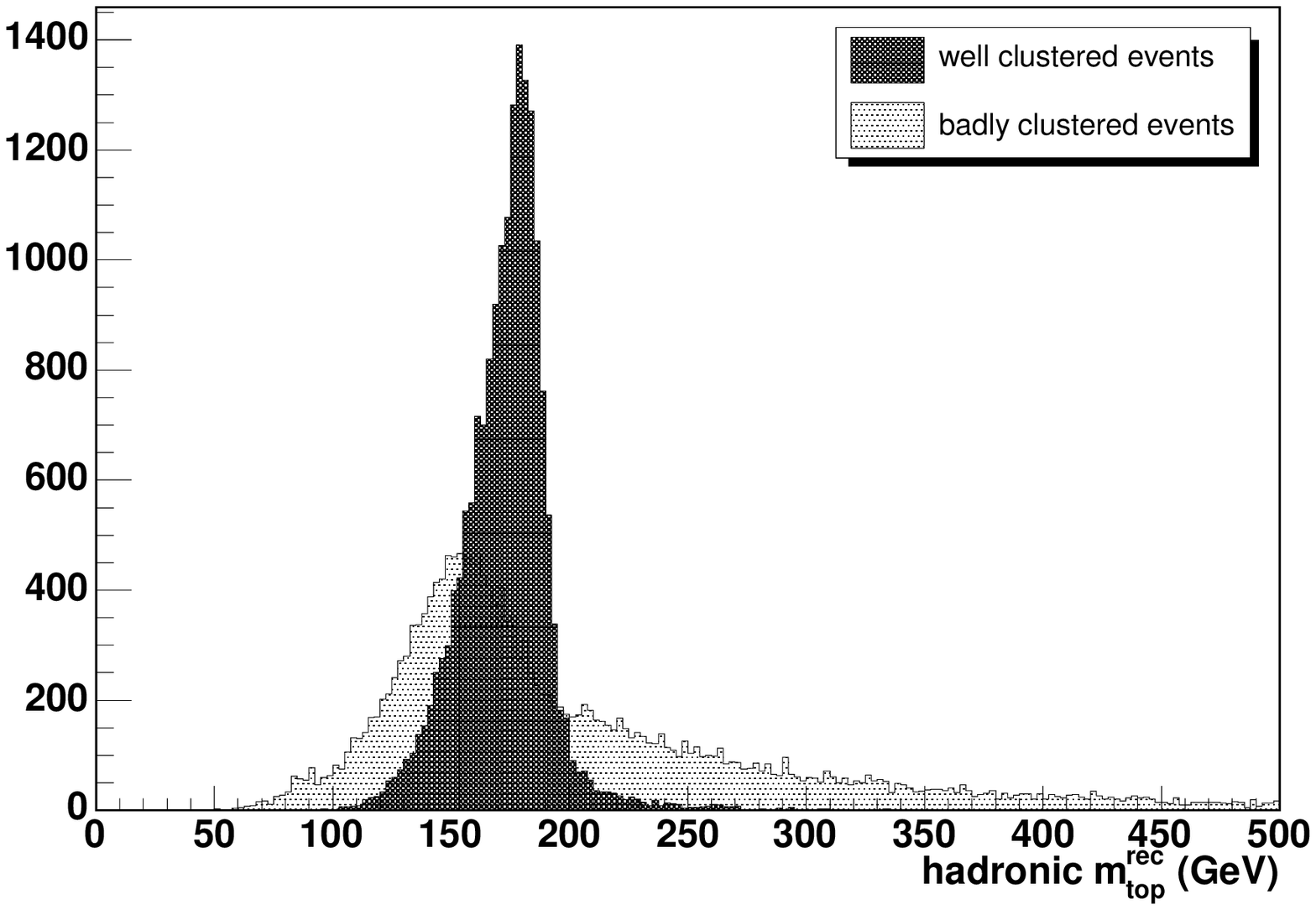} 
%\setcaptionwidth{0.9\textwidth}
    \caption{Distribution of the hadronic top quark mass, using jets clustered with the IC algorithm with a cone radius of 0.4,  applied on a final
        state with four primary quarks.} 
    \label{topmass}
  \end{minipage}    
 \end{figure}

The combined variable is defined as the fraction of events in which both the direction and
the energy of the $n$ primary quarks are well reconstructed using the   
definitions in Section \ref{subsec:alpha} and \ref{efrac}. The
correlation between  $\alpha_{jp}^{max}$ and  $\beta_{jp}^{max}$ is shown in 
Fig.~\ref{exFracIn}, where applying both quality criteria defines a rectangular
area in which the kinematics of the primary quarks are 
well reconstructed. As an illustration of the power of 
this combined variable in identifying well reconstructed events,  
the reconstructed spectrum of the hadronic
top quark mass in the semileptonic $t\bar{t}$ final state is shown in
Fig.~\ref{topmass}. The light grey histogram 
shows the spectrum from events where the  
the kinematics of the primary quarks have been
badly reconstructed according to the combined variable.

\subsubsection{Overall Quality Marker "FracGood"}
The fraction of selected and well reconstructed
events, i.e. the selection efficiency $\epsilon_{s}$, multiplied by the combined variable
Frac($\alpha_{jp}^{max}$+$\beta_{jp}^{max}$) is defined as ``FracGood''.

This last quality marker is interpreted as an estimate for the
reconstruction efficiency of the kinematics of the primary quarks of the complete event, and therefore  used to  
compare different algorithms and their corresponding configurations. Although this variable gives a
powerful overall indication of a reasonable jet definition, it is sometimes useful to consider the partial information of the individual quality markers. Depending on the priorities of a specific physics analysis,
one would be interested in the average number of reconstructed jets, or
the energy resolution for non-overlapping jets, or the efficiency of the angular matching between primary quark and jet.  The
average number of jets gives 
an idea of the sensitivity to pile-up, underlying event, and the rate
of fake jets, while the energy resolution can be linked  
to the issue of jet calibration. 

\subsection{Results}
In this section the most important observations for each jet
clustering algorithm considered are summarized.

\subsubsection{Iterative Cone Algorithm}
\begin{figure}[!htp] 
  \begin{minipage}[c]{0.5\linewidth}
    \centering
    \includegraphics[width=0.6\textwidth]{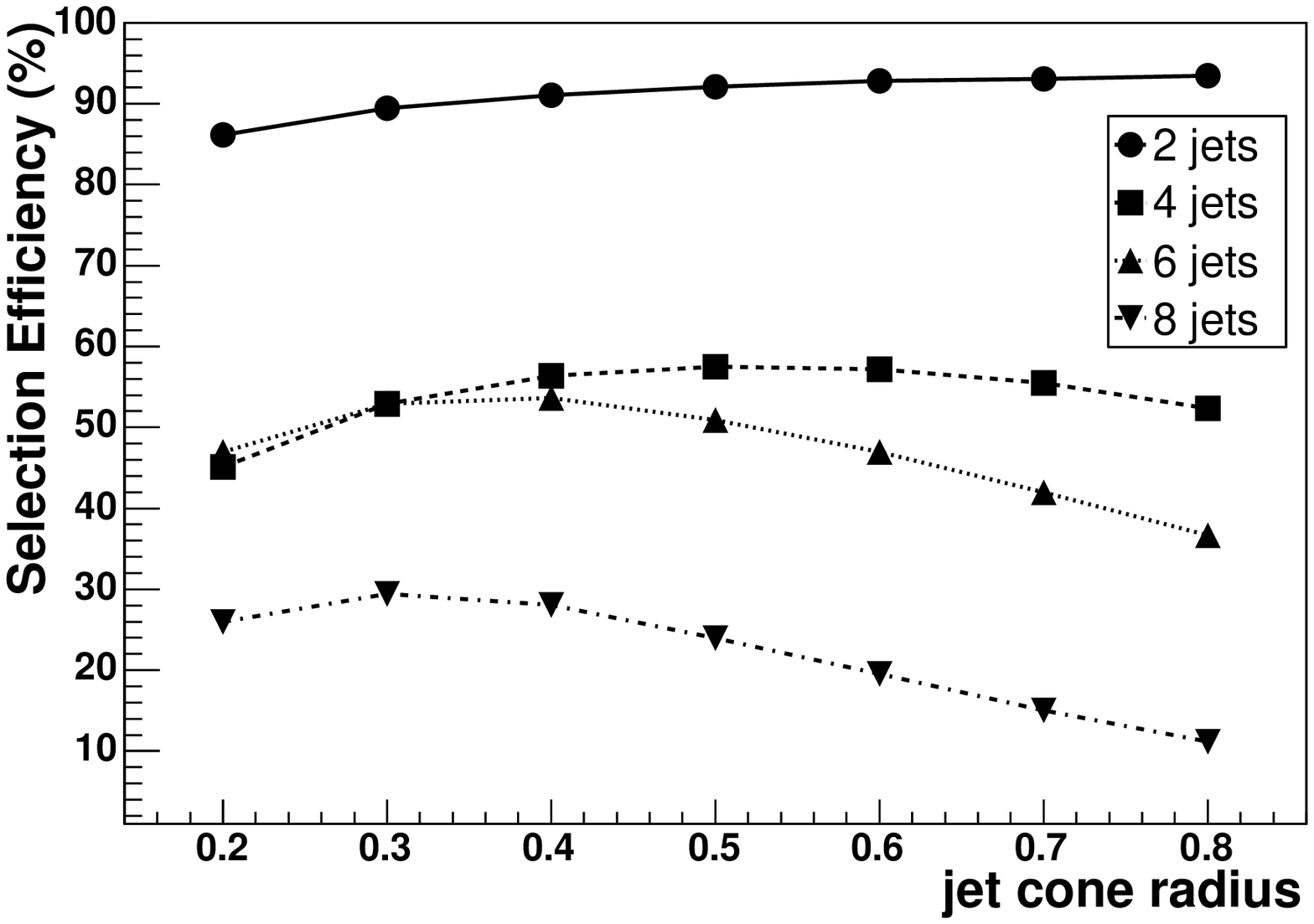} 
%\setcaptionwidth{0.9\textwidth}
    \caption{Fraction of selected events versus the cone radius (IC algorithm).} 
    \label{ICSelEff_OA}
  \end{minipage}    
  \begin{minipage}[c]{0.5\linewidth}
    \centering
    \includegraphics[width=0.6\textwidth]{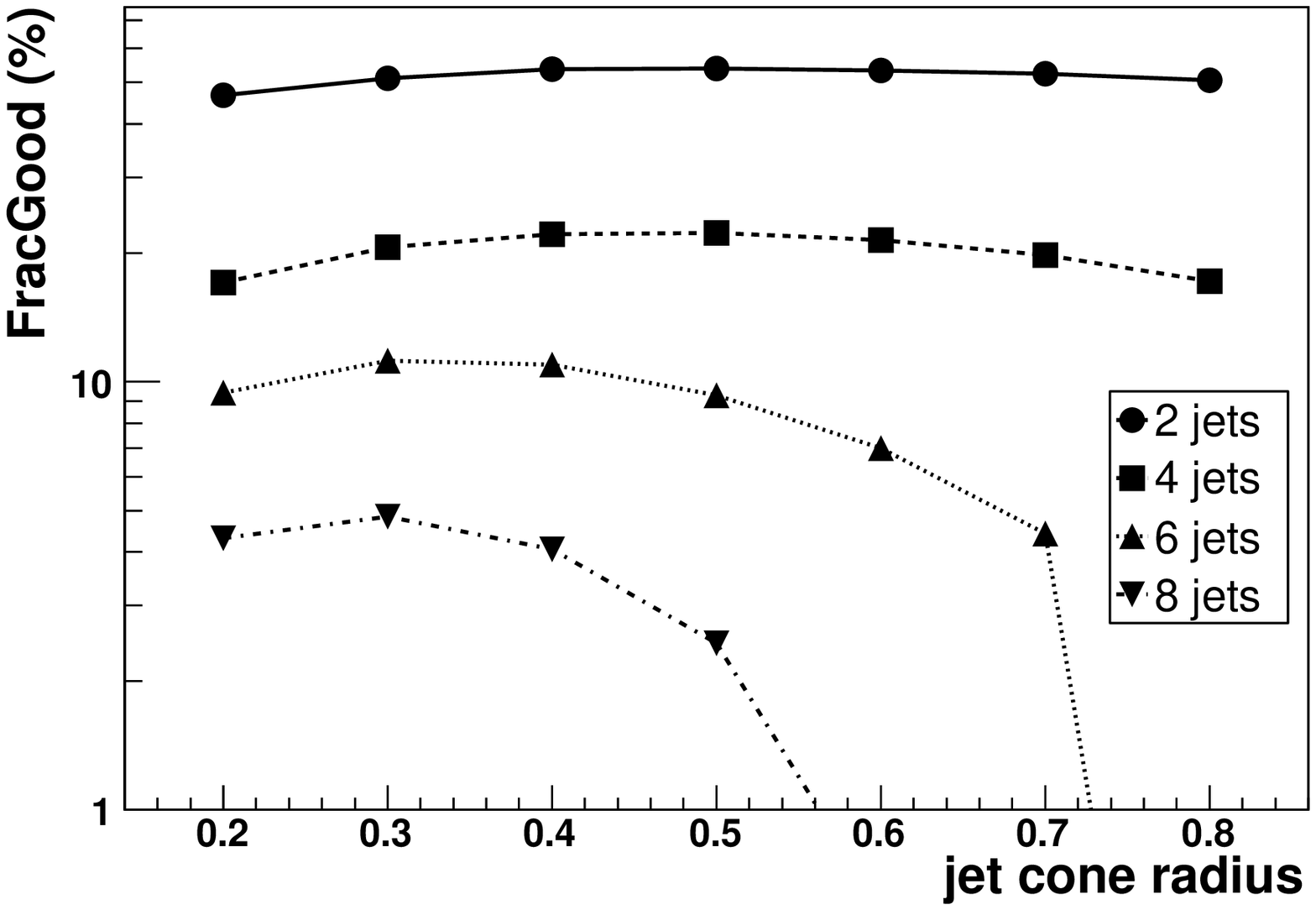} 
%\setcaptionwidth{0.9\textwidth}
    \caption{Fraction of well clustered and selected events versus
        the cone radius (IC algorithm).}
    \label{ICFracGood_OA}
  \end{minipage}
 \end{figure}

Fig.~\ref{ICFracGood_OA} shows the ``FracGood'' variable as a function of the cone radius. 
The maximum fraction
of well reconstructed events is obtained for a cone radius varying from 0.3 to 0.5, 
depending on the event topology.  The dependence of the fraction of well
reconstructed events on the minimal transverse energy of the
jet seed is found  to be negligible. A stronger
dependence as well as a larger optimal cone radius is however expected
when the jet input is changed from simulated to reconstructed particles.  
%Analogue, the addition of pile-up and the bending of particle trajectories 
due to a magnet field, will result in a stronger dependence of the number of 
reconstructed jets with respect to a minimal seed $E_{T}$-value.

%is Although a low seed $E_{T}$ cut seems to be
%favoured, the average number of reconstructed jets increases from xx
%to x when lowering the seed $E_{T}$ cut from 2 to 1 GeV
%(fig.\ref{averagejets}). This is a nice illustration for the fact that
%the ``FracGood'' variable can be interpreted as a rough guideline through
%the possible jet clusering configurations, but that one always should
%check the complete set of quality markers before taking conclusions.

Another important observation is the decrease of the optimal cone radius
for increasing jet multiplicity. This behaviour can be explained by
the higher probability of overlapping jets for higher jet
multiplicities. The generally lower selection efficiency
(Fig.\ref{ICSelEff_OA}) for high multiplicities is interpreted as due to 
higher probability for overlapping jets, the different $p_t$ spectrum of 
the jets, and moreover to the fact that an increase of the
average center of mass energy for $t\bar{t}H$-production compared to
$t\bar{t}$-production will result in extra hard gluon jets.

Furthermore, a lower selection
efficiency $\epsilon_{s}$ is observed for very low jet radii. This can be explained by the
transverse energy cut of 20~GeV which is more severe for small
opening angles. 

Both effects will result in a more
difficult jet clustering task for high jet multiplicities.  Compared
to 55\% of well clustered events in the two quark final state, only
6\% of the events in an eight quark topology pass all the criteria using the optimum cone radius in each case.
%  \begin{center}
%        \includegraphics[width=0.4\textwidth]{s_schmidt/Gen_ICFracGood_seedET.eps} 
%        \caption{Fraction of well clustered and selected events versus
%        the seed \textrm{$E_{T}$} cut, for a jet cone radius of  0.4.} 
%        \label{ICFracGood_seedET}
%  \end{center} 
%\end{figure}

The angular and
energy resolutions for different cone radii are shown in Fig.~\ref{AngRes_vs_EnergyRes}.
The points closest to the origin can be considered to give the best
resolutions and they are in good agreement with the clustering parameters obtained for the optimal reconstruction efficiency.

\subsubsection{Inclusive \textrm {$k_{T}$} Algorithm}
\begin{figure}[!htp] 
  \begin{minipage}[c]{0.5\linewidth}
    \centering
    \includegraphics[width=0.6\textwidth]{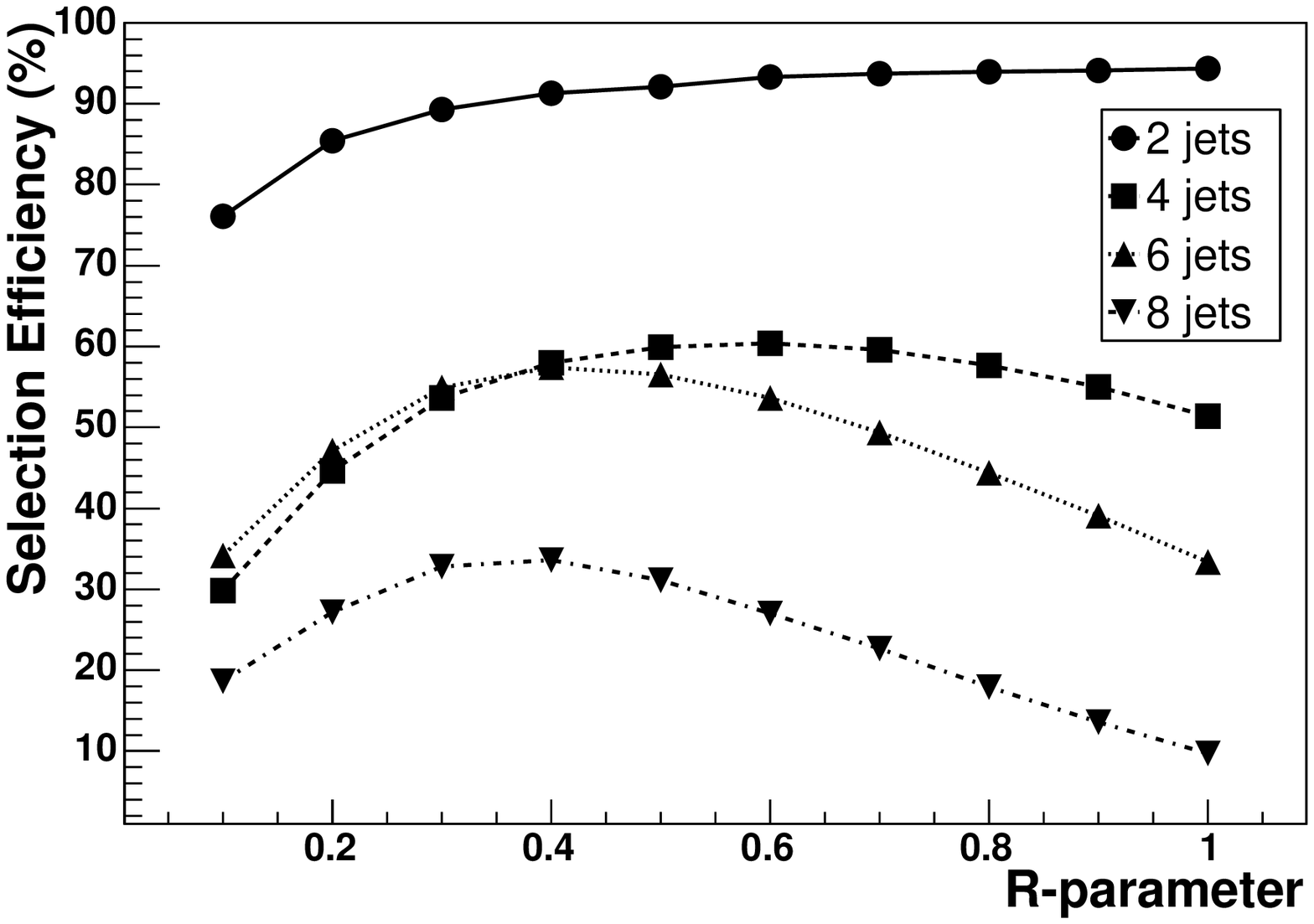} 
%\setcaptionwidth{0.9\textwidth}
    \caption{Fraction of selected events versus
      the  R-parameter ($k_{T}$ algorithm).}
    \label{KTseleff_Rpar}
  \end{minipage}
  \begin{minipage}[c]{0.5\linewidth}
    \centering
    \includegraphics[width=0.6\textwidth]{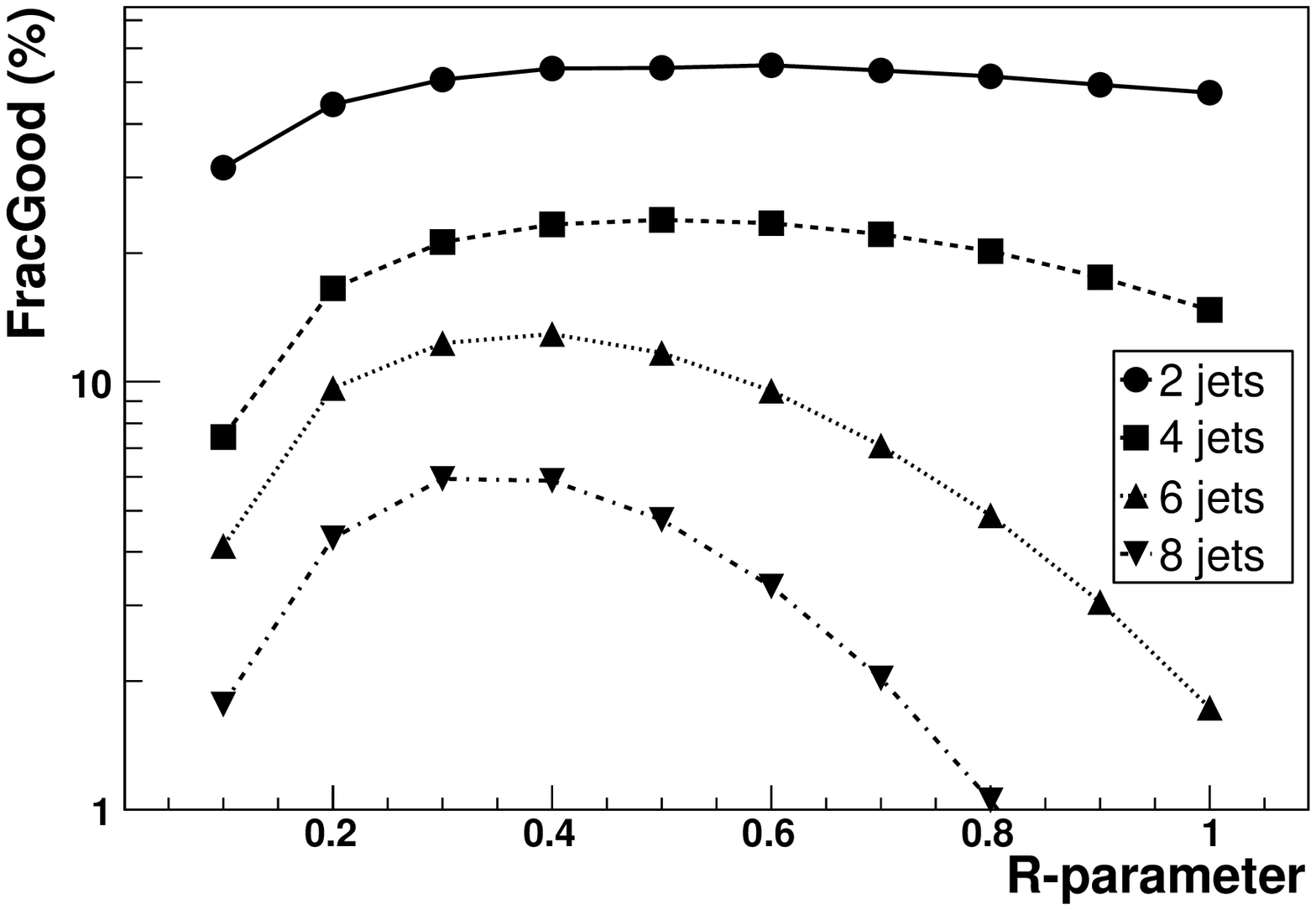} 
%\setcaptionwidth{0.9\textwidth}
    \caption{Fraction of well clustered and selected events versus
      the  R-parameter ($k_{T}$ algorithm).} 
    \label{KTFracGood_Rpar}
  \end{minipage}    
\end{figure}

Fig.~\ref{KTFracGood_Rpar} shows the result for the scan of the
R-parameter of the inclusive $k_{T}$ algorithm. Again, a strong dependence on the
jet multiplicity is observed. For the two quark topology, R=0.6 gives the best performance, 
while this value is reduced to 0.3 for the 8 quark topology. This behaviour is expected 
keeping in mind that the R-parameter
plays a comparable role for the inclusive $k_{T}$ algorithm as the jet radius does
for the {\it Iterative Cone} algorithm. Compared to the optimal
configuration of the {\it Iterative Cone} algorithm, this algorithm
performs almost identical for the two quark case, but is able to get
higher reconstruction efficiencies for larger jet multiplicities.  

\begin{figure}[!htp] 
  \begin{minipage}[t]{0.5\linewidth}
    \centering
    \includegraphics[ width=0.6\textwidth]{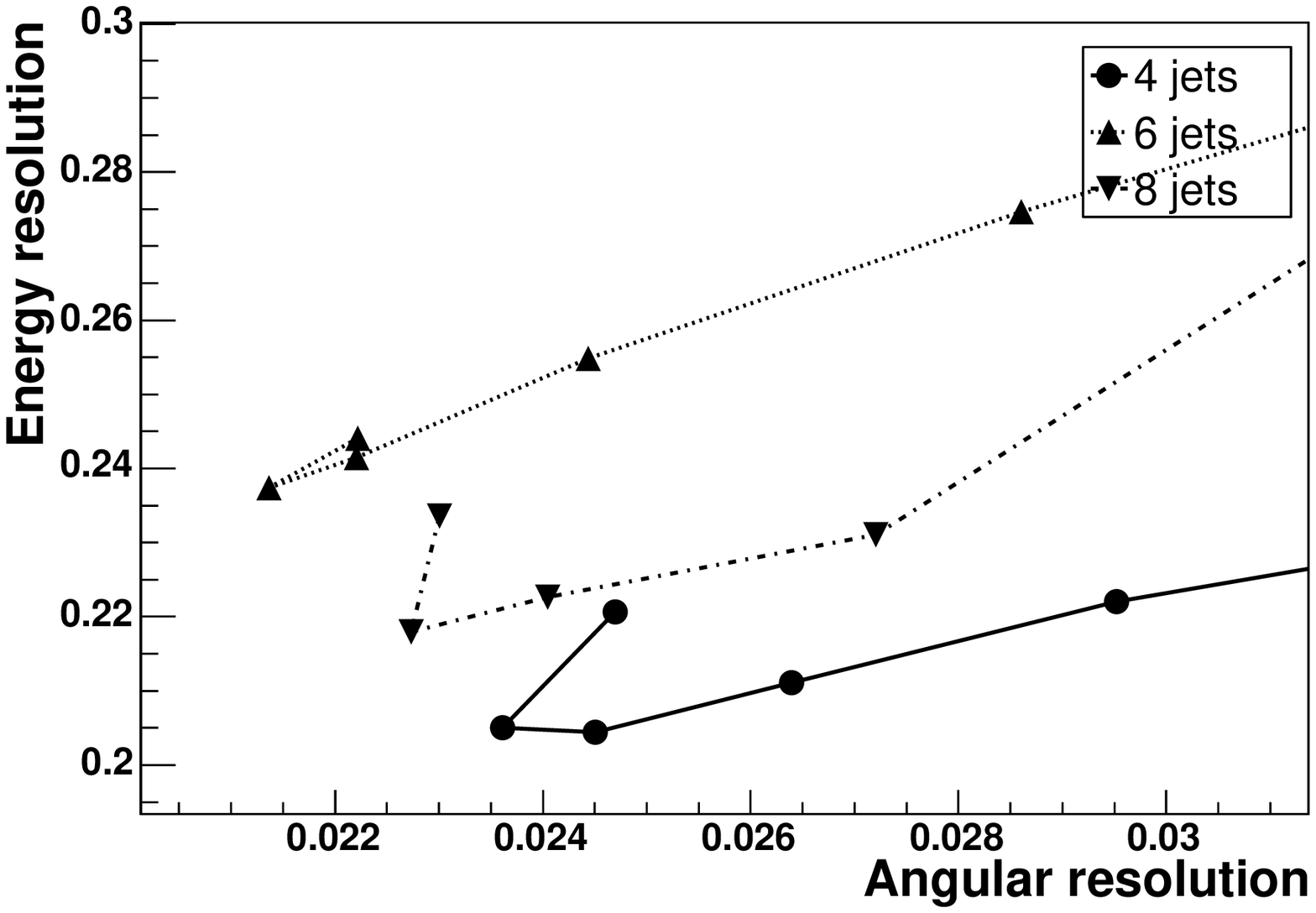} 
  \end{minipage}
  \begin{minipage}[t]{0.5\linewidth}
    \centering
    \includegraphics[ width=0.6\textwidth]{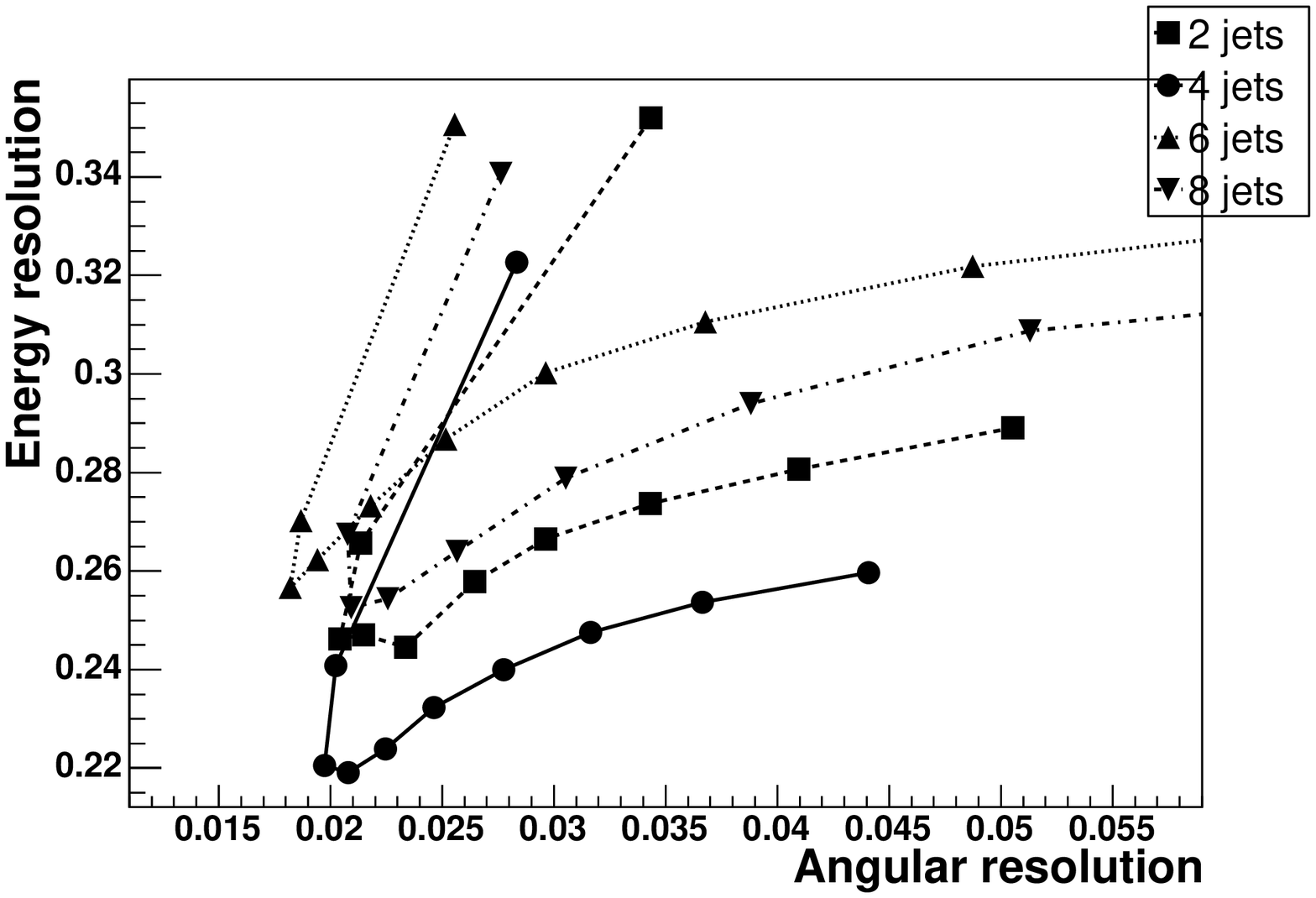} 
  \end{minipage}
  \caption{Relative energy resolution versus angular resolution ($\Delta R$
    distance between jet and quark) for the IC
    algorithm (left) and \textrm{$k_{T}$} algorithm (right). The markers of the same type represent
    the seven 
    different IC cone radii (0.2-0.8) or the ten R-parameter values (0.1-1). The values on the top
    left are the ones with the smallest cone radii or R-parameter values,
    respectively. The energy resolution is defined as the
    RMS divided by the mean value of the $E^{jet}/E^{quark}$ distribution, and the angular
    resolution is defined by the width of a gaussian fit to the
    symmetrized $\Delta R$ distribution. For this plot, the
    two quark-jet pairs with the worst matching (only the worst one in the case of two jets) have been removed to reduce the effect of radiation.}
  \label{AngRes_vs_EnergyRes}  
\end{figure}

The resolution plot in Fig.~\ref{AngRes_vs_EnergyRes} shows a similar
behaviour as for the {\it Iterative Cone} algorithm. The resolution
seems to be optimal for a R-parameter value for which also the fraction of selected
and well clustered events is maximized.

\subsubsection{Midpoint Cone algorithm}
\begin{figure}[!htp] 
  \begin{minipage}[t]{0.50\linewidth}
    \centering
    \includegraphics[width=0.6\textwidth]{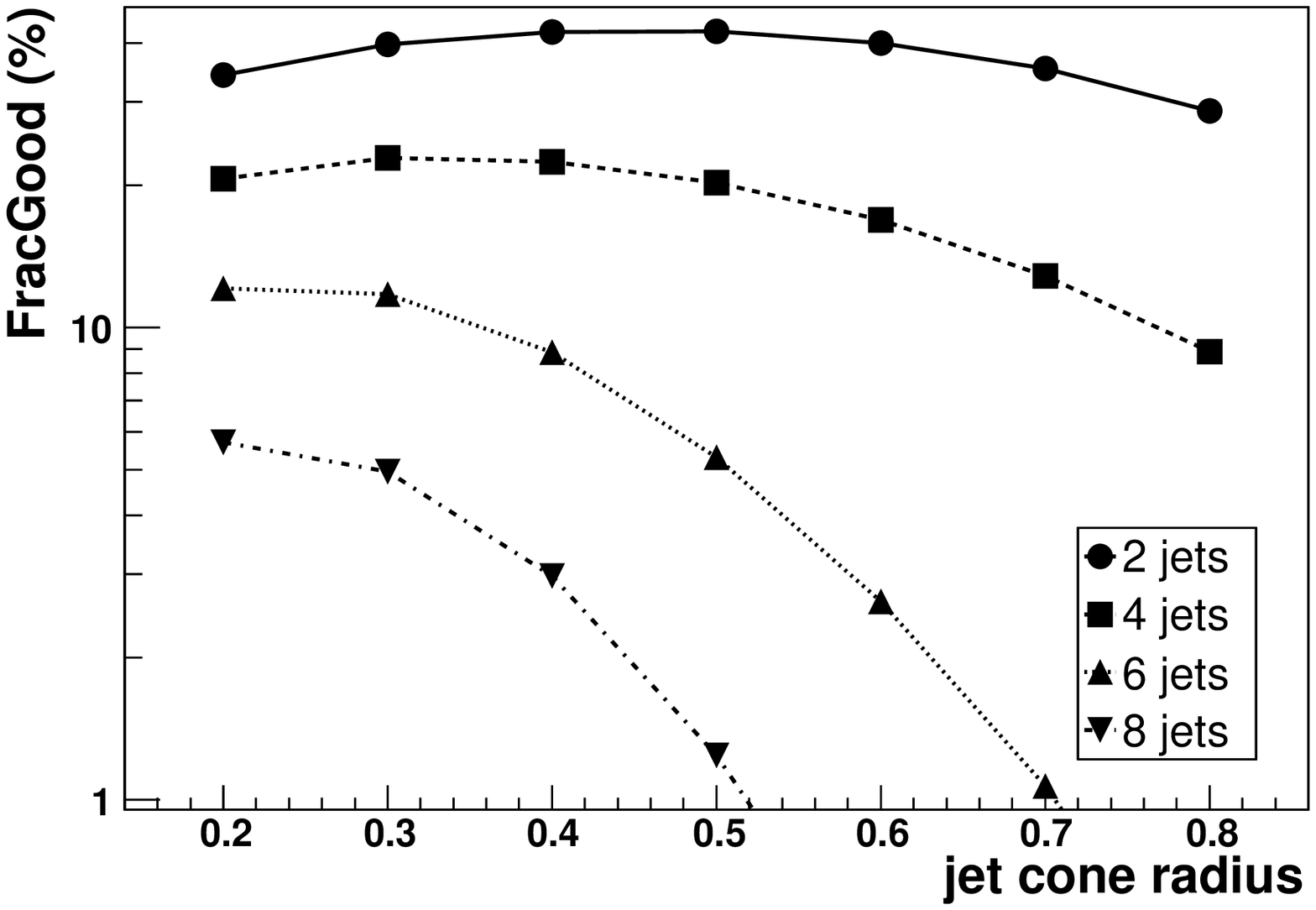} 
    %\setcaptionwidth{0.9\textwidth}
    \caption{Fraction of well clustered and selected events versus the
      cone radius for a merging threshold of 0.5 and a cone area
      fraction of 0.25 (MC algorithm).}
    \label{MCFracGood_coneRad}
  \end{minipage}
  \begin{minipage}[t]{0.50\linewidth}
    \centering
    \includegraphics[width=0.6\textwidth]{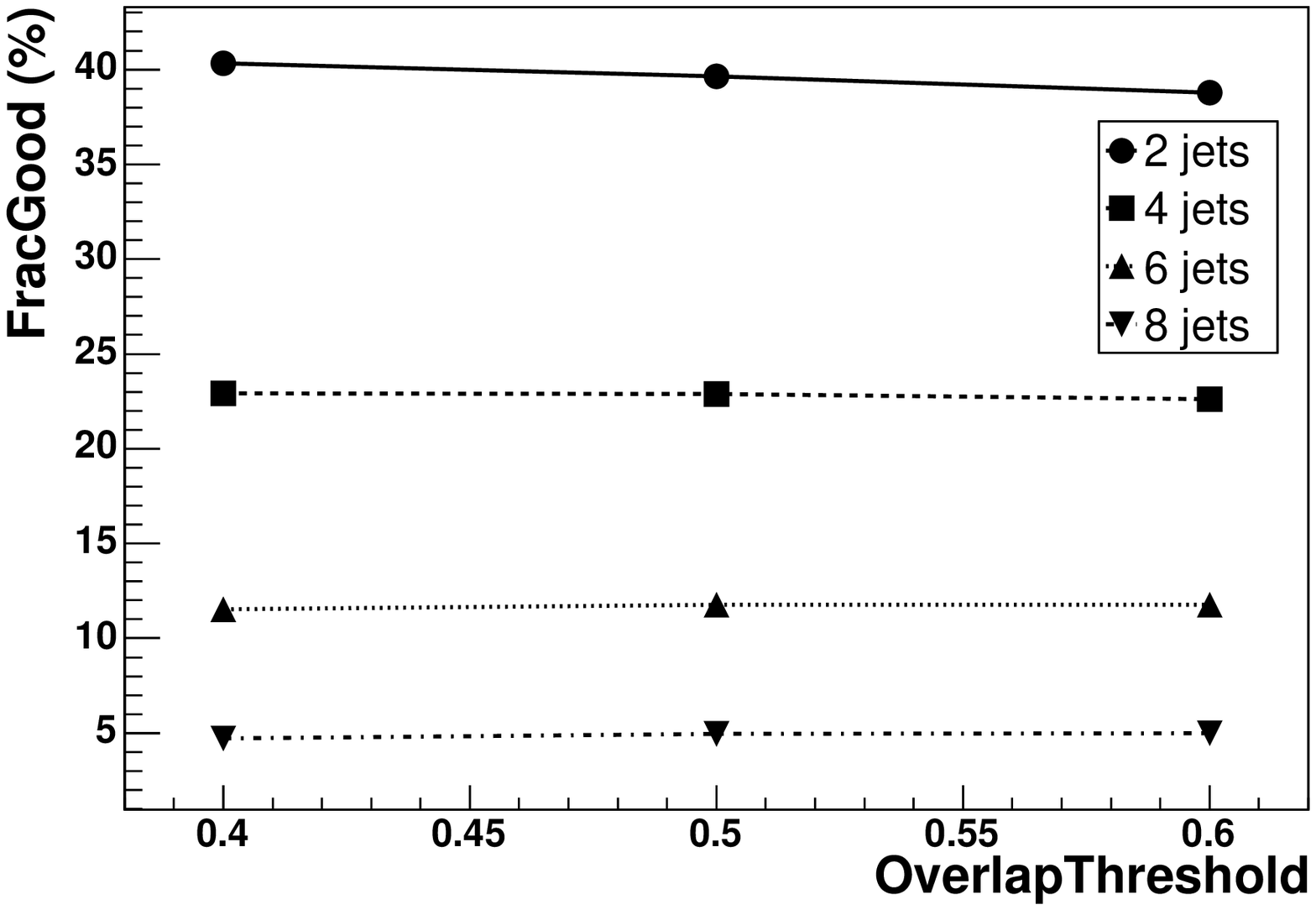} 
    %\setcaptionwidth{0.9\textwidth}
    \caption{Fraction of well clustered and selected events versus the
      threshold for merging for a cone radius of 0.3 and an area
      fraction of 0.25 (MC algorithm).}
    \label{MCFracGood_energyThresh}
  \end{minipage}    
%  \begin{minipage}[t]{0.33\linewidth}
%    \centering
%    \includegraphics[width=0.8\textwidth]{s_schmidt/GenNoMuNu_MCFracGood_ConeAreaFraction} 
%    %%\setcaptionwidth{0.9\textwidth}
%    \caption{Fraction of well clustered and selected events versus the
%      cone area fraction for a cone radius of 0.3 and a threshold for
%      merging of 0.5.}
%    \label{MCFracGood_AreaFraction}
%  \end{minipage}    
 \end{figure}

The scan of the cone radius is shown in Fig.~\ref{MCFracGood_coneRad}
and the dependence on the shared energy fraction threshold for merging is shown in
Fig.~\ref{MCFracGood_energyThresh}.
For high jet multiplicities, the {\it MidPoint Cone} algorithm is able
to reach slightly higher efficiencies than the {\it Iterative Cone}
algorithm. Surprisingly, almost no dependence on the shared energy fraction
threshold for merging has been found. This behaviour might be related to the fact that simulated
particles have been used as input.

\subsubsection{Summary of the Main Observations}
Table \ref{BCDHtable1} summarizes the optimal parameter values for the three jet clustering algorithms, and for each of the 
considered event topologies.
For each optimal jet configuration, the respective estimate of the fraction of well reconstructed events is given.
\begin{table}
\begin{center}
\caption{Overview of the optimal parameter values with their respective estimate of the fraction of well reconstructed events.}
  \vspace*{1mm}
\begin{footnotesize}
\begin{tabular}{|c|c|c|c|c|c|c|c|c|}
\hline
&\multicolumn{2}{c|}{IC}&\multicolumn{2}{c|}{\textrm{$k_T$}}&\multicolumn{4}{c|}{MC}\\
&\multicolumn{2}{c|}{jet radius}&\multicolumn{2}{c|}{R-parameter}&\multicolumn{2}{c|}{jet radius}&\multicolumn{2}{c|}{Overlap 
Threshold}\\
\cline{2-9}
&$Value$&$FracGood$&$Value$&$FracGood$&$Value$&$FracGood$&$Value$&$FracGood$\\
\hline
2 quarks&0.5&53.9&0.6&54.9&0.5&42.4&0.40&40.3\\
4 quarks&0.5&22.3&0.5&23.8&0.3&22.8&0.40-0.50&22.9\\
6 quarks&0.3&11.2&0.4&12.9&0.2&12.1&0.50-0.60&11.8\\
8 quarks&0.3&4.85&0.3&5.93&0.2&5.72&0.60&5.0\\
\hline
\end{tabular}
\label{BCDHtable1}
\end{footnotesize}
\end{center}
\end{table}

\subsubsection{Correlation Between Optimized Configurations}
\begin{figure}[!htp] 
  \begin{minipage}[c]{0.5\linewidth}
    \centering
    \includegraphics[width=0.6\textwidth]{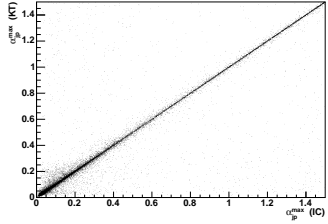} 
    %\setcaptionwidth{0.9\textwidth}
    \caption{Correlation between the IC and \textrm{$k_{T}$} algorithms for the
        $\alpha_{jp}^{max}$ variable in the case of the final state
       with four primary quarks.}
    \label{corr_alpha}
  \end{minipage}
  \begin{minipage}[c]{0.5\linewidth}
    \centering
    \includegraphics[width=0.6\textwidth]{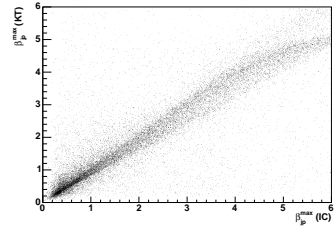} 
    %\setcaptionwidth{0.9\textwidth}
    \caption{Correlation between the IC and \textrm{$k_{T}$} algorithms for the
        $\beta_{jp}^{max}$ variable in the case of the final state
        with four primary quarks.} 
    \label{corr_beta}
  \end{minipage}    
\end{figure}

The correlation between the optimized {\it Iterative
Cone} algorithm and the inclusive $k_{T}$ algorithm for the final
state with four primary quarks is shown in Fig.~\ref{corr_alpha} and
\ref{corr_beta}, for the $\alpha_{jp}^{max}$ and $\beta_{jp}^{max}$
variables.

%The performance of both optimized algorithm configurations is strongly
%correlated. If one selects the same jet candidates with a jet clustering performed with an IC- and a KT algorithm, one has a large possibility that this event was indeed
%well clustered, which is useful information from an analysis
%point of view.

\subsubsection{Robustness Of The Method Against Hard Radiation \label{sec:radiation}}
\begin{figure}[htp] 
  \centering
  \begin{minipage}[t]{0.245\linewidth}
    \centering
    \includegraphics[width=0.99\textwidth]{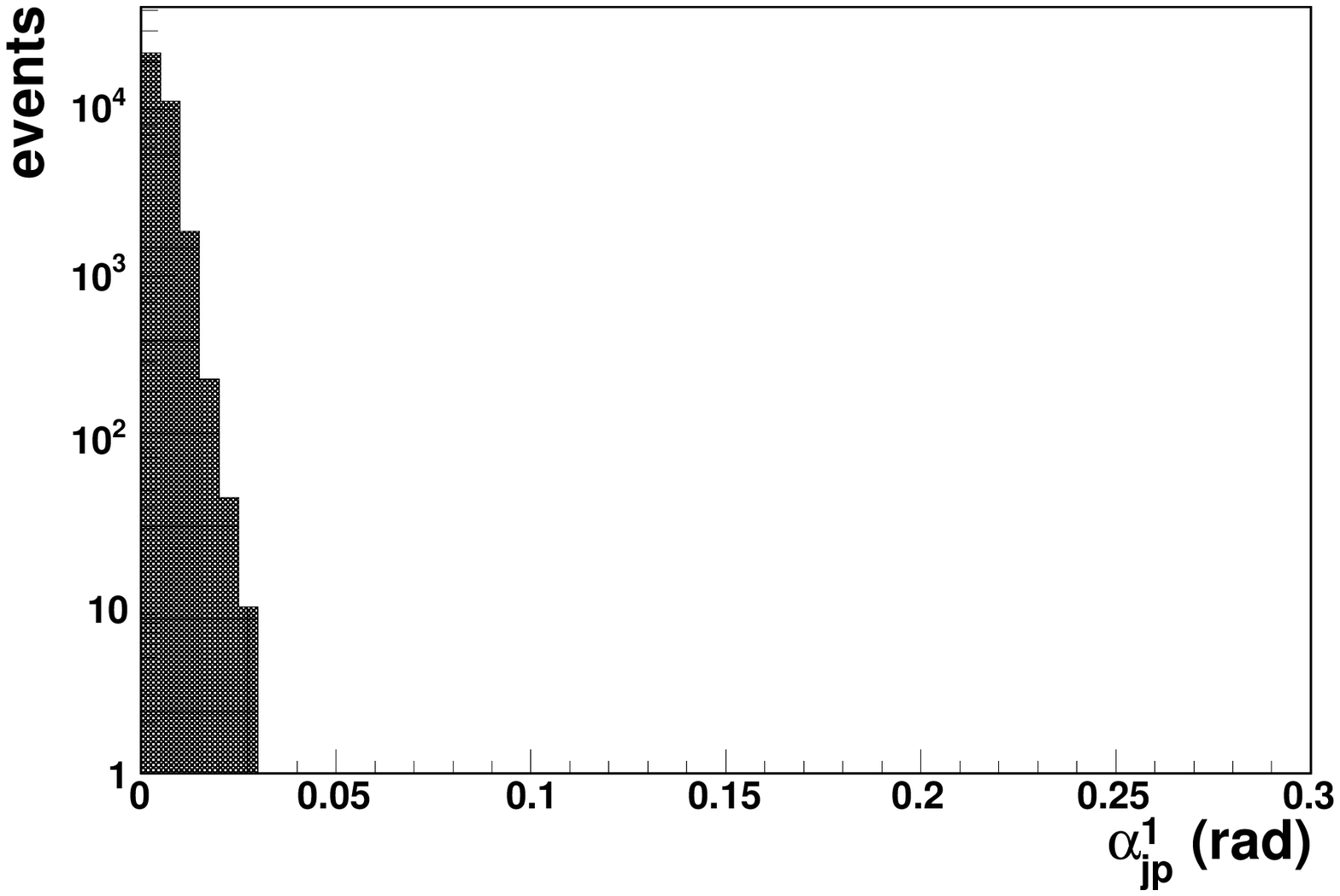} 
  \end{minipage}
  \begin{minipage}[t]{0.245\linewidth}
    \centering
    \includegraphics[width=0.99\textwidth]{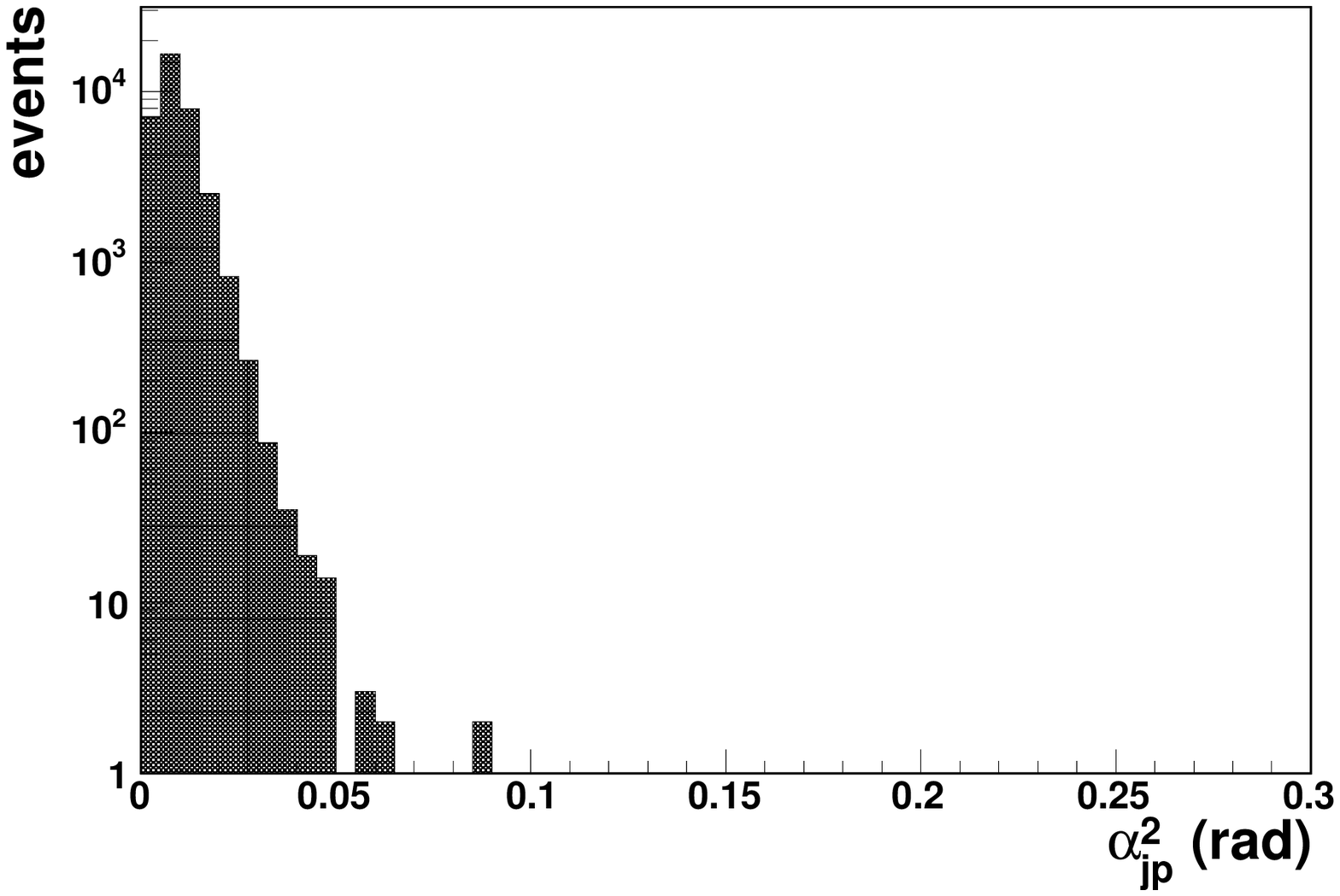} 
  \end{minipage}    
  \begin{minipage}[t]{0.245\linewidth}
    \centering
    \includegraphics[width=0.99\textwidth]{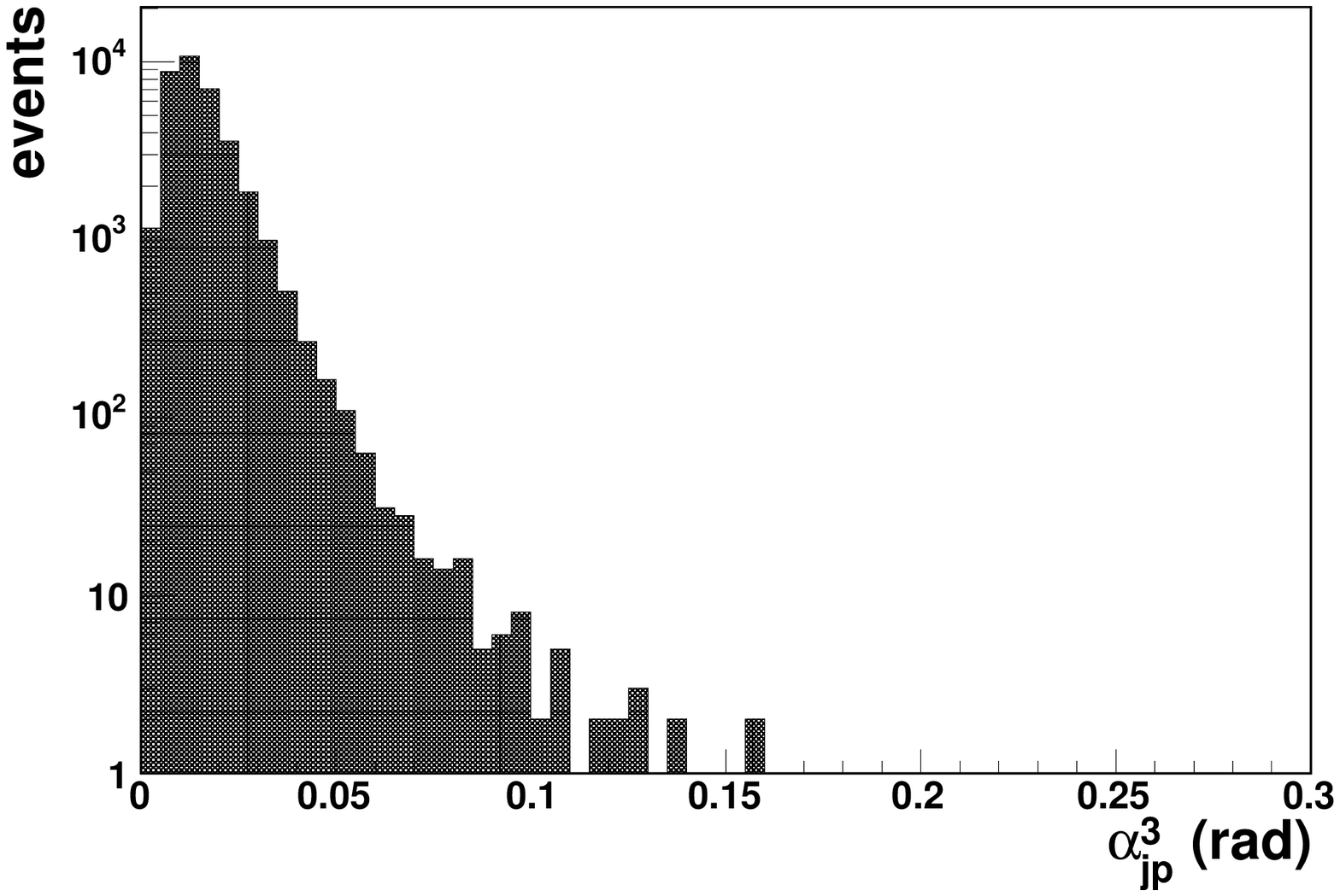} 
  \end{minipage}
  \begin{minipage}[t]{0.245\linewidth}
    \centering
    \includegraphics[width=0.99\textwidth]{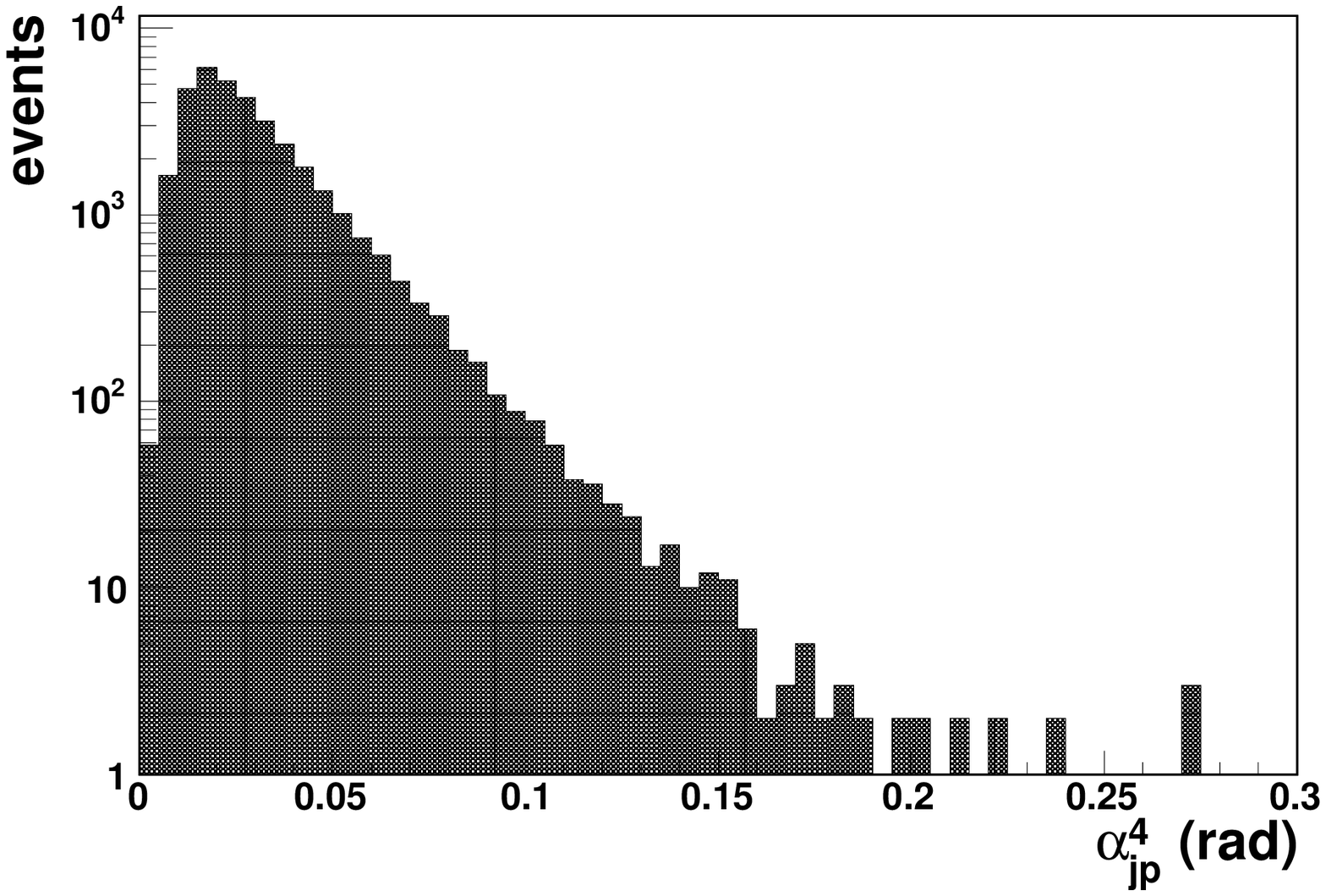} 
  \end{minipage}    
  \caption{Distributions of $\alpha_{jp}^{i}$  in increasing order of magnitude
          for the IC algorithm in the case of a
          final state with four primary quarks which do not radiate hard gluons.}
  \label{maxalphaNOFSR}
 \end{figure}

The sensitivity of the overall observations to
the radiation of gluons with a large transverse momentum relative to their mother quark, or from the initial state proton system, is investigated in the following. The distributions of
the $\alpha_{jp}^{i}$-values ordered by their magnitude within an event are shown in Fig.~\ref{maxalphaNOFSR} for a sample without initial and final state
radiation\footnotemark\footnotetext{PYTHIA parameters $MSTP$ $61$ and $71$ were switched off.}.
%\begin{figure}[hbtp!] 
%  \begin{center}
%        \includegraphics[width=0.4\textwidth]{s_schmidt/MaxAlphaIC041NOFSR} 
%       \caption{Distributions of $\alpha_{jp}$  in increasing order
%          for the {\it Iterative Cone} algorithm in the case of the
%          four jets final state, without final state radiation. } 
%        \label{maxalphaNOFSR}
%  \end{center} 
%\end{figure}
 This has to be compared directly to Fig.~\ref{maxalpha} which shows
the same plots including final state radiation.
Obviously, the long tails are not present in the case without
radiation which indicates that the $\Delta R$ cut of 0.3 for the worst jet is
not expected to have an effect in this case. The observation is
indeed, that the Frac($\alpha_{jp}^{max}$+$\beta_{jp}^{max}$) quality
marker has a flat distribution, but not the selection efficiency and
therefore the ``FracGood'' quality marker.

\begin{figure}[!htp] 
  \begin{minipage}[c]{0.5\linewidth}
    \centering
    \includegraphics[width=0.6\textwidth]{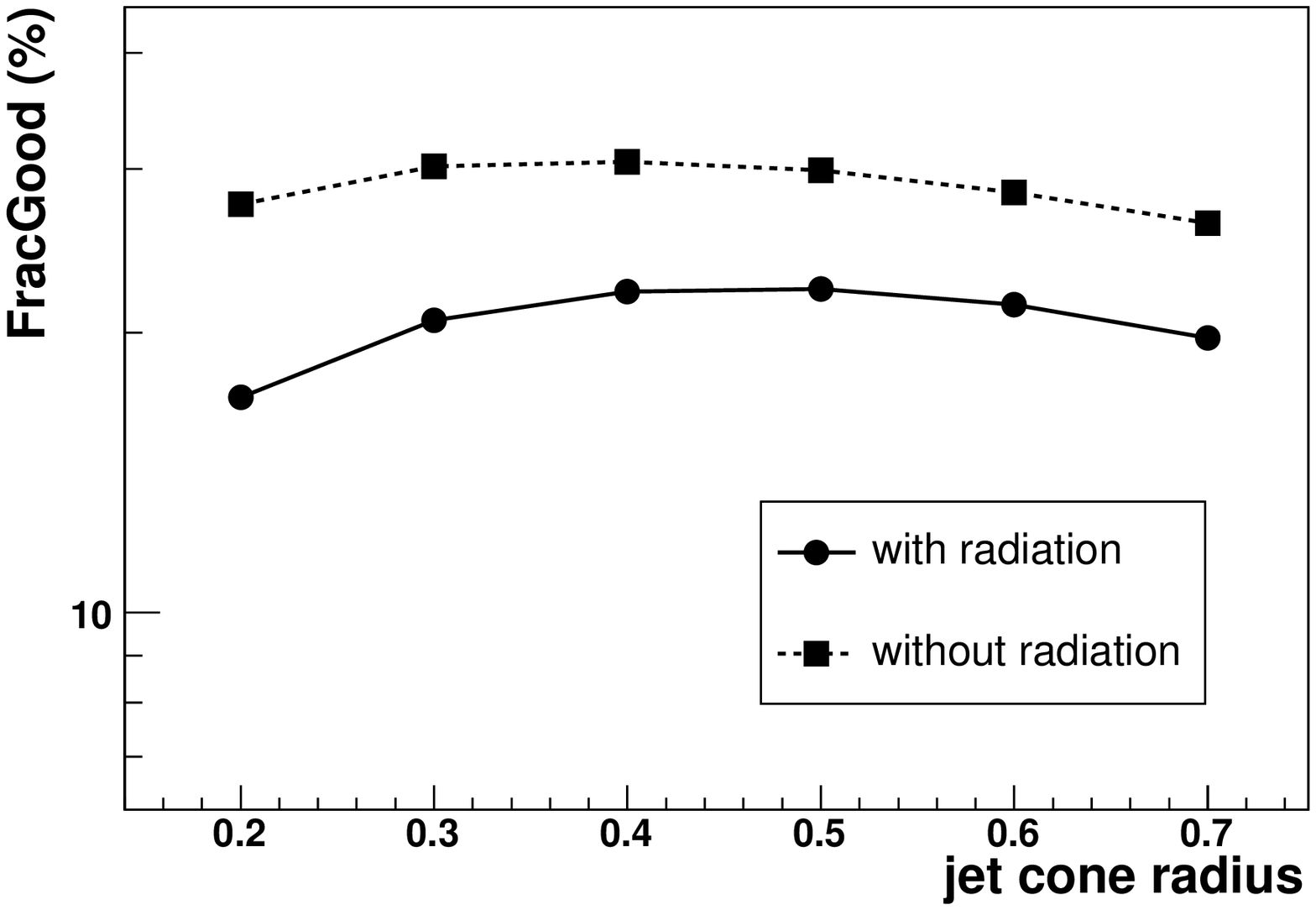} 
    %\setcaptionwidth{0.9\textwidth}
    \caption{Influence of hard gluon radiation on the fraction
        of selected, well clustered events, as a function of the IC  
        cone radius in the case with four primary quarks in the final state.}
    \label{radiationIC}
  \end{minipage}
  \begin{minipage}[c]{0.5\linewidth}
    \centering
    \includegraphics[width=0.6\textwidth]{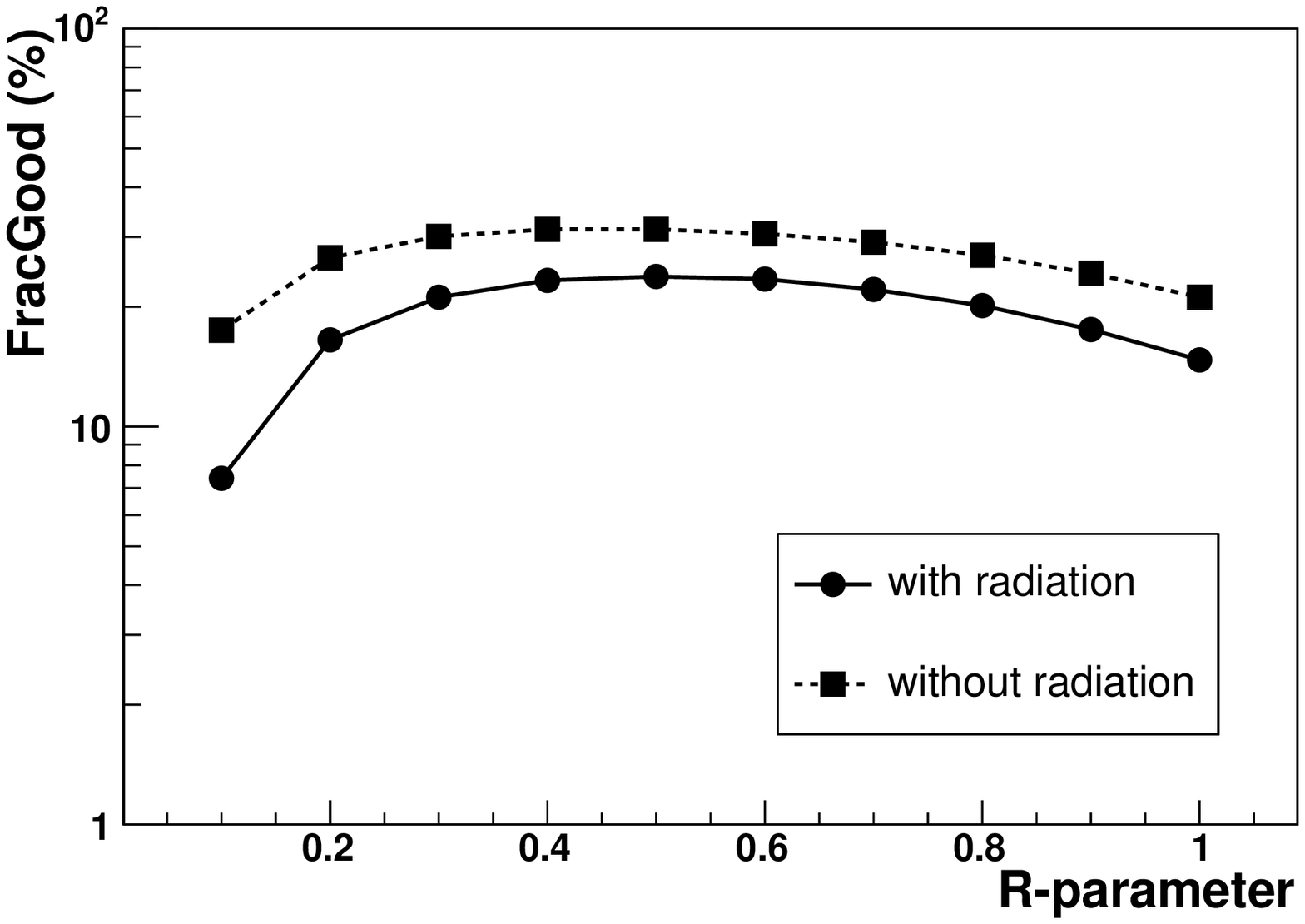} 
    %\setcaptionwidth{0.9\textwidth}
    \caption{Influence of hard gluon radiation on the fraction
        of selected, well clustered events, as a function of the \textrm{$k_{T}$}  
        R-parameter in the case with four primary quarks in the final state.} 
    \label{radiationKT}
  \end{minipage}    
\end{figure}

Fig.~\ref{radiationIC} shows the
fraction of selected, well clustered semileptonic $t\bar{t}$ events with and without
initial and final state radiation for the {\it Iterative Cone}
algorithm. The addition of radiation results in an overall lower
efficiency, but the optimal cone radius and the shape of the curve
are robust. A similar observation was obtained for the inclusive
$k_T$ algorithm in Fig.~\ref{radiationKT}.

\begin{figure}[hbtp] 
  \begin{minipage}[t]{0.5\linewidth}
    \centering
    \includegraphics[width=0.6\textwidth]{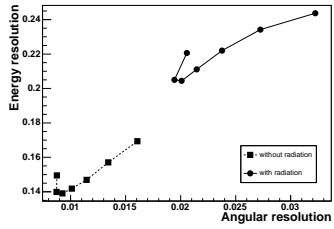} 
  \end{minipage}
  \begin{minipage}[t]{0.5\linewidth}
    \centering
    \includegraphics[width=0.6\textwidth]{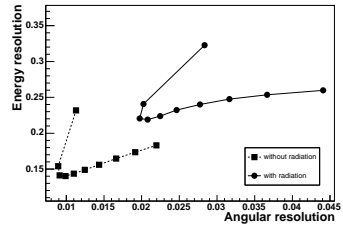} 
  \end{minipage}
  \caption{Energy resolution versus angular resolution ($\Delta R$
    distance between jet and quark) for the IC
    algorithm (left) and \textrm{$k_{T}$} algorithm (right)  in the case of four
    jets in the final state. The definitions are identical to them in
    Fig. \ref{AngRes_vs_EnergyRes}.  } 
  \label{AngRes_vs_EnergyRes_NORAD}  
\end{figure}

In order to quantify the effect of radiation on the resolutions,
Fig.~\ref{AngRes_vs_EnergyRes_NORAD} shows the two cases for the {\it
  Iterative Cone} and the inclusive $k_T$ algorithm.
As expected, the overall resolutions are better in the case without
radiation, but the shape of the curves remains invariant.

%  \begin{minipage}[c]{0.5\linewidth}
%    \centering
%    \includegraphics[width=0.7\textwidth]{} 
%    %%\setcaptionwidth{0.9\textwidth}
%    \caption{}
%    \label{}
%  \end{minipage}
%  \begin{minipage}[c]{0.5\linewidth}
%    \centering
%    \includegraphics[width=0.7\textwidth]{} 
%    %%\setcaptionwidth{0.9\textwidth}
%    \caption{} 
%    \label{}
%  \end{minipage}    
% \end{figure}

\subsection{Conclusions}
In this paper three jet clustering algorithms (the {\it Iterative Cone} algorithm, the
inclusive {\it $k_{T}$} algorithm and the {\it MidPoint Cone} algorithm) are compared according to some pre-defined performance criteria based on reconstruction efficiencies of the kinematics of the primary quarks.
A scan of the most important algorithm parameters is performed in order to determine the
optimal jet clustering from an analysis point of view, i.e. to maximize the
reconstruction efficiency. 

As a proof of concept for the quality definition of the jet clustering, the top quark mass was determined from reconstructed jets. The quality markers were able to isolate the narrow, gaussian top quark mass
peak in a broad distribution.

%Also the robustness of the obtained results against final state
%radiation, as shown in section \ref{sec:radiation}, confirms the usability of
%the presented method. (FIXME: this is not true)

The study was performed on different event samples
with topologies ranging from two primary quarks (fully leptonic and semileptonic
top quark pairs) up to eight ($t\bar t H\rightarrow b q \bar{q}
\bar{b}\bar{q}q b\bar{b}$). As expected, it was found that smaller
opening angles 
are better suited for higher jet multiplicities.

The presented results have been obtained using simulated particle
information at the vertex, and without a magnetic field, but similar results have already been extracted for a full
detector simulation. These results will be published in dedicated
papers for the specific experiments.

%%%%%%%%%%%%%%%%%%%%%%%%%%%%%%%%%%%%%%%%%%%%%%%%%%%%%%%%%%%%%%%%%%%%%%%%%%%%%
\section[Colour annealing --- a toy model of colour reconnections]
{COLOUR ANNEALING --- A TOY MODEL OF COLOUR RECONNECTIONS~\protect
\footnote{Contributed by: M.~Sandhoff, P.~Skands}}
\subsection{Introduction}
Among the central objectives of collider physics is the
precise measurement of the elementary particle masses and couplings. 
Striking recent
examples are the measurements both at LEP and at the Tevatron 
of the mass of the $\skW$ boson to a precision better than one per mille 
\cite{Group:2005di,Abazov:2003sv} --- a precision giving truly
valuable insight into the mechanism of electroweak symmetry breaking
as well as in probing for the quantum effects of New Physics. 

At present, with the top quark in focus at the Tevatron and the
physics programme of the LHC only a few years distant,
the solid understanding of QCD phenomena beyond
leading-order perturbation theory is becoming increasingly more
important, with a large range of both experimental and theoretical
methods and tools being developed. The aim, to achieve
theoretical and systematic uncertainties capable of matching 
the expected statistical precision 
of the large data samples becoming available. 

Apart from developments in flavour physics and lattice QCD, 
essentially all of these approaches focus on the perturbative domain of
QCD --- in brief: including more legs/loops/loga\-rithms
in the calculations. The point we wish to stress here is that, even 
assuming these approaches to one day 
deliver predictions with negligible uncertainties associated with uncalculated
perturbative orders, there still remains the non-perturbative aspects, 
for which current understanding cannot be called primitive, but
certainly not crystal clear either.

Recently, the structure and physics of the underlying event
has received some attention \cite{tunea,Affolder:2001xt,Buttar:2004iy,Field:2005sa}, but again 
the main theoretical thrust, with few exceptions
\cite{Sjostrand:2002ip,Sjostrand:2004pf},  
has taken place in the perturbative modeling, in the
form of more sophisticated models for multiple perturbative
interactions \cite{Sjostrand:1987su,Butterworth:1996zw,Sjostrand:2004ef}. 
While non-perturbative aspects certainly play a significant role,
and enter into the descriptions in the form of various
phenomenological parameters, they 
generally suffer from being hard to quantify, 
hard to calculate, and hard to test. In this study, we shall focus on
precisely such a source of potential uncertainty: colour
reconnection effects in the final state, in particular in the context
of measurements made at hadron colliders. 

In Section \ref{sec:colrec}
we briefly discuss some previous cerebrations on colour reconnections, and in
Section \ref{sec:model} present our own toy model, for use in the present
study. In Section \ref{sec:results} we give a few explicit examples
and show some results for $\skt\sktbar$ events at the
Tevatron. Section \ref{sec:conclusions} contains a summary and
outlook. 

\subsection{Colour Reconnections \label{sec:colrec}}
\begin{figure}
\begin{center}
\begin{fmffile}{fmfww}
\begin{tabular}{cc}
\begin{fmfgraph*}(120,60)
\fmfset{arrow_len}{3mm}
\fmfbottom{b}
\fmfleft{l1,l2}
\fmfright{r2,r1}
\fmf{fermion}{l1,v1}
\fmf{fermion}{v1,l2}
\fmf{boson,tension=2}{v1,c}
\fmf{boson,tension=2}{c,v2}
\fmf{fermion}{r1,v2}
\fmf{fermion}{v2,r2}
\fmfdot{c}
\fmfv{label=$\skq$}{l2}
\fmfv{label=$\skqbar$}{l1}
\fmfv{label=$\skq'$}{r2}
\fmfv{label=$\skqbar'$}{r1}
\fmffreeze
\fmf{dashes,left=0.1}{l2,l1}
\fmf{dashes,left=0.1}{r2,r1}
\fmfv{label=a)}{b}
\end{fmfgraph*}
\hspace*{1cm}
&
\hspace*{1cm}
\begin{fmfgraph*}(120,60)
\fmfbottom{b}
\fmfset{arrow_len}{3mm}
\fmfleft{l1,l2}
\fmfright{r2,r1}
\fmf{fermion}{l1,v1}
\fmf{fermion}{v1,l2}
\fmf{boson,tension=2}{v1,c}
\fmf{boson,tension=2}{c,v2}
\fmf{fermion}{r1,v2}
\fmf{fermion}{v2,r2}
\fmfdot{c}
\fmfv{label=$\skq$}{l2}
\fmfv{label=$\skqbar$}{l1}
\fmfv{label=$\skq'$}{r2}
\fmfv{label=$\skqbar'$}{r1}
\fmffreeze
\fmf{dashes,right=0.1}{l2,r1}
\fmf{dashes,right=0.1}{r2,l1}
\fmfv{label=b)}{b}
\end{fmfgraph*}
\end{tabular}
\end{fmffile}\vskip3mm
\caption{a) the original colour topology in hadronic 
$\ske^+\ske^-\to\skW\skW$ events, and b) a
  reconnected version. Note that these are not Feynman diagrams but 
 rather spatial diagrams depicting the situation after the
  annihilation, with the production point at the origin. Arrows
  pointing against the direction of motion signify antifermions.
\label{fig:WW}}
\end{center}
\end{figure}
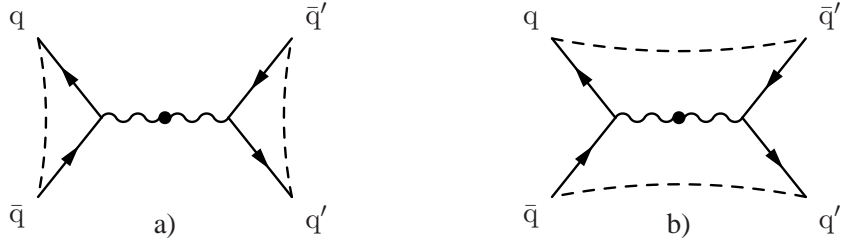

The subject of colour rearrangements was first studied by Gustafson,
Pettersson, and Zerwas (GPZ)  \cite{Gustafson:1988fs}, there in a mainly
qualitative way, and in the
context of rearrangements taking place already at the perturbative level. 
They observed that, e.g.\ in hadronic $\skW\skW\to(\skq_1\skqbar_2)(\skq_3\skqbar_4)$ 
events at LEP, illustrated in Fig.~\ref{fig:WW}a
with colour connections traced by dashed
lines, interference effects and 
gluon exchanges between the decay products could lead to a 
reconfiguration of the colour topology into the one depicted in
Fig.~\ref{fig:WW}b. In the reconnected topology, both the perturbative
QCD cascade and the subsequent hadronisation phase would be
substantially different, leading to very large effects. 

Sj\"ostrand and Khoze (SK) \cite{Sjostrand:1993rb,Sjostrand:1993hi} 
subsequently argued that such large effects
were most likely unrealistic. A reconnection already at the
perturbative level requires at least two perturbative gluon vertices, 
leading to an $\alpha_s^2$
suppression. Moreover, the relevant reconnection diagram is colour 
suppressed by $1/N_c^2$ with respect to the leading (non-reconnected)
$\mathcal{O}(\alpha_s^2)$ diagrams. Finally, for the decay products of
the two $\skW$ bosons to radiate coherently, they must, in the
language of wave mechanics, be in phase, which 
only occurs for radiation at energies smaller than the $\skW$ width. In
other words, gluons with wavelengths smaller than the typical
separation of the two $\skW$ decay vertices will be radiated (almost)
incoherently. For these reasons, SK considered a
scenario where reconnections occur as part of the non-perturbative
hadronisation phase. 

The SK model is based on the standard Lund string
fragmentation model \cite{Andersson:1983ia}, in which 
the chromo-electric flux lines formed between colour
charges separated at distances larger than $\sim 1\skfm$ are represented
by simple massless strings. SK argued that, if two such strings
overlap in space and time, there should be a finite possibility for
them to `cut each other up' and rearrange themselves, much as has been
recently discussed for the case of cosmic and mesonic superstrings
\cite{Jackson:2004zg,Cotrone:2005fr}. However, since we do not
yet know  whether QCD strings behave more like flux tubes 
in a Type II or a Type I superconductor (roughly speaking whether the
topological information is stored in a small core region or not), 
SK presented two distinct models, commonly referred to
as SK-II and SK-I, respectively. As would be expected, both models resulted in 
effects much smaller than in the
GPZ model, leading to a predicted total uncertainty on the $\skW$ mass
from this source of $\sigma_{M_{\skW}}<40~\skMeV$. SK also performed a study of
QCD interconnection effects in $\skt\sktbar$ production 
\cite{Khoze:1994fu}, but only in the context of
$\ske^+\ske^-$ collisions.

Subsequently, a number of alternative models have also been proposed,
most notably the ones proposed by the Lund group, based on QCD dipoles
\cite{Gustafson:1994cd,Lonnblad:1995yk,Friberg:1996xc}, and one based
on clusters by Webber \cite{Webber:1997iw}. Apart from $\skW\skW$ physics,
colour reconnections have also been proposed to model rapidity gaps
\cite{Buchmuller:1995qa,Edin:1995gi,Enberg:2001vq} and quarkonium
production \cite{Edin:1997zb}.

Returning to $\ske^+\ske^-$, experimental investigations at LEP II have
not found conclusive evidence  
of the effect \cite{Abbiendi:1998jb,Abbiendi:2005es}, but were limited
to excluding only the more dramatic scenarios, such as GPZ and 
versions of SK-I with the recoupling strength parameter close to
unity. Hence, while colour reconnection effects cannot be arbitrarily
large, there is room for further speculation. In addition, 
as we shall argue below, it may be possible that the effect is
enhanced in hadron collisions over $\ske^+\ske^-$ --- 
with the added complication that
the environment at hadron colliders is necessarily much less benign
to this sort of measurement than was the case at LEP.

\subsection{Our Toy Model --- Colour Annealing\label{sec:model}} 
In electron--positron annihilation, the two incoming
states carry electromagnetic charge --- giving rise to a dilute cloud of
virtual photons surrounding them --- but no strong charge. From the QCD
point of view, the vacuum state is thus undisturbed in the initial
state, at least up to effects of order $\alpha^2$, i.e.\
  $\ske\to\ske'\gamma^*\to\ske'\skq\skqbar$. 
After the production of, say, a $\skW\skW$ pair, e.g.\ with both $\skW$
bosons decaying hadronically, $\ske^+\ske^- \to \skW^+\skW^- \to
(\skq_1\skqbar_2)(\skq_3\skqbar_4)$, further QCD radiation and hadronisation
then develops, in the background of this essentially pure vacuum
state. As discussed above, the final state colour topology during the
perturbative part of the QCD cascade, at least down to energies of
order the $\skW$ width, in all likelihood is the one depicted in
Fig.~\ref{fig:WW}a. For gluon energies smaller than the $\skW$ width,
however, the question is still relatively open. 

Going to (inelastic, non-diffractive) hadron-hadron collisions, 
the initial state already contains strong charges. Using a simple bag
model for illustration, the vacuum at the collision point and in the space-time
area immediately surrounding it would not be the undisturbed one
above, but would rather correspond to the vacuum \emph{inside}
the hadronic bag. Though detailed modeling is beyond the scope of the
present discussion, we note that soft colour fields living inside this
bag, with wavelengths of order the hadron size $\sim$ hadronisation
length, could impact in a non-trivial way the formation of colour
strings at the time of hadronisation
\cite{Buchmuller:1995qa,Edin:1995gi}, effects that would not
have been present in $\ske^+\ske^-$ collisions.

We are not aware of any detailed studies, neither experimental nor
theoretical at this time. 
Several of the models mentioned above would still be more
or less directly applicable, but the noisier environment of hadron
colliders makes it daunting to attempt to look for any effect. In this
paper, we propose a simple toy model, to give a first indication of
the possible size of the effect, in particular for $\skt\sktbar$
production at the Tevatron. 

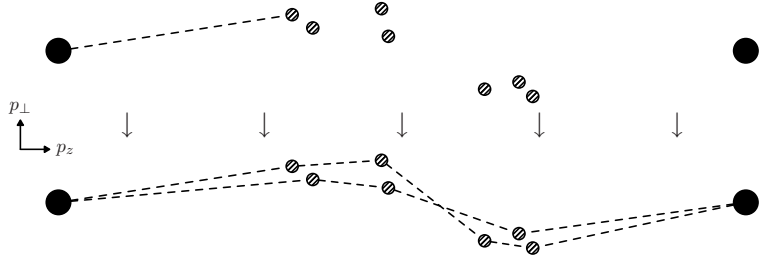
\begin{figure}
%%% Draw ttbar here with (a visualisation of) soft colour field as
%%% background. Include an interaction with background?
\begin{center}
\scalebox{0.65}{
\begin{minipage}{\textwidth}
\vspace*{2.5cm}\begin{fmffile}{fmfaxes}
~~\begin{fmfgraph*}(15,15)
\fmfforce{0.0w,0.0h}{O}
\fmfforce{1.0w,0.0h}{X}
\fmfforce{0.0w,1.0h}{Y}
\fmf{vanilla}{O,X}
\fmf{vanilla}{O,Y}
\fmfv{d.sh=tri,d.siz=4,d.ang=30,lab=$p_z$,l.ang=0}{X}
\fmfv{d.sh=tri,d.siz=4,lab=$p_\perp$,l.ang=90}{Y}
\end{fmfgraph*}\vspace*{-3.6cm}
\end{fmffile}
\begin{center}
\begin{fmffile}{fmfann}
\begin{tabular}{cc}
\begin{fmfgraph*}(400,70)
\fmfleft{brl}
\fmfright{brr}
\fmfforce{0.1w,0.0h}{b1}
\fmfforce{0.3w,0.0h}{b2}
\fmfforce{0.5w,0.0h}{b3}
\fmfforce{0.7w,0.0h}{b4}
\fmfforce{0.9w,0.0h}{b5}
\fmfv{lab=\scalebox{1.5}{$\downarrow$},l.d=1,l.ang=-90}{b1,b2,b3,b4,b5}
\fmfv{d.sh=circ}{brl}
\fmfv{d.sh=circ}{brr}
\fmfforce{0.34w,0.8h}{g2}
\fmfforce{0.37w,0.69h}{g3}
\fmfforce{0.48w,0.62h}{g4}
\fmfforce{0.47w,0.85h}{g5}
\fmfforce{0.62w,0.18h}{g9}
\fmfforce{0.67w,0.24h}{g11}
\fmfforce{0.69w,0.12h}{g12}
\fmfv{d.sh=circ,d.siz=0.1h,d.fi=shaded}{g2}
\fmfv{d.sh=circ,d.siz=0.1h,d.fi=shaded}{g3}
\fmfv{d.sh=circ,d.siz=0.1h,d.fi=shaded}{g4}
\fmfv{d.sh=circ,d.siz=0.1h,d.fi=shaded}{g5}
\fmfv{d.sh=circ,d.siz=0.1h,d.fi=shaded}{g9}
\fmfv{d.sh=circ,d.siz=0.1h,d.fi=shaded}{g11}
\fmfv{d.sh=circ,d.siz=0.1h,d.fi=shaded}{g12}
\fmf{dashes}{brl,g2}
\end{fmfgraph*}\\[5mm]
\begin{fmfgraph*}(400,70)
\fmfleft{brl}
\fmfright{brr}
\fmfv{d.sh=circ}{brl}
\fmfv{d.sh=circ}{brr}
\fmfforce{0.34w,0.8h}{g2}
\fmfforce{0.37w,0.69h}{g3}
\fmfforce{0.48w,0.62h}{g4}
\fmfforce{0.47w,0.85h}{g5}
\fmfforce{0.62w,0.18h}{g9}
\fmfforce{0.67w,0.24h}{g11}
\fmfforce{0.69w,0.12h}{g12}
\fmfv{d.sh=circ,d.siz=0.1h,d.fi=shaded}{g2}
\fmfv{d.sh=circ,d.siz=0.1h,d.fi=shaded}{g3}
\fmfv{d.sh=circ,d.siz=0.1h,d.fi=shaded}{g4}
\fmfv{d.sh=circ,d.siz=0.1h,d.fi=shaded}{g5}
\fmfv{d.sh=circ,d.siz=0.1h,d.fi=shaded}{g9}
\fmfv{d.sh=circ,d.siz=0.1h,d.fi=shaded}{g11}
\fmfv{d.sh=circ,d.siz=0.1h,d.fi=shaded}{g12}
\fmf{dashes}{brl,g2}
\fmf{dashes}{brl,g3}
\fmf{dashes}{brr,g12}
\fmf{dashes}{brr,g11}
\fmf{dashes}{g2,g5}
\fmf{dashes}{g3,g4}
\fmf{dashes}{g5,g9}
\fmf{dashes}{g4,g11}
\fmf{dashes}{g9,g12}
\end{fmfgraph*}\\
\end{tabular}
\end{fmffile}
\end{center}
\end{minipage}}
\end{center}\vspace*{-4mm}
\caption{Type I colour annealing in a schematic $\skg\skg\to\skg\skg$
  scattering. Black dots: beam remnants. Smaller dots: gluons
  emitted in the perturbative cascade. All objects here are colour
  octets, hence each dot must be connected to two string pieces. Upper: the
  first connection made. Lower: the final string topology.\label{fig:ann}}
\end{figure}

Since we do not expect the difference in background vacuum to affect
the short-distance physics, we take the arguments of SK concerning the absence
of colour reconnections at the perturbative level to still be
valid. Though one could still imagine reconnections below the relevant
resonance widths, we shall not consider this. That is, we let
the entire perturbative evolution remain unchanged, and implement our
model at the hadronisation level only. Having no explicit model for
how the presence of soft background fields would affect the collapse
of the colour wave functions at hadronisation time, we consider an extreme case,
where the quarks and gluons completely forget their colour
`history'.  Instead, what determines between which partons hadronising
    strings form is a minimization of the total potential energy stored
    in strings. Specifically, we propose that the partons, 
regardless of their formation history,
    will tend to be colour connected to the partons closest to them in
    momentum space, hence minimizing the string length and thereby the
    average particle multiplicity produced by the configuration, as
    measured by the so-called `Lambda
    measure'\cite{Andersson:1983jt,Sjostrand:2002ip}, here given for
massless partons for simplicity:
\begin{equation}\displaystyle
\Lambda = \Pi_{i=1}^N \frac{m^2_{\skmrm{i}}}{M_0^2}~~~,
\end{equation}
where $i$ runs over the number of colour-anticolour pairs (dipoles) in the
event, $N$, $m_{\skmrm{i}}$ is the invariant mass of the $i$'th
dipole, and $M_0$ is a constant normalisation factor of order
the hadronisation scale.  
The average multiplicity produced by string fragmentation is
proportional to the logarithm of $\Lambda$. 
    Technically, the model implementation starts by erasing the colour
    connections of all final state coloured partons, including ones
    from $\skW$ decays etc. It then begins an
    iterative procedure (which unfortunately can be quite
    time-consuming):
\begin{enumerate}
      \item Loop over all final state coloured partons.
      \item For each such parton with a still unconnected colour or
      anticolour charge, \vskip2mm
	\begin{enumerate}
          \item Compute the $\Lambda$ measure for each possible string
             connection from that parton to other 
             final state partons which have a compatible free colour
             charge. 
%%% Interesting to construct an algorithm that does not require the
%%% colour charge to be free, i.e.\ where already existing connections
%%% can be stolen/snapped/rewired.
         \item Store the connection with the smallest $\Lambda$ measure for 
             later comparison.
	\end{enumerate}\vskip2mm
      \item Compare all the possible `minimal string pieces' found, one for
         each parton. Select the largest of these to be carried out
         physically. That parton is in some sense the one that is
         currently furthest away from all other partons. 
     \item If any `dangling colour charges' are left, repeat from 1.
     \item At the end of the iteration, if the last
         parton is a gluon, and if all other partons already form a 
         complete colour singlet system, the remaining gluon is
         simply attached between the two partons where its presence
         will increase the total $\Lambda$ measure the least.
\end{enumerate}
This procedure will find a local minimum of the $\Lambda$
measure. More aggressive models could still be constructed, most
noticeably by refining the algorithm to avoid being trapped in 
shallow local minima. As a side remark, we note that the above
procedure, which we shall refer to as Type II below, as it stands 
would tend to result in a number of small
closed gluon loops. Hence, we also consider a variant (Type I) where closed
gluon loops are suppressed, if other possibilities exist, see
illustration in Fig.~\ref{fig:ann}. 
Both variants of the annealing algorithm are implemented in
\textsc{Pythia} 6.326, and are carried over to \textsc{Pythia} 6.4, 
where they can be accessed using the
\texttt{MSTP(95)} switch, see also the update notes \cite{update} and
the \textsc{Pythia} 6.4 manual \cite{Sjostrand:2006za}.

\subsection{Results \label{sec:results}}
As a first application of the new models, 
we consider their effects on semileptonic $\skt\sktbar$ events at the
Tevatron. Specifically, whether an effect could be observable in the
light-quark jet system from the hadronic $\skW$ decay. 
This is closely related to the work presented in \cite{Sandhoff:2005}. 

For any fragmentation model, the first step is to make a (re)tune of
the minimum-bias and underlying-event (UE) parameters. Ideally, the
whole range of model parameters should come under scrutiny, however
for the present study we limit ourselves to
a one-parameter retuning of the multiple interactions colour-screening
cutoff in \sktsc{Pythia} (\skttt{PARP(82)}), requiring the retuned models
to agree with the average charged particle multiplicity of Tune A
\cite{tunea}. Below, we compare Tune A to a preliminary tune of the
new UE framework (Old CR) adapted from the Low FSR
tune in \cite{Sjostrand:2004ef}, and to the same model with Type I and
Type II colour reconnections applied. For the 4 models,
\skttt{PARP(82)=2.0}, \skttt{2.1}, \skttt{2.2}, \skttt{1.55}, respectively. 

Next, for each of the tuned models, 
50000 $\skt\sktbar$ events were generated at $E_{CM}=1960~\skGeV$, corresponding
to approximately $8 \skmrm{fb}^{-1}$ of integrated luminosity. Out of the
semi-leptonic fraction of this sample, events with exactly
four charged particle jets were selected (clustered with an
exclusive kT jet algorithm~\cite{ktjet} with $d_{\skmrm{cut}}=150~\skGeV^2$). Finally, 
the jets have to be uniquely identified 
to the correct parton. This was done requiring that 
the (and only the) dedicated jet has a minimal
$\Delta R$ between its axis and the initial parton.

In the undisturbed colour
topology, three string pieces are relevant; one spanned between the
$\skW$ jets, 
one between the $\skb$ quark and the $\skp$ beam remnant, and one between
the $\skbbar$ and the $\skpbar$ remnant. To maximise the overlap of these
strings, and hence create a bias towards situations where colour
reconnections should be enhanced, we reject events that do not fulfill
either condition A) $\eta_{\skq}>\eta_{\skqbar}>\eta_{\skb}$ or
B) $\eta_{\skbbar}>\eta_{\skq}>\eta_{\skqbar}$. 

For each accepted event, we perform a boost to the rest frame of the hadronic
$\skW$, then a polar rotation to line up the decay jets along the $z$
axis (for condition A (B), the quark is rotated to $0^\circ$
($180^\circ$)), and finally an azimuthal rotation to bring the $\skb$
jet from the associated top decay into the $(x,z)$ plane, in the
positive-$x$ hemisphere. We then reject events where the other $\skb$
jet is not also in the positive-$x$ hemisphere, so that the
negative-$x$ hemisphere between the $\skW$ jets should, at least to some extent, 
be free from extraneous hadronic activity. 

\begin{figure}
\vspace*{-5mm}
\begin{center}
\begin{tabular}{cc}
\includegraphics*[scale=0.4]{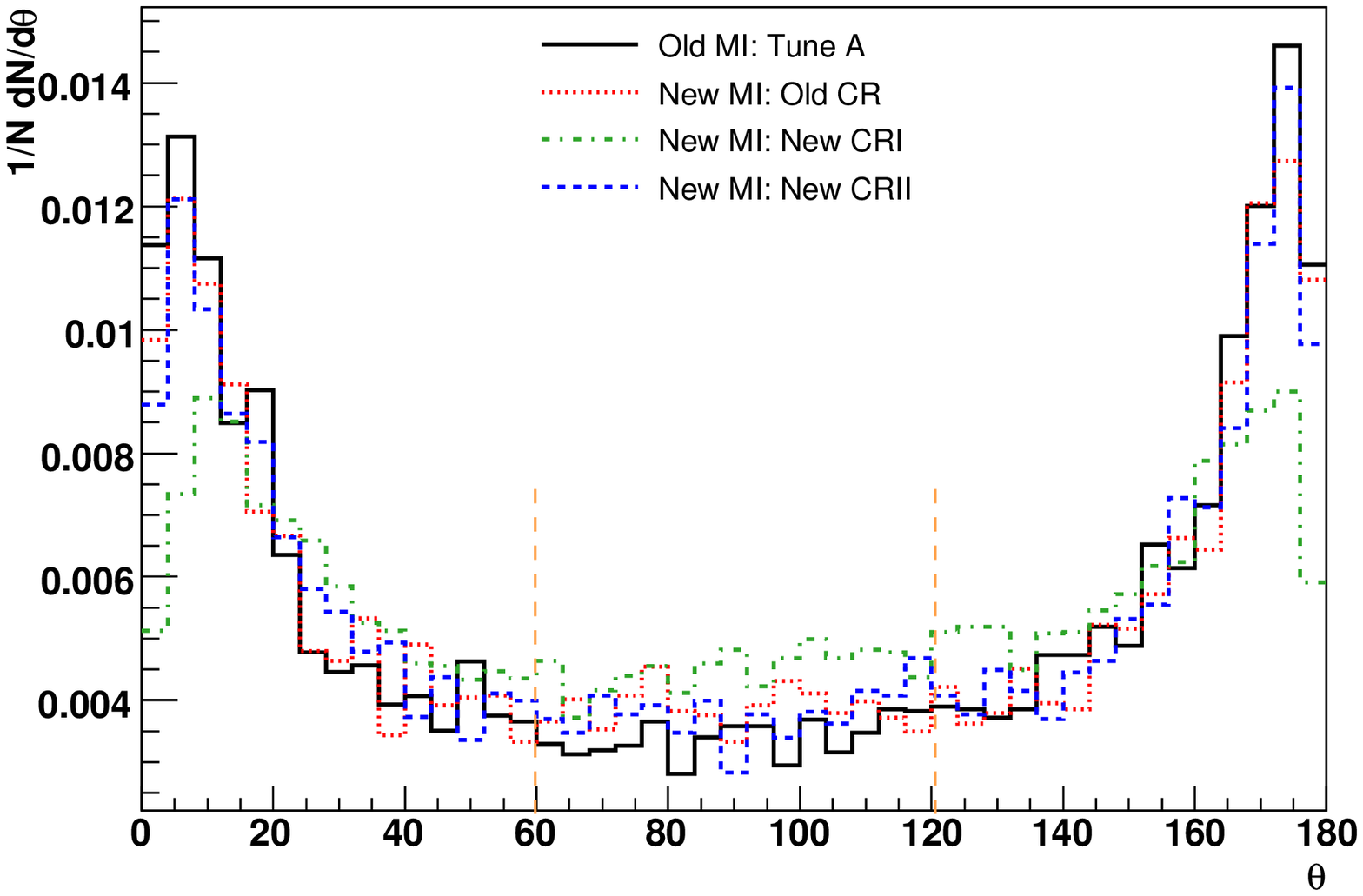} &
\includegraphics*[scale=0.4]{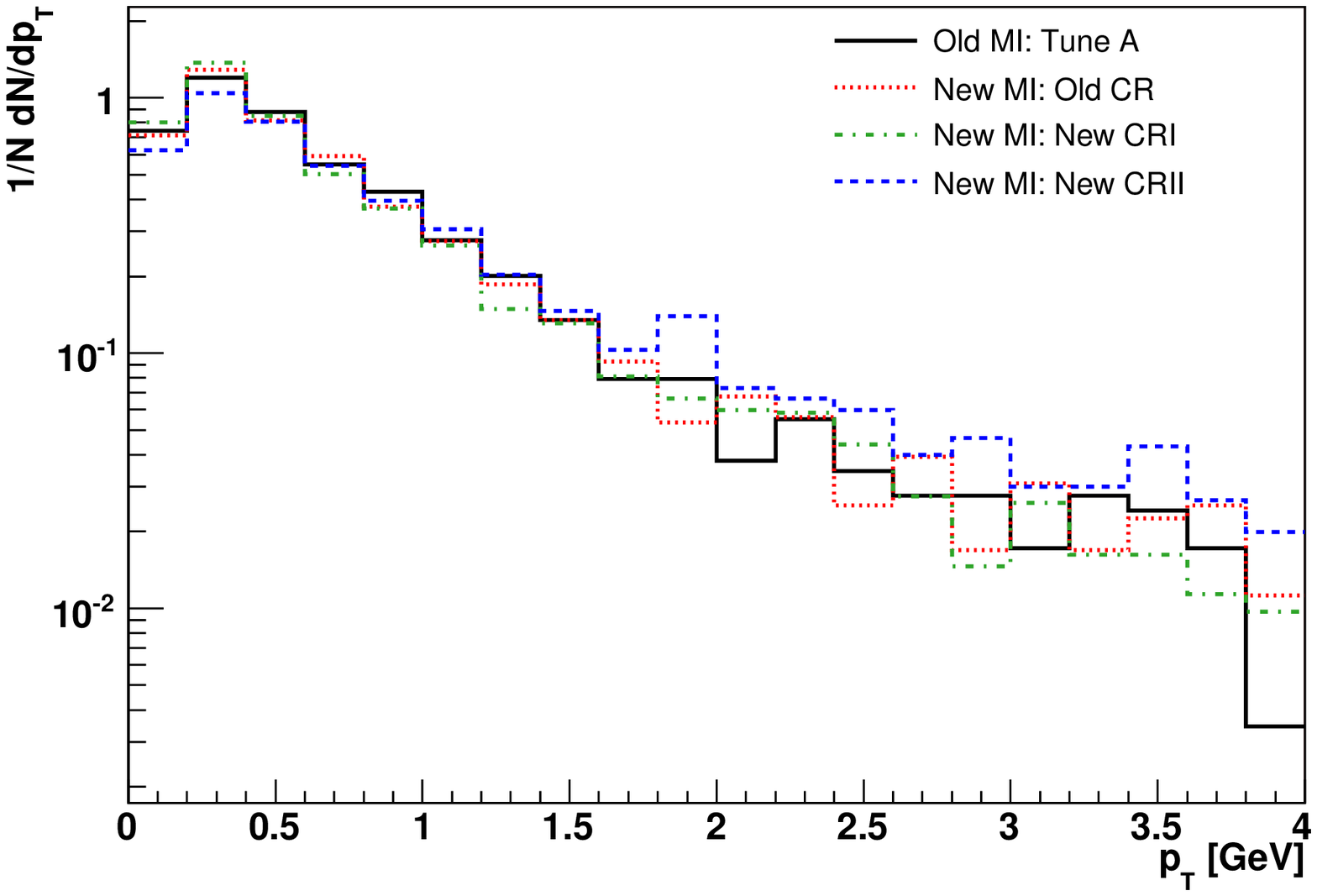}\\[-5mm]
\small\it a) & \small\it b)\end{tabular}\vspace*{-6mm}
\end{center}
\caption{Semi-leptonic top events at the Tevatron (see text). {\it a)} 
Charged particle density between the $\skW$ jets (note the zero
suppression) and
  {\it b)} $p_\perp$ spectra for charged particles in the region $60^\circ <
  \theta < 120^\circ$. 
\label{fig:ill1}
}
\end{figure}
We consider two observables, in both cases only including particles in
the negative-$x$ hemisphere. First, in Fig.~\ref{fig:ill1}a, 
the charged particle multiplicity
between the jets, $1/N_{\skmrm{ch}} \skmrm{d}N_{\skmrm{ch}}/\skmrm{d}\theta$, 
and second, in Fig.~\ref{fig:ill1}b, the transverse momentum
distribution $1/N_{\skmrm{ch}} \skmrm{d}N_{\skmrm{ch}}/\skmrm{d}p_\perp$ for
particles in the inter-jet region, $60^\circ < \theta < 120^\circ$,
indicated in Fig.~\ref{fig:ill1}a by dashed
vertical markers. 

In Fig.~\ref{fig:ill1}a, the asymmetry between the left and right peak
sizes is due to the rapidity constraints and to the way we performed
the rotations; conditions A and B then both force
the associated $\skb$ quark to be closer to the right-hand jet. 
Given the subtle nature of the effect, and the noisy hadronic
environment, the variations in Fig.~\ref{fig:ill1} are quite
large (the distortion of the peak shape at small angles for Type
I is, however, probably too large to be realistic). 
However, notice that the reconnected scenarios do \emph{not} lead to
a significant reduced charged particle density in the inter-jet region, which
would have been the effect we should naively have been looking for, by
comparison to the $\ske^+\ske^-$ studies. We note, however, that the most
agressive of the new models,  
Type II (blue dashed curve), does produce fewer
particles in the fragmentation region than its sister Type I (green
dot-dashed), and also (as shown in figure \ref{fig:ill1}b) that the charged
particles produced in Type II have a higher average $p_\perp$. 

What is going on is that, as for so many aspects of hadron-hadron physics,
the end result is not controlled by one effect alone, but by a
combination of factors. Multiplicity will be increased by allowing
more underlying-event activity and will be decreased by allowing more
colour reconnections. Hence  the same multiplicity can be arrived at through
different mixes of these. By first tuning to the min-bias data we
are to some extent cancelling these effects against each other.
This illustrates an essential
point: in a hadron-hadron environment, 
the multiplicity alone may not be a discriminating variable.
However, the mixes are not completely equivalent. While they may lead
to the same result in one distribution, they will differ for
another. Specifically, by combining the particle flow with the energy
flow, some discriminating power can be gained. One way of realising this is to 
consider that the underlying activity is pumping energy
into the event. To maintain the same multiplicity distribution, the 
particle hardness must then be a function of the underlying
activity, as is illustrated by Fig.~\ref{fig:ill1}. 
While we shall terminate our
discussion here, the subject of disentangling these effects 
certainly merits further
consideration. 

\subsection{Conclusions \label{sec:conclusions}}
We have presented a few simple toy models of colour reconnections,
based on an annealing-like algorithm. These models are quite general
and are directly applicable to any process, 
unlike many previous models for which only
implementations specific to $\skW\skW$ events exist. 

As a first application, we have studied the effects on 
two simple observables in semileptonic $\skt\sktbar$ events at the
Tevatron. We find that, while we cannot discern the presence or
absence of a classical string effect in the multiplicity distributions
alone, it may still be possible to distinguish between different models by
including energy-flow information. The natural next step
would be to consider the extent to which measurements of the top mass 
at the Tevatron and LHC are influenced by these effects. For instance,
an attractive possibility is to use the hadronically reconstructed
$\skW$ mass in these events to set the jet energy scale, hence the degree
to which the hadronic $\skW$ mass reconstruction is affected by the
effects discussed here would be interesting to examine.
 
We intend this study mostly for illustration 
 and for communicating a few
essential remarks. As such, we have freely (ab)used Monte Carlo truth
information and have skipped lightly over a
number of aspects, which would 
have to be more carefully addressed in a real
 analysis. We hope that this work may
 nevertheless serve to stimulate further efforts in this exciting and
 presently little understood field. 

\subsection*{Acknowledgements}
We are grateful to the organisers of Les Houches 2005, ``Physics at
TeV Colliders'' for a wonderful workshop, 
and to T.~Sj\"ostrand for enlightening discussions and comments on the
manuscript. Furthermore we would like to thank P.~M\"attig and T.~Harenberg
for their support. This work was supported by Universities Research Association
Inc.\ under Contract No.\ DE-AC02-76CH03000 with the United States
Department of Energy.

%%%%%%%%%%%%%%%%%%%%%%%%%%%%%%%%%%%%%%%%%%%%%%%%%%%%%%%%%%%%%%%%%%%%%%%%%%%%%
\section[Tuned comparison of electroweak corrections to Drell--Yan-like
$W$- and $Z$-boson production -- a status report]
{TUNED COMPARISON OF ELECTROWEAK CORRECTIONS TO DRELL--YAN-LIKE $W$- AND
$Z$-BOSON PRODUCTION -- A STATUS REPORT~\protect
\footnote{Contributed by: A.~Arbuzov, U.~Baur, S.~Bondarenko, C.~Carloni
        Calame, S.~Dittmaier, M.~Kr\"amer, G.~Montagna, O.~Nicrosini,
        R.~Sadykov, A.~Vicini, D.~Wackeroth}}
\subsection{Introduction}

The basic parton processes for single vector-boson production are
$q\bar{q'} \to W \rightarrow l \nu_{l}$ and $q\bar{q} \to \gamma/Z
\rightarrow l^+l^-$, with charged leptons $l$ in the final state.
Experimental measurements in the vicinity of the W and Z~resonances
allow for a precise determination of the W-boson mass and, from the
forward-backward asymmetry, of the effective weak mixing angle.
Above resonance, the off-shell tails of appropriate distributions
are sensitive to the gauge-boson decay widths.  At high invariant
masses, for example, of the $l^+l^-$ pair, deviations from the
standard cross section and $A_{\mathrm{FB}}$ could indicate new
physics, such as an extra heavy Z$'$ or extra dimensions.  Finally,
the Drell--Yan-like W and Z~production processes may be used as a
luminosity monitor or to further constrain PDFs at the LHC.

Predictions for Drell--Yan processes do not only receive sizeable QCD
corrections, which are known up to
NNLO~\cite{vanNeerven:1991gh,Anastasiou:2003yy}, but also important
corrections of electroweak origin.  Unfortunately, existing Monte
Carlo generators, such as {\sc Horace}
\cite{CarloniCalame:2003ux,CarloniCalame:2005vc}, {\sc Resbos}
\cite{Balazs:1997xd,Landry:2002ix}, {\sc Winhac} \cite{Placzek:2003zg}
and {\sc Wgrad} \cite{Baur:1998kt} / {\sc Zgrad} \cite{Baur:2001ze},
do not yet combine the complete knowledge on strong {\em and}
electroweak higher-order effects.  A first step towards this goal has
been taken in Ref.~\cite{Cao:2004yy} where the combined effect of
soft-gluon summation and final-state photon radiation has been studied
for W-boson production. For more details on predictions for Drell--Yan
processes we refer to the literature
\cite{Haywood:1999qg,Catani:2000jh,Nadolsky:2004vt,Baur:2005rx}.

In this article we focus on electroweak corrections, which are
completely known in ${\cal O}(\alpha)$.  Historically, a first step to
include electroweak corrections was already made in
Ref.~\cite{Berends:1984qa}, where effects of final-state radiation in
the gauge-boson decay stage were taken into account.  For W~production
the approximation of Ref.~\cite{Berends:1984qa} has been improved
later in Ref.~\cite{Baur:1998kt} by the inclusion of ${\cal
  O}(\alpha)$ corrections to resonant W~bosons and in
Refs.~\cite{Dittmaier:2001ay,Baur:2004ig} by the full corrections of
${\cal O}(\alpha)$.  For Z~production the ${\cal O}(\alpha)$ QED
corrections have been presented in Ref.~\cite{Baur:1997wa} and
completed by the corresponding weak contributions in
Ref.~\cite{Baur:2001ze}.  The particular importance of final-state
photon radiation demands a treatment that goes beyond ${\cal
  O}(\alpha)$. Such multi-photon effects have been studied both for W-
and Z-boson production in
Refs.~\cite{CarloniCalame:2003ux,Placzek:2003zg} and
Ref.~\cite{CarloniCalame:2005vc}, respectively. A comparison of these
two calculations can be found in Ref.~\cite{CarloniCalame:2004qw}.

In the following we focus on precision calculations of electroweak
corrections as performed by various groups in recent years and present
a status report of a comparison for a common set of input parameters
and a uniformly tuned setup (input-parameter scheme, PDFs etc.).  In
particular, the recently released ${\cal O}(\alpha)$-improved PDFs
``MRST2004QED'' \cite{Martin:2004dh} are employed.  This set of PDFs
includes a photon distribution function resulting from the ${\cal
  O}(\alpha)$-driven evolution of the PDFs, i.e.\ the Drell--Yan cross
section receives a new type of real correction from photon-induced
processes.

The different approaches that are compared are briefly summarized in
the next section, the precise setup is described in
Section~\ref{se:setup}, and Section~\ref{se:numerics} contains the
numerical results.

\subsection{Different approaches and codes}

The following collaborations have contributed to the tuned comparison
of results on electroweak corrections to Drell--Yan processes:
\begin{itemize}
\item {\sc Dk}: Ref.~\cite{Dittmaier:2001ay} contains a detailed
  description of the calculation of the ${\cal O}(\alpha)$ corrections
  to W~production at hadron colliders and a discussion of results for
  the Tevatron and the LHC. In particular, the full ${\cal O}(\alpha)$
  calculation is compared with a pole approximation for the
  W~resonance.  The case of Z-boson production is not considered.  For
  the present analysis, the calculation of
  Ref.~\cite{Dittmaier:2001ay} has been extended (i) to include
  final-state radiation beyond ${\cal O}(\alpha)$ via structure
  functions and (ii) by implementing the ${\cal O}(\alpha)$-corrected
  PDF set MRST2004QED.  The photon-induced processes $\gamma q\to q' l
  \nu_l$ and $\gamma \bar{q'} \to \bar{q} l \nu_l$ have been
  calculated as described in Ref.~\cite{Diener:2005me}.  The
  evaluation of the $q\bar{q'}$ channel has been technically improved
  by employing a generalization of the dipole subtraction approach
  \cite{Dittmaier:1999mb} to non-collinear-safe observables, as
  partially described in Ref.~\cite{Bredenstein:2005zk}.\\
\item {\sc Horace}~\cite{CarloniCalame:2003ux,CarloniCalame:2005vc} is
  a Monte Carlo generator for single W/Z boson production at hadron
  colliders.  In its published version
  \cite{CarloniCalame:2003ux,CarloniCalame:2005vc} it simulates
  final-state-like multiple photon emission corrections via the QED
  parton-shower algorithm developed in
  Refs.~\cite{CarloniCalame:2000pz,CarloniCalame:2001ny}.  For the
  present study, {\sc Horace} has been extended (i) by including the
  exact $\cal{O}(\alpha)$ electroweak corrections to W production and
  (ii) by implementing the MRST2004QED set of PDFs.  Photon-induced
  processes, as well as the exact $\cal{O}(\alpha)$ corrections to Z
  production, are not taken into account in the present version.  A
  version of the generator, where the exact $\cal{O}(\alpha)$
  corrections and the parton-shower are matched, is in preparation.
  The theoretical and computational details of the calculation of the
  exact $\cal{O}(\alpha)$ corrections to W production and its matching
  with QED radiation beyond $\cal{O}(\alpha)$ will be presented
  elsewhere~\cite{horace_tocome}.\\
\item {\sc Sanc}~\cite{Andonov:2004hi} (see {\tt\small http://sanc.jinr.ru}
  and {\tt\small http://pcphsanc.cern.ch}) is an automated system which
  provides complete parton level results for the electroweak one-loop
  corrections to both neutral- and charged-current Drell-Yan processes.
  {\sc Sanc} is based on the construction of helicity amplitudes and
  electroweak form factors. It automatically generates results in {\sc
    Fortran} format which can be implemented in Monte Carlo event
  generators. The integration over the phase space for hard photon
  emission can be performed either (semi-)analytically or by means of
  a Monte Carlo integrator.  Although the semi-analytical treatment of
  the hard photon contribution does not allow to impose all required
  cuts (i.e.\ the cut on the missing transverse momentum), it provides
  an important check of the Monte Carlo calculation. A detailed
  description of the {\sc Sanc} calculation of the charged-current
  Drell-Yan process can be found in Ref.~\cite{Arbuzov:2005dd}.\\
\item {\sc Wgrad}~\cite{Baur:1998kt,Baur:2004ig} and {\sc
    Zgrad2}~\cite{Baur:2001ze} (see {\tt\small
    http://ubpheno.physics.buffalo.edu/${\tilde{\hphantom{x}}}$dow/})
  are two Monte Carlo programs that include the complete ${\cal
    O}(\alpha)$ electroweak radiative corrections to
  $p\,p\hskip-7pt\hbox{$^{^{(\!-\!)}}$}\to W^{\pm} \to\ell^{\pm} \nu$
  ({\sc Wgrad}) and $p\,p\hskip-7pt\hbox{$^{^{(\!-\!)}}$} \to \gamma,
  Z \to l^+ l^- X (l=e,\mu)$ ({\sc Zgrad2}). Both Monte Carlo programs
  use the phase space slicing method described in
  Refs.~\cite{Baer:1989jg,Baer:1990ra,Harris:2001sx}.  Charged lepton
  mass effects $\propto \ln(\hat s/m_l^2)$ associated with collinear
  final-state photon radiation are included in the calculation while
  very small terms of ${\cal O}(m_l^2/\hat s)$ have been neglected.
  For this comparison the MRST2004QED structure functions have been
  implemented, and the lepton selection cuts and photon-lepton
  recombination procedure have been modified according to the
  specifications given in this report.  Radiative corrections beyond
  ${\cal O}(\alpha)$ that are partially implemented in {\sc Wgrad} and
  {\sc Zgrad2} have been switched off.
\end{itemize}

\subsection{Common setup for the calculations}
\label{se:setup}
\subsubsection{Input parameters and scheme definitions}

The relevant input parameters are
\begin{equation}\arraycolsep 2pt
\begin{array}[b]{lcllcllcl}
\DYGF & = & 1.16637 \times 10^{-5} \DYGeV^{-2}, \quad&
\alpha(0) &=& 1/137.03599911, &
\alpha_{\mathrm{s}} &=& 0.1187,\\
M_\DYPW & = & 80.425\DYGeV, &
\Gamma_\DYPW & = & 2.124\DYGeV, \\
M_\DYPZ & = & 91.1876\DYGeV, &
\Gamma_\DYPZ & = & 2.4952\DYGeV, &
M_\DYPH & = & 115\DYGeV, \\
m_\DYPe & = & 0.51099892\DYMeV, &
m_\mu &=& 105.658369\DYMeV,\quad &
m_\tau &=& 1.77699\DYGeV, \\
m_\DYPu & = & 66\DYMeV, &
m_\DYPc & = & 1.2\DYGeV, &
m_\DYPt & = & 178\;\DYGeV,
\\
m_\DYPd & = & 66\DYMeV, &
m_\DYPs & = & 150\DYMeV, &
m_\DYPb & = & 4.3\DYGeV,
\\
|V_{\DYPu\DYPd}| & = & 0.975, &
|V_{\DYPu\DYPs}| & = & 0.222, \\
|V_{\DYPc\DYPd}| & = & 0.222, &
|V_{\DYPc\DYPs}| & = & 0.975,
\end{array}
\label{eq:SMpar}
\end{equation}
which essentially follow Ref.~\cite{Eidelman:2004wy}.  For the
top-quark mass $m_\DYPt$ the value of Ref.~\cite{Azzi:2004rc} is taken.
The masses of the light quarks are adjusted to reproduce the hadronic
contribution to the photonic vacuum polarization of
Ref.~\cite{Jegerlehner:2001ca}.  The CKM matrix is included via global
factors to the partonic cross sections.

The lowest-order cross section is parametrized in the ``$\DYGF$ scheme''
as defined in Ref.~\cite{Dittmaier:2001ay}, i.e.\ the electromagnetic
coupling $\alpha$ is derived according to $ \alpha_{\DYGF} = \sqrt{2}\DYGF
M_\DYPW^2(1-M_\DYPW^2/M_\DYPZ^2)/\pi$, so that the results are practically
independent of the masses of the light quarks.  Moreover, this
procedure absorbs the corrections proportional to $m_\DYPt^2/M_\DYPW^2$ in
the fermion--W-boson couplings and the running of $\alpha(Q^2)$ from
$Q^2=0$ to the electroweak scale. In the relative radiative
corrections, however, $\alpha(0)$ is used as coupling parameter, which
is the correct effective coupling for real photon emission.  Note that
the ${\cal O}(\alpha)$ corrections in the $\DYGF$ scheme receive a
constant contribution from the quantity $\Delta r$, as described in
Ref.~\cite{Dittmaier:2001ay}.  The W- and Z-boson resonances are
treated with fixed widths without any running effects.

The ${\cal O}(\alpha)$-improved set of PDFs ``MRST2004QED''
\cite{Martin:2004dh} is used throughout.  The factorization of the
photonic initial-state quark-mass singularities is done in the
DIS-like factorization scheme, i.e.\ {\it not} in the
$\overline{\mathrm{MS}}$ scheme as frequently done in the past,
because photon radiation off incoming quarks was ignored in the $F_2$
fit to HERA data \cite{privcom} (see also Ref.~\cite{Diener:2005me}).
The factorization scale is set to the weak boson mass, i.e.\ to
$M_\DYPW$ and $M_\DYPZ$ for W- and Z-boson production, respectively.

\subsubsection{Phase-space cuts and event selection}

In the following the same set of phase-space and event selection cuts
are used as described in Ref.~\cite{Dittmaier:2001ay} for W~production
at the LHC ($\sqrt{s}=14\DYTeV$).  In detail, for the experimental
identification of the process $\DYPp\DYPp\to\DYPW^+\to\nu_l l^+(+\gamma)$
the set of phase-space cuts
\begin{equation}
p_{\mathrm{T},l}>25\DYGeV, \qquad
p_{\mathrm{T,missing}}>25\DYGeV, \qquad
|\eta_l|<1.2,
\label{eq:lcuts}
\end{equation}
is adopted, where $p_{\mathrm{T},l}$ and $\eta_l$ are the transverse
momentum and the rapidity of the charged lepton $l^+$, respectively,
and $p_{\mathrm{T,missing}}=p_{\mathrm{T},\nu_l}$ is the missing
transverse momentum carried away by the neutrino. Note that these cuts
are not ``collinear safe'' with respect to the lepton momentum, so
that observables in general receive corrections that involve large
lepton-mass logarithms of the form $\alpha\ln(m_l/M_\DYPW)$. This is due
to the fact that photons within a small collinear cone around the
charged lepton momentum are not treated inclusively, i.e.\ the cuts
assume a perfect isolation of photons from the charged lepton. While
this is (more or less) achievable for muon final states, it is not
realistic for electrons. In order to be closer to the experimental
situation for electrons, the following photon recombination procedure
is considered:
\begin{enumerate}
\item Photons with a rapidity $|\eta_\gamma| > 2.5$, which are close
  to the beams, are treated as invisible, i.e.\ they are considered as
  part of the proton remnant.
\item If the photon survived the first step, and if the resolution
  $R_{l\gamma} = \sqrt{(\eta_l-\eta_\gamma)^2 + \phi_{l\gamma}^2}$ is
  smaller than 0.1 (with $\phi_{l\gamma}$ denoting the angle between
  lepton and photon in the transverse plane), then the photon is
  recombined with the charged lepton, i.e.\ the momenta of the photon
  and of the lepton $l$ are added and associated with the momentum of
  $l$, and the photon is discarded.
\item Finally, all events are discarded in which the resulting
  momentum of the charged lepton does not pass the cuts given in
  (\ref{eq:lcuts}).
\end{enumerate}
While the electroweak corrections differ for final-state electrons and
muons without photon recombination, the corrections become universal
in the presence of photon recombination, since the lepton-mass
logarithms cancel in this case, in accordance with the KLN theorem.

\subsection{Numerical results}
\label{se:numerics}
\subsubsection{W-boson production}

Table~\ref{tab:W_pTcut} compares results on integrated cross sections
as given in Table~2 of Ref.~\cite{Dittmaier:2001ay}, i.e.\ for
$\DYPp\DYPp\to\DYPW^+\to\nu_l l^+(+\gamma)$ with the different lower cuts on
$p_{\mathrm{T},l}$.  The quantities $\delta_{\DYPep\nu_\DYPe}$ and
$\delta_{\mu^+\nu_\mu}$ correspond to the corrections relative to the
lowest-order prediction $\sigma_0$ for the case that no photon
recombination is applied.
\begin{table}[bth]
\caption{Integrated lowest-order cross sections $\sigma_0$
for W~production at the LHC
for different ranges in $p_{\mathrm{T},l}$ and corresponding
relative corrections $\delta$, as obtained from the various
calculations.}
  \vspace*{1mm}
\centerline{
\def\arraystretch{1.2}
\begin{tabular}{lllllll}
\multicolumn{7}{c}{$\DYPp\DYPp\to\nu_l l^+(+\gamma)$ at $\sqrt{s}=14\DYTeV$}
\\  \hline
$p_{\mathrm{T},l}/\mathrm{GeV}$ &
25--$\infty$ & 50--$\infty$ & 100--$\infty$ & 200--$\infty$ & 500--$\infty$
& 1000--$\infty$
\\  \hline
$\sigma_0/\mathrm{pb}$ &
\\
{\sc Dk} &
2112.2(1) & 13.152(2) & 0.9452(1) & 0.11511(2) & 0.0054816(3) & 0.00026212(1)
\\
{\sc Horace} &
2112.21(4) & 13.151(6) & 0.9451(1) & 0.11511(1) & 0.0054812(4)  & 0.00026211(2)
\\
{\sc Sanc} &
2112.22(2)& 13.1507(2)& 0.94506(1)& 0.115106(1)& 0.00548132(6)& 0.000262108(3)
\\ {\sc Wgrad} &
2112.3(1) & 13.149(1) & 0.94510(5) & 0.115097(5) & 0.0054818(2) & 0.00026209(2)
\\ \hline
$\delta_{\DYPep\nu_\DYPe}/\%$ &
\\
{\sc Dk} &
$-5.19(1)$ & $-8.92(3)$ & $-11.47(2)$ & $-16.01(2)$ & $-26.35(1)$ & $-37.92(1)$
\\
{\sc Horace} &
$-5.23(1)$ & $-8.98(1)$ & $-11.49(1)$ & $-16.03(1)$ & $-26.36(1)$  & $-37.92(2)$
\\
{\sc Wgrad} &
$-5.10(1)$ & $-8.55(5)$ & $-11.32(1)$ & $-15.91(2)$ & $-26.1(1)$  & $-38.2(2)$
\\  \hline
$\delta_{\mu^+\nu_\mu}/\%$ &
\\
{\sc Dk} &
$-2.75(1)$ & $-4.78(3)$ & $-8.19(2)$ & $-12.71(2)$ & $-22.64(1)$ & $-33.54(2)$
\\
{\sc Horace} &
$-2.79(1)$ & $-4.84(1)$ & $-8.21(1)$ & $-12.73(1)$ & $-22.65(1)$ & $-33.57(1)$
\\
{\sc Sanc} &
$-2.80(1)$ & $-4.82(2)$ & $-8.17(2)$ & $-12.67(2)$ & $-22.63(2)$ & $-33.50(2)$
\\
{\sc Wgrad} &
$-2.69(1)$ & $-4.53(1)$ & $-8.12(1)$ & $-12.68(1)$ & $-22.62(2)$ & $-33.6(2)$
\\   \hline
$\delta_{\mathrm{recomb}}/\%$ &
\\
{\sc Dk} &
$-1.73(1)$ & $-2.45(3)$ & $-5.91(2)$ & $-9.99(2)$ & $-18.95(1)$ & $-28.60(1)$
\\
{\sc Horace} &
$-1.77(1)$ & $-2.51(1)$ & $-5.94(1)$ & $-10.02(1) $ & $-18.96(1)$ & $-28.65(1)$
\\
{\sc Sanc} &
$-1.89(1)$ & $-2.56(1)$ & $-5.97(1)$ & $-10.02(1)$& $-18.96(1)$ & $-28.61(1)$
\\
{\sc Wgrad} &
$-1.71(1)$ & $-2.32(1)$ & $-5.94(1)$ & $-10.11(2)$& $-19.08(3)$ & $-28.73(6)$
\\   \hline
$\delta_{\gamma q}/\%$ &
\\
{\sc Dk} &
$+0.071(1)$ & $+5.24(1)$ & $+13.10(1)$ & $+16.44(2)$ & $+14.30(1)$ &
$+11.89(1)$
\end{tabular}
}
\label{tab:W_pTcut}
\end{table}
The corrections for the $\DYPep\nu_\DYPe$ final state are larger in size
compared to the $\mu^+\nu_\mu$ because of the existence of
fermion-mass logarithms originating from collinear final-state
radiation. As explained above, these mass-singular corrections cancel
if the photon is recombined, rendering the corresponding correction
$\delta_{\mathrm{recomb}}$ smaller. At large $p_{\mathrm{T},l}$ the
electroweak corrections are dominated by Sudakov logarithms of the
form $-\alpha/\pi\log^2(\hat{s}/M_{\rm W}^2)$ which are independent of
the lepton species. Comparing the results in Table~\ref{tab:W_pTcut}
we find that the various calculations are, in general, consistent with
each other. More detailed comparisons are in progress to further
improve the agreement.

The corrections originating from the photon-induced processes are not
included in $\delta_{\DYPep\nu_\DYPe}$, $\delta_{\mu^+\nu_\mu}$, and
$\delta_{\mathrm{recomb}}$, but are shown separately as
$\delta_{\gamma q}$ in Table~\ref{tab:W_pTcut}. They are enhanced at
large $p_{\mathrm{T},l}$ because of a new type of contribution where a
W boson is exchanged in the $t$-channel.  The photon-induced processes
could in principle be used to extract information on the photon
content of the proton. However, they are overwhelmed by QCD
corrections and QCD uncertainties which strongly affect the
$p_{\mathrm{T},l}$ spectrum, see e.g.\ Ref~\cite{Cao:2004yy}.  If, on
the other hand, one considers the distribution in the transverse mass
$M_{\mathrm{T},\nu_{l}l}$, which is much less sensitive to QCD
uncertainties, the impact of $\delta_{\gamma q}$ is below the per-cent
level. This is demonstrated in Table~\ref{tab:W_MTcut} where the
${\cal O}(\alpha)$ cross section predictions with cuts on
$M_{\mathrm{T},\nu_{l}l}$ are shown.

\begin{table}
\caption{Integrated lowest-order cross sections $\sigma_0$
for W~production at the LHC
for different ranges in $M_{\mathrm{T},\nu_{l}l}$ and corresponding
relative corrections $\delta$. The transverse mass is defined
by $M_{\mathrm{T},\nu_l l}=
\sqrt{2p_{\mathrm{T},l}p_{\mathrm{T},\nu_l}(1-\cos\phi_{\nu_l l})}$,
where $\phi_{\nu_l l}$ is the angle between the lepton and the
missing momentum in the transverse plane.}
  \vspace*{1mm}
\centerline{
\def\arraystretch{1.2}
\begin{tabular}{lllllll}
\multicolumn{7}{c}{$\DYPp\DYPp\to\nu_l l^+(+\gamma)$ at $\sqrt{s}=14\DYTeV$}
\\  \hline
$M_{\mathrm{T},\nu_{l}l}/\mathrm{GeV}$ &
50--$\infty$ & 100--$\infty$ & 200--$\infty$ & 500--$\infty$ & 1000--$\infty$
& 2000--$\infty$
\\  \hline
$\sigma_0/\mathrm{pb}$ &
\\
{\sc Dk} &
2112.2(1) & 13.152(2) & 0.9452(1)  & 0.057730(5) & 0.0054816(3) & 0.00026212(1) \\
{\sc Horace} &
2112.21(4) & 13.151(6) & 0.9451(1) & 0.057730(2) & 0.0054812(4) & 0.00026211(2) \\
{\sc Wgrad} &
2112.2(1) & 13.150(1) & 0.9450(4)  & 0.057728(2) & 0.0054811(2) & 0.00026210(1)
\\
\hline
$\delta_{\DYPep\nu_\DYPe}/\%$ &
\\
{\sc Dk} &
$-5.20(1)$ & $-7.95(2)$ & $-10.19(2)$ & $-16.69(2)$ & $-24.52(1)$ & $-35.24(1)$\\
{\sc Horace} &
$-5.21(1)$ & $-8.00(1)$ & $-10.20(1)$ & $-16.70(1)$ & $-24.53(1)$ & $-35.25(1)$ \\
{\sc Wgrad} &
$-5.09(1)$ & $-7.73(2)$ & $-10.12(2)$ & $-16.69(3)$ & $-24.50(4)$ & $-35.3(3)$ \\
\hline
$\delta_{\mu^+\nu_\mu}/\%$ &
\\
{\sc Dk} &
$-2.75(1)$ & $-5.03(2)$ & $-7.98(1)$ & $-14.43(1)$ & $-21.99(1)$ & $-32.15(1)$\\
{\sc Horace} &
$-2.77(1)$ & $-5.08(1)$ & $-8.01(1)$ & $-14.44(1)$ & $-21.99(1)$ & $-32.16(1)$ \\
{\sc Sanc} &
$-2.76(2)$ & $-5.06(2)$ & $-7.96(2)$ & $-14.41(2)$ & $-21.94(2)$ & $-32.12(2)$ \\
{\sc Wgrad} &
$-2.69(1)$ & $-4.84(1)$ & $-7.96(1)$ & $-14.48(1)$ & $-22.03(1)$ & $-32.3(1)$
\\  \hline
$\delta_{\mathrm{recomb}}/\%$ &
\\
{\sc Dk} &
$-1.73(1)$ & $-3.43(2)$ & $-6.52(2)$ & $-12.55(1)$ & $-19.51(1)$ & $-28.75(1)$\\
{\sc Horace} &
$-1.75(1)$ & $-3.48(1)$ & $-6.54(1)$ & $-12.57(1)$ & $-19.54(1)$ & $-28.77(1)$\\
{\sc Wgrad} &
$-1.66(1)$ & $-3.27(1)$ & $-6.52(1)$ & $-12.62(2)$ & $-19.60(2)$ & $-29.0(1)$ \\
\hline
$\delta_{\gamma q}/\%$ &
\\
{\sc Dk} &
$+0.0567(3)$ & $+0.1347(1)$ & $+0.2546(1)$ & $+0.3333(1)$ & $+0.3267(1)$ & $+0.3126(1)$
\end{tabular}
}
\label{tab:W_MTcut}
\end{table}

\begin{figure}
%\framebox{
\centerline{
\setlength{\unitlength}{1cm}
\begin{picture}(16,7.6)
\put(-3,-1.1){\includegraphics{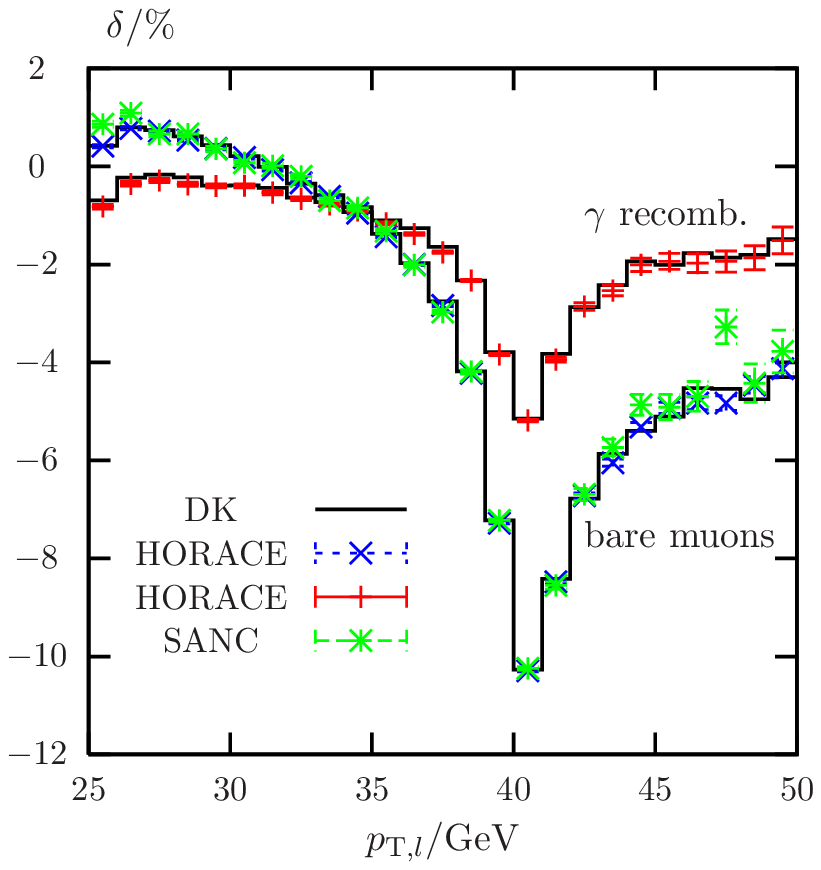}}
\put( 5,-1.1){\includegraphics{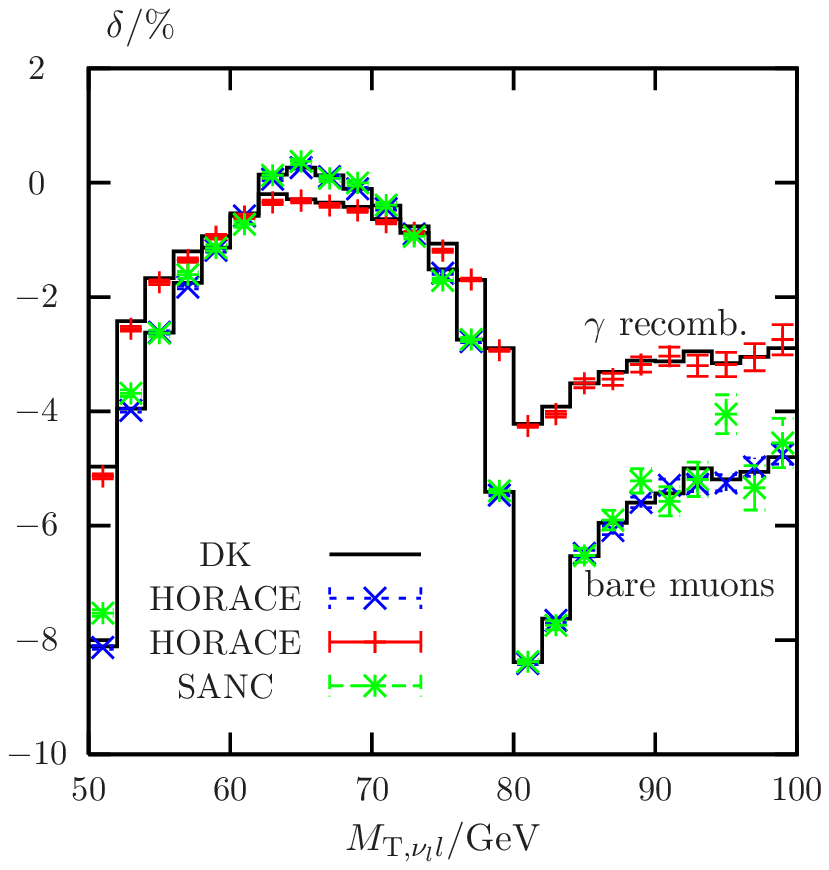}}
\put(1.1,4.75){\small $\DYPp\DYPp\to\nu_l l^+(+\gamma)$}
\put(1.1,4.25){\small $\sqrt{s}=14\DYTeV$}
\put(12.75,6.6){\small $\DYPp\DYPp\to\nu_l l^+(+\gamma)$}
\put(12.75,6.0){\small $\sqrt{s}=14\DYTeV$}
%\put(1.1,4.0){\small $p_{\mathrm{T},l},\dsl{p}_{\mathrm{T}}>25\DYGeV$}
%\put(1.3,4.8){\small $|\eta_l|<1.2$}
\end{picture}
}
\vspace*{-3mm}
\caption{Relative corrections $\delta$ as a function of the
  transverse-momentum $p_{\mathrm{T},l}$ and the transverse mass
  $M_{\mathrm{T},\nu_l l}$, as obtained from the {\sc Dk}, {\sc
    Horace} and {\sc Sanc} calculations. The contributions from the
  photon-induced processes have not been included in this comparison.}
\label{fig:pp_ptl_lhc}
\end{figure}

Figure~\ref{fig:pp_ptl_lhc} shows the relative electroweak correction
$\delta$ as a function of the lepton transverse momentum
$p_{\mathrm{T},l}$, and the transverse mass $M_{\mathrm{T},\nu_l l}$
in $\DYPp\DYPp\to\DYPW^+\to\nu_l l^+(+\gamma)$ for the LHC.  The results
from the {\sc Dk}, {\sc Horace} and {\sc Sanc} collaborations are in
good agreement.  Near $p_{\mathrm{T},l}\approx M_{\DYPW}/2$ and
$M_{\mathrm{T},\nu_l l}\approx M_{\DYPW}$ the correction $\delta$
reaches the order of 10\% for bare muons.  Since these enhanced
corrections originate from collinear final-state radiation, they are
negative for higher $p_{\mathrm{T},l}$ and redistribute events to
lower transverse momenta.  The correction $\delta$ is reduced to a few
per cent after photon recombination, which eliminates the artificial
lepton-mass logarithms.

\subsubsection{Z-boson production}

Table~\ref{tab:Z_Mcut} shows results on integrated cross sections for
$\DYPp\DYPp\to\DYPZ/\gamma\to l^+l^-(+\gamma)$ with the different lower cuts
on $M_{l^+l^-}=\sqrt{(p_{l^+}+p_{l^-})^2}$, as obtained by the {\sc
  Horace} and {\sc Sanc} collaborations. Note that the
ex\-pe\-ri\-men\-tal lepton identification cuts
$p_{\mathrm{T},l}>25\DYGeV$ and $|\eta_l|<1.2$ (cf.
Eq.~(\ref{eq:lcuts})) have been applied.
\begin{table}
\caption{Integrated lowest-order cross sections $\sigma_0$
for Z~production at the LHC for different ranges in $M_{l^+l^-}$ and corresponding
relative corrections $\delta$, as obtained from the {\sc Horace} and {\sc Sanc}
calculations. The ex\-pe\-ri\-men\-tal lepton
identification cuts $p_{\mathrm{T},l}>25\DYGeV$ and $|\eta_l|<1.2$ have
been applied.}
  \vspace*{1mm}
\centerline{
\def\arraystretch{1.2}
\begin{tabular}{lllllll}
\multicolumn{7}{c}{$\DYPp\DYPp\to l^+l^-(+\gamma)$ at $\sqrt{s}=14\DYTeV$}
\\  \hline
$M_{l^+l^-}/\mathrm{GeV}$ &
50--$\infty$ & 100--$\infty$ & 200--$\infty$ & 500--$\infty$ & 1000--$\infty$
& 2000--$\infty$
\\  \hline
$\sigma_0/\mathrm{pb}$ &
\\
{\sc Horace} &
254.64(1) & 10.571(1) & 0.45303(3) & 0.026996(2) & 0.0027130(2) & 0.00015525(1)
\\
{\sc Sanc} &
254.65(2) & 10.572(7) & 0.45308(3) & 0.026996(2) & 0.0027131(2) & 0.000155246(6)
\\  \hline
$\delta_{\mu^+\mu^-}/\%$ &
\\
{\sc Sanc} &
$-3.18(2)$ & $-8.63(2)$ & $-2.62(3)$ & $-5.51(3)$ & $-9.74(3)$ & $-15.26(4)$
\end{tabular}
}
\label{tab:Z_Mcut}
\end{table}
The corrections do not contain contributions from the photon-induced
processes.

\subsection{Conclusions}
We have presented precision calculations for electroweak corrections
to the Drell--Yan-like production of W and Z~bosons at the LHC from
various theoretical collaborations. The calculations have been based
on a common theoretical setup and choice of the input parameters, and
are using the ${\cal O}(\alpha)$-improved MRST2004QED set of parton
distribution functions. We have compared cross section predictions and
differential distributions in the lepton transverse-momentum
$p_{\mathrm{T},l}$ and in the transverse mass $M_{\mathrm{T},\nu_l
  l}$, and find, in general, good agreement between the various
calculations. We have also presented first results for the
photon-induced processes which turn out to be large for large
$p_{\mathrm{T},l}$ but do not contribute significantly to the
$M_{\mathrm{T},\nu_l l}$ distribution. Work is in progress to further
extend and improve the comparison of the various calculations.

\subsection*{Acknowledgements}
The {\sc {\sc Horace}} authors are grateful to Fulvio Piccinini for
stimulating discussions. Three of us (A.A., S.B., R.S.) are grateful
to Dmitri Bardin, Pena Christova, Lidia Kalinovskaya, and all the {\sc Sanc}
collaboration members for help and their contributions to the {\sc Sanc}
system, which were relevant for this study.

%%%%%%%%%%%%%%%%%%%%%%%%%%%%%%%%%%%%%%%%%%%%%%%%%%%%%%%%%%%%%%%%%%%%%%%%%%%%%
\section[Electroweak corrections to large transverse momentum
   production of $Z$ bosons and photons at the LHC]
{ELECTROWEAK CORRECTIONS TO LARGE TRANSVERSE MOMENTUM PRODUCTION OF $Z$
BOSONS AND PHOTONS AT THE LHC~\protect
\footnote{Contributed by: A.~Kulesza, S.~Pozzorini, M.~Schulze}}
\subsection{Introduction}

The Large Hadron Collider (LHC) with its 
high center-of-mass energy and design luminosity
will offer a unique possibility to explore production of gauge bosons with very large
transverse momentum ($\KPSpT$). Being embedded in the
environment of hadronic collisions, the reaction necessarily
involves hadronic physics, like parton distributions, and
depends on the strong coupling constant. In turn, the cross section
for the $Z$-boson or direct photon production at large $\KPSpT$ is an important means
to constrain information on the parton distribution functions (PDFs).
For this class of production processes at the LHC, 
electroweak (EW) corrections from virtual boson exchange become 
important. This is due to the presence of the logarithmic 
terms of the form $\alpha^k\log^{2k-i}(\hat{s}/M_W^2)$ (with $i=0$ for
the leading logarithmic (LL) terms, $i=1$ for the next-to-leading
logarithmic (NLL) terms, etc.) at the $k$-loop level.
These corrections, also known as electroweak Sudakov logarithms, may
well amount to several tens of per cent. A recent survey of the literature on 
logarithmic EW corrections
can be found in Ref.~\cite{Hollik:2004dz}.
Specifically, full one-loop weak corrections to the hadronic $Z$-boson and
prompt photon production at large $\KPSpT$ have been studied 
in Refs.~\cite{Kuhn:2005az,Kuhn:2005gv}. 
Numerical results for the one-loop corrections to these two processes
can be also found in Ref.~\cite{Maina:2004rb}. In the following we
briefly discuss the calculation of Refs.~\cite{Kuhn:2005az,Kuhn:2005gv}
and present compact analytic expressions
for the $Z$-boson and direct photon production in the high energy
approximation, derived from the exact one-loop results
in Refs.~\cite{Kuhn:2005az,Kuhn:2005gv}. Moreover, we also present the NLL contributions
to the two-loop EW corrections,
calculated in Refs.~\cite{Kuhn:2005gv,Kuhn:2004em}. Corresponding numerical
results are discussed.

Of course, to achieve reliable predictions at high $\KPSpT$, the QCD corrections in
next-to-leading order (NLO) need to be taken into
account as they can amount to several tens of per cent correction for both processes.
However, in the following we are interested
only in the EW corrections to the leading order (LO) in the
strong coupling constant $\alpha_\mathrm{S}$  cross section for the  process $pp \to V j$ \mbox{($V=\gamma,Z;\
j=\mathrm{jet}$)}.

\subsection{Calculation}
In the calculation of the one-loop
corrections to the hadronic production of $Z$ bosons~\cite{Kuhn:2005az} and
the hadronic production of photons~\cite{Kuhn:2005gv} at large $\KPSpT$
we consider only weak contributions, i.e. we do not include photonic
corrections.
The quarks are assumed massless and diagrams involving couplings of
quarks to Higgs bosons or would-be-Goldstone bosons are neglected. Moreover,
we omit quark-mixing effects in the calculations. 
The calculation for
the $\bar q q \to V g$ ($V=\gamma,Z$) subprocess is performed at the  
level of matrix elements and allows for full
control over polarization effects. Results for the other contributing subprocesses,
$q\bar q\to V g$,
$g q \to V  q$,
$q g \to V  q$,
$\bar q g  \to V  \bar q$
and
$g \bar q   \to V  \bar q$
are easily obtained from CP symmetry and crossing transformations.
The one-loop amplitude is split
into two parts: Abelian and non-Abelian, 
as the structure of the gauge group generators in front of each term
contributing to the amplitude can
 be classified as either
Abelian (characteristic for Abelian theories) or non-Abelian
(originating from non-commutativity of weak interactions). 
Tensor loop integrals, appearing in
the expressions for one-loop corrections are reduced to scalar integrals
by means of Passarino-Veltman technique. For the $Z$-boson and photon production
the electroweak coupling constants are
renormalized in the $\overline{\mathrm{MS}}$ and on-shell (OS) scheme,
respectively.  The wave functions of all external particles are always
renormalized according to the on-shell prescription.  The weak-boson
masses need not to be renormalized since they do not enter the
tree-level amplitudes.
Further details of the calculation can be found in Refs.~\cite{Kuhn:2005az,Kuhn:2005gv}.

\subsubsection{Analytical results}

The $\KPSpT$ distribution for the partonic reaction $\bar q q \to V g$
($V=\gamma,\;Z$) reads 
\begin{eqnarray}\label{partoniccs1}%\refeq{partoniccs1}
\frac{\mathrm{d} \hat{\sigma}^{\bar q q \to V g}}{\mathrm{d} \KPSpT}
&=&
\frac{\KPSpT}{8\pi N_c^2{\hat s}|{\hat t}-{\hat u}|}
\left[
\overline{\sum}|{\cal {M}}^{\bar q q \to V g}|^2+({\hat t}\leftrightarrow {\hat u})
\right]
,
\end{eqnarray}
where $N_c=3$,
$\overline{\sum}=
\frac{1}{4}
\sum_{\mathrm{pol}} 
\sum_{\mathrm{col}}
$, 
${\hat s}=(p_{\bar q}+ p_q)^2$,
${\hat t}=(p_{\bar q}- p_V)^2$,
${\hat u}=(p_q-p_V)^2$ 
and $p_V^2=M_V^2$.
To  ${\cal O}(\alpha^2\alpha_\mathrm{S})$ for the unpolarized squared matrix element 
we have
\begin{eqnarray}\label{generalresult}%\refeq{generalresult}
\lefteqn{
\overline{\sum}
|{\cal {M}}_1^{\bar q q \to V g}|^2 =
8 \pi^2 \alpha \alpha_\mathrm{S} (N_\mathrm{c}^2-1)\,    
\sum_{\lambda=\mathrm{R,L}}  
\Bigg\{  
\left(I^V_{q_\lambda}\right)^2  \bigg[ 
H_0^V \left( 1+2\delta C^{V,\mathrm{A}}_{q_{\lambda}} \right)} 
\\&&
+ \frac{\alpha}{2\pi}\sum_{V'=\mathrm{Z,W^{\pm}}} \left(I^{V'} I^{\bar{V'}}\right)_{q_{\lambda}}
\,H_1^{V,\mathrm{A}}(M_{V'}^2)
\bigg]
+  \frac{U_{V {W^3}}}{\KPSsw} T_{q_\lambda}^3 I^V_{q_\lambda} 
\bigg[
2 H_0^V\delta C^{V,\mathrm{N}}_{q_{\lambda}}
+ \frac{\alpha}{2\pi}
\frac{1}{\KPSsw^2}H_1^{V,\mathrm{N}}(M_W^2)
\bigg]\Bigg\} , \nonumber
\end{eqnarray}
where $I^V_{q_\lambda}$ represents the 
coupling of an electroweak gauge boson $V$ 
to right-handed ($\lambda=\mathrm{R}$) or left-handed ($\lambda=\mathrm{L}$) quarks.
In terms of the electric charge $Q_q$, the weak isospin $T^3_{q_\lambda}$
and the weak hypercharge $Y_{q_\lambda}$ we have $I^{\gamma}_{q_\lambda}=-Q_q$,
$I^Z_{q_\lambda}=\KPScw/\KPSsw T^3_{q_\lambda}-\KPSsw{Y_{q_\lambda}}/(2\KPScw)$,
with the shorthands $\KPScw=\cos{\KPSthw}$ and  $\KPSsw=\sin{\KPSthw}$
for the  weak mixing angle $\KPSthw$. Moreover the relevant elements of
the Weinberg rotation matrix $U$ are given by $U_{\gamma {W^3}} = -\KPSsw$ and 
$U_{Z {W^3}} = \KPScw$.
The term 
$H_0^V=({{\hat t}^2+{\hat u}^2+2{\hat s} M_V^2})/({{\hat t}{\hat u}})$ in \mbox{Eq.~(\ref{generalresult})}
represents the Born contribution.
$\delta C^{V,\mathrm{A}/\mathrm{N}}_{q_{\lambda}}$
summarize the counterterms associated with the renormalization
of the couplings and the gauge-boson wave function.
The contributions resulting from the bare loop diagrams 
and the fermionic wave-function renormalization correspond to the
functions $H_1^{V,\mathrm{A/N}}$.
The complete analytic expressions for these functions 
and for the counterterms, as well as details concerning the choice of
the renormalization scheme, can be found in Ref.~\cite{Kuhn:2005az}
for the $Z$-boson production and 
in Ref.~\cite{Kuhn:2005gv}
for photon production.
In the following we concentrate on the high-energy asymptotic
behaviour of the corrections, more precisely the
next-to-next-to-leading logarithmic (NNLL) approximation
of the full result.
Formally we consider the limit where $M_W^2/{\hat s}\to 0$ and 
${\hat t}/{\hat s}$, ${\hat u}/{\hat s}$ are constant.
In this limit the one-loop weak corrections are strongly enhanced by 
logarithms of the form $\log({\hat s}/M_W^2)$. 
The functions $H_1^{V,\mathrm{A/N}}$ assume the general form
\begin{equation}\label{nnllstracture}%\refeq{nnllstracture}
H_1^{V,\mathrm{A/N}}(M_{V'}^2) \stackrel{\mathrm{NNLL}}{=}
\mathrm{Re }\,\left[
g_0^{V,\mathrm{A/N}}(M_{V'}^2)\,
\frac{ {\hat t}^2+{\hat u}^2}{{\hat t}{\hat u}}
+g_1^{V,\mathrm{A/N}}(M_{V'}^2)\,
\frac{ {\hat t}^2-{\hat u}^2}{{\hat t}{\hat u}}
+g_2^{V,\mathrm{A/N}}(M_{V'}^2)
\right].
\end{equation}
We find  $g_i^{\gamma,\mathrm{A}}=
g_i^{Z,\mathrm{A}}$ for $i=0,1,2$ and  $g_j^{\gamma,\mathrm{N}}=
g_j^{Z,\mathrm{N}}$ for $j=1,2$. For $V=\gamma,\;Z$ one has: 
\begin{eqnarray}\label{heres1}%\refeq{heres1}
 g_0^{V,\mathrm{A}}(M_{V'}^2)&=& 
-\log^2\left(\frac{-{\hat s}}{M_{V'}^2}\right)
+3\log\left(\frac{-{\hat s}}{M_{V'}^2}\right)
+\frac{3}{2}\Biggl[
\log^2\left(\frac{{\hat t}}{{\hat s}}\right)
+\log^2\left(\frac{{\hat u}}{{\hat s}}\right)
\nonumber\\&&{}
+\log\left(\frac{{\hat t}}{{\hat s}}\right)
+\log\left(\frac{{\hat u}}{{\hat s}}\right)
\Biggr]
+\frac{7\pi^2}{3}-\frac{5}{2}
,\\ 
 g_1^{V,\mathrm{N}}(M_{V'}^2)&=&
- g_1^{V,\mathrm{A}}(M_{V'}^2)+
\frac{3}{2}\Biggl[
\log\left(\frac{{\hat u}}{{\hat s}}\right)
-
\log\left(\frac{{\hat t}}{{\hat s}}\right)
\Biggr]
=
\frac{1}{2}\Biggl[
\log^2\left(\frac{{\hat u}}{{\hat s}}\right)
-
\log^2\left(\frac{{\hat t}}{{\hat s}}\right)
\Biggr]
,\nonumber\\ 
 g_2^{\mathrm{V,N}}(M_{V'}^2)&=&
 -g_2^{\mathrm{V,A}}(M_{V'}^2)=
-2\Biggl[
\log^2\left(\frac{{\hat t}}{{\hat s}}\right)
+\log^2\left(\frac{{\hat u}}{{\hat s}}\right)
+\log\left(\frac{{\hat t}}{{\hat s}}\right)
+\log\left(\frac{{\hat u}}{{\hat s}}\right)
\Biggr]
-4{\pi^2}
. \nonumber
\end{eqnarray}
The non-Abelian function $g_0^{V,\mathrm N}$ is given by
($\delta_{VV'}=1$ for $V=V'$, otherwise $\delta_{VV'}=0$):
\begin{eqnarray}
g_0^{V,\mathrm{N}}(M_W^2)&=& 
2 \left[ \KPSDeltamsbar - \delta_{V \gamma}\log \left({M_W^2 \over M_Z^2}\right)\right]
+\log^2\left(\frac{-{\hat s}}{M_W^2}\right)
-\log^2\left(\frac{-{\hat t}}{M_W^2}\right)
-\log^2\left(\frac{-{\hat u}}{M_W^2}\right)
\nonumber\\&+& \!\!\!\!
\log^2\left(\frac{{\hat t}}{{\hat u}}\right)
-\frac{3}{2}\Biggl[
\log^2\left(\frac{{\hat t}}{{\hat s}}\right)
+\log^2\left(\frac{{\hat u}}{{\hat s}}\right)
\Biggr] - 2\pi^2 + 2 \delta_{VZ} \left( -\frac{\pi^2}{9}
-\frac{\pi}{\sqrt{3}} +2  \right).
\end{eqnarray}
The ultraviolet singularity, 
$\KPSDeltamsbar = 2/(4-D) - \gamma_{\mathrm{E}} +\log(4\pi \mu^2/{M_Z^2})$ 
is cancelled by the counterterms \cite{Kuhn:2005az, Kuhn:2005gv}.

The size of the logarithmically enhanced contributions grows with energy
and for transverse momenta of hundreds of GeV also the 
two-loop logarithms become important. At the NLL accuracy,
our results for the two-loop corrections include contributions with terms of the form  
$\alpha^2\KPSlogar{4}{{\hat s}}$ and $\alpha^2\KPSlogar{3}{{\hat s}}$
where $\KPSlogar{k}{{\hat r}}=\log^k(|{\hat r}|/ M_W^2)$. The expressions
presented here 
have been obtained using results of Refs.~\cite{Denner:2001jv,Denner:2003wi,Melles:2001gw}.
Since at two-loop level the purely weak corrections cannot be isolated 
from the complete electroweak corrections in a gauge-invariant way, we
have to consider the combination of weak and electromagnetic virtual 
corrections. The latter are regularized by means of a fictitious
photon mass\footnote{This approach permits to separate
the finite and infrared-divergent parts of the photonic corrections in
a gauge-invariant way (for a detailed discussion see Sect. 2
of~\cite{Kuhn:2004em}). In our results we include only the finite
part, defined through $\lambda=M_W$. The remaining part, which
contains infrared-divergent logarithms of $\lambda/M_W$, is
gauge-invariant and can thus be treated separately and combined with
real photon radiation.} $\lambda=M_W$.
For the unpolarized squared matrix element we have
$\overline{\sum}|{\cal {M}}_2^{\bar q q \to V g}|^2
-
\overline{\sum}|{\cal {M}}_1^{\bar q q\to V g}|^2 \stackrel{\mathrm{NLL}}{=}
2 {\alpha^3\alpha_\mathrm{S}}(N_\mathrm{c}^2-1)H_0^V
\KPSnewA{V}{2}
$ 
where
\begin{eqnarray}\label{twolooplogs}%\refeq{twolooplogs}
\KPSnewA{V}{2}&=&
\sum_{\lambda=\mathrm{L},\mathrm{R}}\Biggl\{
\frac{1}{2}
\left(
I^V_{q_\lambda}\KPScew_{q_\lambda}
+ S^V_{q_\lambda}
\right)
\Biggl[
I^V_{q_\lambda}\KPScew_{q_\lambda}\left(\KPSlogar{4}{{\hat s}}-6\KPSlogar{3}{{\hat s}}\right)
\nonumber\\&&{}
+
S^V_{q_\lambda}
\left(\KPSlogar{4}{{\hat t}}+\KPSlogar{4}{{\hat u}}-\KPSlogar{4}{{\hat s}}\right) 
\Biggr]
-(\delta_{VZ}-\delta_{V\gamma})\frac{T^3_{q_\lambda}Y_{q_\lambda}}{8\KPSsw^4}
\left(\KPSlogar{4}{{\hat t}}+\KPSlogar{4}{{\hat u}}-\KPSlogar{4}{{\hat s}}\right) 
\nonumber\\&&{}
+\frac{1}{6}I^V_{q_\lambda}
\Biggl[I^V_{q_\lambda}\left(
\frac{b_1}{\KPScw^2}\left(\frac{Y_{q_\lambda}}{2}\right)^2
+\frac{b_2}{\KPSsw^2} C_{q_\lambda}
\right)
+S^V_{q_\lambda}b_2
\Biggr]\KPSlogar{3}{{\hat s}}
\Biggr\}.
\end{eqnarray}
Here $\KPScew_{q_\lambda}=Y_{q_\lambda}^2/(4\KPScw^2)+C_{q_\lambda}/\KPSsw^2$
is the electroweak Casimir operator 
for quarks, with $C_{q_\mathrm{L}}=3/4$  and $C_{q_\mathrm{R}}=0$. $S^V_{q_\lambda}$
is defined as $S^V_{q_\lambda} = C^{\rm ew}_{V V'} I_{q_\lambda}^{V'}/2$ where
$C^{\rm ew}_{V V'}$ is the electroweak Casimir operator in the adjoint
representation~\cite{Denner:2001jv,Pozzorini:2001rs} and $S^{\gamma}_{q_\lambda}= - T^3_
{q_\lambda}/\KPSsw^2,\;S^{Z}_{q_\lambda}= \KPScw T^3_ {q_\lambda}/\KPSsw^3$. 
The  one-loop $\beta$-function coefficients read $b_1=-41/(6\KPScw^2)$
and $b_2=19/(6\KPSsw^2)$. We stress that although the above one- and two-loop results for the
photon and $Z$-boson production can be put in the same form, their
derivation requires separate calculation for each of the processes.

\subsubsection{Numerical results}
The hadronic cross sections are obtained using LO MRST2001 PDFs~\cite{Martin:2002dr}.
We choose $p_T^2$ as the factorization scale
and, similarly as the scale at which the running strong coupling constant
is evaluated. We also adopt the value 
$\alpha_\mathrm{S}(M_Z^2)=0.13$ and use the one-loop running expression for 
$\alpha_\mathrm{S}(\mu^2)$, in accordance with the LO 
PDF extraction method
of the MRST collaboration. 
We use the
following values of parameters~\cite{Eidelman:2004wy}:
$M_Z=91.19$ GeV, $M_W=80.39$ GeV and
$\alpha=1/137.0$, $s^2_{\mathrm{\scriptscriptstyle
W}}=1-c^2_{\mathrm{\scriptscriptstyle W}}=1-M_W^2/M_Z^2$ for the $\gamma$ production
(OS scheme with $\alpha$ in the Thompson limit) or $\alpha=1/128.1$,
$s^2_{\mathrm{\scriptscriptstyle W}}=0.2314$ for the $Z$-boson
production ($\overline{{\rm MS}}$ scheme, as discussed above, 
with $M_Z^2$ as the renormalization scale).

\begin{figure}
\begin{center}
\includegraphics[width=0.5\textwidth]{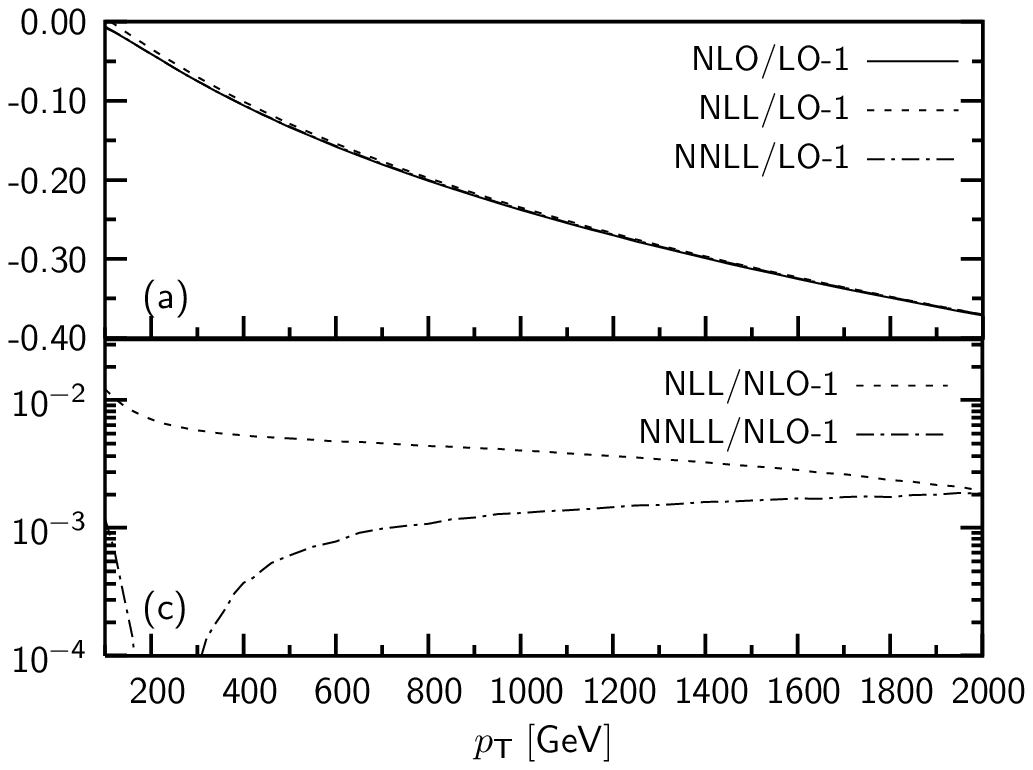}
\includegraphics[width=0.505\textwidth]{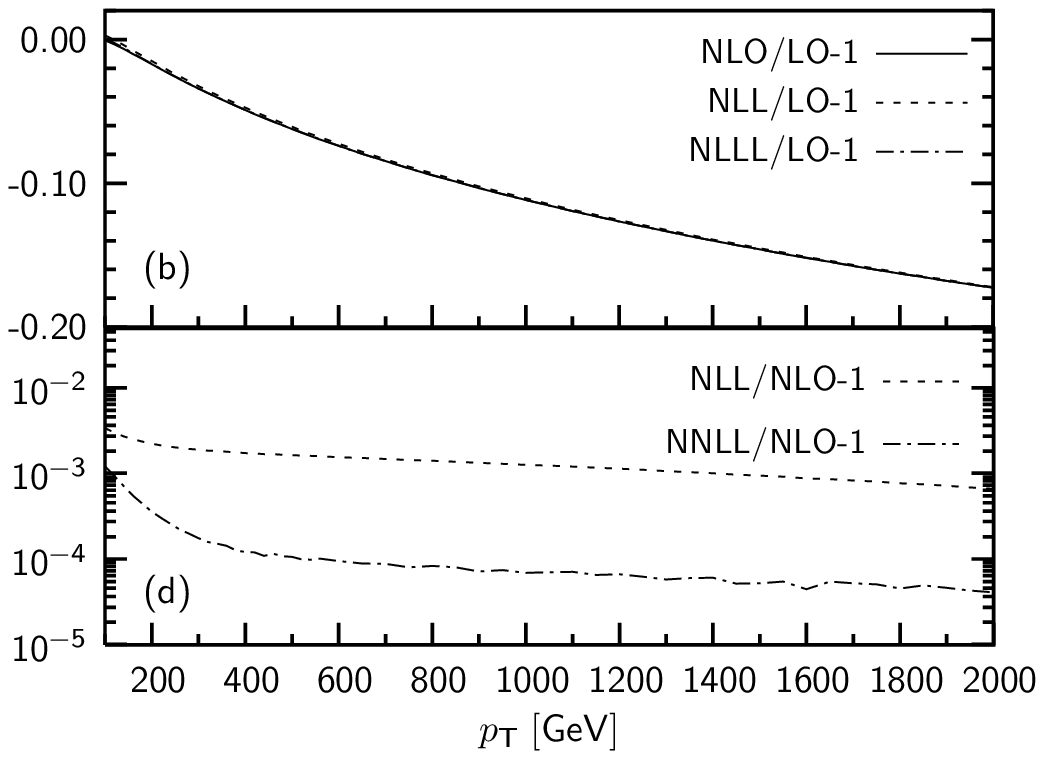}
 \caption{Upper plots: relative NLO (solid), NNL (dotted) and NNLL (dot-dashed) 
weak corrections wrt. the LO $\KPSpT$ distribution for the process (a)
$pp\rightarrow Z j$ and (b) $pp\rightarrow \gamma j$ at
$\sqrt{s}=14$ TeV. Lower plots: NLL (dotted) and NNLL (dot-dashed) approximation
compared to the full NLO result for (c)
$pp\rightarrow Z j$ and (d) $pp\rightarrow \gamma j$ at
$\sqrt{s}=14$ TeV. } 
\label{KPSfig1}
\end{center}
\end{figure}

First, we study the behaviour of the one-loop EW corrections to the transverse
momentum distributions of photons and $Z$ bosons produced at the LHC, see
Fig.~\ref{KPSfig1}. 
The contribution
provided by the NLO correction is negative and increases in size with
$\KPSpT$, reaching -37\%  (-17\%) for the $Z$-boson (photon) production at
$\KPSpT=2 $ TeV. We also conclude that the NLL approximation works good,
at a per cent (or better) level of accuracy both for the photon and $Z$-boson production. 
In comparison, the quality of the NNLL approximation is excellent, at
the level of accuracy of $10^{-3}$ or better in the entire $\KPSpT$ range
for both processes.
%%%%%%%%%%%%%%%
%%%%%%%%%%%%%%

\begin{figure}
\begin{center}
\includegraphics[width=0.5\textwidth]{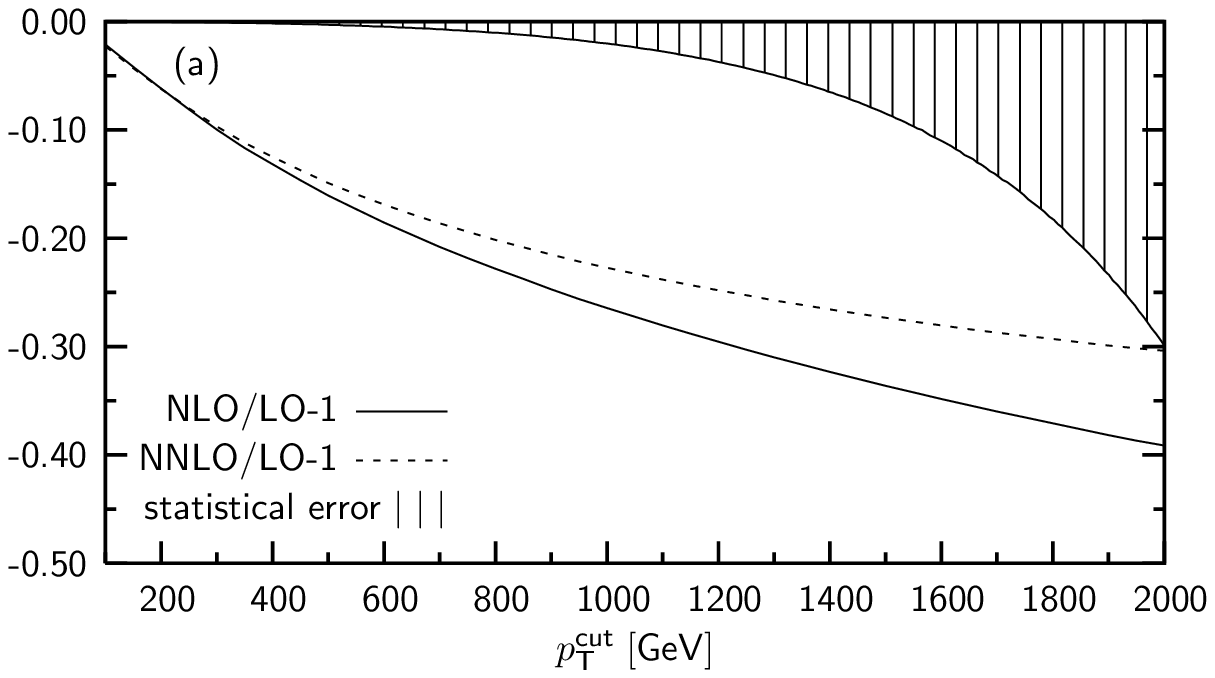}
\includegraphics[width=0.5\textwidth]{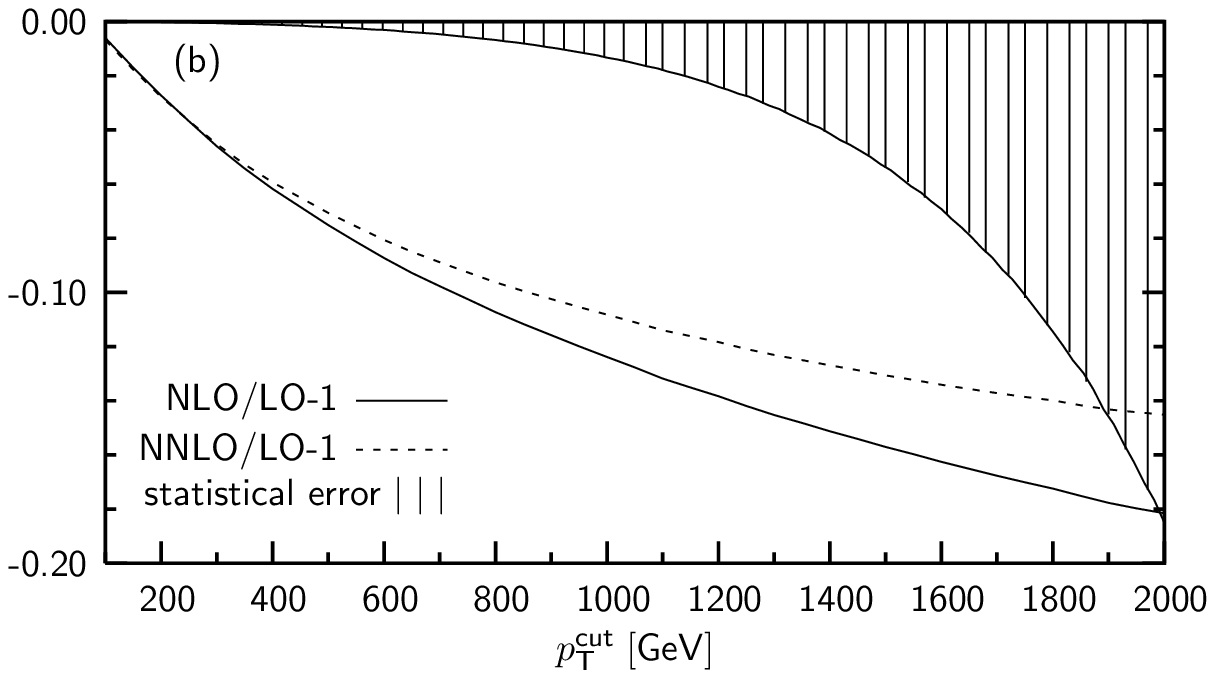}
 \caption{Relative NLO (solid) and NNLO (dotted) corrections 
wrt. the LO prediction  and estimated statistical
  error (shaded area) for the unpolarized integrated cross section for
  (a) $pp\rightarrow Z j$ and (b) $pp\rightarrow \gamma j$ at $\sqrt{s}=14$ TeV as a
  function of $p_T^{\rm cut}$.}
\label{fig2}
\end{center}
\end{figure}

To demonstrate the relevance of the EW effects for the transverse
momentum distributions of the gauge bosons produced at the LHC, in
Fig.~\ref{fig2} we present 
the relative NLO and next-to-next-to-leading order (NNLO)\footnote{Our
NNLO predictions include the
exact NLO contributions combined with the leading and next-to-leading
logarithmic two-loop terms~(\ref{twolooplogs}).} 
corrections for the cross section,
integrated over $\KPSpT$ starting from $p_T = p_T^{\rm cut}$, as a function of
$p_T^{\rm cut}$.
%%%%%%%%%%%%%%%
%%%%%%%%%%%%%%
This is compared with the statistical error, defined as
$\Delta \sigma_{\rm stat} / \sigma = 1 /\sqrt N$ with $N= {\cal L} \times
\sigma_{\rm LO}$. 
We assume a total integrated
luminosity ${\cal L} =300 \;{\rm fb}^{-1}$ for the LHC. 
It is clear from Fig.~\ref{fig2}, that the size of the one-loop
(two-loop logarithmic) corrections 
is much bigger than (comparable to) the statistical error for both the
$Z$-boson and the $\gamma$ production.

\begin{figure}
\begin{center}
\includegraphics[width=0.6\textwidth]{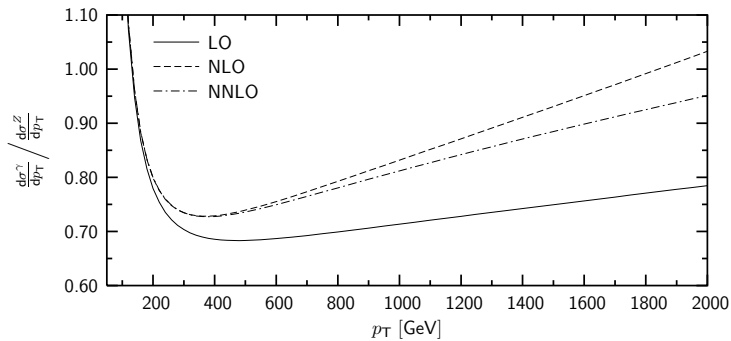}
\caption{Ratio of the transverse momentum distributions
for the processes $pp\to \gamma j$ and  $pp\to Z j$ 
at $\sqrt{s}=14$ TeV for the LO (solid), NLO (dashed) and NNLO
(dotted) predictions.}
\label{KPSfig3}
\end{center}
\end{figure}

In Fig.~\ref{KPSfig3}, we plot the ratio of the $\KPSpT$ distribution for
the $\gamma$ production to the $\KPSpT$ distribution for the $Z$-boson production.
Such ratio is expected to be less sensitive to theoretical errors
than the distributions themselves, since many
uncertainties such as the scale at which
$\alpha_\mathrm{S}$ is calculated or the choice of PDFs 
cancel to a large extent in the ratio. Moreover, due to a similar
cancellation mechanism, the ratio should remain
stable against QCD corrections.
From Fig.~\ref{KPSfig3} we observe that the EW 
corrections modify the production ratio considerably. The effect is
the strongest at high $\KPSpT$. In this region, the LO photon cross
section is smaller than the cross section for $Z$-boson
production by about 25\%. The relatively large NLO corrections for
$Z$-boson production, as compared to $\gamma$ production, cause the full NLO production
rates to become equal at the highest $\KPSpT$ considered here, i.e.
$p_T\sim 2$ TeV. The two-loop corrections modify the ratio and lead
to a few per cent decrease at
high $\KPSpT$.

%%%%%%%%%%%%%%%
%%%%%%%%%%%%%%

\subsection{Conclusions}

At the LHC, the transverse momentum of $Z$ bosons or photons produced in the process 
$pp \to Vj$ ($V=\gamma,Z$) will reach the TeV scale. 
In this $\KPSpT$ regime one-loop weak corrections are negative and large,
of the order of tens of per cent. In the high-energy limit these corrections are 
approximated with an excellent accuracy by the compact analytical expressions presented here. 
We also present expressions for the dominant logarithmic part of the two-loop EW
corrections and point out
that these  corrections are important for the 
correct interpretation of the measurements.
Moreover, we find that the EW corrections modify significantly the
ratio of the $Z$-boson 
and $\gamma$ transverse momentum distributions at high $\KPSpT$.

\subsection{Acknowledgments}
We are grateful to J.~H.~K\"uhn for the collaboration on the project. 
This work was supported in part by 
BMBF Grant No.~05HT4VKA/3
and by the Deutsche Forschungsgemeinschaft 
(SFB/TR-9 
``Computational Particle Physics''). 
M.~S. acknowledges financial 
support from the Graduiertenkolleg "Hochenergiephysik und Teilchenastrophysik".
A.K. would like to thank the organizers for the kind
hospitality and stimulating atmosphere during the workshop.

%%%%%%%%%%%%%%%%%%%%%%%%%%%%%%%%%%%%%%%%%%%%%%%%%%%%%%%%%%%%%%%%%%%%%%%%%%%%%
\section[Impact of weak corrections on LHC jet cross sections]
{IMPACT OF WEAK CORRECTIONS ON LHC JET CROSS SECTIONS~\protect
\footnote{Contributed by: S.~Moretti, M.R.~Nolten, D.A.~Ross}}
\subsection{Weak corrections at TeV scales}
The purely weak (W) component of 
next-to-leading order (NLO) 
Electro-Weak (EW) effects produce leading
corrections of the type $\alpha_{\rm{W}}\log^2({E_T^2}/M_W^2)$, where 
$\alpha_{\mathrm{W}}\equiv \alpha_{\mathrm{EM}}/\sin^2\theta_W$,
with $\alpha_{\mathrm{EM}}$ the Electro-Magnetic coupling
constant and $\theta_W$ the Weinberg angle. In fact,
for large enough $E_T$ values, the jet transverse
energy, such EW effects may be competitive not
only with next-to-NLO (NNLO) (as $ \alpha_{\rm{W}}\approx 
\alpha_{\rm{S}}^2$) but also with NLO QCD corrections (e.g., for
${E_T}=0.5$ TeV, $\log^2({E_T^2}/M_W^2)\approx10$).

These `double logs' are 
due to a lack of cancellation between virtual and real $W$-emission in
higher order contributions. This is in turn a consequence of the 
violation of the Bloch-Nordsieck theorem in non-Abelian theories
\cite{Ciafaloni:2000df}.
The problem is in principle present also in QCD. In practice, however, 
it has no observable consequences, because of the final averaging of the 
colour degrees of freedom of partons, forced by their confinement
into colourless hadrons. This does not occur in the EW case,
where the initial state has a non-Abelian charge,
as in proton-(anti)proton scattering. Besides, these
logarithmic corrections are finite (unlike in
QCD), since $M_W$ provides a physical
cut-off for $W$-emission. Hence, for typical experimental
resolutions, softly and collinearly emitted weak bosons need not be included
in the production cross section and one can restrict oneself to the 
calculation
of weak effects originating from virtual corrections only. 
By doing so, similar
logarithmic effects, $\sim\alpha_{\rm{W}}\log^2({E_T^2}/M_Z^2)$, 
are generated also by $Z$-boson corrections.
Finally, at least in some cases (like the present one),
all these purely weak contributions can  be
isolated in a gauge-invariant manner from EM effects which therefore may not
be included in the calculation. 

In view of all this,  it becomes of crucial importance to assess
the quantitative relevance of such weak corrections
affecting, in particular, key QCD processes at present and future 
hadron colliders, such as jet and heavy quark pair production. 
Published analyses for $b$-jet production
at the Tevatron and Large Hadron Collider (LHC) already exist  \cite{Maina:2003is}
-- see also the previous Les Houches proceedings \cite{Maina:2004ve}
and Refs.~\cite{Maina:2004dm,Hollik:2004dz} -- and those for $t\bar t$
production are in progress 
%\cite{Bernreuther:2005is,Bernreuther:2005ej,Kuhn:2005it}.
\cite{Beenakker:1993yr,Hollik:1997hm,Kao:1997bs,Kao:1999kj,Bernreuther:2005is,Bernreuther:2005ej,Kuhn:2005it}.
(For Standard Model (SM) corrections to heavy quark pairs based on Sudakov leading logarithms only, 
see \cite{Beccaria:2004qg,Beccaria:2005jd}.) 
We show here results for the case of 
jet production at the LHC, a preliminary account of which was given
in \cite{Moretti:2005aa}. (For the case of 
Tevatron, see Ref.~\cite{Moretti:2005ut}.)

\subsection{Corrections to jet production}

It is the aim of this note to report on the computation of the full
one-loop weak effects\footnote{We neglect here
purely EM effects (as well as interferences between these and the
weak ones) through 
${\cal O}(\alpha_{\rm{S}}^2\alpha_{\mathrm{EM}})$, 
as  they are not associated with logarithmic
enhancements either.} entering all possible
`2 parton $\to$ 2 parton' scatterings, through the
perturbative order $\alpha_{\mathrm{S}}^2\alpha_{\mathrm{W}}$. 
(See Ref.~\cite{Baur:1989qt} for
tree-level $\alpha_{\mathrm{S}}\alpha_{\mathrm{EW}}$ interference
effects -- hereafter,
$\alpha_{\mathrm{EW}}$ exemplifies the fact that
both weak and EM effects are included at the given order).
We will instead ignore altogether the contributions
of tree-level $\alpha_{\mathrm{S}}^2\alpha_{\mathrm{W}}$ terms
involving the radiation of real $W$ and $Z$ bosons. Therefore,
apart from $gg\to gg$ (which is not subject to order 
$\alpha_{\mathrm{S}}^2\alpha_{\mathrm{W}}$ corrections), 
there are in total fifteen subprocesses to consider,
\begin{eqnarray}
g g &\to& q \bar q\\
q \bar q &\to& g g\\
q g &\to& q g\\
\bar q g &\to& \bar q g\\
q q &\to& q q\\
\bar q \bar q &\to& \bar q \bar q\\
q Q &\to& q Q ~({\rm{same~generation}})\\
\bar q \bar Q &\to& \bar q \bar Q ~({\rm{same~generation}})\\
q Q &\to& q Q ~({\rm{different~generation}})\\
\bar q \bar Q &\to& \bar q \bar Q ~({\rm{different~generation}})\\
q \bar q &\to& q \bar q \\
q \bar q &\to& Q \bar Q ~({\rm{same~generation}})\\
q \bar q &\to& Q \bar Q ~({\rm{different~generation}})\\
q \bar Q &\to& q \bar Q ~({\rm{same~generation}})\\
q \bar Q &\to& q \bar Q ~({\rm{different~generation}}),
\end{eqnarray}
with $q$ and $Q$ referring to quarks of different flavours
and where the latter are limited to $u$-, $d$-, $s$-, $c$- 
and $b$-type (all taken as massless). While the first four
processes (with external gluons) were already computed 
in Ref.~\cite{Ellis:2001ba}, 
the eleven four-quark processes are new to this study.
Besides, unlike the former, the latter 
can be (soft and collinear) divergent,
so that gluon bremsstrahlung effects ought to be evaluated to obtain
a finite cross section at the considered order. In addition,
for completeness, we have  also included 
the non-divergent `2 parton $\to$ 3 parton'
subprocesses 
\begin{eqnarray}
q g &\to& q q \bar q  \\
\bar q g &\to& \bar q \bar q q\\
q g &\to& q Q \bar Q ~({\rm{same~generation}})\\
\bar q g &\to& \bar q \bar Q Q~({\rm{same~generation}}).
\end{eqnarray}

Notice that in our treatment we identify the jets with the partons
from which they originate and we adopt the cut $|\eta|<2.5$
in pseudorapidity to mimic the LHC detector coverage and
the standard jet cone requirement $\Delta R>0.7$ to emulate their
jet data selection (although we eventually sum the two- and three-jet contributions). Furthermore,
as factorisation and renormalisation scale we use $\mu=\mu_F\equiv\mu_R=E_T/2$ while
we adopt CTEQ6L1 \cite{Pumplin:2002vw} as Parton Distribution Functions. 

\begin{figure}[t!]
\label{fig:Moretti-EWcorrs-LHC}
\begin{center}
\hspace*{0.005truecm}{\epsfig{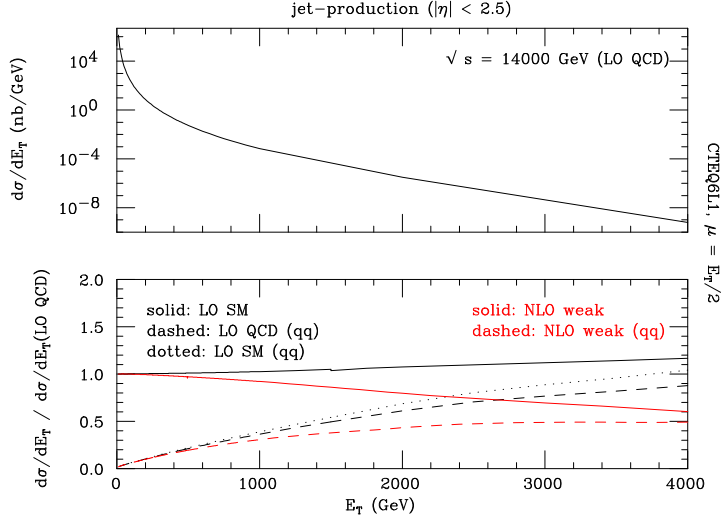}}
\caption{Top: The total single jet inclusive distribution in transverse energy through
${\cal O}(\alpha_{\mathrm{S}}^2)$ at the LHC, via all partonic subprocesses.
Bottom: The  effects of the one-loop
${\cal O}(\alpha_{\mathrm{S}}^2\alpha_{\mathrm{W}})$ and 
tree-level ${\cal O}(\alpha_{\mathrm{S}}\alpha_{\mathrm{EW}})$ 
corrections relative to the spectrum above. The label {\tt (qq)} refers to the case
of subprocesses with no gluons in the initial state.}
\end{center}
\end{figure}

%%%%%%%%%%%%%%%%%%%%%%%%%%%%%%%%
Fig.~\ref{fig:Moretti-EWcorrs-LHC} 
%Fig.~1 
%%%%%%%%%%%%%%%%%%%%%%%%%%%%%%%%
exemplifies the relevance 
that $\alpha_{\mathrm{S}}^2\alpha_{\mathrm{W}}$ effects can have
at the LHC. The $\alpha_{\mathrm{S}}^2\alpha_{\mathrm{W}}$ corrections 
are rather large and grow steadily with the jet transverse energy,
as the total (i.e., via all partonic channels) results through 
${\cal O}(\alpha_{\mathrm{S}}^2+\alpha_{\mathrm{S}}^2\alpha_{\mathrm{W}})$
[labelled {\tt NLO weak}] differ  with respect to
the prediction of total LO QCD through ${\cal O}(\alpha_{\rm{S}}^2)$ [{\tt LO QCD}] 
by up to an astounding $-40\%$, in  the
vicinity of 4 TeV, the highest $E_T$ point that may be reached at the LHC
after full luminosity. In fact, already at $E_T=1$ TeV, the effects amount to $-10\%$.
In the case of subprocesses initiated by (anti)quarks only, one also has LO EW effects through 
${\cal O}(\alpha_{\rm{S}}\alpha_{\mathrm{EW}})$, which can only reach a $3(16)\%$ effect 
at $E_T=1(4)$ TeV,
as shown in the same plot. (Here {\tt LO SM} identifies the sum of 
terms of ${\cal O}(\alpha_{\mathrm{S}}^2)$,
${\cal O}(\alpha_{\mathrm{S}}\alpha_{\mathrm{EW}})$
and ${\cal O}(\alpha_{\mathrm{EW}}^2)$). Between the two
kind of corrections then, are the one-loop ones that dominates over those 
at tree-level. Finally, the plot also presents the contributions of only those 
subprocesses that are not initiated by gluons: it is clear that at large $E_T$ are these channels that dominates. 

\subsection{Conclusions}
In summary, at the LHC, ${\cal O}(\alpha_{\mathrm{S}}^2\alpha_{\mathrm{W}})$ terms are
important contributions to the inclusive jet cross section at
large transverse energy. For the expected highest reach of the machine, $E_T\approx 4$ TeV,
they can be as large as $-40\%$. Therefore, they ought to be included while
comparing experiment {\sl versus} theory. However, particular care
should be devoted to the treatment of real $W$ and $Z$ production and
decay in the
definition of the inclusive jet data sample, as this will determine
whether (positive) tree-level $W$ and $Z$ bremsstrahlung effects have to be included
in the theoretical predictions through ${\cal O}(\alpha_{\mathrm{S}}^2\alpha_{\mathrm{W}})$,
which would tend to counterbalance the negative effects due to the one-loop
$W$ and $Z$ exchange estimated here.  As these were not included in our
calculation, the matter is currently under study.

%%%%%%%%%%%%%%%%%%%%%%%%%%%%%%%%%%%%%%%%%%%%%%%%%%%%%%%%%%%%%%%%%%%%%%%%%%%%%
\section[Search for anomalous couplings in top decay at hadron colliders]
{SEARCH FOR ANOMALOUS COUPLINGS IN TOP DECAY AT HADRON COLLIDERS~\protect
\footnote{Contributed by: S.~Tsuno, I.~Nakano, Y.~Sumino, R.~Tanaka}}
\author{S. Tsuno$^1$, I. Nakano$^1$, Y. Sumino$^2$ and R. Tanaka$^1$}
\subsection{Introduction}

Since the top quark was discovered at the 
Tevatron~\cite{Abe:1995hr,Abachi:1995iq} the number of the observed top quark 
events in the Tevatron Run II experiment is increasing with the integrated 
luminosity, and now reaching of the order of a few hundred. At LHC, of the order of $\sim$10$^7$ top 
quarks will be produced in the first year of 
operation. The study of top quark properties is expected to reveal the 
fundamental structure of top quark interactions currently at the Tevatron, and 
in the future at the LHC and the ILC.

Among various interactions of the top quark, the study of the top quark decay 
properties is particularly interesting. In the Standard Model, the top quark 
decays via electroweak interaction before hadronization, so that the top 
quark's spin information is transferred directly to its decay products, and 
their properties can be predicted reliably using perturbative calculations. 
Thus, the top quark spin can be used as a powerful analyzer to study top 
quark decays.

The anomalous couplings of top decays can be derived indirectly from a constraint 
to the $W$ polarization state from the top quark by using the lepton helicity 
angle~\cite{Abulencia:2005xf,Abazov:2005fk}. The sensitivity of the anomalous 
couplings has also been studied using single top events by looking at 
the differential distributions for signal and background 
events~\cite{Boos:1999dd}. However, single top events have not yet been observed 
at the Tevatron~\cite{Acosta:2004bs,Abazov:2005zz}. Due to the low number of signal events 
and due to the difficult separation of the signal from the backgrounds, the 
experimental results are still poor.

In this article, we propose a new method for reconstructing an effective spin 
direction of the top quark using the $t\bar{t}$ production process~\cite{Sumino:2005pg,Tsuno:2005qb}. In 
particular, the new feature of this technique is that we do not need to 
reconstruct the spin of the anti-top side in a $t\bar{t}$ event, i.e. we do 
not make use of the correlation between the top and anti-top spins. We rather 
make use of the correlation between the top spin and the directions of its decay products. 
So that, even in the $lepton + jets$ channel, the spin direction of the top 
quark can be easily reproduced using the information of one only top quark, and a
high event statistics is obtained with good signal and background separation.
Our method is expected to improve the sensitivity to anomalous couplings 
considerably compared to other measurements.

\subsection{Anomalous couplings in top decay vertices}

The interaction with an anomalous coupling in the top decay vertex can be expressed as 
\begin{equation}
\Gamma_{Wtb}^{\mu} = - \frac{g}{\sqrt{2}} V_{tb} 
\bar{u}(p_{b})\left[ \gamma^{\mu}f_{1}P_{L}
-\frac{i \sigma^{\mu\nu}k_{\nu}}{M_{W}}f_{2}P_{R}\right]
u(p_{t}) \quad ,
\label{eq1}
\end{equation}
where $V_{tb}$ is the CKM (Cabibbo-Kobayashi-Maskawa) matrix 
element, $P_{L,R}$ $=$ $(1 \mp \gamma_{5})/2$ is the 
left-handed/right-handed projection operator, and $k$ is the momentum of the $W$. 
We take the convention in which the energy scale is represented by  
$M_{W}$ (on-shell). For simplicity of the analysis, we assume that the interactions 
preserve the $CP$ symmetry and also neglect the couplings of the right-handed 
bottom quark. Two form factors $f_{1}$ and $f_{2}$ are thus treated as real, 
and then their values are $f_{1}=1$ and $f_{2}=0$ at tree-level in the SM.

From Eq.~{\ref{eq1}}, we may separate the dependence of the decay distribution
on $f_1$ and on $f_2/f_1$. A variation of $f_1$ changes only the normalization 
of the (partial) decay width of the top quark, while a variation of $f_2/f_1$ 
changes both, the normalization and the differential decay distributions. Since 
it is difficult to measure the absolute value of the decay width accurately in 
the near future, our primary goal will be to constrain the value of 
$f_2/f_1$ from the measurement of the differential decay distribution. Since 
the transverse $W$ boson ($W_{T}$) is more sensitive to $f_{2}$ 
than the longitudinal $W$ boson ($W_{L}$), we can enhance the contribution of 
$W_{T}$ using the decay distribution. It is well known that the contribution 
of $W_{T}$ is dominant when the $W$ is emitted opposite to the top spin direction 
in the decay $t \to bW$ and also when the lepton ($l$) is emitted in the opposite direction 
to the $W$ in the decay $W \to l\nu$. Hence, we can select this kinematic 
region in order to enhance the sensitivity to $f_2/f_1$.

\begin{figure}[hptb]
\begin{center}
\begin{tabular}{ccc}
\includegraphics[width=5.0cm]{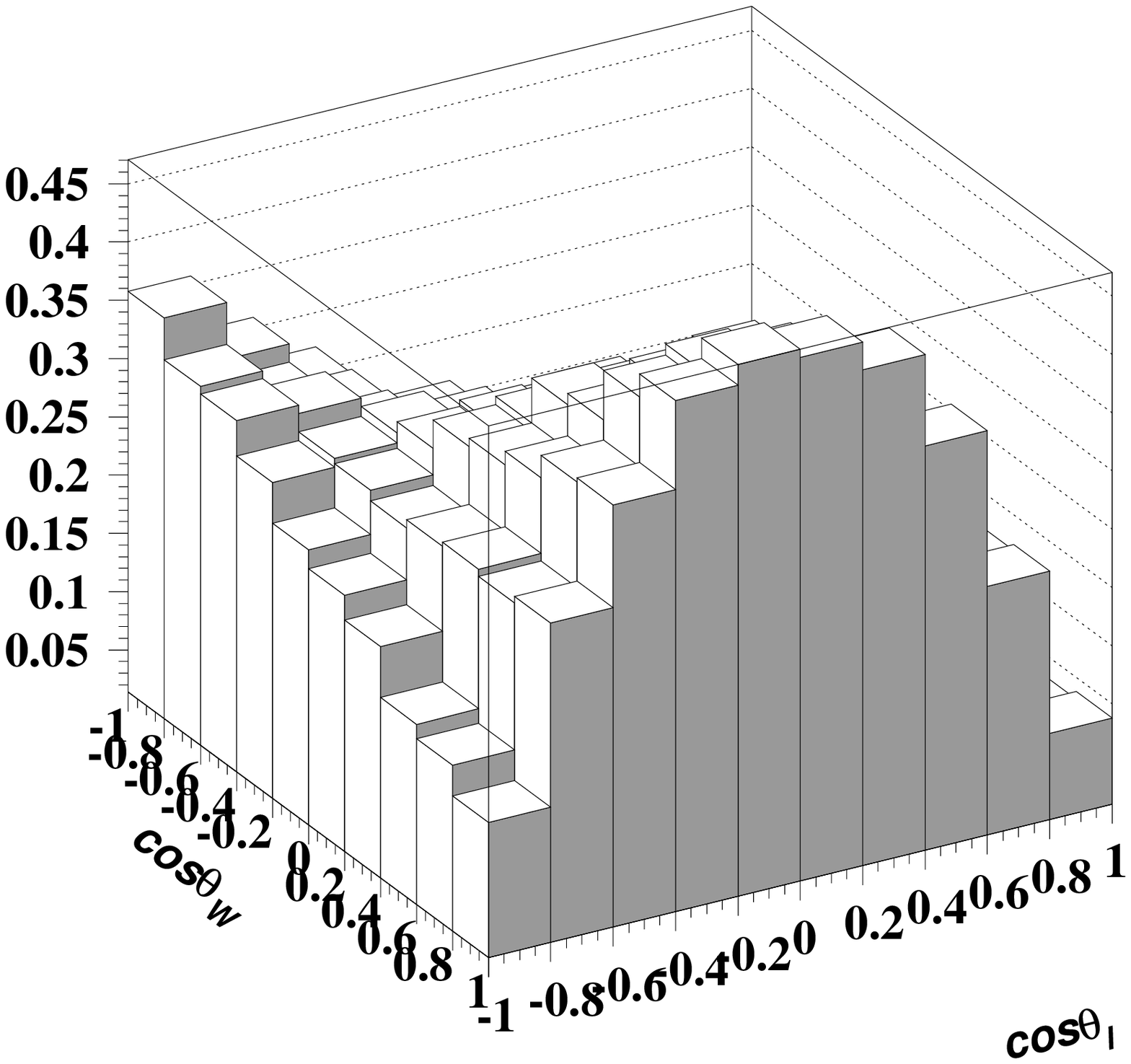} & $\qquad$ $\qquad$ &
\includegraphics[width=5.0cm]{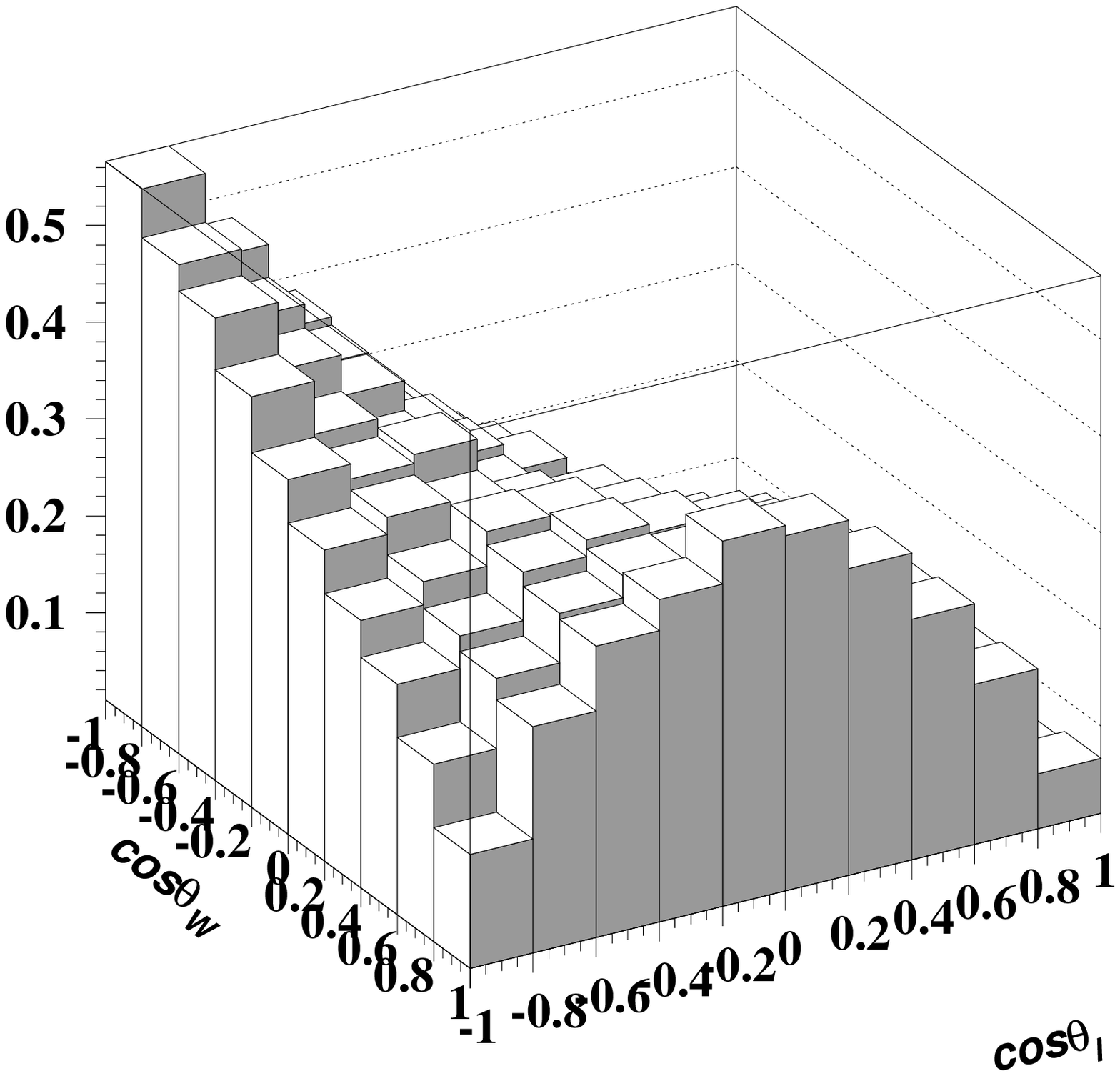}\\
(a) && (b)
\end{tabular}
\caption{ Normalized differential decay distributions (a) for 
$(f_1,f_2) =(1,0)$, and (b) for $(f_1,f_2) =(1,0.3)$. They are normalized to 
unity upon integration.}
\label{spin_sm-f103}
\end{center}
\end{figure}

The differential decay distribution of the $W$ and the $l$ in the semi-leptonic decay 
from a top quark with definite spin orientation, 
$N^{-1}d\Gamma(t \rightarrow bl\nu)/d\cos\theta_{W}d\cos\theta_{l}$, is shown 
in Fig.~\ref{spin_sm-f103} for (a) $(f_1,f_2) =(1,0)$ (tree-level SM) and (b) 
$(f_1,f_2) =(1,0.3)$, respectively. $\theta_{W}$ is defined as the angle 
between the top spin direction and the direction of the $W$ in the top quark rest 
frame. $\theta_{l}$ is defined as the lepton helicity angle, which is the 
angle of the charged lepton in the rest frame of the $W$ with respect to the 
original direction of flight of the $W$. Comparing the two figures, the 
effects of varying $f_2$ are indeed enhanced in the regions 
$\cos\theta_{W} \simeq -1$ and $\cos\theta_{l} \simeq -1$, in accord with the 
enhancement of the $W_{T}$ contributions in these regions. Thus, it is crucial 
to reconstruct the top quark's spin orientation in this method. At hadron 
collider experiments, it is much less trivial to reconstruct the top quark
spin direction, as compared to $e^+e^-$ collider experiments. We discuss how 
to reconstruct the top spin direction in the next section.

\subsection{Effective spin reconstruction}

At hadron colliders, top quarks are produced predominantly through $t\bar{t}$ 
production processes. At the Tevatron, 85\% of the produced $t\bar{t}$ pairs come 
from $q\bar{q}$ initial states, while 15\% come from $gg$ initial states. On 
the other hand, at the LHC, the corresponding fractions are 10\% ($q\bar{q}$) and 
90\% ($gg$), respectively. At these colliders, the polarization of the produced 
top quarks is rather small: at tree level, top quarks are produced 
unpolarized; at NLO, the polarization of top quarks is reported to be very 
small \cite{Bernreuther:2001bx,Bernreuther:2000yn}.

In our analysis of the anomalous couplings, we want to utilize correlations 
between the top quark spin direction and the distribution of its decay 
products. In conventional approaches to reconstruct the top quark spin, the 
top anti-top spin correlation in $t\bar{t}$ events has been 
used~\cite{Mahlon:1995zn,Stelzer:1995gc,Parke:1996pr}. A serious deficit of 
such methods is that they are quite complicated. For instance, the direction 
of the down-type quark in the hadronic decay of anti-top quark is maximally
correlated with the $\bar{t}$ spin. Hence, in order to reconstruct the spin of 
the anti-top quark, we should distinguish the charges of the quarks from the $W$ decay. 
This is a highly non-trivial task and we anticipate that rather large 
systematic errors will be involved before eventually reconstructing the top 
quark spin. On the other hand, if we want to use leptonic decays both, of 
the top and of the anti-top quarks, we suffer from the lack of statistics as well as from 
non-trivial event kinematics due to the two missing momenta of the 
neutrinos.

Here we take another route for reconstructing (effectively) the top quark 
spin. We use the correlation between the top spin and the direction of the 
charged lepton in the top decay for reconstructing the parent top quark's 
spin. Then we use this information to analyze the anomalous couplings in the decay of the same top 
quark. Since we reconstruct the spin and analyze the spin-dependent decay 
distribution using the {\it  same top decay process}, we should make sure that 
we use independent correlations in the former and latter procedures to avoid 
obtaining a meaningless outcome. For this purpose, we take advantage of the 
following facts: (1) Within the SM, the charged lepton is known to be the best 
analyzer of the parent top quark's spin and is produced preferentially in the 
direction of the top spin \cite{Jezabek:1993wk}. (2) The angular distribution 
of the charged lepton with respect to the top spin direction (after all other 
kinematic variables are integrated out) is hardly affected by the anomalous 
couplings of top quark, if the anomalous couplings are small 
\cite{Grzadkowski:1999iq}.\footnote{More precisely, the angular 
distributions of the $\ell^\pm$ are independent of the anomalous couplings up to (and including) 
linear terms in these couplings.} Therefore, we may project the direction of 
the charged lepton onto an appropriate spin basis; then the reconstructed top 
quark spin is scarcely affected by the existence of the anomalous couplings $f_1$ 
and $f_2$, when they are small. We define an effective spin direction by the 
projection of the lepton direction onto the helicity basis:
\begin{equation}
\vec{S}_{\rm SH}  =
\mathrm{sign}(\cos\Theta) \times \frac{\vec{p}_t}{|\vec{p}_t|} 
\quad ,
\label{eq2}
\end{equation}
where $\Theta$ is the angle between the charged lepton and the original 
direction of top quark (opposite of the anti-top direction) in the top rest 
frame; $\vec{p}_t$ is the top quark momentum in the $t\bar{t}$ c.m.\ frame. We 
refer to the effective spin direction above as {\it signed-helicity} (SH) 
direction.

\begin{figure}[htbp]
\begin{center}
\includegraphics[width=6.0cm]{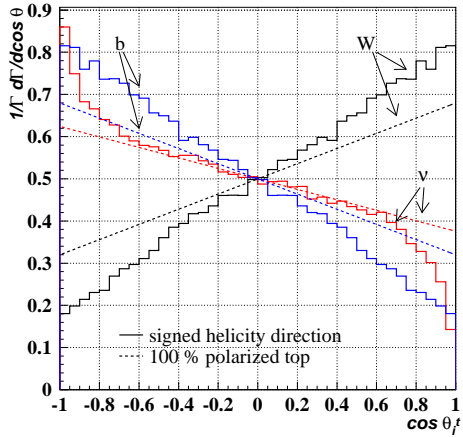}
\caption{ Angular correlations between the directions of
top decay products and the top spin 
direction in the top rest frame, when using the signed-helicity direction 
(solid lines) and when 
using the ideal off-diagonal basis (dashed lines) for the
spin directions, respectively. 
We use the parton information at the generator level, with
the initial state as produced at the Tevatron with $\sqrt{s}=1.96$~TeV.}
\label{helcorr}
\end{center}
\end{figure}

In order to verify how well this effective spin direction reproduces the true 
top spin direction, we demonstrate in Fig.~\ref{helcorr} the angular 
correlations between the directions of the decay products of the top quark and 
the signed-helicity direction (\ref{eq2}). (Obviously, it is tautological to use 
the signed-helicity direction for the analysis of the charged lepton angular 
distribution, so we do not show the lepton angular distribution.) In the same 
figure, the angular correlations using the true top spin direction for 100\% 
polarized top quarks are shown. It is customary to parametrize an angular 
correlation with the linear relation $\frac{1}{2}(1+\alpha\cos\theta)$, where 
$\alpha$ is a correlation coefficient \cite{Mahlon:1995zn}. The correlation 
coefficients $\alpha$ of the $b$ and the $W$ for the signed-helicity direction are 
about twice as large as those for the true top spin direction. This comes from a 
purely kinematic origin, as can be understood as follows: consider a 
hypothetical case, in which no correlation between the true spin direction and 
direction of the $W$ exists (the decay is isotropic). Even in this case, there is 
a positive correlation between the signed-helicity direction and the $W$ in the 
top rest frame, since the charged lepton is emitted more in the direction of 
the $W$ due to the boost of the $W$. The angular correlation of the neutrino with the 
signed-helicity direction does not obey a linear relation.\footnote{This 
dependence is a result of the 100\% anticorrelation of the lepton and the $\nu$ 
directions (back-to-back) in the $l\nu$ c.m.\ frame plus the effect of the boost 
of the $W$.} In this case, the correlation is somewhat stronger than that 
with the true spin direction, too. Thus, the signed-helicity direction reproduces 
qualitatively correct the angular correlations with the decay products, although 
the angular correlations are biased to be somewhat larger than those of the 
true spin direction. In addition, it is important that the dependence of the 
distributions on the anomalous couplings is approximately reproduced in this 
spin reconstruction method.

\subsection{Simulation}

In order to estimate the sensitivity to anomalous couplings in the top 
decay, we perform a Monte Carlo (MC) event generation and a detector simulation. 
The events are produced with both, Tevatron Run II ($p\bar{p}$ collisions with 
$\sqrt{s}$ = 1.96 TeV) and LHC ($pp$ collisions with $\sqrt{s}$ = 14 TeV) 
conditions.

The event generation for the $t\bar{t}$ signal samples is performed by the 
GR@PPA event generator \cite{Tsuno:2002ce} which is interfaced to PYTHIA v6.226, 
a showering MC \cite{Sjostrand:2000wi}. GR@PPA produces the hard process 
based on a $t\bar{t}$ matrix element calculation at tree level. The 
whole decay chain of the top quark is included in the diagram calculation, so 
that the spin correlations in top decays are fully reproduced. The anomalous 
couplings in the top decay are also included. PYTHIA performs 
fragmentation, parton showering, hadronization, and so on and so forth. On the other 
hand, underlying events are produced by PYTHIA alone, using the parameters 
tuned to reproduce the real data from the Tevatron.

The detector simulation is performed by smearing energies for the stable 
particles deposited into a proper segmentation of the calorimeter geometry 
similar to the CDF and ATLAS detectors. Jets are clustered by the cone 
clustering algorithm in PARTHIA ({\tt PYCELL}) with cone size 0.4. We do not 
simulate $b$ tagging. Instead, a $b$-jet is identified as the nearest jet with 
the minimum separation $\Delta R$ between a jet and a $b$-quark at the 
generator level. The separation ($\Delta R$) is defined as 
\begin{equation}
\Delta R \; = \; \sqrt{\Delta \phi^{2} + \Delta \eta^{2}}
\quad ,
\label{eq3}
\end{equation}
where $\Delta \phi$ and $\Delta \eta$ are the separation in the azimuthal 
angle and the pseudorapidity for every pair of a jet and a $b$-quark at 
generator level, respectively. As for leptons, we use the generator level information.

We select the $lepton+jets$ channel in the $t\bar{t}$ production process
by requiring to pass the cuts \\ \\
\hspace{10mm}
\begin{tabular}{lllll}
& Tevatron & & LHC \\
lepton~~~~~~ & $p_{T} \; \geq \; 20 \; \mathrm{GeV}$,
&  $|\eta| \; \leq \; 1.0$  ~~~~~
& $p_{T} \; \geq \; 20 \; \mathrm{GeV}$,
&  $|\eta| \; \leq \; 2.5$ 
\\
$b$--jet & $E_{T} \; \geq \; 15 \; \mathrm{GeV}$,
&  $|\eta| \; \leq \; 1.0$  ~~~~~
& $E_{T} \; \geq \; 30 \; \mathrm{GeV}$,
&  $|\eta| \; \leq \; 2.5$ 
\\
other jet & $E_{T} \; \geq \; 15 \; \mathrm{GeV}$,
&  $|\eta| \; \leq \; 2.0$  ~~~~~
& $E_{T} \; \geq \; 30 \; \mathrm{GeV}$,
&  $|\eta| \; \leq \; 2.5$ 
\\
& ${\not\!\!E}_{T} \, \geq \; 20 \; \mathrm{GeV}$ & & 
${\not\!\!E}_{T} \, \geq \; 20 \; \mathrm{GeV}$ &
\end{tabular}
\label{eq8}
\\ \\
where ${\not\!\!E}_{T}$ is the missing transverse energy calculated by the 
vectorial sum of the selected lepton and the four jets. We require two 
$b$-jets out of least 4 jets in each event.

Although our MC simulation is not fully realistic, we consider it to be 
useful for giving a rough estimate of the sensitivity to anomalous 
couplings before performing a full simulation. In particular, as for the Tevatron 
experiments, our MC simulation would give quite reasonable results. On the 
other hand, for LHC studies, there are some other important ingredients 
that should be taken into account before giving more realistic estimates of 
the sensitivity. Among them, the most important effect would be the presence of events
 with extra jets, i.e.\ $t\bar{t} + n$-jets events, which are not included in 
our event generation. (This effect is expected to be small at the Tevatron.)

The full kinematic event reconstruction for the $lepton + 4 jets$ channel is 
performed by a likelihood fitting reconstruction 
method~\cite{Ikematsu:2003pg} with constrained top and $W$ masses on 
event-by-event basis.\footnote{Ref.\cite{Ikematsu:2003pg} is a dedicated 
study for top quark reconstruction at future $e^+e^-$  linear colliders. In 
order to apply it to hadron collider experiments, some modifications are 
implemented.} This technique has the advantage to choose the correct jet-parton 
assignment by maximizing the likelihood function for each $t\bar{t}$ candidate 
event, as well as a better kinematic reconstruction than the naive 
reconstruction without this likelihood fitting technique.

\subsection{Sensitivity study}

As already explained, we measure the double angular distribution of the $W$ and 
the charged lepton using the effective spin reconstruction method. The top 
quark helicity axis is defined in the top quark rest frame as (the opposite 
of) the direction of the momentum of the hadronically decaying anti-top quark, 
which sequentially decays into three jets. The sign of the top spin is 
defined by the direction of the charged lepton in the top rest frame. The 
reconstructed top quark momentum is also used to measure the helicity angle of 
the charged lepton, since the original direction of the $W$ in the $W$ rest frame 
is equivalent to the opposite of the leptonically decaying top quark direction
in the $W$ rest frame

\begin{figure}[htbp]
\begin{center}
\begin{tabular}{ccc}
\includegraphics[width=5.0cm]{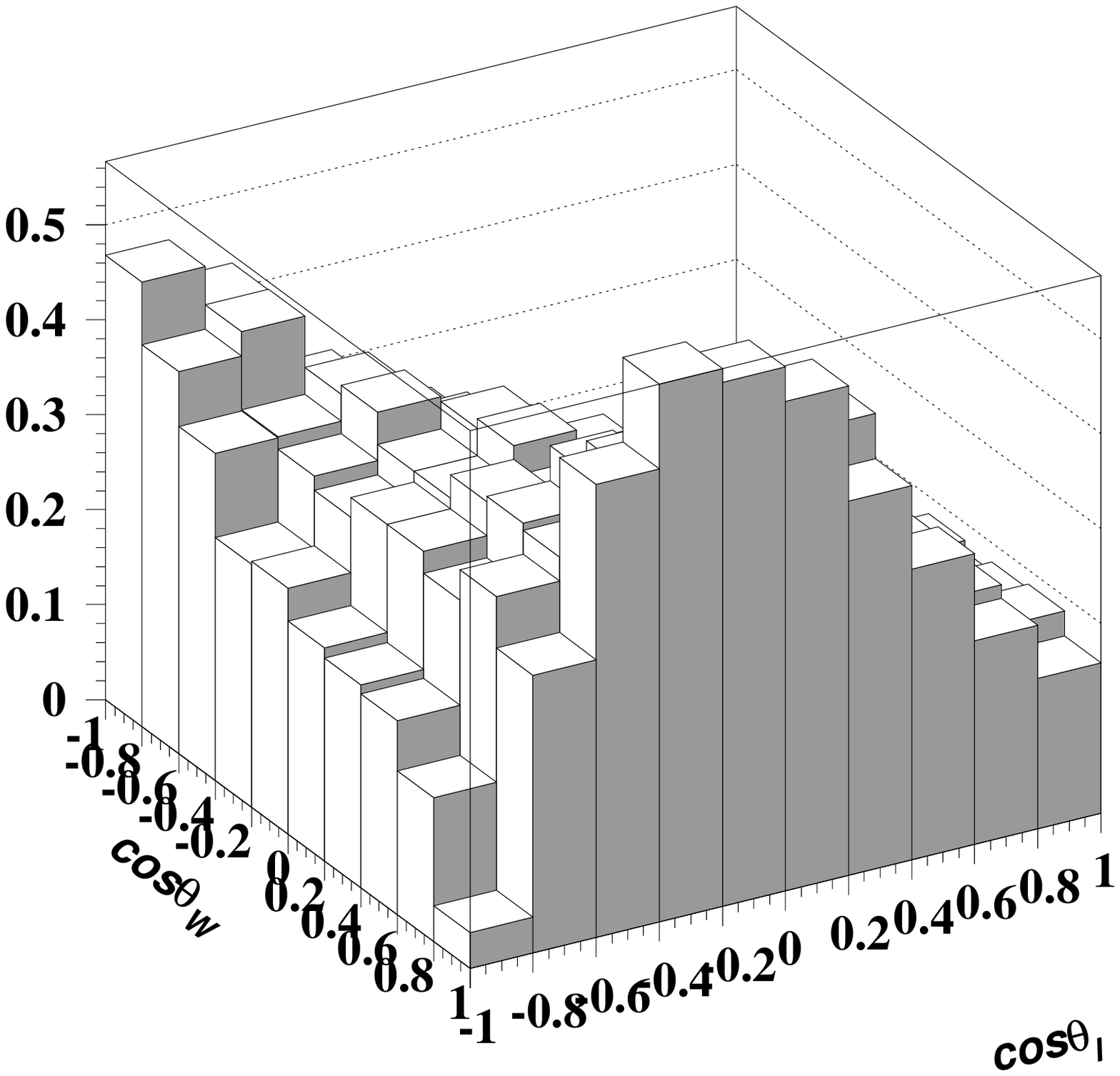} & $\qquad$ $\qquad$ &
\includegraphics[width=5.0cm]{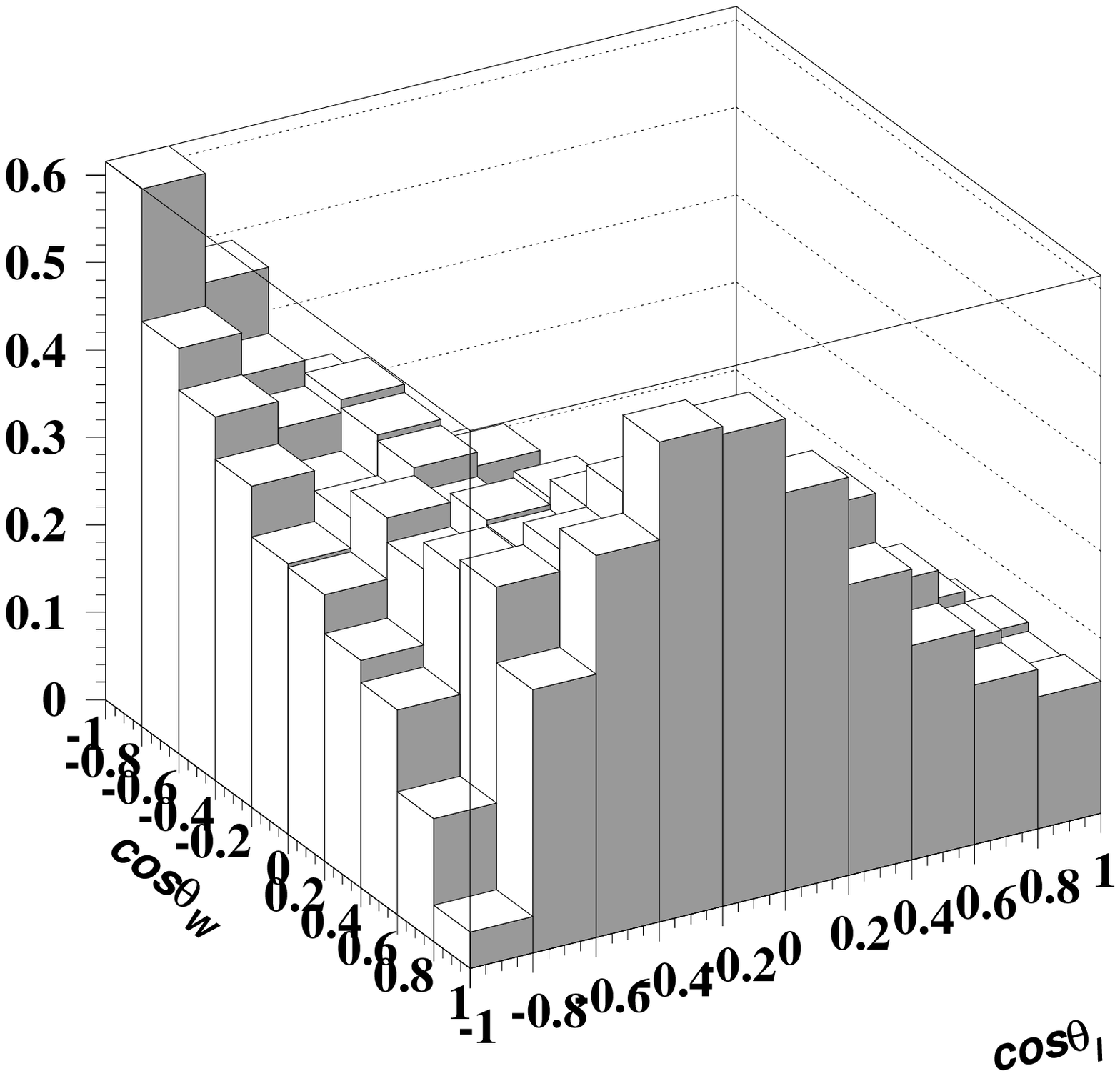}\\
(a) && (b)
\end{tabular}
\caption{ Normalized differential decay distributions using the 
signed-helicity direction {\it after event reconstruction and kinematic 
cuts} (a) for $(f_1,f_2) =(1,0)$, and (b) for $(f_1,f_2) =(1,0.3)$.}
\label{det_spin}
\end{center}
\end{figure}

We show in Fig.~\ref{det_spin} the double angular distributions 
$d\Gamma/d\cos\theta_W\cos\theta_l$ using MC events, after event selection 
and event reconstruction by the kinematic likelihood fitting. Comparing with 
the corresponding parton distributions at generator level in 
Fig.~\ref{spin_sm-f103}, one can see that, even after cuts, the dependence on 
the anomalous couplings remains in the $W_{T}$ region ($\cos\theta_{W}\sim-1$, 
$\cos\theta_{l}\sim-1$). The difference is maximized in the $W_{T}$ region 
($\cos\theta_{W}\sim-1$, $\cos\theta_{l}\sim-1$) and minimized in its diagonal 
opposite region ($\cos\theta_{W}\sim1$, $\cos\theta_{l}\sim1$). The other two 
(diagonal) regions depend weaker on anomalous couplings. 

When the signal statistics is small or the background contribution is not 
well-understood, a simple but not elaborate method to determine the anomalous 
couplings would be practical for a first analysis. Hence, we divide the 
kinematic region into 4 regions and simply count the number of events in 
each region. The regions are defined as follows:
\begin{equation}
\begin{array}{lrccrcl}
\mathrm{Region\ A} : \quad & -1 \; & \leq \; \cos\theta_{W} \; \leq \; 0 
& \quad \mathrm{and} \quad & -1 \; & \leq \; \cos\theta_{l} \; \leq \; 0 & \\
\mathrm{Region\ B} : \quad & -1 \; & \leq \; \cos\theta_{W} \; \leq \; 0
& \quad \mathrm{and} \quad & 0 \; & \leq \; \cos\theta_{l} \; \leq \; 1 & \\
\mathrm{Region\ C} : \quad & 0 \; & \leq \; \cos\theta_{W} \; \leq \; 1 
& \quad \mathrm{and} \quad & -1 \; & \leq \; \cos\theta_{l} \; \leq \; 0 & \\
\mathrm{Region\ D} : \quad & 0 \; & \leq \; \cos\theta_{W} \; \leq \; 1 
& \quad \mathrm{and} \quad & 0 \; & \leq \; \cos\theta_{l} \; \leq \; 1 & \quad ,\\
\end{array}
\label{eq4}
\end{equation}
where $\cos\theta_{l}$ is the lepton helicity angle and $\cos\theta_{W}$ is 
the angle between the $W$ and the signed-helicity direction in the top rest frame. 

\begin{figure}[htbp]
\begin{center}
\includegraphics[width=6.0cm]{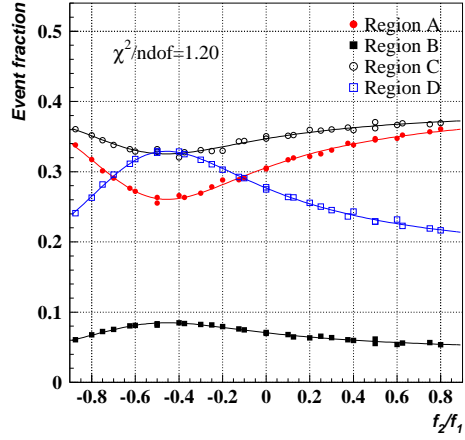}
\caption{ 
MC data and fit results of the event fractions in each region as a functions 
of $f_{2}$/$f_{1}$. All regions are defined in Eq.~(\ref{eq4}). 
}
\label{ratiofit}
\end{center}
\end{figure}

The dependences of the event fractions in these regions on the anomalous 
couplings are shown in Fig.~\ref{ratiofit}. The regions A and D are the 
regions which are most sensitive to anomalous couplings, while the regions B and C 
are less sensitive. We can see that the event fraction in region A 
increases with $f_{2}$/$f_{1}$ when $f_{2}$/$f_{1}$ $>$ 0, and takes a minimum 
value around $f_{2}$/$f_{1} \approx -0.45$, and then increases again if we 
decrease $f_{2}$/$f_{1}$ below $-0.45$. The event fraction in region D has an 
opposite behavior to that of region A. All the event fractions take maximum or 
minimum values around $f_{2}$/$f_{1} = -M_W/M_t \approx -0.45$, where the 
transverse component of the $W$ is canceled. Note that since $f_{1}$ only 
contributes to the normalization of the differential angular distribution, 
which does not affect the shape of the distribution, the event fractions 
depend only on $x=f_{2}/f_{1}$ regardless of the various choices of $f_{1}$ and 
$f_{2}$. 

We fit the MC data as a function of $f_{2}/f_{1}$, shown by discrete points in 
Fig.~\ref{ratiofit}, with analytic functions estimated by the integration over 
each of the regions A--D, where the sum of the event fractions in four regions 
is normalized to one. The fitting results of the event fractions in each 
region are also shown as functions of $f_{2}$/$f_{1}$ in Fig.~\ref{ratiofit}. 
The minimum $\chi^{2}$ per degree of freedom takes a reasonable value 
$\approx 1.20$. The functions, determined by the fit, are used to estimate 
the sensitivity to the anomalous couplings. 

In Table \ref{tablimit}, the expected bounds on the coupling ratio at 
95\% C.L.\ are shown, which are corresponding to 100 and 1000 selected events (after 
cuts) for the Tevatron experiment and 100k selected events (after cuts) for 
the LHC experiment, respectively.\footnote{Using the detection efficiencies, 
100 and 1000 double $b$-tagged events at Tevatron are translated roughly
to 1 and 10~fb$^{-1}$ integrated luminosities, respectively, and 100k events 
to 10 fb$^{-1}$ at LHC.} The input parameters of the MC simulations are taken as 
$(f_1,f_2)=(1,0)$ (tree-level SM values). Only statistical errors are taken 
into account to obtain the allowed regions. For comparison, we present the 
allowed regions using an ideal off-diagonal direction (for the Tevatron), in which 
the spin direction is reconstructed using the off-diagonal basis with the sign 
ambiguity resolved by looking into the information at the generator level; we 
may consider that this ideal off-diagonal direction approximates the true spin 
direction well, so that the corresponding results can be used as references 
(although these include effects of kinematic cuts as well as contamination 
from fake events). We also present the allowed regions using only events 
with a correct assignment of two $b$-jets using the signed helicity direction.

In Table \ref{tablimit}, the bounds using the signed helicity direction are 
not very different from those using the ideal off-diagonal spin direction at 
the Tevatron. Since the latter results can be regarded as references for optimal 
reconstruction of the top spin, it is seen that the signed helicity direction 
is quite efficient for this analysis. In addition, the sensitivities can be 
improved if we can remove the misassignment of the $b$-jets.

\begin{figure}[htbp]
\begin{center}
\includegraphics[width=6.0cm]{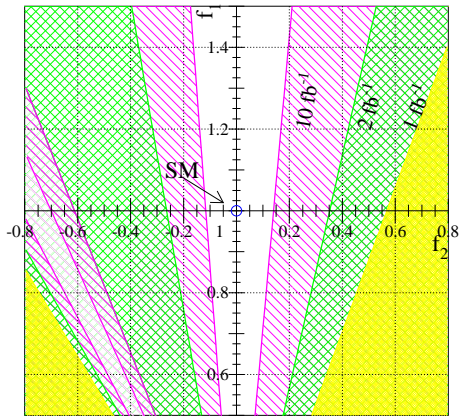}
\caption{ 
Expected excluded regions at 95\% C.L. in the $(f_2,f_1)$-plane at Tevatron. 
The shaded regions correspond to 1~fb$^{-1}$, 2~fb$^{-1}$, and 10~fb$^{-1}$ 
integrated luminosities, respectively. The input SM point is located at 
($f_{2}$,$f_{1}$)=(0,1).}
\label{limit_dsh}
\end{center}
\end{figure}

Finally, we show the expected excluded regions in the $(f_2,f_1)$-plane at 
95\% C.L. for the Tevatron case in Fig.~\ref{limit_dsh}. We thus anticipate 
that our method allows us to cover a wide region in the parameter space even 
in this simplified counting experiment.

\begin{table}
\begin{center}
\caption{ 
Expected bounds at 95\% C.L.\ corresponding to 100 and 1000 events (after 
cuts) for Tevatron and 100k events (after cuts) for LHC. Input parameters of 
the MC simulations are taken as $(f_1,f_2)=(1,0)$. Only statistical errors are 
taken into account. For comparison, the bounds using an ideal off-diagonal 
direction, and those using only the events with correct assignment of the two 
$b$-jets in the signed helicity method are presented.}
  \vspace*{1mm}
\begin{tabular}{c|c|c|c}
& \multicolumn{2}{c|}{Tevatron (1.96 TeV)}
& \multicolumn{1}{c}{LHC (14 TeV)} \\ \cline{2-4}
Number of events
& 100 & 1000 & 100k \\
\hline 
\hline 
Signed-helicity direction
& $-0.93$ $<$ $\frac{f_{2}}{f_{1}}$ $<$ 0.57
& $-0.12$ $<$ $\frac{f_{2}}{f_{1}}$ $<$ 0.14,
& $-0.01$ $<$ $\frac{f_{2}}{f_{1}}$ $<$ 0.01, \\
& 
& $-0.81$ $<\frac{f_{2}}{f_{1}}<$ $-0.70$
& $-0.74$ $<\frac{f_{2}}{f_{1}}<$ $-0.72$ \\ \hline
Ideal off-diagonal direction
& $-0.84$ $<$ $\frac{f_{2}}{f_{1}}$ $<$ 0.50
& $-0.11$ $<$ $\frac{f_{2}}{f_{1}}$ $<$ 0.12,
& Not applicable \\
& 
& $-0.73$ $<\frac{f_{2}}{f_{1}}<$ $-0.61$
&  \\ \hline
Signed-helicity direction
& $-0.29$ $<$ $\frac{f_{2}}{f_{1}}$ $<$ 0.39,
& $-0.09$ $<$ $\frac{f_{2}}{f_{1}}$ $<$ 0.10,
& $-0.01$ $<$ $\frac{f_{2}}{f_{1}}$ $<$ 0.01, \\
with correct $b$ assignment
& $-0.89$ $<\frac{f_{2}}{f_{1}}<$ $-0.59 $
& $-0.80$ $<\frac{f_{2}}{f_{1}}<$ $-0.71 $
& $-0.75$ $<\frac{f_{2}}{f_{1}}<$ $-0.74 $ \\ \hline 
\end{tabular}
\label{tablimit}
\end{center}
\end{table}

\subsection{Conclusions}

We have studied the sensitivities to anomalous couplings $f_1$ 
and $f_2$ in top quark decays at hadron colliders, taking into account realistic experimental 
conditions expected at the Tevatron and the LHC.

We have analyzed a double angular distribution 
$d\Gamma(t\to b l \nu)/d\cos\theta_W d\cos\theta_l$ by using a new method to 
reconstruct the top quark spin direction effectively (referred to as 
{\it signed-helicity direction}). This method does not require the reconstruction 
of the spin of the top quark on the other side, hence it helps to elude possibly 
large systematic uncertainties. These techniques, when used in combination, 
turned out to be quite powerful for the sensitivity study. The $W_T$ region 
$\cos\theta_W \sim -1$, $\cos\theta_l \sim -1$ of the distribution is 
sensitive to the ratio of the anomalous couplings $f_2/f_1$. We confirmed that 
this feature is preserved even after kinematic cuts.

In order to give reliable estimates, we have developed an event generator 
incorporating in the matrix element proper spin correlations of the partons as 
well as the anomalous couplings in the top decay vertices. We also simulate 
the detector effects by assuming a simple geometry and energy resolutions 
based on the CDF and ATLAS detectors for the Tevatron and the LHC colliders, 
respectively. After the event selection, the event kinematics is reconstructed by 
the kinematic likelihood fitting on an event by event basis. It not only 
improves the jet energy scale from the measured jet energy to the 
corresponding parton energy but also helps to select the correct configuration 
of the jets in the top event topology. 

As a first analysis, we have simply counted the event fractions of the double 
angular distribution divided into 4 regions. Then we have performed $\chi^2$--fits 
to these event fractions in order to find the sensitivities to $f_2/f_1$.
The results can be summarized as follows. The bounds obtained at 95\% C.L. 
read
\begin{equation}
\begin{array}{cl}
-0.93 < \frac{f_2}{f_1} < 0.57 & \mbox{for 100 reconstructed events at Tevatron,} \\
-0.81<\frac{f_2}{f_1}<-0.70, ~~ -0.12<\frac{f_2}{f_1}<0.14
& \mbox{for 1000 reconstructed events at Tevatron,} \\
-0.74 < \frac{f_2}{f_1}<-0.72, ~~-0.01 < \frac{f_2}{f_1} < 0.01
& \mbox{for 100k reconstructed events at LHC}. 
\end{array}
\label{eq13}
\end{equation}
We have taken into account only the statistical errors. Due to characteristic 
dependences of the event fractions on $f_2/f_1$, the bound on $f_2/f_1$ 
shrinks quickly as the number of top quark events increases up to a few 
hundred. For more events, the bound scales with $1/\sqrt{N}$, and there 
remains a twofold ambiguity for the allowed ranges of $f_2/f_1$.

Although some simplifications have been made, we consider that our MC study 
for the Tevatron experiment reflects realistic experimental conditions closely 
enough to give reasonable estimates for the sensitivities to anomalous 
couplings. On the other hand, as for the LHC case, some important ingredients 
are still missing in the MC simulation (the most important one would be 
$t\bar{t}+n$-jets events), so our results should be taken as first rough 
estimates.

Since our methods for event reconstruction and effective top spin 
reconstruction are fairly simple, we would expect that they can be applied to 
other analyses, such as precise determination of the $W$ polarization states in top
decays.

\subsection*{Acknowledgments}

S.T. is grateful to the organizers of the Physics at TeV Colliders workshop
(Les Houches 2005) to give an opportunity for fruitful discussions.

%%%%%%%%%%%%%%%%%%%%%%%%%%%%%%%%%%%%%%%%%%%%%%%%%%%%%%%%%%%%%%%%%%%%%%%%%%%%%
\section[Effective NLO approach in the model of single top quark production]
{EFFECTIVE NLO APPROACH IN THE MODEL OF SINGLE TOP QUARK PRODUCTION~\protect
\footnote{Contributed by: E.E~.Boos, V.E.~Bunichev, L.V.~Dudko, V.I.~Savrin,
A.V.~Sherstnev}}
\subsection{Introduction}

The dedicated event generator SingleTop for the  simulation of the
electroweak  production of a single top and its subsequent decays
at the Tevatron and LHC has been achieved with the help of the
CompHEP package~\cite{Boos:2004kh}. Single top is expected to be
discovered at the Tevatron Run II and will be a very interesting
subject of detailed studies at the LHC (see the review
\cite{Beneke:2000hk}).

There are three main processes for single top production at hadron
colliders which could be distinguished by the virtuality $Q^2_W$
of the $W$-boson involved in the process: $t$-channel, $s$-channel
and associated $tW$ mechanisms.

\begin{figure}[hbtp]
\includegraphics[width=0.9\textwidth]{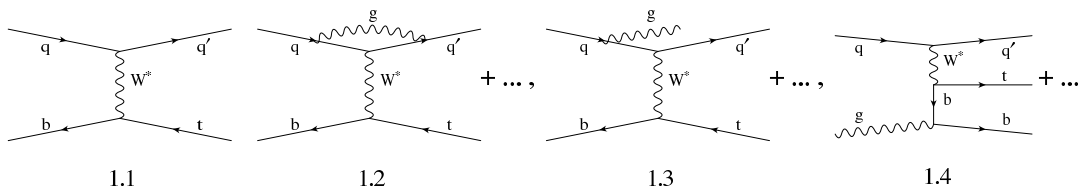}
%\vspace{-15mm}
\caption{LO and representative loop and tree NLO diagrams of the
$t$-channel single top production } \label{SinT}
\end{figure}

The generator SingleTop includes all the three processes and
provides Monte-Carlo unweighted events at the NLO QCD level. In
~\cite{Sullivan:2004ie} it was shown that the NLO distributions
for the $s$-channel process are the same as the LO once rescaled
by a
 known $K$-factor. We discuss shortly here only the
main process with the largest rate, the $t$-channel production.
The representative LO and NLO diagrams are shown in
Fig.\ref{SinT}. The top decay is not shown, however it is included
with all the spin correlations.

\subsection{Overview of the effective NLO approach}

We compute by means of CompHEP the LO  process $2\rightarrow 2$
with the $b$-quark in the initial state including the three-body
decay of the top taking into account spin correlation. This is fed
into PYTHIA~\cite{Sjostrand:2003wg}. We also switch on ISR and
FSR. Then with CompHEP we compute the NLO tree level corrections -
$2\rightarrow 3$ processes with additional $b$- and light quarks
or gluons in the final state including also the top decay with
spin correlations. We split the phase space region in "soft" and
"hard" parts according to the $p_T$ of those additional $b$ and
light jets. The "soft" radiation is taken from PYTHIA radiation
while the "hard" region is a matrix element calculation through
CompHEP. The soft part is normalised in such a way that the value
of the total cross section at NLO, known
from~\cite{Stelzer:1997ns,Harris:2002md}
($\sigma_{NLO}^{t-channel}=242.6 (1.9)$ pb for the LHC
(Tevatron)), is correctly reproduced. The splitting parameters are
tuned based on the requirements that all the distributions become
smooth after  normalization. The performed cross checks show an
agreement with exact NLO calculations where the computed NLO
distributions are correctly reproduced by our method. Therefore,
the generator ``SingleTop'' designed this way does not have a
double counting problem, gives the NLO rate and distributions and
includes all the spin correlations.

The first release of the generator~\cite{Boos:cmsnote} did not include
the hard radiation of the light jets, while the latest version~\cite{Boos:SingleTop}
currently used in the analysis by the Fermilab DO
and the LHC CMS collaborations includes all the mentioned properties.

\subsection{Practical implementation of the method in generator SingleTop}

The generator ``SingleTop''  realizes the effective NLO approach
of event generation by taking into account the main NLO
corrections. It is based on the phase space slicing method.

The cross sections for the  $t$-channel process in the Born
approximation include the  full set of Feynman diagrams where the
top-quark appears with additional $b$ and light quarks in the
final state ($2\to 3$). However, the calculation of the process
$2\to 3$ at  tree level does not include the large logarithmic QCD
corrections (related to the splitting $g\to b\bar b$) that appears
in the "soft" phase space region where the $b$ quark has a small
$P_T$. It is possible to calculate this effect via the standard
renormalization procedure and include it into the partonic
distributions of the $b$-quarks in the proton. In this case the
reaction $2\to 2$ (with $b$-quark in the initial state) would be
the LO approach of the $t$-channel process. In the same way
another $b$-quark should appear also in the final state. It
follows from the fact that $b$-quark can be produced in the proton
only through $b\bar b$ pairs from the virtual gluon. One can
simulate the final b-quark in the process $2\to2$ via ISR. In this
case the $b$-quark could be produced by initial state radiation
and will appear in the final state within a branch of parton
shower, from the splitting function $g\to b\bar b$. One of this
$b$-quarks (from gluon splitting) is the initial hard parton and
the second one goes to the final state.

Calculations of the process $2\to 3$ at the tree level approach
does not include large logarithmic corrections (related to the
process $g\to b\bar b$) but the exact tree level calculations
correctly simulate the behavior of the $b$-quark in the "hard"
phase space region that corresponds to large $P_T$. We will
demonstrate that the combination of the processes $2\to2$ and
$2\to3$ allows us to construct MC samples at "effective" NLO level
approach. We can prepare correct events with "soft" $b$-quark via
ISR simulation. But in this case we loose the significant
contribution of the "hard" $b$-quark. We can probably can get the
appropriate result if we use different strategies of simulation in
the different kinematic regions of phase space. Unfortunately, we
can not naively combine the samples with $2\to2$ and $2\to3$
processes because in this case we get double counting of some
phase space regions. To avoid the problem of double counting we
propose to use different methods of MC simulation in the different
phase space regions and combine them based on some kinematic
parameters.
\begin{figure}[htb]
\begin{minipage}[b]{.46\linewidth}
%\begin{center}
%\epsfxsize=17cm
%\epsfysize=15cm
%\setcaptionmargin{0mm}
%\onelinecaptionsfalse
%\captionstyle{normal}
\includegraphics[width=\textwidth]{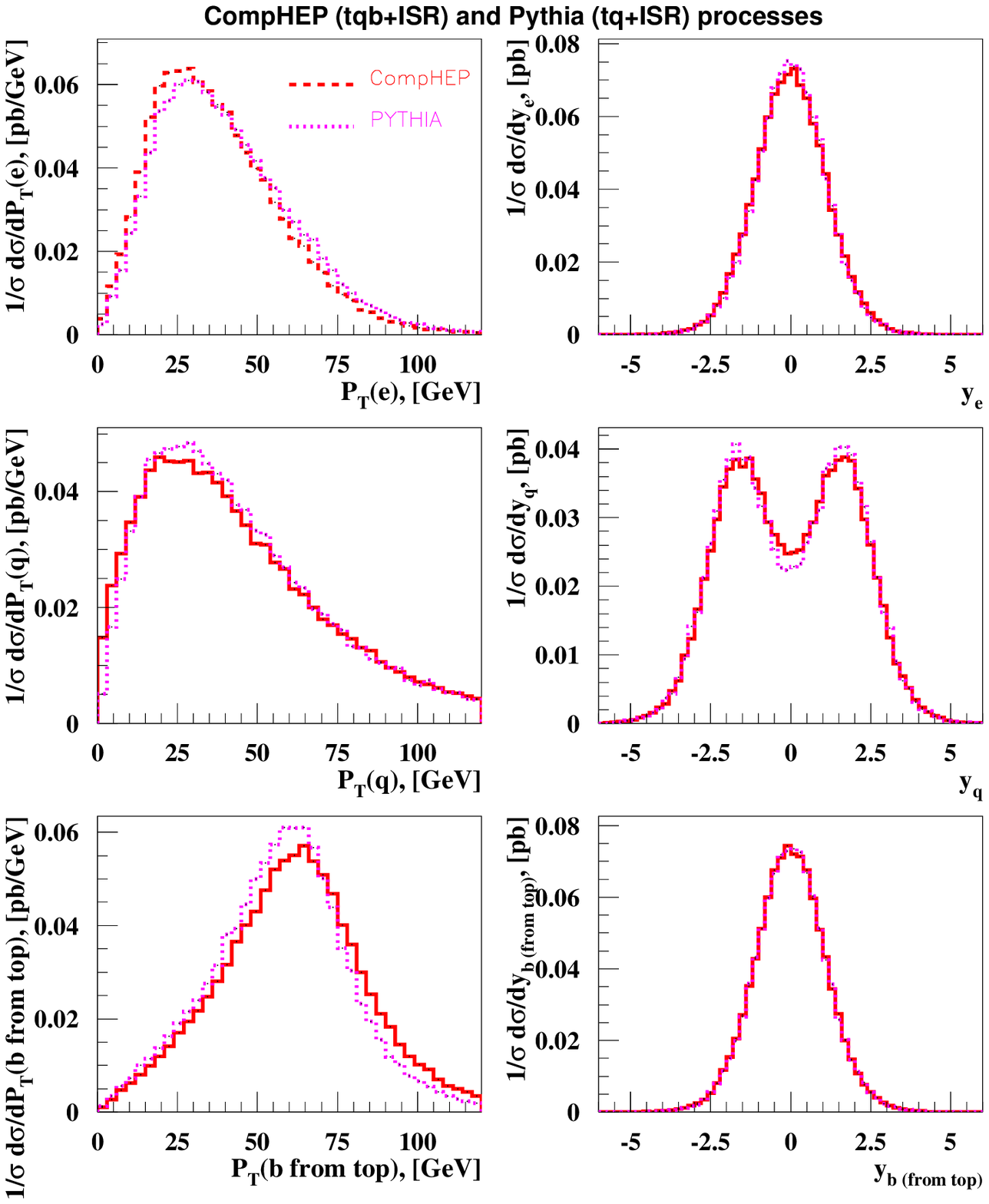}
%\end{center}
\caption{The comparison of $P_T$ and  $\eta$ distributions for the
$pp\to tq+b_{ISR}$~(PYTHIA) and $pp\to tq+b_{LO}$~(CompHEP)
simulations for the Tevatron. The distributions are normalized to
unity and no cuts applied. } \label{fg:tqbtq_tev1}
%\end{figure}
\end{minipage}\hfill
\begin{minipage}[b]{.46\linewidth}
%\begin{figure}[t]
%\begin{center}
%\epsfxsize=12cm
%\epsfysize=15cm
%\setcaptionmargin{0mm}
%\onelinecaptionsfalse
%\captionstyle{normal}
\includegraphics[width=\textwidth]{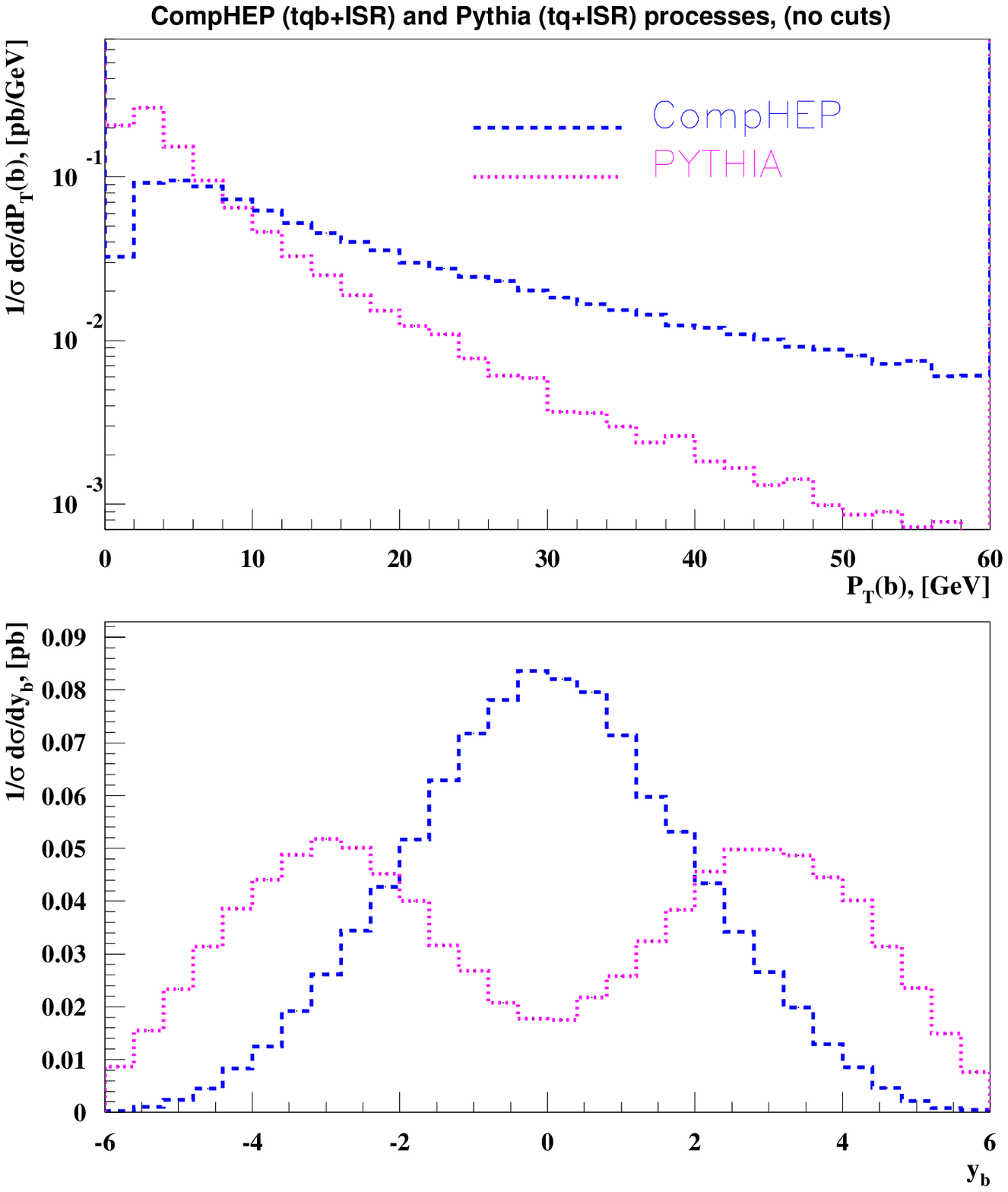}
%\end{center}
\caption{ The comparison of $P_T$ and  $\eta$ distributions for
the $b_{ISR}$ and $b_{LO}$ in the $pp\to tq+b_{ISR}$~(PYTHIA) and
$pp\to tq+b_{LO}$~(CompHEP) simulations for the Tevatron. The
distributions are normalized to unity and no cuts applied. }
\label{fg:tqbtq_tev2}
\end{minipage}
\end{figure}

\begin{figure}[t]
\begin{minipage}[b]{.46\linewidth}
%\begin{center}
%\epsfxsize=12cm
%\epsfysize=15cm
%\setcaptionmargin{0mm}
%\onelinecaptionsfalse
%\captionstyle{normal}
\includegraphics[width=0.95\textwidth]{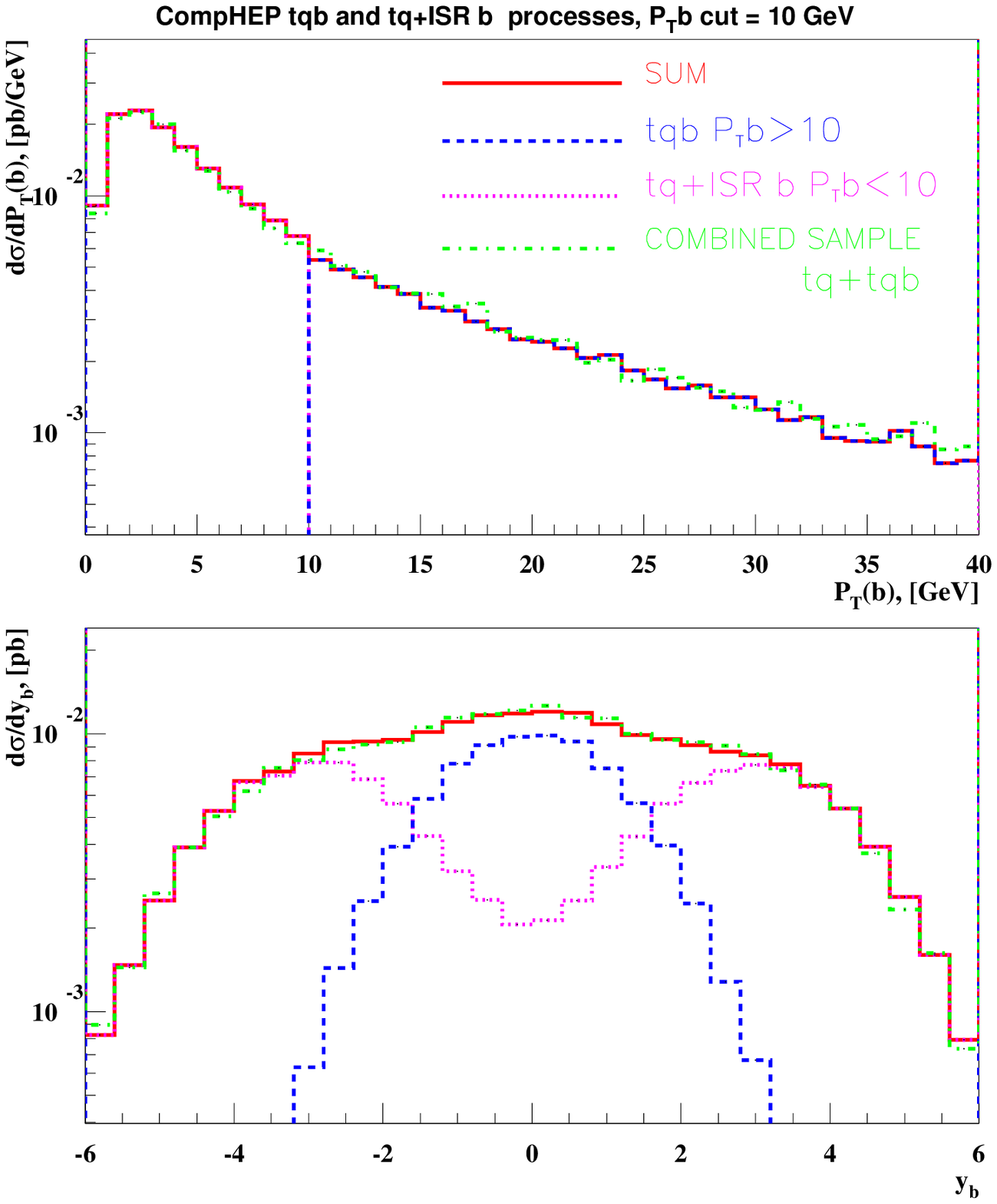}
%\end{center}
\caption{
The combined distributions for the
"soft" $pp\to tq+b_{ISR}$~(PYTHIA) and "hard" $pp\to tq+b_{LO}$~(CompHEP)
regions for the Tevatron collider with $P_T^0(b)=10$ GeV.}
\label{fg:ptd10_tev}
%\end{figure}
\end{minipage}\hfill
\begin{minipage}[b]{.46\linewidth}
%\begin{figure}[t]
%\begin{center}
%\epsfxsize=17cm
%\epsfysize=13cm
%\setcaptionmargin{0mm}
%\onelinecaptionsfalse
%\captionstyle{normal}
\includegraphics[width=\textwidth]{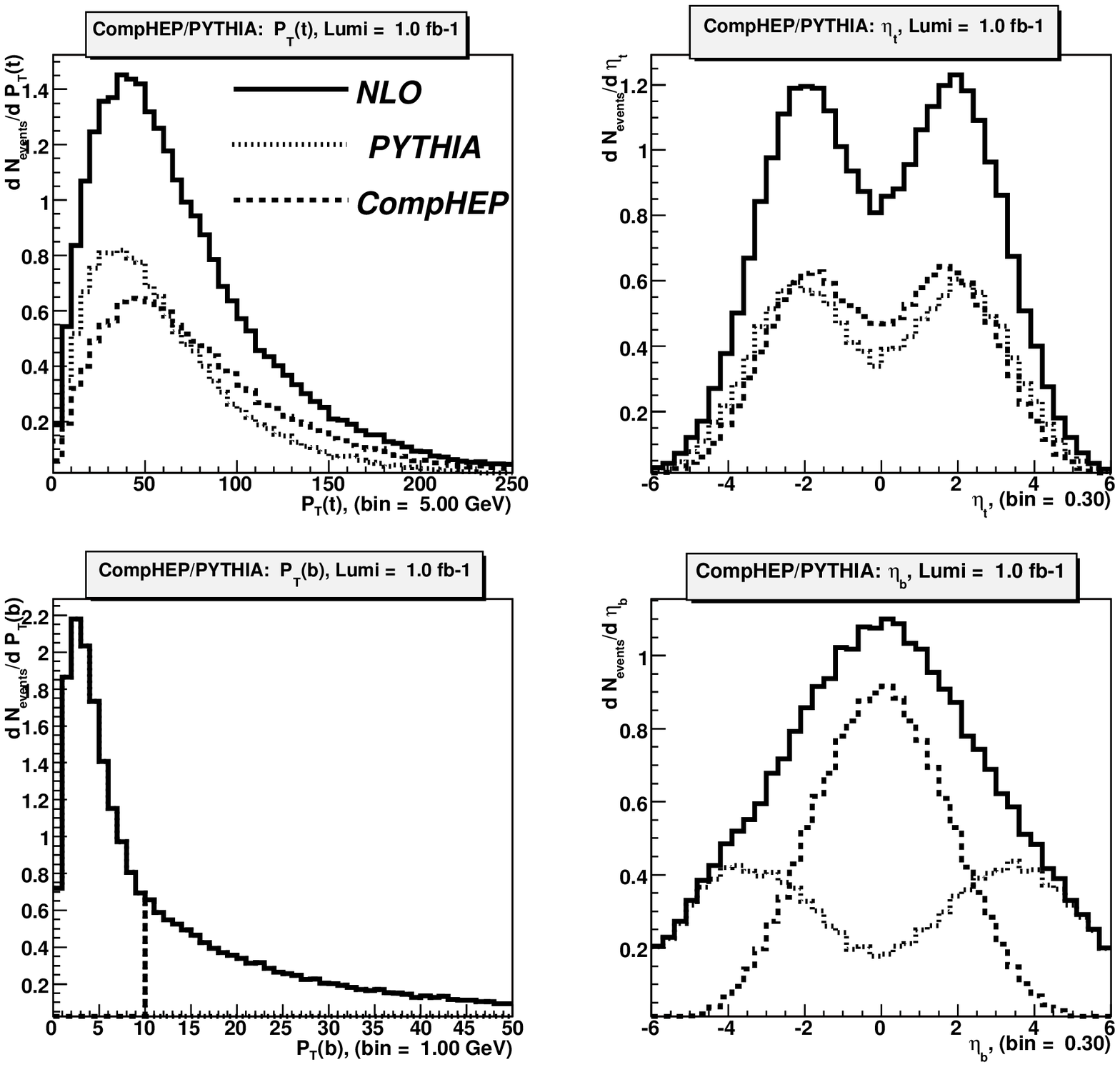}
%\end{center}
\caption{The combined distributions for the
"soft" $pp\to tq+b_{ISR}$~(PYTHIA) and "hard" $pp\to tq+b_{LO}$~(CompHEP)
regions for the LHC collider with $P_T^0(b)=10$ GeV.}
\label{fg:ptd10_lhc}
\end{minipage}
\end{figure}

Figures~\ref{fg:tqbtq_tev1},~\ref{fg:tqbtq_tev2} show the
normalized distributions at Tevatron energies (the distributions
at the LHC energies have almost the same dependencies). On these
plots we can see the distributions for $P_T$ and pseudorapidity of
the top and light quarks looks similar, but the distributions of
additional $b$-quark (that comes not from t-quark decay) differ
significantly. The distribution in pseudorapidity of additional
ISR $b$-quarks, have  pikes at the larger values than the
distributions for processes $2\to 3$ at tree level. Moreover the
$P_T$ spectra for the events derived from  PYTHIA with the ISR
simulation are "softer" than in the tree level calculation. The
main contribution from the large logarithmic appears in the "soft"
region of $P_T(b)$. Therefore, it is reasonable to use the
transverse momentum of the additional $b$-quark as a kinematical
parameter for slicing the phase space between the hard and soft
regions. To prepare events at NLO effective approach we apply the
following procedure: first, we prepare the CompHEP events $2\to 3$
(at tree level) with $P_T(b)$ larger than some critical value
${P^0}_T$. Then we prepare events $2\to2$ in the "soft" region of
the phase space with $P_T(b)<{P^0}_T$. The cross section of
$2\to2$ events in the "soft" region is multiplied by the
$K$-factor. This takes into account loop corrections which do not
change significantly the distributions. The value for the
$K$-factor is derived by normalising the NLO cross section to a
normalised $2 \to 2$ from Pythia and $2 \to 3$ from CompHEP
through
$$
\sigma_{NLO}=K\cdot\sigma_{PYTHIA}(2\to2)|_{P_T(b)<P_T^0}+
                     \sigma_{CompHEP}(2\to3)|_{P_T(b)>P_T^0}.
$$

The $K$-factor here is a function of the slicing parameter
${P^0}_T$. The total NLO cross section we know from exact NLO
calculations~\cite{Stelzer:1997ns,Harris:2002md}.

In case of LHC collider we have:
$$\sigma_{CompHEP}(2\to3)|_{P_T^b>20
   \mbox{\scriptsize\rm GeV}}\approx108.7\;\mbox{\rm pb},$$
$$\sigma_{CompHEP}(2\to3)|_{P_T^b > 10
   \mbox{\scriptsize\rm GeV}}\approx125.7\;\mbox{\rm pb}$$
and K=0.89 for $P_T^{0}=20$ GeV, and k=0.77 for $P_T^{0}=10$ GeV.

In case of the TEVATRON  we have:
$$\sigma_{CompHEP}(2\to3)|_{P_T^b>20
\mbox{\scriptsize\rm GeV}}\approx0.46\;\mbox{\rm pb}$$
$$\sigma_{CompHEP}(2\to3)|_{P_T^b>10
\mbox{\scriptsize\rm GeV}}\approx0.72\;\mbox{\rm pb}.$$
and k=1.32 for $P_T^{0}=20$ GeV, and k=1.21 for
$P_T^{0}=10$ GeV.

The natural requirement for the correct slicing parameter
${P^0}_T$ is a smoothness of the final $P_T$ distribution in the
whole kinematic region for the additional b-quark. After a series
of iterations we  have found that the $P_T$ distribution becomes
smooth enough with ${P^0}_T=10$ GeV. The result is shown in
Figure~\ref{fg:ptd10_tev}. The distributions for the LHC collider
are shown in the figure~\ref{fg:ptd10_lhc} for the same value of
${P^0}_T=10$ GeV. The algorithm described above we have named the
"effective NLO approach".

\subsection{Comparison of the results}

\begin{figure}[hbt]
%\begin{minipage}[b]{.46\linewidth}
%\begin{center}
%\epsfxsize=10cm
%\epsfysize=8cm
%\setcaptionmargin{0mm}
%\onelinecaptionsfalse
%\captionstyle{normal}
\includegraphics[width=\textwidth,height=0.4\textheight]{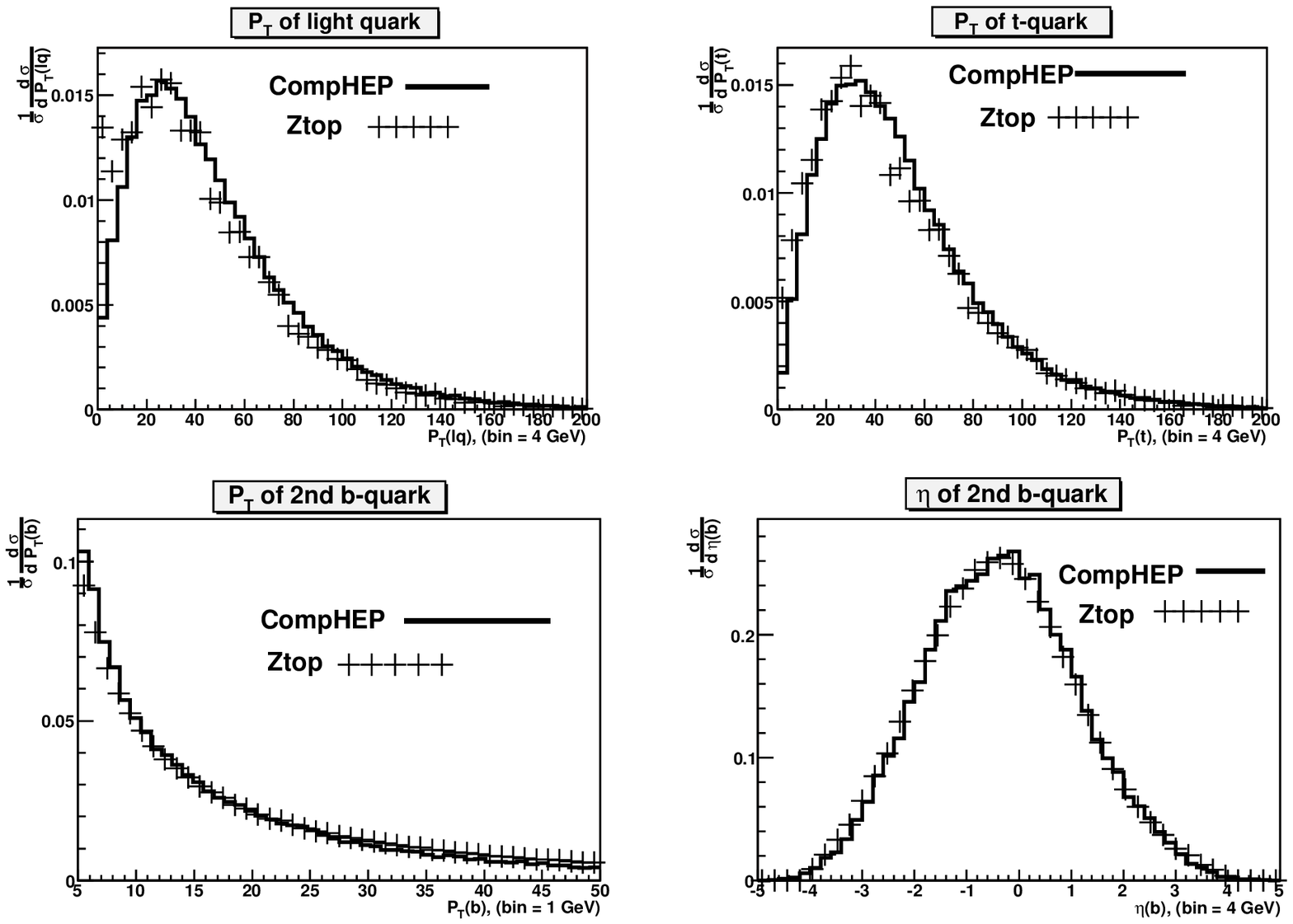}
%\end{center}
\caption{
The $P_T$ and pseudorapidity distributions of final quarks in effective
NLO approach (``SingleTop'') and exact NLO calculations (ZTOP) for the Tevatron collider.
}
\label{fg:ztop_stop}
\end{figure}
%\end{minipage}\hfill
%\begin{minipage}[b]{.46\linewidth}
\begin{figure}[hbt]
%\begin{center}
%\epsfxsize=10cm
%\epsfysize=8cm
%\setcaptionmargin{0mm}
%\onelinecaptionsfalse
%\captionstyle{normal}
\includegraphics[width=\textwidth,height=0.4\textheight]{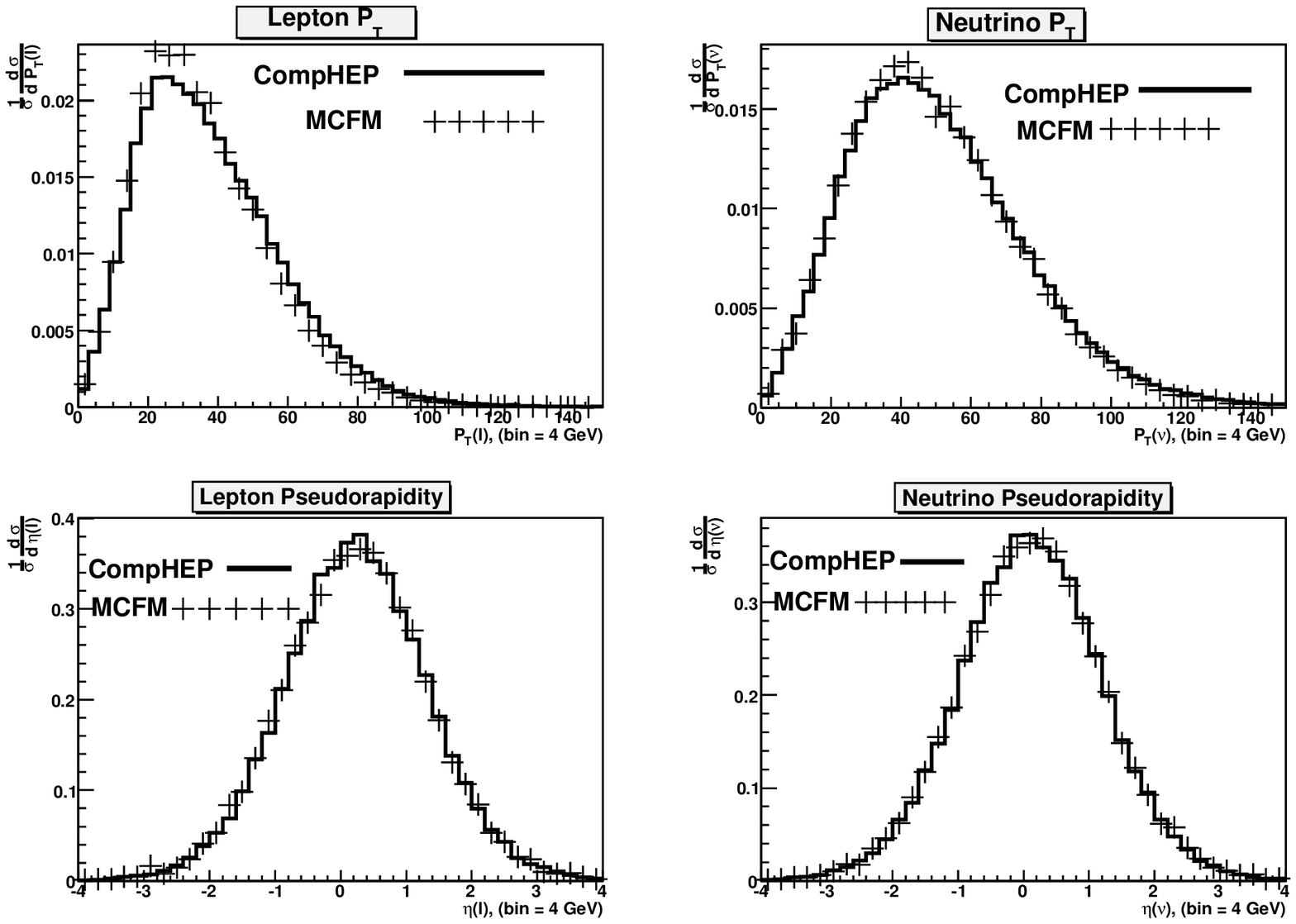}
%\end{center}
\caption{
The $P_T$ and pseudorapidity distributions of final leptons from top-quark decay in effective
NLO approach (``SingleTop'') and exact NLO calculations (MCFM) for the Tevatron collider.
}
\label{fg:mcfm_stop}
%\end{minipage}
\end{figure}

To check the correctness of our approach we compare our results
with two independent NLO calculations. The programs
ZTOP~\cite{Sullivan:2004ie} and MCFM~\cite{Campbell:2004ch}
provides the kinematic distributions at NLO level. The MCFM takes
into account the NLO corrections in the decay of t-quark as well
as in the production. The ZTOP includes NLO corrections only in
the production of top quark. The ZTOP and MCFM programs provide
the possibility to calculate NLO distributions, but do not
simulate events which are important in the real analysis. We
should note, that due to the model of showering for the final
partons, the generator ``SingleTop'' takes into account most of
the part of the NLO corrections in the decay of the  $t$-quark as
well as in the production. We compare the representative
distributions from our effective NLO approach with exact NLO
calculations. The results are shown in
Figures~\ref{fg:ztop_stop},~\ref{fg:mcfm_stop}. We can see how the
events simulated in the effective NLO approach correctly
reproduces the exact NLO distributions which are plotted with the
ZTOP and MCFM programs. The good agreement in the distributions
demonstrates the correctness of the simple approach to model the
most important part of the NLO QCD corrections on the level of
event simulations.

\subsection*{Acknowledgements}
The work is partly supported by RFBR 04-02-16476, RFBR 04-02-17448,
Universities of Russia UR.02.02.503, and Russian Ministry of
Education and Science NS.1685.2003.2 grants.

%%%%%%%%%%%%%%%%%%%%%%%%%%%%%%%%%%%%%%%%%%%%%%%%%%%%%%%%%%%%%%%%%%%%%%%%%%%%%
\section[Progress in ${\rm W^+ W^-}$ production at the LHC]
{PROGRESS IN ${\rm W^+ W^-}$ PRODUCTION AT THE LHC~\protect
\footnote{Contributed by: T.~Binoth, M.~Ciccolini, G.~Davatz, V.~Drollinger,
        M.~D\"uhrssen, A.-S.~Giolo-Nicollerat, M.~Grazzini, N.~Kauer,
        M.~Kr\"{a}mer, M.~Zanetti}}
The production of vector boson pairs in hadron collisions is an important
process within and beyond the Standard Model (SM). Vector boson pairs directly
probe the gauge structure of the $SU(2)\otimes U(1)$ electroweak
theory. Experimentally, various interesting measurements are possible at
hadron colliders. This has been demonstrated at the TeVatron already, for
instance by measuring the ${\rm W^+ W^-}$ cross section or the trilinear
vector boson couplings~\cite{Abazov:2004kc,Acosta:2005mu}.

On the other hand, ${\rm p p \rightarrow W^+ W^-}$ has to be considered as a
background process in many analyses. One of the most prominent examples is the
search for ${\rm h \rightarrow W^+ W^- \rightarrow \ell^+ \nu \ell^-
\bar{\nu}}$, which is the most important SM Higgs search channel in the mass
range between ${\rm 155\ GeV}$ and ${\rm 170\ GeV}$ at the
LHC~\cite{Dittmar:1996ss,Dittmar:1996sp}. Here, W pairs are the irreducible
background to the resonant production of W pairs coming from the Higgs
decay. An accurate theoretical prediction for the ${\rm W^+ W^-}$ background
process is crucial to fully exploit the ${\rm h \rightarrow W^+ W^-}$
discovery channel, in particular as no Higgs mass peak can be reconstructed
from leptonic W decays with two neutrinos in the final state.

In the following, recent progress in the understanding of W pairs is
presented. The ${\rm W^+ W^-}$ cross section is presently known at NLO, and
the contribution from the one-loop ${\rm g g \rightarrow W^+ W^- \rightarrow
\ell^+ \nu \ell^- \bar{\nu}}$ diagrams has been evaluated recently. Although
this is only a part of the (presently unknown) full NNLO contribution, this
calculation is now available also as event generator. Analyses of this process
show that the event properties differ substantially from the LO and NLO
quark-scattering contributions to the ${\rm p p \rightarrow W^+ W^-}$
process. In addition, the event generator for ${\rm g g \rightarrow W^+ W^-}$
has been interfaced to a parton shower program.

For the general case of W pair production soft gluon effects are studied in a
resummed higher order calculation. A solid understanding of soft gluon effects
is important for kinematic properties of ${\rm W^+ W^-}$ events, such as the
transverse momentum of the W pair. Furthermore the results of this calculation
are compared with MC@NLO, in which the spin correlations have been included
quite recently.

After some more comparisons and cross checks, two ${\rm W^+ W^-}$ background
normalization strategies are presented for the Higgs search in the ${\rm h
\rightarrow W^+ W^- \rightarrow \ell^+ \nu \ell^- \bar{\nu}}$ channel, and the
corresponding theoretical uncertainties are evaluated. The uncertainties turn
out to be reduced significantly, when the new achievements are taken into
account.

%\newpage
%==============================================================================
%==============================================================================

\subsection{Soft-gluon effects in ${\rm W^+ W^-}$ production~\protect
\footnote{Author: M.~Grazzini}} 
\label{sec:pp2ww_softGlu}

At present, the WW production cross section is known at NLO accuracy. The NLO
corrections were first obtained with the traditional method by summing over
the W's polarizations \cite{Ohnemus:1991kk,Frixione:1993yp}. Later the
calculation has been extended to fully include spin correlations in the W's
decay \cite{Dixon:1999di,Campbell:1999ah}. The NLO effect increases the cross
section by about $40\%$ at LHC energies.

The fixed-order NLO calculations provide a reliable estimate of ${\rm W^+
W^-}$ cross sections and distributions as long as the scales involved in the
process are all of the same order.  When the transverse momentum of the ${\rm
W^+ W^-}$ pair $p_{T}^{\rm W W}$ is much smaller than its invariant mass
$M_{\rm WW}$ the validity of the fixed-order expansion may be spoiled since
the coefficients of the perturbative expansion can be enhanced by powers of
the large logarithmic terms, $\ln^n M_{\rm WW}/p_{T}^{\rm W W}$.  This is
certainly the case for the $p_{T}^{\rm W W}$ spectrum, which, when evaluated
at fixed order, is even divergent as $p_{T}^{\rm W W}\to 0$, and thus requires
an all-order resummation of the logarithmically enhanced terms.  Resummation
effects, however, can be visible also in other observables, making it
important to study them in detail.

In the following we report on a study of soft-gluon effects in ${\rm W^+ W^-}$
production at hadron colliders \cite{Grazzini:2005vw}.  We use the helicity
amplitudes of Ref.~\cite{Dixon:1998py} and work in the narrow width
approximation (i.e. we only consider double-resonant contributions), but fully
include the decays of the ${\rm W}$ bosons, keeping track of their
polarization in the leptonic decay.  In the large $p_{T}^{\rm W W}$ region we
use LO perturbation theory (${\rm W^+ W^-}$+1 parton); in the region
$p_{T}^{\rm W W}\ll M_{\rm WW}$ the large logarithmic contributions are
resummed to NLL and (almost) NNLL \cite{deFlorian:2000pr,deFlorian:2001zd}
accuracy
\footnote{The inclusion of NNLL terms cannot be complete
\cite{Grazzini:2005vw}, since two-loop corrections to ${\rm W^+W^-}$
production are not yet known.}.

To perform the resummation
we use the formalism of Refs.~\cite{Bozzi:2003jy,Bozzi:2005wk}.
In this approach, the resummation is achieved at the level of the partonic
cross section and the large logarithmic contributions are exponentiated
in a process-independent manner, being constrained to give vanishing
contribution to the total cross section.
Our results have thus uniform NLO accuracy over the entire range of transverse momenta but consistently 
include the all-order resummation of logarithmically enhanced terms in the region $p_{T}^{\rm W W}\ll M_{\rm WW}$.

We present predictions for the transverse momentum spectrum of the ${\rm W^+ W^-}$ pair,
but also for a few leptonic distributions,
comparing our results with those obtained at NLO
with the program MCFM \cite{Campbell:1999ah}, and with the ones from
the general purpose event generator 
MC@NLO~\cite{Frixione:2002ik,Frixione:2003ei} which,
in its latest release \cite{Frixione:2005gz}, partially includes the effect of spin correlations in the ${\rm W}$'s decay.
More details can be found in Ref.~\cite{Grazzini:2005vw}.

To compute the ${\rm W^+ W^-}$ cross section we use MRST2002 NLO densities \cite{Martin:2002aw}
and $\alpha_{\mathrm{S}}$ evaluated at two-loop order.
Our resummed predictions depend on renormalization, factorization and resummation scales.
Unless stated otherwise, the resummation scale is set equal to the
invariant mass $M_{\rm WW}$ of the ${\rm W^+ W^-}$ pair, whereas
renormalization and factorization scales are set to $2M_W$.
The latter choice allows us to exploit our unitarity constraint and to exactly recover
the total NLO cross section when no cuts are applied.
At NLO we consistently use $\mu_F=\mu_R=2M_W$ as default choice,
whereas in MC@NLO $\mu_F$ and $\mu_R$ are set to the default choice, the average transverse mass of the ${\rm W}$ bosons.

We start by considering the inclusive cross sections.
Our NLL+LO result is 115.6 pb,
and agrees with the NLO one (116.0 pb) to better than $1\%$.
The cross section from MC@NLO is instead lower, about 114.7 pb.
The above difference is due to the different choice
of the scales, and to the different convention in the choice of the electroweak couplings adopted in MC@NLO.
%==========================================
\begin{figure}
\begin{center}
\begin{tabular}{cc}
\epsfysize=5.5truecm
\epsffile{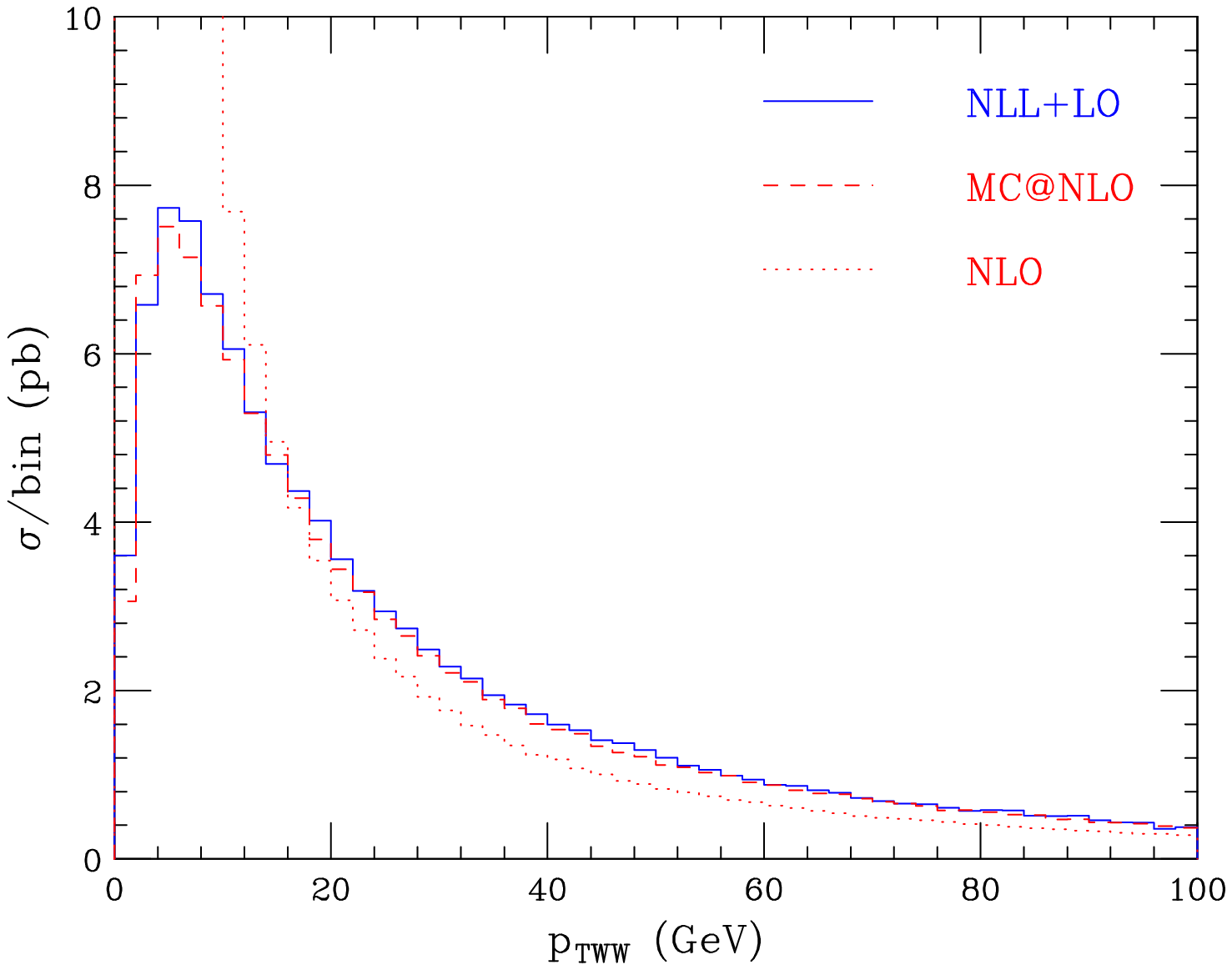} & \epsfysize=5.5truecm\epsffile{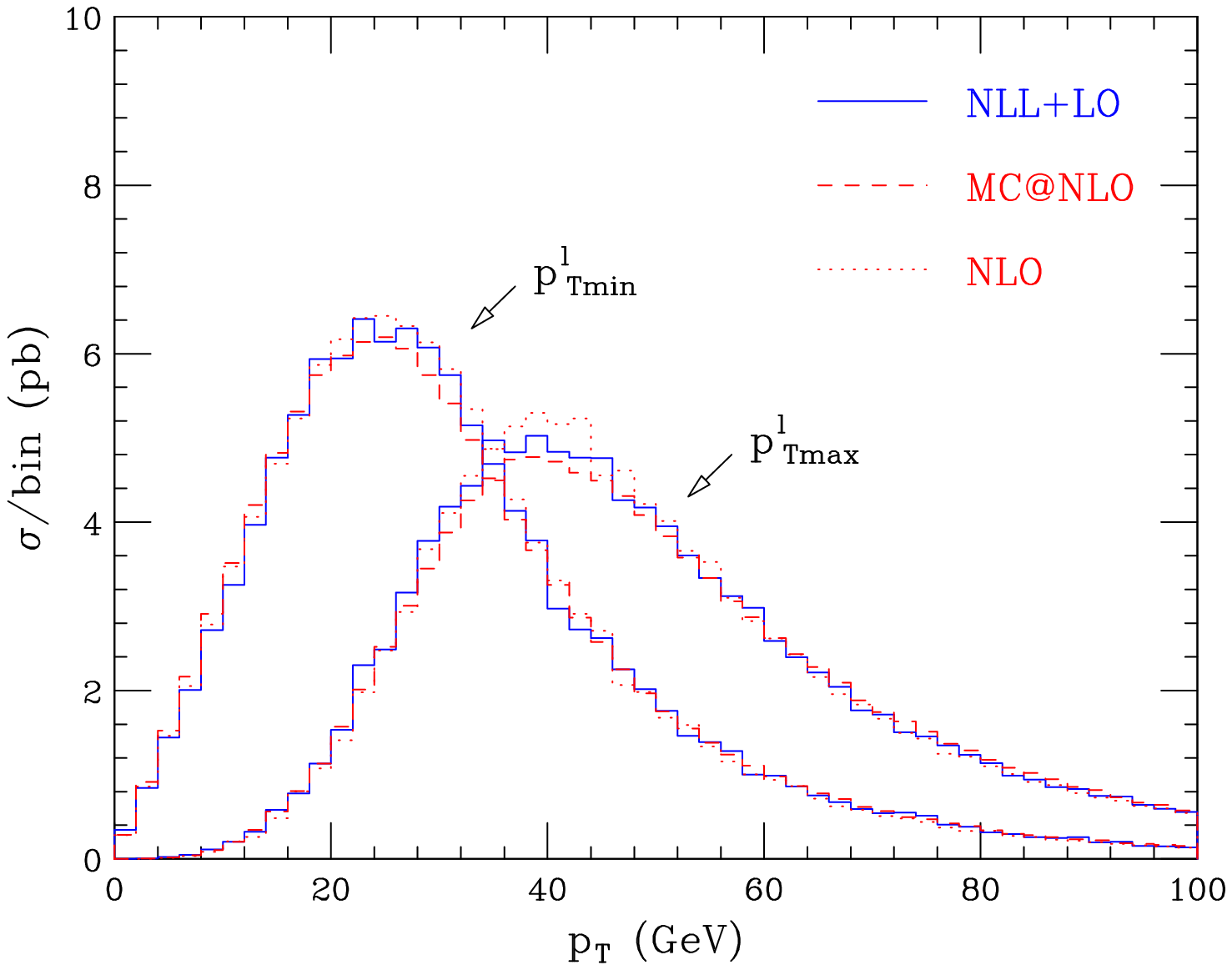}\\
\end{tabular}
\end{center}
\vspace*{-3mm}
\caption{\label{fig1}{\em Left: Comparison of the transverse momentum spectra of the ${\rm W^+W^-}$ pair obtained at NLL+LO, NLO and with MC@NLO. No cuts are applied. Right: corresponding predictions for the transverse momentum spectra of the lepton with minimum and maximum $p^l_T$.}}
\end{figure}
%=========================================

In Fig.~\ref{fig1} (left) we show the $p_{T}^{\rm W W}$ distribution,
computed at NLO (dotted), NLL+LO (solid) and with MC@NLO (dashed).
We see that the NLO result diverges to $+\infty$ as $p_{T}^{\rm W W}\to 0$.
The NLL+LO and MC@NLO results are instead finite as $p_{T}^{\rm W W}\to 0$ and are in good agreement,
showing a kinematical peak around $p_{T}^{\rm W W}\sim 5$ GeV.

We now consider the $p_T$ spectra of the leptons.
For each event, we classify the transverse momenta of the two charged leptons
into their minimum and maximum values, $p^l_{T{\rm min}}$ and $p^l_{T{\rm max}}$.
In Fig.~\ref{fig1} (right) we plot the corresponding $p_T$ spectra,
computed at NLL+LO (solid), NLO (dotted) and with MC@NLO (dashes).
All the three predictions are clearly in good agreement:
the effect of resummation, which is essential in the $p_{T}^{\rm W W}$ spectrum,
is hardly visible in the leptonic spectra.

To further assess the effect of resummation in the leptonic observables,
we consider the application of the following cuts,
suggested by the study of Ref.~\cite{Davatz:2004zg}:

\begin{itemize}
\item For each event, $p^l_{T{\rm min}}$ should be larger than $25$ GeV and $p^l_{T{\rm max}}$ should be between $35$ and $50$ GeV.
\item The invariant mass $m_{ll}$ of the charged leptons should be smaller than $35$ GeV.
\item The missing $p_T$ of the event should be larger than $20$ GeV.
\item The azimuthal charged lepton separation in the transverse plane $\Delta\phi$ should be smaller than $45^o$.
\item A jet veto is mimicked
by imposing that the transverse momentum of the ${\rm W^+ W^-}$ pair should be smaller than $30$ GeV.
This cut is perfectly legitimate in our resummed calculation and is exactly equivalent to a jet veto at NLO.
\end{itemize}
These cuts, designed for the search of a Higgs boson with $M_{\rm h} = 165$ GeV,
strongly select the small $\Delta\phi$ region. The jet veto is usually applied
in order to reduce
the $t{\bar t}$ contribution, which is expected to produce large-$p_T$ $b$-jets from the decay of the top quark.

The NLL+LO (MC@NLO) accepted cross section is 0.599 pb (0.570 pb) which should be contrasted with the NLO result,
which is 0.691 pb, about $20\%$ higher.
This relative large difference is due to the fact that these cuts
enhance the relevance of the small-$p_{T}^{\rm W W}$ region,
where the NLO calculation is not reliable.

\begin{figure}[htb]
\begin{center}
\begin{tabular}{c}
\epsfysize=7truecm
\epsffile{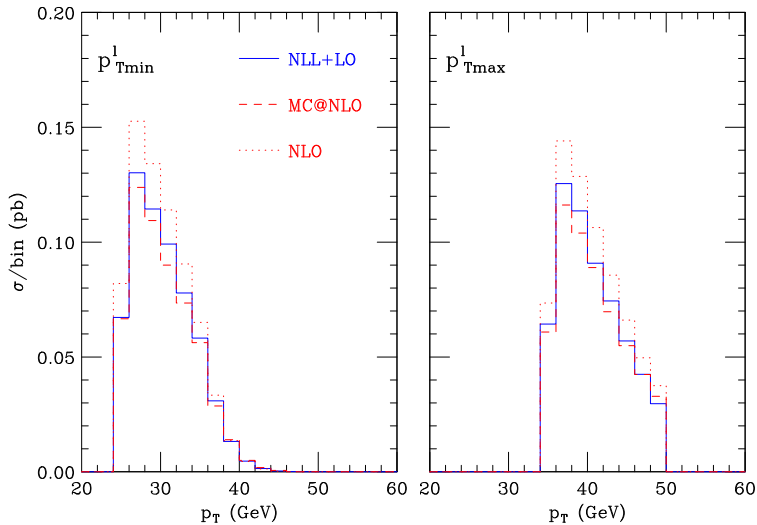}\\
\end{tabular}
\end{center}
\vspace*{-3mm}
\caption{\label{WWfig2}{\em Distributions of $p^l_{T{\rm min}}$ and $p^l_{T{\rm max}}$ when cuts are applied.}}
\end{figure}

In Fig.~\ref{WWfig2} the $p^l_{T{\rm min}}$ and $p^l_{T{\rm max}}$ distributions are presented.
We see that although the three predictions are
in reasonable agreement in shape,
differences are now evident.
In particular, the $p^l_{T{\rm min}}$ distribution at NLO is steeper than the other two.
Comparing NLL+LO and MC@NLO spectra, we see that
the former are steeper than the latter:
with the application of strong cuts
the differences between NLL+LO and MC@NLO predictions are enhanced.

\begin{figure}[htb]
\begin{center}
\begin{tabular}{cc}
\epsfysize=5.5truecm
\epsffile{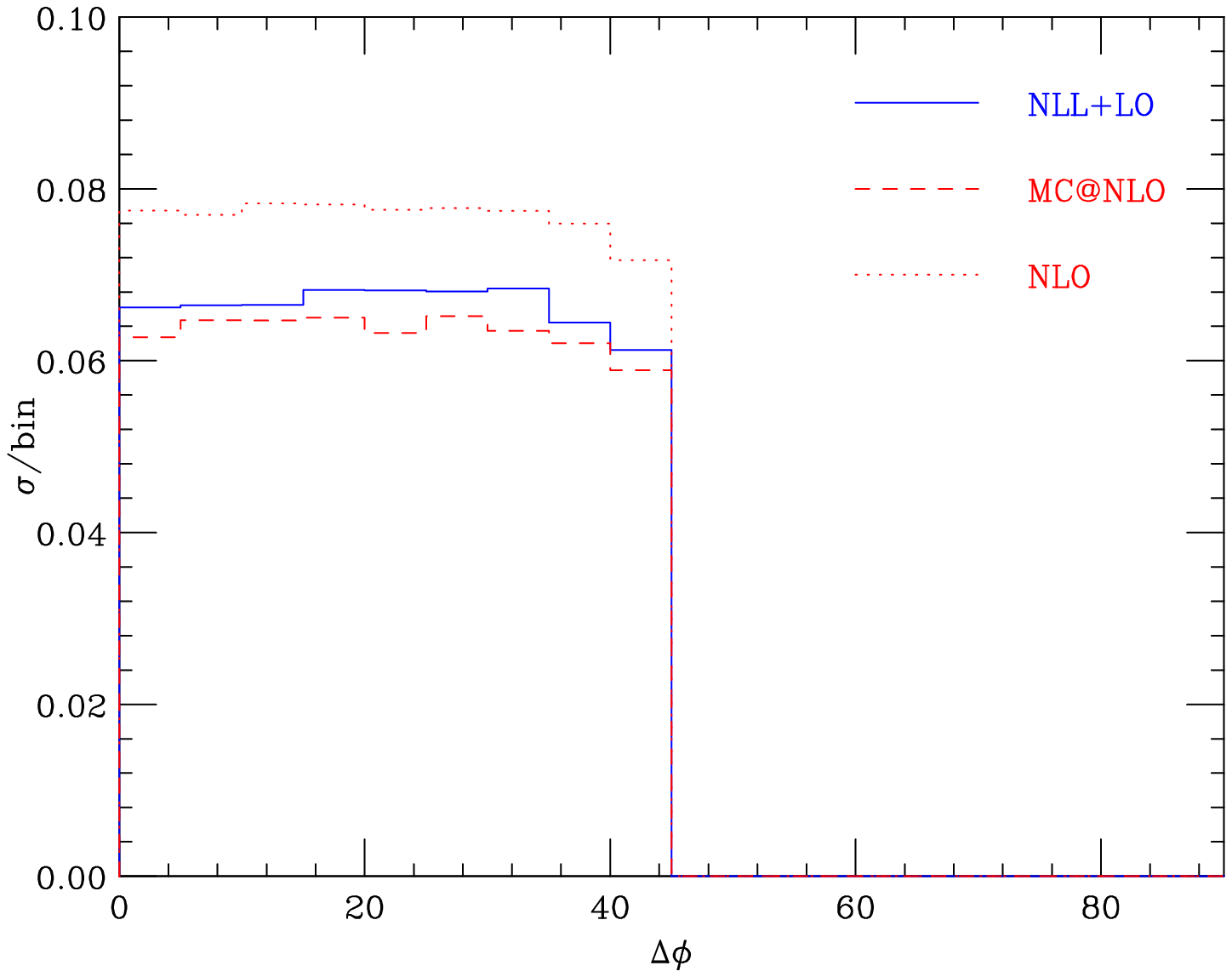} &\epsfysize=5.5truecm\epsffile{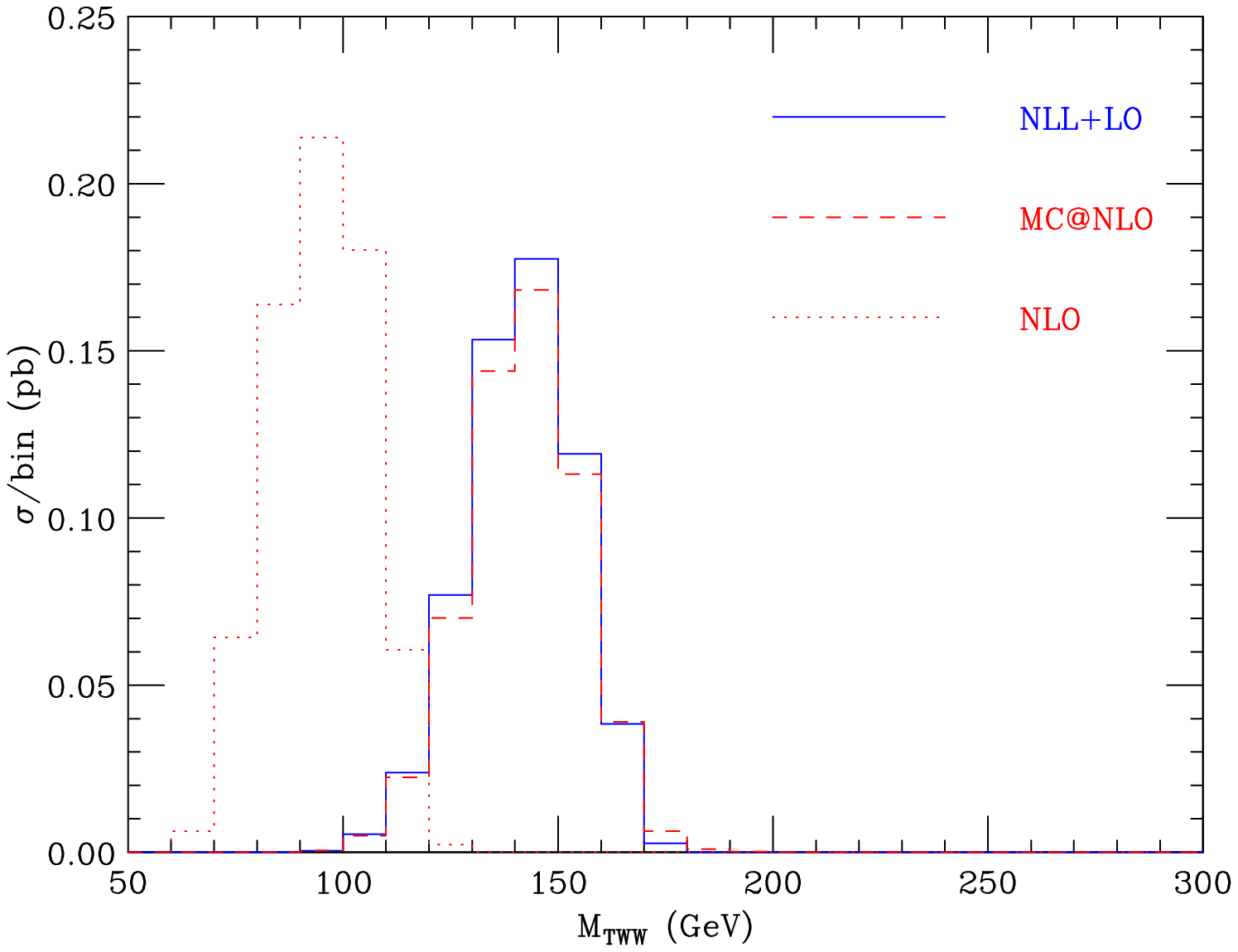}\\
\end{tabular}
\end{center}
\vspace*{-3mm}
\caption{\label{fig3}{\em Left: $\Delta\phi$ distribution when cuts are applied. Right: Transverse-mass distribution. }}
\end{figure}
In the search for the Higgs boson in the ${\rm h \rightarrow W^+ W^- \rightarrow \ell^+ \nu \ell^- \bar{\nu}}$ channel
an important difference between the signal and the background is found in the $\Delta\phi$ distribution.
Since the Higgs is a scalar, the charged leptons tend to be produced quite close in angle.
As a consequence, the signal is expected to be peaked at small values of $\Delta\phi$,
whereas the $\Delta\phi$ distribution for the background is expected to be reasonably flat.
It is thus important to study the effect of resummation on this distribution,
which is also known to be particularly sensitive to spin correlations.
In Fig.~\ref{fig3} (left) the $\Delta\phi$ distribution is displayed.
We see that the shapes of the three results
are in good agreement with each other, although
a slightly different slope of the
NLL+LO result with respect to MC@NLO and NLO appears.
We remind the reader that the NLO and NLL+LO calculations exactly include spin correlations, whereas
MC@NLO neglects spin correlations in the finite (non-factorized) part of the one-loop contribution.

In Fig.~\ref{fig3} (right) we finally consider the transverse-mass distribution of the ${\rm W^+ W^-}$ system, defined as in Ref.~\cite{Rainwater:1999sd}.
The NLO result (dotted) is compared to the NLL+LO one (solid) and to MC@NLO (dashes).
We see that the effects of soft-gluon resummation are dramatic for this distribution:
the NLL+LO result is shifted towards larger values
of $M_{TWW}$ by about 50 GeV, which is mainly due to the divergence of the NLO curve, shown in Fig.~\ref{fig1}.
We find that this big difference is
mainly due to the leptonic cuts: removing the jet veto the shift in
the transverse-mass distribution is basically unchanged.
Comparing the shapes of the histograms
we see that at NLO the shape
is fairly different with respect to NLL+LO and MC@NLO.
Also the NLL+LO and MC@NLO distributions now show clear differences:
the position of the peak is the same, but
the NLL+LO result is steeper and softer than the MC@NLO one.

In this contribution we have examined soft-gluon effects in ${\rm W^+ W^-}$ production at the LHC.
We find that resummation has a mild impact on inclusive leptonic distributions.
On the other hand, when stringent cuts are applied,
the effects of resummation are strongly enhanced.
The most significant effect is seen in the transverse mass distribution, for which the NLO calculation
is clearly not reliable.
Our resummed predictions are generally in good agreement with those of the MC@NLO event generator.

%==============================================================================
%==============================================================================

\subsection{Gluon-induced ${\rm W^+W^-}$ background to Higgs boson 
searches~\protect\footnote{Authors: T.~Binoth, M.~Ciccolini, N.~Kauer, 
M.~Kr\"{a}mer}}

\subsubsection{Introduction\label{intro-section}}

The hadronic production of W pairs has been studied extensively in
the literature (see e.g.\ Ref.~\cite{Haywood:1999qg}). In this short
note we focus on the gluon-induced loop process ${\rm gg \to
W^{\ast}W^{\ast} \to \ell\bar{\nu}\bar{\ell'}\nu'}$. Although
suppressed by two powers of $\alpha_{\rm s}$ relative to
quark-antiquark annihilation, the importance of the gluon-gluon
induced background process is enhanced by the experimental Higgs
search cuts which exploit the longitudinal boost and the spin
correlations of the ${\rm W^+W^-}$ system to suppress W pair continuum
production through quark-antiquark annihilation. We present the first
complete calculation of the gluon-fusion process ${\rm gg \to
W^{\ast}W^{\ast} \to \ell\bar{\nu}\bar{\ell'}\nu'}$, including spin and
decay angle correlations and allowing for arbitrary invariant masses
of the W bosons. This work extends our previous
calculation~\cite{Binoth:2005ua}, which did not include the
contribution from the intermediate top-bottom massive quark loop.

Our calculation demonstrates that the gluon-fusion contribution to
on-shell W pair production only provides a $5\%$ correction to the
inclusive W pair production cross section at the LHC. However, after
taking into account realistic experimental cuts, the process ${\rm gg \to
W^{\ast}W^{\ast} \to \ell\bar{\nu}\bar{\ell'}\nu'}$ becomes sizeable
and enhances the theoretical ${\rm W^+W^-}$ background estimate for Higgs
searches by about $30\%$.  In the following we will present a brief
discussion of the numerical results. Details of the calculation can be
found in Ref.~\cite{Binoth:2005ua} and in a forthcoming article. We
note that an independent calculation of the ${\rm gg \to W^+W^-}$ background has
been performed in Ref.~\cite{Duhrssen:2005bz}. A comparison of the two
calculations Refs.~\cite{Binoth:2005ua} and \cite{Duhrssen:2005bz} is
in progress.

% ------------------------------------------------------------------------

\subsubsection{Results \label{results-section}}

In this section we present numerical results for the process ${\rm pp \to
W^{\ast}W^{\ast}\to \ell\bar{\nu}\bar{\ell'}\nu'}$ at the LHC, where $\ell
= e~\mbox{or}~\mu$.  We tabulate the total cross section and the cross
section for two sets of experimental cuts.  In addition, we show
various differential distributions. The experimental cuts include a
set of ``standard cuts''~\cite{Haywood:1999qg}, motivated by the
finite acceptance and resolution of the detectors, where we require
all charged leptons to be produced at $p_{T,\ell} > 20$ GeV and
$|\eta_\ell| < 2.5$, and a missing transverse momentum $\not\!p_T >
25$~GeV. Cross sections calculated with this set of cuts will be
labeled $\sigma_{\rm std}$.  We also present results after imposing
Higgs search cuts following a recent experimental
study~\cite{Davatz:2004zg}.  In addition to the ``standard cuts''
defined above, we require that the opening angle between the two
charged leptons in the plane transverse to the beam direction must
satisfy $\Delta\phi_{T,\ell\ell} < 45^{\circ}$ and that the dilepton
invariant mass $M_{\ell\ell}$ be less than $35$~GeV.  Furthermore, the
larger and smaller of the lepton transverse momenta are restricted as
follows: $25~\mbox{GeV} < p_{T,{\rm min}}$ and $35~\mbox{GeV} < p_{T,
  {\rm max}} < 50~\mbox{GeV}$.  Finally, a jet-veto is imposed that
removes events with jets where $p_{T,{\rm jet}} > 20$~GeV and
$|\eta_{\rm jet}| < 3$.  Cross sections evaluated with the Higgs
selection cuts will be labeled $\sigma_{\rm bkg}$.

To obtain numerical results we use the following set of input
parameters:
$M_W = 80.419~{\rm GeV}$, 
$M_Z = 91.188~{\rm GeV}$, 
$G_\mu  = 1.16639 \times 10^{-5}~{\rm GeV}^{-2}$,
$\Gamma_W  = 2.06~{\rm GeV}$,
$\Gamma_Z  = 2.49~{\rm GeV}$, and 
$V_{\rm CKM}  = \mathbbm{1}$.
The weak mixing angle is given by $c_{\rm w} = M_W/M_Z,\ s_{\rm w}^2 =
1 - c_{\rm w}^2$.  The electromagnetic coupling is defined in the
$G_\mu$ scheme as $\alpha_{G_\mu} = \sqrt{2}G_\mu M_W^2s_{\rm
  w}^2/\pi$.  The masses of external fermions are neglected. The
values of the heavy quark masses in the intermediate loop are set to
$m_{\rm top} = 178$~GeV and $m_{\rm b} = 4.4$~GeV.  The $pp$ cross
sections are calculated at $\sqrt{s} = 14$~TeV employing the CTEQ6L1
and CTEQ6M \cite{Pumplin:2002vw} parton distribution functions at
tree- and loop-level, corresponding to $\Lambda^{\rm LO}_5 = 165$ MeV
and $\Lambda^{\overline{{\rm MS}}}_5 = 226$ MeV with one- and two-loop
running for $\alpha_s(\mu)$, respectively.  The renormalization and
factorization scales are set to $M_W$.  Fixed-width Breit-Wigner
propagators are used for unstable gauge bosons.

We compare results for ${\rm W^+W^-}$ production in gluon scattering with LO and
NLO results for the quark scattering processes. Since we are
interested in ${\rm W^+W^-}$ production as a background, the ${\rm gg\to h \to W^+W^-}$
signal amplitude is not included. The LO and NLO quark scattering
processes are computed with MCFM \cite{Campbell:1999ah}, which
implements helicity amplitudes with full spin correlations
\cite{Dixon:1998py} and includes finite-width effects and
single-resonant corrections.  Table~\ref{tbl:xsections} shows gluon
and quark scattering cross sections for the LHC.
\begin{table}[hbt]
\caption{\label{tbl:xsections} 
  Cross sections for the gluon and quark scattering contributions to
  ${\rm pp \to W^{\ast}W^{\ast}\to \ell\bar{\nu}\bar{\ell'}\nu'}$ at the LHC
  ($\sqrt{s} = 14$ TeV) without selection cuts (tot), with standard
  LHC cuts (std: $p_{T,\ell} > 20$ GeV, $|\eta_\ell| < 2.5$,
  $\not\!p_T > 25$ GeV) and Higgs search selection cuts (bkg, see
  main text) applied. The integration error is given in brackets. We also 
  show the ratio of the NLO to LO cross sections and the ratio of the 
  combined NLO+$gg$ contribution to the NLO cross section.}
\vspace*{1mm}
\centering{
\def\arraystretch{1.4}
\begin{tabular}{|c|c|cc|c|c|}
 \cline{2-6}
\multicolumn{1}{c|}{} & \multicolumn{5}{c|}{${\rm \sigma(pp \to W^{\ast}W^{\ast}\to
   \ell\bar{\nu}\bar{\ell'}\nu')}$~[fb]} \\ \cline{2-6}
\multicolumn{1}{c|}{} & & 
\multicolumn{2}{c|}{\raisebox{1ex}[-1ex]{$q\bar{q}$}}
& \multicolumn{1}{c|}{} &\multicolumn{1}{c|}{} \\[-1.5ex]
\cline{3-4}
\multicolumn{1}{c|}{} & 
\multicolumn{1}{c|}{\raisebox{2.7ex}[-2ex]{$gg$}} & 
\raisebox{0.9ex}{LO} & \raisebox{0.9ex}{NLO} 
& \raisebox{2.25ex}[-2ex]{$\frac{\sigma_{\rm NLO}}{\sigma_{\rm LO}}$} & 
  \raisebox{2.25ex}[-2ex]{$\frac{
 \sigma_{{\rm NLO}+gg}}{\sigma_{\rm NLO}}$}
\\[-1.5ex]
\hline
 $\sigma_{\rm tot}$ & $60(1)$ & $875.8(1)$ & $1373(1)$ & 1.57 & 1.04 \\
 \hline
 $\sigma_{\rm std}$ & $29.8(6)$ & $270.5(1)$ & $491.8(1)$ & 1.82 & 1.06 \\
 \hline
 $\sigma_{\rm bkg}$ & $1.41(3)$ & $4.583(2)$ & $4.79(3)$ & 1.05 & 1.29 \\
 \hline
\end{tabular}}
\end{table}
% ------------------------------------------------------------------------
Total cross sections ($\sigma_{\rm tot}$) are compared with cross
sections when standard LHC cuts ($\sigma_{\rm std}$) and selection
cuts optimized for Higgs boson searches ($\sigma_{\rm bkg}$) are
applied (see above for the definition of the cuts).  The $gg$ process
only yields a 5\% correction to the total ${\rm W^+W^-}$ cross section
calculated from quark scattering at NLO QCD.  When realistic Higgs
search selection cuts are applied the correction increases to 30\%.
Note that the experimental Higgs search cuts include a jet veto which
suppresses large contributions from gluon-quark scattering at NLO and
thereby reduces the K-factor for ${\rm q\bar{q}\to W^+W^-}$ from 1.6 to 1.1.
For the ${\rm gg \to W^+W^-}$ process we find a renormalization and
factorization scale uncertainty of approximately $25\%$. The scale
uncertainty of the ${\rm q\bar{q}\to W^+W^-}$ process is approximately
5\%~\cite{Binoth:2005ua}.

The massive top-bottom loop increases the result based on intermediate
light quarks~\cite{Binoth:2005ua} by 12\% and 15\% for the inclusive
cross section, $\sigma_{\rm tot}$, and the cross section with standard
cuts, $\sigma_{\rm std}$, respectively.  After imposing Higgs search
cuts, however, the contribution of the massive quark loop is reduced
to 2\% only.  This reduction can largely be attributed to the cut on
$\Delta\phi_{T,\ell\ell}$ as can be seen in Fig.~\ref{fig:etal_delphill} (right)
below.  We note that the impact of the massive quark loop contribution
is mainly due to the interference with the massless loops.

\begin{figure}[htb]
\begin{center}
\epsfig{file=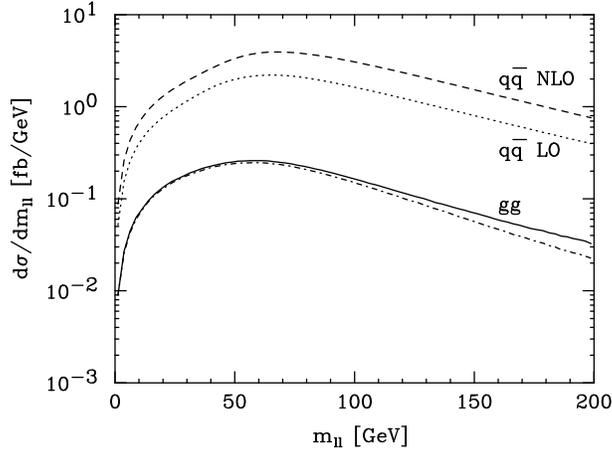,angle=90,width=0.5\textwidth} 
\end{center}
\vspace*{-3mm}
\caption{\label{fig:mll} Distributions in the charged lepton-pair invariant
  mass $m_{\ell\ell}$ for the gluon scattering process (solid) and the
  quark scattering process in LO (dotted) and NLO QCD (dashed) of ${\rm pp
  \to W^{\ast}W^{\ast}\to \ell\bar{\nu}\bar{\ell'}\nu'}$ at the LHC.
  The dashed-dotted line shows the gluon scattering process without
  the intermediate top-bottom loop~\cite{Binoth:2005ua}. The input
  parameters are defined as in the main text. Standard LHC cuts have been
  applied (see main text and Table \protect\ref{tbl:xsections}).}
\end{figure}

\begin{figure}[htb]
\begin{center}
\epsfig{file=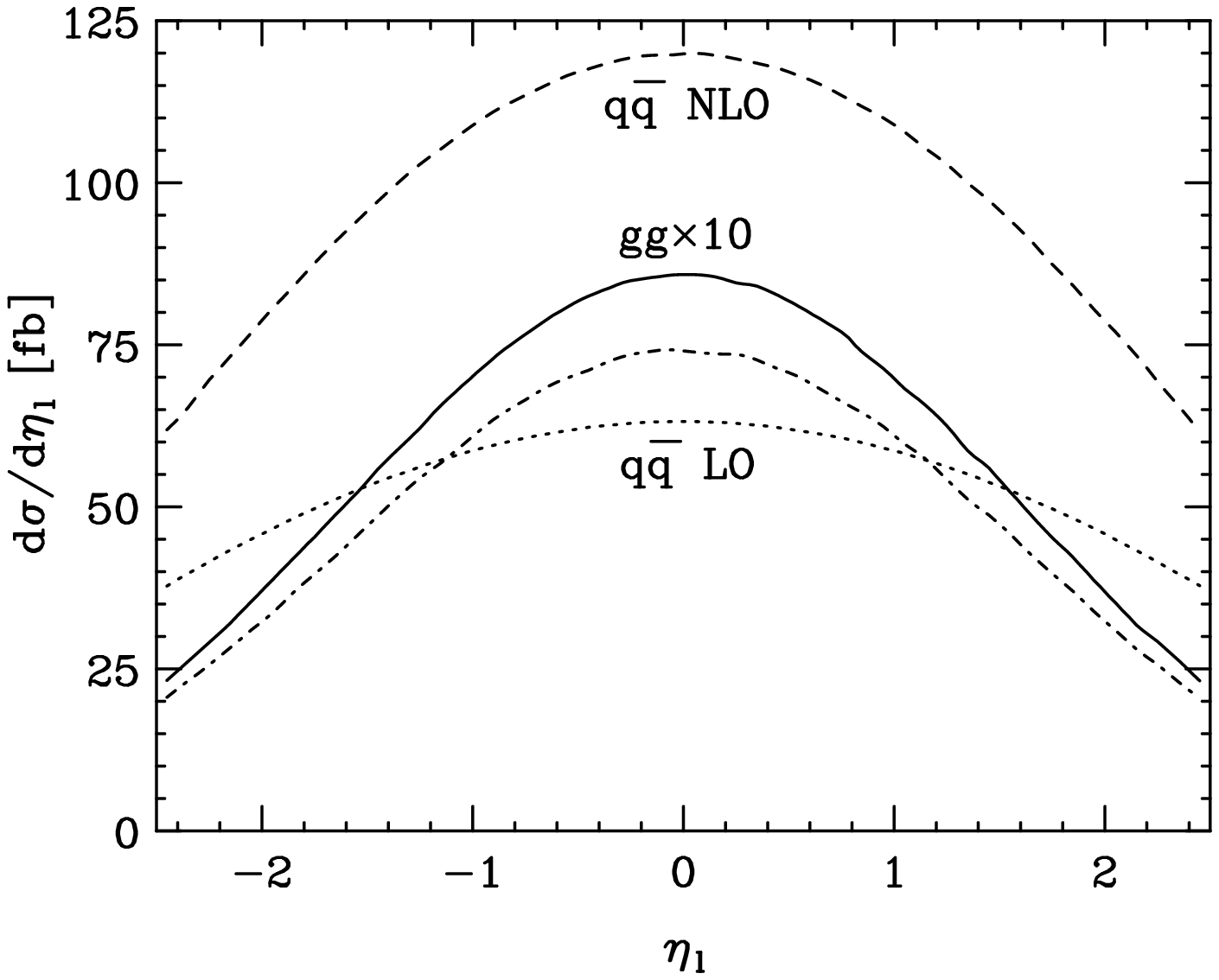,angle=90,width=0.49\textwidth} 
\epsfig{file=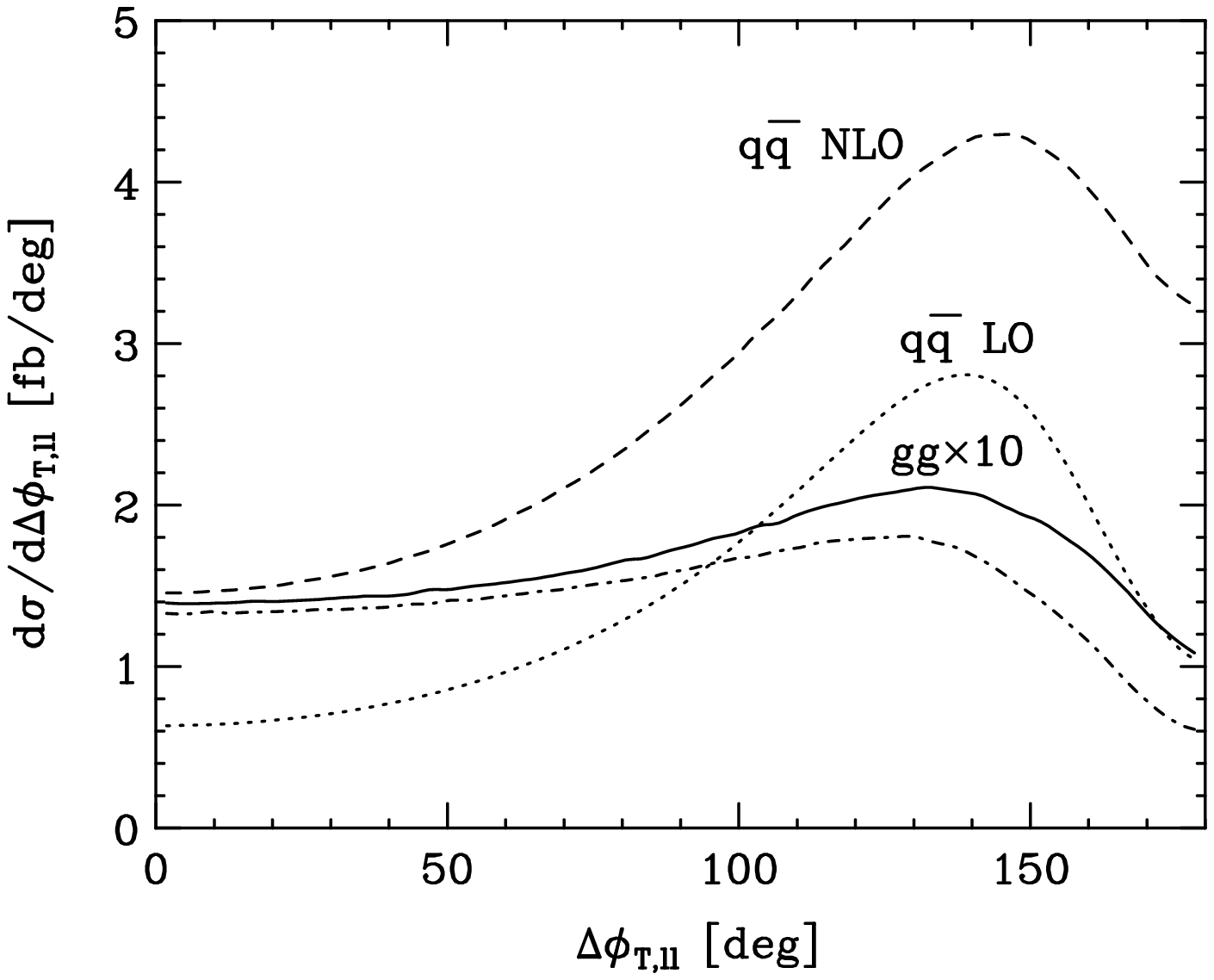,angle=90,width=0.49\textwidth}
\end{center}
\vspace*{-3mm}
\caption{\label{fig:etal_delphill}
  Left: distributions in the pseudorapidity $\eta_{\ell^-}$ of the
  negatively charged lepton. 
  Right: distributions in the transverse-plane opening
  angle of the charged leptons $\Delta\phi_{T,\ell\ell}$.
  Details as in Fig.~\protect\ref{fig:mll}.
  The $gg$ distribution is displayed after multiplication with a
  factor 10.  The dashed-dotted line shows the gluon scattering
  process without the intermediate top-bottom
  loop~\cite{Binoth:2005ua}.}
\end{figure}

Selected differential distributions for ${\rm pp \to W^{\ast}W^{\ast}\to
\ell\bar{\nu}\bar{\ell'}\nu'}$ at the LHC are shown in
Figs.~\ref{fig:mll} and \ref{fig:etal_delphill}. The
standard set of cuts defined above has been applied throughout.
Figure~\ref{fig:mll} shows the distribution in the invariant mass of
the pair of charged leptons. We compare the gluon-gluon induced
contribution with the quark scattering process in LO and NLO. In order
to facilitate the comparison with Ref.\cite{Binoth:2005ua}, the
gluon-fusion cross section is shown with and without the top-bottom
intermediate loop.  We observe that the invariant mass distribution of
the gluon-gluon induced process is similar in shape to the quark
scattering contributions and suppressed by more than one order of
magnitude in normalization.

W-boson pairs produced in quark-antiquark scattering at the LHC are
in general strongly boosted along the beam axis. Gluon induced
processes on the other hand result in ${\rm W^+W^-}$ events at more central
rapidities. This feature is born out by the distribution in the
pseudorapidity of the negatively charged lepton shown in
Fig.~\ref{fig:etal_delphill} (left). In order to distinguish the shapes of the various
contributions we have chosen a linear vertical scale and plot the
gluon-gluon contribution multiplied by a factor~10. Compared to LO
quark-antiquark scattering, the lepton distribution of the gluon-gluon
process shows a more pronounced peak at central rapidities. We also
observe an enhancement of the NLO corrections at central rapidities
which is due to the substantial contribution of gluon-quark processes
at NLO.

Figure~\ref{fig:etal_delphill} (right) finally shows the distribution in the
transverse-plane opening angle of the charged leptons. This observable
reflects the spin correlations between the ${\rm W^+W^-}$ pair and allows one to
discriminate W bosons originating from scalar Higgs decays and ${\rm W^+W^-}$
continuum production. Note that the importance of the gluon-gluon
process is enhanced by the Higgs search selection cuts which require a
small opening angle $\Delta\phi_{T,\ell\ell} < 45^{\circ}$. This
selection cut, on the other hand, reduces the contribution of the
intermediate top-bottom loop to the gluon-fusion cross section.

% ------------------------------------------------------------------------

\subsubsection{Conclusions \label{concl-section}}
We have calculated the loop-induced gluon-fusion process ${\rm gg \to
W^{\ast}W^{\ast} \to \ell\bar{\nu}\bar{\ell'}\nu'}$ which provides an
important background for Higgs boson searches in the ${\rm h \to W^+W^-}$
channel at the LHC. We have presented numerical results for the total
cross section, the cross section with two sets of experimental cuts
and various differential distributions. The results extend our
previous calculation \cite{Binoth:2005ua} by including the
intermediate top-bottom loop. Our calculation demonstrates that the
gluon-fusion contribution to on-shell W pair production only yields
a $5\%$ correction to the inclusive W pair production cross section
at the LHC.  However, after imposing realistic Higgs search selection
cuts, the process ${\rm gg \to W^{\ast}W^{\ast} \to
\ell\bar{\nu}\bar{\ell'}\nu'}$ becomes the dominant higher-order
correction to the ${\rm W^+W^-}$ background estimate and enhances the
theoretical prediction from quark-antiquark scattering at NLO by
approximately $30\%$. We conclude that gluon-gluon induced W pair
production is essential for a reliable description of the background
and has to be taken into account to exploit the discovery potential of
Higgs boson searches in the ${\rm pp\to h\to W^+W^- \to leptons}$ channel
at the LHC.

%==============================================================================
%==============================================================================

\subsection{Effect of parton showering on gluon-induced ${\rm W^+W^-}$ 
production~\protect\footnote{Authors: G.~Davatz, A.-S.~Giolo-Nicollerat, 
M.~Zanetti}}
The main background for the Higgs search decaying in ${\rm W^+ W^-
\rightarrow \ell^+ \nu \ell^- \bar{\nu}}$ is the continuum ${\rm W^+ W^-}$
production, ${\rm q \bar{q}\to W^+W^-}$. Recently a NNLO correction to this
process was calculated, the gluon-induced ${\rm W^+ W^-}$ production, ${\rm g
g \to W^+ W^-}$~\cite{Binoth:2005ua,Duhrssen:2005bz}. This process represents
only a 4\% correction to the inclusive ${\rm W^+ W^-}$ production cross
section at NLO. However, when the selection cuts specific to Higgs search in
the ${\rm W^+ W^-}$ channel are applied, this fraction increases to 30\%. This
is due to the fact that ${\rm g g \to W^+ W^-}$ tends to have leptons emitted
more centrally than continuum ${\rm W^+ W^-}$ production, rendering the Higgs
selection cuts less efficient against this background.

So far ${\rm g g \to W^+ W^-}$ was only studied using a parton-level
generator. In the following the effects of adding a parton shower to this
process will be investigated. The ${\rm g g \to W^+W^-}$ parton-level program
provided by N. Kauer was linked to PYTHIA for the showering step. The W bosons
were then forced to decay into leptons.  The addition of a parton shower is
expected to have mainly an effect on the lepton isolation requirement and on
the jet veto efficiency. In order to study this effect, the initial state radiation
was switched on and off, and the distributions of characteristic variables were compared
 after specific cuts have been applied. 
 The same selection and
reconstruction as in the ``Top background generation in the ${\rm h\to
W^+W^-}$ channel'' chapter of these proceedings was used.

\begin{figure}[htb]
\centering
\includegraphics{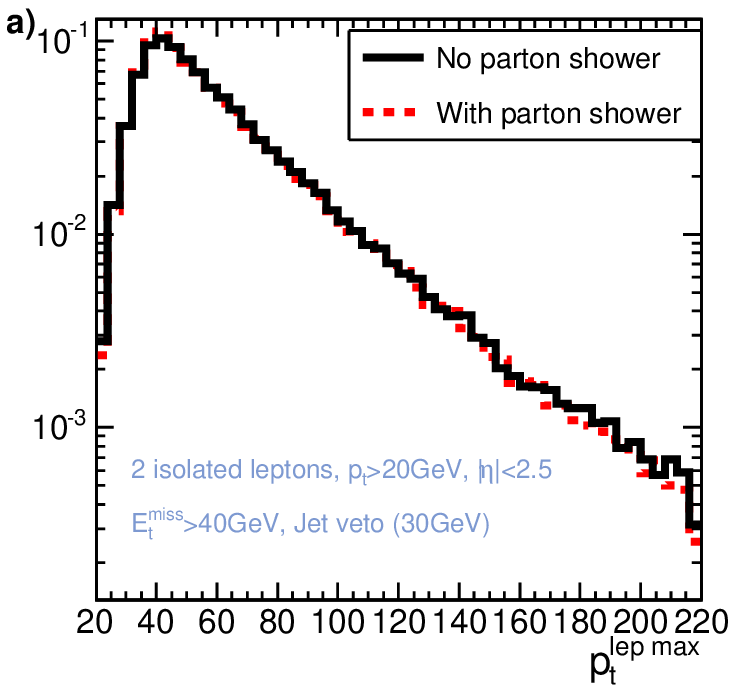}
\includegraphics{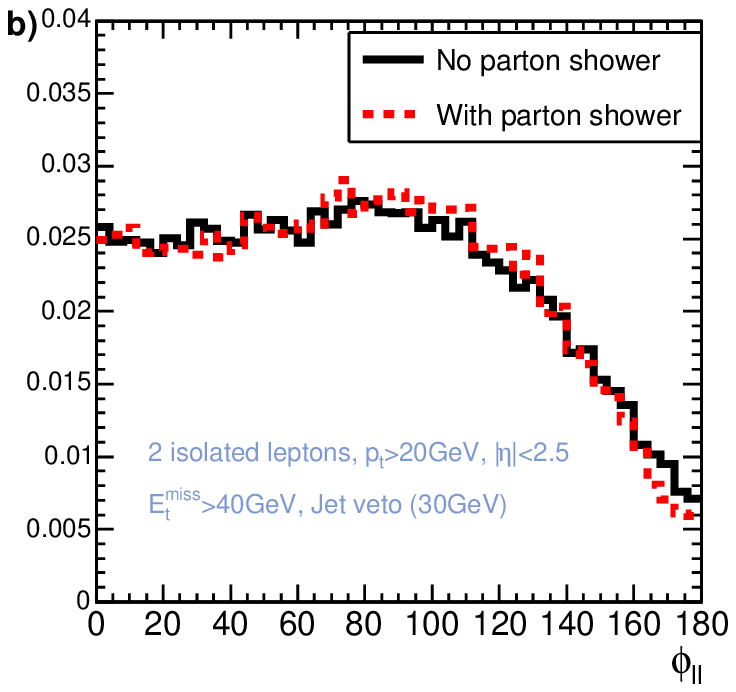}
 \vspace*{-3mm}
\caption{The $p_t$ spectrum of
the lepton with the highest $p_t$ (a) and
the opening angle between the leptons in the transverse plane (b)
for a sample simulated without initial state radiation (black solid line) and with
initial state radiation (red dotted line) for events with two isolated
leptons, a missing energy higher than 40~GeV and no jets with $\rm E_T>$30~GeV.}
\label{phill}
\end{figure}

The addition of the parton shower reduces 
the efficiency of finding two isolated leptons in the final state by 20\%.
Adding a jet veto after all other selection cuts are applied reduces
the total efficiency by 10\%. The changes to the jet veto efficiency
due to the addition of parton shower is thus smaller than its effect
on the lepton isolation.
The shapes of the other cut variables
remain similar with or without initial state radiation. 
Figure~\ref{phill} shows a comparison of the $p_t$
spectrum of the lepton with the highest $p_t$ and the angle
between the leptons in the transverse plane
for a ${\rm g g \to W^+ W^-}$ sample produced
without (black solid line) and with (red dashed line) initial state
radiation. 

Since the gluon-induced ${\rm W^+ W^-}$ production is known at
LO only, it is not possible to know if the parton shower will describe
accurately the inclusion of higher order QCD corrections. Moreover when applying a jet
veto, essentially only the leading order part of this process
remains. For further simulation of this process with parton showering we
recommend to apply a jet veto in order to be in the correct kinematic
region but to set its efficiency to one to take into account the fact
that the contribution from ${\rm g g \to W^+ W^-}$ is only known at LO,
 where no additional jets are expected.

An analysis with detailed CMS detector simulation of this process has been performed.
After all selection cuts for the ${\rm h\to W^+W^-}$ analysis~\cite{Davatz:2006xx},
${\rm g g \to W^+ W^-}$ still represents a contribution of about 30\% to
the continuum W pair production cross section at NLO.

%==============================================================================
%==============================================================================

\subsection{Modeling the production of $W$ pairs~\protect
\footnote{Authors: V.~Drollinger, M.~D\"uhrssen}}

%==============================================================================
\subsubsection{Introduction}

In order to measure the contribution from the ${\rm W^+ W^-}$ background in
the Higgs signal region it is necessary to extrapolate the number of ${\rm W^+
W^-}$ events from an almost pure background region to the signal region. In
general, it is favorable to study the behavior of such an extrapolation from a
clean subsample of the corresponding background events. In this particular
case, no method is known how to isolate ${\rm W^+ W^-}$ background events in a
clean way. The only option left, is to rely on theoretical predictions.

The scope of this work is to study effects which could lead to systematic
uncertainties in the measurement of ${\rm W^+ W^-}$ events. After the events
are generated with various programs, a basic selection of the events is
applied which follows the selection proposed for the ${\rm h \rightarrow W^+
W^- \rightarrow \ell^+ \nu \ell^- \bar{\nu}}$ \cite{Dittmar:1996ss}
analysis. For the selected events a set of kinematic distributions is
compared. In particular, the main attention is drawn to the
$\Delta\phi(\ell^+,\ell^-)$ distribution which is characteristic for both,
signal and background, and therefore is a good candidate for the normalization
of the ${\rm W^+ W^-}$ background.

%------------------------------------------------------------------------------
\paragraph{Event Generators}

There are many programs available to generate ${\rm W^+ W^-}$ events. For
simplicity, only the final state with two muons and two neutrinos is
generated. Different event generators have been employed in order to compare
the kinematic distributions: PYTHIA, CompHEP, MC@NLO, and GG2WW. All
generators can be run with different parameters and options and have different
strengths and weaknesses.

PYTHIA \cite{Sjostrand:2000wi} is a well known program which generates
everything starting from the hard interaction at LO until the complete final
state, including showers and the underlying event, in a self consistent
way. PYTHIA is used to study the effect of spin correlations and scale
dependence.

CompHEP \cite{Boos:2004kh} is an event generator which allows to generate the
hard processes of almost any tree digram. Among the programs considered, it is
the only one which calculates  the full ${\rm 2 \rightarrow 4}$ (two particles in the
initial state and four particles in the final state) matrix elements. The
showering is done in a separate step with PYTHIA.

MC@NLO \cite{Frixione:2002ik,Frixione:2003ei} 
is used to evaluate the effect of higher order
corrections to ${\rm W^+ W^-}$ production as well as the effect of spin
correlations which have been included recently and are available in version
3.10 \cite{Frixione:2005gz}. The events are weighted with constant weights
which differ only in the sign. The showering is performed with HERWIG
\cite{Corcella:2000bw}.

GG2WW~\cite{Binoth:2005xx} is an event generator that generates the hard
process of ${\rm g g \rightarrow W^+ W^-}$ at LO and decays the W bosons. It
has all important features: all six quarks (top and bottom quarks massive), W
bosons are allowed to be off shell, and all spin correlations are taken into
account. Higher order corrections, which are expected to be similar to other
gluon induced processes at the LHC, are unknown at present. On the other hand,
${\rm g g \rightarrow W^+ W^-}$ can be considered as a higher order partonic
sub-process to ${\rm p p \rightarrow W^+ W^-}$ production in general, where
the sub-process ${\rm q\bar{q} \rightarrow W^+ W^-}$ represents the lowest
order.

%------------------------------------------------------------------------------
\paragraph{Event Selection}

In order to be able to compare events in a phase space region that is typical
for an analysis, the pre-selection cuts of the ${\rm h \rightarrow W^+ W^-
\rightarrow \ell^+ \nu \ell^- \bar{\nu}}$ analysis, as suggested in
Ref.~\cite{Dittmar:1996ss} (cuts 1 to 6), are applied. The lepton isolation
(cut 3) is omitted because the leptons from W decays are typically isolated
anyway. The cuts, which will be applied in all cases, are $p_T(\mu^\pm_1) >
{\rm 20\ GeV/c}$ and $p_T(\mu^\pm_2) > {\rm 10\ GeV/c}$, $|\eta(\mu^\pm)| < 2$
for both muons, $m(\mu^+,\mu^-) < {\rm 80\ GeV/c^2}$, $p_T(\mu^+ + \mu^-) >
{\rm 20\ GeV/c}$, and $\Delta\phi(\mu^+,\mu^-) < {\rm 2.4\ rad}$.

The following naming conventions are used: $\mu_1$ is the muon with the
highest $p_T$ in the event, $\mu_2$ is the muon with the second highest $p_T$,
${\rm W_1}$ is the W boson that decays to $\mu_1$, and ${\rm W_2}$ is the W
boson that decays to $\mu_2$. Kinematic distributions for W boson pairs and
muon pairs are compared in the following in case the events have passed the
selection described above. The distributions of W pairs are not accessible
experimentally, but important to understand some event properties. All
distributions are normalized to unit area in order to be able to compare the
shapes of the distributions more easily.

%==============================================================================
\subsubsection{Results and discussion}

In the following, various kinematic properties of W pair events are compared
for the different event generators in the first part. The
$\Delta\phi(\mu^+,\mu^-)$ distribution, which turns out to be the most
sensitive observable, is discussed in more detail in the second and the third
part.

%------------------------------------------------------------------------------
\paragraph{Comparison of Generators}

\begin{figure}[htb]
\centering
\includegraphics[width=0.99\textwidth]{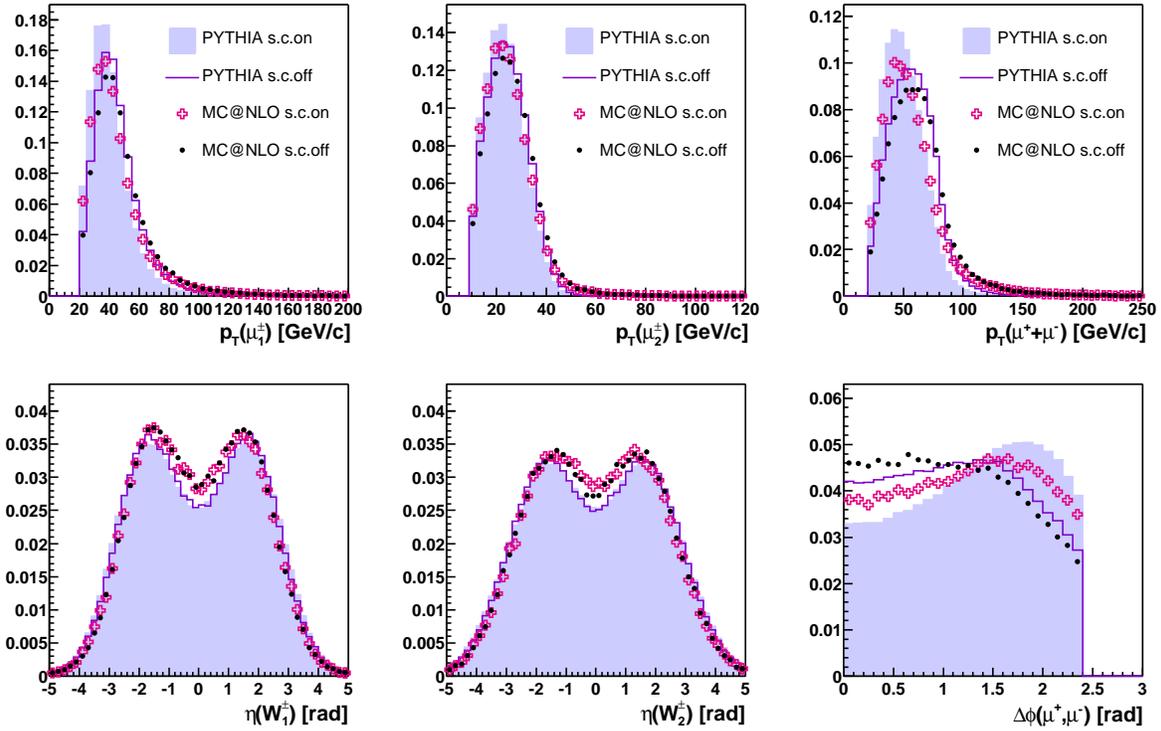}
 \vspace*{-3mm} \caption{Comparison of W boson and muon distributions: there
 is a clear difference visible between the simulation of the hard process at
 LO and NLO. The effect of spin correlations (labelled ``s.c.'') plays an
 important role, too, at LO and at NLO.}
\label{fig:pp2ww_kine}
\end{figure}

In order to get an idea about the importance of HO corrections and spin
correlations, six typical distributions of W pair production are compared in
Fig.~\ref{fig:pp2ww_kine}. Both effects are clearly visible, when MC@NLO and
PYTHIA are compared with spin correlations switched ``on'' and ``off'' in both
cases. There is also difference possible due to the different underlying
events of MC@NLO (HERWIG) and PYTHIA. This difference is expected to
disappear, after both event generators have been tuned consistently to LHC
data. Recently, the soft gluons have been resummed up to NNLL for W pair
production at NLO in Ref.~\cite{Grazzini:2005vw}, summarized in section \ref{sec:pp2ww_softGlu}.
 The result is in good agreement with the prediction of MC@NLO.
The approximation of generating W pairs in a $2 \rightarrow 2$ process and
then performing the W decays in an separate step has no visible effect on the
kinematic distributions, when PYTHIA is compared with CompHEP which computes
the full $2 \rightarrow 4$ matrix element. Even the fact that contributions
from Feynman diagrams with only one W boson are not present in case of PYTHIA
does not lead to recognizable differences, because the contribution from
processes with only one W boson is strongly suppressed.

In this study, no emphasis is placed on the comparison of the cross sections
which are summarized in Table~\ref{tab:wqeff} before and after the event
selection. The selection efficiencies give an idea about the quantitative
differences of the various simulations. These differences should not be
understood as the uncertainty of the W pair production process, because it
becomes clear that a proper choice of the event generator, in this case MC@NLO
(with spin correlations), can describe all important features with a better
accuracy than the differences between the scenarios investigated.
\renewcommand{\arraystretch}{1.1}
\begin{table}[htb]
 \caption{\sl Cross sections and selection efficiencies for the scenarios considered.\rm}
  \vspace*{1mm}
 \centering
 \begin{tabular}{l|c|c|c}
   program and setup &  total ${\rm \sigma \times BR}$ & ${\rm \sigma \times BR}$ after selection & selection efficiency \\
   \hline
   PYTHIA: spin correlations ``on''                               &  ~~828 fb  &  122 fb  &  14.7\% \\
   PYTHIA: spin correlations ``off''                              &  ~~828 fb  &  137 fb  &  16.5\% \\
   CompHEP: ${\rm qq \rightarrow WW \rightarrow 2\mu 2\nu}$       &  ~~900 fb  &  127 fb  &  14.1\% \\
   CompHEP: ${\rm qq \rightarrow W\mu\nu \rightarrow 2\mu 2\nu}$  &  ~~900 fb  &  127 fb  &  14.1\% \\
   MC@NLO: spin correlations ``on''                               &   1287 fb  &  206 fb  &  16.0\% \\
   MC@NLO: spin correlations ``off''                              &   1287 fb  &  212 fb  &  16.5\% \\
   \hline
   GG2WW: six quarks, ${\rm W}$s off shell                        & ~~~~60 fb  & ~~19 fb  &  31.4\% \\
 \end{tabular}
\label{tab:wqeff}
\end{table}
\renewcommand{\arraystretch}{1.0}

%------------------------------------------------------------------------------
\paragraph{Theoretical Uncertainties at NLO}
\label{sec:pp2ww_UncertNLO}

\begin{figure}[htb]
\centering
\includegraphics[width=0.49\textwidth]{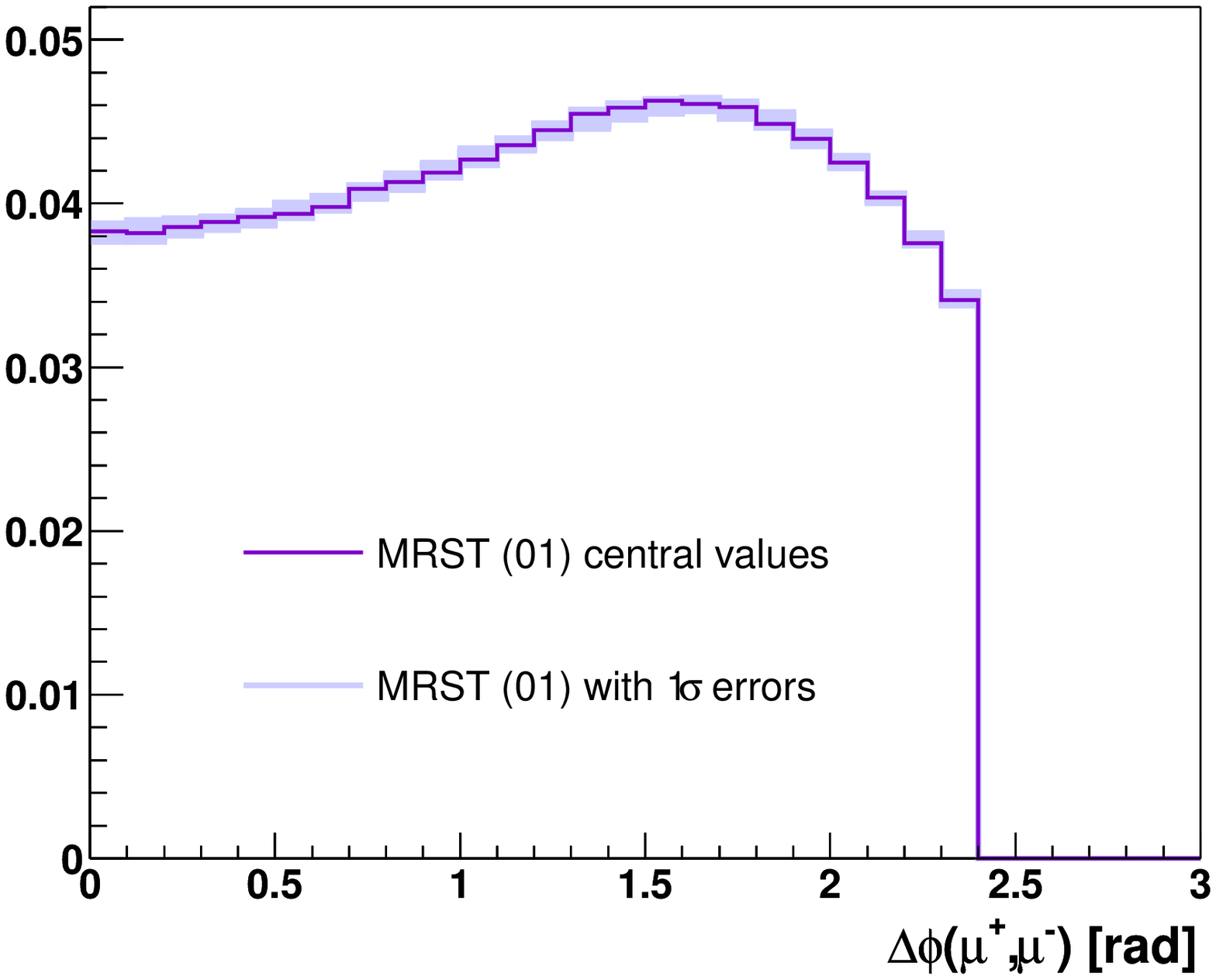}
\includegraphics[width=0.49\textwidth]{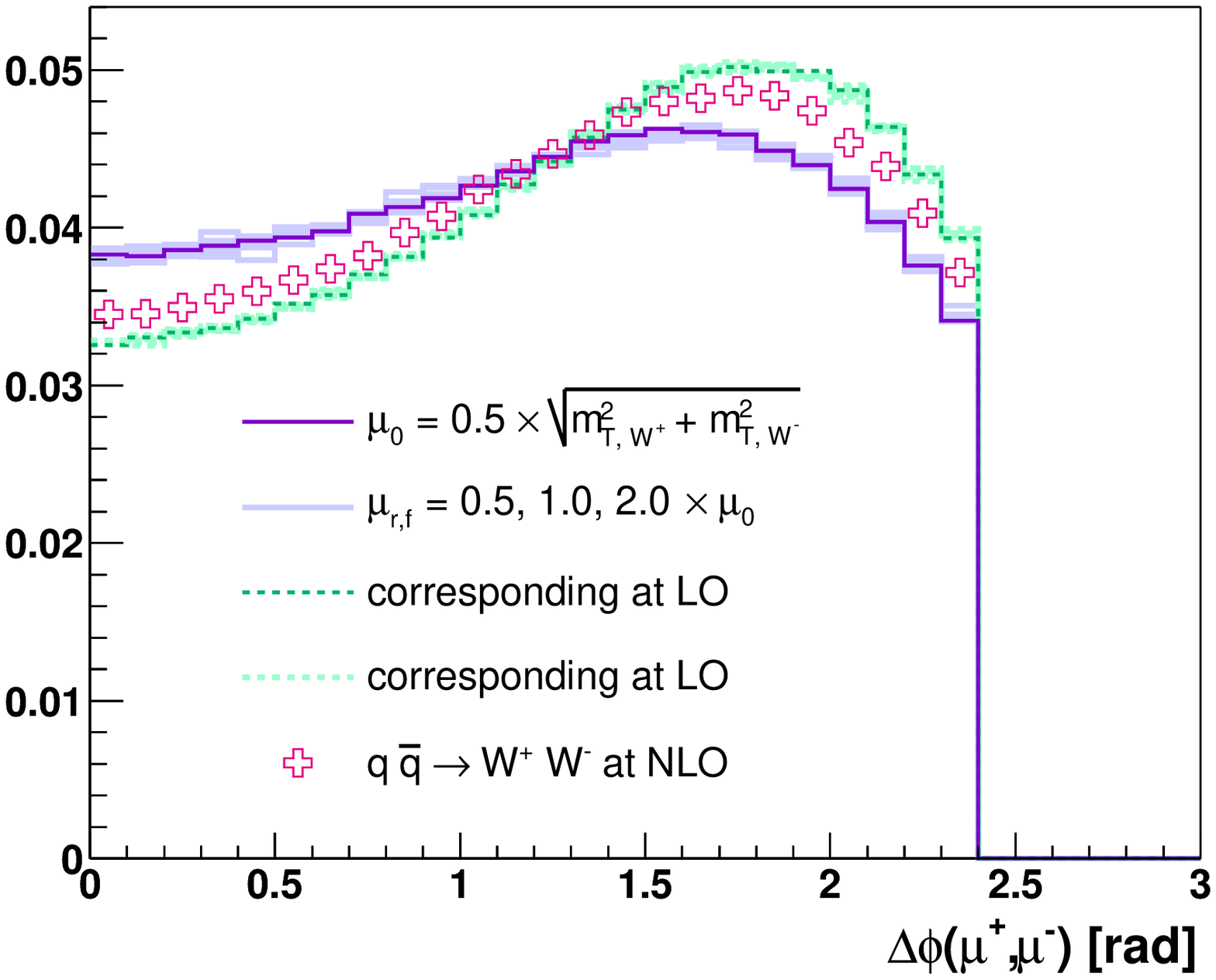}
 \vspace*{-3mm}
 \caption{Left: PDF uncertainties with MC@NLO: shown are the nominal MRST
structure functions with one set of nominal parameters and with 30 sets of
parameters where one parameter is varied by one standard deviation at a time.
Right: Scale uncertainties with MC@NLO: the scales ${\rm \mu_r}$ and ${\rm
\mu_f}$ are varied independently by factors of two with respect to the nominal
scale $\mu_0$. The scale dependence at LO is studied correspondingly with
PYTHIA, which yields shapes close to ${q \bar{q} \rightarrow W^+ W^-}$ at
NLO. Note that there are small statistical fluctuations which contribute to
the width of the error bands, too.}
\label{fig:pp2ww_sys}
\end{figure}

Once all features are included in an event generator - in this case MC@NLO -
the shape of $\Delta\phi(\mu^+,\mu^-)$ is still not perfectly known. There are
two theoretical uncertainties than can be potentially large in this particular
example. In order to study the PDF uncertainty, each PDF parameter is varied
independently by one standard deviation. In case of the structure function
MRS, this leads to thirty error PDFs. The $\Delta\phi(\mu^+,\mu^-)$ shape
variation, shown in Fig.~\ref{fig:pp2ww_sys} (left), turns out to be small.

In order to study the uncertainties due to higher order corrections, often the
scale dependence of the renormalization scale ${\rm \mu_r}$ and factorization
scale ${\rm \mu_f}$ are varied independently by a factor of two in both
directions, which results in nine different scales including the nominal
scale. The shape of the $\Delta\phi(\mu^+,\mu^-)$ distribution of MC@NLO is
stable against the variation of the scales as shown in
Fig.~\ref{fig:pp2ww_sys} (right). But this is not only the case at NLO. The
shape variation is similar at LO, even though both shapes are different. What
is the reason for this discrepancy? Whereas initial states at LO is always
${\rm q \bar{q}}$, the possible initial states at NLO are ${\rm q \bar{q}}$,
${\rm q g}$ and ${\rm g \bar{q}}$. The events with gluons in the initial state
have different spin properties which are visible in the
$\Delta\phi(\mu^+,\mu^-)$ distribution. A cross check for this hypothesis is
also shown in Fig.~\ref{fig:pp2ww_sys} (right), where the NLO events with
${\rm q \bar{q}}$ initial state are shown separately. This shape is remarkably
close\footnote{There is a difference visible which is due to the different
underlying events, and the somewhat different $p_T$ spectra of the ${W^+ W^-}$
system, respectively.} to the shape at LO. In conclusion, the new partonic
process coming in at NLO is not reflected by the scale variations, which means
that the method of changing the scales does not work in this particular
case. In order to study the scale uncertainties in a reliable way, all
partonic sub-processes have to be taken into account, which means that in case
of W pair production the contribution from the ${\rm g g}$ initial state has
to be added.

%------------------------------------------------------------------------------
\paragraph{Contribution from ${\rm g g \rightarrow W^+ W^-}$}

\begin{figure}[htb]
\centering
\includegraphics[width=0.49\textwidth]{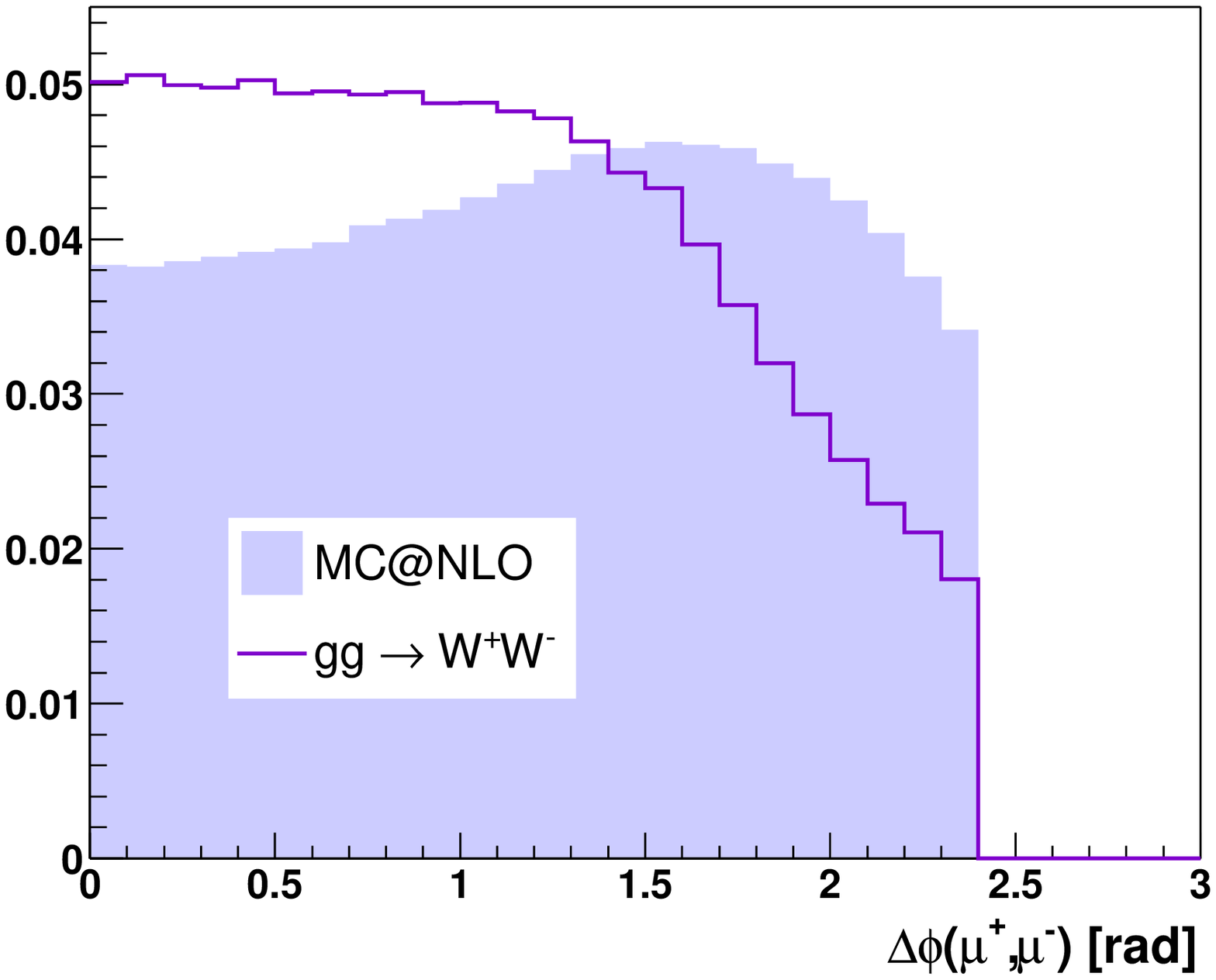}
\includegraphics[width=0.49\textwidth]{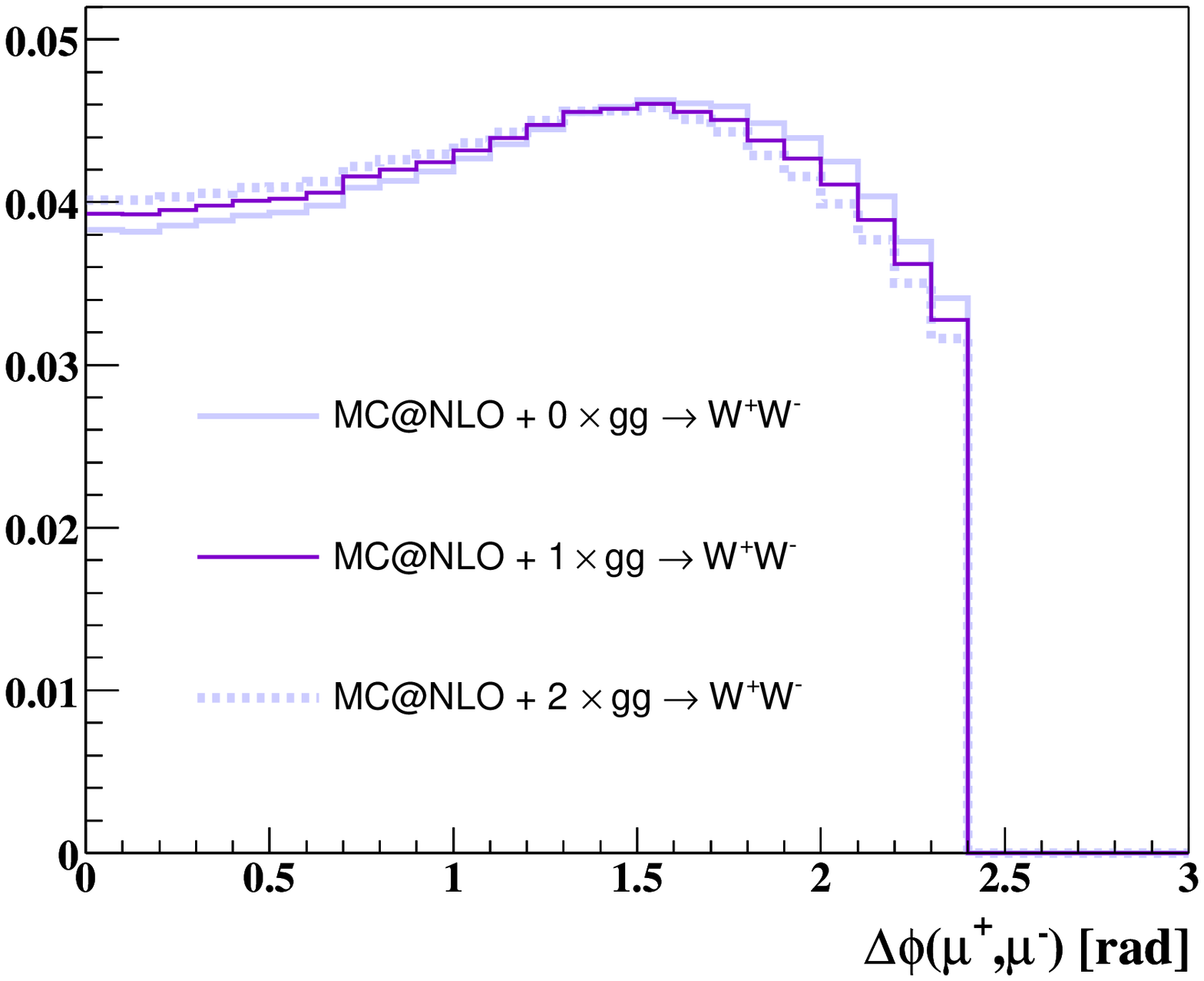}
 \vspace*{-3mm}
 \caption{Left: Shape comparison of $g g \rightarrow W^+ W^-$ (GG2WW) events
with MC@NLO, which includes the partonic sub-processes $q \bar{q} \rightarrow
W^+ W^-$, $q g \rightarrow W^+ W^-$ and $\bar{q} g \rightarrow W^+ W^-$.
Right: $p p \rightarrow W^+ W^-$ with all partonic sub-processes. Three
scenarios for the partonic sub-process $g g \rightarrow W^+ W^-$ are assumed:
zero times the nominal LO cross section, one times the nominal LO cross
section, and two times the nominal LO cross section.}
\label{fig:pp2ww_sum}
\end{figure}

Compared to the process ${\rm q \bar{q} \rightarrow W^+ W^-}$ the cross
section for ${\rm g g \rightarrow W^+ W^-}$ is more than an order of magnitude
smaller. On the other hand, the latter process has a higher selection
efficiency, as shown in Table~\ref{tab:wqeff}, and becomes more important
after cuts relative to ${\rm q \bar{q} \rightarrow W^+ W^-}$. In
Fig.~\ref{fig:pp2ww_sum} (left) the shape of ${\rm g g \rightarrow W^+ W^-
\rightarrow \mu^+ \nu \mu^- \bar{\nu}}$ events is compared with the
corresponding distribution from MC@NLO. There is a clear difference visible,
which means that the addition of the ${\rm g g \rightarrow W^+ W^-}$
contribution changes the shape of the sum of all partonic sub-processes.

In order to get the whole picture the ${\rm g g \rightarrow W^+ W^-}$ events
are added to the events, generated with MC@NLO, taking into account the
different cross sections after cuts. Three scaling factors for ${\rm g g
\rightarrow W^+ W^-}$ are shown in Fig.~\ref{fig:pp2ww_sum} (right), namely
${\rm k_{g g \rightarrow W^+ W^-} = 0, 1, 2}$. The largest deviation from the
nominal (${\rm k_{g g \rightarrow W^+ W^-} = 1}$) shape is the shape without
contribution from ${\rm g g \rightarrow W^+ W^-}$. In order to get a feeling
for the uncertainty due to the missing NLO corrections for ${\rm g g
\rightarrow W^+ W^-}$, this sub-process is multiplied with a factor of two,
which is roughly the k-factor of ${\rm g g \rightarrow h}$, another process
with two gluons in the initial state. In a comparable case, ${\rm g g
\rightarrow \gamma \gamma}$ at NLO~\cite{Bern:2002jx}, the k-factor turns out
to be even lower than two. The change of the shape of the
$\Delta\phi(\mu^+,\mu^-)$ distribution due to the ${\rm g g \rightarrow W^+
W^-}$ contribution is small for this particular selection, but it has to be
kept in mind that the fraction of this sub-process can be enhanced further
with additional cuts.

%------------------------------------------------------------------------------
\subsubsection{Analyzing the transverse Mass Distribution of W Pairs}

Besides the transverse opening angle $\Delta\phi(\ell^+,\ell^-)$ also the
distribution of the pseudorapidity of the sum of the two leptons $\eta(\ell^+
+ \ell^-)$ and the transverse mass $M_T$ promise some discrimination power
between a Higgs boson signal and the $\rm W^+W^-$ background (The invariant
mass $m(\ell^+ + \ell^-)$ of the two leptons is strongly correlated to
$\Delta\phi(\ell^+,\ell^-)$ and not used here). For simplicity only relaxed
pre-selection cuts are applied: two leptons ($e^\pm$ or $\mu^\pm$) with
\begin{itemize}
  \item $p_T(\ell^\pm_1) > {\rm 20\ GeV/c}$ and $p_T(\ell^\pm_2) > {\rm 10\ GeV/c}$
  \item $|\eta(\ell^\pm)| < 2.5$ 
  \item $p_T^{\rm{miss}}=p_T(\nu_1+\nu_2) > {\rm 20\ GeV/c}$ .
\end{itemize} 
The usual definition of the transverse mass is $M_T=\sqrt{2 \cdot p_T(\ell^+ +
\ell^-) \cdot p_T^{\rm{miss}}\cdot (1-\cos\phi_T)}$ for the $\rm h \to W^+W^-$
analysis, where $\phi_T$ is the transverse opening angle between the vector
sum of the two leptons and $p_T^{\rm{miss}}$. This definition works best if
the two leptons are almost collinear. For increasing opening angle between the
leptons, $M_T$ decreases since both $p_T(\ell^+ + \ell^-)$ and
$p_T^{\rm{miss}}$ get smaller. To compensate for this the definition
$M^\prime_T=\sqrt{M_T^2+m^2(\ell^+ + \ell^-)}$ is used rather than $M_T$ .

%------------------------------------------------------------------------------
\paragraph{Defining signal and normalization regions}

\begin{figure}[htb]
\centering
\includegraphics[width=0.325\textwidth]{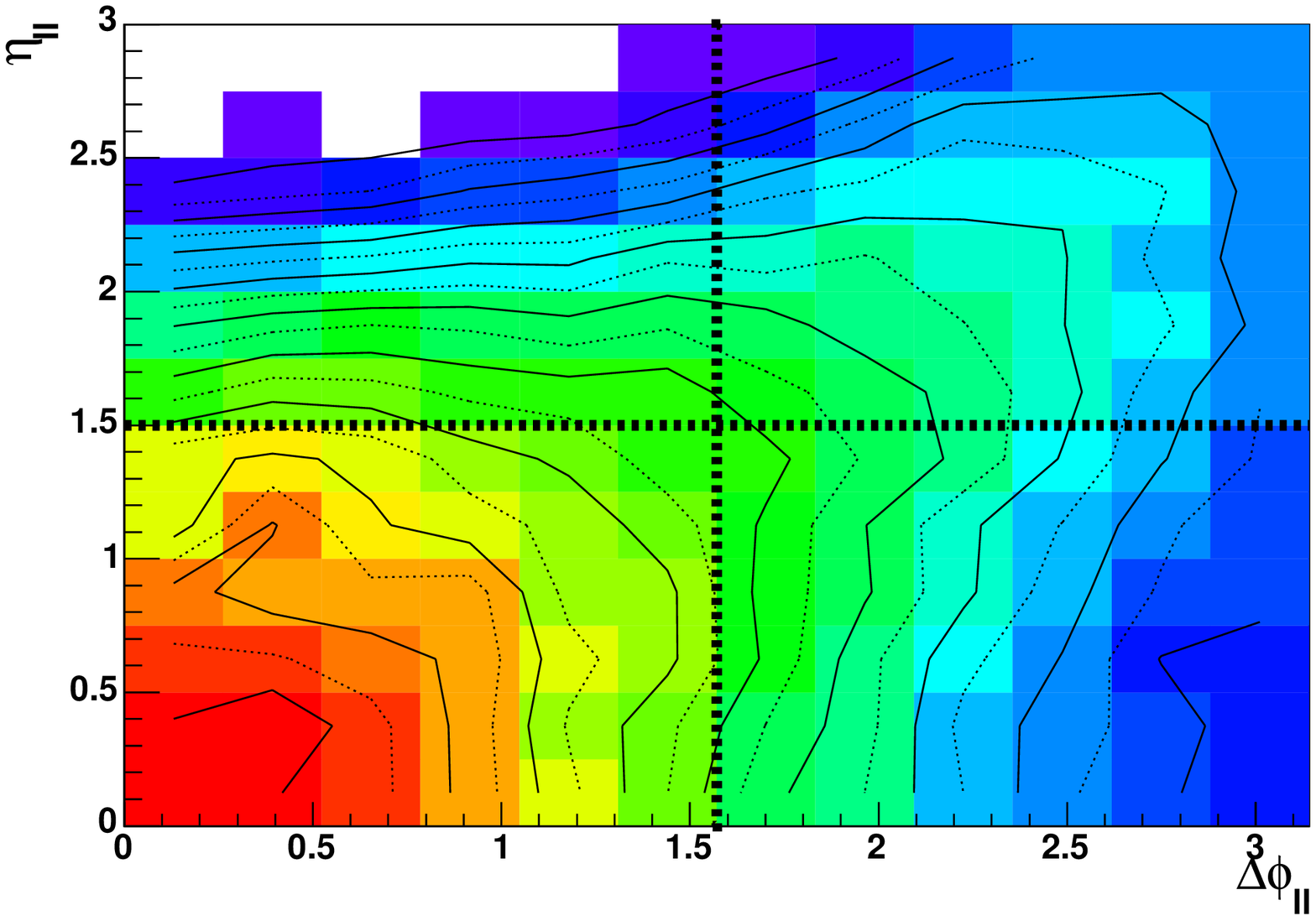}
\includegraphics[width=0.325\textwidth]{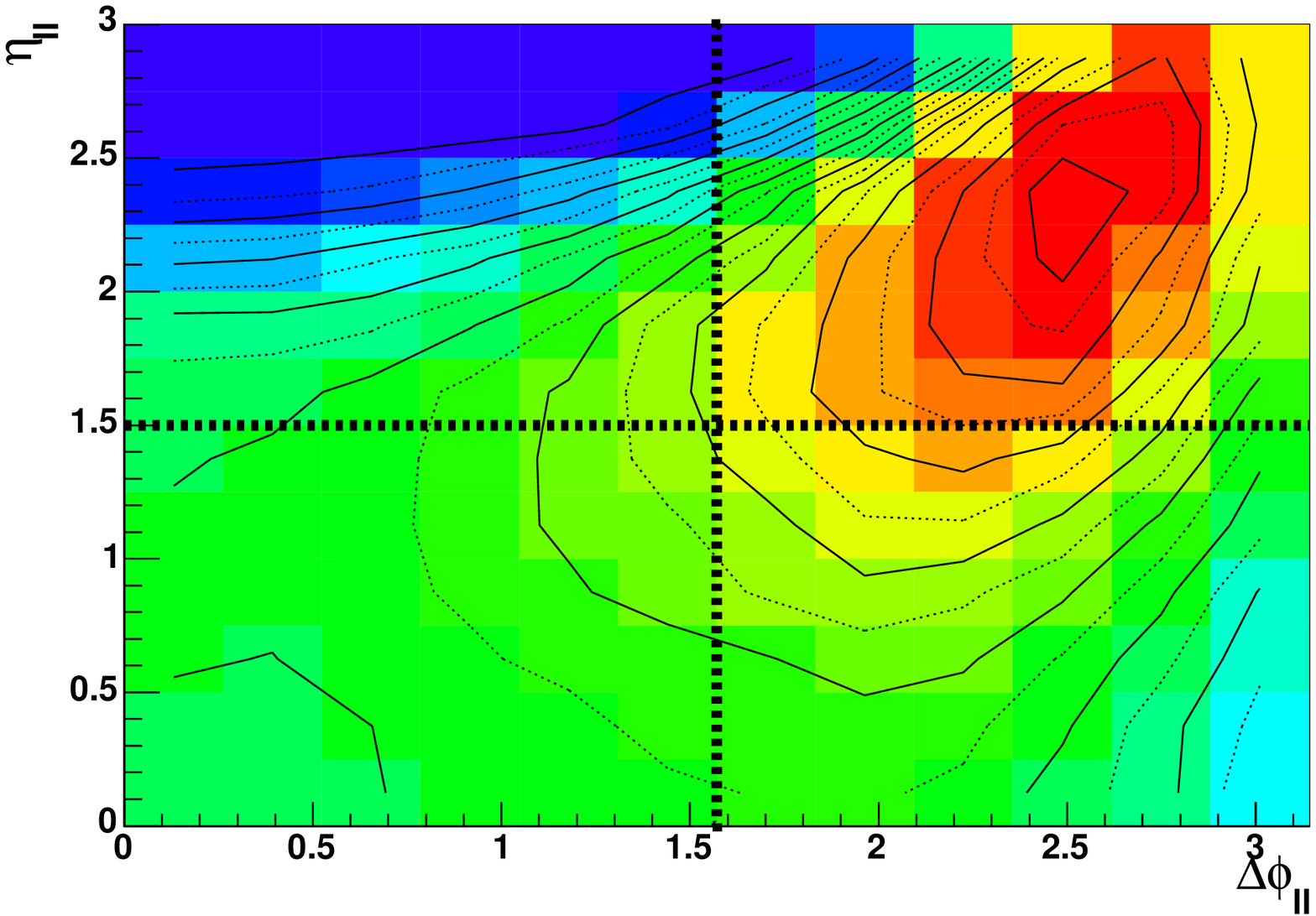}
\includegraphics[width=0.325\textwidth]{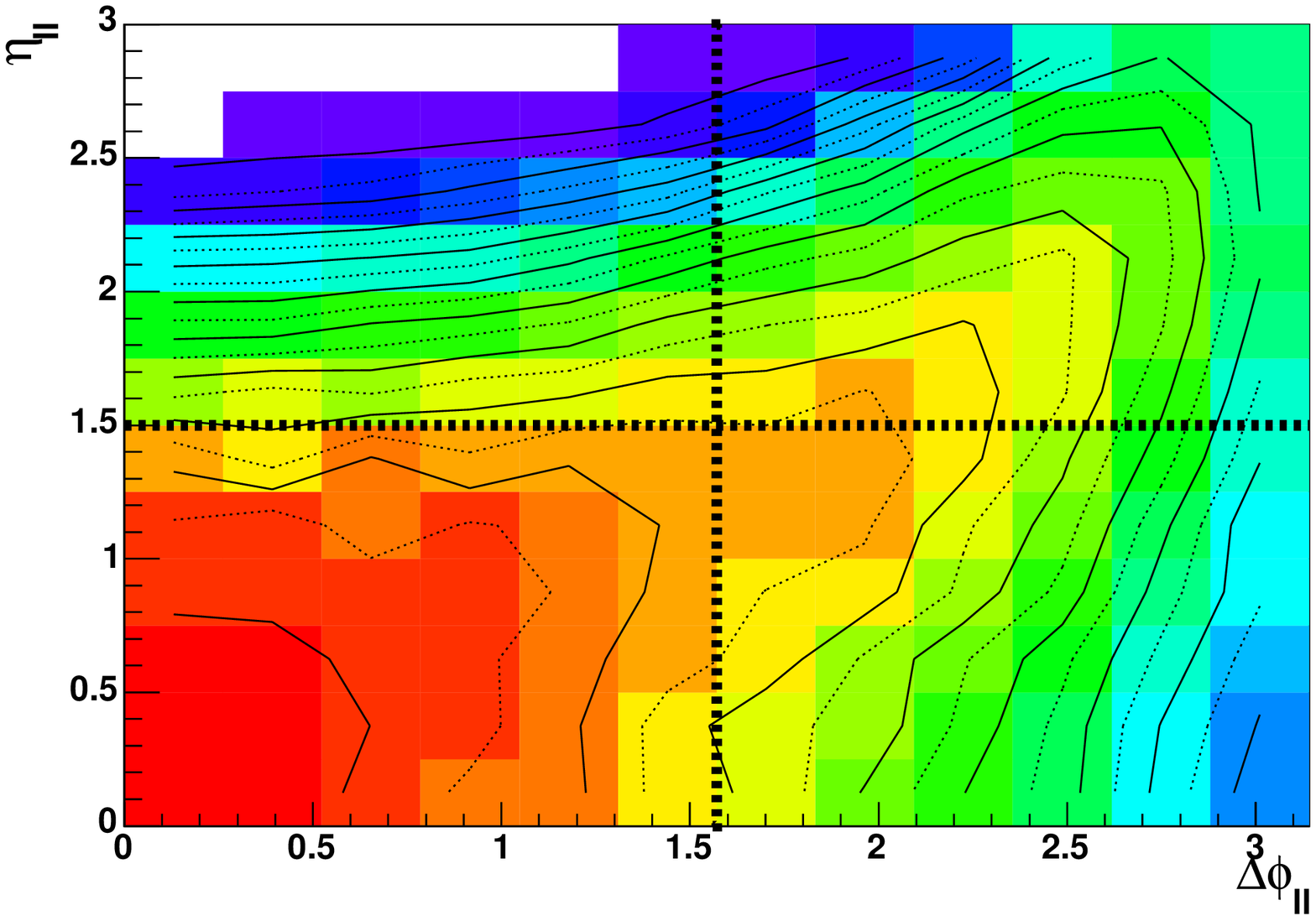}
 \vspace*{-3mm}
 \caption{\label{fig:pp2ww_etaphi} Event distribution for $\rm W^+W^- \to
          \ell^+\nu \, \ell^-\bar{\nu}$ in the plane of $\eta_{ll}=\eta(\ell^+
          + \ell^-)$ and $\Delta\phi_{ll}=\Delta\phi(\ell^+,\ell^-)$. For the
          normalization of the $M^\prime_T$ shape the plane is split into four
          regions at $\eta(\ell^+ + \ell^-)=1.5$ and
          $\Delta\phi(\ell^+,\ell^-)=\frac{\pi}{2}$.\newline Left: $\rm gg \to
          h \to W^+W^-$ (MC@NLO).  Center: $\rm q\bar{q} \to W^+W^-$ (MC@NLO).
          Right:$\,\,gg \to W^+W^-$ (GG2WW).}
\end{figure}

In Fig. \ref{fig:pp2ww_etaphi} the distribution of events in the plane of
$\eta(\ell^+ + \ell^-)$ vs. $\Delta\phi(\ell^+,\ell^-)$ is shown for a Higgs
boson of 170 GeV (left), $\rm q\bar{q} \to W^+W^-$ (center) and $\rm gg \to
W^+W^-$ (right).  A potential Higgs boson signal would appear dominantly in
the signal region $\eta(\ell^+ + \ell^-)<1.5$ and
$\Delta\phi(\ell^+,\ell^-)<\frac{\pi}{2}$ (very similar to $\rm gg \to
W^+W^-$).  On the other hand the dominant $\rm q\bar{q} \to W^+ W^-$
background is found in the normalization regions $\eta(\ell^+ + \ell^-)>1.5$
or $\Delta\phi(\ell^+,\ell^-)>\frac{\pi}{2}$.

The aim is to normalize the shape of $M^\prime_T$ for $\rm q\bar{q} \to
W^+W^-$ in the signal region using the shape of $M^\prime_T$ in the
normalization region. Such a shape normalization to data has the advantage
that experimental uncertainties that enter $M_T$ due to the missing transverse
momentum can be reduced in the ratio.

For a systematic comparison of the $M^\prime_T$ shapes the $\eta(\ell^+ +
\ell^-)$ vs. $\Delta\phi(\ell^+,\ell^-)$ plane is split into four regions at
$\eta(\ell^+ + \ell^-)=1.5$ and $\Delta\phi(\ell^+,\ell^-)=\pi/2$.

%------------------------------------------------------------------------------
\paragraph{Normalizing the ${M^\prime_T}$ shape for ${\rm q\bar{q} \to W^+W^-}$}

\begin{figure}[htb]
\centering
\includegraphics[width=0.49\textwidth]{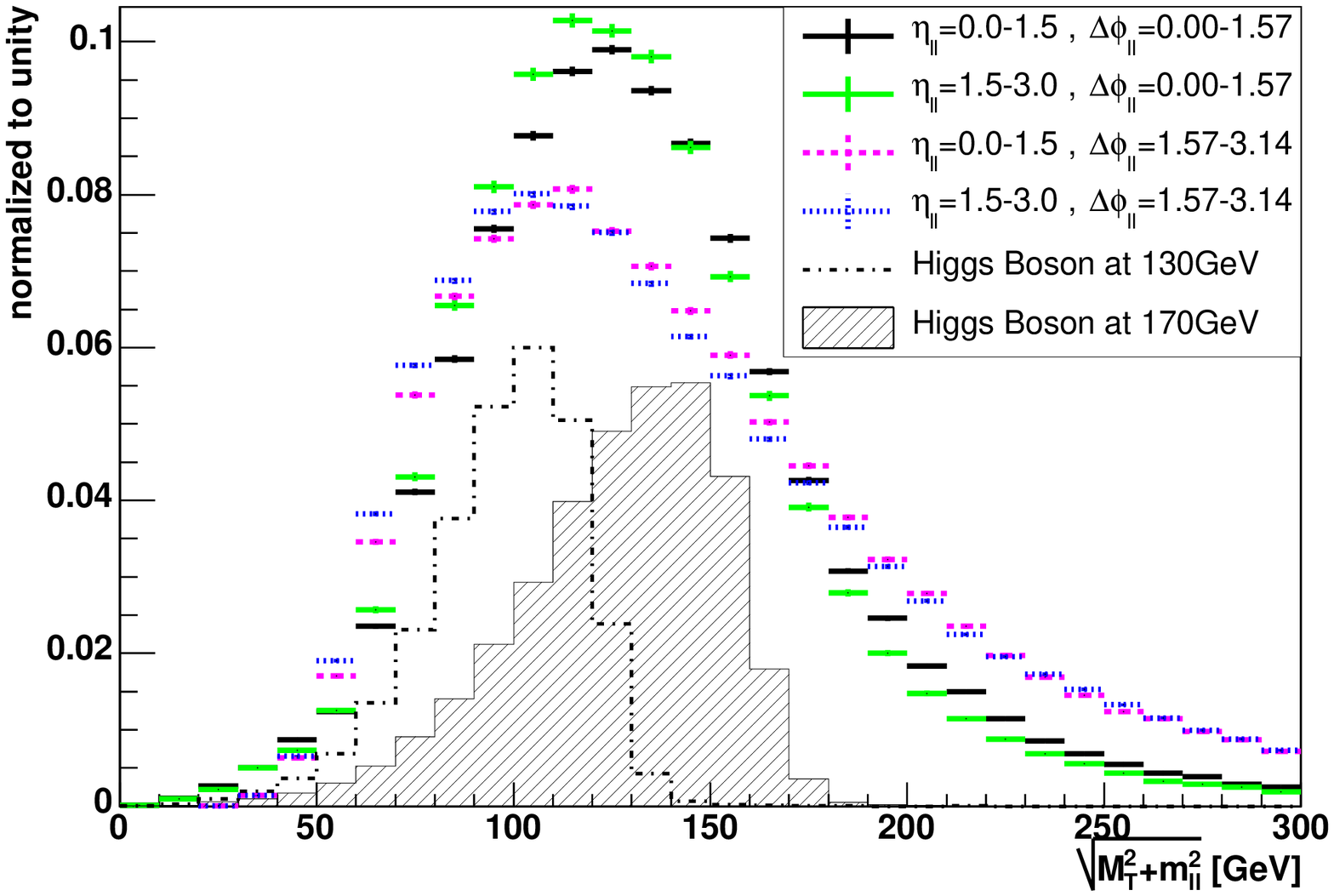}
\includegraphics[width=0.49\textwidth]{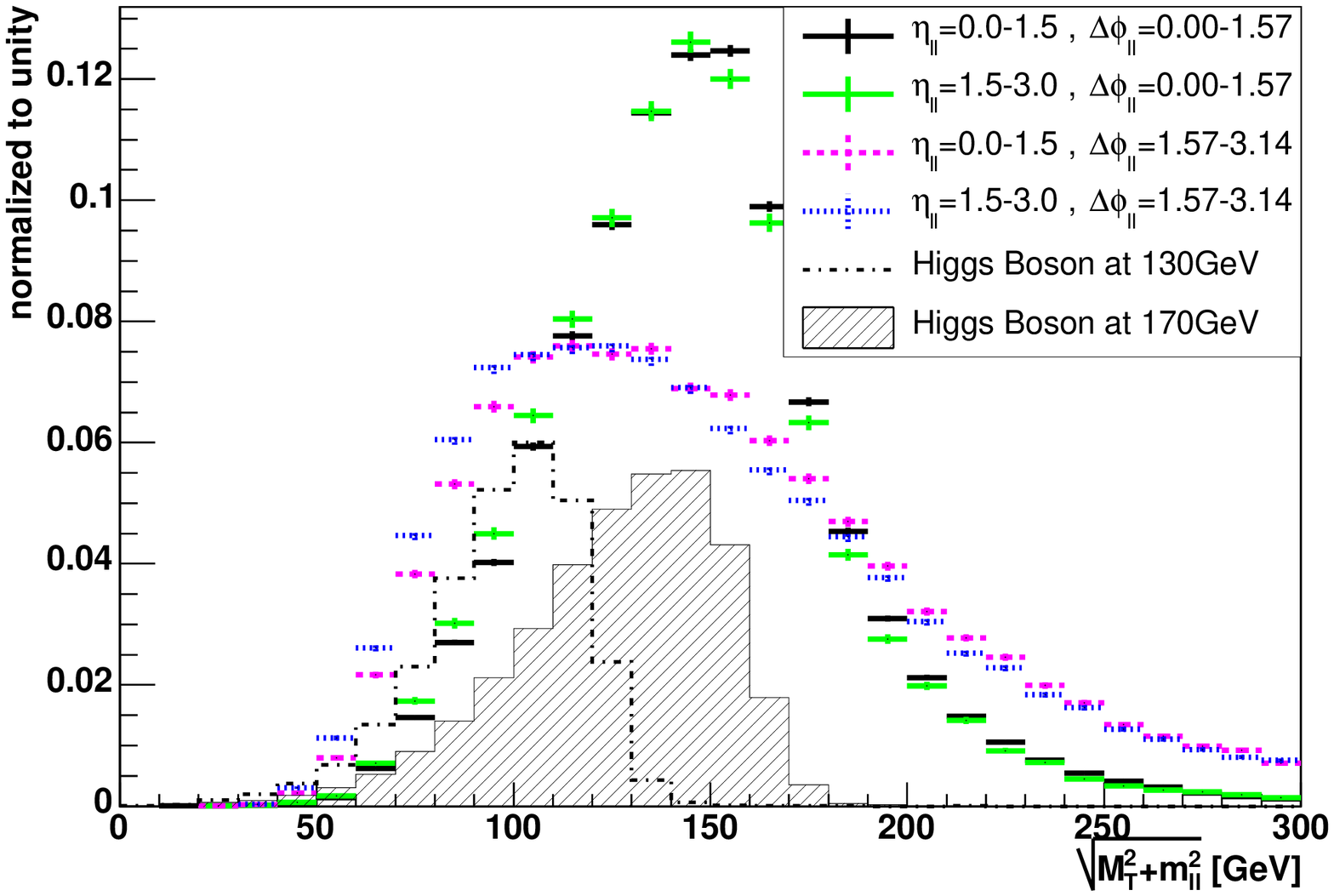}
 \vspace*{-3mm}
 \caption{Distribution of $M^\prime_T$ for $\rm q\bar{q}\to W^+W^-$ (left) and
 $\rm gg\to W^+W^-$ (right) in four different regions of the $\eta(\ell^+ +
 \ell^-)$ vs. $\Delta\phi(\ell^+,\ell^-)$ plane of Fig. \ref{fig:pp2ww_etaphi}
 (normalized to unity). The error bars give the Monte Carlo statistical error
 on the points. For comparison the dashed-dotted and the hashed histogram give
 the shape of $\rm h \to W^+W^-$ events in the signal region (~$\eta(\ell^+ +
 \ell^-)<1.5$, $\Delta\phi(\ell^+,\ell^-)<\frac{\pi}{2}$~), arbitrary
 normalization).}
\label{fig:pp2ww_MT}
\end{figure}

The shape of $M^\prime_T$ for $\rm q\bar{q} \to W^+W^-$ and $\rm gg \to
W^+W^-$ is shown in Fig. \ref{fig:pp2ww_MT} for each region of
Fig. \ref{fig:pp2ww_etaphi} (histograms normalized to unity). For additional
comparison the shape of $M^\prime_T$ in the signal region is shown for Higgs
boson events. Also in this shape $\rm gg \to W^+W^-$ is very similar to a
Higgs boson with a mass close to 170 GeV.

\begin{figure}[htb]
\centering
\includegraphics[width=0.325\textwidth]{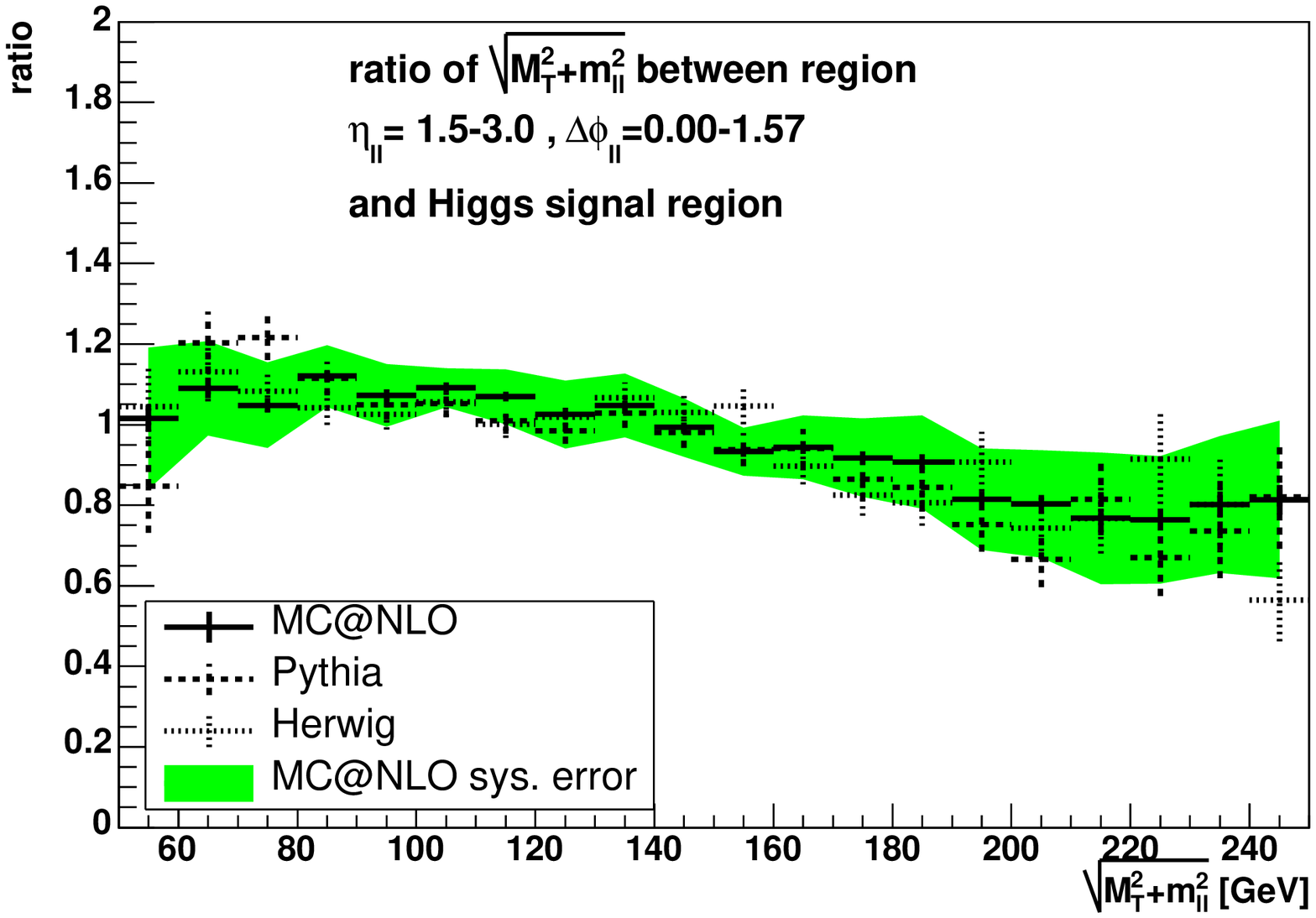}
\includegraphics[width=0.325\textwidth]{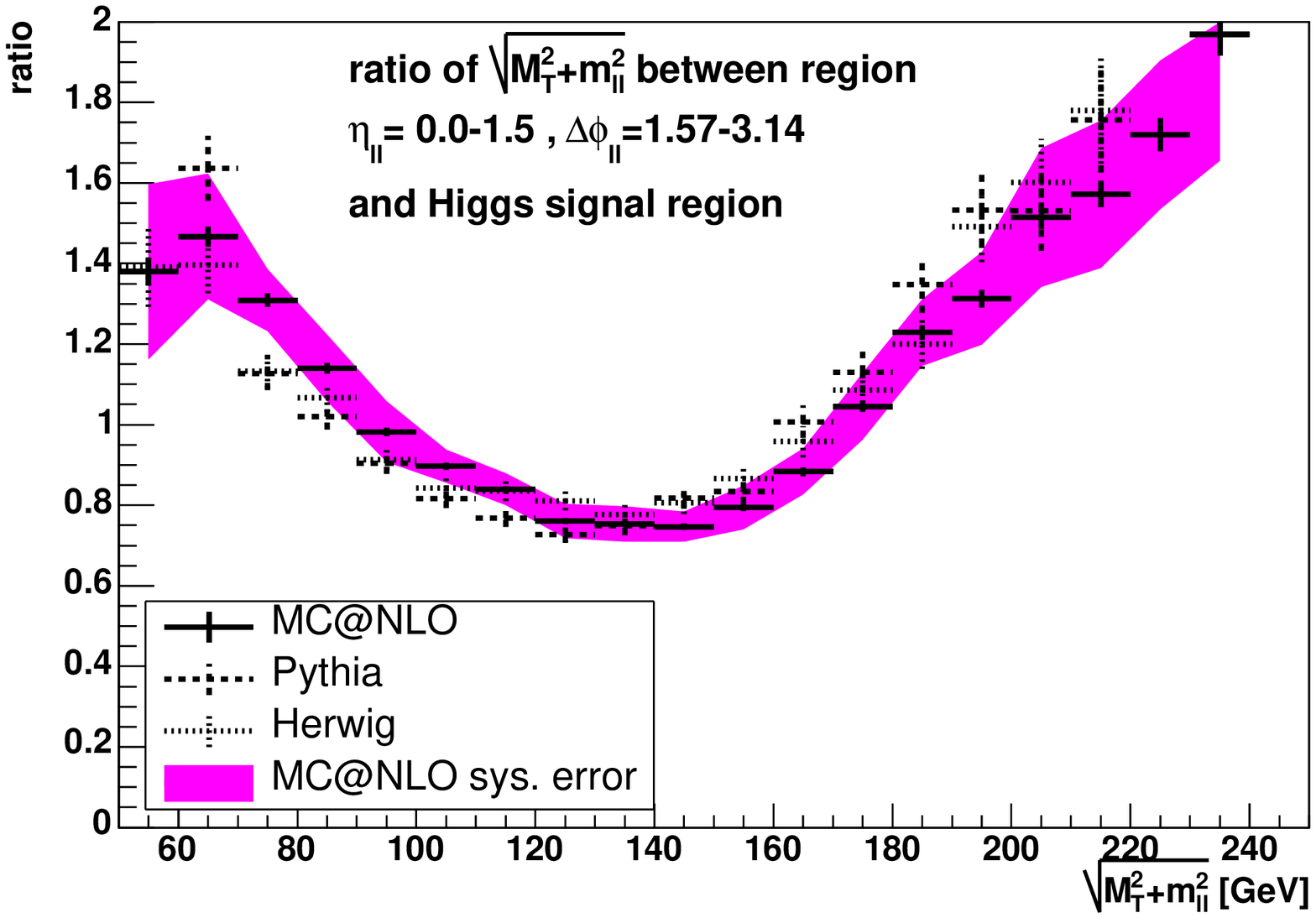}
\includegraphics[width=0.325\textwidth]{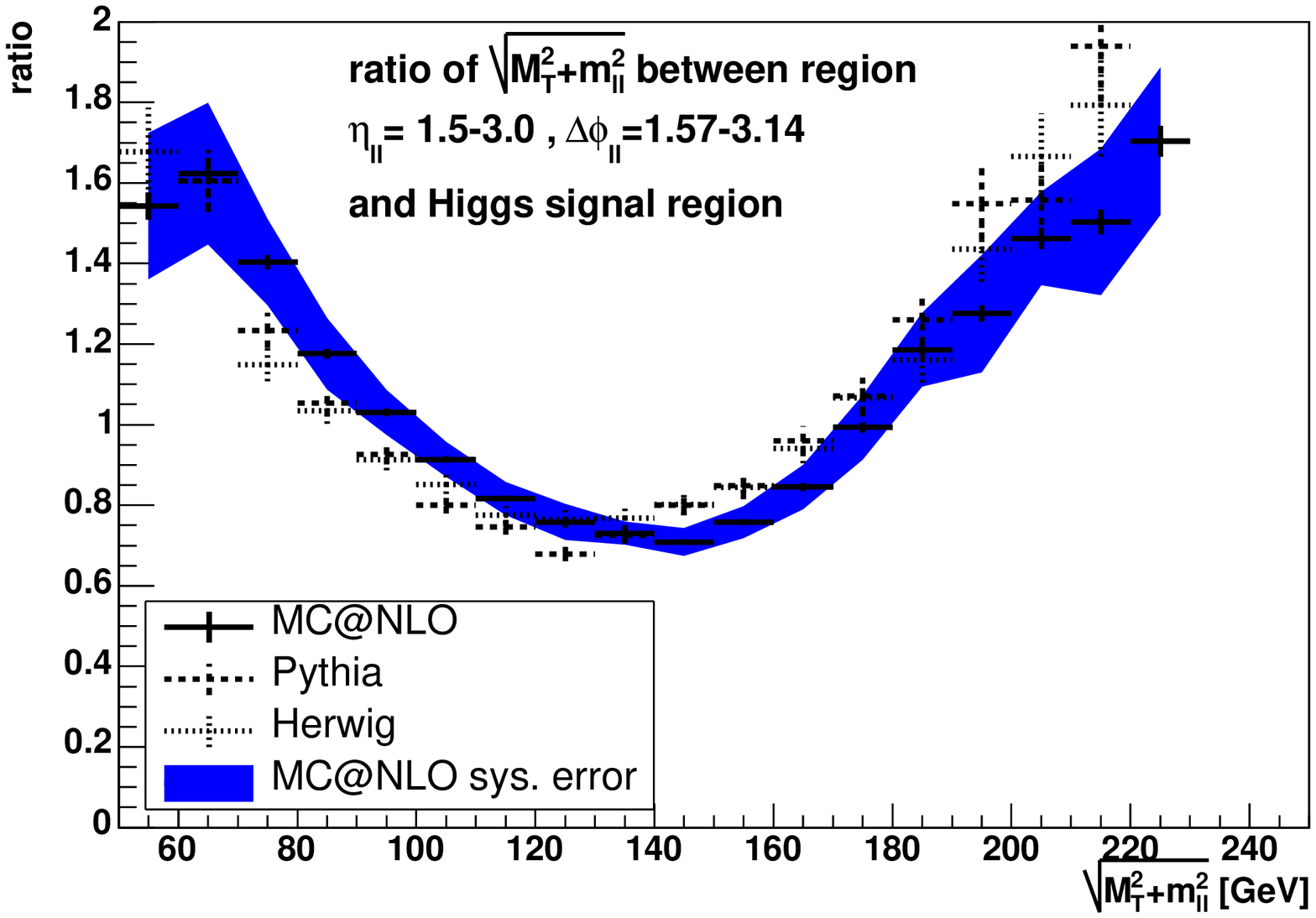}
 \vspace*{-3mm}
 \caption{Ratio between the $M^\prime_T$ distribution in the three
 normalization regions and the signal region for MC@NLO. The error bars give
 the Monte Carlo statistical error on the shape. The colored error band gives
 an upper limit on the systematic uncertainty on this shape from QCD
 factorization and renormalization scale variation (each within a factor 2)
 and one sigma error PDF variation. The width of this error band is dominated
 by statistical fluctuations, the true uncertainty should be much smaller. For
 comparison also the leading order shape from PYTHIA and HERWIG is shown.}
\label{fig:pp2ww_MTratio}
\end{figure}

In Fig. \ref{fig:pp2ww_MTratio} the ratio of the $M^\prime_T$ shape in the
signal region and the $M^\prime_T$ shape in the normalization regions is shown
for MC@NLO. The colored error band gives the systematic uncertainty on the
shape of MC@NLO from scale and PDF variations. The QCD factorization and
renormalization scale inside MC@NLO is varied independently within factors of
2. PDF uncertainties are evaluated by scanning through the CTEQ6
\cite{Pumplin:2002vw} error PDFs. The width of the error band is actually
consistent with the statistical fluctuations expected from the independent
Monte Carlo samples and therefore gives an upper limit on the systematic
uncertainty on the ratio. The ratio given by PYTHIA \cite{Sjostrand:2003wg}
and HERWIG \cite{Corcella:2002jc} shows a slight shift to smaller values of
$M^\prime_T$ (the main reason is the missing $gq$ and $qg$ initial state as
shown in section \ref{sec:pp2ww_UncertNLO}). In the relevant mass range of 100
GeV $<M^\prime_T<$ 200 GeV a systematic uncertainty of less than 10\% on these
ratios can be expected (not including experimental uncertainties from
$p_T^{\rm{miss}}$).

The most promising way of normalizing the $M^\prime_T$ shape in the signal
region is from the normalization region $\eta(\ell^+ + \ell^-)>1.5$ and
$\Delta\phi(\ell^+,\ell^-)<1.5$ (Fig. \ref{fig:pp2ww_MTratio} left), which has
the same $\Delta\phi(\ell^+,\ell^-)$ distribution as the signal region. Since
the ratio is relatively flat, systematic shifts in $M^\prime_T$ are
uncritical. However, this region suffers from low experimental event
statistics and has already some Higgs boson signal contribution as can be seen
in Fig. \ref{fig:pp2ww_etaphi}.

The normalization regions with $\Delta\phi(\ell^+,\ell^-)>\frac{\pi}{2}$
should be cleanly measurable, but the ratio to the signal region is not flat
and systematic uncertainties on $p_T^{\rm{miss}}$ might affect the ratio (this
needs further experimental studies).

Once the $M^\prime_T$ shape of the background is measured, one could go beyond
number counting in the signal region and use the information contained in the
$M^\prime_T$ distribution. One could follow two approaches: By using
$\Delta\phi(\ell^+,\ell^-)$ to normalize the $\rm W^+W^-$ background one could
subtract the extrapolated $M^\prime_T$ shape in the signal region and look for
the Jacobian Higgs peak. Alternatively one could use the extrapolated
$M^\prime_T$ shape to normalize the $\rm W^+W^-$ background directly by using
the sideband $M^\prime_T>200$ GeV. This would be an independent method of
normalizing the background directly in the Higgs boson signal region.

Using both methods it should also be possible to measure the Higgs boson mass
from the position of the Jacobian peak. Without any background determination
from data, this might prove difficult, since the peak of the background and
the peak of the signal are very close together.

The shape of the ratios in Fig. \ref{fig:pp2ww_MTratio} is very similar for
$\rm gg \to W^+W^-$ and also for $t\bar{t}$, which contributes as additional
background to $h \to W^+W^-$ searches. This reduces the dependency of the
extrapolation on the relative normalization of the various backgrounds.

%------------------------------------------------------------------------------
\subsubsection{Conclusions}

The modeling of W pair production at the LHC has been investigated by
comparing several event generators with different features. MC@NLO turns out
to be the most reliable program available. The prediction of ${\rm p p
\rightarrow W^+ W^-}$ is improved further by combining ${\rm g g \rightarrow
W^+ W^-}$ events with the events from MC@NLO. More results and details can be
found in Ref.~\cite{Drollinger:2005ug}.

After the event selection, which is borrowed from the ${\rm h \rightarrow W^+
W^-}$ analysis, the shape of the $\Delta\phi$ distribution can be used to
extrapolate the number of W pair events from the background control region
(${\rm 1.4\ rad} < \Delta\phi < {\rm 2.4\ rad}$) into the Higgs signal region
($\Delta\phi < {\rm 0.8\ rad}$). For this particular example the theoretical
uncertainties of the $\Delta\phi$ shape are ${\rm \delta_{PDF} < 0.7\% }$ for
the PDF uncertainty of MC@NLO, ${\rm \delta_{scale} < 2.0\% }$ for the scale
uncertainty of MC@NLO, and ${\rm \delta_{g g \rightarrow W^+ W^-} \approx
3.8\% }$ for the uncertainty of the ${\rm g g \rightarrow W^+ W^-}$
contribution. For comparison, ignoring either the spin correlations or the
higher order corrections would lead to uncertainties of the order of 30\%.

By measuring the distribution of the transverse mass $M^\prime_T$ at large
$\Delta\phi(\ell^+,\ell^-)$ or $\eta(\ell^+ + \ell^-)$ an extrapolation to the
transverse mass distribution into the Higgs boson signal region seems
feasible. The theoretical uncertainty on this extrapolation estimated from PDF
and scale uncertainties of MC@NLO is less than 10\%. Using this extrapolation
an independent normalisation of $\rm W^+W^-$ and an observation of the
Jacobian Higgs boson peak should be possible.

\subsubsection*{Acknowledgements}
Special thanks to Stefano Frixione for including the important spin
correlations in MC@NLO, and many thanks to Alexandre Sherstnev for the nice
introduction to the use of CompHEP and to Edward Boos for giving further
details. Many thanks to Massimiliano Grazzini for his very useful comments
about higher order uncertainties.  Work supported in part by the European
Community's Human Potential Programme under contract HPRN-CT-2002-00326,
[V.D.].

%%%%%%%%%%%%%%%%%%%%%%%%%%%%%%%%%%%%%%%%%%%%%%%%%%%%%%%%%%%%%%%%%%%%%%%%%%%%%
\section[Top background generation for the $H\to WW$ channel]
{TOP BACKGROUND GENERATION FOR THE $H\to WW$ CHANNEL~\protect
\footnote{Contributed by: G.~Davatz, A.-S.~Giolo-Nicollerat, M.~Zanetti}}
\subsection{Introduction}

The $\rm t\bar{t}$ production is known as an important background for many
processes at the LHC. Large uncertainties can be expected from the different
Monte Carlo simulations. We will study the $\rm t\bar{t}$ background in the
phase space specific for the SM Higgs channel $\rm H \to WW \to \ell \nu \ell
\nu$ by comparing four different Monte Carlo event generators:
HERWIG~\cite{Corcella:2002jc}, MC@NLO~\cite{Frixione:2002ik,Frixione:2003ei},
PYTHIA~\cite{Sjostrand:2003wg} and TopReX~\cite{Slabospitsky:2002ag}.\\
The Higgs decay into two $\rm W$ bosons and subsequently into two charged leptons is expected to be the main
discovery channel for intermediate Higgs mass, between
$\rm 2 m_W$ and $\rm 2 m_Z$~\cite{Dittmar:1996ss}.
The signature of this decay is characterized by two
leptons and high missing transverse energy. 
However, since no narrow mass peak can be reconstructed, 
a good control of the background, together with a high
signal to background ratio, is needed. 
The most important backgrounds, which give similar signature as the
signal (i.e. two leptons and missing energy),
are the continuum WW production and the $\rm t\bar{t}$ production.
In order to separate the signal from the backgrounds, one has to require a
small opening angle between the leptons in the plane transverse to the beam against the continuum WW production 
and apply a jet veto against $\rm t\bar{t}$ production This implies a restriction to a very specific region
of the phase space. \\
First, we estimate
how well Leading Order (LO) Monte Carlos generate top production in the phase
space relevant for Higgs search with respect to NLO Monte Carlos
by comparing MC@NLO with HERWIG.
Then by comparing PYTHIA and HERWIG we determine the effect of
using different parton shower models.
Finally, we estimate the effect of
spin correlations between the t and the $\rm \bar{t}$. 
More details about this study can be found in Ref. ~\cite{Davatz:2005}.

\subsection{Generating top background}
For each Monte Carlo program, one million $\rm pp\to t\bar{t} \to
WbWb\to \ell\nu\ell\nu bb$ events are generated ($\rm \ell = e$, $\rm
\mu$ and $\rm \tau$). 
The PDF chosen for HERWIG, PYTHIA and TopReX is CTEQ5L and for MC@NLO
CTEQ5M1. No underlying event is generated. 
The top mass is set to 175~GeV. The events are reconstructed using stable detectable  particles. 
First, a pre-selection requires
two isolated opposite charged leptons with $\rm p_t>20$~GeV 
and $\rm |\eta|<2$, cuts which can easily be satisfied by CMS and ATLAS. This pre-selection is always applied in the following. The final selection requires: \label{higgs_sel}

\begin{itemize} 
\item rejecting all events including a
jet\footnote{The jets are reconstructed using an iterative cone algorithm with a
cone size, $\rm \Delta$R, of 0.5. A jet is kept if its
$\rm p_t$ is higher than 20 GeV and $\rm |\eta|<4.5$.} with
$\rm p_t$ higher than 30~GeV and $\rm |\eta|<$~2.5 (jet veto)

\item $\rm E_t^{miss}>$~40~GeV ($\rm E_t^{miss}$ is formed with the sum
of isolated leptons and jets transverse momenta)
\item $\rm {\phi}_{\ell\ell}<~45^{\circ}$ (angle between the leptons in the
transverse plane) 
\item 5~GeV~$\rm <m_{\ell\ell}<$~40~GeV (the invariant mass of the two leptons)
\item 30~GeV~$\rm <p_{\ t\ lep\ max}<$~55 GeV (lepton with the maximal
$\rm p_t$)
\item $\rm p_{\ t\ lep\ min}>$~25~GeV (lepton with the minimal $\rm p_t$)
\end{itemize}

\subsection{Comparison between HERWIG and MC@NLO}

To estimate the effect of an accurate inclusion of NLO matrix elements, HERWIG~6.508 and MC@NLO~2.31 were compared \footnote{HERWIG~6.508 was also used for the showering step in MC@NLO. HERWIG~6.508 is an update of the HERWIG~6.507 version with a bug fixed concerning the top decay.}. The spin correlations between $\rm t$ and $\rm \bar{t}$ are not included in MC@NLO. HERWIG events were therefore also simulated without these spin correlations. As the same showering model is used, the difference between the two simulations should be mostly due to the additional NLO matrix elements in MC@NLO. 

\begin{figure}[h!]
\begin{center}
\includegraphics[scale=0.8]{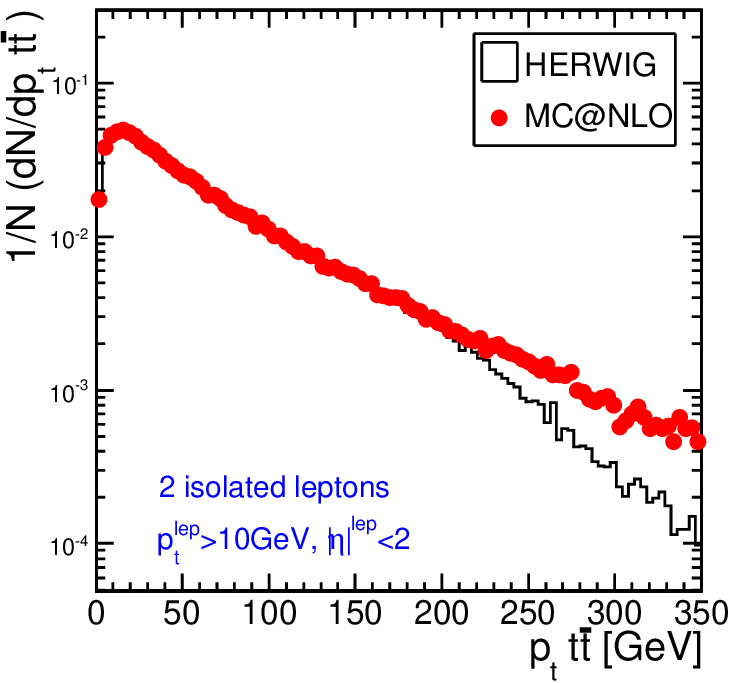}
\includegraphics[scale=0.8]{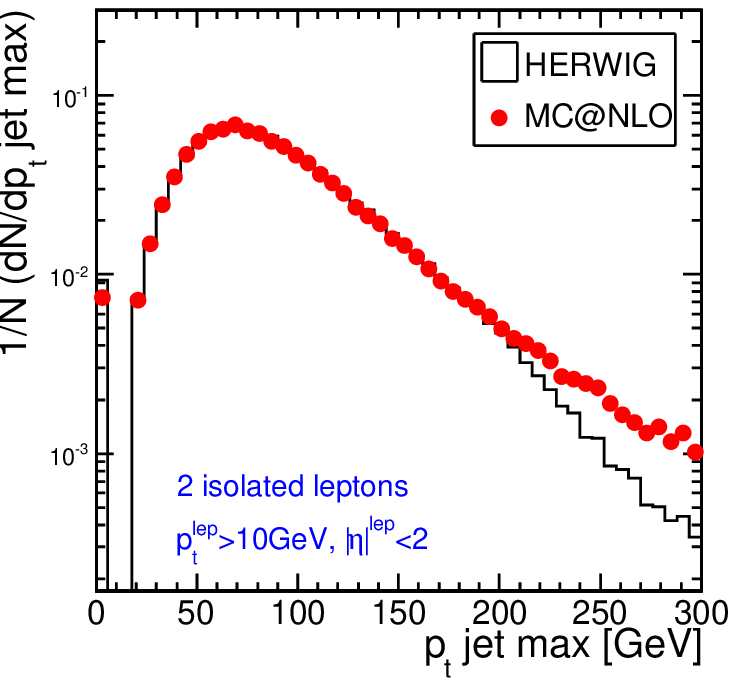}
\end{center}
\caption{The $\rm p_t$ distribution of the $\rm t\bar{t}$ system (left) and the leading jet (right) in HERWIG and MC@NLO.}
\label{h_m_ptjet}
\end{figure}

Figure~\ref{h_m_ptjet} (left) shows the transverse momentum of the $\rm t\bar{t}$ system for HERWIG and MC@NLO. At low $\rm p_t$, the two Monte Carlos are very similar, as the soft and collinear emissions are generated by HERWIG in both simulations. 
The $\rm t\bar{t}$ system is balanced by gluon emissions from the initial state radiation. MC@NLO produces in addition to the hard process up to one hard jet accurate to NLO. Therefore, the high $\rm p_t$ region of the $\rm t\bar{t}$ system is harder in MC@NLO.\\
In Fig.~\ref{h_m_ptjet} (right), the $\rm p_t$ of the hardest jet is
shown, taking into account all jets in the final state (from
the hard process and from the gluon emission).
$\rm p_{\ t\ jet\max}$ equal to zero means that there is no reconstructed jet with $\rm p_t$ higher than 20 GeV and $\rm |\eta|<4.5$ in the event. In the high $\rm p_t$ region the leading jet is harder in MC@NLO, but again at low $\rm p_t$, the two simulations are very similar. 
The region relevant for the $\rm  H\to WW\to \ell \nu \ell \nu$ signal selection is the very low $\rm p_t$ region, where HERWIG and MC@NLO agree very well. In addition, the shapes of all the other cut
variables are very similar in MC@NLO and HERWIG without spin
correlations.
After comparing the relative efficiencies of the different cuts, the differences between the two Monte Carlos are essentially due to the jet veto cut and smaller than 10\%. 
Since there are already two $\rm b$-jets in the $\rm t\bar{t}$ final state, the
jet veto tends to be less sensitive to additional jet activity.
From this comparison one can conclude that implementing accurately the NLO
contribution in the simulation has a small
effect on the shapes of the variables considered
and the selection efficiencies for the phase space relevant for the
$\rm H\to WW$ search. The region where NLO makes a
difference is at very high $\rm p_t$, whereas the bulk of the selected events is in
the low $\rm p_t$ region. \\

\subsection{Effect of showering models, differences between HERWIG and PYTHIA}
In the following, we study how different showering models influence the variable shapes and
selection efficiencies. For this, PYTHIA
6.325, based on the Lund hadronization model, was compared with
HERWIG without spin correlations, based on the cluster model for hadronization. 
Furthermore, we simulate two PYTHIA samples, one with the default $\rm
Q^2$-ordered parton shower model (so-called 'old showering') and one
with the $\rm p_t$-ordered parton showering model (so-called 'new
showering'). For all three simulations, default scales are chosen.

\begin{figure}[h!]
\begin{center}
\includegraphics[scale=0.8]{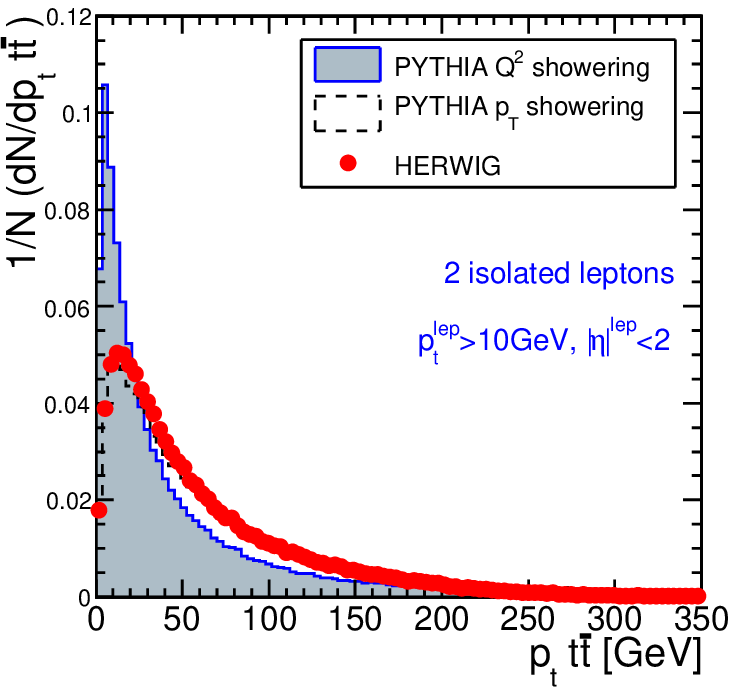}
\includegraphics[scale=0.8]{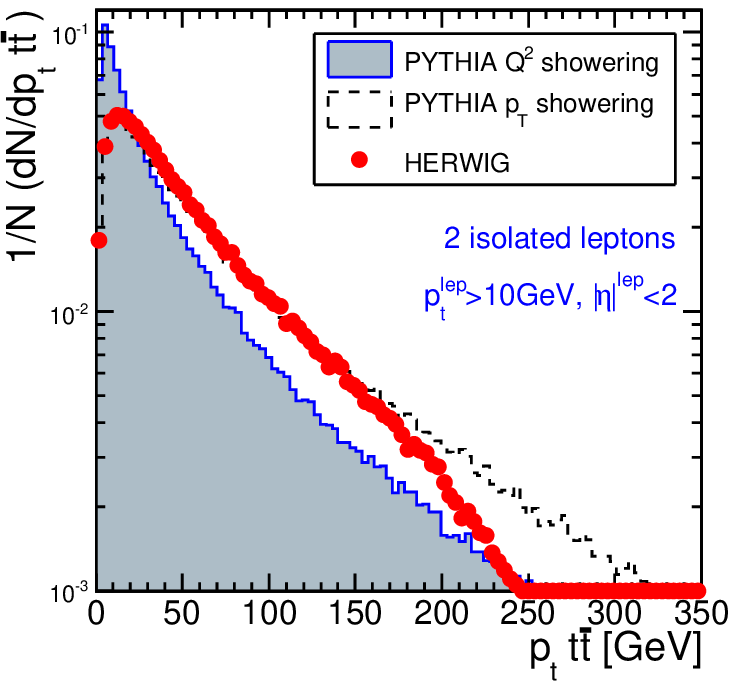}
\end{center}
\caption{The pt spectrum of the $\rm t\bar{t}$ system in HERWIG, PYTHIA new ($\rm p_t$-ordered)
and old ($\rm Q^2$-ordered) showering.}
\label{h_p_p_ptt}
\end{figure}

\begin{figure}[h!]
\begin{center}
\includegraphics[scale=0.8]{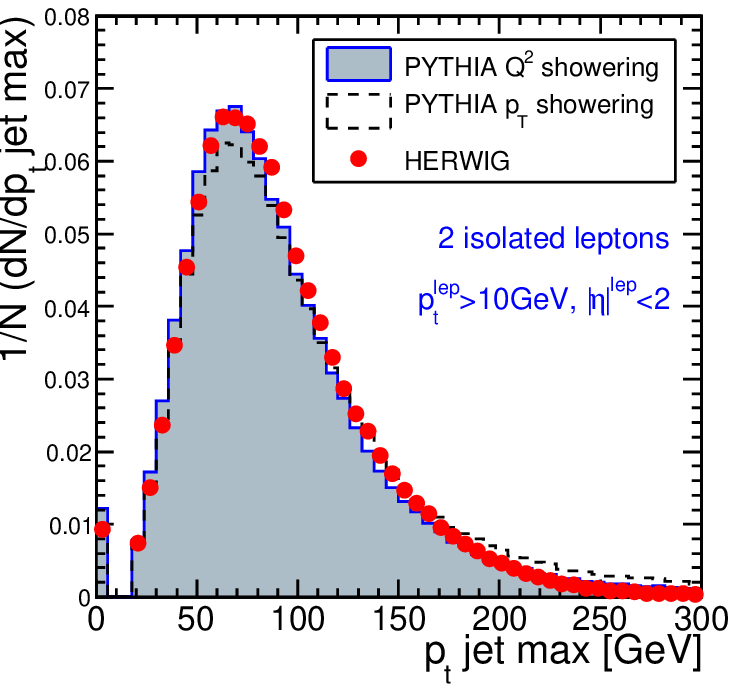}
\includegraphics[scale=0.8]{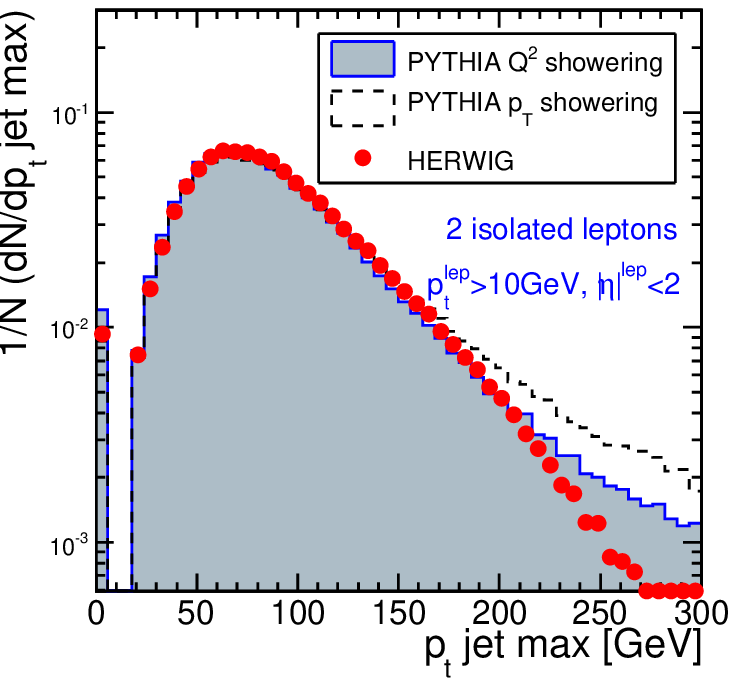}
\end{center}
\caption{The pt spectrum of the leading jet in HERWIG, PYTHIA new ($\rm p_t$-ordered) and old ($\rm Q^2$-ordered) showering.}
\label{h_p_p_ptjet}
\end{figure}

\begin{figure}[h!]
\begin{center}
\includegraphics[scale=0.8]{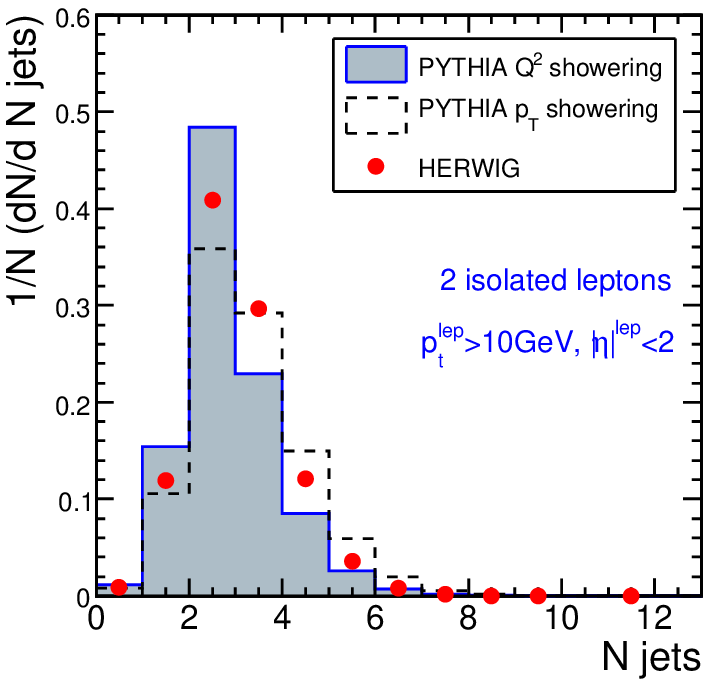}
\includegraphics[scale=0.8]{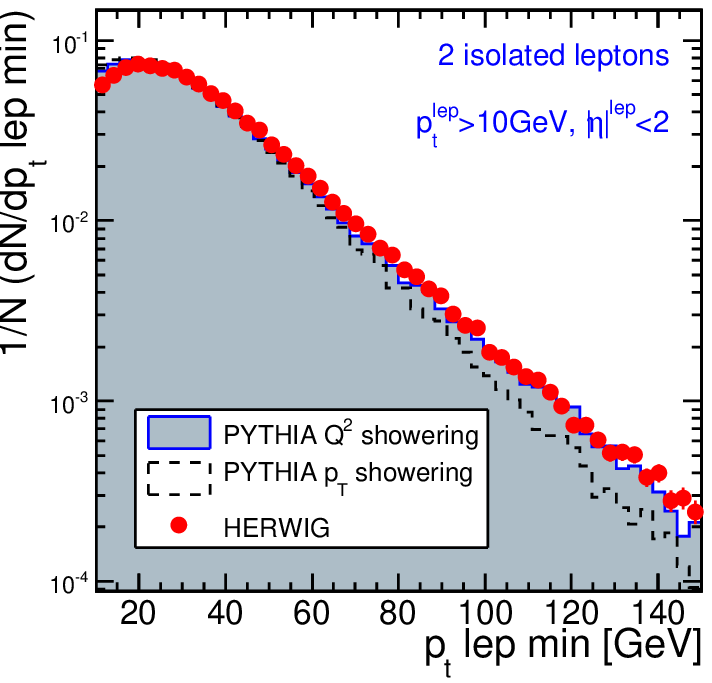}
\end{center}
\caption{The number of jets (left) and the $\rm p_t$ distribution of the softest
lepton (right) in HERWIG, PYTHIA new ($\rm p_t$-ordered)
and old ($\rm Q^2$-ordered) showering.}
\label{h_p_p_njets}
\end{figure}

In Fig.~\ref{h_p_p_ptt}, the $\rm p_t$ spectrum of the $\rm
t\bar{t}$ system is shown. The PYTHIA old showering tends to produce a
much softer spectrum than HERWIG and the PYTHIA new showering model. 
The $\rm p_t$ spectrum of the $\rm t\bar{t}$ system in HERWIG and the
new showering model in PYTHIA agree very well, except in the high $\rm
p_t$ region, which is due to the fact that HERWIG applies no matrix element
corrections at all.

Figure~\ref{h_p_p_ptjet} shows the $\rm p_t$ spectrum of the hardest jet for PYTHIA and
HERWIG. The leading jets in the new showering program are harder than in the old one. 
As the jets are harder, the number of jets increase with respect to
the old showering model, as can be seen in Fig.~\ref{h_p_p_njets}.
If one compares the other cut variables, the old showering model in
PYTHIA and HERWIG agree well, whereas the new showering model in
PYTHIA produces softer leptons, as shown in Fig.~\ref{h_p_p_njets}. 
Comparing the relative efficiencies after the selection cuts were
applied, the biggest differences come from the jet
veto and the lepton isolation cut efficiencies. While PYTHIA with the old showering model and
HERWIG have about the same isolation efficiency, PYTHIA with the new
showering model has a 20\% lower efficiency for the isolation of the
leptons. This is due to the fact that particles
from the new shower have on average higher $\rm p_t$ than the ones from
the old shower, making the leptons less isolated.\\
The jet veto efficiencies from HERWIG and the new showering model in PYTHIA are very similar, whereas the veto is less effective in the old showering model due to the fact that the jets are softer and therefore more events pass the jet veto. This leads to a difference in the jet veto efficiency of about 20\%. In order to get lower uncertainties from the use of different Monte Carlos, it will be very important to tune the Monte Carlos with data.

\subsection{Effect of the spin correlations}

\begin{figure}[h!]
\begin{center}
\includegraphics[scale=0.8]{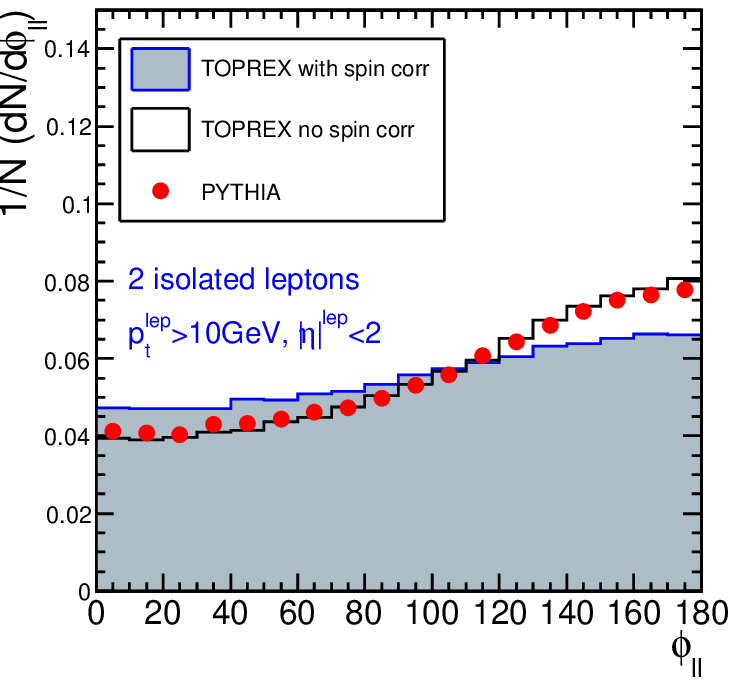}
\includegraphics[scale=0.8]{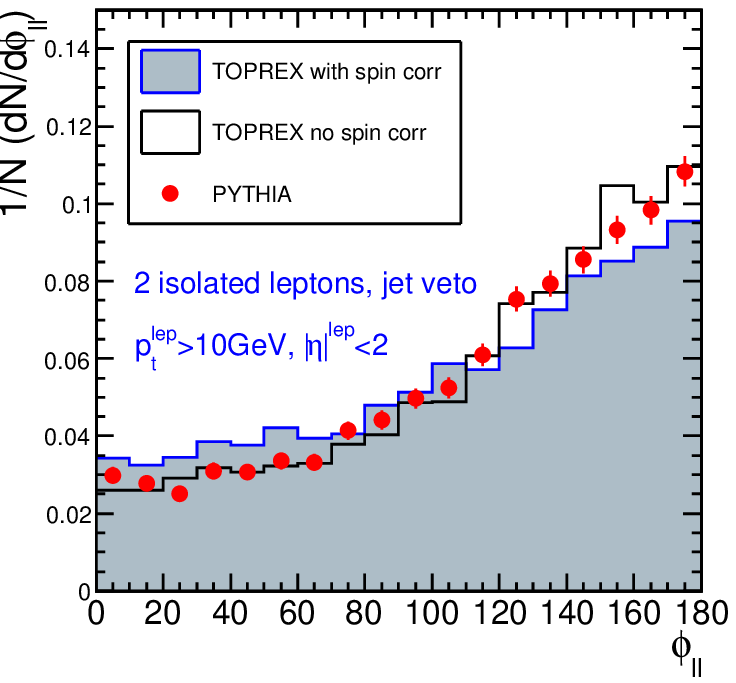}
\end{center}
\caption{$\rm \phi_{\ell\ell}$ is the angle between the leptons in the plane transverse to the beam. TopReX with and without spin correlations is shown, as well as PYTHIA. On the left, only very basic cuts are applied, whereas on the right a jet veto is applied in addition. The region important for the Higgs signal search is the low $\rm \phi_{\ell\ell}$ region.}
\label{t_p_phill}
\end{figure}

In the H $\rm \to$ WW Higgs search, a cut has to be
applied on the opening angle between the leptons in the transverse
plane ($\rm \phi_{\ell\ell}$) in order 
to differentiate the signal from continuum WW background. The variable $\rm \phi_{\ell\ell}$, as much as the mass of the di-lepton system $\rm m_{\ell\ell}$, are sensitive to spin correlations. In the following, the influence of the inclusion of spin correlations in the
$\rm t\bar{t}$ process is studied.
PYTHIA does not include the spin correlation between $\rm t$ and $\rm \bar{t}$. Thus we use TopReX with and without spin correlations, interfaced to PYTHIA for the showering
step\footnote{The
difference between PYTHIA and TopReX without spin correlation is
mostly due to the fact that 
the top quarks are not allowed to radiate gluons in TopReX, and the
different treatment of $\rm m_{top}$. For this comparison, the old showering model is chosen in PYTHIA.}. 
Figure~\ref{t_p_phill} shows the angle $\rm \phi_{\ell\ell}$ between the
leptons for the simulations with and without spin correlations. On the left, the only requirements are two isolated leptons with
$\rm p_t>$~10~GeV and $\rm |\eta|<$~2. On the right, a jet veto is applied in addition. As $\rm \phi_{\ell\ell}$ and $\rm m_{\ell\ell}$ are correlated, we only show the $\rm \phi_{\ell\ell}$ distribution.
PYTHIA and TopReX without spin correlations show the same
$\rm \phi_{\ell\ell}$ distribution. Including spin correlations leads to a flatter $\rm \phi_{\ell\ell}$ distribution. The same is studied with HERWIG with spin correlations, compared to HERWIG and MC@NLO without spin correlations. The difference due to the inclusion of spin correlations is slightly bigger in the comparison of TopReX and PYTHIA. Again, HERWIG without spin correlations has the same $\rm \phi_{\ell\ell}$ distribution as MC@NLO.
After a jet veto is applied, the distributions with and without spin correlations look more similar in both cases.

The difference of the relative efficiencies with and without spin
correlations in HERWIG is about 10\% and the same difference can be observed in TopReX. However, 
the relative efficiency for the $\rm \phi_{\ell\ell}$ cut in TopReX is slightly higher than in HERWIG.

In conclusion, the following approach could be used to generate the $\rm t \bar{t}$ background:
since the difference between MC@NLO and HERWIG without spin
correlations is rather small in our region of interest, HERWIG with
spin correlations could be used, re-weighted to the NLO cross section with an inclusive K-factor. \\
The new showering model of PYTHIA predicts similar shapes for the jets
and the $\rm t \bar{t}$ system as HERWIG, but the isolation of the leptons
leads to a difference of about 20\% and the other lepton variables are softer. 
On the other hand, the old showering model of PYTHIA is more similar to HERWIG in the lepton isolation and the lepton variable shapes, but has much softer jets.
This needs to be studied further. \\
When data is available, these
uncertainties can be reduced by tuning the different Monte Carlos to
data. In any case, it will be important to estimate the $\rm t \bar{t}$ background contribution for the Higgs search using data by defining normalization regions.

\subsection*{Acknowledgements}
We would like to thank Gennaro Corcella, Torbj\"orn Sj\"ostrand, Peter
Richardson and Stefano Frixione for
the useful discussions and help on the topic.

%%%%%%%%%%%%%%%%%%%%%%%%%%%%%%%%%%%%%%%%%%%%%%%%%%%%%%%%%%%%%%%%%%%%%%%%%%%%%
\section[Estimation of $t\bar{t}$ background for $H\to WW$ channel]
{ESTIMATION OF $t\bar{t}$ BACKGROUND FOR $H\to WW$ CHANNEL~\protect
\footnote{Contributed by: G.~Davatz, A.-S.~Giolo-Nicollerat, M.~Zanetti}}
\subsection{$\rm t \bar{t}$ normalization from data}

The presence of two neutrinos in the final state of the $H\DGZra W^+ W^-\DGZra\ell^+\nu\ell^-\bar{\nu}$ 
decay chain does not allow the reconstruction of a narrow invariant mass peak. 
Moreover, the rejection needed to reduce the different background processes 
is very high, in the specific case of $\rm t\bar{t}$ being ${\cal O}(10^{-5})$.
The precise understanding of the backgrounds is the most critical issue
concerning this Higgs discovery channel. The most reliable approach to address 
this problem is to measure the different sources of background directly
from the data. The commonly used method consists on selecting a signal-free 
phase space region (control region), where a given background process is enhanced. 
The contribution of that background in the signal region is then extrapolated
from the control region taking into account the observed amount of events. 
This procedure relies on the relation:
\begin{equation} \label{normalization_relation}
  N_{signal\_reg}=\frac{N_{signal\_reg}^{Monte Carlo}}{N_{control\_reg}^{Monte Carlo}} N_{control\_reg}
  =\frac{\sigma_{signal\_reg}\cdot\epsilon_{signal\_reg}}{\sigma_{control\_reg}\cdot\epsilon_{control\_reg}} N_{control\_reg}
\end{equation}
where $N_{signal\_reg}^{Monte Carlo}$ and $N_{control\_reg}^{Monte Carlo}$ are the numbers 
of events predicted by the Monte Carlo simulation in the signal and control region.
Each of this two numbers can be expressed as a product of the theoretical cross
section in that phase space area, $\sigma_{signal_rec,control_rec}$, 
and the experimental efficiency of reconstructing
events in the same region, $\epsilon_{signal_rec,control_rec}$\footnote{
  The experimental uncertainties could modify the boundaries 
  defining the phase space where the cross section is calculated theoretically. 
  This is the case in particular when the selections involve jets. 
  The ``$\epsilon$'' terms in relation
  (\ref{normalization_relation}) are assumed to account also for this effect.
}.
% As a consequence ``$\epsilon$'' could also be greater than $1$}. 
This will allow to better point out the different sources of systematic uncertainties. 
In particular the theoretical predictions enter the procedure only via 
the ratio $\sigma_{signal\_reg}/\sigma_{control\_reg}$, leading to a much smaller scale dependency
and thus to smaller theoretical uncertainties. \\
The theoretical issues concerning the $\rm t\bar{t}$ normalization have been deeply studied 
in \cite{Kauer:2004fg}, following the work done in the Les Houches Workshop in the year 2003.
The primary goal of this note is to provide a reliable description of the experimental aspects,
specifically the ones related to the CMS detector. For this study a full detector simulation 
has then been exploited. \\
The main requirement from the experimental side on the choice of the control 
region is to limit as much as possible the error due to the ``$\epsilon$'' terms
in relation (\ref{normalization_relation}). This implies to use similar
selections as for the signal region. Moreover the contamination from
other physical and instrumental backgrounds should be negligible. \\
In order to estimate the $\rm t\bar{t}$ contribution in the signal region, we 
exploit the presence of two additional high $E_{t}$ jets coming form the top quark 
decay. Two procedures are proposed: the first based on the
tagging of the two jets as originating from $b$ quarks, and the other is 
requiring simply the $E_t$ of the jets to be above a certain threshold. 
Both control regions will be defined by the same selections on the leptons 
as for the signal region. \\
The cuts used to define the signal region together with the 
corresponding number of events expected for 1~fb$^{-1}$ for the fully simulated 
signal (for a Higgs mass of 165 GeV), $\rm t\bar{t}$ and $\rm Wt$ are summarized 
in Table~\ref{signal_selections}.

\begin{table}[h!]
\begin{center}
\caption{The expected number of events for a luminosity of 1~fb$^{-1}$
for the signal with a Higgs mass of 165~GeV and the  $\rm t\bar{t}$ and
$\rm tWb$ background. The relative efficiency with respect to
the previous cut is given inside the brackets.}
\label{signal_selections}
  \vspace*{1mm}
\begin{tabular}{|l|l|c|c|c|}
\hline
& & $\rm H\to WW$ ($\rm m_H=$~165~GeV) &  $\rm t\bar{t}$ & $\rm tWb$ \\
\hline
&$\rm \sigma \times BR(e,\mu,\tau)$ [fb] &2360 & 86200 & 3400\\
\hline
1) & Trigger & 1390 (59\%) &  57380 (67\%) & 2320 (68\%)  \\
\hline
2) & lepton ID & 393 (28\%) &15700 (27\%) & 676 (29\%)  \\
%& \small{$\rm \sigma_{IP}>$~3, $\rm |\Delta z_{lep}|<$~0.2~cm} &&&\\
\hline
3) & $\rm E_t^{miss}>$~50~GeV & 274 (70\%) &9332 (59\%) & 391 (58\%) \\
\hline
4) & $\rm \phi_{\ell\ell}<$~45 & 158 (58\%) &1649 (18\%) & 65 (17\%)\\
\hline
5) & 12~GeV~$<m_{\ell\ell}<$~40~GeV & 119 (75\%) & 661 (40\%) &
28 (43\%)\\
\hline
6) & 30~GeV$\rm <p_{t}^{\ell\, max}<$55~GeV & 88 (74\%) &304 (46\%)
& 13 (46\%)\\
\hline
7) & $\rm p_{t}^{\ell\, min}>$25~GeV & 75 (85\%) & 220 (73\%) & 9.2 (71\%)\\
\hline
8) & Jet veto & 46 (61\%) &  9.8 (4.5\%) & 1.4 (15\%)\\
%\hline \hline
%& $\rm \varepsilon_{tot}$ & $(1.92\pm0.06)$\% & $(0.011\pm 0.002)$\% &
%$(0.041\pm 0.005)$\%\\
\hline
\end{tabular}
\end{center}
\end{table}

The main cut to reject the $\rm t\bar{t}$ is the jet veto. An event is rejected, if there
is at least one reconstructed jet with $E_t>15$ GeV within $|\eta|=2.5$. 
In order to reduce the fake jets, when the measured jet $E_t$ is between 15 and 20 GeV,
the ratio of the sum of the $\rm p_t$ of all tracks inside the jet 
over the transverse jet energy deposited in the calorimeter, referred to as ``$\alpha$'',
is required to be greater than $0.2$.

\subsubsection{b-tagging jets based $\rm t\bar{t}$ normalization}

\begin{figure}[h]
  \begin{center}
    \includegraphics[scale=0.93]{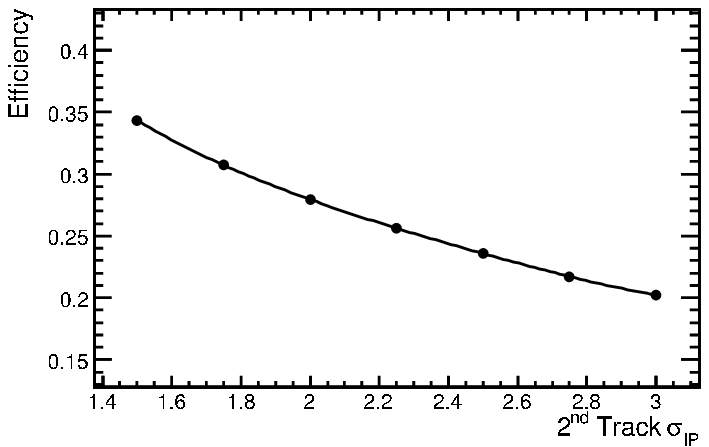}
    \includegraphics[scale=0.93]{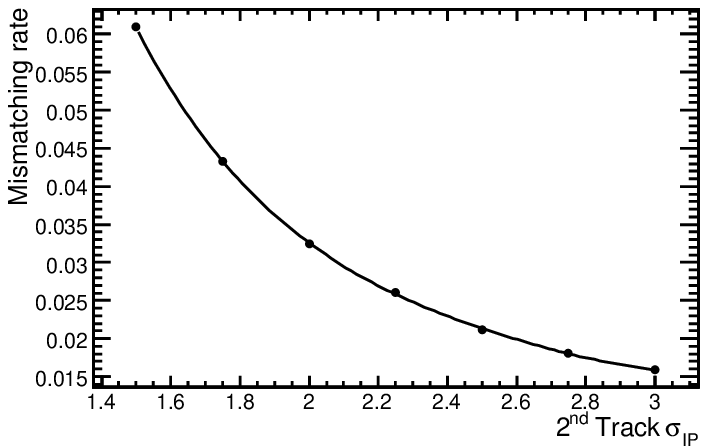}
  \end{center}
  \caption{Tagging efficiency and mistagging rate as a function of the discriminator. 
    Left plot shows the efficiency of b-tagging both the jets originated
    from b quarks in $\rm t\bar{t}$ events, whereas right plot shows the percentage
    of mistagged jets.} 
  \label{BTagDiscriminatorPlots}
\end{figure}

The presence of two b-tagged jets together with two isolated leptons is a striking evidence for $\rm t\bar{t}$
events. In addition to the requirement of two b-jets, the control region 
for $\rm t\bar{t}$ extrapolation is defined by all the cuts in
Table ~\ref{signal_selections} but the jet veto. \\
The algorithm, used to discriminate whether a jet is originated from
a $b$ quark. is based on the impact parameters of charged particle tracks
associated to the jet~\cite{BTagging:2005an}.
The parameter, in the following called ``discriminator'', 
that characterizes the efficiency and the mistagging rate of the 
algorithm, is the impact parameter significance $\sigma_{IP}$ 
of a minimum number of tracks associated to the jet. 
In this study, a jet is tagged as a $b$-jet if its measured $E_t$ is greater
than $20~GeV$ and if there are at least 2 tracks with 
$\sigma_{IP}$ above a given threshold.
%The choice of a suitable value for this threshold is a compromise
%between the efficiency of selecting $\rm t\bar{t}$ and the purity
%of the events selected.
%EVENTUALLY COMMENT THE FOLLOWING PART OUT || || ||
%                                          \/ \/ \/
The dependence of the efficiency of selecting $\rm t\bar{t}$ and the purity
of the events selected on the discriminator value is shown
in the plots of Fig.~\ref{BTagDiscriminatorPlots}. \\
A discriminator value of 2 for jet b-tagging is used in this analysis. In this
case the double b-tagging efficiency is ${\cal O}(30\%)$ while the mistagging rate is 
${\cal O}(3\%)$. Table~\ref{doublebTag_table} 
summarizes the number of events expected for 10~fb$^{-1}$ in the
control region for $\rm t\bar{t}$, $\rm Wt$ and the signal in the case of 
$\rm 2\mu$, $\rm 2e$ and $\rm e\mu$ final states.

%The left plot shows
%the efficiency of b-tagging at least 2 jets  
%when 2 $\rm b$'s are present within $\rm |\eta|<$~2.5 (the CMS
%pixel vertex detector acceptance) %CHECK THIS!! 
%The right plot represents the fraction of the mistagged jets   
%calculated out of a sample with jets with $E_t>20$ GeV. 
 
%                                         /\ /\ /\
%                                         || || ||

Not all the processes with $2\ell+2b+E_t^{miss}$ as final state have been fully simulated
for this analysis. Nevertheless, general considerations and fast Monte Carlo
level checks can lead to exclude other relevant sources of backgrounds.  \\
The more natural concurrent process is $W^{+}W^{-}b\bar{b}\DGZra2\ell 2\nu b\bar{b}$ which is anyway
$\alpha_{weak}^2$ suppressed with respect to $\rm t\bar{t}$. Its cross section is indeed
expected to be smaller than $1~pb$.   
Assuming the same efficiency for the kinematic selections as for the $W^{+}W^{-}\DGZra2\ell$,
i.e. ${\cal O}(10^{-3})$, less than $10$ events are expected for $10~fb^{-1}$ in the control 
region even without taking into account the double-b tagging efficiency.  
In the case of same flavor leptons in the final state, $\gamma^{*}/Z^{*}\DGZra 2\ell + b\bar{b}$ 
(the vector boson mass being away from the $Z$ peek, i.e. $m_{\ell\ell}<40~ GeV$) 
could also contribute as an instrumental background, when an high value of $E_t^{miss}$ is
yielded by the imperfect hermeticy of the detector and due the finite resolution
of the calorimeters, respectively. 
Anyway for a fully simulated sample of $\gamma^{*}/Z^{*}\DGZra 2\ell + 2jets$ with jets' $E_t$ 
grater than $20$ GeV, the fraction of events with $E_t^{miss}>50~GeV$ 
(the actual cut applied for the signal selection) is smaller than 1\%.
Applying the same kinematic selection, but the $E_t^{miss}$ cut on a
$pp\DGZra \gamma^{*}/Z^{*}\DGZra 2\ell + b\bar{b}$ sample generated with MadGraph
\cite{Maltoni:2002qb},
$200$ events are expected for $10~fb^{-1}$ which vanish if
the rejection due to a realistic $E_t^{miss}$ selection is included.

\subsubsection{Two high $E_t$ jets based $\rm t\bar{t}$ normalization}

Although very powerful, the method proposed above for the estimation of the $\rm t\bar{t}$ background from data
relies entirely on jet b-tagging which is known to be a sophisticated 
procedure from the hardware as well as from the algorithmic point of view. 
In order to avoid the systematics due to this method, 
it is then worth to have alternative methods to estimate  
the $\rm t\bar{t}$ background from data. \\
Each of the two $b$ quarks in the $\rm t\bar{t}$ final state come from a $175~GeV$ 
central object; their $E_t$ spectra are then rather hard. An alternative
method to define a $\rm t\bar{t}$ control region is thus to require simply two hard jets in the detector 
in addition to the signal cuts listed in Table~\ref{signal_selections}.\\
In analogously to the normalization, based on b-tagging, Drell Yan can
be a dangerous background. In this case, the general $2\ell+ 2j$ final state has a much
higher cross section with respect to the $2\ell+ 2b$ one. As a fully simulated sample
based on matrix element calculations was not available, 
a Monte Carlo level analysis has been performed, leading to the result that,
 after applying the ${\cal O}(10^{-2})$ reduction due to the $E_t^{miss}$ cut,
the contribution of this process in the control region can not be neglected.
In order to get rid of this additional background, only the $e\mu$ final state has
been considered. \\
The thresholds on the jet transverse energy that maximize the signal ($\rm t\bar{t}$) 
over the background (Wt$+$signal) ratio and minimize the statistical error 
have been found to be $50$ and $30$ GeV. The number of events expected events
for $\rm 10~fb^{-1}$ for $\rm t\bar{t}$, Wt and the signal are summarized 
in Table~\ref{doublebTag_table}.

A background process, not considered in the full simulation analysis, is
$W^{+}W^{-}\DGZra\mu\nu_{\mu}+e\nu_{e}+2j$. The cross section, after geometrical acceptance cuts,
is  $0.4~pb$, whereas the signal selection cut efficiency
is smaller than $5\cdot 10^{-4}$ (with an uncertainty of
$\sim{8}\%$ from the limited Monte Carlo statistics). The contribution of this background can then be assumed to
be at maximum of the order as the signal. \\
In case one jet is misidentified as an electron, $W^{\pm}\DGZra\mu\nu_{\mu}+3j$
%providing the same final state topology we are treating in this section,
could be a source of background, too.
For the CMS detector, the probability of electron misidentification is 
estimated to be  ${\cal O}(10^{-4})$\footnote{
  The muon misidentification rate is at least one order of magnitude
  smaller}. 
Given its cross section, calculated to be $\sim{200}$ pb after the geometrical  
acceptance cuts, the latter rejection factor together
with the kinematic selection efficiency, estimated again from a generator level
study is ${\cal O}(10^{-4})$, lead to neglect this process as a source 
of contamination of the $\rm t\bar{t}$ control region.

\begin{table}
  \begin{center}
  \caption{Number of events of $\rm t\bar{t}$, signal and $Wt$ expected for $10~fb^{-1}$ 
    in the two control regions, described in the text, and in the signal region.
    The results are shown for $2\mu$, $2e$, $e\mu$ final states.
    \label{doublebTag_table}}
  \vspace*{1mm}
    \begin{tabular} {|c|c|c|c|c|c|c|c|c|c|}
      \hline
      & \multicolumn{3}{c|}{``b-tagging'' control region} 
      & \multicolumn{3}{c|}{``hard jets'' control region} 
      & \multicolumn{3}{c|}{signal region} \\
      \cline{2-10}
      & $2\mu$ & $2e$ & $e\mu$ & $2\mu$ & $2e$ & $e\mu$& $2\mu$ & $2e$ & $e\mu$ \\
      \hline
      $\rm t\bar{t}$ & 194 & 107 & 245 & - & - & 411 & 33 & 22 & 44\\
      \hline
      $Wt$  & 1 & $\rm <1$ & 2 & - & - & 6 & 5 & 3 & 6\\
      \hline
      $Signal~(m_H=165)$  & $\rm <1$ & $\rm <1$ & 1 & - & - & 11 & 156 & 89 & 214\\
      \hline
    \end{tabular}
  \end{center}
\end{table}

\subsection{$\rm t\bar{t}$ normalization procedure uncertainties}

\subsubsection{Systematics uncertainties}

Our proposed procedure to estimate the number of $\rm t\bar{t}$ events 
in the signal phase space region exploits relation (\ref{normalization_relation}).
In order to compute the systematic uncertainties on the final result we 
consider separately those related to each term present in the formula.

\begin{itemize}

\item {\bf Theoretical uncertainty} 

  Taking the ratio of the
  $t\bar{t}$ cross sections in the signal and control region avoids much of the 
  theoretical systematic uncertainties.  This is in fact the main 
  justification of rel. (\ref{normalization_relation}), first proposed in Ref.
  \cite{Kauer:2004fg}. In that paper the theoretical uncertainty 
  on the ratio $\sigma_{signal\_reg}/\sigma_{control\_reg}$
  has been studied at parton level with LO precision 
  by varying the renormalization and factorization scales. 
  The error has been estimated to range 
  between $3\%$ to $10\%$, mostly due to the choice of the PDF. \\
  In Ref. \cite{ttMCStudy:2005lh}, the NLO effects on $\rm t\bar{t}$ simulation have been studied,
  with the result that the shapes of the distributions involved in the normalization
  procedure, i.e. the $E_t$ spectra of the jets and the jet multiplicity are not
  affected by higher orders contributions.
  However, the comparison of different showering models shows some discrepancies
  either in the jet multiplicity or the jets $E_t$ spectra, introducing a further
  uncertainty with respect to the one due to the PDF set. \\
  For what concerns the proposed normalization procedure, the dependence on the 
  showering model has been studied in this analysis.
  Nevertheless, the Monte Carlo predictions concerning 
  $\rm t\bar{t}$ will be intensively compared and tuned directly with the data, also
  considering the very high $\rm t\bar{t}$ rate at the LHC. 
  A 10\% systematical error due to theoretical uncertainty will be assumed as
  reported in Ref. \cite{Kauer:2004fg}, although baring in 
  mind that this could be an optimistic estimate.

\item {\bf Jet energy scale uncertainty} 

  In the background normalization procedures, we propose, the jet energy scale (JES) 
  uncertainty
  is particularly important since it affects in opposite manners the signal region, 
  defined by vetoing the jets, and the control region where the presence of two jets is 
  required. To take into account this sort of anti-correlation of $\epsilon_{signal\_reg}$ 
  and $\epsilon_{control\_reg}$, we estimate the effect of the JES uncertainty directly
  on their ratio by rescaling the measured jet four momentum by an amount corresponding
  to the fractional uncertainty (i.e. $P^{\mu}_{jet} = (1+\lambda)P^{\mu}_{jet}$). 
  
  In the plot of Fig.~\ref{JESUncert}  the relative 
  variation of $\frac{\epsilon_{signal\_reg}}{\epsilon_{control\_reg}}$ for various values of 
  $\lambda$
  %\footnote{Actually the dependency of the JES uncertainty from the
  %  jet $E_t$ is taken into account by dividing $\lambda$ by $2$ for jets above
  %  $50~GeV$} 
  is shown. In the plot the triangles represent the control region, defined by
  requiring two jets with $E_t$ greater than 50 and 35 GeV,
  whereas the squares stand for the control region defined
  by requiring two b-tagged jets\footnote{
    The reason, why the ratio $\epsilon_{signal\_reg}/\epsilon_{control\_reg}$ in the latter
    case is less sensitive to the JES uncertainty is that the $E_t$ threshold for
    the b-jets candidates is $20~GeV$ and the fraction of $\rm t\bar{t}$ events 
    with b-tagged jets with $E_t$ close to that threshold is very small.
  }
  
  A realistic estimation of the JES uncertainty of CMS after integrating
  $10~fb^{-1}$ of LHC is ${\cal O} (5\%)$. The corresponding relative variation
  of $\rm \epsilon_{signal\_reg}/\epsilon_{control\_reg}$ is  
  $\rm \sim8\%$ for the double b-tagging defined control region 
  and $\rm \sim10\%$ for the two high $\rm E_t$ jets control region.

%  corresponding relative variation
%  of $\epsilon_{signal\_reg}/\epsilon_{control\_reg}$ is respectively 
%  $\sim8\%$ and $\sim10\%$ for the two control regions.

  \begin{figure}[htbp]
    \begin{center}
      \includegraphics[scale=0.99]{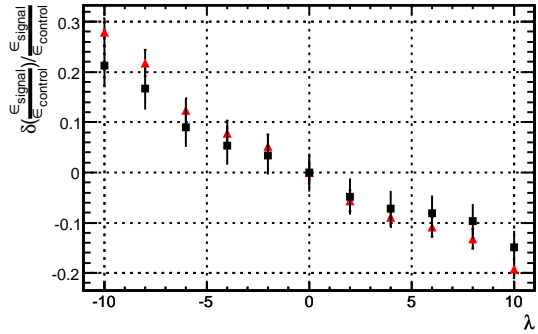}
    \end{center}
    \caption{Relative variation of $\frac{\epsilon_{signal\_reg}}{\epsilon_{control\_reg}}$
      as a function the jet momentum rescaling factor $\lambda$. The red triangles
      represent the control region defined by two hard jets whereas the black squares
      correspond to the two b-tagged jets phase space area.
    }
  \label{JESUncert}
  \end{figure}

\item {\bf $\alpha$ criterion uncertainty} 

In order to prevent the contamination from fakes when vetoing jets down to a 
raw transverse energy of 15 GeV, it is useful to cut on the track content of the jets. 
For jets with $E_t$ between 15 and 20 GeV the $\alpha$ 
criterion is then exploited, as explained before.   
In order to estimate the systematic uncertainty due to this criterion, the cut on $\rm \alpha$ 
has been varied from 0.15 to 0.25. Moreover, different values of the
minimum $\rm p_t$ for a track, to be included in the sum, have been
tried from 2 to 3~GeV. The consequent variation of the jet veto 
efficiency ($\rm \epsilon_{signal\_reg}$) is relatively small, i.e. of the order of 4\%.

\item {\bf b-tagging uncertainty} 

  In Ref. \cite{BTagSys:2005cn} the precision, with which the b-tagging efficiency of CMS
  will be known at CMS, is expected to be $11\%$ for $1~fb^{-1}$ integrated luminosity
  and it is foreseen to improve till $7\%$ with $10~fb^{-1}$. These values
  represent directly the uncertainty on $\epsilon_{control\_reg}$ in the case
  of the control region defined by requiring two b-tagged jets.

\item {\bf Uncertainties on $N_{control\_reg}$} 

  It has been shown in the previous section that $\rm t\bar{t}$ is plainly the 
  dominant process in both control regions. With the cuts used for
  selecting these control regions, i.e. the signal kinematic cuts plus 
  two b-tagged jets or two high $E_t$ jets, we expect to identify almost
  purely $\rm t\bar{t}$ events. In the worst case, i.e. when the control region
  is defined by two high $E_t$ jets, the fraction of events coming form
  other processes is smaller than 4\%.
  Provided that this fraction is small, it is safe to simply 
  neglect this source of systematic uncertainty.

\end{itemize}

For $10~fb^{-1}$, the experimental uncertainties listed above account for a systematic
error of $\sim{11}\%$ for both $\rm t\bar{t}$ control regions. Including
the theoretical uncertainty this error does not exceed $16\%$.

\subsubsection{Statistical uncertainties}

The statistical precision with which the number of $\rm t\bar{t}$ events in the 
signal region can be known depends on the expected number of $\rm t\bar{t}$ 
events in the control region. From the numbers quoted in Table~\ref{doublebTag_table} 
 and assuming a Poissonian behavior it is clear that 
the error due to systematic uncertainties is predominant with
respect to the statistical ones for both of the proposed normalization procedures.

%%%%%%%%%%%%%%%%%%%%%%%%%%%%%%%%%%%%%%%%%%%%%%%%%%%%%%%%%%%%%%%%%%%%%%%%%%%%%
\section[Single resonant top production as background to the $H\to WW$ search]
{SINGLE RESONANT TOP PRODUCTION AS BACKGROUND TO THE $H\to WW$ SEARCH~\protect
\footnote{Contributed by: J.~Campbell, G.~Davatz, A.-S.~Giolo-Nicollerat,
    F.~Maltoni, S.~Willenbrock, M.~Zanetti}}

At leading order, the inclusive double resonant top production process,
$\rm pp\to t\bar{t} \to WbWb\to \ell\nu\ell\nu bb$, where $\rm
\ell=e,\mu,\tau$, has a cross section times branching ratio of about
52~pb. Single resonant top production $\rm pp\to Wtb$ represents a
contribution about ten times smaller. After applying a jet veto, the
singly resonant top contribution is increased with respect to the
doubly resonant one, since the $\rm b$-jet is typically produced at
much lower transverse momentum. It is this contribution which we
will study in detail here.

\begin{figure}[htbp]
\begin{center}
\includegraphics[scale=0.8]{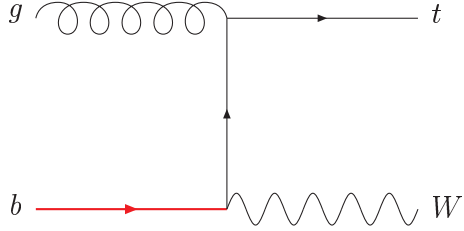}
\end{center}
\caption{A lowest order diagram for the singly resonant top production
process. A $\rm b$ quark is produced inside the proton via gluon
splitting and the resulting anti-b is unobserved.}
\label{wtdiag}
\end{figure}

In order to resum large logarithms of the form $\log [(m_t+m_W)/m_b]$, it
is preferable to view the singly resonant process as one in which a
$\rm b$ quark is probed directly inside the proton. In this case, the
single resonant leading order process is $\rm gb \to Wt$, as depicted
in Fig.~\ref{wtdiag}.
Starting from this process one can calculate NLO corrections, which
naively include the doubly resonant diagrams in the real radiation
contribution. Previous attempts to remove these contributions have
either relied on subtracting the doubly resonant cross
section~\cite{Tait:1999cf} or on  applying a mass window
cut~\cite{Belyaev:2000me}, both of which suffer from ambiguities
related to the interference between the singly and doubly resonant
graphs. However, by applying a veto on the presence of an extra $\rm b$
quark, the interference effect is greatly suppressed and the
contribution from the doubly resonant diagrams can be unambiguously
removed~\cite{Campbell:2005bb}.

Therefore we can estimate the singly resonant top production rate at
NLO in a region where a jet veto is applied, which in the case of the
Higgs search corresponds to the signal region. Clearly, the NLO
prediction for the rate depends on the region of phase space which is
probed, in particular on the definition of the jet veto. In the
following, we will study the sensitivity to NLO corrections of the
different kinematic variables used for $\rm H\to WW\to \ell\nu\ell\nu$.

The NLO corrections to $\rm Wt$ production, implemented using a veto on
an extra $\rm b$ quark, are calculated using the Monte Carlo program
MCFM~\cite{Campbell:2005bb,mcfm}. The factorization and renormalization
scales chosen to be equal to the jet veto value used, i.e. 40~GeV.
We have calculated the efficiencies obtained
for the Higgs selection cuts, which are defined in the chapter, `Top
background generation in the $\rm H\to WW$ channel' of these
proceedings. A comparison between the  LO and NLO  results is shown in
Table~\ref{mcfm_wt}. We note that, since MCFM is a parton level
generator, the jet veto actually corresponds to a veto directly on the
partons, requiring no parton with $\rm p_t>$ 40~GeV. Moreover, no
requirement on the lepton isolation is added.  Without selection cuts,
the effect of the NLO corrections is to increase the LO cross section
by a factor of about 1.4. After all selection cuts, this factor drops
to approximately 0.7 almost entirely due to the effect of the jet
veto. This is expected since the presence of an extra parton in
the NLO calculation means that a jet is vetoed more frequently. The
efficiency for the other selection cuts are very similar at LO and NLO.
In order to account for the difference in the jet veto efficiency between NLO
and LO, the K-factor that will be used to approximate the NLO cross
section is determined in the signal region by the ratio of the NLO to
LO cross sections of MCFM.

\begin{table}[htb]
\begin{center}
\caption{Higgs selection cut efficiencies for the singly resonant Wt
process at LO and NLO, simulated with MCFM (parton level)~\cite{mcfm} 
and TopReX (LO and parton shower). Here a veto is applied on the $\rm
p_t$ of the generated b and is set to 40~GeV. The cross section is
given after the following branching ratio has been included,
[$\rm W^+\to e^-\nu$][$\rm t\to e^-\nu \bar{b}$].
  \vspace*{1mm}
\label{mcfm_wt}}
\begin{tabular}{|l||c|c|c|c||c|}
\hline 
& \multicolumn{4}{c||}{MCFM} & TopReX\\
\hline
 & \multicolumn{2}{c|}{LO} & \multicolumn{2}{c||}{NLO}& LO \\
\hline
Selection cuts & $\rm \sigma\times BR$ & rel. eff & $\rm \sigma\times BR$ & rel. eff& rel. eff \\
& (fb) &  & (fb) & & \\
\hline
No cuts & 271 & & 377 & &\\
\hline
2 lep, \small{$\rm |\eta|<$~2, $\rm \rm p_t>$~20~GeV} & 204 & 0.75 & 277 & 0.73 & \\
\hline
$\rm E_t^{miss}>$~40 & 148 & 0.73 & 209 & 0.75 & 0.75\\
\hline
$\rm \phi_{\ell\ell}<$~45 & 20.8 & 0.14 & 34.4 & 0.16 & 0.17 \\
\hline
5~GeV~$\rm <m_{\ell\ell}<$~40~GeV & 10.6 & 0.51 & 15.6 & 0.45 & 0.50 \\
\hline
Partonic jet veto, 40~GeV & 1.55 & 0.15 & 1.12 & 0.07 & 0.16\\
\hline
30~GeV$\rm <p_{t}^{\ell\, max}<$55~GeV & 1.08 & 0.70 & 0.73 & 0.65 & 0.63\\
\hline
$\rm p_{t}^{\ell\, min}>$25~GeV & 0.73 & 0.68 & 0.49 & 0.67 & 0.67 \\
\hline
\end{tabular}
\end{center}
\end{table}

The cut selection efficiency obtained with MCFM is then compared to a
simulation performed using TopReX~\cite{Slabospitsky:2002ag}, in which the effects
of a parton shower are included. The cut efficiencies obtained using
this approach are shown in the third column of Table~\ref{mcfm_wt}.
TopReX and MCFM lead to very similar results, with the exception of
the jet veto.

\begin{figure}[htbp]
\begin{center}
\includegraphics[scale=0.8]{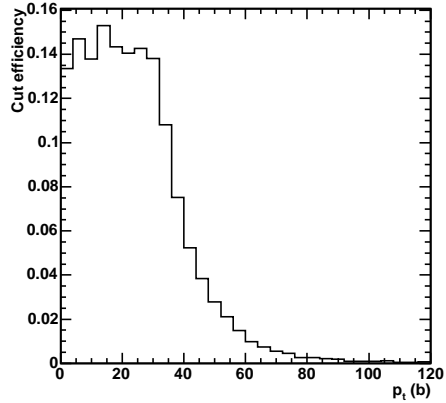}
\end{center}
\caption{Cut efficiency as a function of the transverse momentum of the b quark, after requiring two isolated leptons with $\rm p_t>$~20~GeV, $\rm |\eta|<$~2 and no reconstructed clustered jet with $\rm p_t>$~30~GeV for a simulation with TopReX.}
\label{ptb_eff}
\end{figure}

The difference between the efficiencies of the jet veto is a direct
consequence of the limitations of the parton level generator, MCFM.
Whereas MCFM includes no showering and thus applies the jet veto
directly at the parton level, the events produced by TopReX can be
vetoed according to jets produced by the shower\footnote{For this
study, as before, the jets are reconstructed using a cone algorithm on
the stable particles from the MC tree.}. It is clear that the
transverse momentum of the jet produced by the shower is not the same
as the $\rm p_t$ of the parton that is produced in the hard
interaction. We find that, at leading order, requiring no parton with
$\rm p_t>$~40~GeV has a similar efficiency as requiring no jets with
$\rm p_t>$~30~GeV. Thus a parton cut at 40~GeV will correspond to a jet
cut at 30~GeV. Fig.~\ref{ptb_eff} shows the selection efficiency as a
function of the $\rm p_t$ of the b for finding two leptons with $\rm
p_t>$ ~20~GeV and vetoing all clustered jets with $\rm p_t>$~30~GeV, for the
TopReX sample.
In this case, 85\% of the events have $\rm p_t(b)<$~40~GeV and  94\%
have $\rm p_t(b)<$~60~GeV.

A leading order parton shower Monte Carlo should thus already provide a
good approximation of this process.
A NLO K-factor of 0.7 has been determined in the
signal region for the Higgs search in the WW channel. In
particular, the veto on additional jet activity occurs at a transverse
momentum of 30~GeV, corresponding to a parton-level veto of 40~GeV. 
The theoretical error on the Wt cross section is estimated to be of the
order of 10-20\%, including PDF and scale variation~\cite{Campbell:2005bb}. 
Therefore a
conservative estimate of the Wt background in this region could be
obtained by simply using the LO cross section without any additional
K-factor, since it is predicted to be slightly less than unity at most.

%%%%%%%%%%%%%%%%%%%%%%%%%%%%%%%%%%%%%%%%%%%%%%%%%%%%%%%%%%%%%%%%%%%%%%%%%%%%%
\section[Study of PDF and QCD scale uncertainties in $pp \rightarrow ZZ
    \rightarrow 4\mu$ events at the LHC]
{STUDY OF PDF AND QCD SCALE UNCERTAINTIES IN $pp \rightarrow ZZ
\rightarrow 4\mu$ EVENTS AT THE LHC~\protect
\footnote{Contributed by: S.~Abdullin, D.~Acosta, P.~Bartalini,
        R.~Cavanaugh, A.~Drozdetskiy, A.~Korytov, G.~Mitselmakher,
    Yu.~Pakhotin, B.~Scurlock, A.~Sherstnev, H.~Stenzel}}
\subsection{Introduction}

The ${\rm q\bar{q}\rightarrow ZZ \rightarrow 4\mu}$ process is the main
irreducible background in searches for the Higgs boson via its ${\rm H
\rightarrow ZZ \rightarrow 4\mu }$ decay mode.
Theoretical uncertainties affect the current estimation of the physics reach for the search analysis and may turn into 
contributions to the total systematic errors on significance estimators, as background evaluation on a specific 
4-muon mass range often relies on extrapolations from regions with larger background statistics, 
which are based on Monte Carlo Models.
Normalization to higher rate processes like single $Z$ production may help to reduce these uncertainties.
This work concentrates on the estimation of current errors in the calculations for total and differential 
cross sections for the process ${\rm q\bar{q}\rightarrow ZZ \rightarrow 4\mu}$ arising from PDF and perturbative
uncertainties, following the guidelines described in~\cite{Bartalini:2005zu} for the evaluation of theoretical 
uncertainties in LHC analyses.

\subsection{Event Generation}
All results are obtained at NLO with MCFM~\cite{mcfm} version 4.0 interfaced to 
the standard Les Houches accord PDF package LHAPDF~\cite{lhapdf}.
%for a convenient selection of various PDF families and evaluation of their intrinsic uncertainties. 
The cross sections are evaluated within a typical experimental acceptance and for momentum cuts summarised
in Section~\ref{sec:selection}
%, optimized for three different higgs mass scenarios.
%The muonic decays of Z is considered (massless leptons).
The calculations with MCFM are carried out for a given fixed set of electroweak input parameters
using the effective field theory approach. The PDF family CTEQ61 provided by the CTEQ collaboration~\cite{Stump:2003yu} 
is taken as nominal PDF input.
Quantitative error analysis is performed following the prescription of reference~\cite{Pumplin:2001ct} using the 40 sets of CTEQ61.
Errors are propagated via re-weighting to the final observables.
MRST2001E given by the MRST group~\cite{Martin:2002aw} is considered as an additional cross check. 
The value of the strong coupling $\alpha_s$ is not a free input parameter for the cross section calculation but taken 
from the corresponding value in the PDF. 

%Important input parameters are renormalization and factorization scales. 
The dependence of the observables on the choice for renormalization and factorization scales 
is unphysical and should be regarded as one important contribution to the total uncertainty 
in the theoretical predictions accounting for missing higher orders in QCD calculations.
The reference cross sections and distributions  are obtained with $\mu_R = \mu_F = 2 M_Z$. 
Missing higher orders are estimated by independent variations of the two scales in the range $M_Z < \mu < 4 M_Z$,
following prescriptions applied to other processes~\cite{Jones:2003yv}.

\subsection{Definition of observables and event selection \label{sec:selection}}

In order to perform a generator-level study with MCFM, we select events fulfilling acceptance
and momentum cuts very much along the lines of the ones optimized for full simulation-level analysis (in progress). 
General pre-selection cuts and three different sets of selection cuts are defined, the latter
being driven by the Higgs search in four muon final states at low, average and high Higgs masses
($\rm M_H =$ 150, 250, and 500 GeV respectively).

The pre-selection cuts are: 
\begin{itemize}
\item There should be at least four such muons (2 opposite sign muon
      pairs) for an event to be considered.
\item ${\rm PT > 7 }$ GeV for all the four muons.
%      (for the barrel, ${\rm |\eta| < 1.1}$) or
%      ${\rm P > 9 }$ GeV (for the endcaps, ${\rm |\eta| > 1.1}$) for
%      all considered muons. These cuts correspond to a muon
%      reconstruction efficiency of 80-90\%.
\item Selected opposite sign muon pairs arising from Z/$\gamma$ decays should have
      invariant mass ${\rm M_{\mu+\mu-} > 12 }$ GeV. 
      This cut on ${\rm M_{\mu^+\mu^-} }$ removes low-mass resonances.
\end{itemize}

\noindent
The selection cuts are obtained from the pre-selection cuts, increasing the lower PT threshold on the four muons
to 10, 16 and 25 GeV for $\rm m_H =$ 150, 250, and 500 GeV respectively.

The notations we use in this work include:
\begin{itemize}
\item ${\rm M_{4\mu} }$ is the invariant mass of the four selected muons.
\item ${\rm PT_{4\mu} }$ is the transverse momentum of the four muons system.
\item ${\rm Z1 }$ (${\rm M_{\mu^+\mu^-} }$) refers to the muon pair
  with invariant mass closest to the ${\rm Z^0}$ mass and ${\rm Z2 }$ refers to the second muon pair selected from the rest of the muons with the highest PT.
%\item ${\rm \mu_1, ..., \mu_4 }$ are the four selected muons when they
%  are sorted by PT, largest to smallest.
\end{itemize}

\begin{figure}
    \begin{tabular}{p{.47\textwidth}p{.47\textwidth}} 
      \resizebox{\linewidth}{0.65\linewidth}{\includegraphics{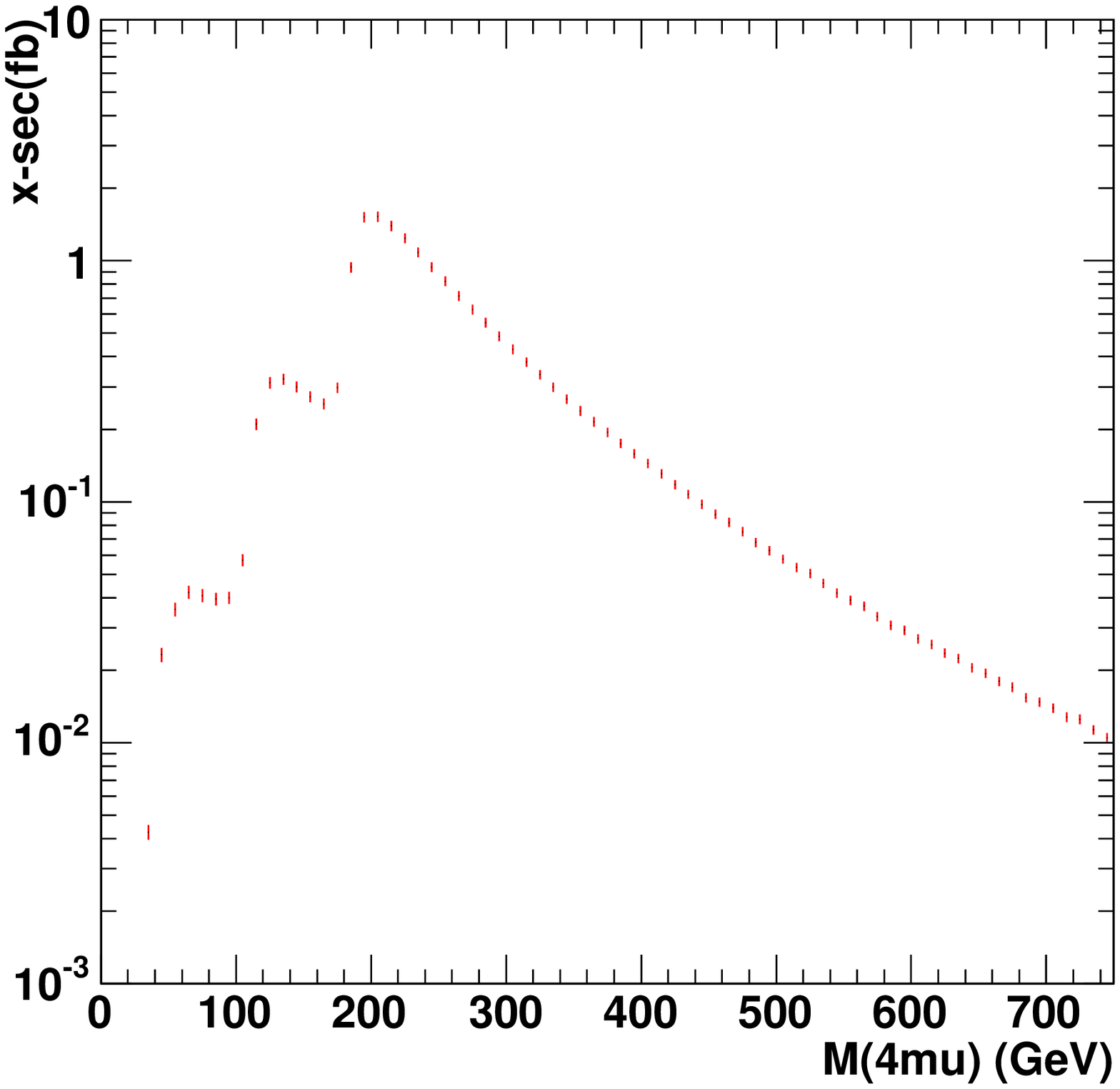}} &
      \resizebox{\linewidth}{0.65\linewidth}{\includegraphics{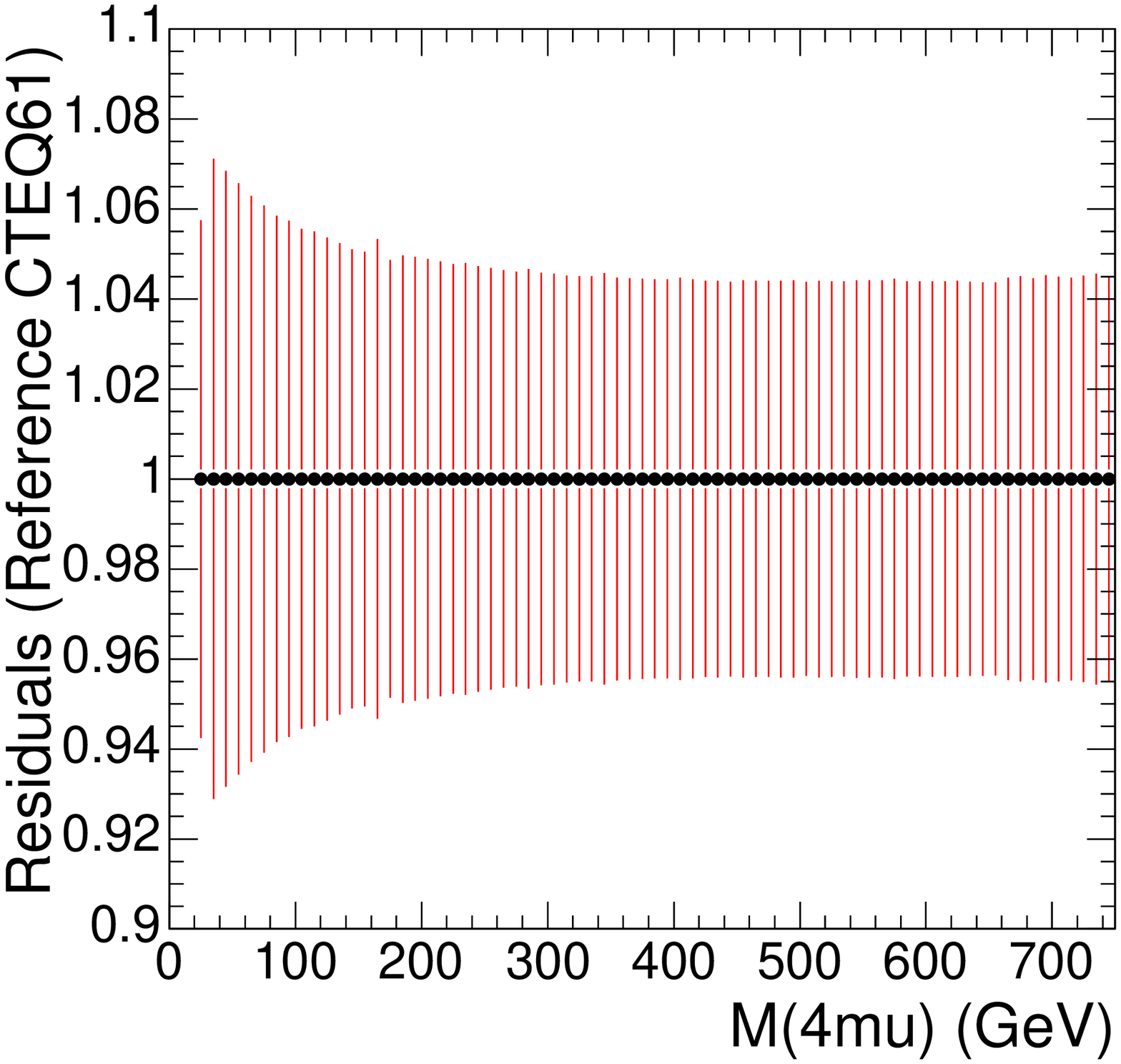}} \\
      \caption{${\rm M_{4\mu}}$ = four muon invariant mass distribution, normalized to femtobarns per 10 GeV bins.
      PDF = CTEQ61; $\mu_F = \mu_R = 2 * M_Z$. Symmetric error bars result from full error analysis with the CTEQ61 error 
      sets: they are reported as relative uncertainties in Fig.~\ref{m4mu_pdf}.}
      \label{m4mu_n} &
      \caption{${\rm M_{4\mu}}$ distribution: symmetric relative uncertainties from full error analysis with the CTEQ61 error sets.}
      \label{m4mu_pdf} \\
      \resizebox{\linewidth}{0.65\linewidth}{\includegraphics{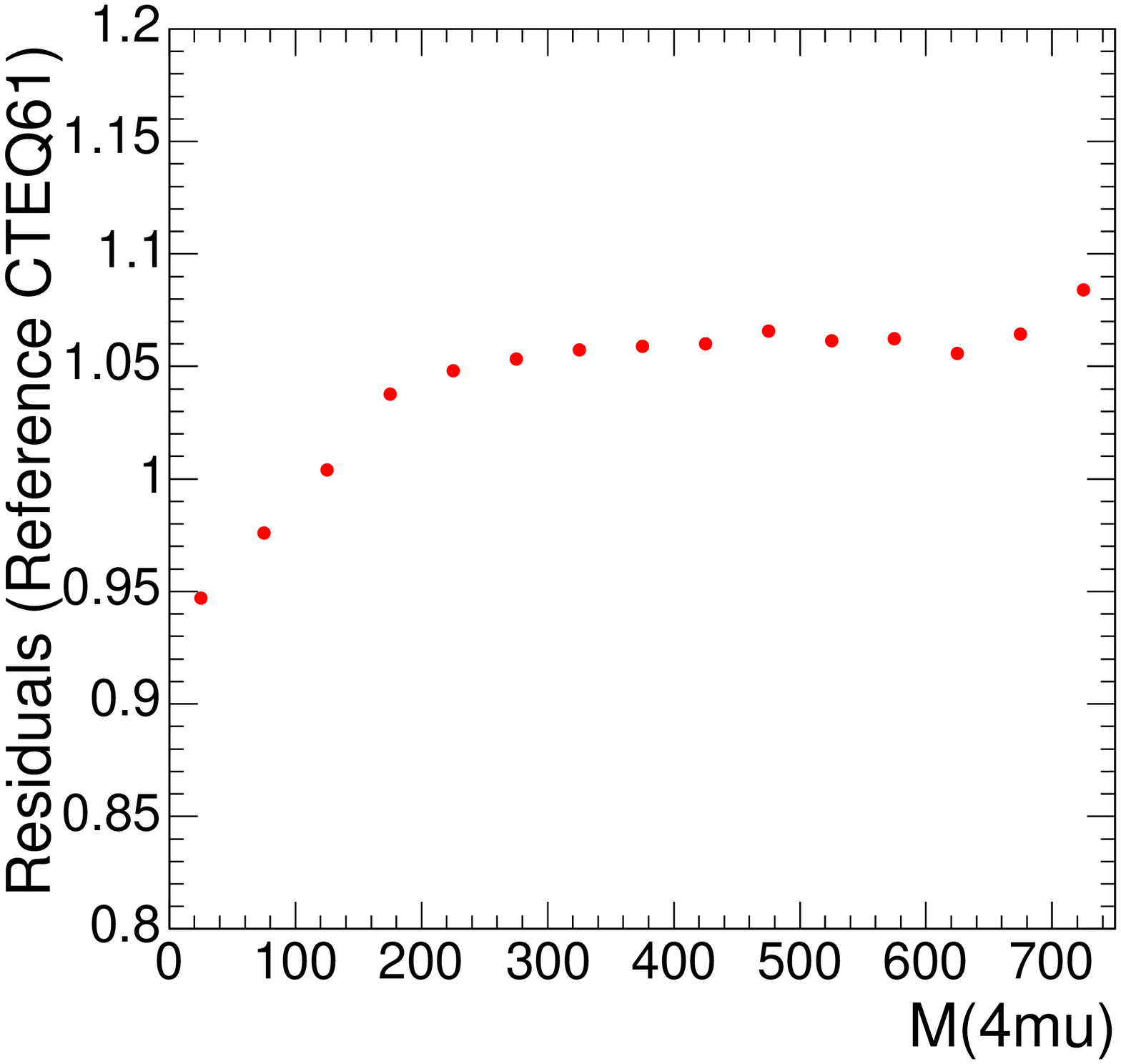}} &
      \resizebox{\linewidth}{0.65\linewidth}{\includegraphics{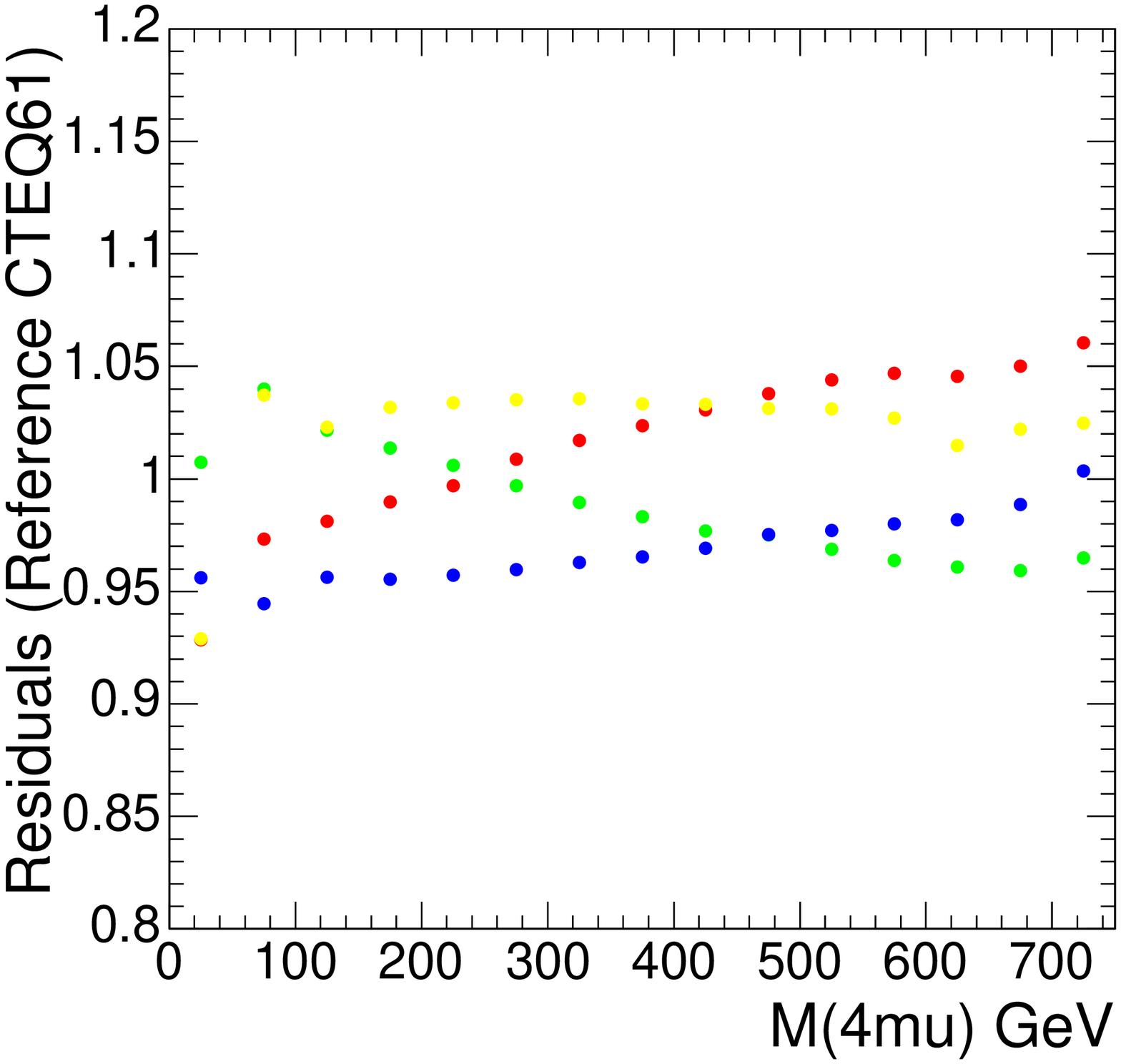}} \\
      \caption{Ratio between ${\rm M_{4\mu}}$ distributions obtained with PDF = MRST2001 and PDF= CTEQ61 respectively.}
      \label{m4mu_mrst} &
      \caption{${\rm M_{4\mu}}$ distribution according to four different renormalization and factorization scale settings with respect to the reference one ($\mu_F = \mu_R = 2 M_Z$):  $\mu_F = M_Z, \mu_R = M_Z$ (red); $\mu_F = 4 M_Z, \mu_R = M_Z$ (yellow); $\mu_F = M_Z, \mu_R = 4 M_Z$ (green); $\mu_F = 4 M_Z, \mu_R = 4 M_Z$ (blue).}
      \label{m4mu_s} \\
    \end{tabular}
\end{figure}
\begin{table}
  \small
  \caption{Relative uncertainty on total cross section
 $\sigma$(${\rm q\bar{q}\rightarrow ZZ \rightarrow 4\mu}$) with pre-selection cuts 
and on d$\sigma$/d${\rm M_{4\mu}}$ evaluated for three values of ${\rm M_{4\mu}}$ with selection cuts. 
Reference figures correspond to CTEQ61 PDF set and $\mu_F = \mu_R = 2 * M_Z$. 
Asymmetric errors arising from the choice of the QCD scales are obtained adopting independent variations of $\mu_F$ and $\mu_R$ in the range $M_Z < \mu < 4 M_Z$.
Symmetric errors from PDF parameterization are obtained using the CTEQ61 error sets. Comparison with reference MRST2001E predictions is also reported.
}
  \label{tab:ANA}
  \vspace*{1mm}
  \centering
    \begin{tabular}{||c|c|c|c|c||} \hline
                                    & $\Delta$($\sigma$) & $\Delta$(d$\sigma$/d${\rm M_{4\mu}}$) & $\Delta$(d$\sigma$/d${\rm M_{4\mu}}$) & $\Delta$(d$\sigma$/d${\rm M_{4\mu}}$) \\ \hline
                                & (pre-selection cuts)   & (${\rm M_{H}}$=150 GeV)               & (${\rm M_{H}}$=250 GeV)               & (${\rm M_{H}}$=500 GeV)               \\ \hline
    $\mu_F$ and $\mu_R$             & +3.2\%             & +2.3\%     & +3.4\%         & +3.8\%      \\
    scales                          & -4.0\%             & -4.4\%     & -4.3\%         & -2.5\%      \\ \hline
    PDF (CTEQ61)                    & $\pm$4.8\%         & $\pm$5.1\% & $\pm$4.7\%     & $\pm$4.4\%  \\ \hline
    $\Delta$(MRST2001E)             & +4.6\%             & +0.4\%     & +4.8\%         & +6.6\%      \\ \hline
    \end{tabular}
\end{table}
\begin{figure}
    \begin{tabular}{p{.47\textwidth}p{.47\textwidth}} 
      \resizebox{\linewidth}{0.65\linewidth}{\includegraphics{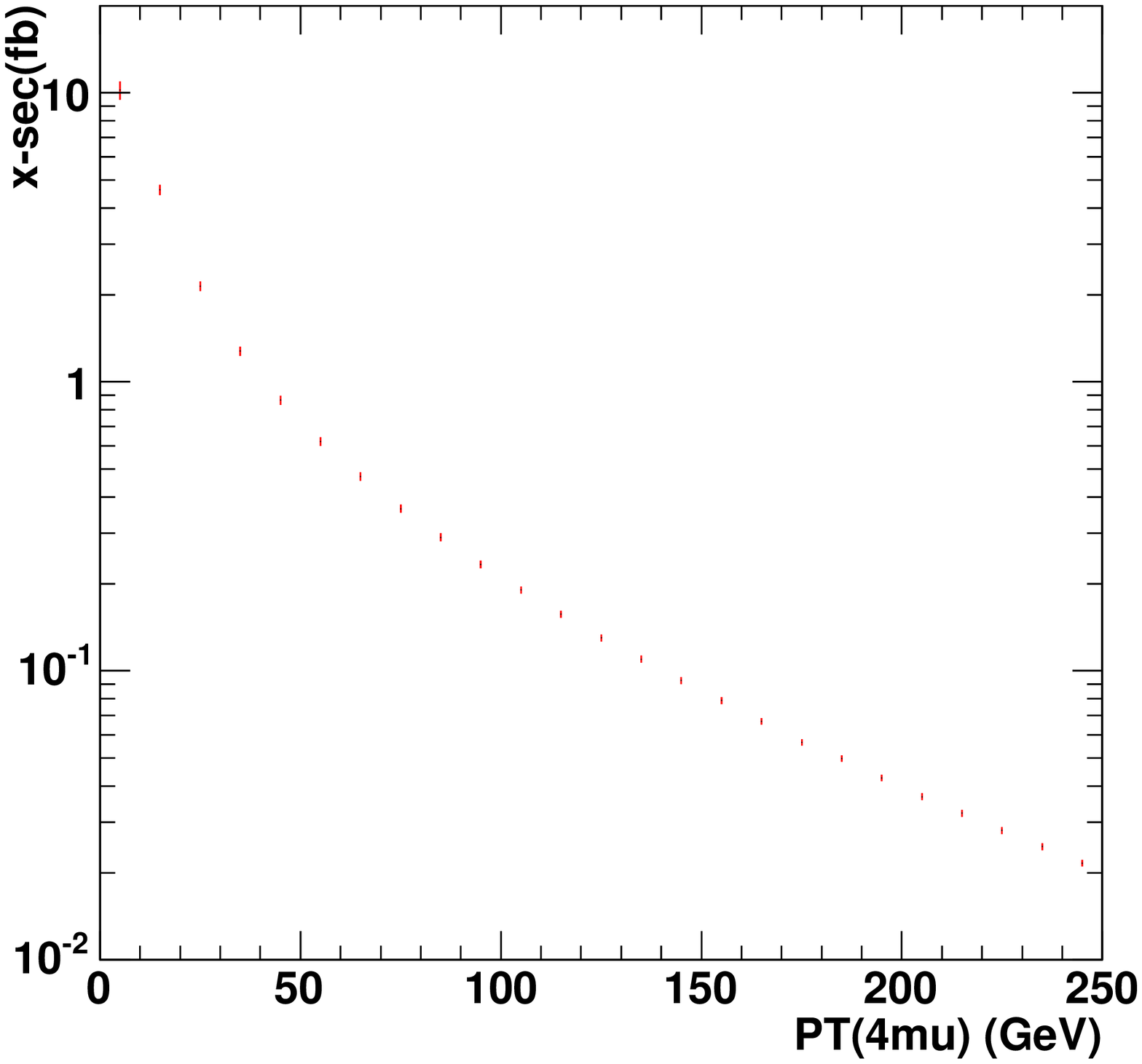}} &
      \resizebox{\linewidth}{0.65\linewidth}{\includegraphics{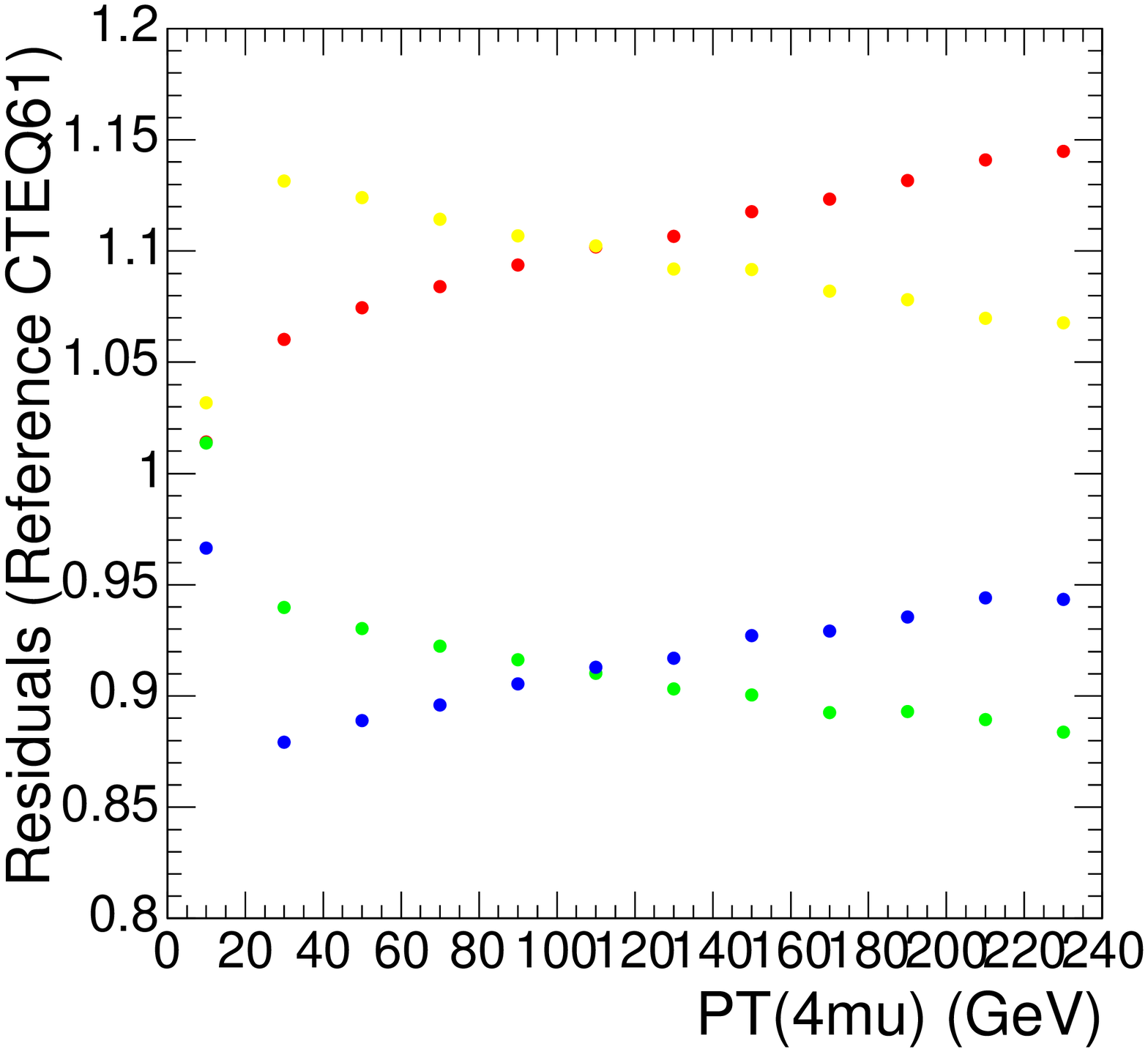}} \\
      \caption{PT(4$\mu$) = transverse momentum of the 4$\mu$ system. PDF = CTEQ61; $\mu_F = \mu_R = 2 * M_Z$. Symmetric error bars result from full error analysis with the CTEQ61 error sets.}
      \label{pt4mu_n} &
      \caption{PT(4$\mu$) distribution according to four different renormalization and factorization scale settings with respect to the reference one ($\mu_F = \mu_R = 2 M_Z$): $\mu_F = M_Z, \mu_R = M_Z$ (red); $\mu_F = 4 M_Z, \mu_R = M_Z$ (yellow); $\mu_F = M_Z, \mu_R = 4  M_Z$ (green); $\mu_F = 4 M_Z, \mu_R = 4 M_Z$ (blue).}
      \label{pt4mu_s} \\
    \end{tabular}
\end{figure}
\begin{figure}
    \begin{tabular}{p{.47\textwidth}p{.47\textwidth}} 
      \resizebox{\linewidth}{0.65\linewidth}{\includegraphics{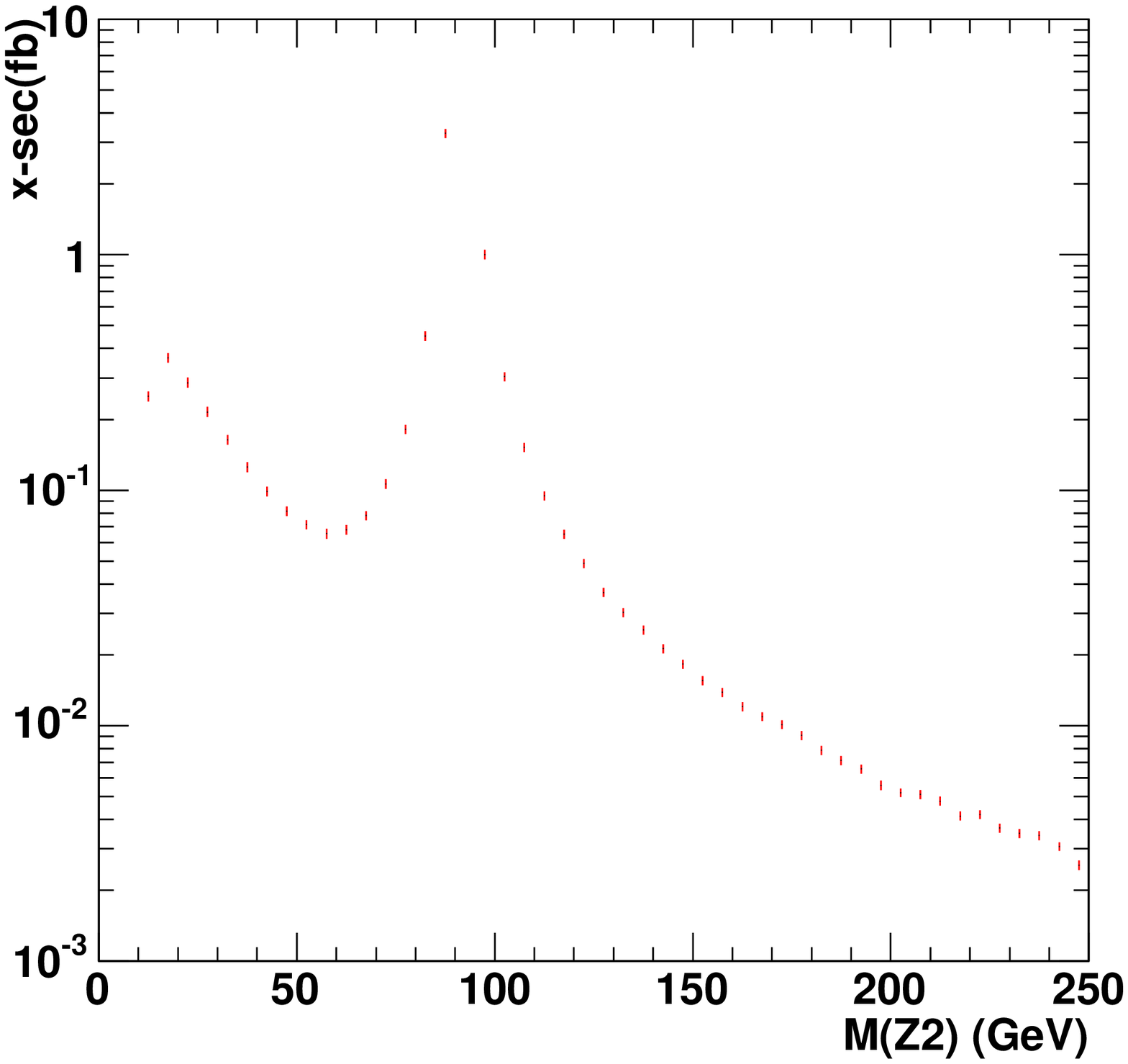}} &
      \resizebox{\linewidth}{0.65\linewidth}{\includegraphics{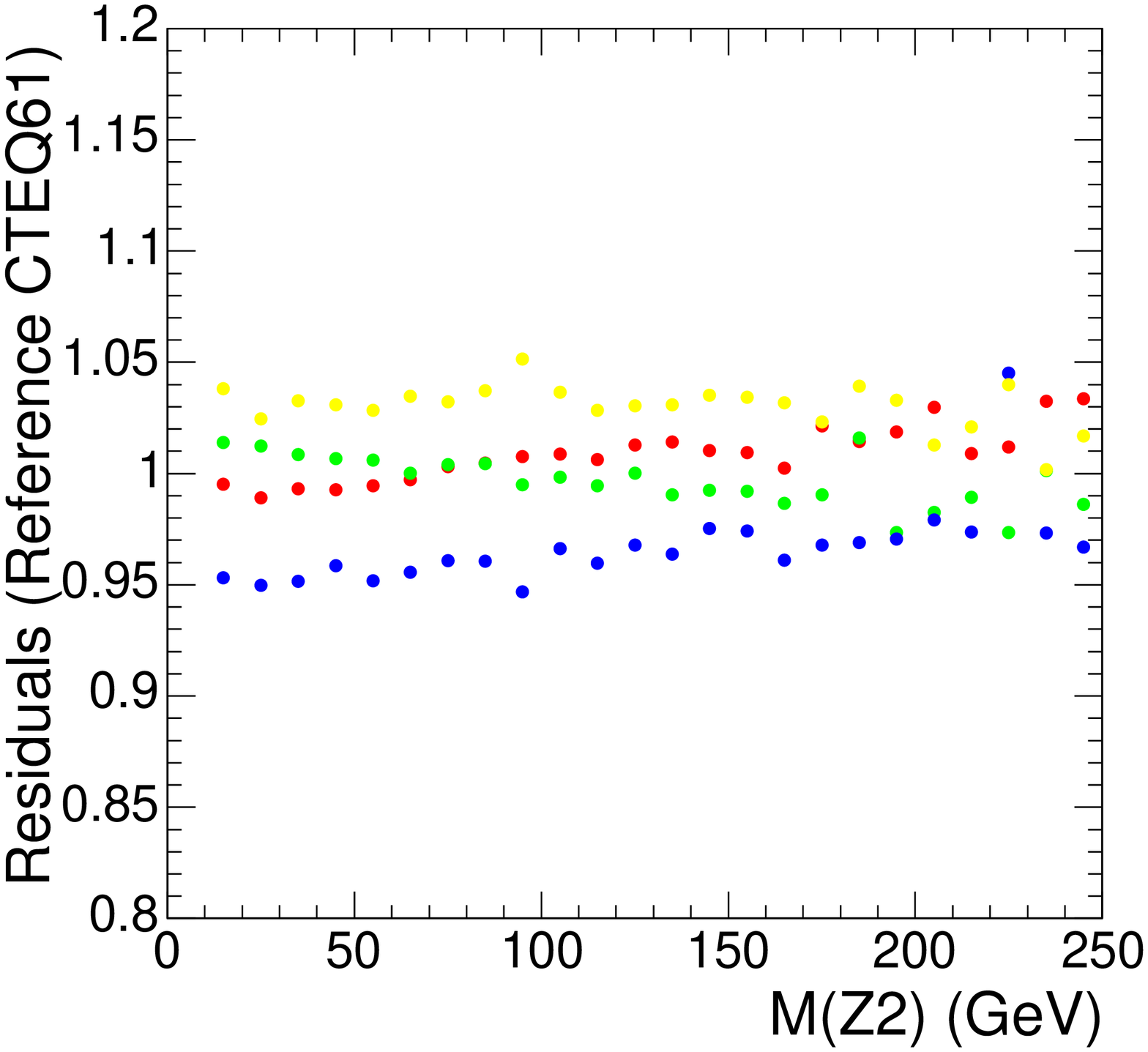}} \\
      \caption{Invariant mass of the Z2 candidate. PDF = CTEQ61; $\mu_F = \mu_R = 2 * M_Z$. Symmetric error bars result from full error analysis with the CTEQ61 error sets.}
      \label{mZ2_n} &
      \caption{Invariant mass of the Z2 candidate according to four different renormalization and factorization scale settings with respect to the reference one ($\mu_F = \mu_R = 2M_Z$): $\mu_F = M_Z, \mu_R = M_Z$ (red); $\mu_F = 4M_Z, \mu_R = M_Z$ (yellow); $\mu_F = M_Z, \mu_R = 4 M_Z$ (green); $\mu_F = 4 M_Z, \mu_R = 4 M_Z$ (blue).}
      \label{mZ2_s} \\
    \end{tabular}
\end{figure}
\begin{figure}
  \begin{center}
\includegraphics[width=0.35\textwidth]{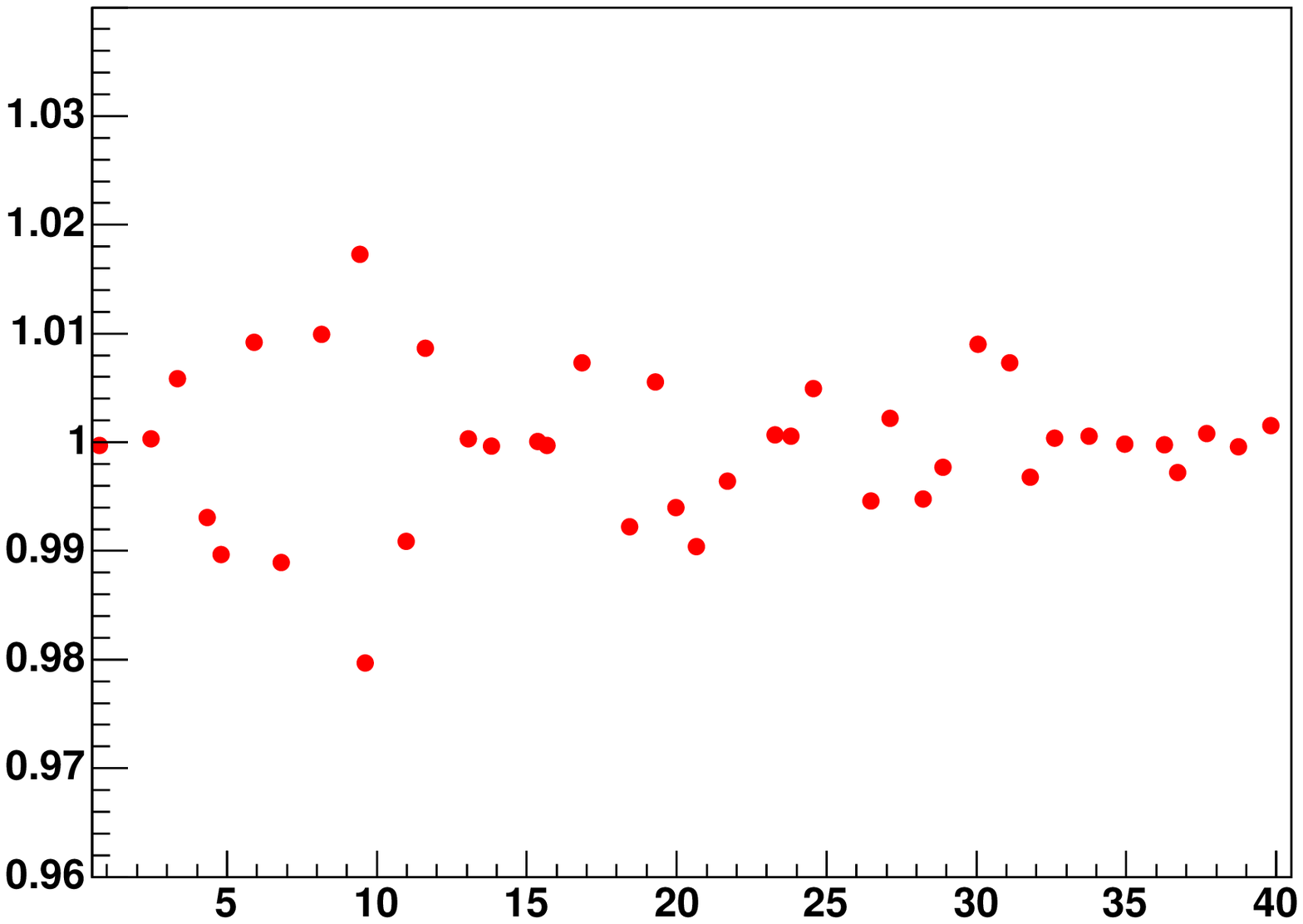}
\caption[]{Normalized ratio between $ZZ \rightarrow 4\mu$ and $Z \rightarrow 2\mu$ cross sections according to the 40 members of the CTEQ61 sets.} 
\label{norm}
  \end{center}
 \end{figure}

\subsection{Study of uncertainties from PDF and QCD scales}
The total effective cross section $\sigma$(${\rm q\bar{q}\rightarrow ZZ \rightarrow 4\mu}$) with pre-selection cuts
for CTEQ61 PDF set and $\mu_F = \mu_R = 2 * M_Z$ turns out to be 18.6~fb. 
The ${\rm M_{4\mu} }$ distribution is given in Fig.~\ref{m4mu_n}, along with uncertainties from CTEQ61
error analysis; the corresponding relative uncertainties are also reported in Fig.~\ref{m4mu_pdf}, which indicates a flat
behaviour for ${\rm M_{4\mu} } >$ 150 GeV.
An additional cross check is made in Fig.~\ref{m4mu_mrst}, which reports the comparison between the predictions of CTEQ61 
and MRST2001E PDFs. 

The effect of $\mu_F$ and $\mu_R$ variations on ${\rm M_{4\mu} }$ is shown in Fig.~\ref{m4mu_s}; one may notice that each of the four different combinations turns out to be  dominant as lower (or upper) error boundary in a given ${\rm M_{4\mu} }$ region, with an overall effect which results in flat boundaries. Adopting just $\mu_F$ - $\mu_R$ correlated variations would  underestimate the contribution of QCD scales to the total theoretical uncertainty.

All these results are summarized in Table~\ref{tab:ANA}. 
We quote 3-4\% effects arising from the variation of the QCD scales and 4-5\% effects from CTEQ61 error analysis, while MRST2001 predictions turn out to be consistent with CTEQ61 error boundaries. No sensitive dependency of the error boundaries with ${\rm M_{4\mu}}$ is observed.

In general, CTEQ61 error analysis achieves similar results for all the single muon, di-muons and four-muons kinematic distributions in ${\rm q\bar{q}\rightarrow ZZ \rightarrow 4\mu}$ events. 
QCD scale variations also achieve similar results for single muon distributions. 
However, more sensitive relative uncertainties of around 10-15\% 
are observed on four-muons and di-muons PT and pseudorapidity distributions
in ${\rm q\bar{q}\rightarrow ZZ \rightarrow 4\mu}$ events. See for example
Fig.~\ref{pt4mu_n}, which reports the ${\rm PT_{4\mu}}$ distribution and Fig.~\ref{pt4mu_s}, which shows the effect of 
of $\mu_F$ and $\mu_R$ variations on ${\rm PT_{4\mu}}$.

Fig.~\ref{mZ2_n} reports the invariant mass distribution of the Z2  along with uncertainties from CTEQ61
error analysis (corresponding to a relative error of around 4\% on the overall Z2 mass spectrum), while Fig.~\ref{mZ2_s} shows the effect of $\mu_F$ and $\mu_R$ variations on the same distribution, which turns out to be slightly more pronounced on the nominal mass of the $Z$ resonance.

The effects of the complete logarithmic electroweak $O(\alpha)$ corrections on
the production of vector-boson pairs at the LHC have been studied
in \cite{Accomando:2004de}. These corrections, that we don't take into account,
turn out to be relevant for $M_{4\mu}$ of the order of several 100GeV lowering
the Born level predictions by more than 10\% for $M_{4\mu} > 500 GeV$.

\subsection{Normalization to Drell-Yan}

Normalization to higher rate processes involving $q\bar{q}$ initial state may provide us with an experimental methodology to absorb part of the theoretical uncertainties arising from PDF and QCD scales. Single $Z$ boson events decaying to $\mu^+\mu^-$ events are generated with MCFM with pre-selection cuts (applied to di-muons final states) described in section~\ref{sec:selection} 
The total effective cross section $\sigma$(${\rm q\bar{q}\rightarrow Z \rightarrow \mu^+\mu^-}$) turns out to be 924~pb.
Fig.~\ref{norm} shows the prediction for the normalized ratio 
$\sigma$(${\rm q\bar{q}\rightarrow ZZ \rightarrow 4\mu}$)/$\sigma$(${\rm q\bar{q}\rightarrow Z \rightarrow 2\mu}$)
with pre-selection cuts according to the 40 members of CTEQ61.
PDF uncertainty on the ratio reduces to $\pm$3.2\%, against $\pm$4.8\% which is the value quoted for $\sigma$(${\rm q\bar{q}\rightarrow ZZ \rightarrow 4\mu}$) (Table~\ref{tab:ANA}).
A similar approach is followed for the uncertainty on the ratio arising from QCD scale variations, which gives an asymmetric error of +3.5 \% and -2.8\%.

As mentioned in the previous section, this study doesn't take into account electroweak corrections.
Although the size of these corrections turns out to be similar~\cite{Accomando:2005ra} between for single and double boson production, corrections to the ratio might be sensitive in the high $M_{4\mu}$ region.

\subsection{Acknowledgments}

We would like to thank 
M.~Aldaya,
P.~Arce, 
D.~Bourilkov,
J.~Caballero, 
J.~Campbell,
B.~Cruz,
S.~Ferrag,
U.~Gasparini,
P.~Garcia,
J.~Hernandez,
I.~Josa,
E.R.~Morales,
N.~Neumeister,
A.~Nikitenko,
T.~Sjostrand
for their active participation in the analysis discussions and comments on this letter.

%%%%%%%%%%%%%%%%%%%%%%%%%%%%%%%%%%%%%%%%%%%%%%%%%%%%%%%%%%%%%%%%%%%%%%%%%%%%%
\section[Relative contributions of $t$- and $s$-channels to the
    ${ZZ \rightarrow 4 \mu }$ process]
{RELATIVE CONTRIBUTIONS OF $t$- and $s$-CHANNELS TO THE
${ZZ \rightarrow 4 \mu }$ PROCESS~\protect
\footnote{Contributed by: S.~Abdullin, D.~Acosta, P.~Bartalini,
        R.~Cavanaugh, A.~Drozdetskiy, A.~Korytov, G.~Mitselmakher,
    Yu.~Pakhotin, B.~Scurlock, A.~Sherstnev}}
\subsection{Introduction}

\begin{figure}
    \begin{tabular}{p{.47\textwidth}p{.47\textwidth}} 
      \resizebox{\linewidth}{0.65\linewidth}{\includegraphics{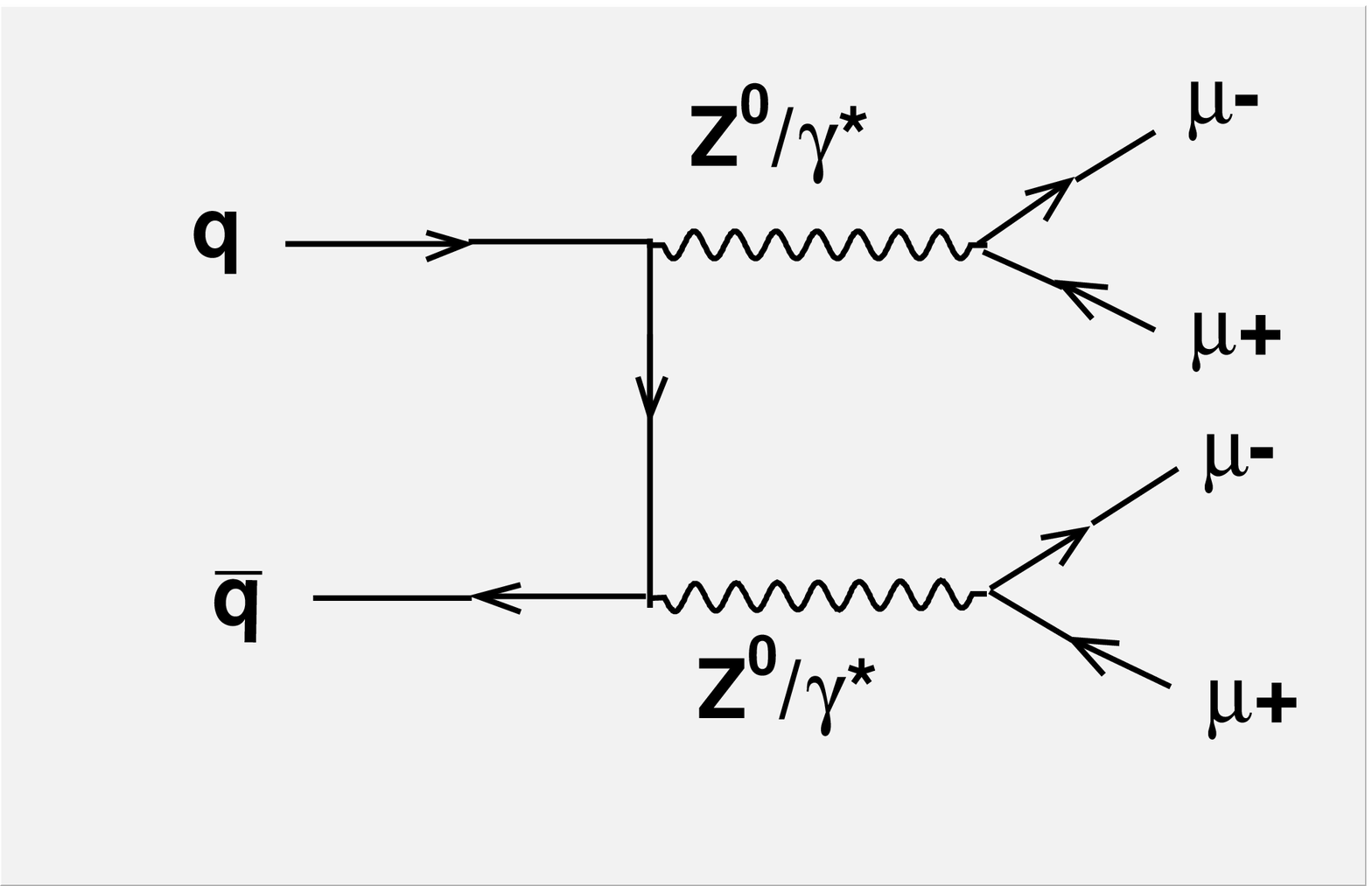}} &
      \resizebox{\linewidth}{0.65\linewidth}{\includegraphics{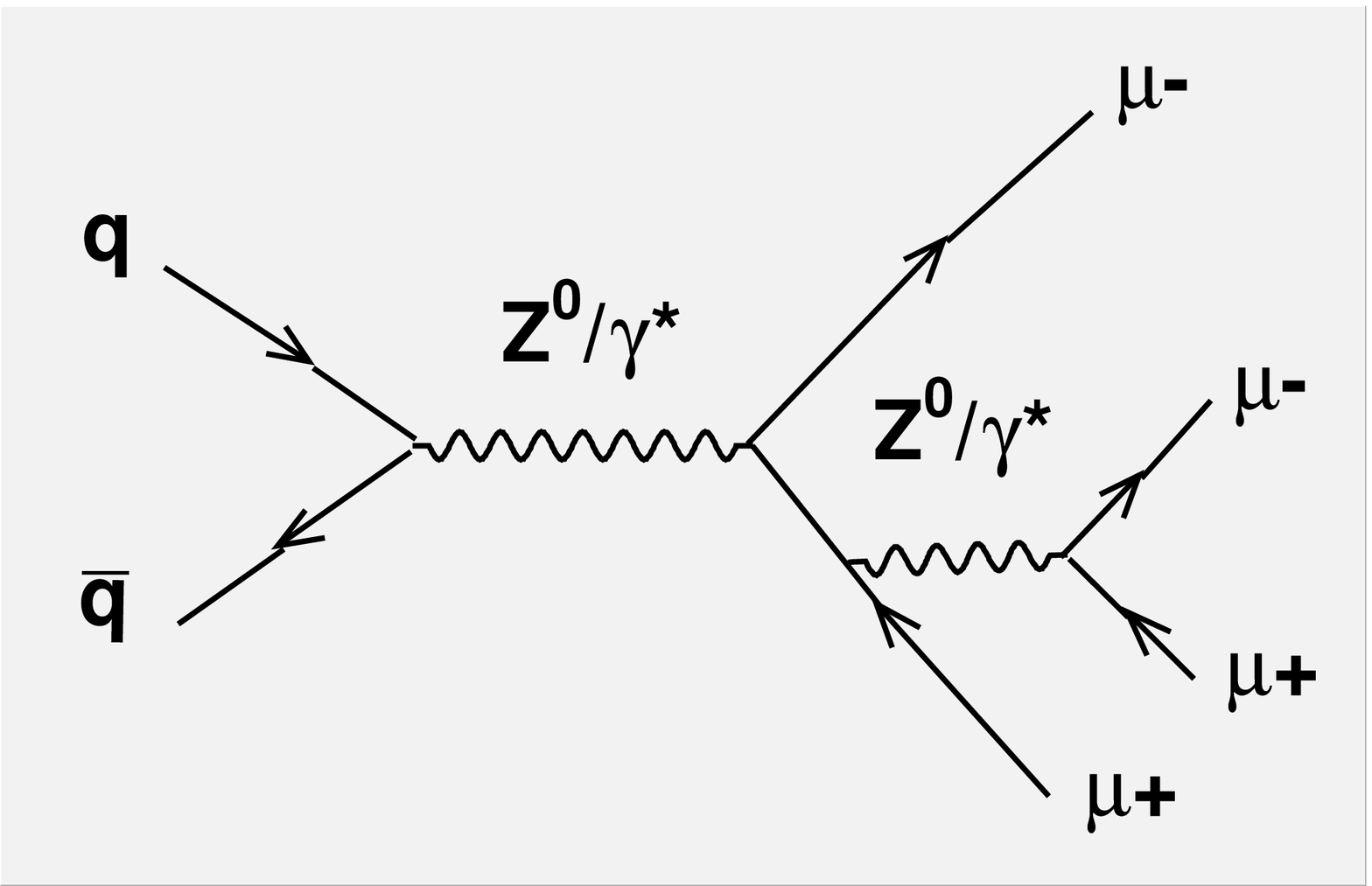}} \\
      \caption{ZZ background: t-channel diagram.} 
      \label{fig:tch} &
      \caption{ZZ background: s-channel diagram.}
      \label{fig:sch} \\
    \end{tabular}
\end{figure}

The ${\rm q\bar{q}\rightarrow ZZ \rightarrow 4\mu}$ process is the main
irreducible background in searches for the Higgs boson via its ${\rm H
\rightarrow ZZ \rightarrow 4\mu }$ decay mode. Figures
~\ref{fig:tch} and ~\ref{fig:sch} show the t- and s-channel contribution
diagrams. PYTHIA \cite{Sjostrand:2000wi}, an event generator commonly used for
simulation of this process at the LHC is unfortunately missing the
s-channel contribution. In this note, we show that the s-channel
sub-process and its interference with the t-channel cannot be
neglected if one aims to simulate the ZZ-background with a precision
of 10\% or better.

One may notice that very different kinematics are expected for the s-
and t-channel events. For example, the invariant mass of the four
muons for the s-channel contribution will tend to have a peak around
the ${\rm Z^0}$ mass with a tail to high invariant masses, because a Z
is radiated from one of the muon legs in the decay of the first Z,
whereas the t-channel has a more complicated structure with at least
two distinct peaks around the ${\rm Z^0}$ mass and twice the ${\rm
Z^0}$ mass, with a tail to even higher values.

\subsection{Event Generation}
\label{sec:presel}

For this study we used event samples of ZZ (by Z in the ZZ process
here and below we mean ${\rm Z/Z^*/\gamma^*}$) background produced with
PYTHIA only (PYTHIA parameters: MSEL = 0, MSUB 22 = 1) and with
CompHEP-PYTHIA. The latter uses the CompHEP \cite{Boos:2004kh} matrix
element (ME) generator interfaced to PYTHIA, which is used for
showering and hadronization in the same way and with the same
parameters as for the pure-PYTHIA sample. Below, we refer to
CompHEP-PYTHIA samples simply as CompHEP samples. The main
subprocesses resulting in a ${\rm 4\mu}$ final state are:

${\rm ZZ \rightarrow 4\mu }$ 

${\rm ZZ \rightarrow 2\tau 2\mu \rightarrow 4\mu }$ (not used in this
analysis)

${\rm ZZ \rightarrow 2b 2\mu \rightarrow 4\mu }$ (not used in this
analysis)

For normal analysis cuts, which select a region of ${\rm 4\mu}$ invariant
masses between about 110 and 170 GeV, we expect about 33 events from
the first process, about 4 events for the second and about 3 events
for the third. The latter one will become negligible after isolation
cuts. All event numbers in this note are normalized to ${\rm 30fb^{-1} }$ of
integrated luminosity.

We used the CTEQ5L PDF \cite{Stump:2002yv} and the ${\rm \hat{s} }$ ${\rm Q^2 }$ scale parameter
\cite{Sjostrand:2000wi} in both CompHEP and PYTHIA (the ${\rm \hat{s} }$ ${\rm Q^2 }$ scale is not
a default in PYTHIA 6.223). Generator-level ``pre-selection'' cuts are: ${\rm PT > 3 }$ GeV,
${\rm |\eta| < 2.5 }$ for all four muons. The PYTHIA sample's
generation-level ``pre-selection'' cuts: ${\rm PT > 3 }$ GeV,
${\rm |\eta| < 2.5 }$ for the four selected muons. Additional cuts on the
invariant masses of any two pairs of selected opposite sign muons are:
${\rm 5 < M_{\mu+\mu-} < 150 }$ GeV (the cross sections, especially for the
s-channel, are sensitive to the lower limit; the upper limit, once it
is sufficiently higher than ${\rm m_{Z^0}}$, is not important).

\subsection{Event selection and analysis cuts}
\label{sec:selec}

To perform a generator-level study, we select events as for the full
simulation-level analysis in progress (selection cuts).

The selection cuts are: 
\begin{itemize}
\item ${\rm PT > 7 }$ GeV (for the barrel, ${\rm |\eta| < 1.1}$) or
      ${\rm P > 9 }$ GeV (for the endcaps, ${\rm |\eta| > 1.1}$) for
      all considered muons. These cuts correspond to a muon
      reconstruction efficiency of 80-90\%.
\item There should be at least four such muons (2 opposite sign muon
      pairs) for an event to be considered.
\item All four permutations of opposite sign muon pairs should have
      invariant mass ${\rm M_{\mu+\mu-} > 12 }$ GeV (for the four
      muons selected). This cut on ${\rm M_{\mu^+\mu^-} }$ removes
      low-mass resonances.
\end{itemize}

We also use in this study an example of analysis cuts optimized for
small Higgs boson masses (${\rm m_H < 160 }$ GeV for the full
simulation-level analysis in progress), as listed in
Table~\ref{tab:ANAxyz}.

The notations we use for the analysis-level cuts include:
\begin{itemize}
\item ${\rm Z1 }$ (${\rm M_{\mu^+\mu^-} }$) refers to the muon pair
  with invariant mass closest to the ${\rm Z^0}$ mass and ${\rm Z2 }$
  refers to the second muon pair selected from the rest of the muons
  with the highest PT.
\item ${\rm \mu_1, ..., \mu_4 }$ are the four selected muons when they
  are sorted by PT, largest to smallest.
\item ${\rm M_{4\mu} }$ is the invariant mass of the
four selected muons.
\end{itemize}

\begin{table}[htb]
  \small
  \caption{Analysis-level cuts (example of cuts optimized for the
  small Higgs boson mass region, ${\rm M_{4\mu} < 160 }$ GeV).}
  \vspace*{1mm}
  \label{tab:ANAxyz}
  \centering
    \begin{tabular}{||c|c||} \hline
      parameter & cut, GeV \\ \hline
      PT ${\rm \mu_1 }$ & 14 \\ \hline
      PT ${\rm \mu_2 }$ & 10 \\ \hline
      PT ${\rm \mu_3 }$ & 10 \\ \hline
      PT ${\rm \mu_4 }$ & 7 \\ \hline
      Z1 (${\rm M_{\mu+\mu-} }$) & $>60$ \\ \hline
      Z1 (${\rm M_{\mu+\mu-} }$) & $<110$ \\ \hline
      Z2 (${\rm M_{\mu+\mu-} }$) & $>12$ \\ \hline
      Z2 (${\rm M_{\mu+\mu-} }$) & $<60$ \\ \hline
      ${\rm M_{4\mu} }$ & $>110$ \\ \hline
    \end{tabular}
\end{table}

\subsection{CompHEP vs. PYTHIA: comparison of t-channel only samples}

Before making a comparison of events for which the s- and t-channel
diagrams are included (CompHEP) with t-channel diagram events only
(pure-PYTHIA), we compare t-channel CompHEP and t-channel pure-PYTHIA
events. This cross check is necessary to be sure that the effect, if
it exists, is not due to a difference in internal cuts, model
parameters or something similar, but indeed is a consequence of taking
the s-channel into account, as well as interference between the s- and
t-channels.

\begin{figure}
    \begin{tabular}{p{.47\textwidth}p{.47\textwidth}} 
      \resizebox{\linewidth}{0.65\linewidth}{\includegraphics{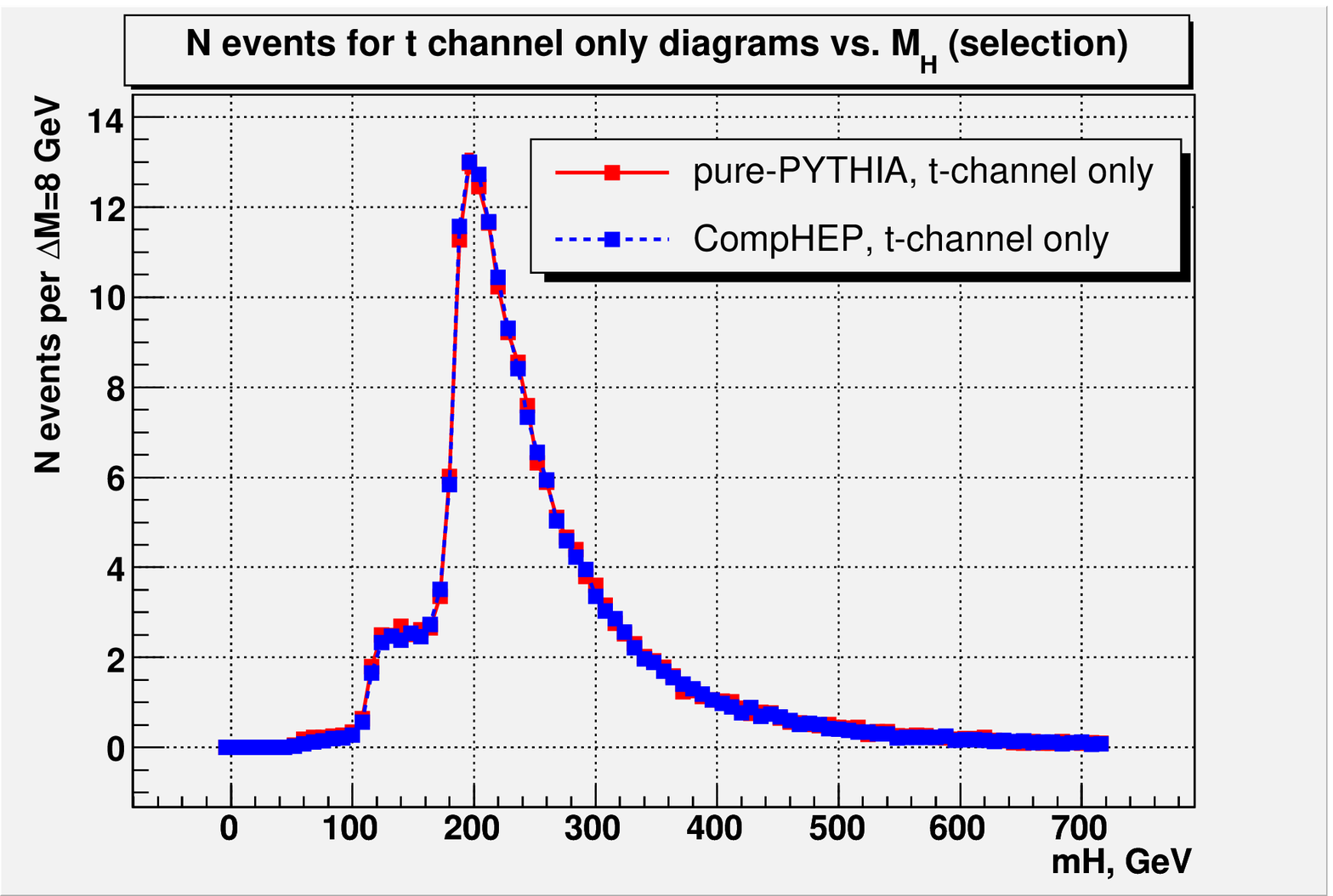}} &
      \resizebox{\linewidth}{0.65\linewidth}{\includegraphics{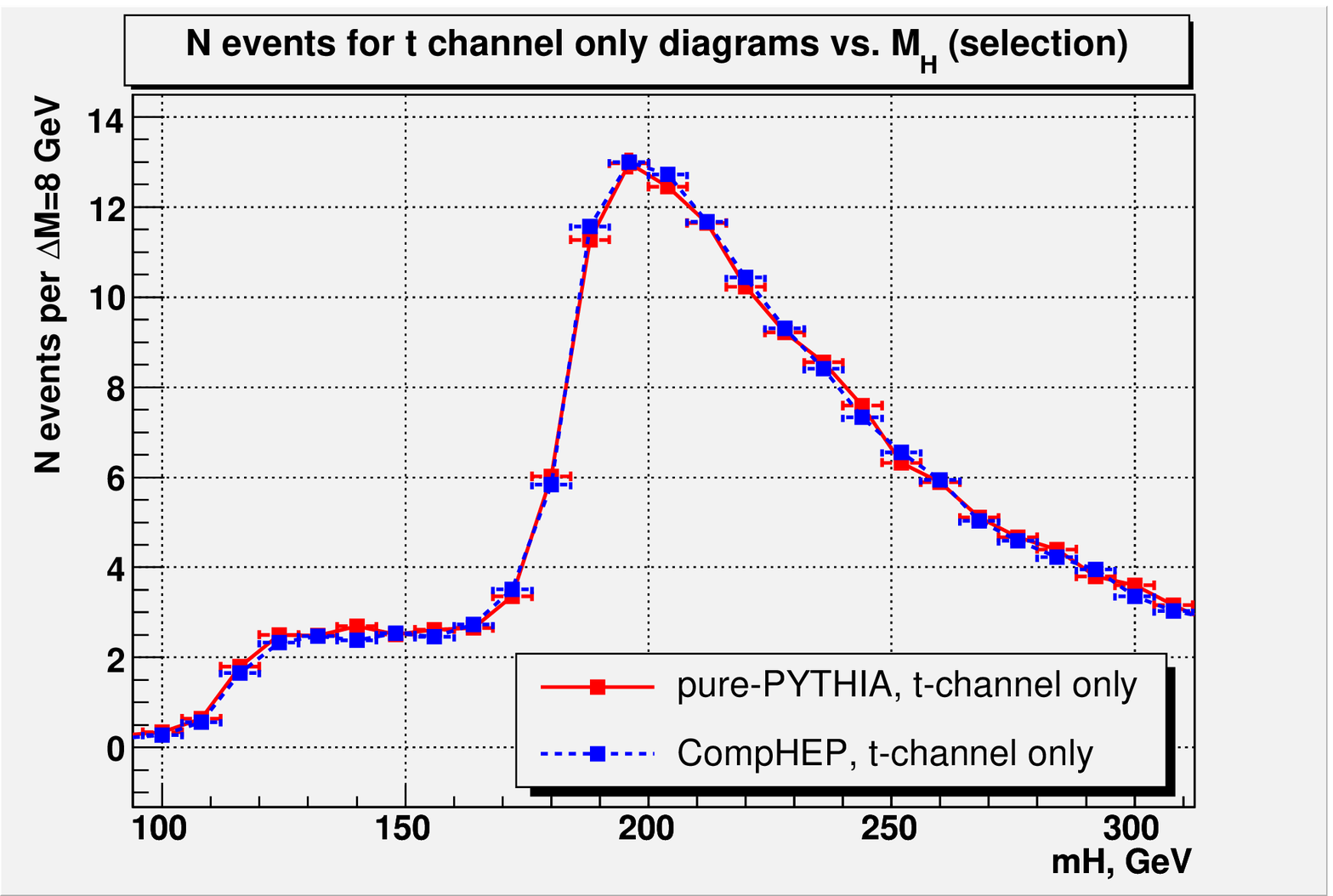}} \\
      \caption{${\rm 4\mu}$ invariant mass distribution after selection
	cuts, ${\rm L = 30 ~fb^{-1} }$. Comparison of t-channel CompHEP
	generated events and t-channel PYTHIA ones. Error bars include the
	MC statistical contribution only.}
      \label{fig:tt} &
      \caption{Enlarged part of Fig.~\ref{fig:tt}, ${\rm 100 < M_{4\mu} <
	  300 }$ GeV.}
      \label{fig:tt1} \\
    \end{tabular}
\end{figure}

\begin{figure}
    \begin{tabular}{p{.47\textwidth}p{.47\textwidth}} 
      \resizebox{\linewidth}{0.65\linewidth}{\includegraphics{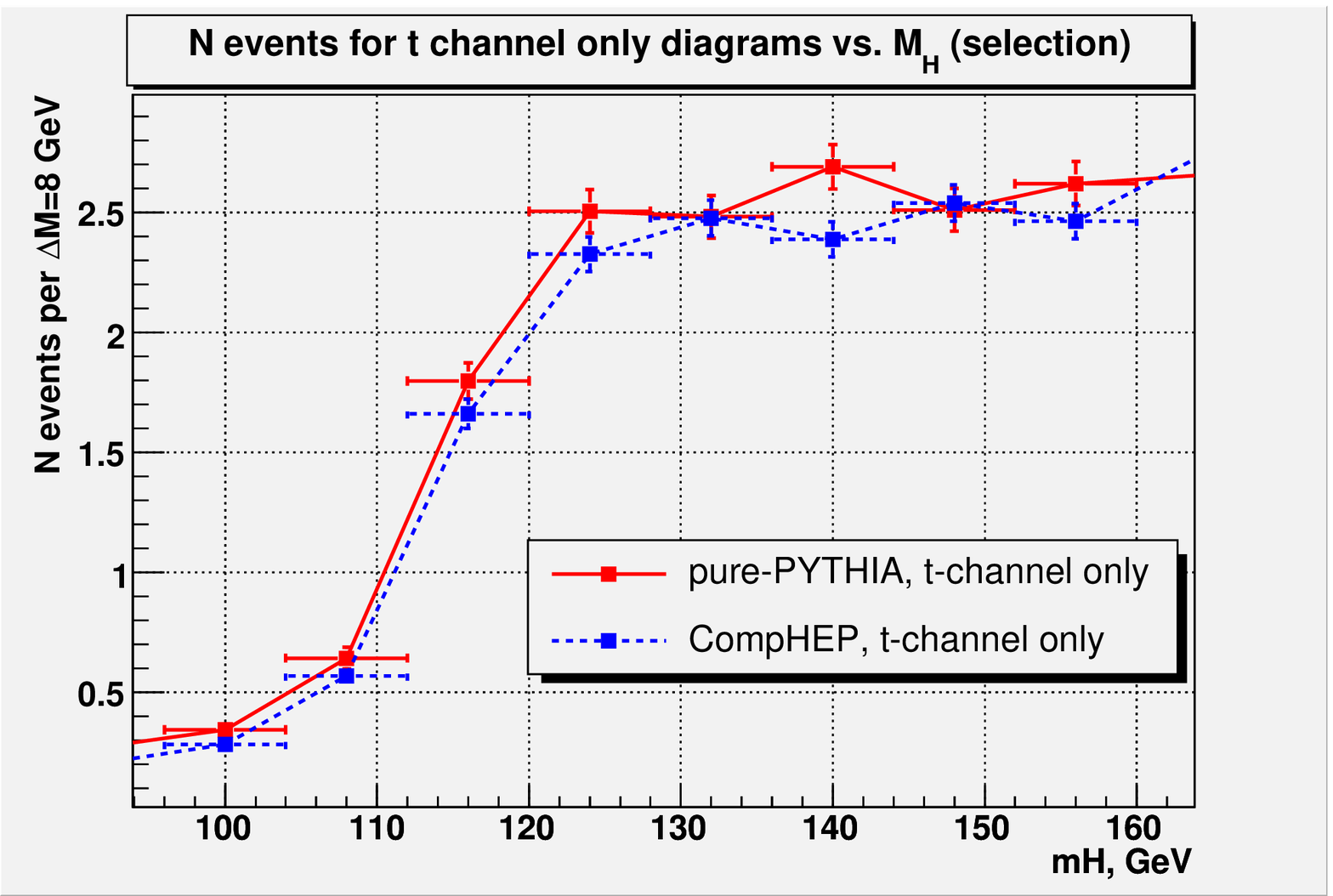}} &
      \resizebox{\linewidth}{0.65\linewidth}{\includegraphics{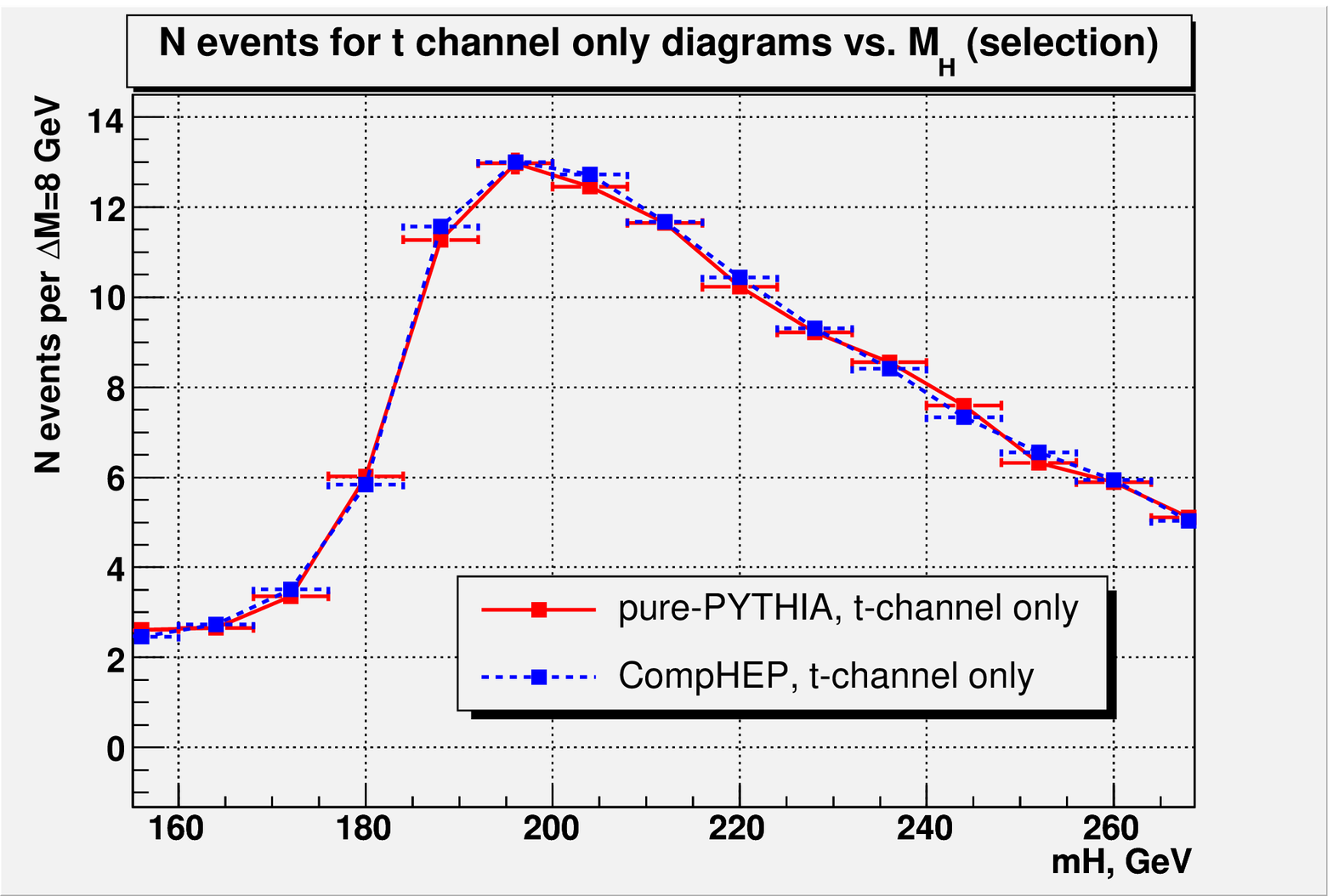}} \\
      \caption{Enlarged part of Fig.~\ref{fig:tt}, ${\rm 100 < M_{4\mu} <
	  160 }$ GeV. }
      \label{fig:tt2} &
      \caption{Enlarged part of Fig.~\ref{fig:tt}, ${\rm 160 < M_{4\mu} <
	  260 }$ GeV.}
      \label{fig:tt3} \\
    \end{tabular}
\end{figure}

Figures \ref{fig:tt},~\ref{fig:tt1},~\ref{fig:tt2} and \ref{fig:tt3}
show the results of the t-channel only comparison. Figure \ref{fig:tt}
shows the entire ${\rm M_{4\mu} }$ interval of interest, and
Figs. \ref{fig:tt1},~\ref{fig:tt2} and \ref{fig:tt3} show different
sub-intervals for better comparison. It is clear, that the t-channel
only samples generated with PYTHIA and with CompHEP have almost
identical ${\rm M_{4\mu} }$ spectra (up to the level of the
statistical precision of the results).

\subsection{Comparison of t- and s-channel sample (CompHEP)
  vs. t-channel sample only (pure-PYTHIA)}

We now compare the s- and t-channel CompHEP events to t-channel only
PYTHIA events.

\begin{figure}
    \begin{tabular}{p{.47\textwidth}p{.47\textwidth}} 
      \resizebox{\linewidth}{0.65\linewidth}{\includegraphics{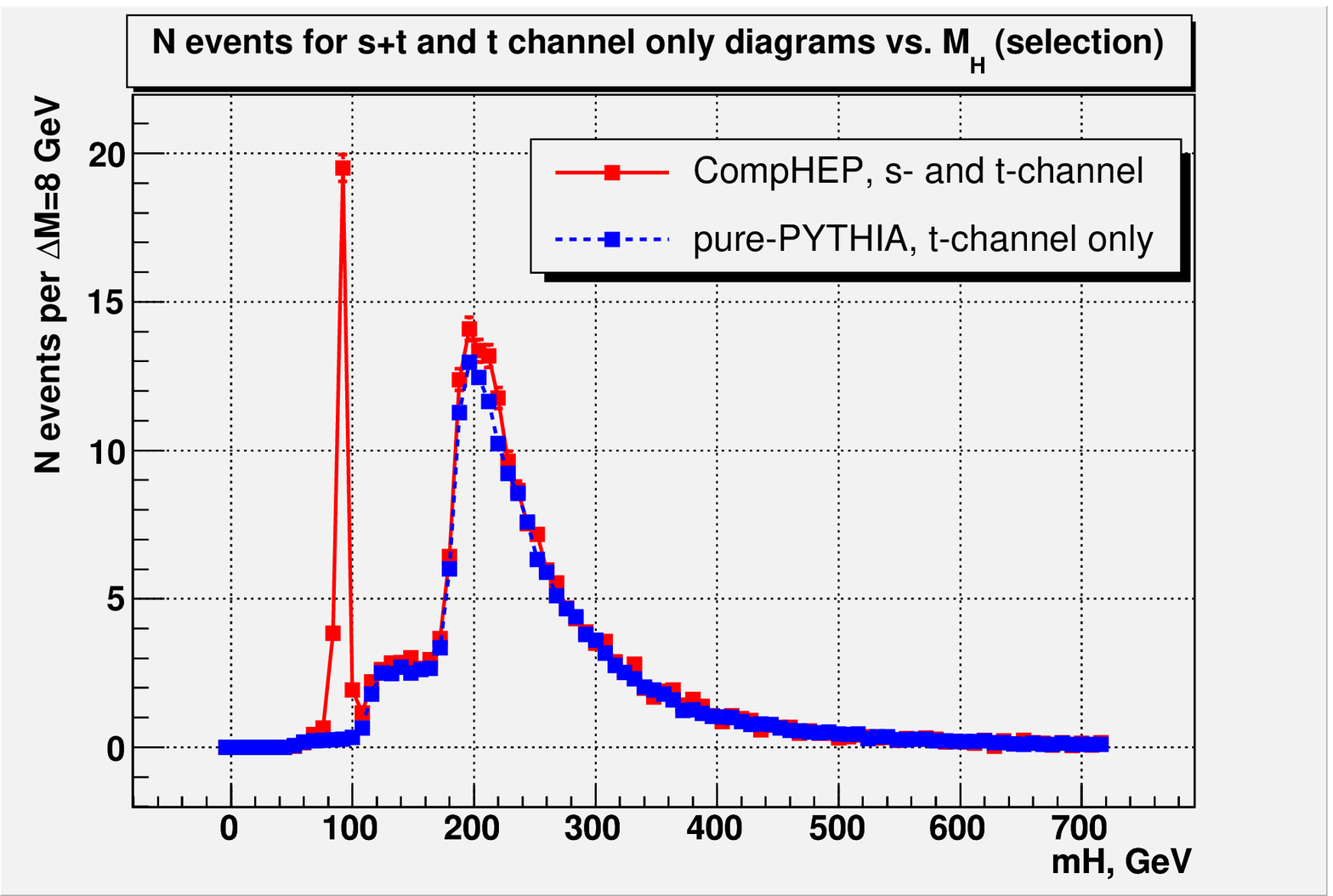}} &
      \resizebox{\linewidth}{0.65\linewidth}{\includegraphics{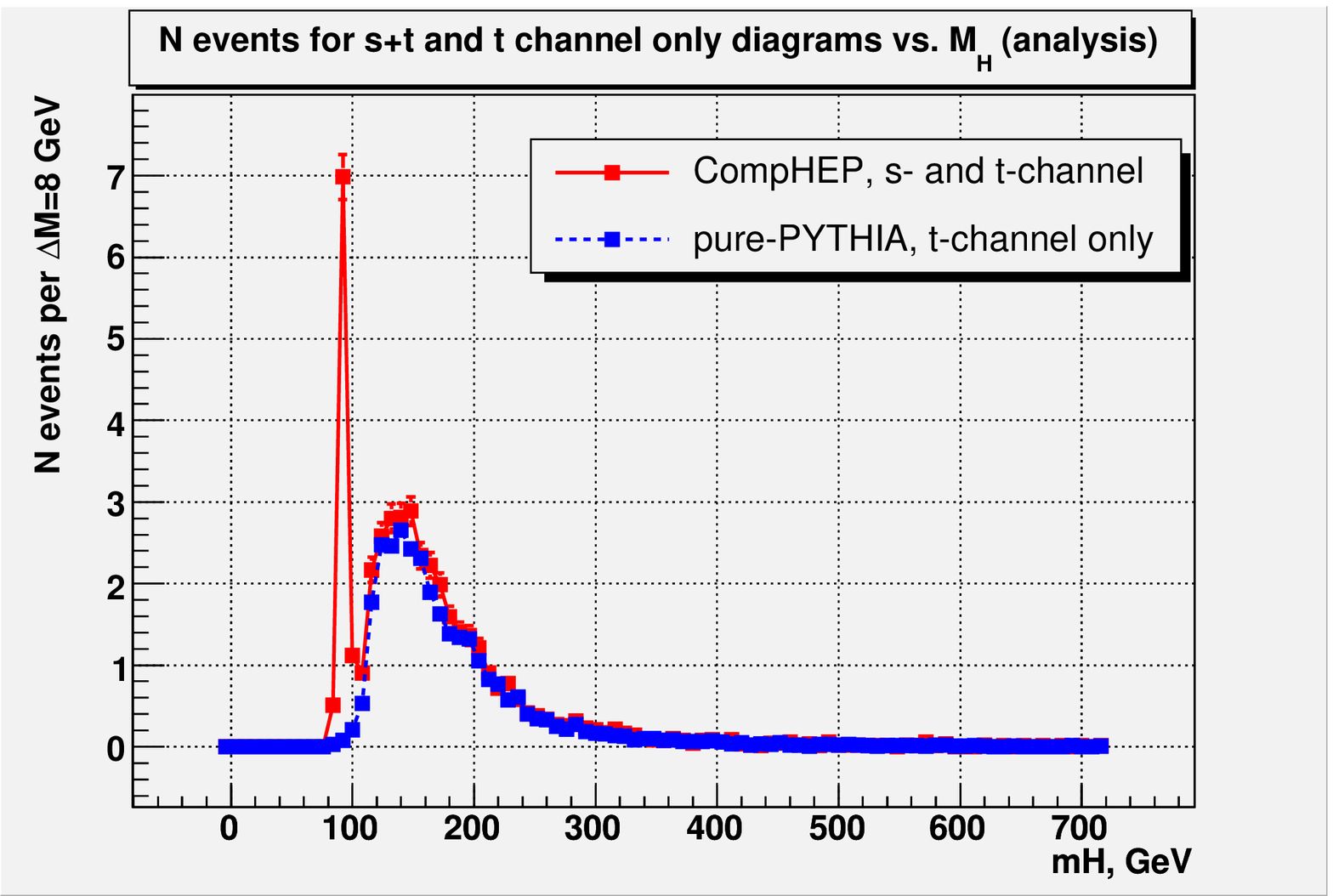}} \\
      \caption{${\rm 4\mu}$ invariant mass distribution after selection
	cuts, ${\rm L = 30 ~fb^{-1} }$. Comparison of s- plus t-channel
	CompHEP generated events and t-channel only PYTHIA ones. Error bars
	include MC statistical contribution only.}
      \label{fig:sel} &
      \caption{Same as Fig.~\ref{fig:sel} but after analysis cuts (see
	Table~\ref{tab:ANAxyz}).}
      \label{fig:ana} \\
    \end{tabular}
\end{figure}

There are three regions of interest in the ${\rm 4\mu }$ invariant
mass (${\rm M_{4\mu} }$). The first one is near the ${\rm Z^0}$
mass. Because of the s-channel, in particular, this region has a peak.
The peak is clearly seen after both selection and analysis cuts
(optimized for small ${\rm m_H }$ region), see Figs.~\ref{fig:sel} and
\ref{fig:ana}.

\begin{figure}
  \begin{tabular}{p{.47\textwidth}p{.47\textwidth}} 
    \resizebox{\linewidth}{0.65\linewidth}{\includegraphics{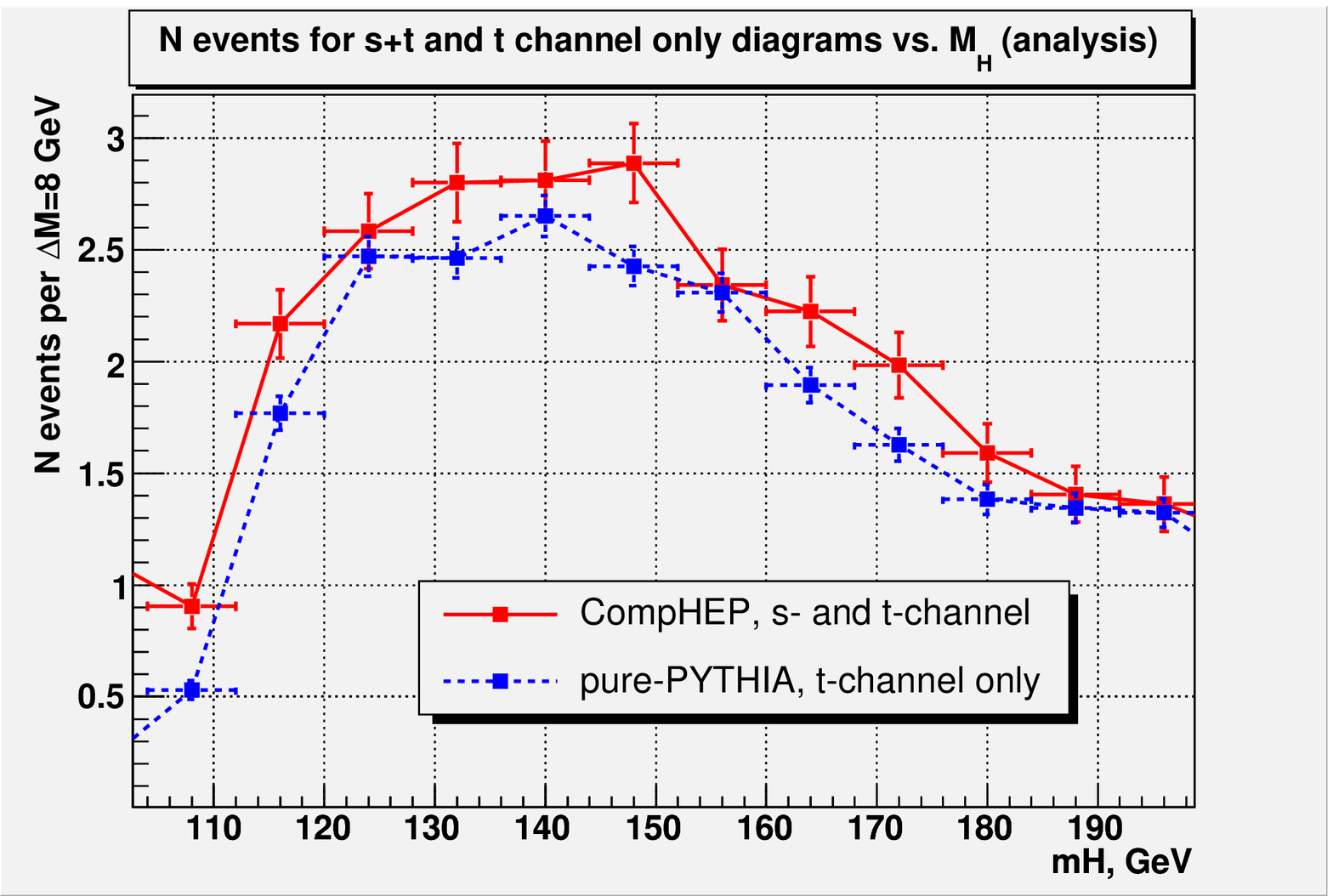}} &
    \resizebox{\linewidth}{0.65\linewidth}{\includegraphics{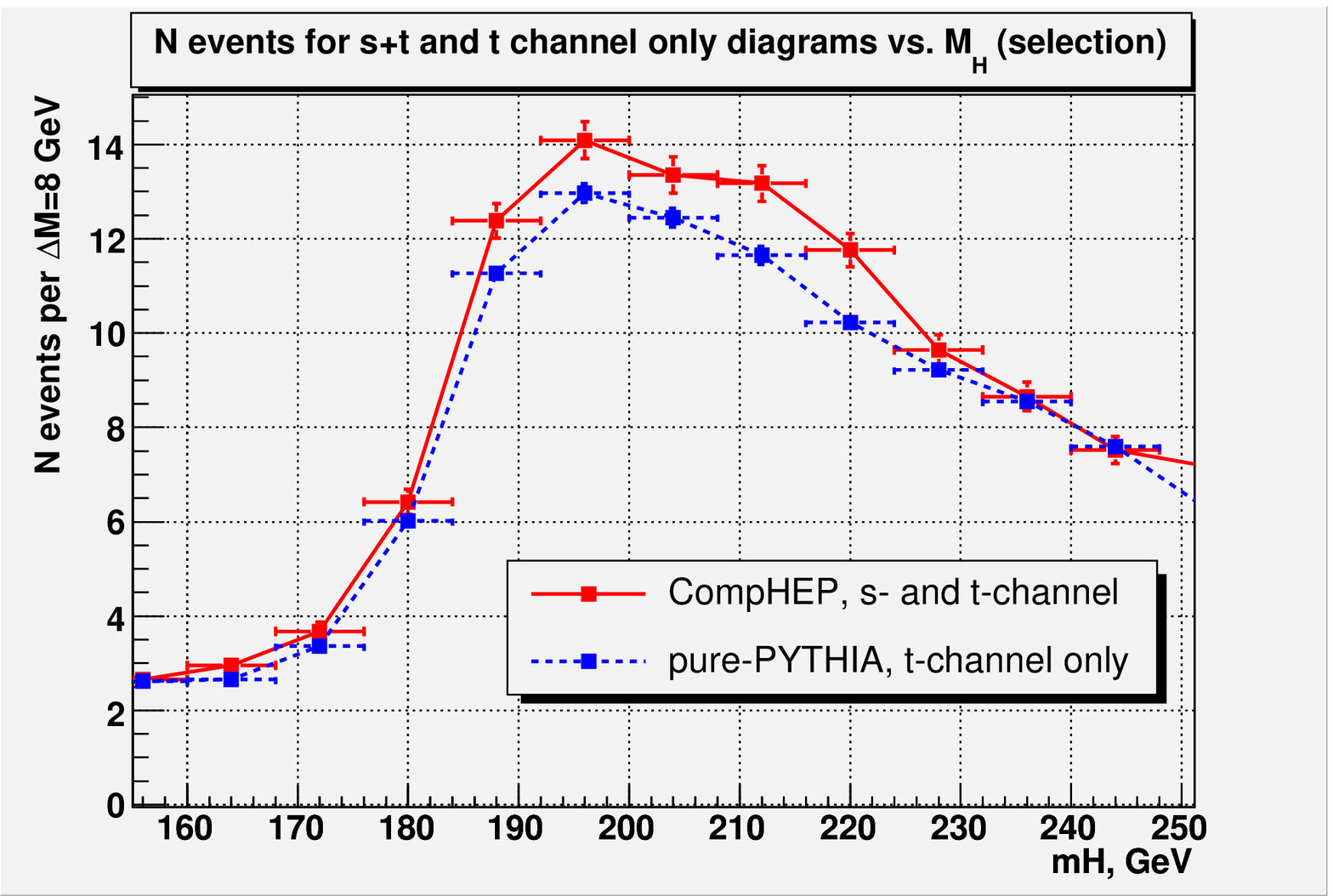}} \\
    \caption{Enlarged part of Fig.~\ref{fig:ana}.}
    \label{fig:sel2} &
    \caption{Enlarged part of Fig.~\ref{fig:sel} for ${\rm M_{4\mu} >
	160 GeV}$.}
    \label{fig:sel3} \\
  \end{tabular}
\end{figure}

Another region of interest is the low mass region with ${\rm M_{4\mu} <
160 }$ GeV. This is where we applied our example set of analysis cuts
(optimized for small ${\rm m_H }$ region). In this region, due to the
s-channel presence and interference between the t- and s-channels, we
see an excess of events over the t-channel-only case at the level of
10-15\% (even after the analysis cuts), see Fig.~\ref{fig:sel2}.

Even in the third region of interest (${\rm M_{4\mu} > 160 GeV}$), the
s-channel contribution still is not negligible, ${\rm \sim 5-7\% }$
(Fig.~\ref{fig:sel3}).

The overall numbers of ${\rm 4\mu }$ events after different cuts are
shown in Table~\ref{tab:CSNEV}. "Pre-selection" cuts are defined in
Section~\ref{sec:presel}, "selection" and "analysis" cuts are defined
in Section~\ref{sec:selec}. The numbers for the t-channel
contributions for the PYTHIA- and CompHEP-produced samples in the
first two columns ($\sigma$ (pre-selection) in fb and the
corresponding N (pre-selection) of expected events) are different
because of different pre-selection cuts for these two generators. Once
the cut on the invariant mass of all four permutations of ${\rm
\mu^+\mu^-}$-pairs is introduced (and other cuts are the same as
well), the expected event numbers for the t-channel contribution for
the PYTHIA and CompHEP samples are the same up to the level of
statistical precision.

\begin{table}[htb]
  \small
  \caption{Cross section values for the t- and s-channel CompHEP
  sample, the t-channel CompHEP and PYTHIA samples and the
  corresponding expected numbers of events for ${\rm 30 ~fb^{-1} }$
  integrated luminosity with MC statistical errors (numbers and
  corresponding statistical errors are scaled according to
  cross-section and integrated luminosity from a much larger number of
  MC generated events).}
  \label{tab:CSNEV}
  \vspace*{1mm}
  \centering
    \begin{tabular}{||c|c|c|c|c||} \hline
      process & $\sigma$ (pre-selection), fb & N (pre-selection) & N
      (selection) & N (analysis) \\ \hline
      CompHEP, s- and t-channel     & 65.0 & $1950 \pm 4.6$ & $224 \pm 1.6$  & $42.6 \pm 0.68$  \\ \hline
      CompHEP, t-channel only  & 18.8 & $565 \pm 1.1$  & $184 \pm 0.64$ & $29.3 \pm 0.25$  \\ \hline
      pure PYTHIA              & 9.93 & $298 \pm 0.98$ & $186 \pm 0.77$ & $30.4 \pm 0.31$  \\ \hline
    \end{tabular}
\end{table}

\subsection{Summary}

PYTHIA does not include the s-channel (and its interference with the
t-channel) in ZZ background generation. 

We have shown that the s-channel contribution to the ZZ background in
the analysis of ${\rm H \rightarrow 4\mu }$ with ${\rm M_{4\mu} >
115}$ GeV (area of interest for the Standard Model Higgs boson search
at LHC: lower limit is from LEP studies) is non-negligible and remains
non-negligible after applying analysis cuts optmised for a low mass
Higgs search.

\subsection*{Aknowledgments}

We would like to thank 
M.~Aldaya, 
P.~Arce, 
J.~Caballero, 
B.~Cruz, 
T.~Ferguson,
U.~Gasparini,
P.~Garcia,
J.~Hernandez,
I.~Josa, 
P.~Moisenz, 
E.R.~Morales, 
N.~Neumeister,
A.~Nikitenko, 
F.~Palla,
T.~Sjostrand and
I.~Vorobiev
for their active participation in the analysis discussions and
comments on this letter.

%%%%%%%%%%%%%%%%%%%%%%%%%%%%%%%%%%%%%%%%%%%%%%%%%%%%%%%%%%%%%%%%%%%%%%%%%%%%%
\section[Sensitivity of the muon isolation cut efficiency to the underlying
event uncertainties]
{SENSITIVITY OF THE MUON ISOLATION CUT EFFICIENCY TO THE UNDERLYING
EVENT UNCERTAINTIES~\protect
\footnote{Contributed by: S.~Abdullin, D.~Acosta, P.~Bartalini,
        R.~Cavanaugh, A.~Drozdetskiy, A.~Korytov, G.~Mitselmakher,
    Yu.~Pakhotin, B.~Scurlock, A.~Sherstnev}}
\subsection{Introduction}

In future searches for the Higgs boson at the LHC via its 4-muon decay
channel, ${\rm H \rightarrow ZZ \rightarrow 4 \mu }$, the muon
isolation cut plays a key role in suppressing many otherwise dominating
backgrounds where all or some muons originate from hadronic decays
(${\rm t\bar{t} }$ and ${\rm Zb\bar{b} }$ are the most important
processes in this category). In reducing the ${\rm
t\bar{t} }$ and ${\rm Zb\bar{b} }$ backgrounds to a negligible level,
the ZZ background and signal is also suppressed. Therefore, one must
worry about the efficiency of the muon isolation cut with respect to
the ZZ background and Higgs boson signal and, even more, about the
sensitivity of this efficiency to the large theoretical uncertainties
associated with a poor understanding of the underlying event (UE)
physics. The UE is defined as \cite{PAOLO} all the remnant activity
from the same proton-proton interaction.  The goal of the studies
presented in this letter was not to optimize the muon isolation cut in
order to maximize the signal-over-background significance, but rather
to understand how well can we predict the isolation cut efficiency using the
current Monte Carlo generators, and to determine how to
measure the isolation cut efficiency using the
experimental data themselves.

In these generator-level studies, we looked only at the
tracker-based isolation cut.
%; we believe the relative sensitivity of
%the calorimeter-based muon isolation cut to the UE uncertainties must
%be correlated with that for the tracker-based muon
%isolation. Calorimeter-based muon isolation by itself depends more
%than the tracker-based one on an accurate detector realization and
%that was the reason we avoided its implementation for this
%generator-level analysis.

The analysis presented in this letter is done in accordance with CMS
guidelines described in \cite{PAOLO} for estimating uncertainties arising due to the UE. 

%%%%%%%%%%%%%%%%%%%%%%%%%%%%%%%%%%%%%%%%%%%%%%%%%%%%%%%%%%%%%%%%%%%%%%%%%%%%%%%%%%

\subsection{Event generation parameters for PYTHIA}

Higgs boson, ${\rm t\bar{t} }$ and Z-inclusive data samples were generated with
PYTHIA 6.223 \cite{Sjostrand:2000wi}. The ZZ data sample was generated at the
matrix-element level with CompHEP \cite{Boos:2004kh} and, then, PYTHIA was
used to complete the event simulation (parton shower development, UE,
hadronization, and particle decays). The PYTHIA parameters that drive
the UE simulation were consistently chosen to match those selected for
the Data Challenge 2005 (DC05) CMS official production (see
Table~\ref{tab:DC05}). Detailed discussion of the associated
phenomenology and the corresponding references can be found elsewhere
\cite{PAOLO}.

\begin{table}[htb]
  \small
  \caption{Parameters in PYTHIA for multi-parton interactions (MI) and
  UE for CDF, ATLAS and CMS.}
  \vspace*{1mm}
  \label{tab:DC05}
  \centering
    \begin{tabular}{||c|c|c|c|c|c||} \hline
      parameter & CDF & ATLAS & CMS (DC04) & CMS (DC05) & comment \\ \hline
      PARP(82) & 2    & 1.8  & 1.9 & 2.9 & regularization scale of PT spectrum for MI \\ \hline
      PARP(84) & 0.4  & 0.5  & 0.4 & 0.4 & parameter of matter distribution inside hadrons \\ \hline
      PARP(85) & 0.9  & 0.33 & 0.33 & 0.33 & probability in MI for two gluons with color connections \\ \hline
      PARP(86) & 0.95 & 0.66 & 0.66 & 0.66 & probability in MI for two gluons (as a closed loop) \\ \hline
      PARP(89) & 1800 & 1000 & 1000 & 14000 & reference energy scale \\ \hline
      PARP(90) & 0.25 & 0.16 & 0.16 & 0.16 & power of the energy-rescaling term \\ \hline
      ${\rm pt_{\textrm{cut-off}} }$ & 3.34 & 2.75 & 2.90 & 2.90 & final ${\rm pt_{\textrm{cut-off}} }$ \\ \hline
    \end{tabular}
\end{table}

The most critical parameter affecting the UE activity is
${\rm pt_{\textrm{cut-off}} }$, the lowest PT allowed for multi-parton interactions.
The smaller ${\rm pt_{\textrm{cut-off}} }$ is, the larger is the number of tracks
associated with the underlying event. The ${\rm pt_{\textrm{cut-off}} }$ value and its
evolution with the center of mass energy of proton-proton collisions
are defined via the following formula:

\begin{center}
${\rm pt_{\textrm{cut-off}} = PARP(82)*(14000/PARP(89))^{PARP(90)} }$
\end{center}

The three parameters, PARP(82,89,90), have meaning only in this
combination.  
The parameters PARP(89) and PARP(90) are fixed at 14,000 and 0.16,
correspondingly. We decided to vary ${\rm pt_{\textrm{cut-off}} }$ by
${\rm \pm 3\sigma }$, or ${\rm \pm 0.5 ~GeV }$, which seems to be a
sensible estimation of theoretical uncertainties arising from UE
modeling \cite{Nason:1999ta}. Note that ${\rm pt_{\textrm{cut-off}}=3.34 ~GeV
}$, as extracted from CDF's Tune A of PYTHIA MI parameters, differs
from the default values used by ATLAS (${\rm 2.75 ~GeV }$) and CMS
(${\rm 2.9 ~GeV }$) by ${\rm \sim 0.5 ~GeV }$ because it was done
using a different PYTHIA parameter tuning model and is listed for
completeness only in Table~\ref{tab:DC05}.

%%%%%%%%%%%%%%%%%%%%%%%%%%%%%%%%%%%%%%%%%%%%%%%%%%%%%%%%%%%%%%%%%%%%%%%%%%%%%%%%%%

\subsection{Monte Carlo sample production}

Processes used in these studies were: ${\rm t\bar{t} }$ (PYTHIA parameter
MSEL = 6); Higgs boson signal (${\rm m_H = 150 }$ GeV, PYTHIA parameters MSEL = 0,
MSUB(102,123,124) = 1 with H allowed to decay to ${\rm Z/\gamma* }$ only,
${\rm Z/\gamma* }$ allowed to decay to ${\rm e/\mu/\tau }$ pair only and ${\rm \tau }$
allowed to decay to ${\rm e/\mu }$ only); ZZ (PYTHIA parameters MSEL = 0,
MSUB(1) = 22 with ${\rm Z/\gamma* }$ allowed to decay to ${\rm e/\mu/\tau }$ pair
only and ${\rm \tau }$ allowed to decay to ${\rm e/\mu }$ only); Z-inclusive (PYTHIA
parameters MSEL = 0, MSUB(1) = 1 with Z allowed to decay to muon
pair only). For Higgs boson signal, we used PHOTOS as a generator of
bremsstrahlung photons.

Generator-level cuts:

\begin{itemize}

\item ${\rm t\bar{t} }$: at least four muons with ${\rm PT > 7 ~GeV }$ and ${\rm |\eta| <
  2.4 }$;

\item Higgs boson signal: at least four muons with ${\rm PT > 7 ~GeV
}$ and ${\rm |\eta| < 2.4 }$; ${\rm 5 < M_{inv}(\mu^+\mu^-) < 150 ~GeV
}$ for 2 intermediate resonances (${\rm Z/\gamma* }$);

\item ZZ-sample: same as for signal;

\item Z-inclusive: no user defined cuts.

\end{itemize}

\subsection{Event selection}
\label{sec:goodmu}

Event-selection cuts were further imposed on the produced Monte Carlo samples.
These cuts were chosen to mimic those optimized for the future data analysis.
There are two distinct sets of such cuts.

First, only ''good muons'' were selected. A muon was considered to be
''good'' if it had ${\rm PT > 7 }$ GeV in the barrel region (${\rm |\eta| < 1.1 }$)
or ${\rm P > 9 }$ GeV in the endcaps (${\rm 1.1 < |\eta| < 2.4 }$). This ensures
that the muon reconstruction efficiencies are flat with respect to PT or P, which
helps minimize systematic uncertainties on the muon reconstruction
efficiency.

Then, event-selection cuts similar to the full analysis cuts were
applied. They are:

\begin{itemize}

\item At least 2 opposite sign muon pairs with invariant masses for
   all ${\rm \mu^+\mu^- }$ pair permutations being greater than 12 GeV (this
   cut suppresses heavy-quark resonances).

\item PT of all four selected muons must be greater than 10 GeV
   (signal-over-background optimization).

\item invariant mass of the four muons must be greater than 110 GeV
   and less than 700 GeV (Higgs boson with ${\rm M<114.4 ~GeV }$ is
   excluded at LEP, Higgs boson with mass over 700 GeV is strongly
   disfavored by theory and, also, would have too low a production
   cross section).

\item ${\rm ISOL = \sum{PT_i} }$ (PT with respect to the beam
   direction) should be less or equal to 0, 0, 1, 2 GeV for the four
   muons when the muons are sorted by the ISOL parameter. The sum runs
   over only charged particle tracks with PT greater then 0.8 GeV and
   inside a cone of radius ${\rm R = \sqrt{(\Delta\phi)^2 +
   (\Delta\eta)^2}=0.3 }$ in the azimuth-pseudorapidity space. A PT
   threshold of 0.8 GeV roughly corresponds to the PT for which tracks
   start looping inside the CMS Tracker. Muon tracks were not included
   in the calculation of the ISOL parameter.

\end{itemize}

\subsection{Tracker-based muon isolation cut efficiency}

\begin{figure}
    \begin{tabular}{p{.47\textwidth}p{.47\textwidth}} 
      \resizebox{\linewidth}{0.65\linewidth}{\includegraphics{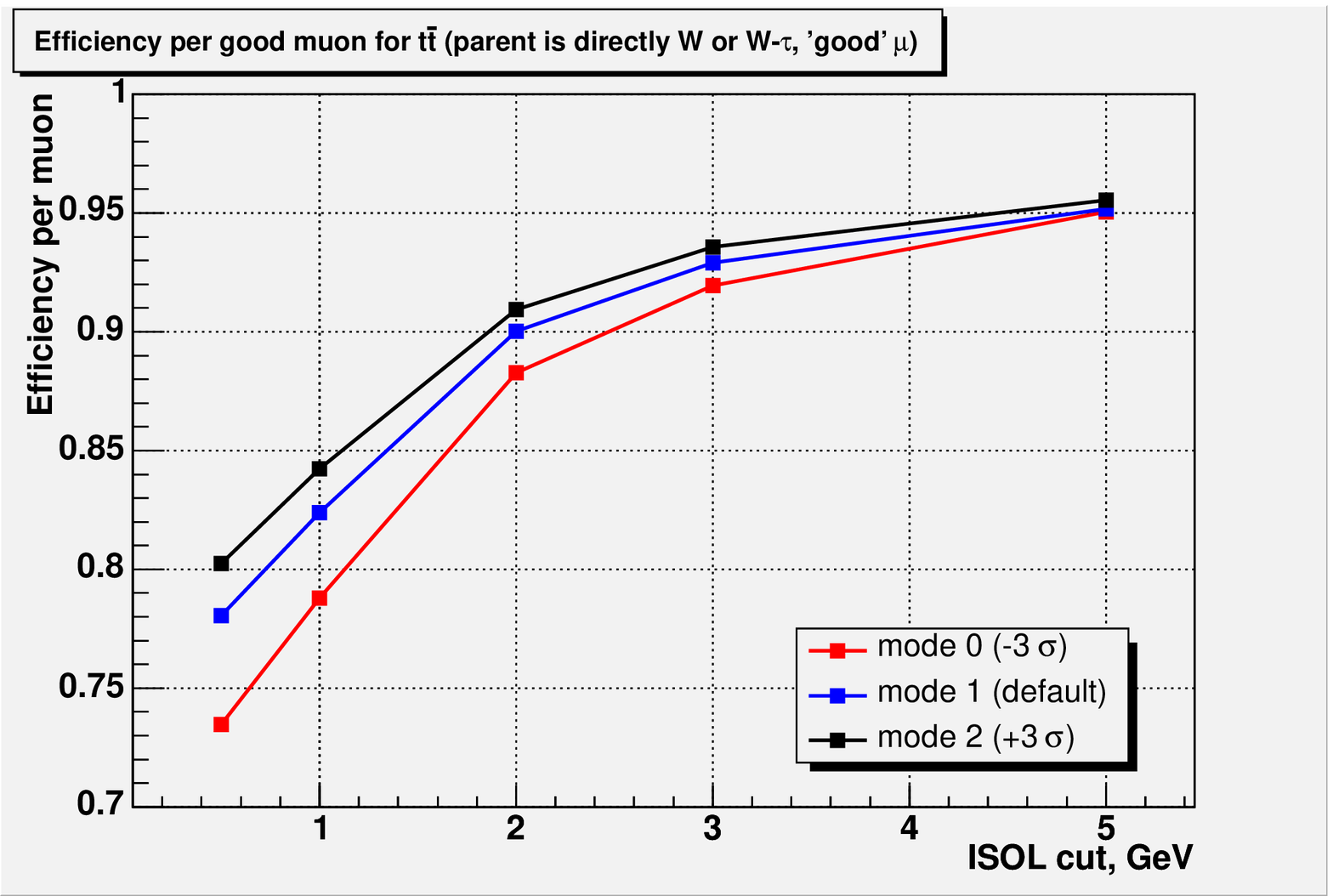}} &
      \resizebox{\linewidth}{0.65\linewidth}{\includegraphics{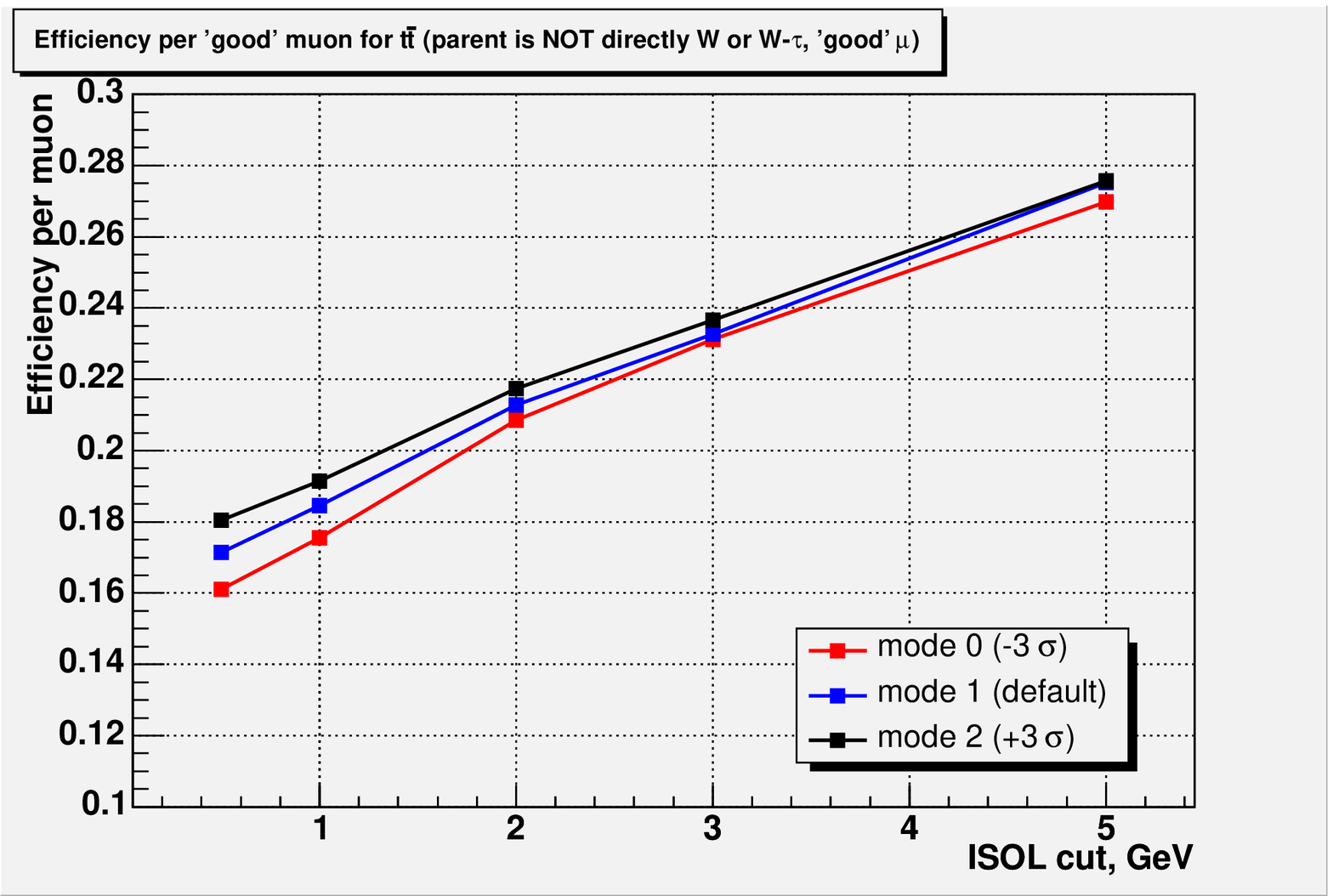}} \\
      \caption{Muon isolation cut efficiency averaged over selected muons
	whose parents are W bosons (${\rm t\bar{t} }$ events). The blue middle line
	is for the default MI ${\rm pt_{\textrm{cut-off}} }$, the black upper line is for
	downward ${\rm -3\sigma }$ variation of ${\rm pt_{\textrm{cut-off}} }$ value, the red lower
	line is for upward ${\rm +3\sigma }$ variation.}
      \label{fig:GoodBckgW} &
      \caption{Similar to Fig.~\ref{fig:GoodBckgW} for muons from hadronic
	decays (${\rm t\bar{t}}$ events).}
      \label{fig:GoodBckgB} \\
      \resizebox{\linewidth}{0.65\linewidth}{\includegraphics{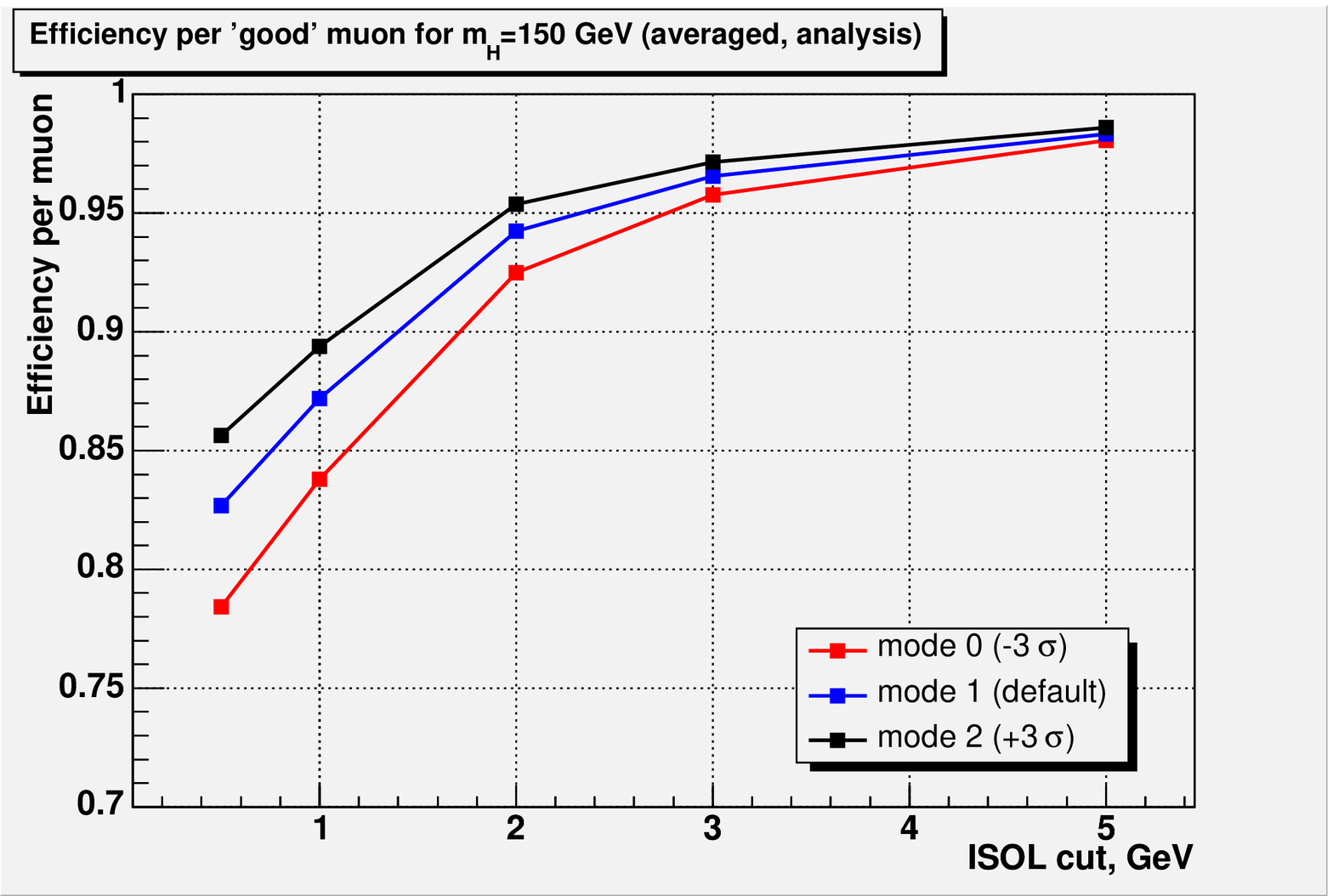}} &
      \resizebox{\linewidth}{0.65\linewidth}{\includegraphics{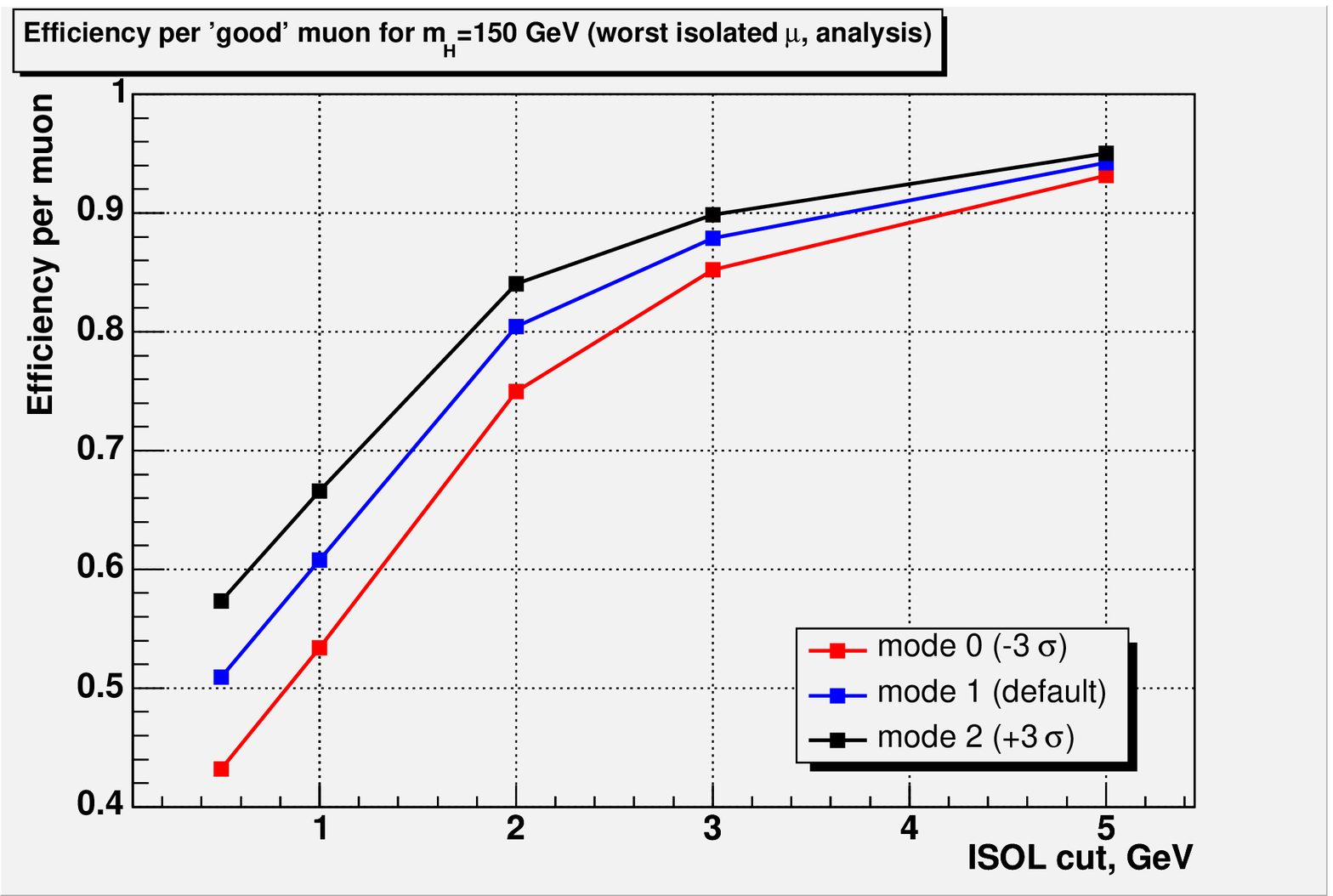}} \\
      \caption{Similar to Fig.~\ref{fig:GoodBckgW} for Higgs boson events.}
      \label{fig:GoodSig} &
      \caption{Muon isolation cut efficiency for the least isolated muon
	from 4 selected ones in Higgs boson events.}
      \label{fig:SigWorst} \\
    \resizebox{\linewidth}{0.65\linewidth}{\includegraphics{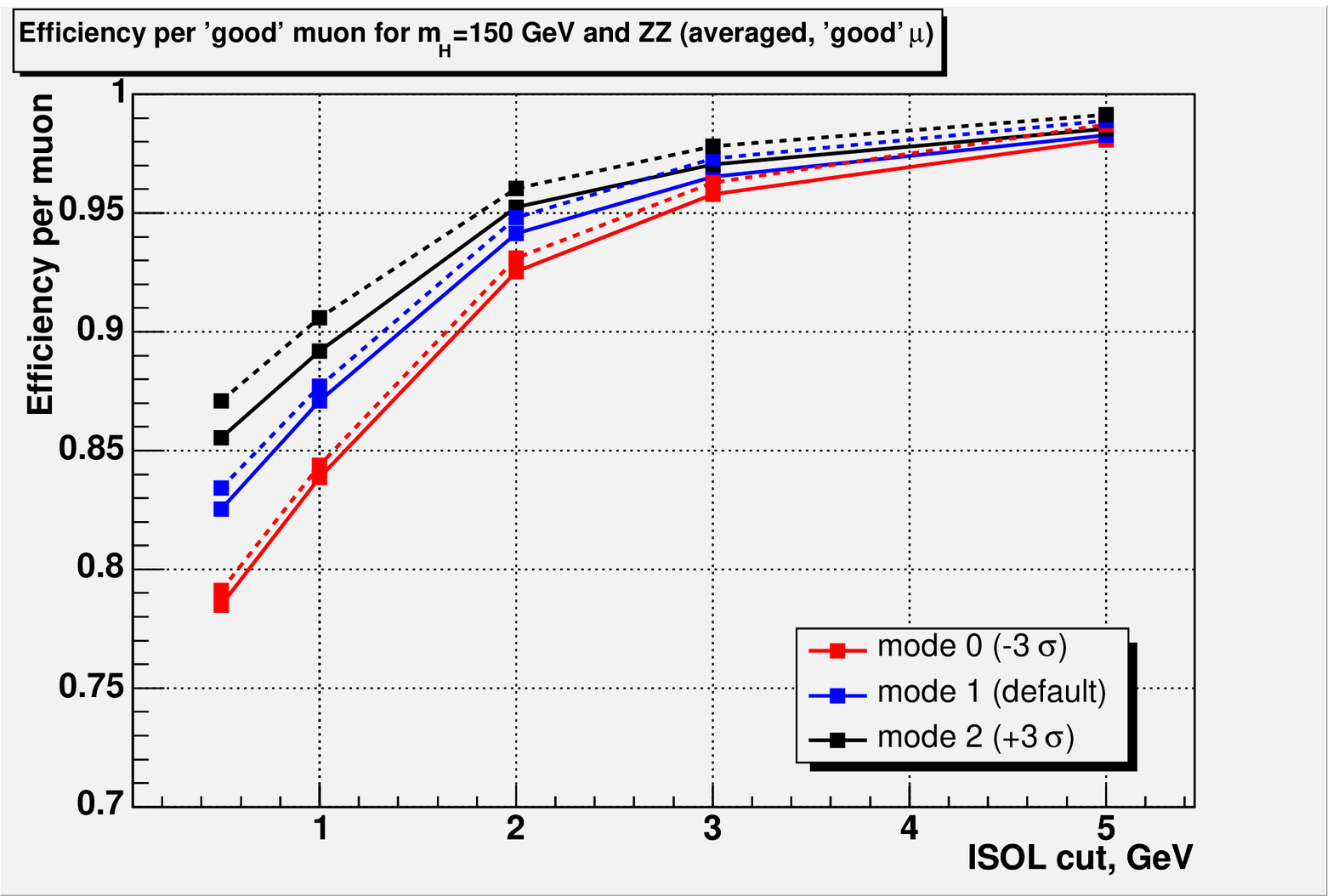}} &
    \resizebox{\linewidth}{0.65\linewidth}{\includegraphics{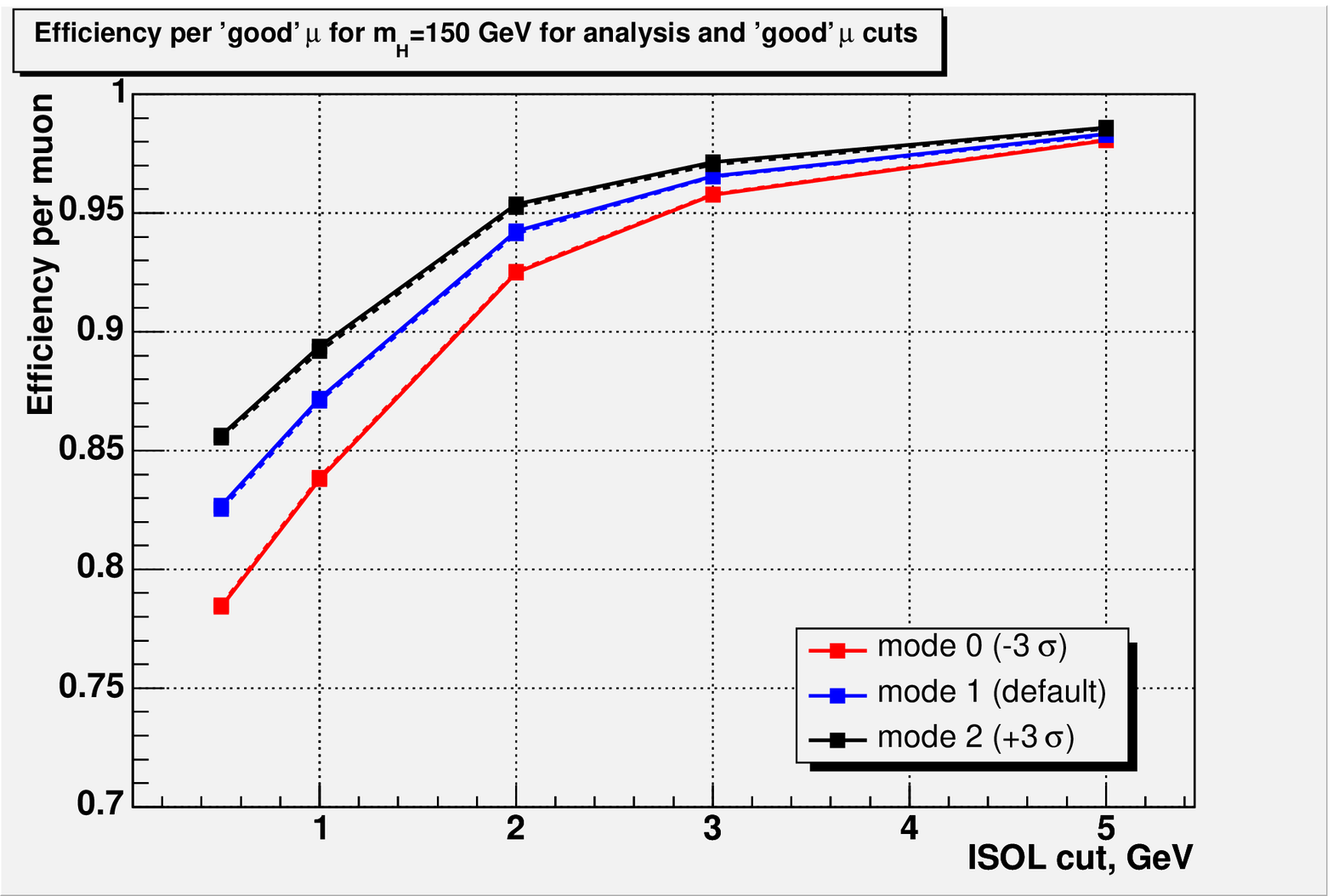}} \\
    \caption{Muon isolation cut efficiency averaged over 4 selected
      muons for signal events (solid lines, Fig.~\ref{fig:GoodSig})
      and ZZ background (dashed lines). The blue middle line is for
      the default MI ${\rm pt_{\textrm{cut-off}} }$, the black upper line is for
      downward ${\rm -3\sigma }$ variation of ${\rm pt_{\textrm{cut-off}} }$ value, the red
      lower line is for upward ${\rm +3\sigma }$ variation.}
    \label{fig:IsolZZsig} &
    \caption{Muon isolation cut efficiency averaged over 4 selected
      muons for signal events. Solid lines are for good muons from
      events after analysis cuts (same as Fig.~\ref{fig:GoodSig}) and
      dashed lines are for good muons from events before analysis
      cuts. There is no difference at statistical precision level for
      two graph sets. Color notations are the same as for Fig.~\ref{fig:IsolZZsig}.}
    \label{fig:GoodBeforeAfter} \\
    \end{tabular}
\end{figure}

Figures \ref{fig:GoodBckgW},~\ref{fig:GoodBckgB} and \ref{fig:GoodSig}
show the muon isolation cut efficiency averaged over all "good" muons
(see section~\ref{sec:goodmu}) for the ${\rm t\bar{t} }$ sample and the Higgs
boson. For ${\rm t\bar{t} }$ background, we show two plots: one for
muons originating from ${\rm W \rightarrow \mu\nu }$ and ${\rm W
\rightarrow \tau\nu \rightarrow \mu\nu\nu\nu }$ decays and the other
for muons originating from hadronic decays (typically, the former
would tend to be isolated and the latter non-isolated). The average
isolation efficiency per "good" muon is calculated as the ratio of the
number of "good" muons with the isolation parameter ISOL below a
particular threshold to the total number of "good" muons.  Figure
\ref{fig:SigWorst} shows the isolation cut efficiency for the least
isolated muon out of four (Higgs boson sample). We use a cut at ISOL=2
GeV for such muons. One can see that this cut alone will have ${\rm
\sim 80\% }$ efficiency with ${\rm \pm 5\% }$ uncertainty due to the UE model.
%\begin{figure}
%    \begin{tabular}{p{.47\textwidth}p{.47\textwidth}} 
%    \end{tabular}
%\end{figure}
    
    Figure \ref{fig:IsolZZsig} compares the muon isolation cut efficiency
curves for the main irreducible ZZ background and for the Higgs boson
events. Clearly, these efficiencies are very similar.

\subsubsection{Sensitivity to kinematical cuts}

Figure \ref{fig:GoodBeforeAfter} demonstrates another very important
feature of the tracker-based muon isolation cut: its efficiency is not
very sensitive to the kinematical analysis cuts. The figure has two
sets of efficiency curves: one is obtained for "good" muons and
another for "good" muons passing further event selection cuts as
described in section~\ref{sec:goodmu}. One can hardly see any
difference. Therefore, the conclusions of this analysis will not
depend on the choice of the final event selection cuts.
\begin{figure}
  \begin{tabular}{p{.47\textwidth}p{.47\textwidth}} 
    \resizebox{\linewidth}{0.65\linewidth}{\includegraphics{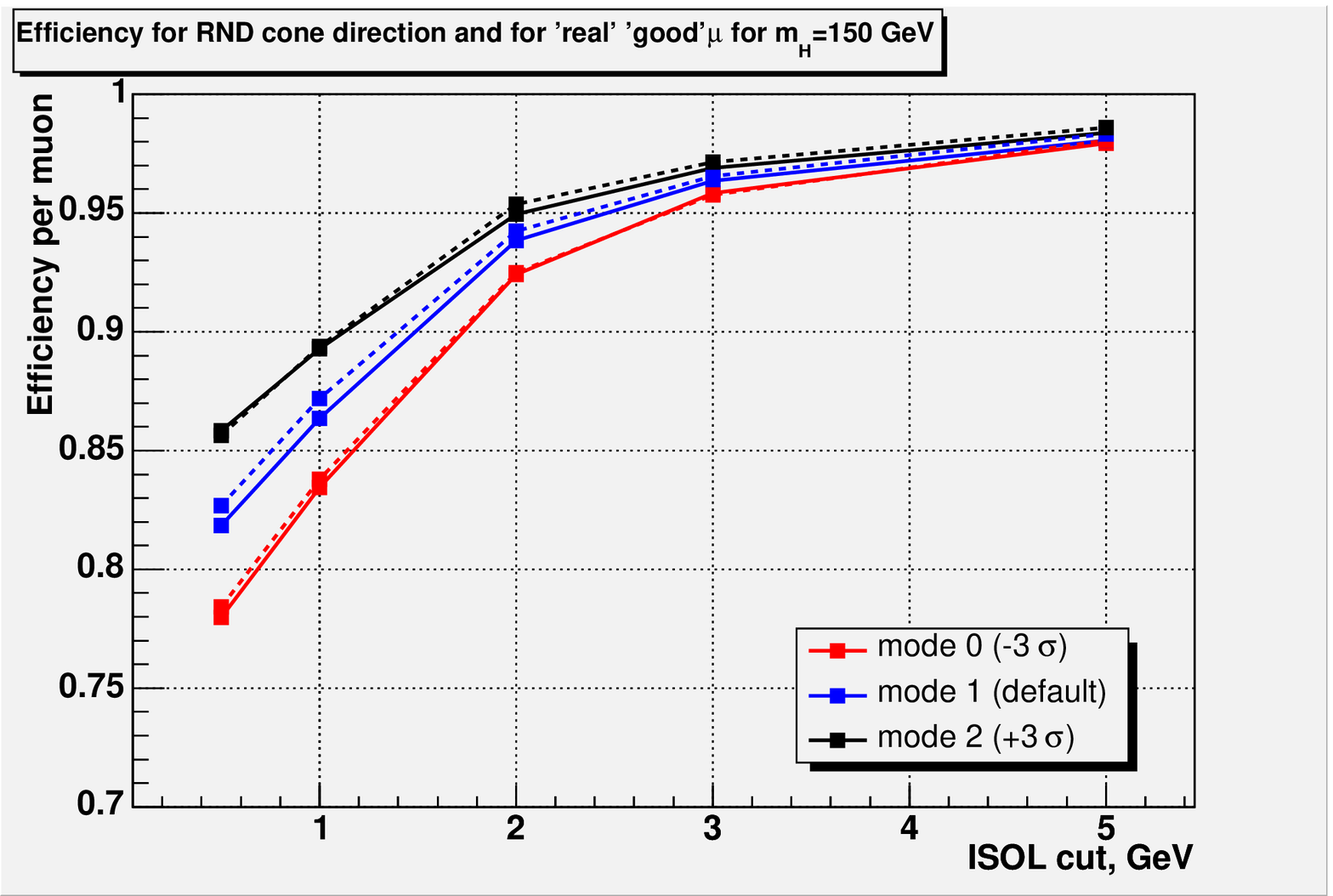}} &
    \resizebox{\linewidth}{0.65\linewidth}{\includegraphics{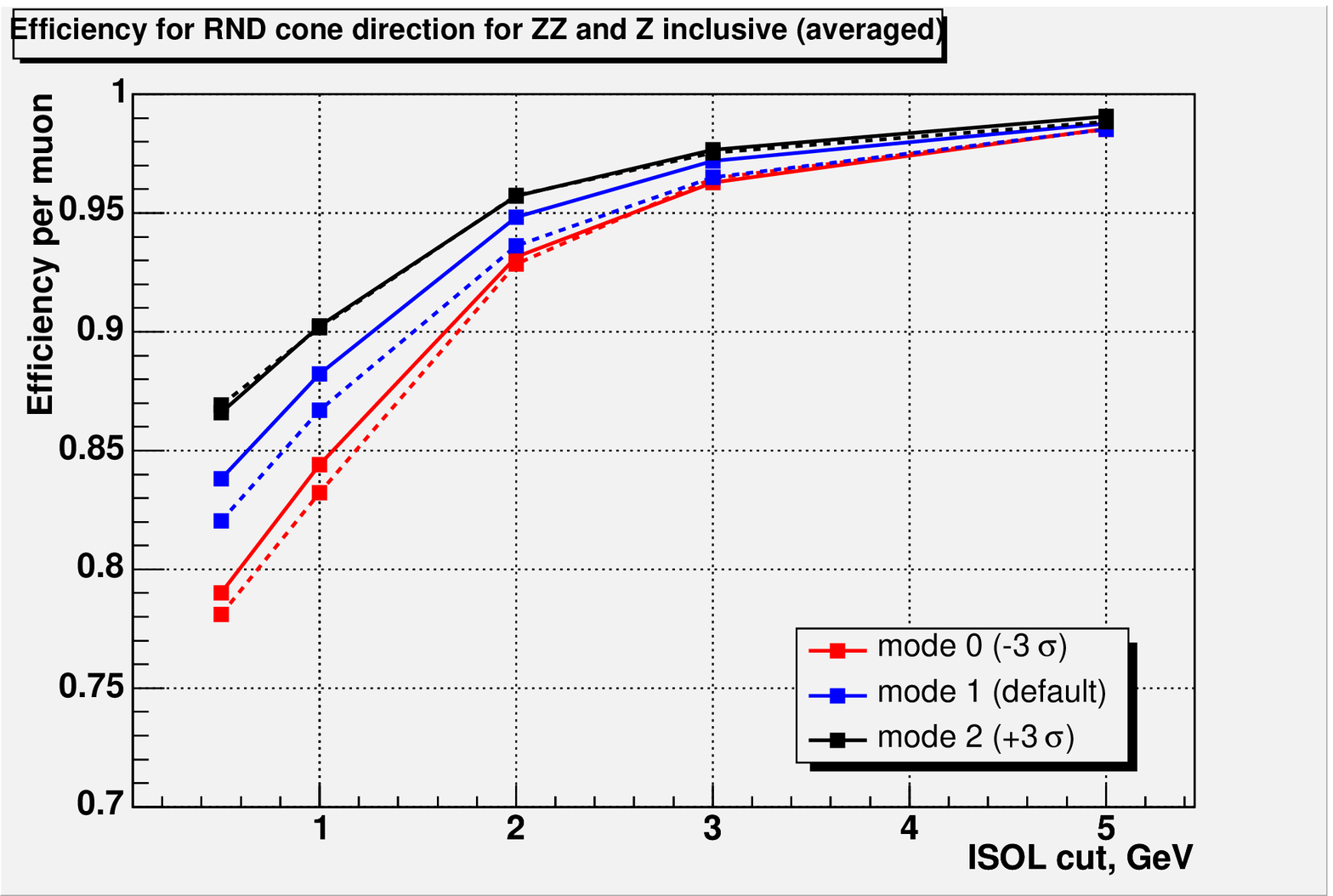}} \\
    \caption{Muon isolation cut efficiency for random-cone directions
      (solid lines) and for muons (dashed lines) for signal events. The
      blue middle lines are for the default MI ${\rm pt_{\textrm{cut-off}}
      }$, the black upper lines are for downward ${\rm -3\sigma }$
      variation of ${\rm pt_{\textrm{cut-off}} }$ value, the red lower
      lines are for upward ${\rm +3\sigma }$ variation.}
    \label{fig:SigRND} &
    \caption{Muon isolation cut efficiency for random-cone directions for
      Z-inclusive (dashed lines) and for ZZ (solid lines) events. The blue
      middle lines are for the default MI ${\rm pt_{\textrm{cut-off}} }$, the black upper
      lines are for downward ${\rm -3\sigma }$ variation of ${\rm pt_{\textrm{cut-off}} }$ value,
      the red lower lines are for upward ${\rm +3\sigma }$ variation.}
    \label{fig:ZincRND} \\
  \end{tabular}
\end{figure}

\subsubsection{Evaluation of the muon isolation cut efficiency 
from data using random-cone directions}

Figure \ref{fig:SigRND} 
%and \ref{fig:IsolGENrnd} 
shows the isolation cut efficiency 
%and the isolation cut parameter distribution 
as calculated for random directions uniformly distributed in ${\rm \eta-\phi }$
space (${\rm |\eta|<2.4 }$). The algorithm of the ISOL parameter calculation
is the same as for ``real'' MC muons, except that now the ISOL
parameter is calculated from the sum of PT for tracks around random
directions in the acceptance region. The Higgs boson Monte Carlo sample was
used to make these plots. We see that the graphs obtained for the
random cone (solid lines) and for ``real'' muons (dashed line;
identical to Figures \ref{fig:GoodSig} and \ref{fig:GoodBeforeAfter})
look very similar. In fact, they agree within statistical uncertainties.
This observation motivated us to investigate whether we can measure
the isolation cut efficiency by using some distinct reference data
sample and applying the random-cone technique. The reference data
sample must have a large cross section (to provide good statistics),
be relatively clean from backgrounds, and have a similar underlying
structure to ZZ events. Inclusive ${\rm Z \rightarrow \mu\mu }$ seems
to be just what we need. The cross section is ${\rm \sim1.6 }$ nb,
${\rm Z \rightarrow \mu\mu }$ has a very clean signature.
%\begin{figure}
%  \begin{tabular}{p{.47\textwidth}p{.47\textwidth}} 
%  \end{tabular}
%\end{figure}

Figure \ref{fig:ZincRND} shows the isolation cut efficiencies computed
for random-cone directions in Z-inclusive Monte Carlo sample.  One can
see that the isolation cut efficiencies for muons in the ZZ sample
are very well mimicked by the efficiencies calculated for random cones
in the Z-inclusive sample. The variations in the UE
${\rm pt_{\textrm{cut-off}} }$ have nearly identical effects on both data
samples.

%Just for the record, we also show the random-cone-based efficiencies
%for the ${\rm t\bar{t} }$ sample (Fig.~\ref{fig:ttRND}). Clearly, the
%${\rm t\bar{t} }$-sample is not good at mimicking the UE of the ZZ
%sample: calibration with ${\rm t\bar{t} }$-sample would give similarly
%large error as UE theoretical uncertainty. 
%\begin{figure}
%  \begin{tabular}{p{.47\textwidth}p{.47\textwidth}} 
%    \resizebox{\linewidth}{0.65\linewidth}{\includegraphics{s_abdullin2/ue-ttrnd-per-mu.eps}} &
%    \resizebox{\linewidth}{0.65\linewidth}{\includegraphics{s_abdullin2/mu_trk_iso_rnd-150-15000-plus-default-signal.1.eps}} \\
%    \caption{Distribution of the tracker-based muon isolation parameter
%      ISOL for random-cone directions (blue solid line) and for muons (red
%      dashed line). Higgs boson events with the default UE simulation
%      parameters are used.}
%    \label{fig:IsolGENrnd} &
%    \caption{Similar to Fig.~\ref{fig:ZincRND} for ${\rm t\bar{t} }$
%      background events. }
%    \label{fig:ttRND} \\
%  \end{tabular}
%\end{figure}

\subsubsection{${\rm 4\mu}$ Isolation cut efficiency per event}

Efficiencies per event are listed in Table~\ref{tab:perEvnt}.  We
observe that the values for Signal, ZZ-background, and Z-inclusive
using random-cone technique samples are in agreement with each other
for all three tested UE scenarios. The range of efficiencies for the
ZZ-background spans from ${\rm \sim0.72 }$ to ${\rm \sim0.84 }$. This
range of ${\rm \pm 6\%}$ absolute of the central value can be
associated with the uncertainties on the 4-muon isolation cut
efficiency arising from theoretical uncertainties on considered UE
parameters in PYTHIA.

\begin{table}[htb]
  \small
  \caption{Efficiency per event using different events samples: Higgs
  boson signal with ${\rm m_H=150 }$ GeV, ZZ background, Z-inclusive
  (4 RND muons), ${\rm t\bar{t} }$ background. ``4 RND muons'' means
  that for a particular process in each event 4 random cone directions
  were used to calculate the ISOL parameter and the corresponding
  values were treated as ones for ``real'' muons.}
  \label{tab:perEvnt}
  \vspace*{1mm}
  \centering
    \begin{tabular}{||c|c|c|c|c||} \hline
      process/case & efficiency (default) & efficiency (${\rm -3\sigma }$) & efficiency (${\rm +3\sigma }$) \\ \hline
      signal, ${\rm m_H = 150 }$ GeV         & $0.775 \pm 0.004$ & $0.707 \pm 0.005$ & $0.812 \pm 0.004$ \\ \hline
      ZZ background                   & $0.780 \pm 0.004$ & $0.721 \pm 0.005$ & $0.838 \pm 0.004$ \\ \hline
      4 RND muons, Z-inclusive events & $0.762 \pm 0.007$ & $0.706 \pm 0.007$ & $0.821 \pm 0.006$ \\ \hline
      ${\rm t\bar{t} }$ background           & $0.016 \pm 0.001$ & $0.013 \pm 0.001$ & $0.015 \pm 0.001$ \\ \hline
    \end{tabular}
\end{table}

On the other hand, it appears possible to use the Z-inclusive sample
to gauge the UE activity and evaluate the 4-muon isolation cut
efficiency experimentally.  There might be a small systematic shift of
the order of ${\rm \sim 2\% }$ in efficiencies between the ZZ and
Z-inclusive samples, and this is a shift can be evaluated from data, 
and the result is then to a large degree independent from a
particular UE scenario.
For the three different UE simulations we used in these studies, we
obtain the following offsets: $0.018 \pm 0.008$, $0.015 \pm 0.009$,
$0.017 \pm 0.007$. Much larger Monte Carlo samples would be needed to
pin it down more accurately. However, conservatively, one may
ignore this correction and assign a $2\%$ systematic uncertainty on
the Z-sample-based estimate of the 4-muon isolation cut efficiency for
ZZ-background and Higgs boson signal events as it is
already much smaller in comparison to the other systematics such as
theoretical uncertainties associated with the choice of PDF's and QCD
scale, NLO/NNLO corrections, etc.

The efficiency for accepting ${\rm t\bar{t} }$-events is of the order of
0.015 $\pm$ 0.001. Its sensitivity to the UE could not be studied due to
lack of statistics, but it is not expected to be too large as it is
dominated by the jet activity. In fact, if the reducible ${\rm t\bar{t} }$-
and ${\rm Zb\bar{b} }$-backgrounds could not be suppressed well below the
ZZ-background, one would need to study their sensitivity to the UE
physics, as well as to the jet fragmentation modeling.

%%%%%%%%%%%%%%%%%%%%%%%%%%%%%%%%%%%%%%%%%%%%%%%%%%%%%%%%%%%%%%%%%%%%%%%%%%%%%%%%%%

\subsection{Summary}

The isolation cut efficiency per muon due to uncertainties in the
UE can vary as much as $\pm 5$\% (the efficiency
itself and its uncertainty strongly depend on how tight the ISOL cut
is). The 4-muon isolation cut efficiency per event for ${\rm ZZ
\rightarrow 4 \mu }$ background is measured to be $\sim (78 \pm 6)\%$.

To decrease these large uncertainties to a negligible level with
respect to other systematic uncertainties, one can calibrate the
isolation cut efficiency from data using Z-inclusive events (${\rm Z
\rightarrow 2 \mu }$) and the random-cone technique. We show that this
indeed significantly decreases uncertainties associated with the poor
understanding of the UE physics. There might be $\sim 2\%$ systematic
shift in the 4-muon isolation cut efficiencies obtained this way. In
principle, one can correct for this shift, but it does not appear to
be necessary as this uncertainty is already very small.

The results and described techniques in this letter may be of interest
for all analyses relying on lepton isolation cuts.

\subsection{Acknowledgments}

We would like to thank 
M.~Aldaya, 
P.~Arce, 
J.~Caballero, 
B.~Cruz, 
G.~Dissertori,
T.~Ferguson,
U.~Gasparini,
P.~Garcia,
J.~Hernandez,
I.~Josa, 
M.~Konecki,
P.~Moisenz, 
E.R.~Morales, 
N.~Neumeister,
A.~Nikitenko,
F.~Palla and
I.~Vorobiev
for their active participation in the analysis discussions and
comments on this letter.

%%%%%%%%%%%%%%%%%%%%%%%%%%%%%PART%%%%%%%%%%%%%%%%%%%%%%%%%%%%%
\part[HIGGS PHYSICS]{HIGGS PHYSICS}
%%%%%%%%%%%%%%%%%%%%%%%%%%%%%PART%%%%%%%%%%%%%%%%%%%%%%%%%%%%%

%%%%%%%%%%%%%%%%%%%%%%%%%%%%%%%%%%%%%%%%%%%%%%%%%%%%%%%%%%%%%%%%%%%%%%%%%%%%%
\section[$gg\rightarrow H$ at the LHC: uncertainty due to a jet veto]
{$gg\rightarrow H$ AT THE LHC: UNCERTAINTY DUE TO A JET VETO~\protect
\footnote{Contributed by: G.~Davatz, A.~Nikitenko}}
\subsection{Overview}

The experimental cross section $\sigma$$_{meas}$ of the Higgs signal and other
final states is given by
\begin{equation}
\sigma_{meas}= N_{s} / (\epsilon_{sel} \times  L_{pp})
\end{equation}
with N$_{s}$ being the number of signal events, $\epsilon_{sel}$ the
efficiency after all signal selection cuts are applied and $L_{pp}$ the integrated
proton-proton luminosity.  In order to get an estimation of the cross section
uncertainty, the statistical and systematic uncertainties have to be
determined.  The systematic uncertainties come from the experimental
selection, background and luminosity uncertainties.

 In the Higgs mass range
between 155 and 180 GeV, H$\rightarrow$WW$\rightarrow \ell \nu\ell \nu$ is
considered to be the main Higgs discovery channel
\cite{Dittmar:1996ss,Davatz:2004zg}. The signal consists of two isolated
leptons with a small opening angle and large missing $\rm E_T$. In order to
reduce the top background, a jet veto has to be applied. The signal over
background ratio is found to be around 2:1 at a Higgs mass of 165 GeV. For
lower and higher Higgs masses, this signal over background ratio decreases
slightly \cite{Davatz:2004zg}.

 As the signal over background ratio is small
in this channel, the systematic uncertainties should be known very well. This
study concentrates on the uncertainty of the signal efficiency due to the jet
veto. The systematics were obtained using different Monte Carlo simulations.

Three different Monte Carlo generators are compared: PYTHIA 6.319, HERWIG
6.507 and MC@NLO 2.31\cite{Sjostrand:2004ef, Corcella:2000bw,
Frixione:2002ik,Frixione:2003ei}.  
All three are so-called parton shower Monte Carlos.

PYTHIA is a general purpose Leading Order (LO) Monte Carlo, based on LO matrix
elements and Lund hadronization. HERWIG is also a Leading Order Monte Carlo
based on the Cluster model for hadronization. MC@NLO matches Next-to-Leading
Order (NLO) calculations to a parton shower Monte Carlo. Its total cross
section is calculated with NLO accuracy. In MC@NLO, HERWIG is used for the
showering. 

 The three Monte Carlos treat the high transverse momentum ($\rm
p_T$) region in different ways: PYTHIA includes matrix element corrections in
the $m_{top}$ $\rightarrow$ $\infty$ limit, whereas HERWIG has no hard
matrix element corrections included in gg $\rightarrow$ H so far. \ MC@NLO includes
the NLO matrix elements in an exact way. 

 For PYTHIA, two different samples
have been generated for comparison: One with the default $\rm Q^2$ ordered
showering model and one with the new $\rm p_T$ ordered showering model. In the
beginning, we make the comparisons with the default $\rm Q^2$ ordered
showering and then also include the new $\rm p_T$ ordered showering model.

In the following, the pdf set chosen for HERWIG and PYTHIA is CTEQ5L, while
for MC@NLO CTEQ5M is taken. Jets are reconstructed using an iterative cone
algorithm with cone size 0.5. The leading particle (seed) of the jet has to
have a $\rm p_T$ higher than 1 GeV. The $\mid \eta \mid$ of the jet should be
smaller than 4.5 (here the CMS detector acceptance is chosen \cite{tdr}). An
event is rejected if it contains a jet with a $\rm p_T$ higher than 30
GeV. The Higgs mass for this study is chosen to be 165 GeV, which is the
region where the best signal over background ratio can be found. The top mass
is set to 175 GeV. First, all events are studied without considering the
underlying event. Finally, PYTHIA is also studied including different underlying
event schemes.

 A similar study was done in the content of the HERA/LHC
workshop, with the CASCADE program included in the comparison \cite{DeRoeck:2005rf}
\footnote{CASCADE is a full hadron level Monte Carlo generator for ep and pp
scattering at small x built according to the CCFM evolution equation
\cite{Jung:2000hk}.}.

 At LO, the $\rm p_T$ of the Higgs is
zero. However, parton shower Monte Carlos emit soft gluons which balance the
Higgs boson and introduce a transverse momentum in LO parton shower Monte Carlos. As
the Higgs is balanced by jets, the transverse momentum is very sensitive to
the jet veto and therefore also the efficiency of a jet veto depends strongly
on $\rm p_T$ Higgs.

\begin{figure}[!htb]
\begin{center}
\includegraphics*[scale=0.37]
{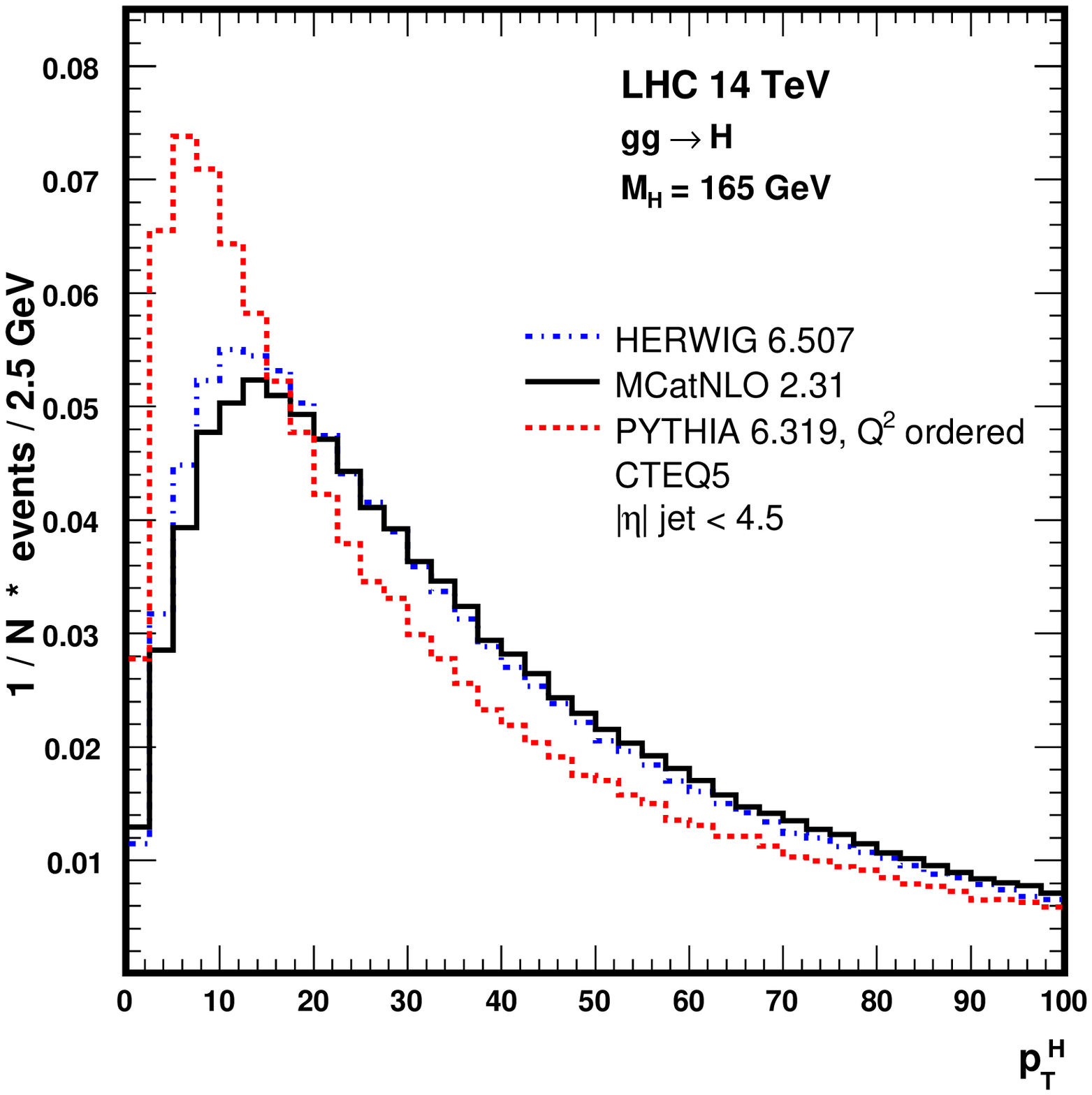}
\includegraphics*[scale=0.37]
{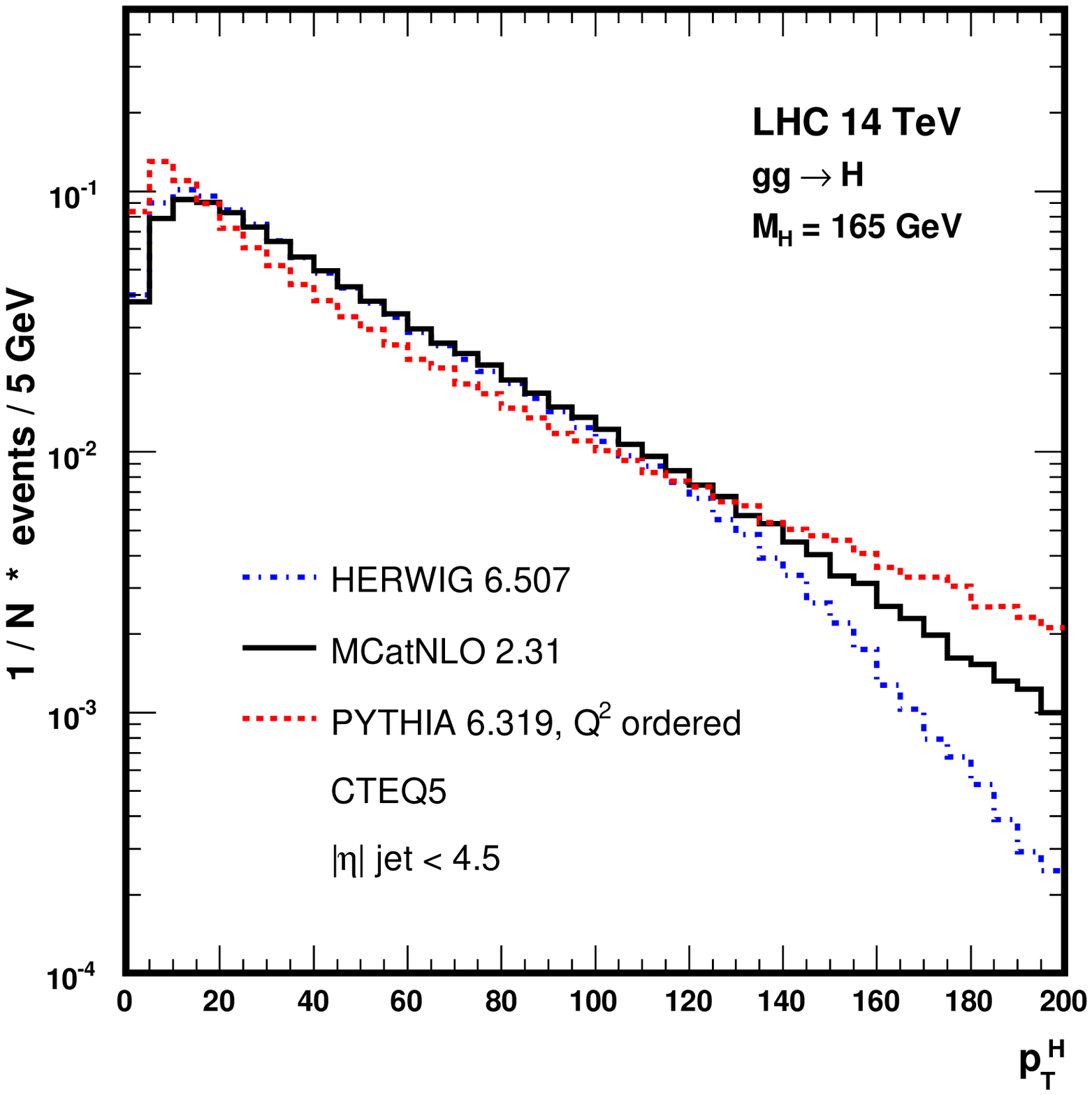}
\end{center}
\caption[fig1]{$\rm p_T$ Higgs spectra for PYTHIA, HERWIG and MC@NLO in linear and logarithmic scale.}
\label{pth}
\end{figure}

\subsection{Comparing PYTHIA with $\rm Q^2$ ordered showering, HERWIG and MC@NLO}
\begin{figure}[!htb]
\begin{center}
\includegraphics*[scale=0.37]{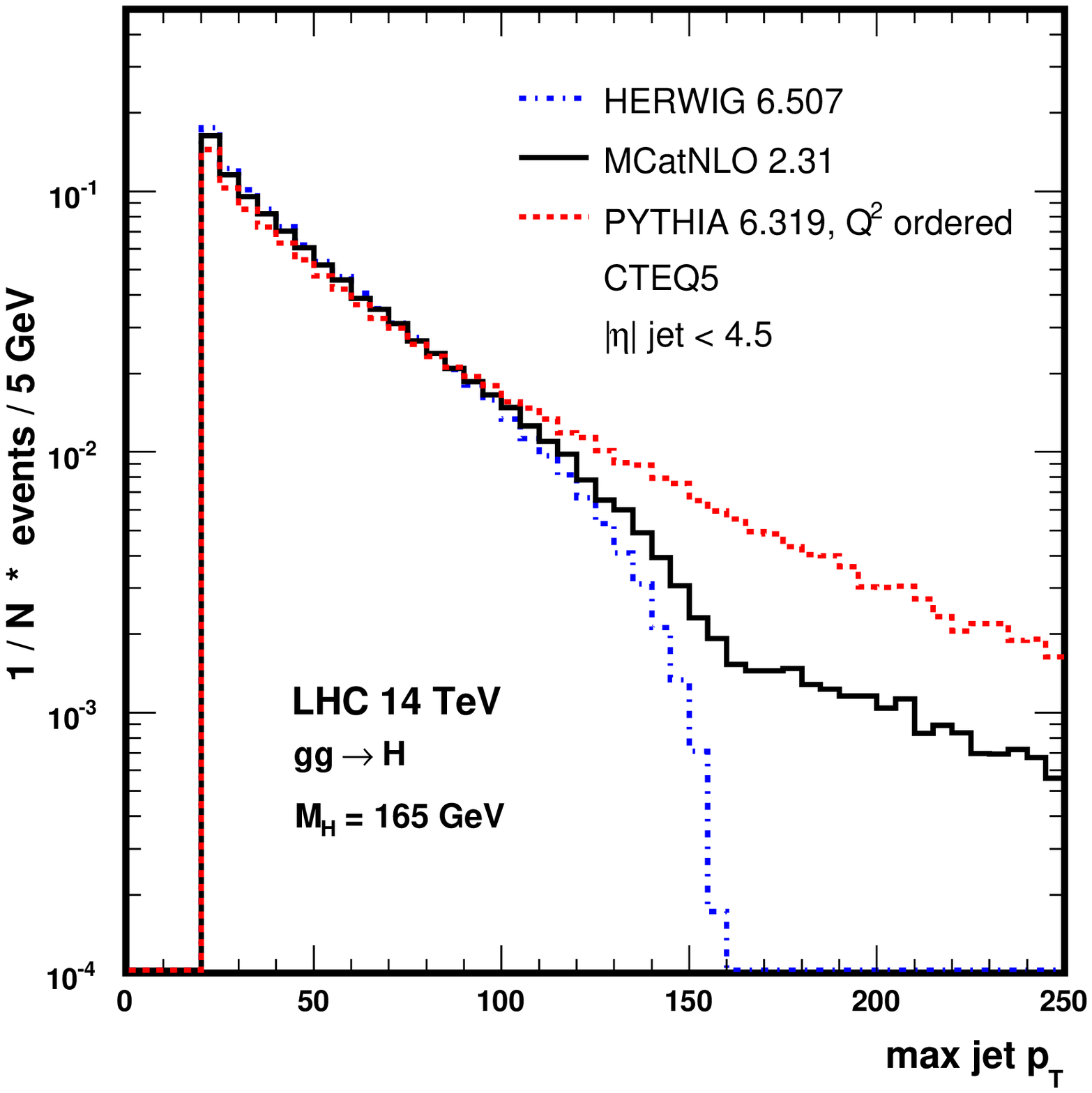}
\includegraphics*[scale=0.37]{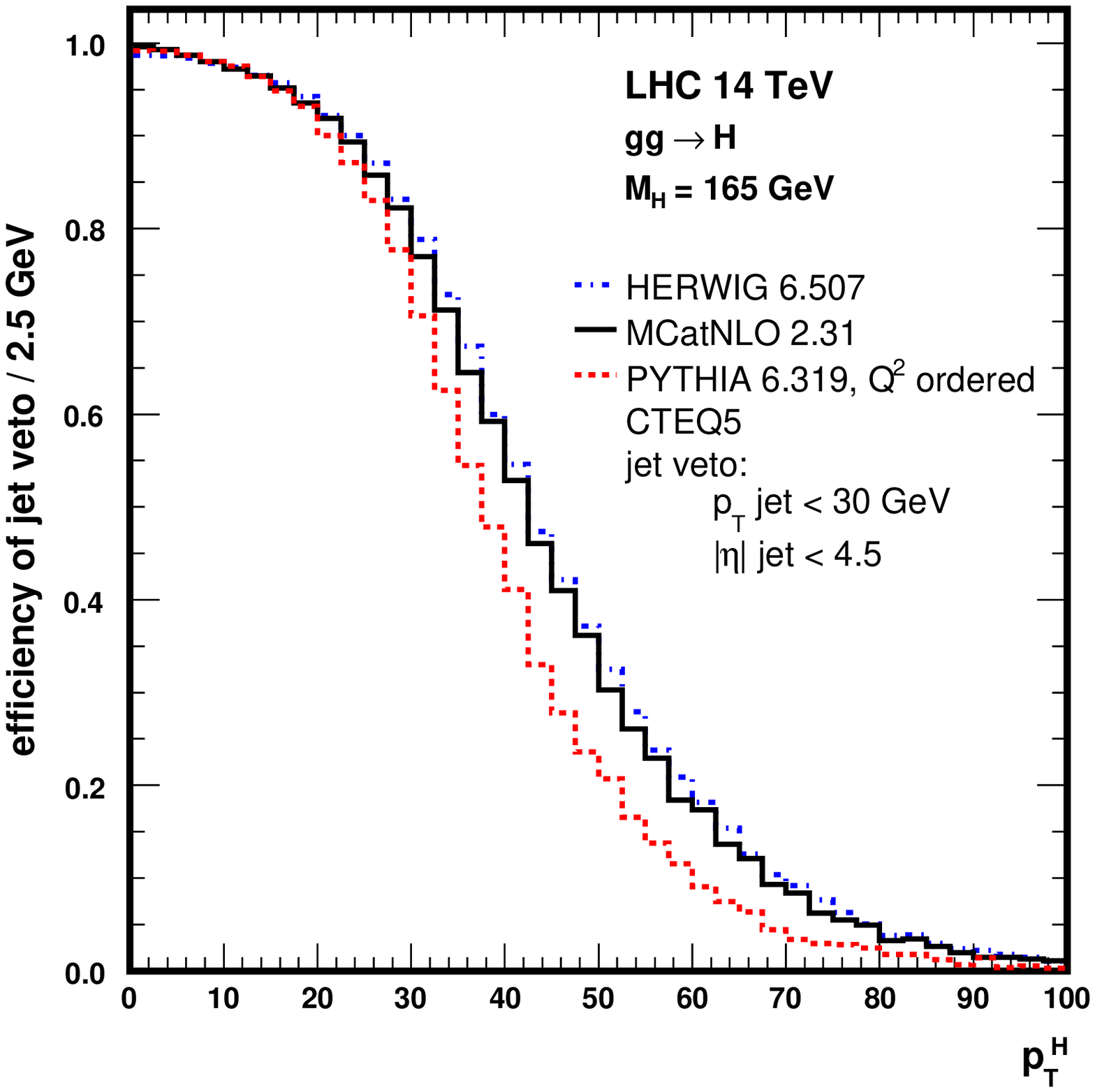}
\caption[fig2]{$\rm p_T$ of the leading jet for PYTHIA, HERWIG and MC@NLO
(left), and efficiency of the jet veto with 30 GeV as a function of $\rm p_T$
Higgs (right). }
\label{maxjetpt}
\end{center}
\end{figure}

In this section, PYTHIA with the default $\rm Q^2$ ordered showering, HERWIG
and MC@NLO is compared.  In Fig.~\ref{pth}, the normalized $\rm p_T$
Higgs spectra are shown for the three Monte Carlos. In the linear scale, one
can see that at low $\rm p_T$, HERWIG and MC@NLO are very similar. This can be
expected as the soft and collinear emissions of MC@NLO are treated by
HERWIG. In the low $\rm p_T$ region, PYTHIA predicts a softer leading jet
spectrum than HERWIG and therefore also a softer $\rm p_T$ Higgs spectrum in
this region. At high $\rm p_T$ however, PYTHIA is harder than HERWIG. 
Figure~\ref{maxjetpt}(left) shows the leading jet spectrum in the logarithmic
scale. HERWIG implements angular ordering exactly and thus correctly sums the
$LL$ (Leading Log) and part of the $N^kLL$ (Next-to..Leading Log)
contributions. However, the current version of HERWIG, available on the web,
does not treat hard radiations in a consistent way. Hence the spectrum drops
quickly at high $\rm p_T$ Higgs (Fig.~\ref{pth}(right)) and high $\rm p_T$
of the leading jet (Fig.~\ref{maxjetpt}(left)). In contrast, PYTHIA does not
treat angular ordering in an exact way, but includes hard matrix element
corrections. Therefore, PYTHIA looks more similar to MC@NLO at high $\rm
p_T$. MC@NLO on the other hand correctly treats the hard radiation up to NLO,
combining the high $p_T$ spectrum with the soft radiation of HERWIG. 

 In Fig.~\ref{maxjetpt}(right), the efficiency of the jet veto is shown for the
three different Monte Carlos as a function of $\rm p_T$ Higgs.  One observes a
strong dependency of the $\rm p_T$ Higgs on the jet veto. Once a jet veto is
defined, the efficiency starts to drop quickly as soon as the $\rm p_T$ of the
Higgs is close to the $\rm p_T$ used to define a jet veto. However, as the
Higgs $\rm p_T$ can be balanced by more than one jet, the efficiency is not
zero above this value.

\subsection{HERWIG + matrix element corrections and PYTHIA with new $\rm p_T$ ordered shower model}
\begin{figure}[!htb]
\begin{center}
\includegraphics*[scale=0.26]{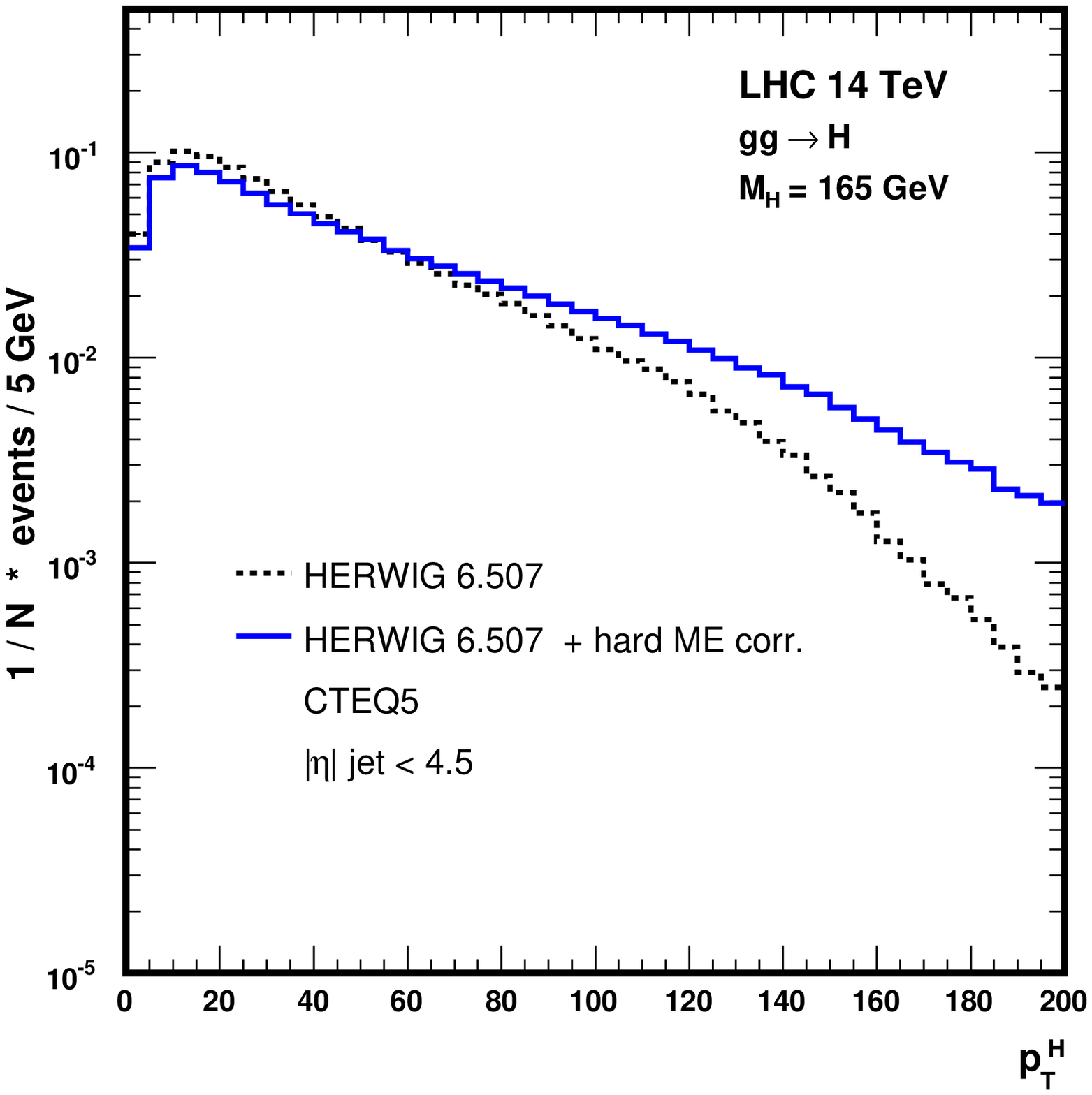}
\includegraphics*[scale=0.26]{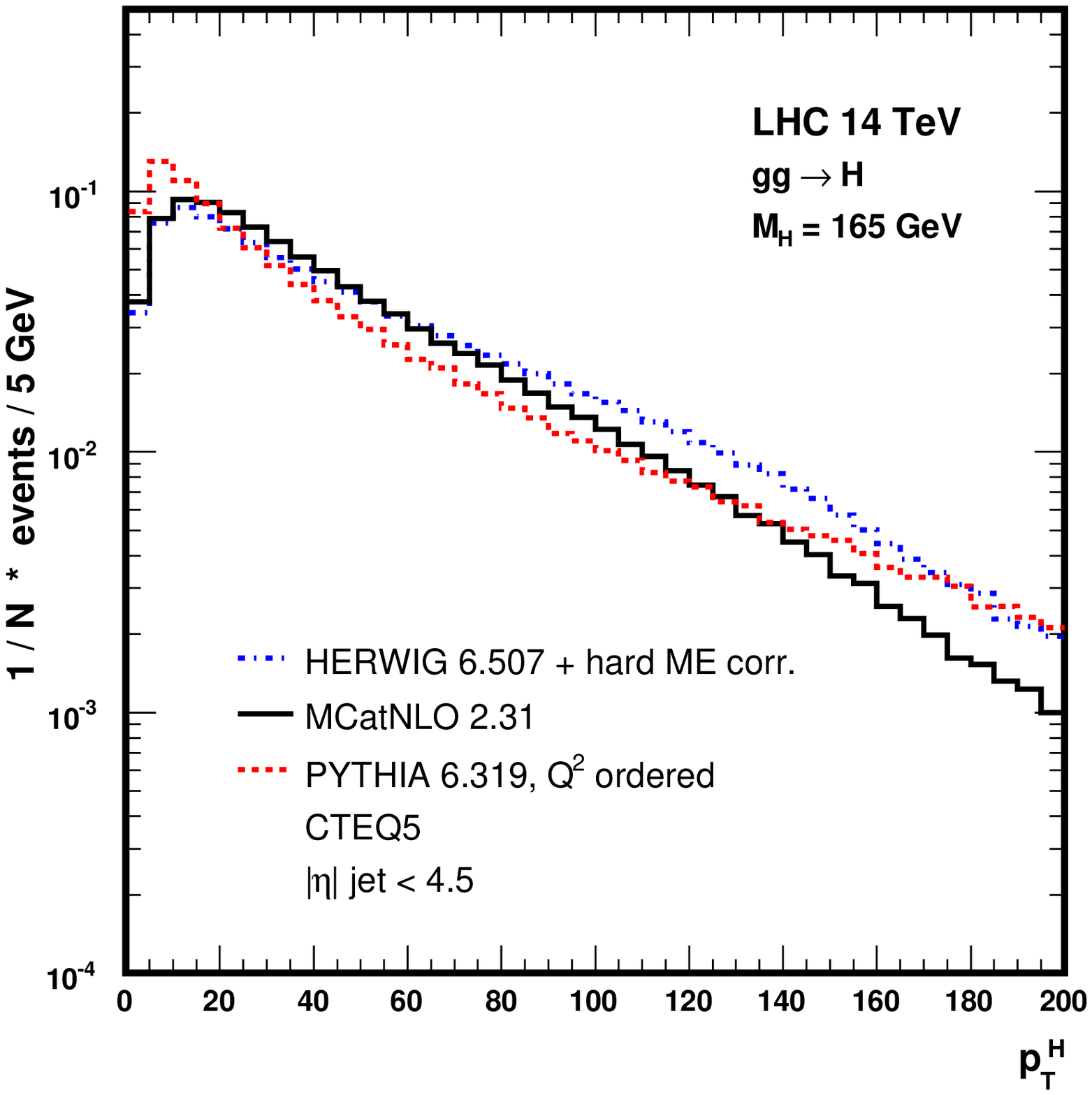}
\includegraphics*[scale=0.26]{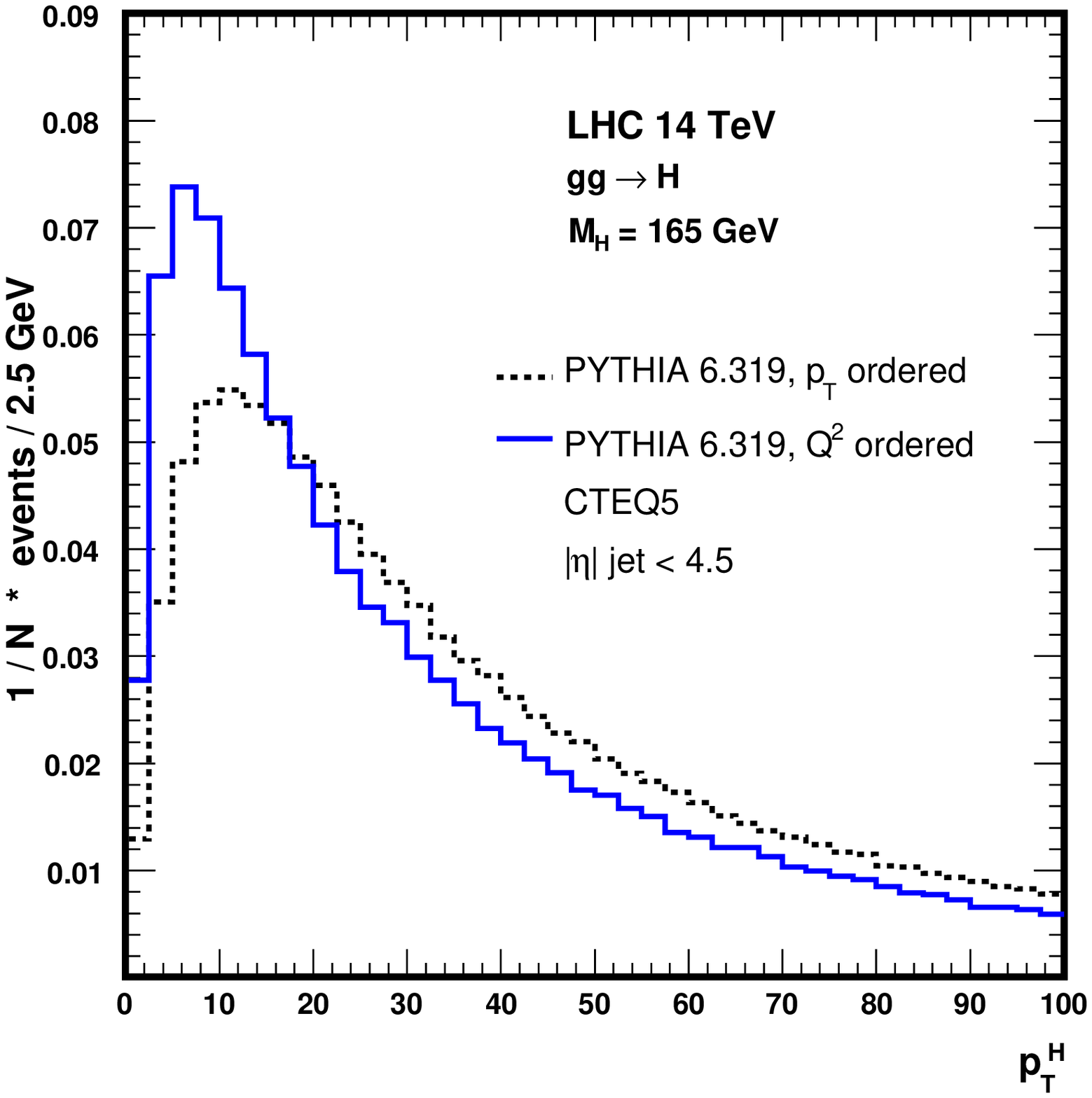}
\caption[fig3]{The $\rm p_T$ Higgs spectrum for HERWIG with and without hard
matrix element corrections (left) and HERWIG with matrix element corrections
in comparison with PYTHIA and MC@NLO (middle). On the right, the $\rm p_T$
Higgs spectrum for the default $Q^2$ ordered and the new $\rm p_T$ ordered
showering models is shown.}
\label{herwmo}
\end{center}
\end{figure}

G.~Corcella provided a preliminary version of HERWIG including hard matrix
element corrections for gg $\rightarrow$H \cite{Corcella:2004fr}. The hard
matrix element corrections lead to harder jets (Fig.~\ref{herwmo}(left)) and
therefore the jet veto is more effective. At high $\rm p_T$, PYTHIA and HERWIG
become now very similar (Fig.~\ref{herwmo}(middle)). 

\begin{figure}[!h]
\begin{center}

\includegraphics*[scale=0.37]{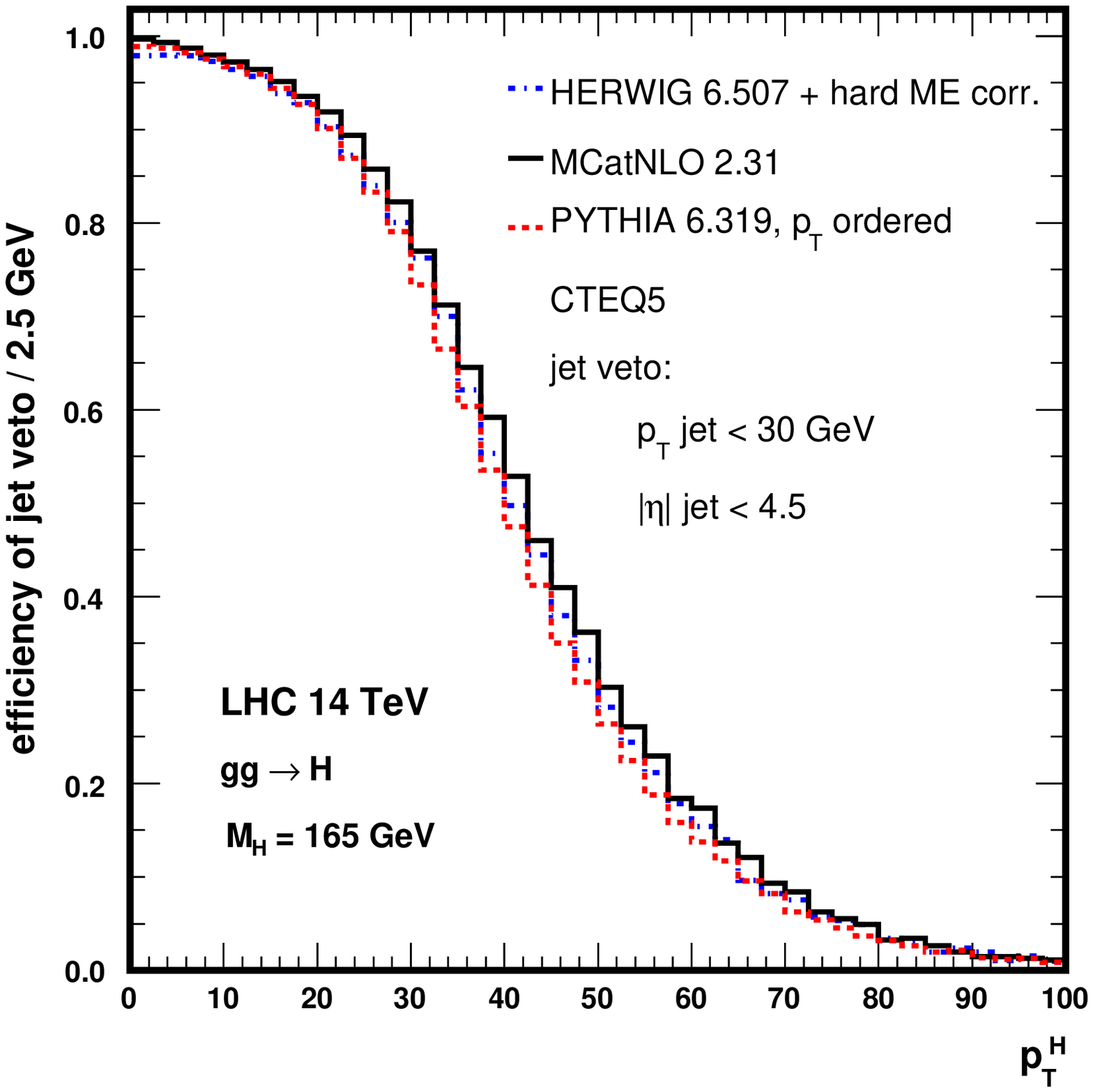}
\includegraphics*[scale=0.37]{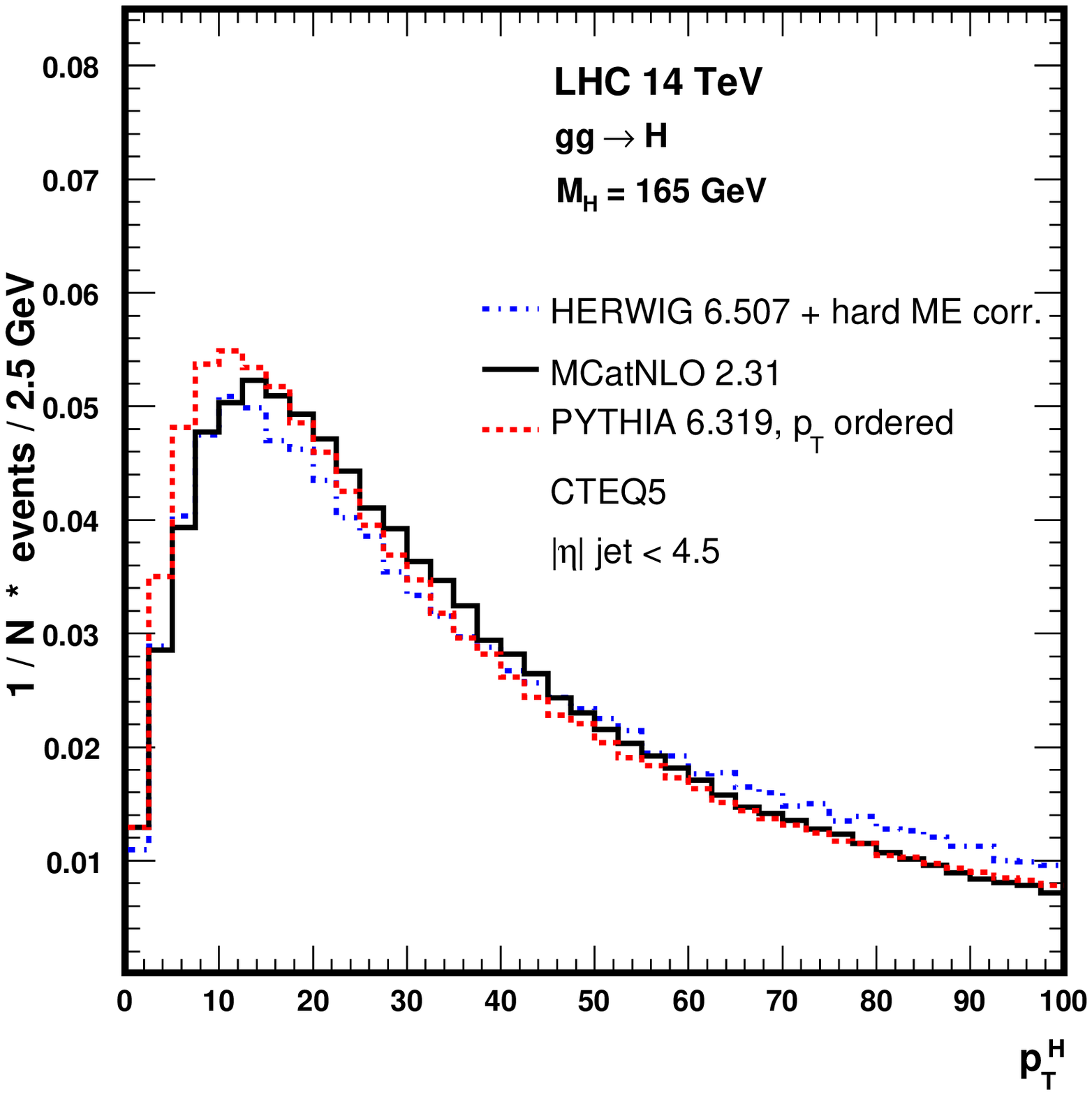}
\caption[fig4]{Efficiency after a jet veto is applied and $p_T$ Higgs distribution for HERWIG with matrix element corrections, PYTHIA with new $p_T$ ordered shower model and MC@NLO. }
\label{effptnew}
\end{center}
\end{figure}

 Also the new $\rm
p_T$ ordered showering model in PYTHIA is tested. Figure~\ref{herwmo}(right)
shows the $\rm p_T$ Higgs spectrum for the default $\rm Q^2$ ordered and the
new $\rm p_T$ ordered showering models. The jets from the new showering model
are shifted to higher $\rm p_T$ in the low $\rm p_T$ region and therefore also
the $\rm p_T$ of the Higgs boson is more similar to HERWIG and MC@NLO in this
region. In Fig.~\ref{effptnew}, the efficiency after a jet veto is applied
(left) and the $p_T$ Higgs distribution (right) for HERWIG with matrix element
corrections, PYTHIA with new $p_T$ ordered shower model and MC@NLO is shown.


Table \ref{DNtable1} shows the number for the efficiency of a jet veto of 30 GeV
for MC@NLO, PYTHIA and HERWIG with and without matrix element corrections.  In
the first row, the number of the efficiency for a $\rm p_T$ Higgs between 0
and 80 GeV is shown. The second row shows the inclusive efficiency over all
events. One has to keep in mind that after all selection cuts are applied,
only the low $\rm p_T$ Higgs region is important \cite{Davatz:2004zg}. One can
see that in the region important for the Higgs search in the WW channel (first
row), the difference between the new $\rm p_T$ ordered PYTHIA version, HERWIG
with matrix element corrections and MC@NLO is only around 1\%. The overall
uncertainty between all these different simulations is around 10\%.

\begin{table}[!ht]
\centering
\caption{Efficiency of the jet veto for MC@NLO, PYTHIA with $\rm Q^2$- and $\rm p_T$- ordered shower models, HERWIG with and without matrix element corrections.}
\vspace*{1mm}
\begin{tabular}{|c|c|c|}
\hline
 & Efficiency for events with a  & Inclusive efficiency  \\
 & $\rm p_T$ Higgs between 0 and 80 GeV & (all events) \\
\hline
MC@NLO 2.31  & 0.69&0.58  \\
\hline
PYTHIA 6.319, $\rm Q^2$ ordered  & 0.73 &0.62 \\  
\hline
PYTHIA 6.319, $\rm p_T$ ordered  & 0.68 &0.53 \\ 
\hline
HERWIG 6.507 & 0.70 &0.63 \\
\hline
HERWIG 6.507 + ME Corrections  &  0.68 &0.54\\
\hline
\end{tabular}
\label{DNtable1}
\end{table}

In Ref.~\cite{DeRoeck:2005rf}, the effect of including a realistic detector resolution,
NNLO calculations (described in Ref.~\cite{Davatz:2004zg}) and different tunings
for the underlying event were studied in addition. As a result, the effect on
the jet veto efficiency, when smearing the $\rm E_T$ of a jet with the jet
resolution of e.g. CMS \cite{tdr}, is less than 1\%. The uncertainty of the jet
veto efficiency does not change significantly including higher order
corrections with the re-weighting method described in Ref.~\cite{Davatz:2004zg}. The
biggest part of the events is at low $p_T$, while the effect of higher order
corrections occurs mostly at very high $p_T$. PYTHIA with $\rm Q^2$ showering
model was studied with different underlying event tuning schemes, which are
the ATLAS Tune \cite{Dawson:2004th}, CDF Tune A \cite{Field:2005sa} and PYTHIA default
(MSTP(81)=1, MSTP(82)=3 \cite{Sjostrand:2004ef}). The different tunings lead
to about the same efficiency, and also the difference in the efficiency with
and without underlying event is less than 1\%.

\subsection{MC@NLO: Effect of varying the factorization and renormalization scale}
\begin{figure}[!h]
\begin{center}
\includegraphics*[scale=0.37]{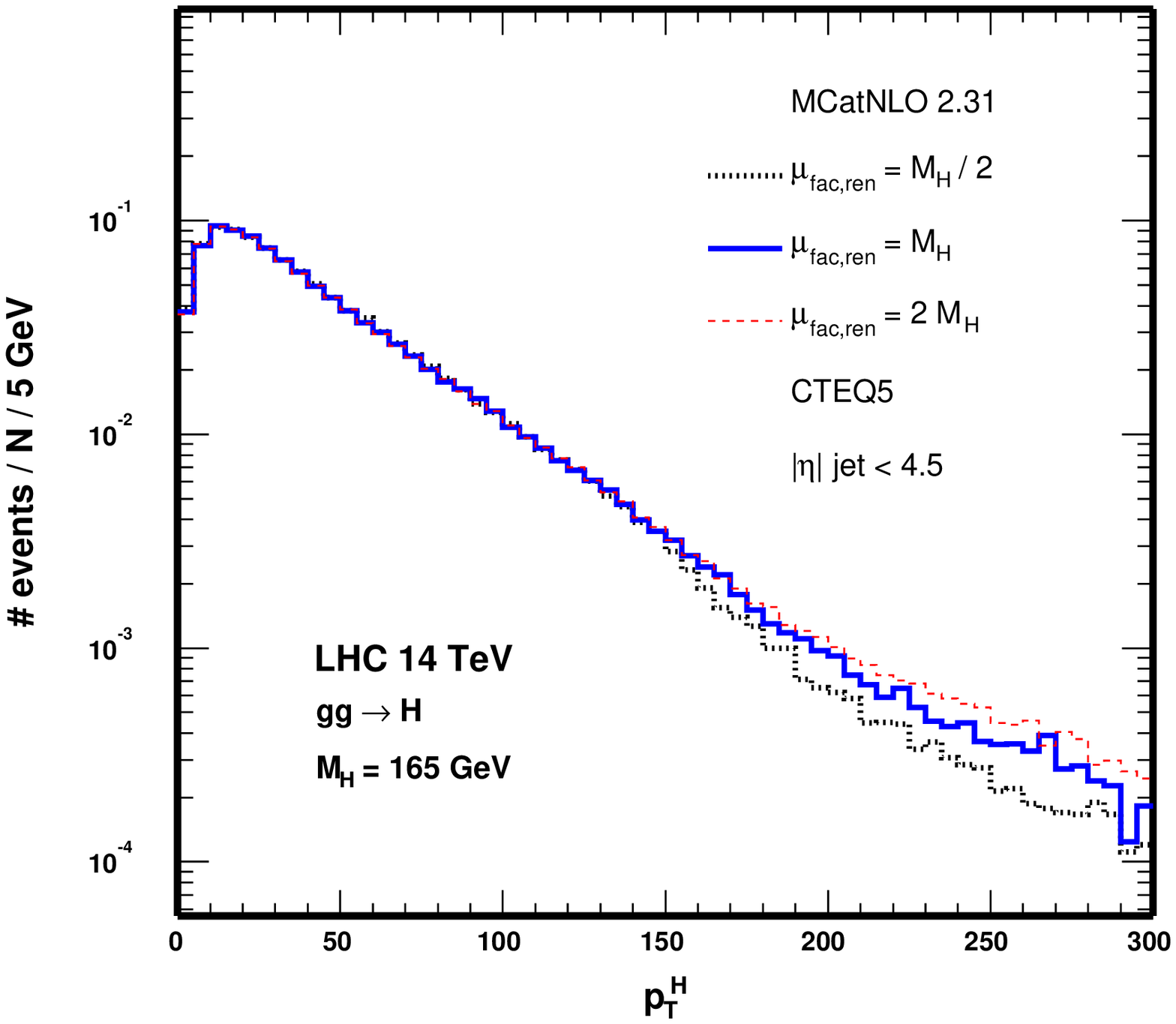}
\includegraphics*[scale=0.37]{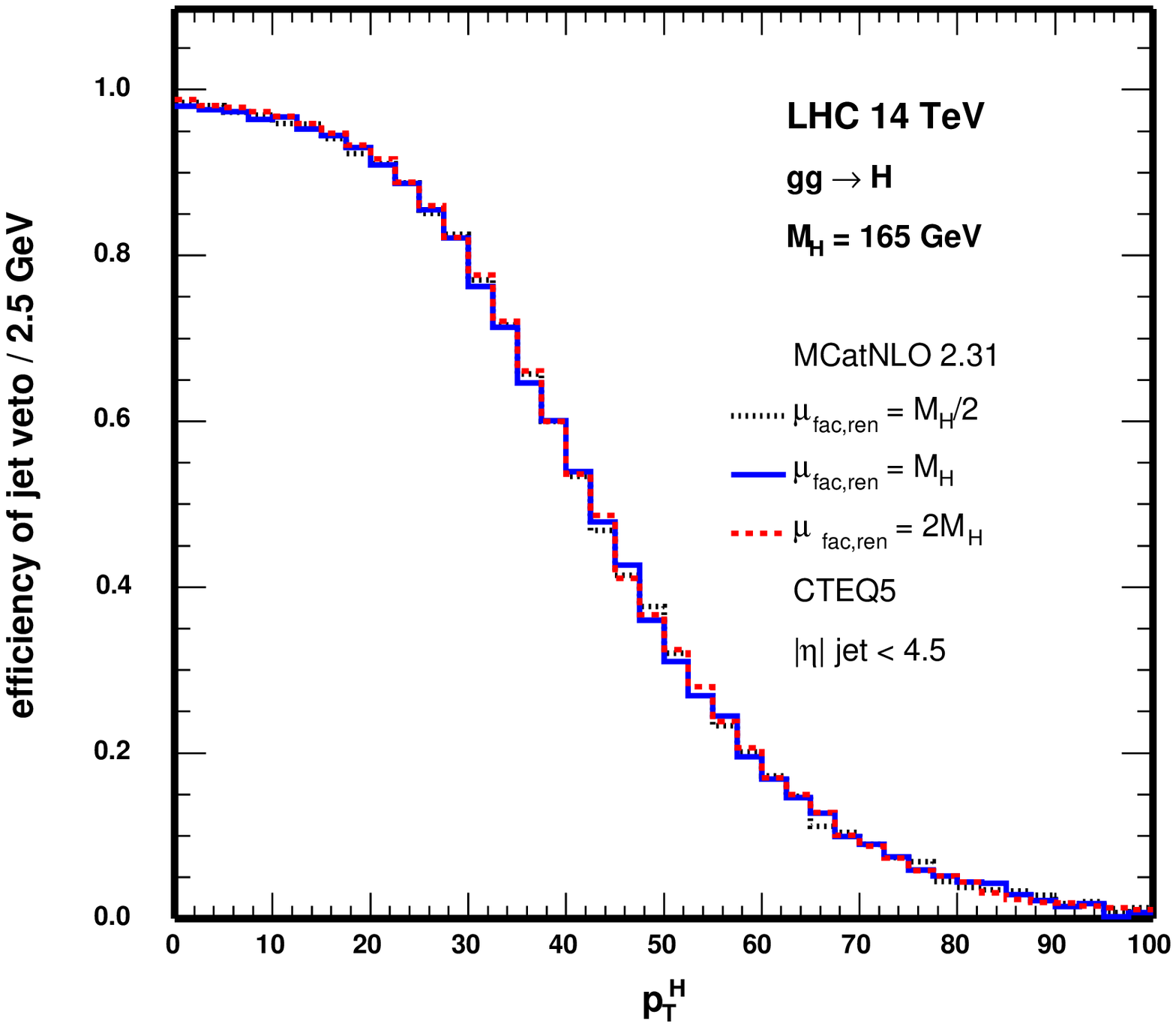}
\caption[fig5]{Number of events and efficiency after a jet veto 30 GeV is applied for MC@NLO with different scale choices.}
\label{effptmu}
\end{center}
\end{figure}

To get an estimate of the uncertainty due to different factorization and
renormalization scales, three MC@NLO samples were produced with scales $\rm
\mu_{fac, rec}$ between $\rm M_{H}/2$ and $2 \rm M_{H}$. In
Fig.~\ref{effptmu}(left), the $\rm p_T$ Higgs spectrum and in
Fig.~\ref{effptmu}(right) the efficiency after a jet veto of 30 GeV is
applied are shown for these three samples. The only difference is at very high
$p_t$, whereas the bulk of the events is at low $\rm p_T$. Therefore, as can
be seen also in Table~\ref{mutable}, the effect of different scales on the jet
veto efficiency is negligible.

\begin{table}[!ht]
\centering
\caption{Efficiency of the jet veto for MC@NLO with different scale choices.}
\vspace*{1mm}
\begin{tabular}{|c|c|c|}
\hline
 & Efficiency for events with a  & Inclusive efficiency  \\
 & $\rm p_T$ Higgs between 0 and 80 GeV & (all events) \\
\hline
$\rm \mu_{fac, rec} = M_{H}/2$  & 0.685&0.585  \\
\hline
$\rm \mu_{fac, rec} = M_{H}$  & 0.692 &0.583 \\  
\hline
$\rm \mu_{fac, rec} = 2 M_{H}$ & 0.687 &0.582 \\
\hline
\end{tabular}
\label{mutable}
\end{table}

\subsection{Conclusions}
We have studied the uncertainty of the jet veto efficiency due to the use of
different Monte Carlo generators in the gg$\rightarrow$H channel. The
uncertainty between PYTHIA, HERWIG and MC@NLO without underlying event lies
within 10\%. Including higher order QCD corrections does not increase this
uncertainty significantly. Also the effect of including a realistic jet $E_T$
resolution (for this study we took the CMS jet $E_T$ resolution) is very
small. We also studied the effect of the underlying event with different
tuning models in PYTHIA (PYTHIA default, ATLAS Tune and CDF Tune A). The
tuning models considered lead to about the same efficiency and the effect of
including underlying events or not is smaller than 1\%. Taking into account
the new $p_T$ ordered showering model of PYTHIA and a preliminary version of
HERWIG with matrix element corrections reduces the uncertainty in the region
which is relevant for the Higgs to WW signal search to 1\%.

\subsection*{Acknowledgements}
We would like to thank G.~Corcella, M.~Dittmar, S.~Frixione, M.~Seymour and
T.~Sjostrand for useful discussions and comments.

%%%%%%%%%%%%%%%%%%%%%%%%%%%%%%%%%%%%%%%%%%%%%%%%%%%%%%%%%%%%%%%%%%%%%%%%%%%%%
\section[Comparison between MCFM and PYTHIA for the $gb\rightarrow bh$ and
$gg\rightarrow bbh$ processes at the LHC]
{COMPARISON BETWEEN MCFM AND PYTHIA FOR THE $gb\rightarrow bh$ and
$gg\rightarrow bbh$ PROCESSES AT THE LHC~\protect
\footnote{Contributed by: J.M.~Campbell, A.~Kalinowski, A.~Nikitenko}}
\subsection{Introduction}

  An accurate generation of the $\rm gb \rightarrow \rm bh$ and $\rm gg \rightarrow \rm b \bar{\rm b} \rm h$ 
  processes is crucial both for the measurement of the MSSM $\rm gg \rightarrow \rm b \bar{\rm b} \rm h$, 
  $\rm h \rightarrow$2$\tau$ cross section and for constraining tan($\beta$) from event-counting 
  at the LHC \cite{Kinnunen:2004ji}. The production of a MSSM Higgs boson in association with b quarks is the dominant
  production process at high tan($\beta$) and for $\rm M_{\rm h}>$ 150-200 GeV/$c^2$. The CMS experimental selections
  include single b-tagging, a veto on the other jets in the event (excluding $\tau$ jets), a cut on the angle between 
  the two $\tau$ leptons in the transverse plane and a cut on the reconstructed mass of the $\tau$-lepton pair using the 
  missing transverse energy. Thus, the correct generation of the pseudorapidity and $\rm p_{\rm T}$ of the b 
  quarks and the Higgs boson is very important.

  In PYTHIA \cite{Sjostrand:2000wi}, both the $\rm gb \rightarrow \rm bh$ (2$\rightarrow$2) and 
  $\rm gg \rightarrow \rm b \bar{\rm b} \rm h$ (2$\rightarrow$3) processes are 
  available, each of which produces a $\rm b \bar{\rm b} \rm h$ final state. 
  In the $\rm gb \rightarrow \rm bh$ process the second b quark ($\bar{\rm b}$) comes from the 
  gluon splitting ($\rm g \rightarrow \rm b\bar{\rm b}$) in the initial state parton shower
  and is always present in the PYTHIA event. 

  In this paper we compare the kinematics of the PYTHIA 2$\rightarrow$2 and 2$\rightarrow$3 processes  
  with the next-to-leading order (NLO) calculations implemented in the MCFM 
  program \cite{Campbell:2002zm}. The NLO calculations in MCFM start from the leading order (LO) 
  $\rm gb \rightarrow \rm bh$ process, with the LO $\rm gg \rightarrow \rm b \bar{\rm b} \rm h$ contribution
  included as part of the NLO calculation. The LO MCFM calculations were also compared with the PYTHIA 2$\rightarrow$2 
  process when both the initial and final state radiation was switched off.   

\subsection{Simulation setup}

The kinematic distributions were compared for two values of the Higgs boson mass, $\rm m_{\rm h}$=200 and 500 GeV/$c^{2}$. 
PYTHIA 6.227 was used to generate the processes $\rm gg \rightarrow \rm b \bar{\rm b} \rm h$ 
(MSUB(121)=1, KFPR(121,2)=5) and $\rm gb \rightarrow \rm bh$  (MSUB(32)=1 with gluon and b quark as incoming partons). 
The CTEQ6L1 PDF was used with the renormalization and the factorization scales equal and set to 
$\mu_{\rm R} = \mu_{\rm F} = (\rm m_{\rm h} + 2 \rm m_{\rm b})/4$. The primordial parton $\rm k_{\rm T}$  
was switched off in PYTHIA (MSTP(91)=0). To reduce the CPU time, the fragmentation, decays and multiple interactions 
were switched off in PYTHIA (MSTP(111)=0, MSTP(81)=0). 
For the $\rm gb \rightarrow \rm bh$ process, a lower cut of 20 GeV/$c$ was set on the $\rm p_{\rm T}$ of the  outgoing
partons in the rest frame of the hard interaction (CKIN(3)=20 in PYTHIA).  The jets were reconstructed from the partons
using the simple cone algorithm with a cone size of 0.7.

\subsection{Comparison of PYTHIA and MCFM at leading order}

The distributions for the $\rm gb \rightarrow \rm bh$ process in PYTHIA and LO MCFM were compared. The initial and the
final state radiation in PYTHIA was switched off, so that a direct comparison of the LO matrix element implementation
in PYTHIA and MCFM could be performed. The distributions of the b quark $\rm p_{\rm T}$ and the Higgs boson 
mass are shown in Figures~\ref{ptb1_Fix} and~\ref{higgsMass500} respectively, for $\rm m_{\rm h}$=500 GeV/$c^{2}$. 
The dashed line shows the PYTHIA distributions, whereas the dotted line shows the MCFM distributions. 
There is a clear difference between the PYTHIA and MCFM curves. The dominant reason is that, in PYTHIA the matrix
elements make use of the kinematic relation $\rm s + \rm t+ \rm u = \rm m_{\rm h}^2$. In contrast, MCFM uses
$\rm s + \rm t + \rm u = \rm Q^2$, where $\rm Q^2$ is the virtuality of the Higgs boson. This is the
appropriate form to use when the Higgs boson is allowed off-shell using the Breit-Wigner approximation; it gives rise 
to a large discrepancy when the Higgs boson is very far off-shell (for instance, $\rm Q^2 \gg \rm m_{\rm h}^2$).
Corrections to the PYTHIA matrix elements 
were made by substituting $\rm Q^2$ for $\rm m_{\rm h}^2$ where appropriate\footnote{
Thanks to T. Sj$\ddot{\rm o}$strand for providing the  fixed matrix element.} and
the solid lines in Figures~\ref{ptb1_Fix} and~\ref{higgsMass500} reflect the PYTHIA results after 
this change. With the corrected matrix elements the discrepancy between
PYTHIA and MCFM is significantly reduced. The remaining difference in the Higgs boson mass distribution is due to the 
different treatment of the Higgs boson propagator. MCFM uses the fixed width approach, whereas PYTHIA uses a
width which is dependent on $\rm Q^2$. In particular, the drop near 160 GeV/$c^{2}$ corresponds to the closure of
the WW decay channel for the Higgs boson. This calculation is most useful in the resonance region. 
Away from the resonance peak, once the decay of the Higgs boson is included contributions from other 
interfering diagrams (such as ones in which the Higgs is replaced by a Z boson) can change the shape of the
prediction.

\subsection{Comparison of next-to-leading order MCFM and PYTHIA}

The comparison between the MCFM NLO predictions and PYTHIA was made when the initial and the final state radiation in
PYTHIA was switched on. In all figures shown below the solid line represents the distribution for the PYTHIA 
$\rm gb \rightarrow \rm bh$ process generated with the corrected matrix element and $\hat{\rm p}_{\rm T}>$20 GeV/$c$, 
the dashed line shows the distribution for the
PYTHIA $\rm gg\rightarrow \rm b\bar{\rm b} \rm h$ process and the dotted line corresponds to the MCFM 
$\rm gb \rightarrow \rm bh$ process at NLO. 

The $\rm p_{\rm T}$ distribution of the highest $\rm p_{\rm T}$ b jet with $|\eta|<2.4$ is shown in Figure~\ref{bjet1_pt_200} 
for $\rm m_{\rm h}$=200 GeV/$c^{2}$ and Figure~\ref{bjet1_pt_500} for $\rm m_{\rm h}$=500 GeV/$c^{2}$. Each of the
histograms is normalized to unity in the region $\rm p_{\rm T}>$ 20 GeV/$c$. One sees that both PYTHIA processes show 
good agreement with MCFM.

The efficiency of the central jet veto (after single b tagging) depends, in particular, on the $\rm p_{\rm T}$ 
and  $\eta$ distributions of the second (less energetic) b jet.  The $\rm p_{\rm T}$ distribution of the second b jet within 
$|\eta|<2.4$ is shown in Figure~\ref{bjet2_pt_200} for $\rm m_{\rm h}$=200 GeV/$c^{2}$ and Figure~\ref{bjet2_pt_500} for 
$\rm m_{\rm h}$=500 GeV/$c^{2}$, after requiring that the first (most energetic) b jet be in the tagging range 
$\rm p_{\rm T}^{\rm b ~ \rm jet}>$20 GeV/$c$ and $|\eta^{\rm b ~ \rm jet}|<$2.4. 
Once again, the histograms are normalized to unity in the region $\rm p_{\rm T}>$20 GeV/$c$. 
One can see that the second b jet in the PYTHIA $\rm gb \rightarrow \rm bh$ process is much softer
than in NLO MCFM, while this calculation agrees well with the PYTHIA 
$\rm gg\rightarrow \rm b\bar{\rm b} \rm h$ process. This is to be expected since the second b quark ($\bar{\rm b}$)
in the $\rm gb \rightarrow \rm bh$ process is produced by the parton shower in the initial state. At high
$p_T$ one expects the 2$\rightarrow$3 process, which is included as a NLO effect in MCFM, to provide a better description
and one sees that this is indeed the case.

Figures~\ref{bjet1_eta_200}, \ref{bjet1_eta_500}, \ref{bjet2_eta_200} and \ref{bjet2_eta_500} show the pseudorapidity 
distributions for the first and the second b jets for Higgs boson masses of 200 and 500 GeV/$c^2$. The content of
the histograms is normalized to unity in the $\eta$ interval between -2 and +2. The PYTHIA distributions for the leading 
b jet for the Higgs boson mass of 200 GeV/$c^{2}$ agree well with the MCFM result (Figure~\ref{bjet1_eta_200}), whereas 
for $\rm m_{\rm h}$=500 GeV/$c^{2}$ the MCFM $\eta$ distribution is less central than in PYTHIA (Figure~\ref{bjet1_eta_500}).
The second b jet in the $\rm gb \rightarrow \rm bh$ process is distributed in the forward/backward direction more in PYTHIA 
than in MCFM (Figures~\ref{bjet2_eta_200} and \ref{bjet2_eta_500}). This is again due to the aforementioned reason that
the second b quark is produced in PYTHIA from the parton shower. The $\eta$ distribution of the second 
b jet in the PYTHIA $\rm gg\rightarrow \rm b\bar{\rm b} \rm h$ process is close to MCFM, but there is still some 
difference which is more pronounced for $\rm m_{\rm h}$=200 GeV/$c^{2}$ than for $\rm m_{\rm h}$=500 GeV/$c^{2}$.

The experimental selections include cuts on the visible $\tau$-lepton energy, on the angle between the 
two $\tau$ leptons in the transverse plane and on the mass reconstructed from the missing transverse energy. 
Therefore the selection efficiency depends, in particular, on the $\rm p_{\rm T}$ spectrum of the Higgs boson. 
Figures~\ref{higgs_pt_200} and \ref{higgs_pt_500} show the Higgs boson $\rm p_{\rm T}$ spectrum after cuts
which imitate the experimental selections of single b tagging and a jet veto. These cuts require that:
\begin{itemize}
\item the first b jet must lie in the tagging range, $\rm p_{\rm T}^{\rm b ~ \rm jet}>$ 20 GeV/$c$
  and $|\eta^{\rm b ~ \rm jet}|<$2.4;
\item no other jets should be observed in the central region, $\rm p_{\rm T}^{\rm other ~ \rm jet}<$ 20 GeV/$c$
 or $|\eta^{\rm other ~ \rm jet}|>$2.4.
\end{itemize}
Since MCFM includes the b quark as a massless particle, predictions are only available when applying a cut
on the b quark $\rm p_{\rm T}$. By momentum balance, this means that the Higgs boson transverse momentum is constrained
at LO to be greater than the jet cut of 20 GeV/$c$. However, when moving to NLO, the region below this begins
to be populated. This feature means that the NLO calculation does not provide reliable predictions in the
close vicinity of the jet cut.
Therefore we perform the comparison only for $\rm p_{\rm T}>$30 GeV/$c$ and normalize the histograms in 
Figures~\ref{higgs_pt_200} and \ref{higgs_pt_500} to unity in the $\rm p_{\rm T}$ interval between 30 and 
200 GeV/$c$. One can see that the Higgs boson $\rm p_{\rm T}$ spectrum calculated to NLO in MCFM is slightly softer
than either PYTHIA prediction. The effect on the selection efficiency requires further study but is expected to be small.

\subsection{Conclusions}

A comparison of the shapes of the kinematic distributions of b quarks and the Higgs boson was performed for the PYTHIA 
$\rm gb \rightarrow \rm bh$ and $\rm gg\rightarrow \rm b \bar{\rm b} \rm h$ processes and the
$\rm gb \rightarrow \rm bh$ process implemented at LO and NLO in MCFM.
The study was performed for two masses of the Higgs boson, 200 and 500 GeV/$c^2$, which lie at either end
of the interesting analysis region.

It was observed that the $\rm p_{\rm T}$ spectrum of the leading b jet in the PYTHIA $\rm gg\rightarrow \rm b\bar{\rm b} \rm h$ 
process is in good agreement with the one obtained from the NLO MCFM $\rm gb \rightarrow \rm bh$ process.
The PYTHIA $\rm gb \rightarrow \rm bh$ process leads to the second b jet being produced with a softer $\rm p_{\rm T}$
spectrum, due to the parton shower.
Neither of the two PYTHIA processes agrees exactly with the $\eta$ spectrum of the b jets in the NLO MCFM 
$\rm gb \rightarrow \rm bh$ process, but the PYTHIA $\rm gg\rightarrow \rm b\bar{\rm b} \rm h$ process shows
better agreement. The $\rm p_{\rm T}$ spectrum of the Higgs boson in the PYTHIA processes is slightly harder than in 
NLO MCFM. 

The $\rm p_{\rm T}$ shapes for the b jet and the Higgs boson were compared for 
$\rm p^{\rm b ~ \rm jet}_{\rm T}>$ 20 GeV/$c$ and $\rm p^{\rm H}_{\rm T}>$ 30 GeV/$c$. 
Since the experimental jet energy resolution for 20 GeV jets in CMS is of the order of 40\%, it would be very desirable
to make a comparison with NLO calculations using a much lower cut-off, for instance $\simeq$ 5 GeV/$c$.
However, such an exercise would require further theoretical input, namely a calculation which extends the MCFM
treatment to include effects due to the mass of the final state b quark.

\subsection*{Acknowledgments}

We would like to thank S. Willenbrock, F. Maltoni, M. Spira, M. Kr$\ddot{\rm a}$mer and J. Collins for very
useful discussions.

\begin{BBH2figures}{!hbtp}
  \resizebox{\linewidth}{!}{\includegraphics{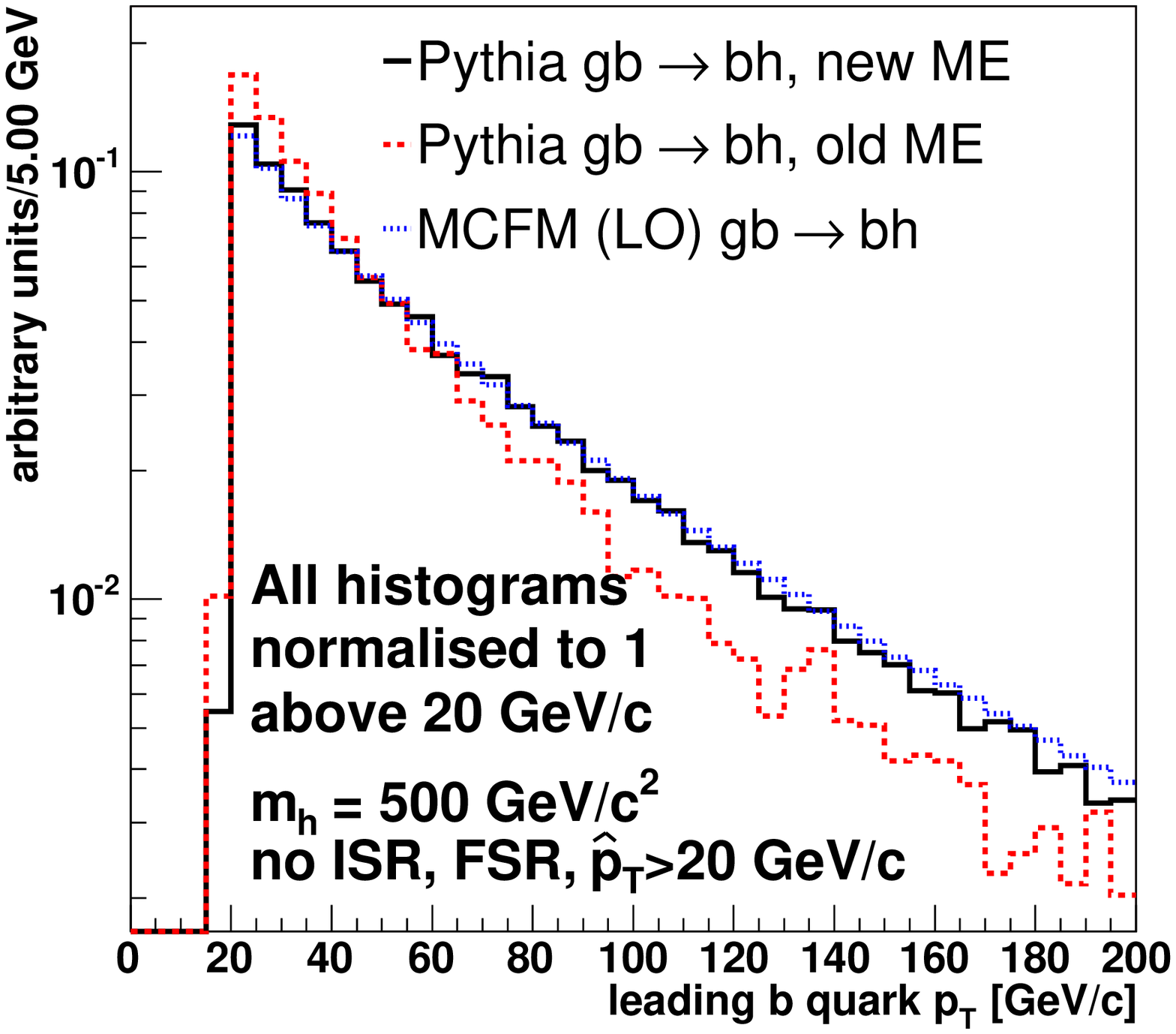}} &
  \resizebox{\linewidth}{!}{\includegraphics{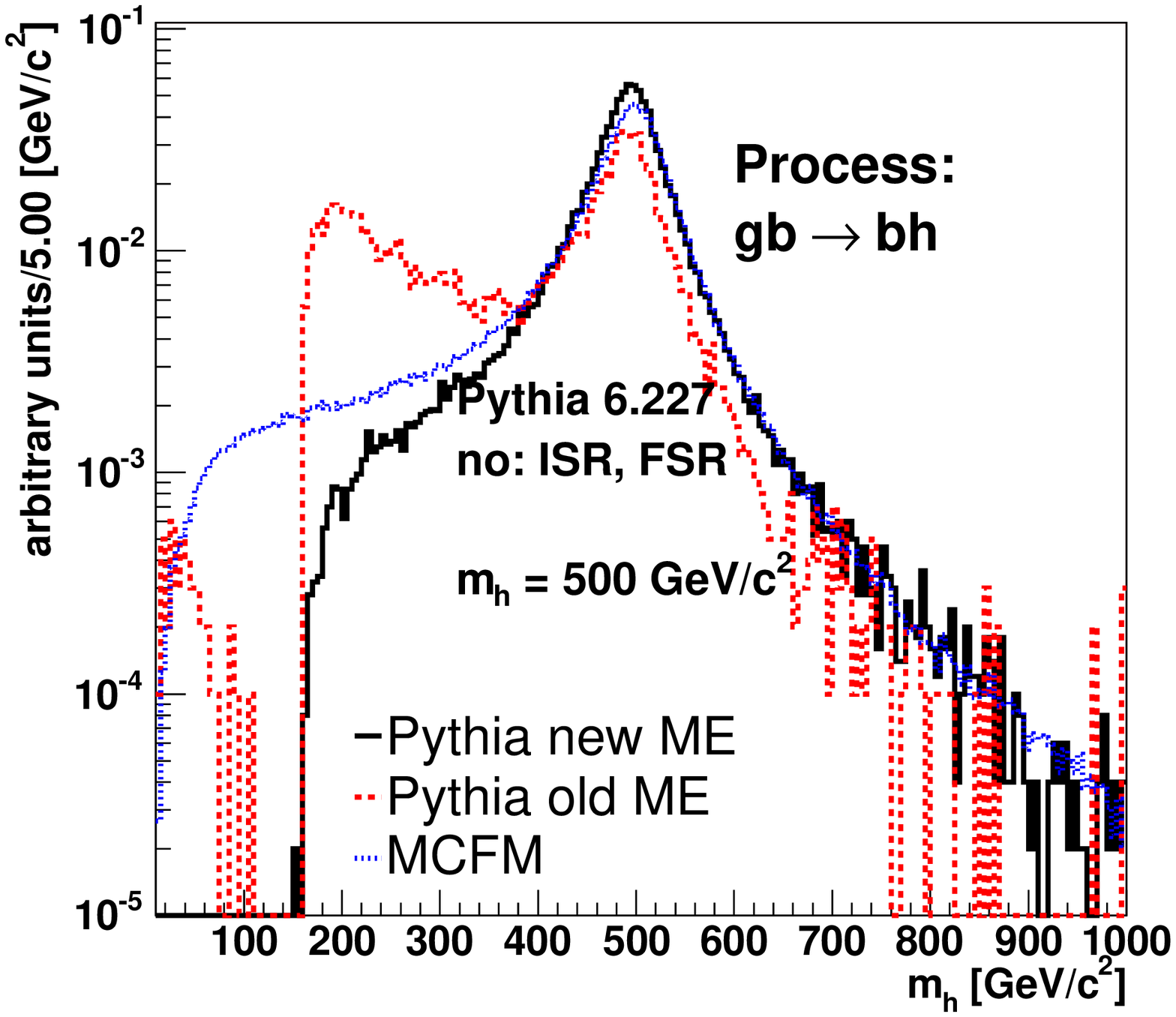}} \\
   \caption{The $\rm p_{\rm T}$ of the b quark in PYTHIA and LO MCFM for $\rm gb \rightarrow \rm bh$ process, 
            $\rm m_{\rm h}$=500 GeV/$c^{2}$. See more explanations in the text.} \label{ptb1_Fix} &  
   \caption{The Higgs boson mass distribution in PYTHIA and MCFM for $\rm m_{\rm h}$=500 GeV/$c^{2}$.} \label{higgsMass500}
\end{BBH2figures}

\begin{BBH2figures}{!hbtp}
  \resizebox{\linewidth}{!}{\includegraphics{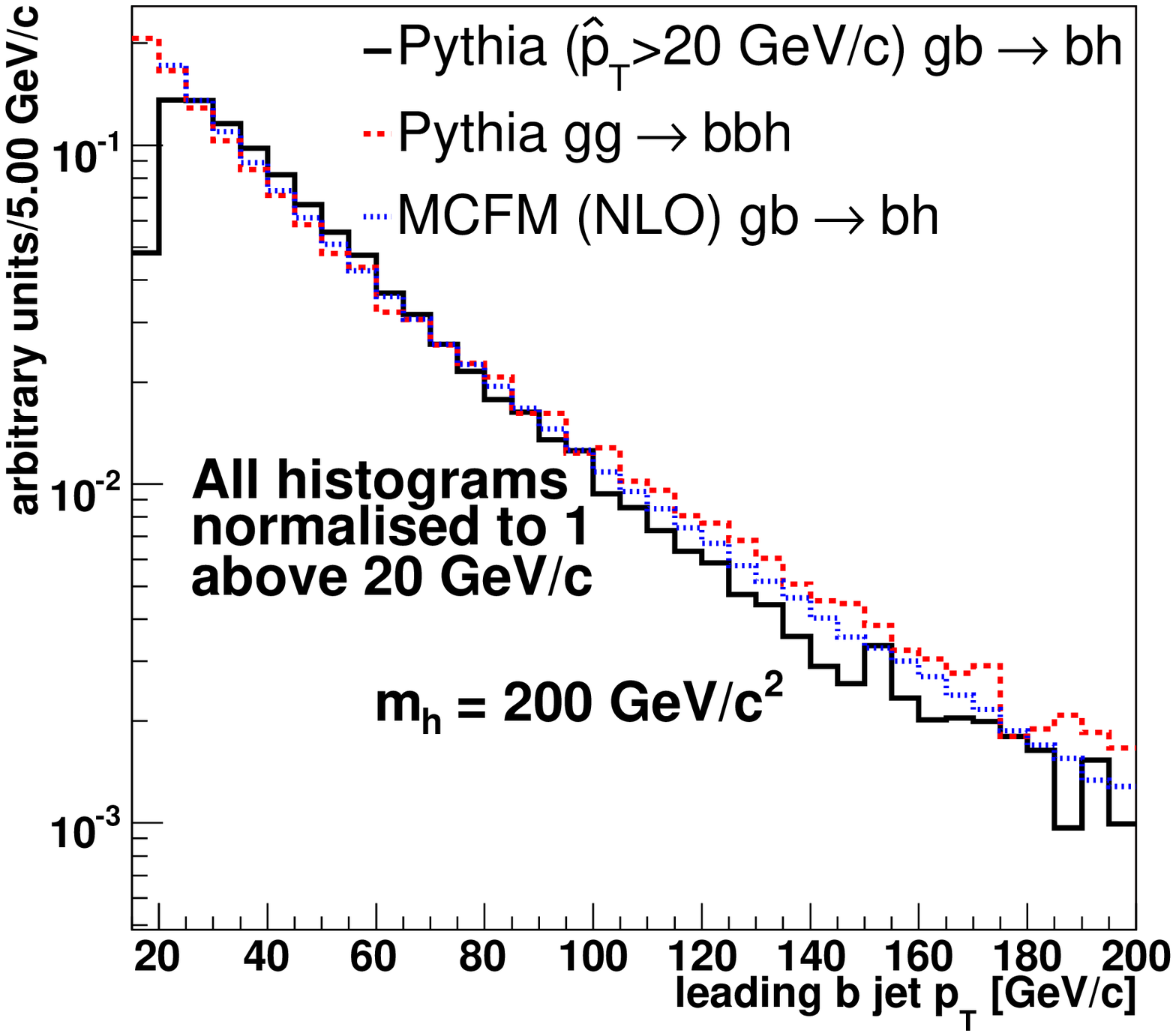}} &
  \resizebox{\linewidth}{!}{\includegraphics{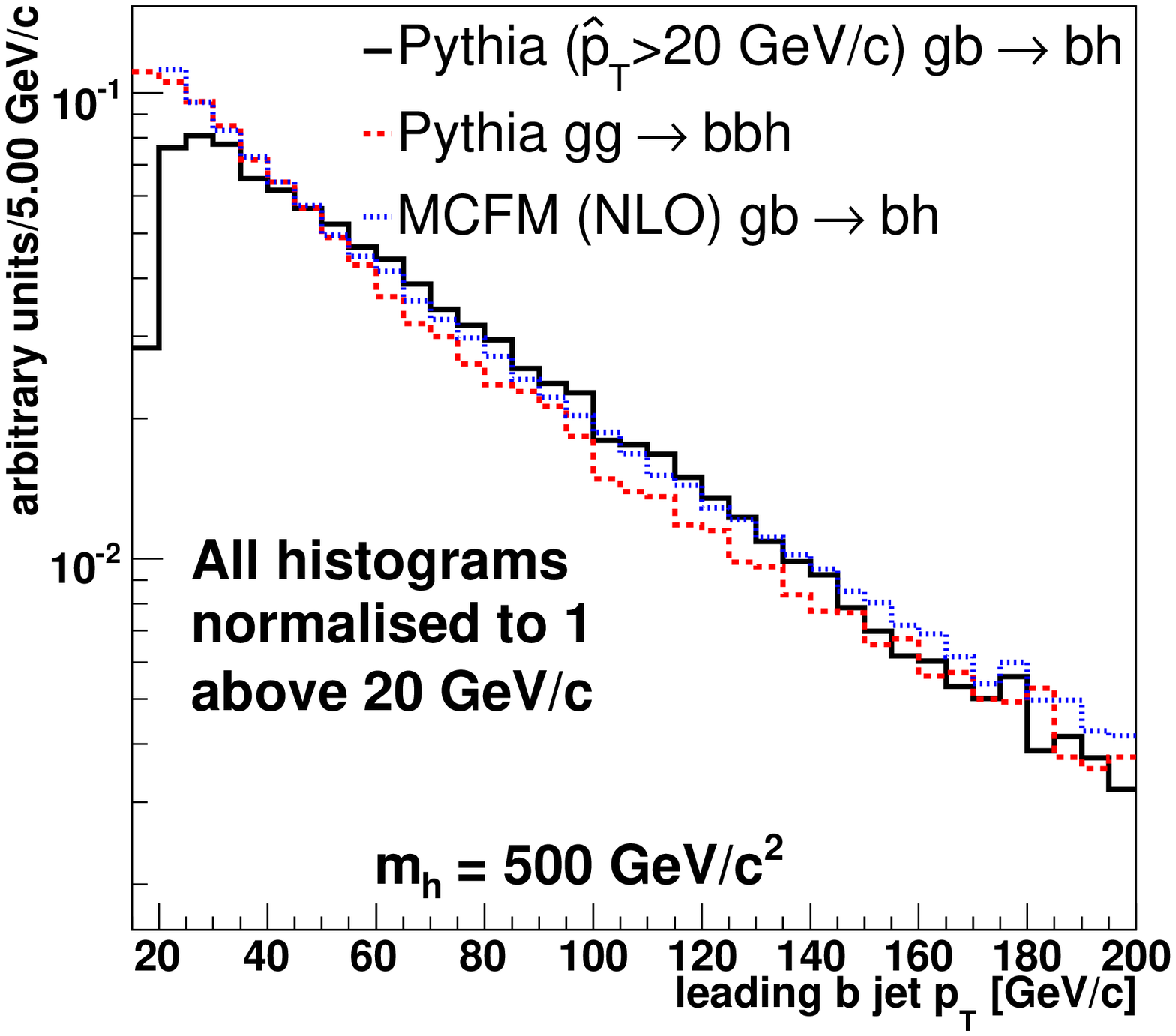}} \\
   \caption{The $\rm p_{\rm T}$ of the leading b jet in PYTHIA and in MCFM for 
            $\rm m_{\rm h}$=200 GeV/$c^{2}$.} \label{bjet1_pt_200} &  
   \caption{The $\rm p_{\rm T}$ of the leading b jet in PYTHIA and in MCFM for 
            $\rm m_{\rm h}$=500 GeV/$c^{2}$.} \label{bjet1_pt_500}
\end{BBH2figures}

\begin{BBH2figures}{!hbtp}
  \resizebox{\linewidth}{!}{\includegraphics{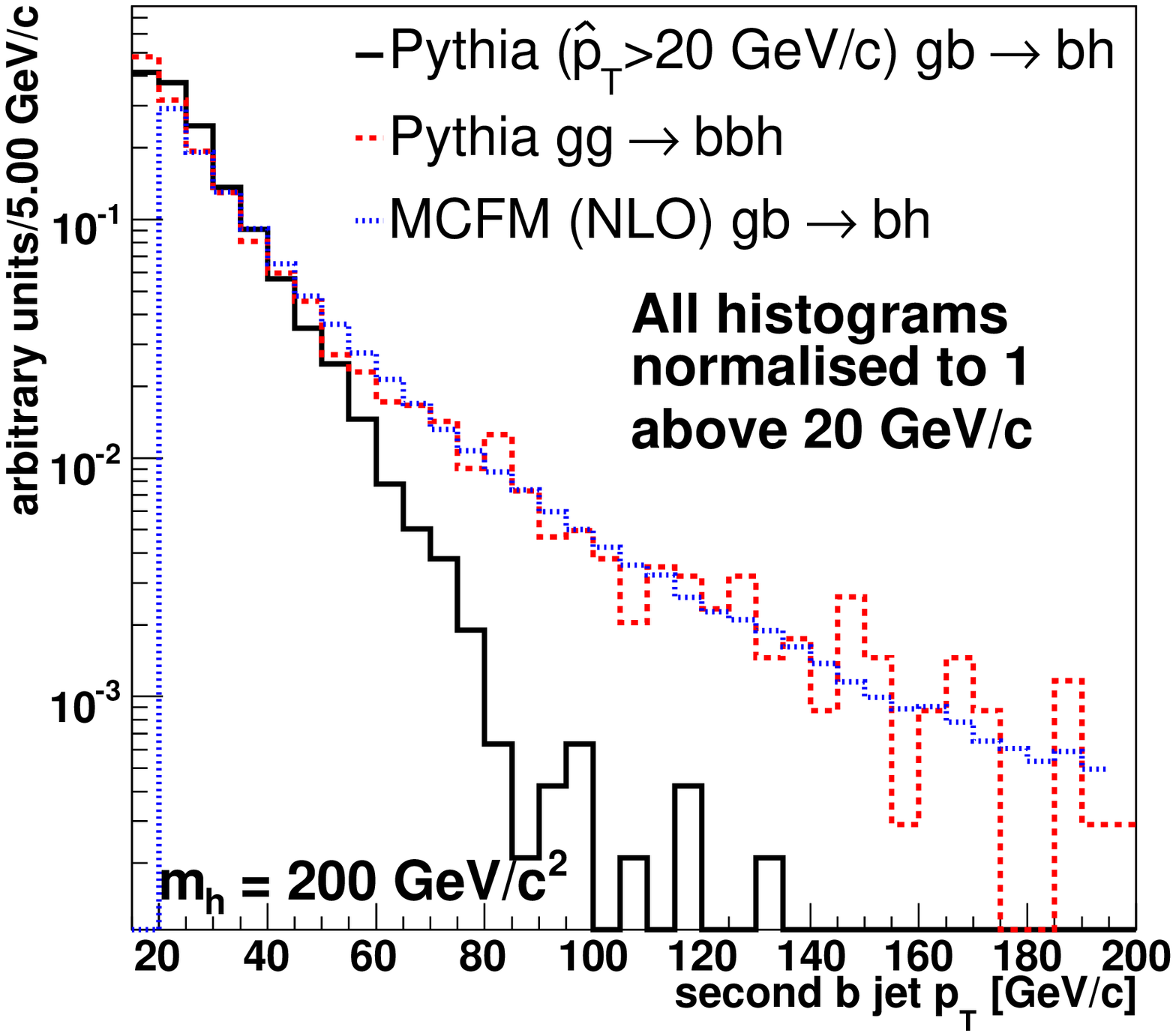}} &
  \resizebox{\linewidth}{!}{\includegraphics{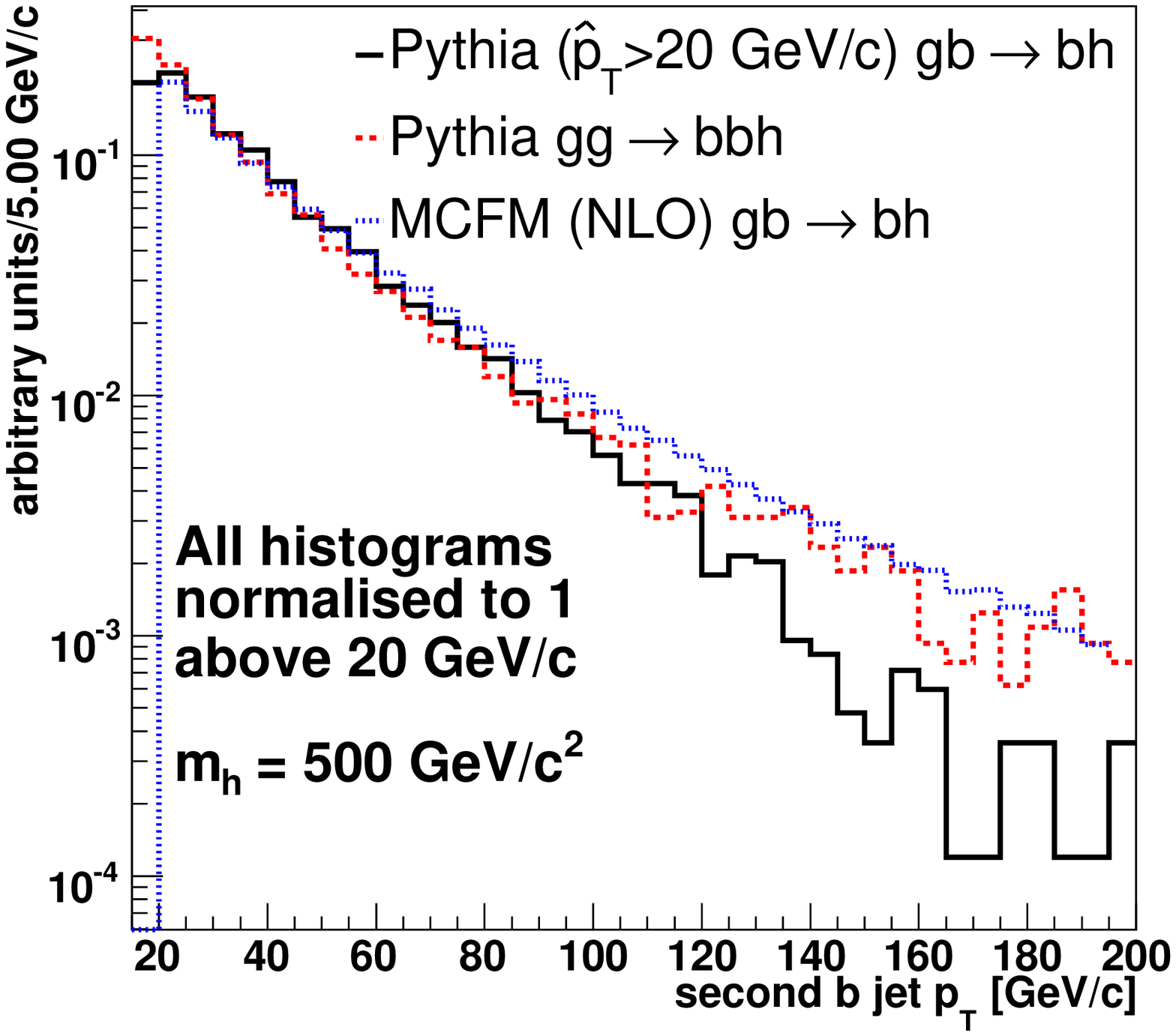}} \\
   \caption{The $\rm p_{\rm T}$ of the second b jet in PYTHIA and in MCFM for 
            $\rm m_{\rm h}$=200 GeV/$c^{2}$.} \label{bjet2_pt_200} &  
   \caption{The $\rm p_{\rm T}$ of the second b jet in PYTHIA and in MCFM for 
            $\rm m_{\rm h}$=500 GeV/$c^{2}$.} \label{bjet2_pt_500}
\end{BBH2figures}

\begin{BBH2figures}{!hbtp}
  \resizebox{\linewidth}{!}{\includegraphics{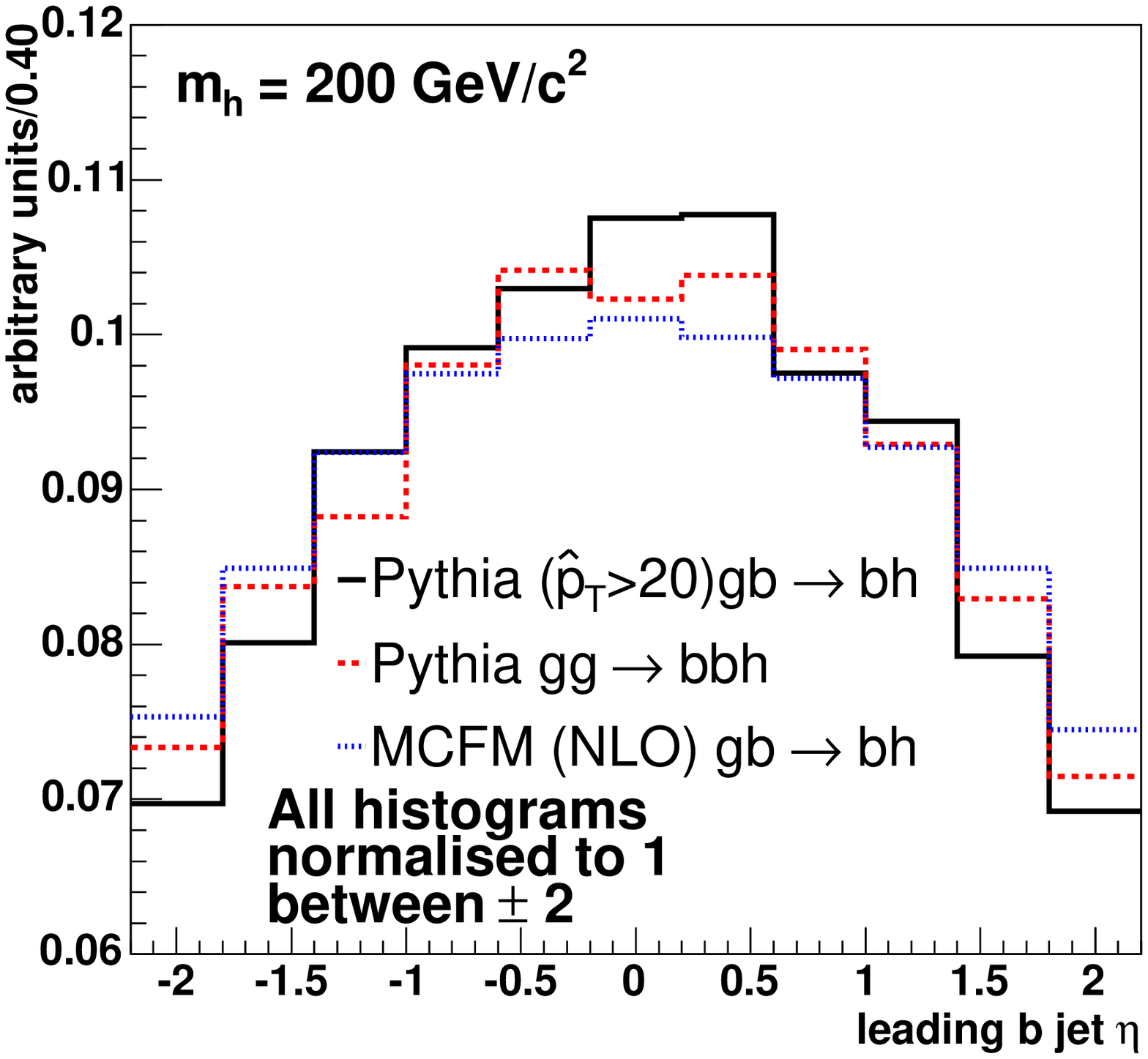}} &
  \resizebox{\linewidth}{!}{\includegraphics{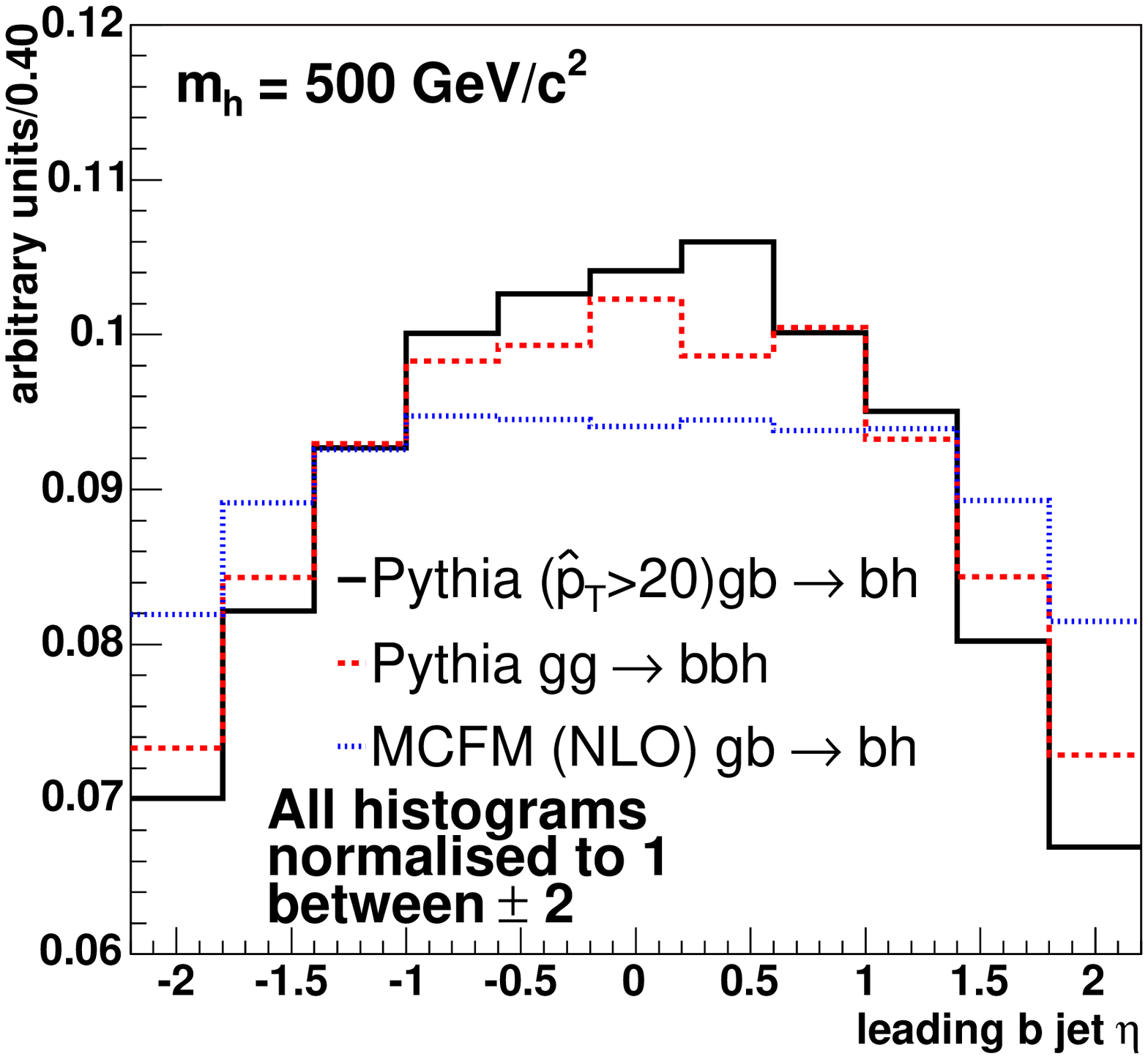}} \\
   \caption{The $\eta$ of the leading b jet in PYTHIA and in MCFM for 
            $\rm m_{\rm h}$=200 GeV/$c^{2}$.} \label{bjet1_eta_200} &  
   \caption{The $\eta$ of the leading b jet in PYTHIA and in MCFM for 
            $\rm m_{\rm h}$=500 GeV/$c^{2}$.} \label{bjet1_eta_500}
\end{BBH2figures}

\begin{BBH2figures}{!hbtp}
  \resizebox{\linewidth}{!}{\includegraphics{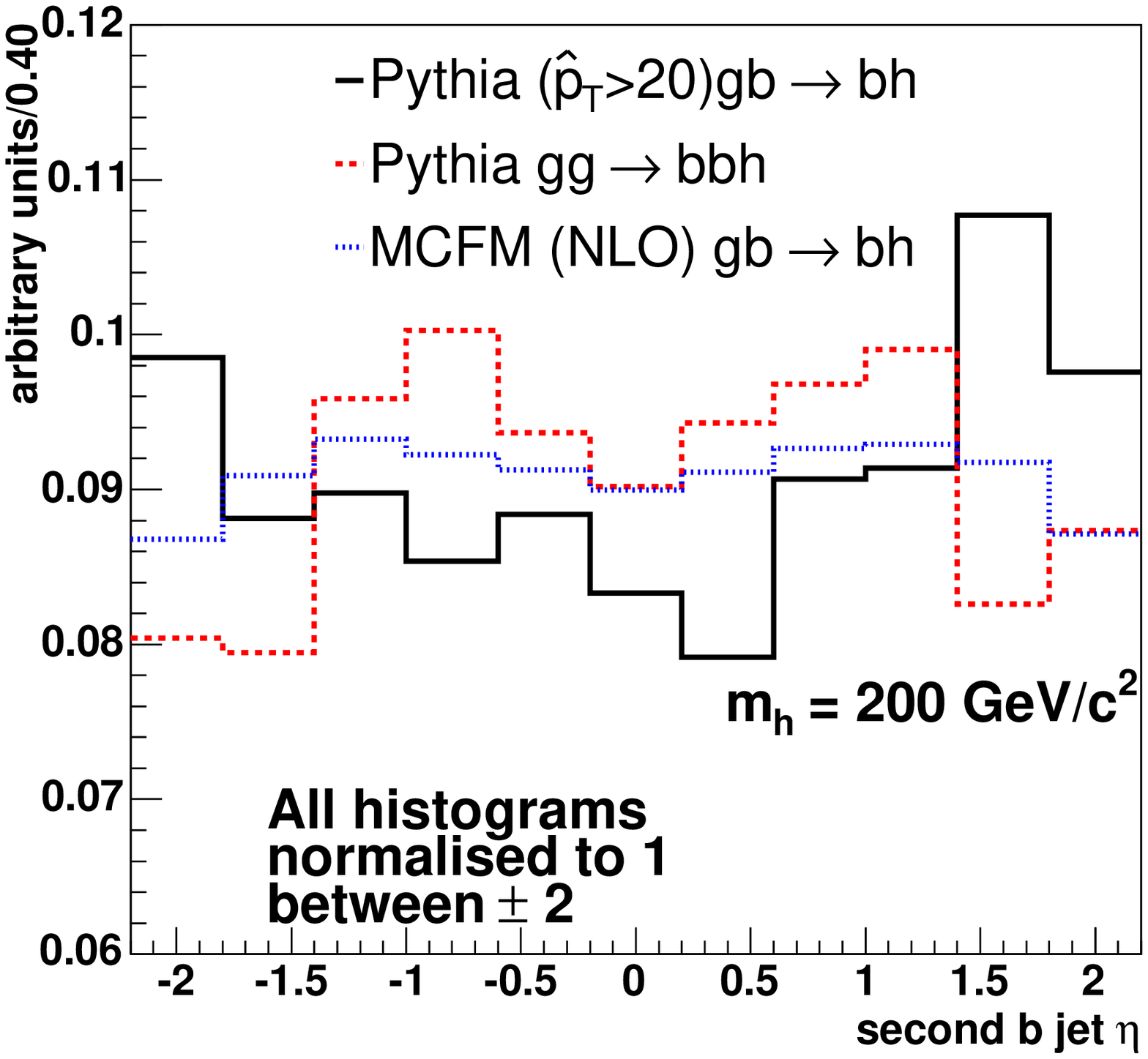}} &
  \resizebox{\linewidth}{!}{\includegraphics{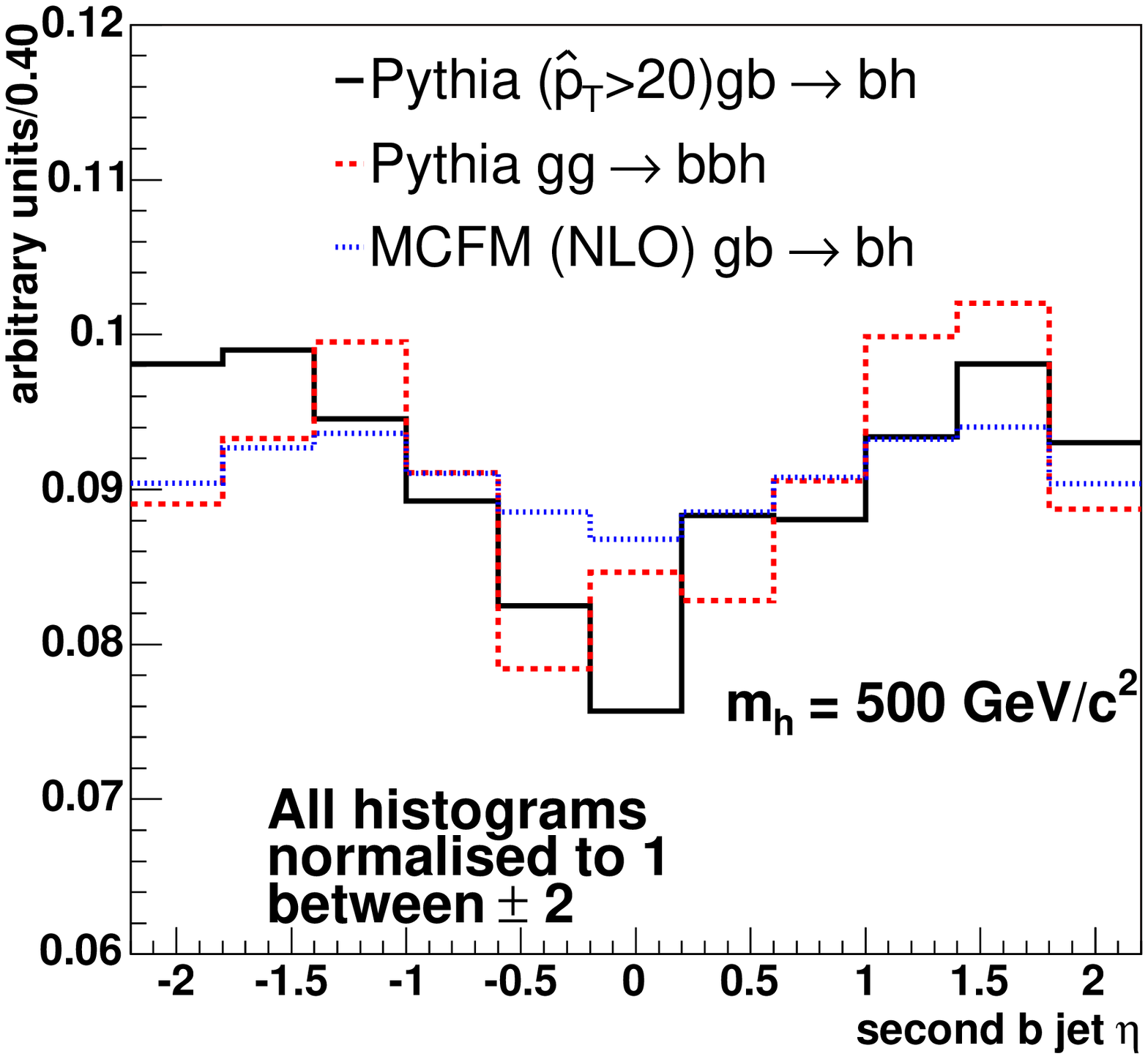}} \\
   \caption{The $\eta$ of the second b jet in PYTHIA and in MCFM for 
            $\rm m_{\rm h}$=200 GeV/$c^{2}$.} \label{bjet2_eta_200} &  
   \caption{The $\eta$ of the second b jet in PYTHIA and in MCFM for 
            $\rm m_{\rm h}$=500 GeV/$c^{2}$.} \label{bjet2_eta_500}
\end{BBH2figures}

\begin{BBH2figures}{!hbtp}
  \resizebox{\linewidth}{!}{\includegraphics{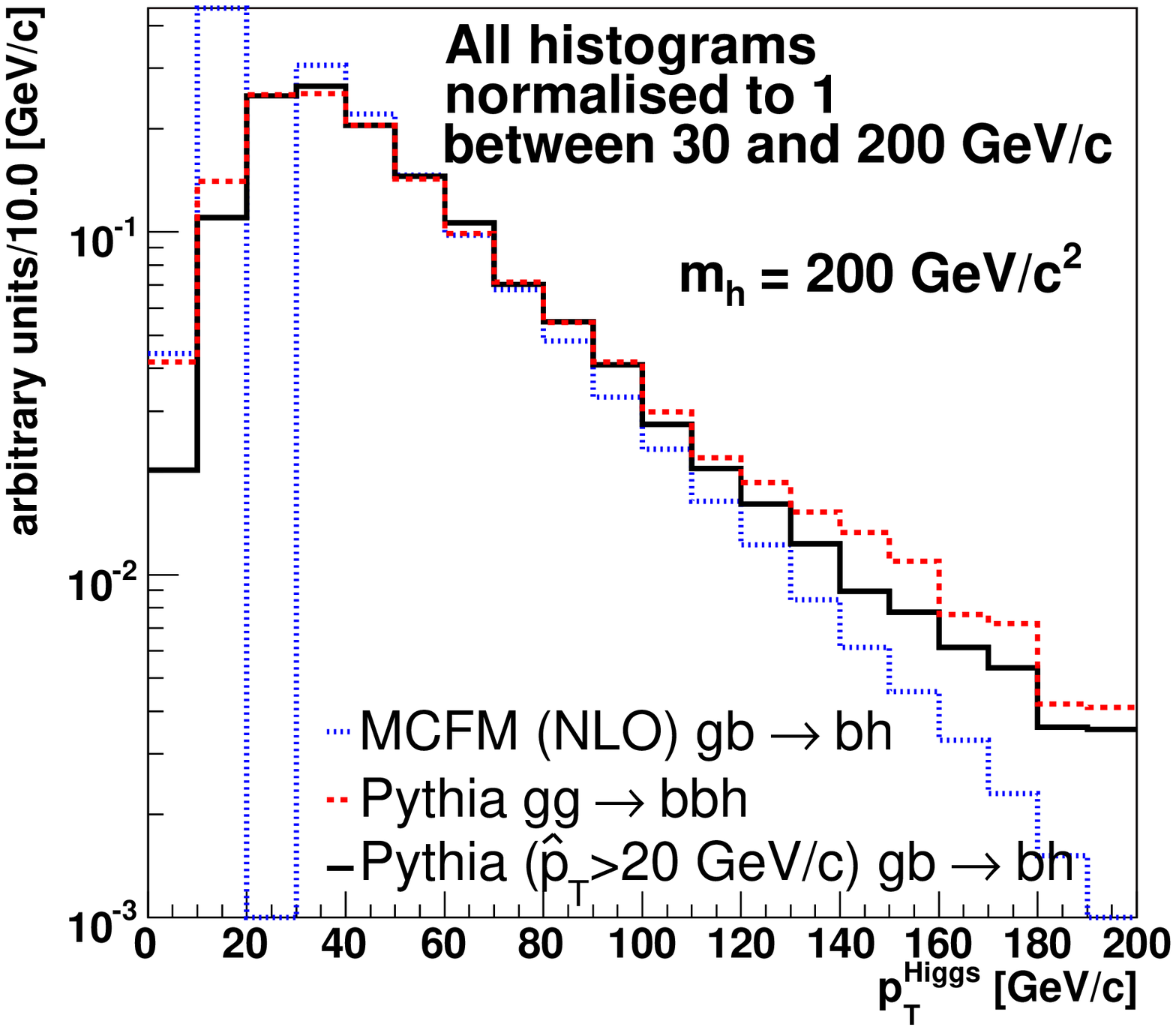}} &
  \resizebox{\linewidth}{!}{\includegraphics{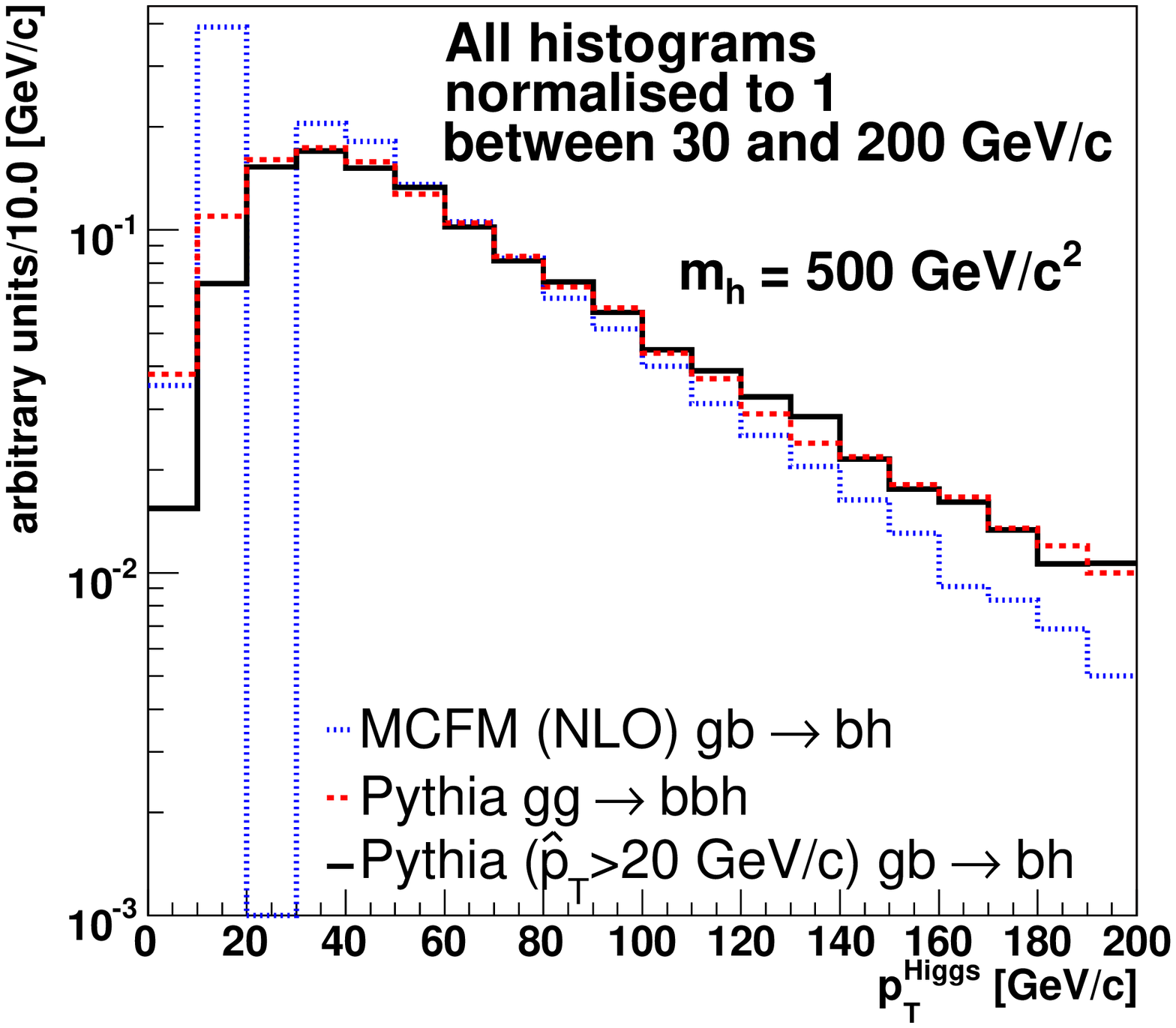}} \\
   \caption{The $\rm p_{\rm T}$ of the Higgs boson for the leading b jet in the tagging range and 
            no other jets in the central region, $\rm m_{\rm h}$=200 GeV/$c^{2}$.} 
   \label{higgs_pt_200} &  
   \caption{The $\rm p_{\rm T}$ of the Higgs boson for the leading b jet in the tagging range and 
            no other jets in the central region, $\rm m_{\rm h}$=500 GeV/$c^{2}$.} 
   \label{higgs_pt_500}
\end{BBH2figures}

%%%%%%%%%%%%%%%%%%%%%%%%%%%%%%%%%%%%%%%%%%%%%%%%%%%%%%%%%%%%%%%%%%%%%%%%%%%%%
\section[Higgs production in association with bottom quarks]
{HIGGS PRODUCTION IN ASSOCIATION WITH BOTTOM QUARKS~\protect
\footnote{Contributed by: J.~Campbell, S.~Catani, J.~Collins,
S.~Dittmaier, S.~Frixione, R.~Harlander, W.~Kilgore, M.~Kr\"amer,
L.~Magnea, F.~Maltoni, S.~Moretti, P.~Nason, F.~Olness, S.~Schumann,
J.~Smith, M.~Spira, S.~Willenbrock}}
\subsection{Introduction}

At large values of $\tan\beta$, some or all of the MSSM Higgs bosons
have enhanced couplings to bottom quarks.  The neutral MSSM Higgs
bosons may therefore be copiously produced in association with
bottom quarks.  There are two different formalisms that have been
employed to calculate the cross sections for such processes. The
four-flavor scheme begins with $gg\to b\bar bh$ as the leading-order
(LO) process. The cross sections with zero, one, or two high-$p_T$
$b$ jets are known at next-to-leading order (NLO) in QCD
\cite{Dittmaier:2003ej,Dawson:2003kb,Dawson:2004sh,Dawson:2005vi}.
In contrast, the five-flavor scheme uses a bottom-quark distribution
function in the initial state.  The inclusive cross section begins
with $b\bar b\to h$ at LO, and has been calculated at NLO
\cite{Dicus:1998hs,Balazs:1998sb,Maltoni:2003pn} and NNLO
\cite{Harlander:2003ai}.  The cross section with one high-$p_T$ $b$
jet begins with $gb\to hb$ at LO and is known at NLO
\cite{Campbell:2002zm}.  The cross section with two high-$p_T$ $b$
jets can only be calculated in the four-flavor scheme.

The five-flavor scheme has two advantages with respect to the
four-flavor scheme.  Collinear logarithms, proportional to powers of
$\alpha_s\ln(\mu_F/m_b)$ ($\mu_F$ is the factorization scale), that
appear in the four-flavor scheme are resummed to all orders in the
five-flavor scheme.  Thus one expects a more convergent perturbation
series in the five-flavor scheme.  The second advantage is that the
LO process in the five-flavor scheme is simpler, and makes
higher-order corrections tractable.  For example, the inclusive
cross section for Higgs production in association with $b$ quarks is
known at NNLO in the five-flavor scheme \cite{Harlander:2003ai}, but
only at NLO in the four-flavor scheme
\cite{Dittmaier:2003ej,Dawson:2005vi}.

Comparisons between calculations of Higgs production in the two
schemes have been carried out in
Refs.~\cite{Dawson:2004sh,Dawson:2005vi,Campbell:2004pu}. Generally
speaking, the two calculations agree within their respective
uncertainties. However, there are various ways in which the
comparisons can and should be improved.

\begin{figure}[htb]
\begin{center}
\vspace*{5mm}
\includegraphics[bb=50 250 580 600,scale=0.4]{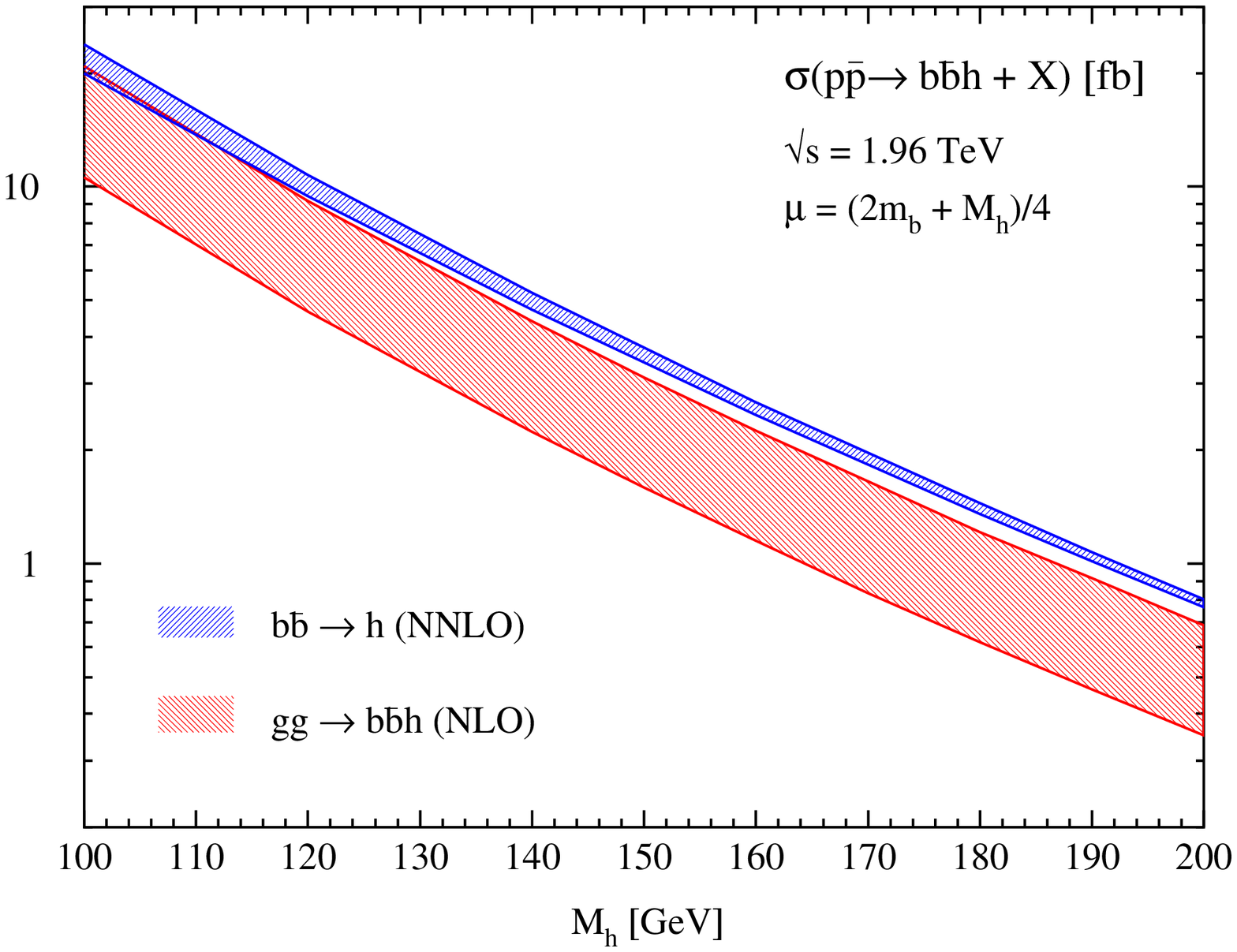}
\includegraphics[bb=50 250 580 600,scale=0.4]{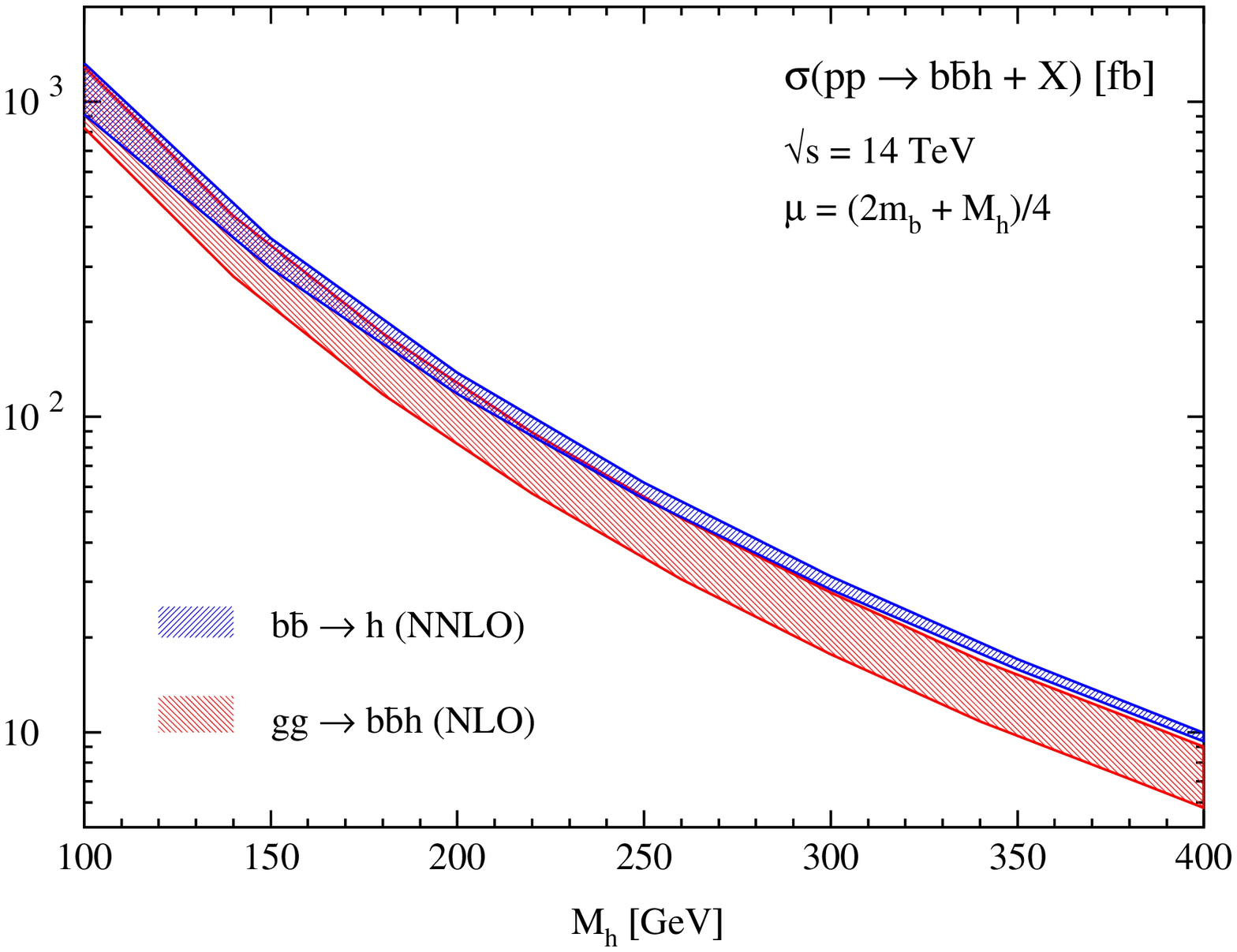}
\caption[]{Inclusive cross sections for $p{\overline p} (pp)
\rightarrow b {\overline b} h+X$ at the Tevatron and the LHC as a
function of the Higgs mass $M_h$. The error bands correspond to
varying the scale from $\mu_R=\mu_F=(2m_b+M_h)/8$ to
$\mu_R=\mu_F=(2m_b+M_h)/2$.  The NNLO curves are from
Ref.~\cite{Harlander:2003ai}.} \label{fg:0b_sigma}
\end{center}
\end{figure}

Let us focus on the inclusive cross section for Higgs production in
association with bottom quarks.  A comparison of the four- and
five-flavor calculations, taken from the 2003 Les Houches
proceedings \cite{Campbell:2004pu}, is shown in
Fig.~\ref{fg:0b_sigma}. The five-flavor calculation has a smaller
uncertainty since it is NNLO, while the four-flavor calculation is
NLO.  Although they are consistent with each other, the five-flavor
cross section lies near the top of the uncertainty band of the
four-flavor cross section. This may be due in part to the fact that
the five-flavor calculation is one order higher, and that it also
resums collinear logarithms. However, there are also ways in which
the comparison could be made more fairly.

In this review, we discuss some of the ways that the comparison
between the four- and five-flavor scheme calculations could be
improved.  After a review of the formalism, we discuss the effect of
a finite $b$ mass; top-loop diagrams; four- and five-flavor parton
distribution functions; and NNLO parton distribution functions.  We
also estimate the effect of the resummation of collinear logarithms.
We conclude with a summary of our results.

\subsection{Formalism}

If the characteristic energy scale $\mu$ is small compared to the
$b$-quark mass, $m_b$, then the $b$ quark decouples from the
dynamics and {\it does not} appear as a partonic constituent of the
hadron; that is, $b(x,\mu<m_b)=0$ and we are working in a
four-flavor scheme. In such a scheme, the Higgs is produced in the
${\cal O}(\alpha_s^2)$ process $gg \to b\bar bh$. Calculations in
the four-flavor scheme have the advantage that they do not need to
introduce the $b$ distribution function.

If instead we consider energy scales much larger than the $b$-quark
mass ($\mu\gg m_b$), then we work in a five-flavor scheme where the
$b$ quark {\it does} appear as a partonic constituent of the hadron,
$b(x,\mu>m_b)>0$.  In this regime, the $b$-quark mass enters as
powers of $\alpha_s \ln(\mu^2/m_b^2)$ which are resummed via the
DGLAP equations. This scheme has the advantage that it involves
lower-order Feynman graphs, and the $\alpha_s \ln(\mu^2/m_b^2)$
terms are resummed.

Ideally, there is an intermediate region where the 4-flavor and
5-flavor schemes are both a good representation of the physics; in
this region we can make a transition from the low-energy 4-flavor
scheme to the high-energy 5-flavor scheme thereby obtaining a
description of the physics that is valid throughout the entire
energy range from low to high scales.\footnote{We label the
four-flavor and five-flavor schemes as ``fixed-flavor-number'' (FFN)
schemes since the number of partons flavors is fixed.  The hybrid
scheme which combines these FFS is a ``variable-flavor-number''
(VFN) scheme since it transitions from a four-flavor scheme at low
energy to a five-flavor scheme at high energy
\cite{Aivazis:1993pi,Aivazis:1993kh}.}

When we evolve the $b$ distribution function in the context of the
DGLAP evolution equation $db \sim  P_{b/i} \otimes f_i$, we have the
option to use splitting kernels which are either mass-dependent
[$P_{b/i}(m_b\not=0)$] or mass-independent [$P_{b/i}(m_b=0)$]. While
one might assume that using $P_{b/i}(m_b\not=0)$ yields more
accurate results, this is not the case. The choice of
$P_{b/i}(m_b\not=0)$ or $P_{b/i}(m_b=0)$ is simply a choice of
scheme, and both schemes yield identical results up to high-order
corrections \cite{Olness:1997yn}. For simplicity, it is common to
use the mass-independent scheme since the $P_{b/i}(m_b=0)$ coincide
with the $\overline{\rm MS}$ kernels.

When the factorization proof of the ACOT scheme was extended to
include massive quarks, it was realized that fermion lines with an
initial or internal ``cut'' could be taken as massless
\cite{Kramer:2000hn}.  This simplification, referred to as the
simplified-ACOT (S-ACOT) scheme, is {\it not} an approximation; it
is again only a choice of scheme, and both the results of the ACOT
and S-ACOT schemes are identical up to high-order corrections
\cite{Collins:1998rz}.  The S-ACOT scheme can lead to significant
technical simplifications by allowing us to ignore the heavy quark
masses in many of the individual Feynman diagrams.  Furthermore, the
mass of the heavy quark in the initial state must be set to zero in
order to avoid infrared divergences that appear starting at NNLO
\cite{Doria:1980ak,Di'Lieto:1980dt,Catani:1985xt,Catani:1987xy,Catani:2002hc}.

\subsection{Finite $b$ mass}

\begin{figure}
  \begin{center}
    \leavevmode
    \begin{tabular}{c}
      \includegraphics[bb=110 265 465 560,width=.6\textwidth]{%
    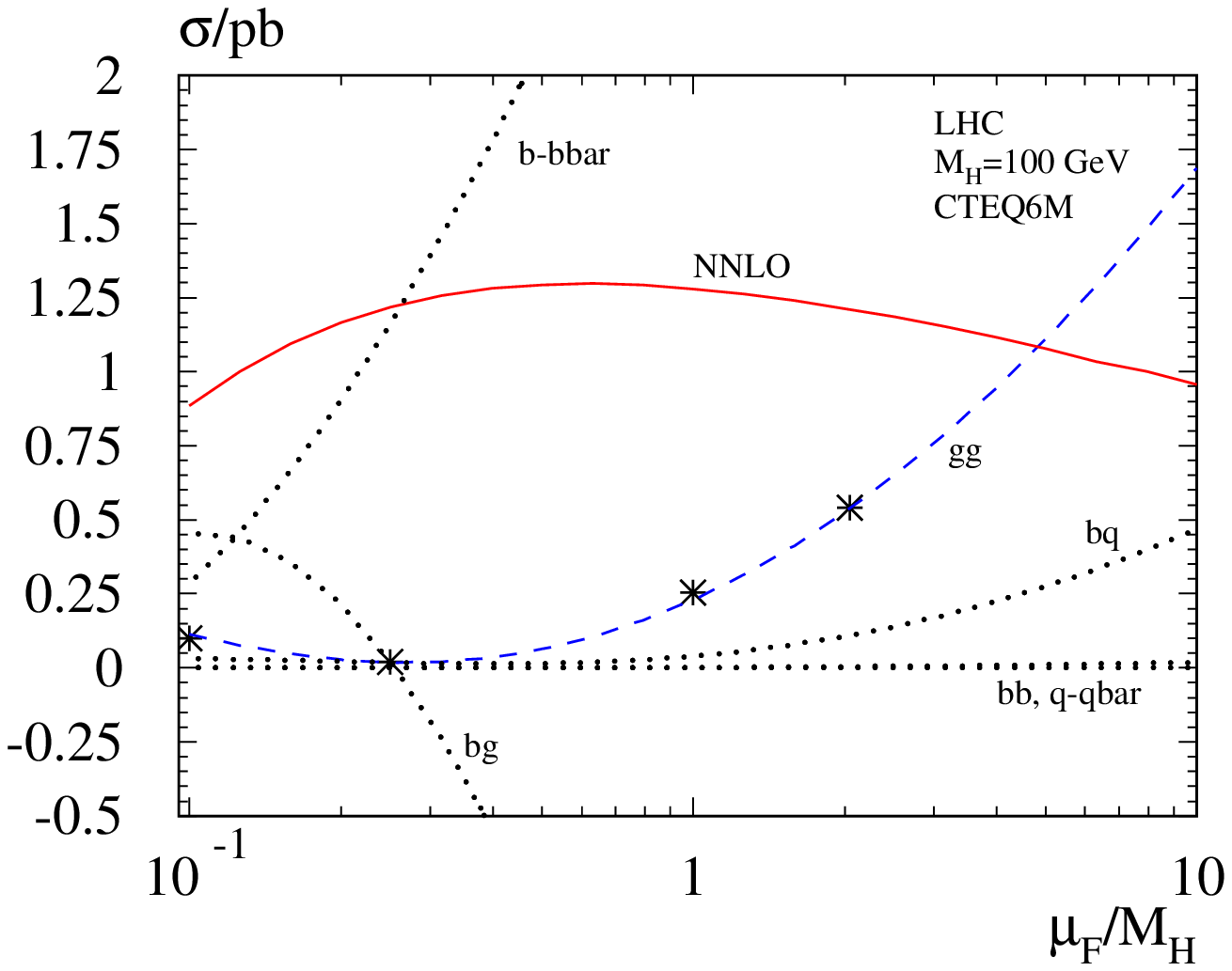}
    \end{tabular}
    \parbox{14.cm}{
      \caption[]{\label{fig::ggmuf100.2}\sloppy The cross section for
    the process $b\bar b\to h+X$ (LHC, $m_h=100$ GeV) at NNLO
    (solid line), split into the individual sub-processes (dashed
    and dotted). The markers denote the contribution from the $gg$
    sub-processes, with mass divergences subtracted minimally, and
    bottom quark mass terms kept in the matrix element calculation.
    }}
  \end{center}
\end{figure}

The NNLO calculation of $b\bar b\to h$ was carried out with $m_b=0$
throughout \cite{Harlander:2003ai}.  This is not an approximation at
LO or NLO, since all diagrams have at least one $b$ quark in the
initial state.  However, at NNLO the process $gg\to bbh$ arises, and
here the bottom-quark mass may be kept finite.  The numerical impact
of neglecting the $b$-quark mass can be determined by extracting
this contribution from the fully massless result of
Ref.~\cite{Harlander:2003ai} and comparing it to the terms denoted
by ``LO+$1/\ln$+$1/\ln^2$'' in Ref.~\cite{Maltoni:2003pn}, where a
finite $b$-quark mass was used. In both cases, the mass divergences
are subtracted in the $\overline{\rm MS}$ scheme, and the difference
between them is expected to be of order $(m_b/m_h)^2$.

The results for the LHC are shown in Fig.~\ref{fig::ggmuf100.2},
where the dashed line denotes the massless result, and the markers
represent individual values read off of the relevant curve in Fig.~7
of Ref.~\cite{Maltoni:2003pn}. As expected, the markers hardly
deviate from the curve, thus showing that the bottom quark mass
effects are indeed negligible.

Another observation is that the $gg$ channel in the $\overline{\rm
MS}$ scheme almost vanishes at a factorization scale of
$\mu_F=m_h/4$; in fact, this is true for {\it all} sub-processes
(dash-dotted lines), except for the $b\bar b$ channel. This supports
$\mu_F\approx m_h/4$ as the factorization scale for this process, as
argued in Ref.~\cite{Maltoni:2003pn} based on the collinear behavior
of the NLO correction. The solid line is the sum of all
sub-processes and thus represents the NNLO result (note, however,
that we used a NLO parton density set to make these curves).

\begin{figure}
  \begin{center}
    \leavevmode
    \begin{tabular}{c}
      \includegraphics[bb=110 265 465 560,width=.6\textwidth]{%
    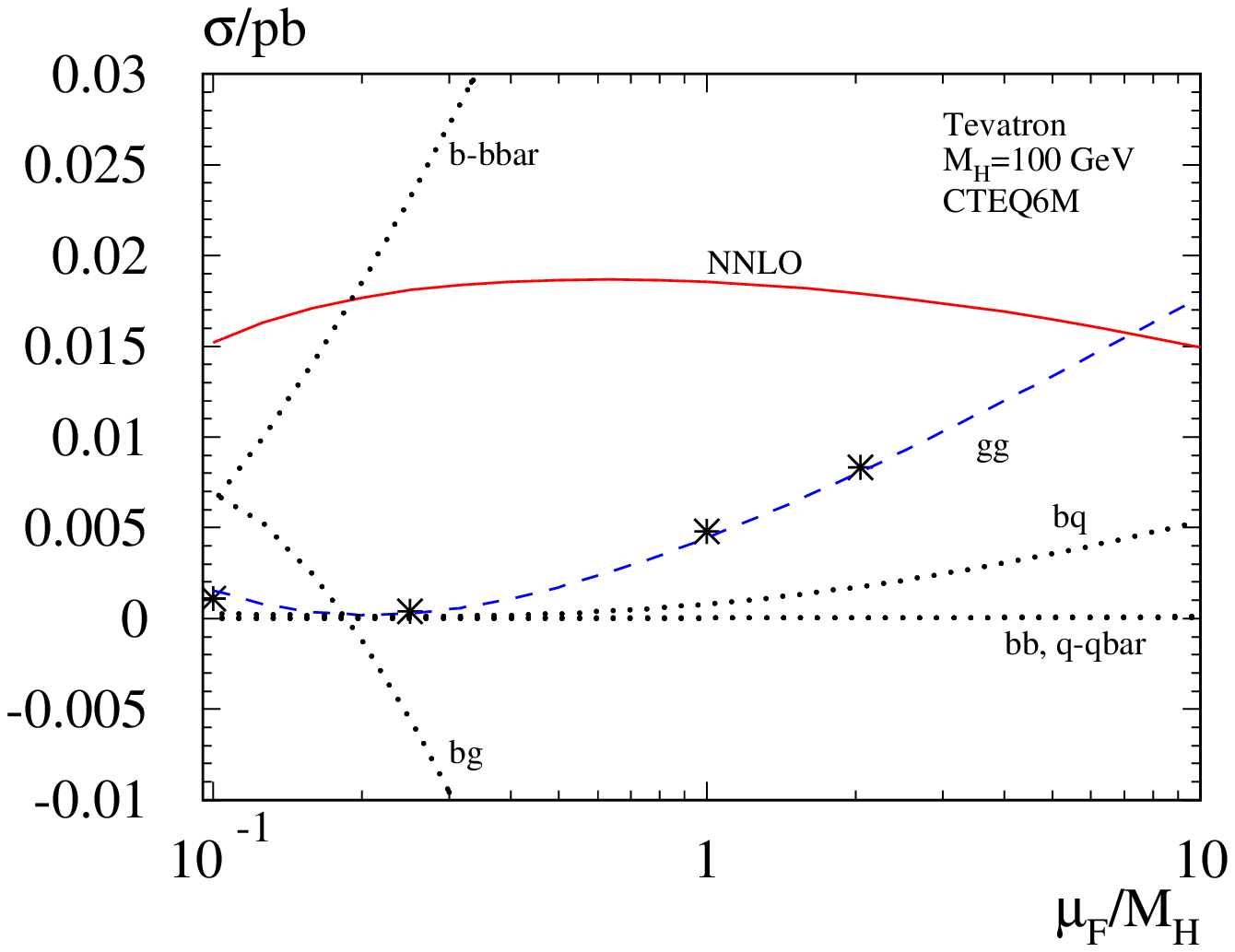}
    \end{tabular}
    \parbox{14.cm}{
      \caption[]{\label{fig::ggmuf100tev.2}\sloppy
    Same as Fig.\,\ref{fig::ggmuf100.2}, but for the Tevatron
    (1.96\,TeV).
        }}
  \end{center}
\end{figure}

The analogous plot for the Tevatron is shown in
Fig.~\ref{fig::ggmuf100tev.2}. Qualitatively, one observes the same
behavior as for the LHC, only the factorization scale at which only
the $b\bar b$ curve contributes to the rate is a little lower
($\mu_F\approx m_h/5$).

\subsection{Top loop}

\begin{figure}[b]
\begin{center}
    \leavevmode
    \begin{tabular}{c}
      \includegraphics
      %[bb=110 265 465 560,width=.6\textwidth]
      [width=.3\textwidth]
      {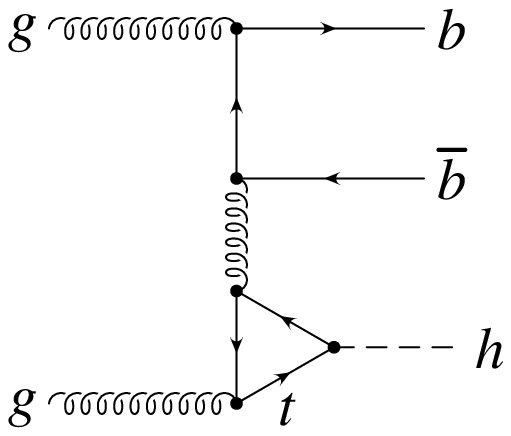}
    \end{tabular}
    \parbox{14.cm}{
      \caption[]{\label{fig:gg_bbxh_toploop_3}\sloppy Higgs
      producton via a top-quark loop.   This diagram interferes with the
      tree diagram for $gg\to bb\bar h$.
    }}
\end{center}
\end{figure}

In both the four- and five-flavor schemes, one encounters
higher-order diagrams where the Higgs boson couples to a top-quark
loop, not to the bottom quark.  Such a diagram is shown in
Fig.~\ref{fig:gg_bbxh_toploop_3}.  In both schemes it contributes
via its interference with the tree diagram $gg\to b\bar bh$, which
is proportional to $m_b$ due to the chiral structure of the
diagrams. In the comparison of the inclusive cross sections, shown
in Fig.~\ref{fg:0b_sigma}, this class of diagrams was included in
the four-flavor scheme calculation but not in the five-flavor
scheme. This contribution is negative, and accounts for about 4\% of
the difference in the two schemes at the Tevatron, and about 9\% at
the LHC, for $m_h=120$ GeV \cite{Campbell:2004pu}.

Upon further reflection, it seems more appropriate to regard the
class of diagrams involving a top loop as being associated with the
process $gg\to h$, which also contributes to inclusive Higgs
production. The cross sections shown in Fig.~\ref{fg:0b_sigma} do
not include this process.  From this point of view, the top-loop
contribution discussed above is not really a radiative correction to
$gg\to b\bar bh$ (four-flavor scheme) or $b\bar b\to h$ (five-flavor
scheme), but rather an interference between these processes and
$gg\to h$.  It is common to find that two different LO processes
interfere at higher order.

The most systematic way to organize the calculation is in powers of
the Yukawa couplings $y_b$ and $y_t$.  The inclusive cross section
contains terms proportional to $y_b^2$ and $y_t^2$, as well as
interference terms proportional to $m_by_by_t$.

Regardless of one's point of view, a fair comparison of the four-
and five-flavor schemes should treat the class of diagrams
containing a top-quark loop identically.  These diagrams are treated
consistently in Refs.~\cite{Dawson:2004sh,Dawson:2005vi}.

\subsection{Four- and five-flavor parton distribution functions}

A four-flavor calculation should use a four-flavor set of parton
distribution functions, that is, one in which there is no $b$
distribution function.  Unfortunately, no such set is available in
the standard parton distribution sets.  Here we estimate the
numerical impact that a four-flavor set would make on the
calculation of $gg\to b\bar bh$.

%%%%%%%%%%%%%%%%%%%%%%%%%%%%%%%%%%%%%%%%%%%%%%%%%%%%%%%%%%%%%%%%%%%%%%%%%%
% DOUBLE FIGURE: MOMENTUM FRACTION AND LUMINOSITY
\begin{figure}[b]
\begin{center}
\leavevmode \vbox{
 \hbox{
 \epsfxsize=0.45\hsize \epsfbox{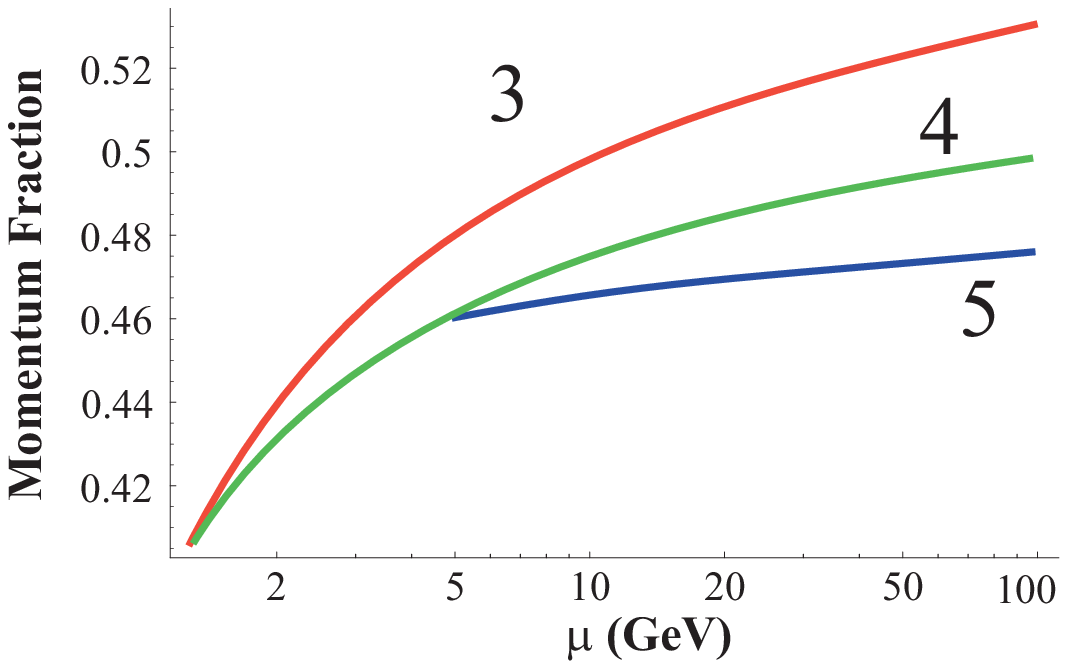}
 \hfill
 \epsfxsize=0.45\hsize \epsfbox{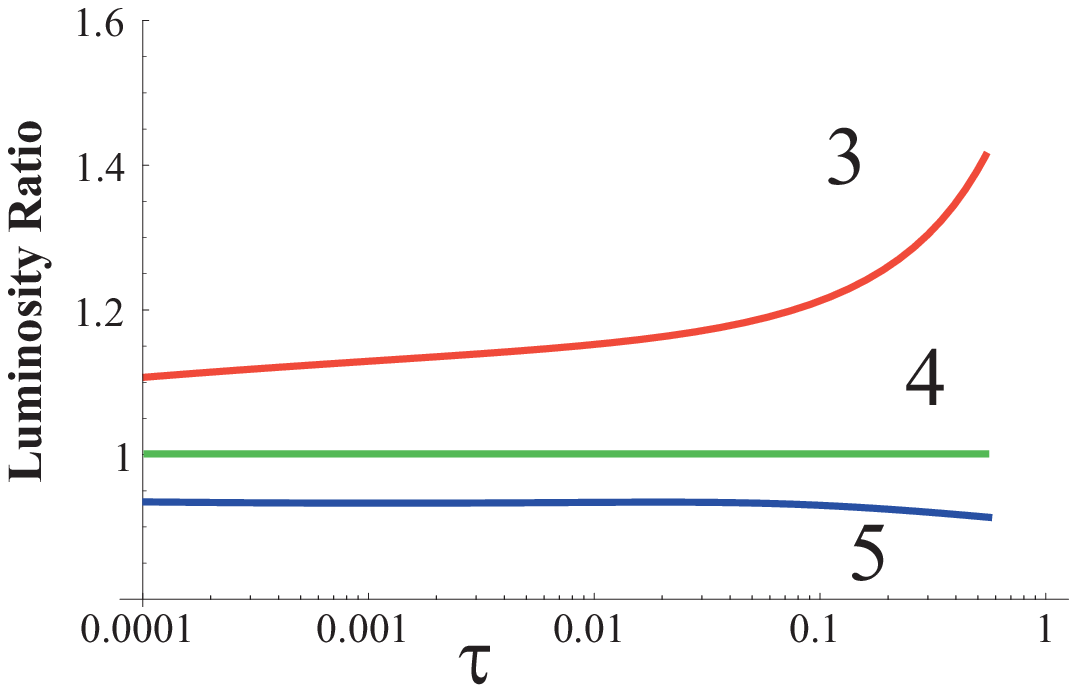}
 }
} \vskip -00mm
\end{center}
 \caption{
(a) Integrated momentum fraction, $\int_0^1 x f_g(x,\mu) \, dx$ vs.
$\mu$ of the gluon for $N_F=\{3,4,5\}$ = \{Red, Green, Blue\}.
\protect\\
(b) The ratio of the gluon-gluon luminosity  ($d{\cal
L}_{gg}/d\tau$) vs.~$\tau$ for $N_F=\{3,4,5\}$ = \{Red, Green,
Blue\} as compared with $N_F=4$ at $\mu=30$~GeV. }
 \label{fig:fredi}
\end{figure}
%%%%%%%%%%%%%%%%%%%%%%%%%%%%%%%%%%%%%%%%%%%%%%%%%%%%%%%%%%%%%%%%%%%%%%%%%%

To illustrate how the active number of ``heavy'' flavors affects the
``light'' partons, in Fig.~\ref{fig:fredi}(a) we show the momentum
fraction of the gluon vs. the factorization scale $\mu$. We have
started with a single PDF set at $\mu=1.3$~GeV, and evolved from
this scale invoking the ``heavy'' flavor thresholds as appropriate
for  the specified number of flavors. While all three PDF sets start
with the same initial momentum fraction, once we go above the charm
threshold ($m_c=1.3$~GeV) the $N_F=\{4,5\}$ momentum fractions are
depleted by the $g\to c\bar{c}$ process. In a similar fashion, the
momentum fraction for $N_F=5$ is depleted compared to $N_F=4$ by the
$g\to b\bar{b}$ channel above the bottom threshold ($m_b=5$~GeV).

To gauge the effect of the different number of flavors on the cross
section, we compute the gluon-gluon luminosity which is defined as $
d{\cal L}_{gg}/d\tau = g \otimes g $.  We choose a scale of
$\mu\approx m_h/4=30$~GeV which is characteristic of a Higgs of mass
120 GeV. In terms of the luminosity, the cross section is given as
$d\sigma/d\tau \sim [d{\cal L}_{gg}/d\tau] \,
[\hat{\sigma}(\hat{s}=\tau s)]$ with $\tau = \hat{s}/s = x_1 x_2$.

To highlight the effect of the different $N_F$ PDF's, we plot the
ratio of the luminosity as compared to the $N_F=4$ case in
Fig.~\ref{fig:fredi}(b). We see that the effects of
Fig.~\ref{fig:fredi}(a) are effectively squared, as expected, when
examining the curves of Fig.~\ref{fig:fredi}(b).

The blue (bottom-most) curve in Fig.~\ref{fig:fredi}(b) shows that
by using a five-flavor set in the four-flavor calculation of $gg\to
b\bar bh$, one is underestimating the cross section by about 7\%.
While this is not a very large effect, it does go in the correct
direction to improve the agreement between the four- and five-flavor
calculations of inclusive Higgs production.

%While this simple qualitative calculation gives us a general idea
%how the actual cross sections might vary, a full analysis is
%required to properly balance all the competing factors.

\subsection{NNLO parton distribution functions}\label{sec:NNLO}

The production of heavy quarks in deep-inelastic scattering (DIS)
and the incorporation of heavy quarks into parton densities are
related and interesting topics.  The fixed-flavor NLO QCD
corrections to charm quark electro-production were calculated in
Ref.~\cite{Laenen:1992zk} in the three-flavor scheme.  At high
energies, the three-flavor scheme should be replaced by a
four-flavor scheme, and eventually a five-flavor scheme.  In the
intermediate region, a variable flavor number scheme should provide
a smooth switch from the three-flavor scheme to the four-flavor
scheme.\cite{Tung:2001mv}

The treatment of the heavy quark as a parton density requires the
identification of the large logarithmic terms $\log(Q^2/m^2)$, which
was done in Ref.~\cite{Buza:1995ie} through next-to-next-leading
order (NNLO).  Then based on a two-loop analysis of the heavy quark
structure functions from an operator point of view, it was shown in
Refs.~\cite{Buza:1997nv}, \cite{Buza:1996wv} and
\cite{Matiounine:1998ky} how to incorporate these large logarithms
into charm (and bottom) densities. Two different NNLO variable
flavor number schemes were defined in Refs.~\cite{Chuvakin:1999nx}
and \cite{Chuvakin:2000jm}, where it was shown how they could be
matched to the three-flavor scheme at small $Q^2$, the four-flavor
scheme at large $Q^2$, and the five-flavor scheme at even larger
$Q^2$.

This NNLO analysis yielded two important results. One was the
complete set of NNLO matching conditions for massless parton
evolution between $N$ and $N+1$ flavor schemes.  Unlike the NLO
case, the NNLO matching conditions are discontinuous at these flavor
thresholds. Such matching conditions are necessary for any NNLO
calculation at the LHC, and have already been implemented in parton
evolution packages by \cite{Chuvakin:2001ge}, \cite{Vogt:2004ns} but
unfortunately not yet in the programs which make global fits to
experimental data. Note that the NNLO matching conditions on the
running coupling $\alpha_s(N_F,Q^2)$ as $Q^2$ increases across
heavy-flavor flavor thresholds have been calculated in
\cite{Bernreuther:1981sg,Bernreuther:1983zp} and
\cite{Larin:1994va,Chetyrkin:1997sg}. Furthermore, the NNLO two-loop
calculations above explicitly showed that the heavy quark structure
functions in variable flavor approaches are not infrared safe. A
precise definition of the heavy-flavor content of the deep inelastic
structure function requires one to either define a heavy quark-jet
structure function, or introduce a fragmentation function to absorb
the uncanceled infrared divergence.  Similar issues arise for
inclusive $Z$ production in association with heavy quarks
\cite{Maltoni:2005wd}. In either case, a set of contributions to the
inclusive light parton structure functions must be included at NNLO.

A dedicated analysis \cite{Chuvakin:2000zj} for charm
electro-production showed that even at relatively large $Q^2$ one
could not distinguish between the fixed order NLO calculation of
\cite{Laenen:1992zk} and the NNLO VFNS calculation of
\cite{Buza:1996wv}, given the large error bars on the experimental
data then available in the year 2000. This demonstrates that terms
in $\ln(Q^2/m^2)$ in fixed flavor number schemes are proportional to
the convolution of small terms and therefore do not necessarily make
a large contribution to the deep-inelastic cross section. To
quantify this statement one requires more precise data from the HERA
collider on charm and bottom quark electro-production analyzed in
both fixed-flavor and variable flavor schemes. Since there is an
increasing use of variable flavor schemes with massless charm and
bottom parton densities in hadronic collisions it is important to
clarify this topic.

\subsection{Resummation}

%%%%%%%%%%%%%%%%%%%%%%%%%%%%%%%%%%%%%%%%%%%%%%%%%%%%%%%%%%%%%%%%%%%%%%%%%%
% DOUBLE FIGURE: COMPARE EVOLVED AND PERT PDF
\begin{figure}[ht]
\begin{center}
\leavevmode \vbox{
 \hbox{
 \epsfxsize=0.45\hsize \epsfbox{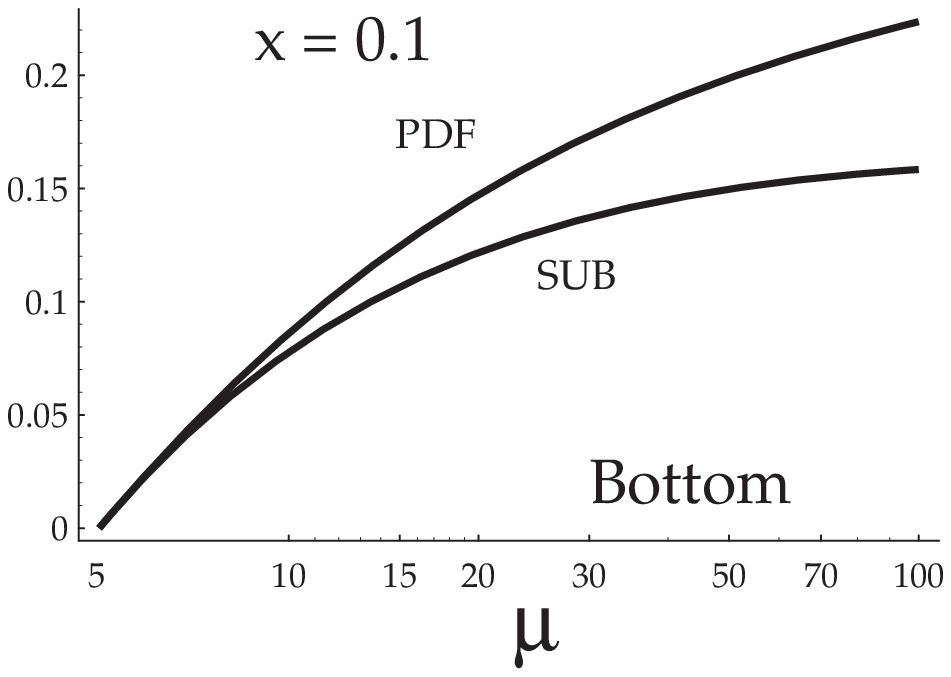}
 \hfill
 \epsfxsize=0.45\hsize \epsfbox{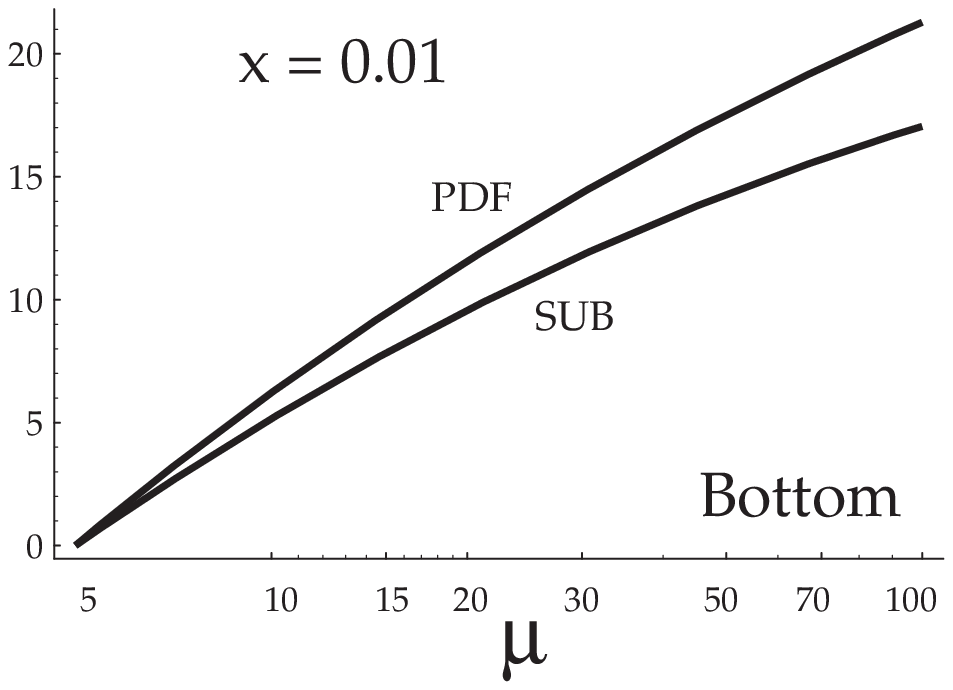}
 }
} \vskip -00mm
\end{center}
 \caption{
Comparison of the evolved PDFs, $b(x,\mu)$ (labeled PDF), and
 perturbative PDFs,
$\tilde{b}(x,\mu)\sim P_{b/g}\otimes g$ (labeled SUB),
 as a function of the renormalization scale $\mu$
 for bottom at a) $x=0.1$  and b) $x=0.01$.
}
 \label{fig:fredii}
\end{figure}
%%%%%%%%%%%%%%%%%%%%%%%%%%%%%%%%%%%%%%%%%%%%%%%%%%%%%%%%%%%%%%%%%%%%%%%%%%

The fundamental difference between the $gg\to b\bar bh$ process and
the $b\bar{b}\to h$ process amounts to whether the radiative
splittings ({e.g.}, $g\to b\bar{b}$) are computed by the DGLAP
equation as a part of the parton evolution, or whether they are
external to the hadron and computed explicitly.  In essence, both
calculations are represented by the same perturbation theory with
two different expansion points; while the full perturbation series
will yield identical answers for both expansion points, there will
be differences in the truncated series.

To understand the source of this difference, we examine the
contributions which are resummed into the $b$-quark distribution
function by the DGLAP evolution equation, $df \sim P\otimes f$.
 Solving this equation perturbatively in the region of the $b$-quark
threshold, we obtain $\tilde{b} \sim P_{b/g}\otimes g$. This term
simply represents the first-order $g\to b\bar{b}$ splitting that
is fully contained in the ${\cal O}(\alpha_s^2)$ $gg\to b\bar bh$
calculation.

In addition to this initial splitting, the DGLAP equation resums an
infinite series of such splittings into the perturbatively-evolved
PDF, $b$. Both $b$ and $\tilde{b}$ are shown in
Fig.~\ref{fig:fredii} for two choices of $x$ \cite{Olness:1997yc}.
Near threshold, we expect $b$ to be dominated by the single
splitting contribution, and this is verified in the figure. In this
region, $b$ and $\tilde{b}$ are comparable, and we expect the
four-flavor $gg\to b\bar bh$ calculation should be reliable in this
region.

As we move to larger scales, we see $b$ and $\tilde{b}$ begin to
diverge at a few times $m_b$ since $b$ includes higher-order
splitting such as $\{P^2, P^3, P^4, ...\}$ which are not contained
in $\tilde{b}$.  We expect the five-flavor $b\bar{b}\to h$
calculation should be most reliable in this region since it resums
the iterative splittings.

Fig.~\ref{fig:fredii} shows that the resummation is a bigger
effect at larger values of $\mu$, as expected.  This may explain
why the five-flavor curves in Fig.~\ref{fg:0b_sigma} deviate more
from the four-flavor curves at larger values of the Higgs mass.
Fig.~\ref{fig:fredii} also shows that the resummation is a bigger
effect at larger values of $x$, which may explain why the Tevatron
curves deviate more from each other than the LHC curves.

This analysis also explains a puzzling difference between the two
formalisms. The $b \bar{b} \to h$ calculation appears to be affected
by Sudakov logarithms due to soft and collinear gluon emission,
yielding for example terms of the form $\alpha_s^2\log^{4} N$ at
NNLO, with $N$ the Mellin variable conjugate to $m_h^2/\hat{s}$. The
presence of these logarithms would suggest the need to perform a
resummation.  In the $g g \to b \bar{b} h$ calculation, however,
Sudakov logarithms are subleading because each final-state quark
provides a suppression factor, roughly $1 - m_h^2/\hat{s}$ as
$\hat{s} \to m_h^2$, corresponding to a $1/N$ suppression of the
Mellin transform. However, this is not a real difference, just a
rearrangement the perturbative series: in the $b \bar{b} \to h$
calculation, the $1/N$ suppression factors are provided by the
$b$-quark distribution, which is evaluated perturbatively, and
acquires a $1/N$ factor through the Mellin transform of the
splitting kernel $P_{b/g}$.

\subsection{Conclusions}

In this review we have discussed a few improvements that could be
made in the calculation of Higgs production in association with
bottom quarks in the four- and five-flavor schemes.  A consistent
treatment of top-quark loop diagrams would improve the agreement
between the inclusive cross sections, shown in
Fig.~\ref{fg:0b_sigma}, by 4\% at the Tevatron and 9\% at the LHC
for $m_h=120$ GeV.  We estimate that using a four-flavor set of
parton distribution functions in the four-flavor calculation would
improve the agreement by about another 7\%.  We showed that using a
finite $b$ mass in the NNLO five-flavor calculation has no numerical
impact. The NNLO parton distributions used in that calculation could
be improved by implementing a proper matching at heavy-flavor
thresholds, but we cannot estimate what the numerical impact will
be.  Finally, we discussed how resummation could explain some of the
features of the comparison between the four- and five-flavor
calculations.

%%%%%%%%%%%%%%%%%%%%%%%%%%%%%%%%%%%%%%%%%%%%%%%%%%%%%%%%%%%%%%%%%%%%%%%%%%%%%
\section[Associated $t\bar{t}H$ production with $H\rightarrow\gamma\gamma$
at the LHC]{ASSOCIATED $t\bar{t}H$ PRODUCTION WITH $H\rightarrow\gamma\gamma$
AT THE LHC~\protect\footnote{Contributed by: S.~Dittmaier, R.~Frazier,
S.~Gascon-Shotkin, M.~Kr\"{a}mer, F.~Maltoni, D.~Mercier, M.~Moretti,
A.~Nikitenko, F.~Piccinini, R.~Pittau, M.~Spira}}
\subsection{Introduction}
%***Check standard (SMH) SM and 2-d, MSSM refs
 A Standard Model
%~\cite{Allanach:2002ph,Allanach:2001hx,Allanach:2001mv}
or two-doublet
%~\cite{Allanach:2002ph,Allanach:2001hx,Allanach:2001mv}
neutral Higgs boson produced in
association with a
 $\rm t \bar{\rm t}$ pair, with $\rm H(h^0)\rightarrow \gamma \gamma$
would share
a fully reconstructible mass peak with the inclusive
$\rm H(h^0)\rightarrow \gamma \gamma$ signature.  But like the other
associated-production channels
$\rm W \rm H$ and $\rm Z \rm H$
~\cite{Assamagan:2004mu},
% channels discussed in section XXXXX,
the signature could contain an
isolated high-transverse-momentum charged lepton which can be used both to discriminate
against
QCD background and
reconstruct
 the primary vertex; the associated production channels could hence be
less dependent on
 photon energy resolution. In particular, the presence of two top quarks
would tend to
 produce high-multiplicity events, which could offer additional
discriminating power
 against light jet QCD background.  And in the case of the two-doublet
MSSM
%~\cite{Allanach:2002ph,Allanach:2001hx,Allanach:2001mv},
the gluon
 fusion Higgs production channel could in fact be subject to suppression
with respect to
 the associated production channels in the case of top-stop degeneracy
("maximal mixing")~\cite{Belanger:1999pv}.
%(ref??).
 Prior generator-level studies for the detection of the SM
(~\cite{:1997du}) and MSSM~\cite{:1997ki}
% (ref. CMS Note 1997/057)
Higgs bosons in CMS~\cite{:1994pu} via this
channel have indicated a signal-to-background ratio of approximately 1.
A full simulation study in the ATLAS Physics
Technical Design Report~\cite{unknown:1999fr} has predicted a
signal significance of
 $S/\sqrt{B}=${4.3-2.8} for m$_H=${100--140} GeV with a signal efficiency of $\sim$30\%.
% CERN-ATL-COM-PHYS-2004-056 par Beauchemin, P and Azuelos, Georges
%"Search for the SM Higgs Boson in the gamma gamma + ETmiss channel"
A more recent, related ATLAS study involving a 2-photon signature
accompanied by missing energy~\cite{Beauchemin:2004zi} has
indicated, for 100 fb$^{-1}$ a signal-to-background ratio of $\sim$2
%For 100fb-1, for ttbarh channel,
for m$_H=$120 GeV.
%(10.2 signal events for 5.4 background events).
%***Also ATLAS notes and results (see presentations)

\subsection{Signal production cross-sections and event rates}
Production cross-sections for $\rm t \bar{\rm t} \rm H$ have been calculated at
next-to-leading order~\cite{Beenakker:2001rj,Beenakker:2002nc,Dawson:2002tg}.
Taking the
branching ratio for
 $\rm H \rightarrow \gamma \gamma$ from HDECAY~\cite{Djouadi:1997yw} and
assuming in
 addition that the decay of one of the top quarks yields a lepton
(electron or muon) from
 $\rm W^\pm\rightarrow \rm l + \nu_l$ (including the possibility of tau
lepton decays to
muons or electrons), we estimate for several Higgs boson masses the
 number of signal events for 30 and 100 fb$^{-1}$
(Table~\ref{ttg2gamma:tab:nosig}):

\begin{table}[!h]
\centering
\caption{\footnotesize{Estimated number of signal events for
$\rm t \bar{\rm t} \rm H, \rm H \rightarrow \gamma \gamma$, assuming
NLO production cross sections~\cite{Beenakker:2002nc}, Higgs
branching ratios to 2 photons~\cite{Djouadi:1997yw}, and
1 electron or muon from top decay (including from tau lepton decays).}}
  \vspace*{1mm}
\footnotesize{
\begin{tabular}{|c|c|c|}
\hline
Higgs Mass (GeV)  & After 30 fb$^{-1}$ & After 100 fb$^{-1}$  \\ \hline
115               & 20.75       &       69.18             \\
120               & 19.53           &   65.10             \\
130       & 15.92       &       53.05             \\
140           & 11.18           &   37.28              \\
\hline
\end{tabular}
}
\label{ttg2gamma:tab:nosig}
\end{table}

\begin{figure}[!Hhtb]
\centering
\includegraphics[width=0.50\textwidth]{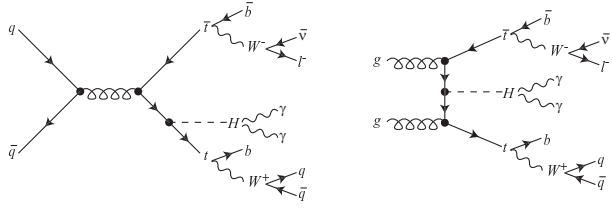}
%\caption{Reconstructed mass of Higgs boson for
%$\rm M_{\rm H}$=130 GeV/$c^{2}$}
\caption{Typical s- and t-channel diagrams for $\rm t \bar{\rm t}
 \rm H$ production with $\rm H \rightarrow \gamma \gamma$ and at least one
charged
lepton from the decay of a top or antitop quark.}
\label{higgs_sm_tth_2gamma_fig1}
\end{figure}

 Figure~\ref{higgs_sm_tth_2gamma_fig1}  shows typical Feynman diagrams of
the signal process.
%(see Sasha's example, and should use distiller to make pdfs)
%Figure~\ref{higgs_sm_h_zz_2e2mu_fig1} shows reconstructed Higgs boson mass
%for $\rm M_{\rm H}$=130 GeV/$c^{2}$.

 Signal events were generated with both the MADGRAPH~\cite{Maltoni:2002qb,Stelzer:1994ta,Murayama:1992gi} and
ALPGEN~\cite{Mangano:2002ea,Mangano:2001xp,Caravaglios:1998yr} LO exact matrix
element generators, for each of the Higgs masses shown in Table 1 (at least
30000 events per mass value with statistical error below 1\%). Events from
both generators were found to
yield comparable LO cross-section results and
 kinematical distributions.  The LO cross-sections were also found to
agree with those
from the program HQQ~\cite{Spira:1997dg} at the percent level.
%(*** check it from my logs of mails).

It should be noted that for the current study all signal events have been generated
such that exactly one of the two W bosons from the two top quarks decays leptonically.
It can be assumed, however, that the event selection which will be described below
will also have some efficiency for events where both W bosons yield leptons, thus
potentially increasing the total number of signal events expected to be observed. This will
be evaluated in a later study.

\subsection{Identified background processes and event generation}
\begin{figure}[!Hhtb]
\centering
\includegraphics[width=0.50\textwidth]{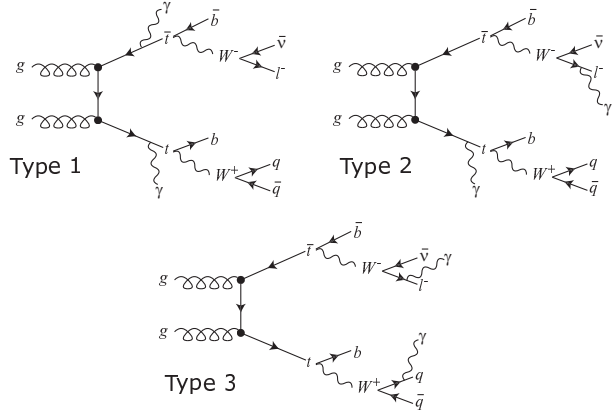}
%\caption{Reconstructed mass of Higgs boson for
%$\rm M_{\rm H}$=130 GeV/$c^{2}$}
\caption{A subsample of the relative Feynman graphs illustrating the three types of
$t\bar t\gamma\gamma$ processes.}
\label{higgs_sm_tth_2gamma_fig2}
\end{figure}

Standard Model processes resulting in both irreducible and
reducible backgrounds have been identified. A background is called
irreducible if it is capable of giving rise to the same
signature on the particle level as that searched for in a signal event, that
is to say, a lepton and two photons ($\rm l \gamma \gamma$).
Among the irreducible
processes, special care has been taken to properly treat the
$\rm t \bar{\rm t} \gamma \gamma $ background.
Feynman diagrams of three possible types of $\rm t \bar{\rm t} \gamma \gamma $
processes
considered are shown in Figure~\ref{higgs_sm_tth_2gamma_fig2}. In the
first case, called ``Type 1'', both photons are radiated from either outgoing
top quark or incoming parton lines. In the third case, called ``Type 3'', both
are radiated from top quark decay products. The second case, ``Type 2''
combines one photon radiated according to ``Type 1'' with the second radiated
according to ``Type 3''. (A fourth process arises from both photons being
radiated from different decay products of the {\em same} top quark; for the relevant
event selection (see pertinent section below) with $m_{\gamma\gamma}>$70 GeV we have
verified that this contribution is completely negligible).
Since at the time of undertaking the study no
matrix element generator included either the Types 2 or 3 processes, a
collaboration was begun with the authors of ALPGEN to add them. Also added to ALPGEN
was the process W$\gamma\gamma +$ 4 jets.  The performance of this sample versus
an inclusive W$\gamma\gamma$ sample (with all possible extra jets coming from
parton showering), also considered, is evaluated in subsequent sections.
Where applicable in the ALPGEN samples, top quarks and W bosons are decayed within
ALPGEN which assures
preservation of spin correlation information which could impact kinematical
distributions.
%Figure~\ref{higgs_sm_h_zz_2e2mu_fig1} shows reconstructed Higgs boson mass
%for $\rm M_{\rm H}$=130 GeV/$c^{2}$.

 Table~\ref{ttg2gamma:tab:genirred} lists the considered irreducible
background processes,
the
generators used to either generate or cross-check event samples,
the LO cross-section with statistical errors, the number of
events expected for 30 (100) fb$^{-1}$ of integrated luminosity, the
number of events generated and the statistical weight of
each generated event for 30 (100) fb$^{-1}$ of integrated luminosity. The cross-sections
reflect preselection criteria imposed at generator-level which are described
in the next section. In the processes involving real top quarks as well
as in the W$\gamma\gamma +$ 4j process, one top quark/the W boson was forced
to decay leptonically, and the stated cross-section therefore implicitly
includes the relevant branching ratio. It can be seen that the effect of the
inclusion of background Types 2 and 3
is to augment the total initial contribution (before selection) from $\rm t \bar{\rm t} \gamma \gamma $
by approximately one order of magnitude.
%%%This table needs to have update completed (weights)updated to latest results!
\begin{table}[!h]
\centering
 \caption{\footnotesize{Cross-sections at leading order (statistical
errors in parentheses), number
 of events generated, numbers of events and statistical weight/generated event for
30 and 100 fb$^{-1}$ of
integrated luminosity for the irreducible backgrounds considered.
}}
  \vspace*{1mm}
\footnotesize{
\begin{tabular}{|c|c|c|c|c|c|c|c|}
\hline
Process   &  $\sigma\times$ BR &&&&&& \\
          &   (1 W$\rightarrow l\nu$)&    Ngen &    N 30 fb$^{-1}$ & Wgt
30 fb$^{-1}$  & N 100 fb$^{-1}$ & Wgt 100 fb$^{-1}$ &   Generator \\ \hline
 tt$\gamma\gamma$ 1    &  1.6 fb ($\leq$ 1/mil)& 9296 & 48  &    .0052
 &      160 &     .0172 &   AL,MG \\
 tt$\gamma\gamma$ 2    & 6.1 fb ($\leq$ 1\%) & 2310 &   183 &    .0792
 &      610 &     .2641 &    AL \\
 tt$\gamma\gamma$ 3    &     4.9 fb ($\leq$ 1\%) &  914 &   147 &
.1608   &      490  &    .5361 &     AL  \\
 bb$\gamma\gamma$       &      318.1 fb     & 159829 &  9543 &   .0597  &
 31810  &  0.1990   &      MG  \\
 W$\gamma\gamma$ 4j    &    11.5 fb (1.2\%) & 4587 &    345 &
 0.0752 &         1150   &    .2507  &       AL \\
 Z$\gamma\gamma$        &  29.0 fb &    50005   & 870  &
 0.0174        & 2900 &  0.0580 &       MG  \\ \hline
 W$\gamma\gamma$        &  23.6 fb & 112000 & 708       &  0.0063
&    2360 &   0.0211 &       MG
\\
\hline
\end{tabular}
}
\label{ttg2gamma:tab:genirred}
\end{table}
%XXX some stat errors still missing
%XXX check on bbgg force decay to lepton NO it is not forced decay to lepton

A background is called reducible if at least one element of the
final-state signature is mistakenly identified due to incomplete detector
coverage or other instrumental effects.  This could arise if one or more
electrons or jets are misidentified as photons, or a jet as an electron
or a muon. Therefore possible background processes can be grouped into
the following signature categories: $\rm l \rm l \gamma$, $\rm l \rm l \rm j$,
 $\rm l \rm j \rm j$, $\rm l \gamma \rm j$, $\gamma \gamma \rm j$, $\gamma
\rm j \rm j$,
 $\rm j \rm j \rm j$, where $\rm l$ is a lepton and $\rm j$ is a jet.
Table~\ref{ttg2gamma:tab:catred} lists
the reducible
background processes to be considereed for each category.
% along with the generators to be used
% to either
% generate or cross-check event samples.
It should be noted that several
processes
could contribute to more than one signature category.

During the time horizon of the workshop, due to the implementation of the many new generator
processes, it has been possible to study only the irreducible backgrounds with acceptable
statistics, so only these will be presented in this report. Low-statistics tests on most of the
processes in Table~\ref{ttg2gamma:tab:catred} have been performed, and as many of these processes
as possible will be included with high statistics in a definitive study now in progress
with events fully simulated and reconstructed in the CMS detector.
\begin{table}[!h]
\centering
\caption{\footnotesize{Identified reducible background processes to be considered for each
signature category.
% along with the generators to be used to either generate or
%cross-check event samples.
}}
  \vspace*{1mm}
\footnotesize{
\begin{tabular}{|c|c|}
\hline
Signature & Process  \\ \hline
$\gamma\gamma$j & m$\gamma$ + njets \\
ll$\gamma$ & ll$\gamma$ \\
l$\gamma$ j & W(Z)$\gamma$ + njets      \\
            & $\rm b \bar{\rm b}\gamma$ + njets \\
            & $\rm t \bar{\rm t}\gamma$ + njets \\
$\gamma$jj & m$\gamma$ + njets \\
           & $\rm b \bar{\rm b}\gamma$ + njets \\
            & $\rm t \bar{\rm t}\gamma$ + njets \\
           & W(Z)$\gamma$ + njets      \\
ljj, llj & W(Z) + $\rm t \bar{\rm t}$ + njets \\
         & W(Z) + $\rm b \bar{\rm b}$ + njets \\
         & kW+mZ+njets                        \\
         & t$\bar{\rm b}$ (W)+njets          \\
        &  t + njets, Wt$\bar{\rm b}$ + njets \\
        & $\rm b \bar{\rm b} \rm t \bar{\rm t}$ + njets\\
        & $\rm b \bar{\rm b} \rm b \bar{\rm b}$ + njets \\
        &  $\rm t \bar{\rm t} \rm t \bar{\rm t}$ + njets \\
jjj     & W(Z) + $\rm t \bar{\rm t}$ + njets \\
        & W(Z) + $\rm b \bar{\rm b}$ + njets \\
         & kW+mZ+njets                        \\
         & t$\bar{\rm b}$ (W)+njets           \\
        &  t + njets, Wt$\bar{\rm b}$ + njets \\
        & $\rm b \bar{\rm b} \rm t \bar{\rm t}$ + njets \\
        & $\rm b \bar{\rm b} \rm b \bar{\rm b}$ + njets \\
        &  $\rm t \bar{\rm t} \rm t \bar{\rm t}$ + njets \\
\hline
\end{tabular}
}
\label{ttg2gamma:tab:catred}
\end{table}

All generated signal and background events
were fragmented and hadronized with PYTHIA~\cite{Sjostrand:2003wg,Sjostrand:2000wi} version 6.227.
%\begin{figure}
%\begin{center}
%\includegraphics[width=0.5\textwidth]{Fig1}
% \caption{Search reach for the $\mu \gamma {\not\!\!E}_{T}$ signal
%(as defined in the
%   text) for
%   300 fb$^{-1}$ integrated luminosity  at the LHC.
%}
%\label{search}
%\end{center}
%\end{figure}
%We show the region of parameter space corresponding to
%\footnote{The
%  statistical uncertainties
%  on fitted $a$ and $b$ parameters make a negligeable difference to the
%final numerical results.}

\subsection{Description of preselections}
%***Check these again to see consistency, also with Damien_unweighted.txt (dr_gamma_l)
No generator-level preselections were made on signal events.
For the irreducible background events, the following preselection
was made:
%Kine preselections on background events:
\begin{itemize}
 \item m$_{\gamma\gamma}\geq$80 GeV + where applicable:
 \item p$_{T{\gamma}}\geq$ 20 GeV, $|\eta_{\gamma}|\leq$ 2.5 or
p$_{T{\gamma}}\geq$ 15 GeV, $|\eta_{\gamma}|\leq$ 2.7
 \item p$_{T{j,l,b}}\geq$ 15 GeV, $|\eta_{j,l,b}|\leq$ 2.7, $\Delta$R(l,j or j,j or
b,b or $\gamma$,j or $\gamma,\gamma)\geq$ 0.3
\end{itemize}

where p$_T$ refers to the transverse momentum of the particle, $\eta$ its
rapidity and $\Delta R=\sqrt{(\Delta{\eta}^2+\Delta{\phi}^2})$ where
$\phi$ is the azimuthal angle.

The logical .OR. of the above
 generator-level criteria were then imposed on all signal event samples
at the particle level as well as the following fiducial acceptances on signal
as well as on background events:
\begin{itemize}
 \item $|\eta_{\gamma, e}|\leq$ 2.5,$|\eta_{\mu}|\leq$ 2.1,
 \item $\Delta R_{\gamma_{1},\gamma_{2}}\geq$ 0.3 where
$\gamma_{1}$ and $\gamma_{2}$ are p$_T$-ordered
\end{itemize}

\subsection{Description of preliminary particle-level selection}
%***From newest results, use july presentation as template
After the preselection, the selection imposed on all signal and
background events includes the following criteria: first, that the
two photons from the Higgs boson decay as well as the lepton
coming from one of the top quarks will have significant p$_T$:
\begin{itemize}
\item p$_{T{\gamma 1, \gamma 2, lepton}}\geq$50,25,20 GeV
\item p$_{T{\gamma 1}} + p_{T{\gamma 2}}\geq$120 GeV
\end{itemize}

Second, that the two Higgs photons and the lepton from a top quark
will be isolated:
\begin{itemize}
\item $\Delta R_{\gamma_{1},lepton},\Delta R_{\gamma_{2},lepton}\geq$ 0.4,0.6
\item The $\Delta R$ of the closest charged particle with p$_T >$1 GeV to $\gamma_1$
($\gamma_2$) must be greater than or equal to 0.2 (0.15)
\item The $\Delta R$ of the closest charged particle with p$_T >$1 GeV to the lepton
must be greater than or equal to 0.15 and less than or equal to 2
\item The absolute value of the scalar product of the $\gamma_1$ and
lepton momenta must be greater than or equal to 300 GeV$^2$
\end{itemize}

Third, that the scalar nature of the Higgs boson will assure a flat
distribution of the variable $\cos\theta^{*}$ for signal events,
where
$\tan\theta^{*}=\frac{|\vec{p_i}|\sin\theta_i}{\gamma(|\vec{p_i}|\cos\theta_i
-\beta E_i)}$, and $E_i$ and $\theta_i$ refer respectively to the
energy of and the 3-space angle between either of the two Higgs
photon directions and the direction of their joint 4-vector, in the
laboratory frame. The same distribution should be peaked in the
forward and backward directions for background events. The
requirement imposed is : $\cos\theta^{*}\leq$ 0.9.

Fourth, that the presence of a real $t\bar{t}$ pair in signal events
should result in a multiplicity significantly greater than for
background events from processes not containing such a pair.
Events must therefore contain at least eight particle-level jets as
constructed with the PYCELL algorithm of the PYTHIA
package.

Finally, the invariant mass of the two photons selected as coming
from a Higgs boson must lie within a 3 GeV-wide window around the putative
Higgs boson mass corresponding to the signal event sample considered.

\subsection{Preliminary particle-level results for the
Standard Model Higgs boson}
%***template fromjuly pres table or other tables here
%***don't forget statistical errors in table

Table~\ref{ttg2gamma:tab:perfdet30} shows, for each of the Standard Model Higgs boson
masses considered, the signal selection efficiency and the number of signal (N$_S$) and background
events expected, from each irreducible background process, for 30 fb$^{-1}$ of
integrated luminosity (corresponding to approximately
the first three years of LHC running at 10$^{33}$cm$^{-2}$s$^{-1}$) after application of the
selection described in the previous section. It can be seen that the leading-order
W$\gamma\gamma$ sample seems to strongly underestimate the contribution to the total
background, as compared to the W$\gamma\gamma +$4 jet sample. Therefore, for the current
study we include the W$\gamma\gamma +$4 jet contribution instead of the
W$\gamma\gamma$ contribution when calculating the total number of expected background
events (N$_B$) and the signal significance as reflected by the
quantity N$_S$/$\sqrt{N_B}$, both of which are also shown in
Table~\ref{ttg2gamma:tab:perfdet30}. It should be noted however that, by the same argument,
the leading-order Z$\gamma\gamma$ sample considered probably also represents an
underestimated contribution relative to that of a hypothetical Z$\gamma\gamma +$4 jet
sample, not available at the time of the study. This contribution may be of the
same order as that from the W$\gamma\gamma +$4 jet sample, though perhaps slightly reduced
in analogy with the relative importances of the leading-order W$\gamma\gamma$ and
Z$\gamma\gamma$ contributions. This will be evaluated with a soon-to-be-available
Z$\gamma\gamma +$4 jet sample.
\begin{table}[!h]
\centering
\caption{\footnotesize{Estimated number of signal and background events, signal
selection efficiency and
signal significance for
$\rm t \bar{\rm t} \rm H, \rm H \rightarrow \gamma \gamma$, after 30 fb$^{-1}$ of
integrated luminosity.}}
  \vspace*{1mm}
\footnotesize{
\begin{tabular}{|c|c|c|c|c|}
\hline Higgs Mass (GeV)  & 115  & 120 & 130 & 140 \\ \hline\hline
Signal Selection Efficiency (\%) & 19.09  & 20.78 & 24.65 & 25.58
\\ \hline Number Signal Evts (N$_S$)         & 3.96   & 4.06  & 3.92
& 2.86    \\ \hline $t\bar{t}\gamma\gamma$ Type 1    & 0.17   & 0.11
& 0.14  & 0.16 \\ \hline $t\bar{t}\gamma\gamma$ Type 2    & 0.08 &
0.16  & 0.08  & 0.16 \\ \hline $t\bar{t}\gamma\gamma$ Type 3    &
$<$0.2   & 0.2   & $<$0.2  & $<$0.2 \\ \hline Z$\gamma\gamma$ & 0.23
& 0.21  & 0.24  & 0.16       \\ \hline W$\gamma\gamma$4j & 0.4    &
0.9   & 1.9   & 1.4      \\ \hline bb$\gamma\gamma$ & $<$ 0.06 &
0.06  & 0.06  & $<$ 0.06   \\ \hline \hline Total
Number Background Evts.(N$_B$)& 0.88  & 1.63  & 2.42  & 1.88 \\
\hline Signal Significance              & 4.22   & 3.18  & 2.52  &
2.09    \\ \hline \hline W$\gamma\gamma$                  & 0.37   &
0.42  & 0.37  & 0.39    \\ \hline
\end{tabular}
}
\label{ttg2gamma:tab:perfdet30}
\end{table}

Table~\ref{ttg2gamma:tab:perfdet100} shows the same results for 100
fb$^{-1}$ of integrated luminosity (corresponding to approximately
one year of LHC running at 10$^{34}$ cm$^{-2}$s$^{-1}$).

\begin{table}[!h]
\centering
\caption{\footnotesize{Estimated number of signal and background events, signal
selection efficiency and
signal significance for
$\rm t \bar{\rm t} \rm H, \rm H \rightarrow \gamma \gamma$, after 100 fb$^{-1}$ of
integrated luminosity}}
  \vspace*{1mm}
\footnotesize{
\begin{tabular}{|c|c|c|c|c|}
\hline Higgs Mass (GeV)  & 115  & 120 & 130 & 140 \\ \hline\hline
Signal Selection Efficiency (\%) & 19.09  & 20.78 & 24.65 & 25.58
\\ \hline Number Signal Evts (N$_S$)         & 13.2   & 13.5  & 13.1
& 9.5    \\ \hline $t\bar{t}\gamma\gamma$ Type 1    & 0.57   & 0.38
& 0.48  & 0.53 \\ \hline $t\bar{t}\gamma\gamma$ Type 2    & 0.3    &
0.5   & 0.3   & 0.5  \\ \hline $t\bar{t}\gamma\gamma$ Type 3    &
$<$0.5   & 0.5   & $<$0.5  & $<$0.5 \\ \hline Z$\gamma\gamma$ & 0.8
& 0.7   & 0.8   & 0.5       \\ \hline W$\gamma\gamma$4j & 1.5    &
3.0   & 6.2   & 4.7      \\ \hline bb$\gamma\gamma$ & $<$0.2   & 0.2
& 0.2   & $<$ 0.2   \\ \hline \hline Total Number Background
Evts.(N$_B$)& 3.17  & 5.28  & 7.98  & 6.23 \\ \hline Signal
Significance              & 7.41   & 5.88  & 4.64  & 3.81
\\ \hline \hline W$\gamma\gamma$                  & 1.25   & 1.35  &
1.23  & 1.27    \\ \hline
\end{tabular}
}
\label{ttg2gamma:tab:perfdet100}
\end{table}

For both cases it can be seen that the contributions to the total
survivng background from the $t\bar{t}\gamma\gamma$ Type 2 and, in
the limit of generated statistics, Type 3 processes are of the same
order of magnitude as the Type 1 process.

The limited-statistics samples of the reducible background processes
considered have not resulted in a significant contribution to the number of
surviving background events, in the context of this particle-level study.

The above results would indicate a possibility of signal observability
 in excess of 3$\sigma$ for Higgs boson masses below 120 GeV after 30 fb$^{-1}$, with
approximately four signal events observed corresponding to the signal selection
efficiencies of approximately 20\%.

For 100 fb$^{-1}$, there would be discovery potential in excess of 5$\sigma$ and ranging
as high as over 7$\sigma$ for Higgs boson masses up to 120 GeV. For masses up to 140 GeV
there would be a possibility of signal observability in excess of 3$\sigma$.

\subsection{Conclusions and future work}
%Addition of

The preliminary particle-level selection presented above has not yet been optimized.
Furthermore, there is a possibility to enhance its performance via the inclusion
of variables involving missing energy and/or momentum, or the identification of b quark
jets, which may be effective
in vetoing background events not including real top quarks but nontheless having
high jet multiplicity. The present study would indicate that it is this type of process
(for example W + $\gamma\gamma$ + N jets) which will prove to be the most challenging background.

\begin{figure}[!Hhtb]
\centering
\includegraphics[width=0.50\textwidth]{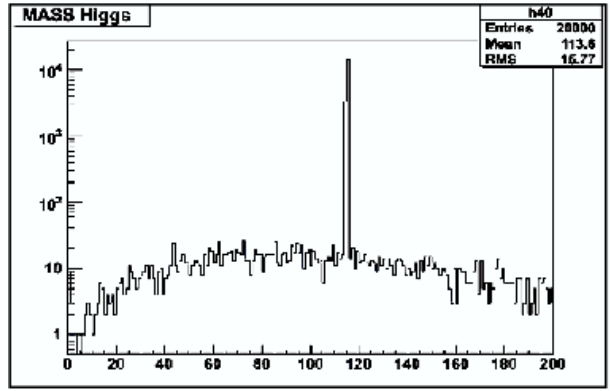}
%\caption{Reconstructed mass of Higgs boson for
%$\rm M_{\rm H}$=130 GeV/$c^{2}$}
\caption{Invariant $\gamma\gamma$ mass from events from the signal process
$\rm t \bar{\rm t}
 \rm H$ production with $\rm H \rightarrow \gamma \gamma$ and at least one
charged
lepton from the decay of a top or antitop quark (m$_H=$115 GeV), where the two photons with
the highest values of p$_T$  have been identified as the Higgs photons.}
\label{higgs_sm_tth_2gamma_fig3}
\end{figure}

The method used to select the two putative Higgs photons in each event may also have an
effect on
the selection's performance.  In some prior studies of associated Higgs
production~\cite{Assamagan:2004mu},
the two photons with the highest values of p$_T$ have been assumed to have come from
the Higgs boson. However we have observed (see Figure~\ref{higgs_sm_tth_2gamma_fig3})
that the use of this criterion results in considerable sidebands in the two-photon invariant
mass distribution, at the level of approximately 10\% for $\rm t \bar{\rm t} \rm H$
events with $\rm
H \rightarrow
\gamma \gamma$.
  Investigation has shown that this faulty combinatorial choice concerns
overwhelmingly the photon with the second-highest p$_T$, since the
photon with the highest p$_T$ is not a Higgs photon only at the
level of approximately 1/mil, as calculated from signal events. The
origin of these `fake' second Higgs photons is approximately 80\%
from $\pi^0$'s, 10\% from $\eta$'s, a few percent from $\omega$'s,
and the remainder from other particles. Fully 80\% of these fake
Higgs photon `mother' particles appear to come from parton showers
whose origin is one of the two final-state top quarks, and as such
are peculiar to the $\rm t \bar{\rm t}  \rm H$ process.  The other
20\% come from showering from the initial-state protons and hence
are common to all the associated production channels. In this way we
can estimate that the level of the effect on the WH and ZH processes
would be approximately 2-3\%.

\begin{figure}[!Hhtb]
\centering
\includegraphics[width=0.90\textwidth]{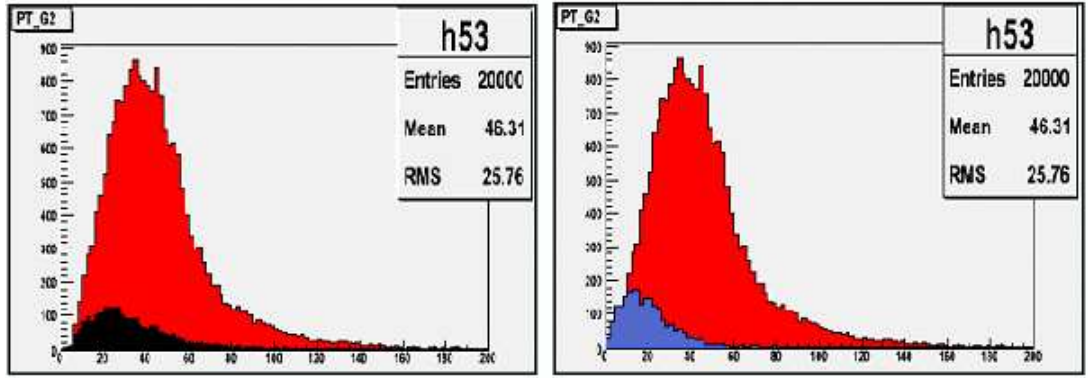}
%\caption{Reconstructed mass of Higgs boson for
%$\rm M_{\rm H}$=130 GeV/$c^{2}$}
\caption{Distribution of the transverse momentum for the second-highest
p$_T$ photon in each event. (Left) The shaded superimposed distribution corresponds to
events
where this photon is not from Higgs boson decay. (Right) The shaded superimposed
distribution corresponds to the true Higgs photon in the same events.}
\label{higgs_sm_tth_2gamma_fig4}
\end{figure}

In addition to the pure combinatorial effect discussed above, which
in itself would be damaging to signal selection efficiency, the
selection of `fake' Higgs photons could result in biased kinematical
distributions used to construct the selection itself, whether this
last is composed of the mere sequential imposition of criteria or of
a more sophisticated nature such as likelihood or neural
network-based selections. An example of this possibility is shown in
Fig.~\ref{higgs_sm_tth_2gamma_fig4}, where the total distribution of
the photon with the second-highest p$_T$ in each event is plotted,
with that for those events in which the second photon is not a Higgs
photon superposed on it, on the left-hand side.
 The plot on the right-hand side  shows the same total distribution,
but the superimposed sub-distibution is that of the true Higgs photons in the same events.
Techniques are currently under investigation to try to improve the Higgs photon selection
so as to correct this problem, both at the particle level as well as in the context of a detailed
CMS simulation and reconstruction study, which is now ongoing.
In this detailed simulation and reconstruction study we will study the contributions of
the reducible as well as the irreducible backgrounds. We will interpret the results in
the context of models with two Higgs doublets as well as attempt to estimate the sensitivity of
the $\rm t \bar{\rm t}  \rm H$ process to the CP nature of the Higgs boson, as has been
suggested by some authors~\cite{Gunion:1996xu}.

%impact of power showers.

%Resonant slepton production and its decays into $l \gamma {\tilde G}$ or $l
%{\tilde G}$ can be discovered at the LHC for slepton masses into the multi-TeV
%region, depending upon the $R_p$ violating coupling and provided that the gravitino is
%ultra-light (with a mass less than 0.1 eV). Various $M_T$ distributions will
%allow the accurate measurement of sparticle masses involved.

\subsection*{Acknowledgements}
%ask authors to add their acknowledgements
The authors would like to thank the organizers and session convenors of
the Les Houches Workshop. S. G.-S. is indebted to Peter Skands for his invaluable help
on the ALPGEN/PYTHIA interface for the new ALPGEN processes.
R.P. acknowledges the financial support of MIUR, under contract
n. 2004021808 009.
%K Sridhar would like to thank CERN and LAPTH for hospitality offered
%during which some of the work contained herein was performed.

%%%%%%%%%%%%%%%%%%%%%%%%%%%%%%%%%%%%%%%%%%%%%%%%%%%%%%%%%%%%%%%%%%%%%%%%%%%%%
\section[Study of $bbZ$ as a benchmark for MSSM $bbH$]
{STUDY OF $bbZ$ AS A BENCHMARK FOR MSSM $bbH$~\protect
\footnote{Contributed by: S.~Lehti}}
\subsection{Introduction}
The Z boson production process with associated b jets 
$\rm gg/qq\rightarrow b\bar{\rm b}Z$ is topologically similar to 
$\rm gg/qq\rightarrow b\bar{\rm b}H$. In the MSSM the associated Higgs production
dominates at large values of tan$\beta$. At large tan$\beta$ the most 
important decay modes for
the Higgs boson are $\rm H\rightarrow b\bar{\rm b}$ and 
$\rm H\rightarrow \tau\tau$. Here we concentrate on Higgs decaying into 
$\tau$'s with $\tau$ decaying to an electron or muon. 

The $\rm b\bar{\rm b}Z$ production at the LHC can be used as a benchmark for testing the
Higgs boson reconstruction methods~\cite{Maltoni:2005wd}. The Z mass is known with a good precision,
which can be used to verify the mass reconstruction method. It is also possible
to measure the $\rm b\bar{\rm b}Z$ cross section to verify NLO calculations,
and b jet and Z transverse momentum spectra to verify the kinematics.

The aim of this study is to show that the Higgs boson mass peak can be reconstructed
by reconstructing and understanding the Z mass peak, and to show that it is
possible to extract the signal from the background and to measure the cross 
section.
If that can be done, the same method should work
similarly for the Higgs boson in $\rm gg/qq\rightarrow b\bar{\rm b}H$.

\subsection{Cross sections}

The signal consists of $\rm Z/\gamma^*$ events produced in association with b quarks.
The Z boson and $\gamma^*$ are allowed to decay to electron, muon or tau pairs, tau decaying
leptonically. Two possibilities exist, either to select any two lepton final state, or to 
select e+$\mu$ final states only. The former has a larger cross section, but the latter
has a significantly smaller background~\cite{note2002_035}.

The signal ($\rm b\bar{\rm b}Z/\gamma^*$) is generated,
and the signal cross sections calculated with CompHEP~\cite{Boos:2004kh}.
The LO signal cross section for any two-lepton
final state is 58 pb. No $\rm p_{\rm T}$ and $\eta$ cuts are applied on massive
b quarks in
$\rm b\bar{\rm b}Z/\gamma^*$ process generation.
%Massive b quark is assumed with no cuts on b quark $\rm p_{\rm T}$ or $\eta$.
The background comes
mainly from two sources, $\rm Z/\gamma^*$ associated with light quark and gluon
jets, generated with PYTHIA~\cite{Sjostrand:2003wg}, and $\rm t\bar{\rm t}$, tW events, 
generated with TopReX~\cite{Slabospitsky:2002ag}. 
An NLO cross section of 1891~pb~\cite{DY_cross_section}, calculated with 
MCFM~\cite{mcfm}, is used for Drell-Yan 
$\rm Z/\gamma^*\rightarrow LL$ ($\rm LL = ee$, $\mu\mu$ or $\tau\tau$) events
with $\rm m_{\rm LL} >$ 80 GeV/${c^2}$.
For $\rm t\bar{\rm t}$ and tW a cross section of 840 and 60 pb is used,
respectively~\cite{Beneke:2000hk}.
The $\rm Z/\gamma^*$ background sample
%, generated with PYTHIA,
consists also of events with two associated b quarks, but to prevent double counting,
those events are removed using the available generation level information of the events.

As shown in Table~\ref{table:cross_section}, the cross section for
$\rm gg/qq\rightarrow b\bar{\rm b}Z/\gamma^*, 
Z/\gamma^*\rightarrow\tau\tau\rightarrow\ell\ell+X$
is quite small, in fact of the order of the Higgs boson cross section at 
m$_{\rm A}$ = 200 GeV/$c^2$, tan$\beta\sim$ 25. 
As the mass of the Z boson is lower than that of the Higgs boson, the leptons and
jets have lower p$_{\rm T}$,
and the selection efficiency for the associated 
$\rm Z/\gamma^*\rightarrow\tau\tau\rightarrow\ell\ell$ events is lower.
Therefore, studying $\rm gg/qq\rightarrow b\bar{\rm b}Z/\gamma^*, 
Z/\gamma^*\rightarrow\tau\tau\rightarrow\ell\ell+X$
as a benchmark for the Higgs boson in e+$\mu$ final states is not feasible.
However, it is possible to study the mass reconstruction using inclusive
$\rm Z/\gamma^*\rightarrow\tau\tau$ events.

Other potential backgrounds are $\rm b\bar{\rm b}$,
WW, WZ and ZZ events. The contribution from these backgrounds turn out, however, to be 
negligible after the selection.
%, and they are not taken into account. 
The cross sections for signal and main
background processes are shown in Table~\ref{table:cross_section}.

\begin{table}[h]
  \vskip 0.1 in
  \centering
  \caption{Cross sections for signal and background processes.}
  \vspace*{1mm}
  \begin{tabular}{|ll|c||ll|c|}
  \hline
Signal $\rm b\bar{\rm b}Z/\gamma^*$ && pb & Background && pb \\
  \hline
$\rm \tau\tau b\bar{\rm b}$ &{\tiny (60$<\rm m_{\tau\tau}<100$GeV/$c^2$)} & 3.29 & 
$\rm Z/\gamma^*\rightarrow\tau\tau\rightarrow\ell\ell$ &{\tiny (80$<\rm m_{\tau\tau}<100$GeV/$c^2$)} & 223.2\\
$\rm \tau\tau b\bar{\rm b}$ &{\tiny ($\rm m_{\tau\tau}>100$GeV/$c^2$)} & 0.132 &
$\rm Z/\gamma^*\rightarrow\tau\tau\rightarrow\ell\ell$ &{\tiny ($\rm m_{\tau\tau}>100$GeV/$c^2$)} & 10.1\\
$\rm \mu\mu b\bar{\rm b}$ &{\tiny (60$<\rm m_{\mu\mu}<100$GeV/$c^2$)} & 26.2 &
$\rm Z/\gamma^*\rightarrow\mu\mu$ &{\tiny ($\rm m_{\mu\mu}>80$GeV/$c^2$)} & 1891 \\
$\rm \mu\mu b\bar{\rm b}$ &{\tiny ($\rm m_{\mu\mu}>100$GeV/$c^2$)} & 1.05 &
$\rm Z/\gamma^*\rightarrow ee$ &{\tiny ($\rm m_{ee}>80$GeV/$c^2$)} & 1891\\
$\rm eeb\bar{\rm b}$ &{\tiny (60$<\rm m_{ee}<100$GeV/$c^2$)} & 26.3 &
$\rm t\bar{\rm t}\rightarrow ee/\mu\mu/\tau\tau$ && 86.2 \\
$\rm eeb\bar{\rm b}$ &{\tiny ($\rm m_{ee}>100$GeV/$c^2$)} & 1.05 &
$\rm tW\rightarrow ee/\mu\mu/\tau\tau$ && 6.16 \\
  \hline
  \end{tabular}
  \label{table:cross_section}
\end{table}

\subsection{Detector simulation}
The detector simulation is done using full GEANT~\cite{geant} simulation in the 
ORCA~\cite{orca} framework. Version ORCA\_8\_7\_4 of CMS OO reconstruction software is used.
The CMS detector is simulated with the complete ideal detector, no staging
%\cite{CMSstaging}
and no misalignment of the detector elements is assumed. The ORCA 
reconstruction is based on official CMS digitized datasets with pile-up 
included (3.4 minimum bias events superimposed per event crossing for a 
luminosity of $\rm 2\times 10^{33}cm^{-2}s^{-1}$).

\subsection{Event selection}
\subsubsection{Trigger}
The events are triggered with a single and double electron and muon trigger.
The p$_{\rm T}$ threshold for single muons is 19 GeV/c, for single electrons 29 GeV/c,
for double muons 7 GeV/c and for double electrons 17 GeV/c~\cite{Sphicas:2002gg}.
The Level 1 trigger efficiency for the signal is 0.914. 
%The HLT efficiency for 
%single muons, single electrons, double muons and double electrons is 
%0.9*0.97*0.97, x,x, 0.95*0.872*0.946, respectively.
The overall trigger efficiency for signal is found to be 0.826.
Stronger trigger thresholds and lower efficiency on single and double 
electrons suppress the electron 
final states with respect to the muon final states. An offline cut on lepton 
$\rm p_{\rm T}>$ 20 GeV/c balances the different thresholds for events passing 
the two electron and two muon trigger,
but the events triggered with single electron trigger are still suppressed.
Therefore, there are more muon events than electron events from the signal and
the background passing the trigger.

\subsubsection{Offline selection}
The basic event selection is a requirement of two isolated leptons 
p$_{\rm T}>$ 20 GeV/c in the central detector acceptance region $|\eta|<$ 2.5 coming 
from a reconstructed primary vertex. These cuts reduce efficiently the
backgrounds with soft leptons 
($\rm pp\rightarrow b\bar{\rm b}$,$\rm c\bar{\rm c}$,..).
The leptons are defined isolated
when there are no other tracks from the primary vertex with p$_{\rm T}>$ 1 GeV/c within a cone
$\rm\Delta R = \sqrt{\Delta\varphi^2 + \Delta\eta^2}\leq 0.4 $ around the lepton. Other methods,
which are used to suppress the backgrounds, are b tagging and central jet veto.
The missing transverse energy, reconstructed from high p$_{\rm T}$ objects, such as leptons and jets,
is used to suppress $\rm t\bar{\rm t}$ events, and it is also
needed in the $H \rightarrow\tau\tau$ mass reconstruction method due to 
neutrinos in the final state.

B jets, associated with the Z boson, provide a powerful tool to separate
the $\rm b\bar{\rm b}Z/\gamma^*$ events from the Z/$\gamma^*$ background.
The Z/$\gamma^*$ events are mostly produced with no significant jet activity,
and the associated jets are mostly light quark and gluon jets.
Therefore, the Z/$\gamma^*$ background can be suppressed by requiring
reconstructed jets to be present in the event, and even further by requiring that
the associated jets are identified as b jets. There are two possibilities
available, either to require one b tagged jet per event and veto other
jets, or to require two b tagged jets in the event. Here the 2b-tagging option
is used in order to have a more pure sample of $\rm b\bar{\rm b}Z/\gamma^*$ 
events. A E$_{\rm T}$ threshold of 20 GeV is used for both jets.

\begin{figure}[h]
  \centering
\includegraphics[width=0.5\textwidth,height=0.3\textwidth]{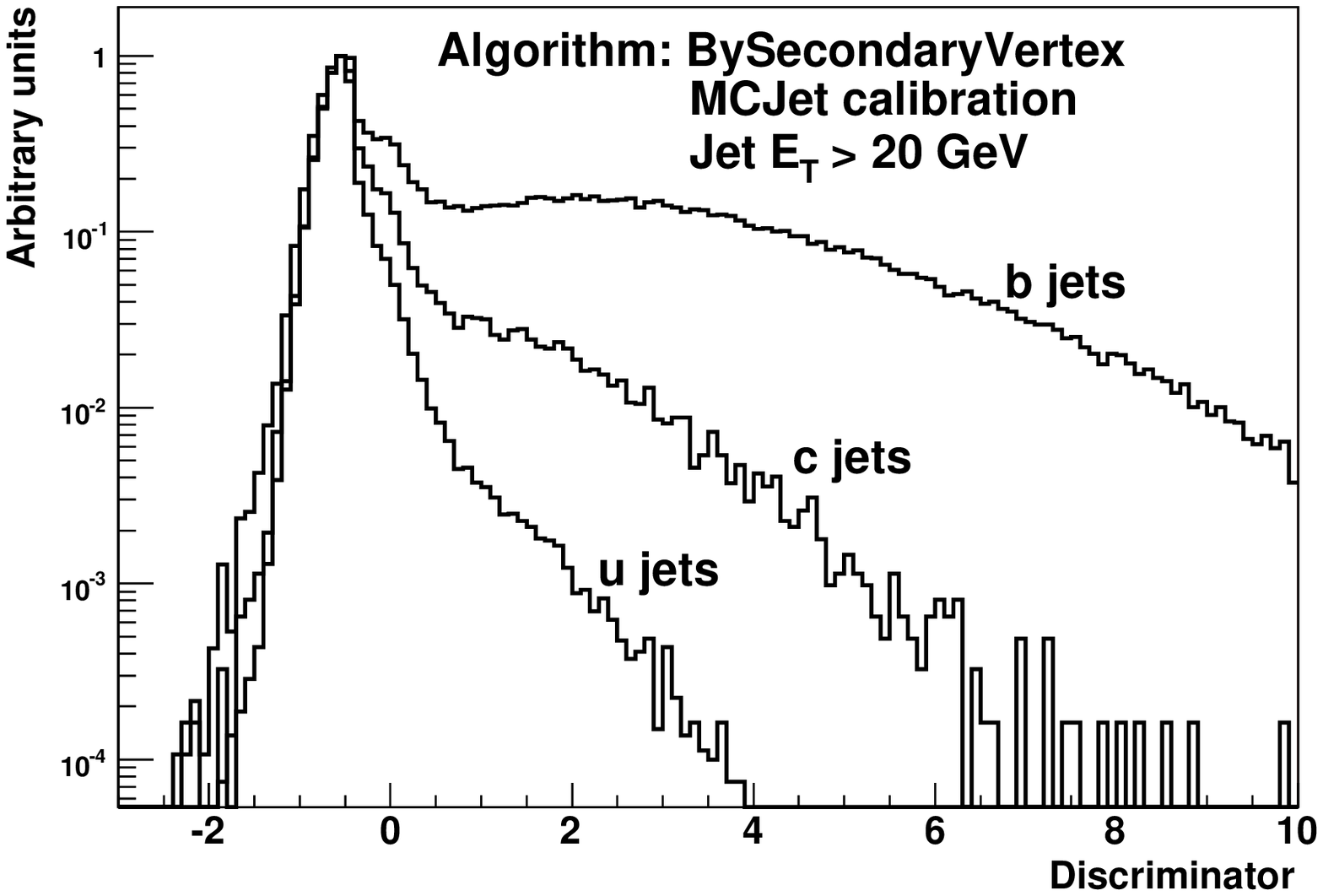}
    \caption{The output of the b tagging algorithm for b-, c- and
light quark and for gluon jets in $\rm t\bar{\rm t}$ events.
  \label{fig:btagging}}
\end{figure}

B jets associated with the Higgs and Z bosons are generally very soft, which
makes their tagging a challenging task. In a low E$_{\rm T}$ jet, the track
multiplicity and momenta tend to be low, and many jets do not have enough significant
tracks to be identified as b jets. As a consequence, the b tagging efficiency
is not very high.
%, $\sim$ X \%, if the mistagging rate is to be kept under 
%1 \% level. 
%To have good purity of the $\rm b\bar{\rm b}Z/\gamma^*$ events stronger
%b tagging cut is used here. 
In this study, a b tagging algorithm based on the reconstruction
of the secondary decay vertex of the decaying B hadron~\cite{pTDR1}
is chosen. The discriminator of that algorithm is shown in 
Fig.~\ref{fig:btagging} for b, c and light quark and gluon jets. 
A cut of discriminator $>2$ gives on average 22\% b tagging
efficiency per jet ($\rm b\bar{\rm b}Z/\gamma^*$) with 0.091\% mistagging 
rate ($\rm jjZ/\gamma^*$). A cut stronger than this 
suppresses the signal 
too much with respect to $\rm t\bar{\rm t}$ events, for which the b jets are 
more energetic, more central, and therefore easier to reconstruct and to b tag. 

The $\rm t\bar{\rm t}$ events have more
jet activity than $\rm b\bar{\rm b}Z/\gamma^*$, and a jet veto is used
to suppress the $\rm t\bar{\rm t}$ background. Events with
jets $\rm E_{\rm T}>$ 20 GeV within the tracker 
acceptance region in addition to the two b jets are rejected.

In $\rm t\bar{\rm t}$ events the leptons come from W decays, so there are 
always neutrinos in the final state. For the signal there are no neutrinos in 
the final state ($\rm b\bar{\rm b}Z/\gamma^*\rightarrow 
b\bar{\rm b}\tau\tau\rightarrow b\bar{\rm b}\ell\ell+X$ represents only a 
tiny fraction of the signal events) and missing transverse energy is expected only
due to measurement error. The missing $\rm E_{\rm T}$ (MET) is reconstructed 
from
the high $\rm E_{\rm T}$ objects in the event: the two leptons, and the jets 
coming from the primary vertex. A jet is defined to be coming from the primary 
vertex, if at least half of its tracks are coming from the primary vertex.
A MC jet correction~\cite{pTDR1} on jet energy scale is used.
A MET cut MET $<$ 30 GeV is applied, which is 
already close to the detector MET resolution. 

A strong method to separate the Z boson events from the $\rm t\bar{\rm t}$
background is to reconstruct the invariant Z mass. The two leptons are measured
with a good accuracy, and the invariant mass distribution in 
Fig.~\ref{fig:invmass} 
shows a clear peak at the nominal Z boson mass. Events with invariant a mass of 
85$\rm < m_{\ell\ell} <$ 95 GeV/$c^2$ are chosen for further analysis.

After the selection described above, from the total number of passed events 
(1065 events for 30~fb$^{-1}$) the fraction of signal events 
($\rm b\bar{\rm b}Z/\gamma^*$) is 70\%, the fraction of $\rm Z/\gamma^*$
with no associated b jets is 11\%, and the fraction of $\rm t\bar{\rm t}$
events is 19\%. All other backgrounds are negligible.

\subsection{Results}
\subsubsection{Mass reconstruction}

In the $\rm H_{\rm SUSY}\rightarrow\tau\tau$ analysis the
Higgs boson mass is reconstructed using a
collinear approximation method. Due to neutrinos in the final state a
precise mass reconstruction is impossible. In the collinear approximation the
neutrinos are assumed to be emitted along the leptons, which is a valid 
assumption for the signal events due to large Lorentz boosts of the two $\tau$'s.
The missing transverse energy is projected along the lepton transverse momentum
directions, giving an estimate for the neutrino momentum including
the z component of the neutrino momentum. The reconstructed mass is the invariant 
mass of the summed lepton and neutrino 4-momenta.

\begin{figure}[t]
  \centering
  \vskip 0.1 in
  \begin{tabular}{cc}
  \begin{minipage}{7.5cm}
    \centering
  \includegraphics[height=70mm,width=80mm]{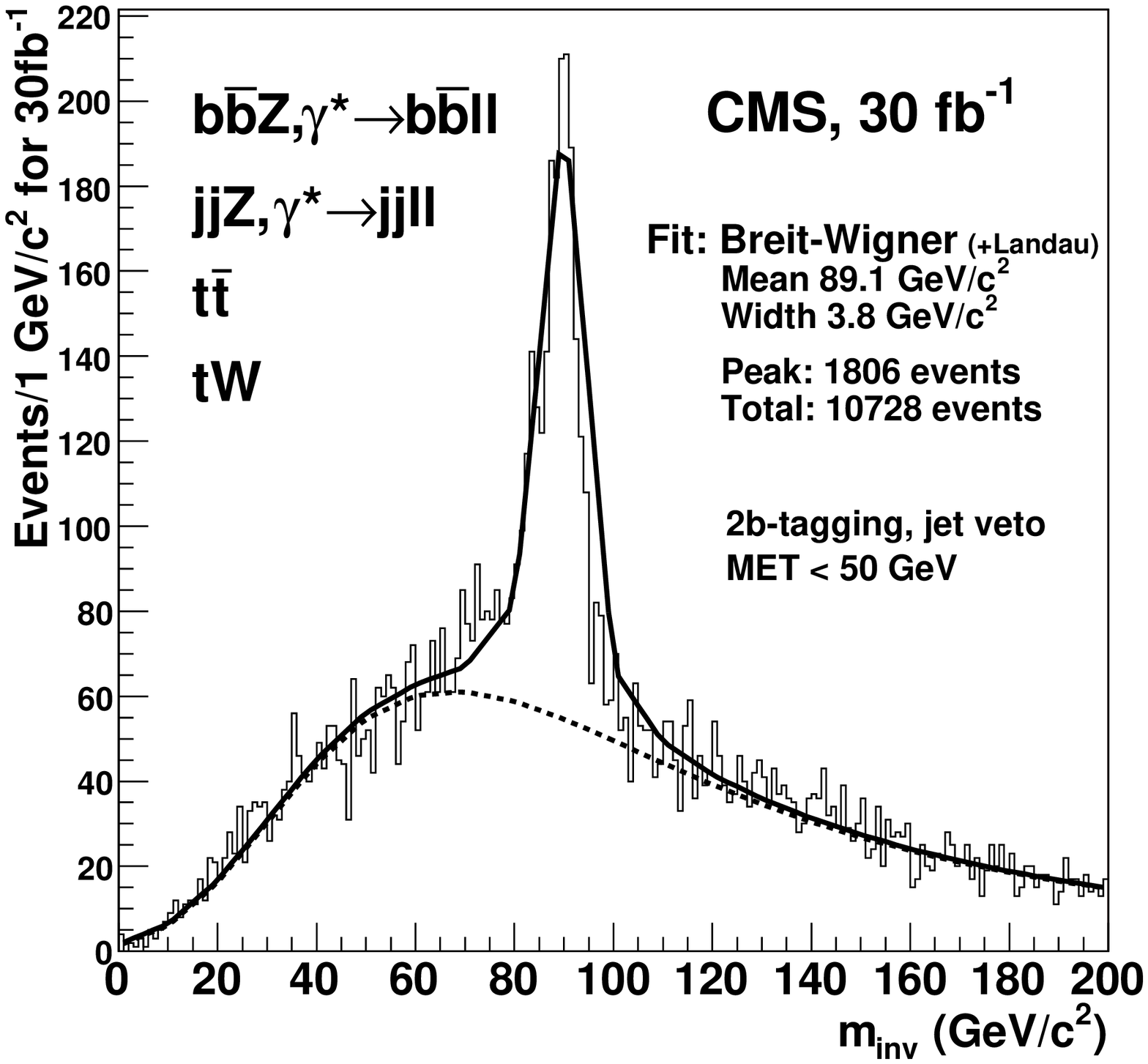}
      \caption{Invariant mass of the two leptons.}
  \label{fig:invmass}
  \end{minipage}
  &
  \begin{minipage}{7.5cm}
    \centering
  \includegraphics[height=70mm,width=80mm]{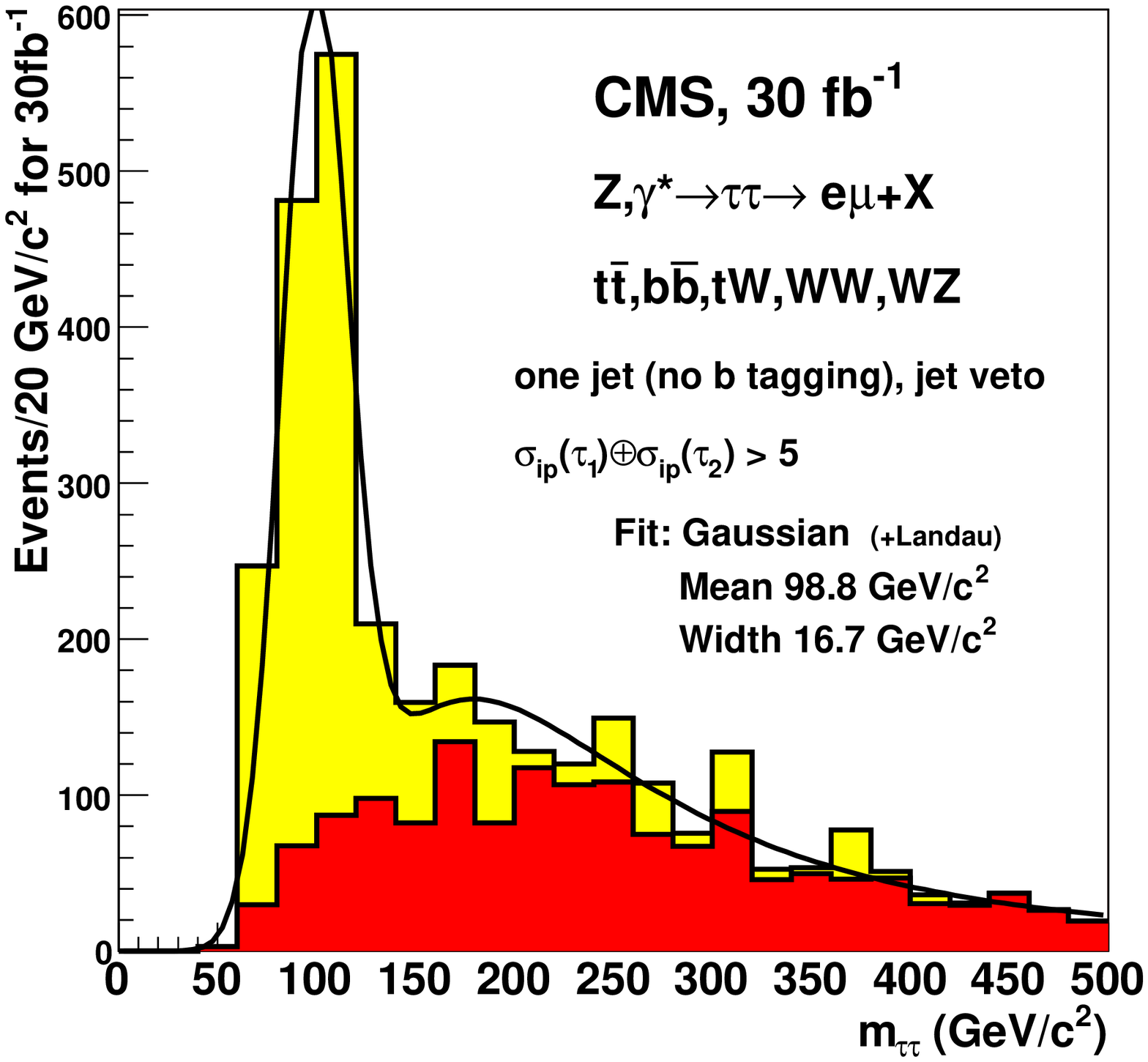}
      \caption{Mass reconstructed using collinear approximation.}
  \label{fig:effmass}
  \end{minipage}
  \end{tabular}
\end{figure}

Mass reconstruction using the collinear approximation is possible, when the two 
leptons are not in a back-to-back configuration.
Events in back-to-back configuration are removed with a cut $\Delta\varphi(\ell_1,\ell_2)<$
175$^o$, where $\Delta\varphi(\ell_1,\ell_2)$ is the angle between the two 
leptons in the transverse plane. The mass reconstructed using the collinear 
approximation is shown in Fig.~\ref{fig:effmass}. 
The e+$\mu$ final states are chosen to select events with intermediate $\tau$'s. 
In the Higgs boson studies one b jet is required present in the event with a veto
on additional jets. Similar events are chosen here, one associated jet is required
in the event, but no b tagging is used.
Since the leptons are generally well 
measured, the mass peak position and width are highly dependent on the 
quality of the missing transverse energy measurement. 

\subsubsection{Verification of Monte Carlo}

The verification of the Monte Carlo for $\rm b\bar{\rm b}Z/\gamma^*$ events 
includes the verification of the cross section, the associated b jet 
E$_{\rm T}$ and $\eta$ distributions, and the Z p$_{\rm T}$ distribution.
These has been studied in Ref.~\cite{Campbell:2003hd}.
Each of these distributions consists of both, signal and background events,
the measured distribution is a convolution of different signal and background
distributions. The shapes of the background distributions can be measured
from the data, both $\rm Z/\gamma^*$ cross section with associated light quark and 
gluon jets
and the $\rm t\bar{\rm t}$ cross section are large compared to the signal
cross section. 
%The $Z/\gamma^*$ and $\rm t\bar{\rm t}$ cross sections can
%also be measured using data. Therefore one can assume that the fraction
%of signal events, $Z/\gamma^*$ events and $\rm t\bar{\rm t}$ events can be 
%measured from data. 
%The fraction of background events can also be measured from the data.
This information can be used to estimate the shape of
the convoluted distributions, which can then be compared with the measured
distributions.

\begin{figure}[h]
  \centering
  \vskip 0.1 in
  \begin{tabular}{ccc}
  \begin{minipage}{5cm}
    \centering
  \includegraphics[height=70mm,width=50mm]{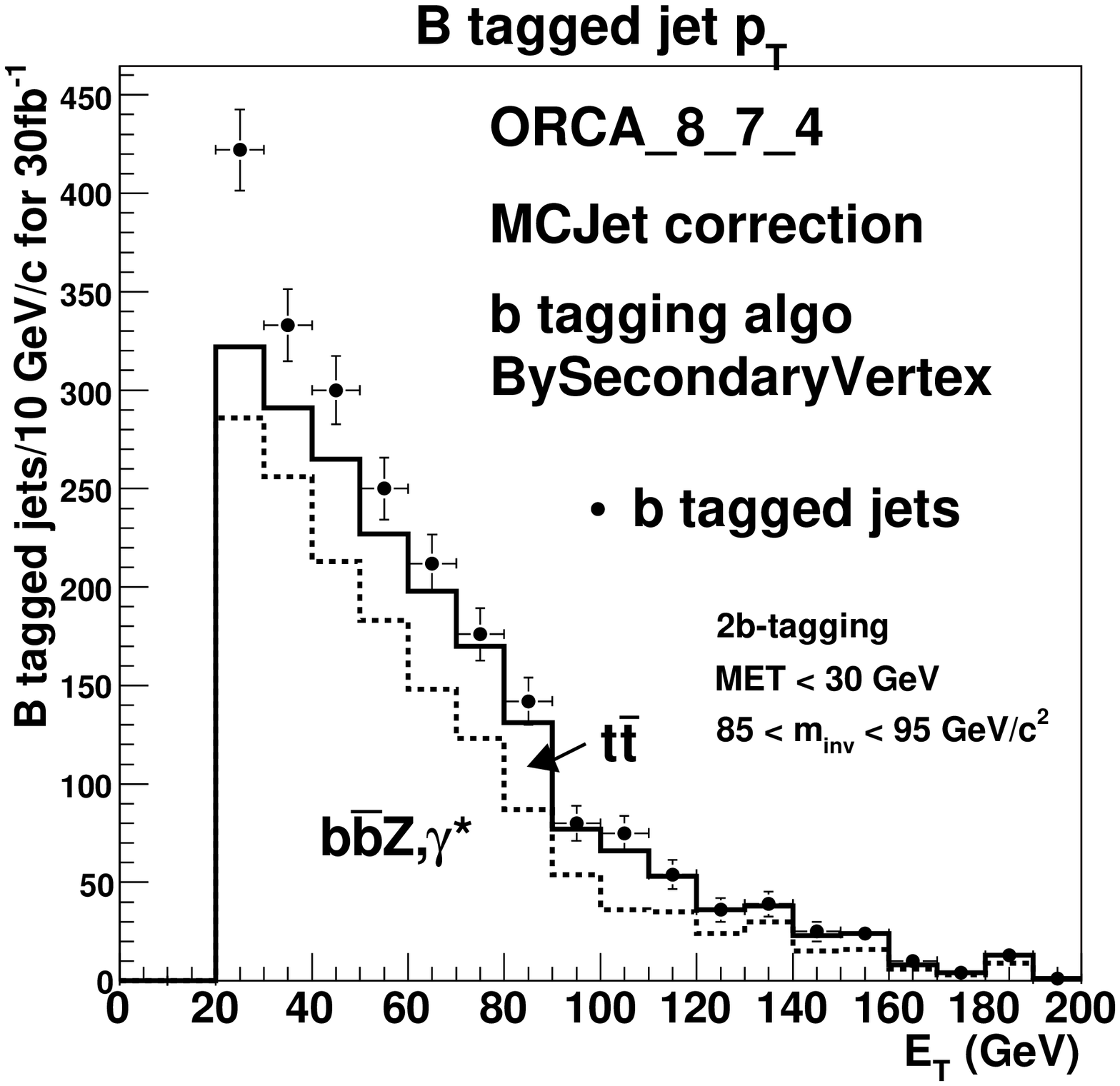}
      \caption{Reconstructed and b tagged jet p$_{\rm T}$ (points) and the
fraction of $\rm b\bar{\rm b}Z/\gamma^*$ (dashed histogram) and 
$\rm t\bar{\rm t}$ events (solid histogram).}
  \label{fig:bjet_et}
  \end{minipage}
  &
  \begin{minipage}{5cm}
    \centering
  \includegraphics[height=70mm,width=50mm]{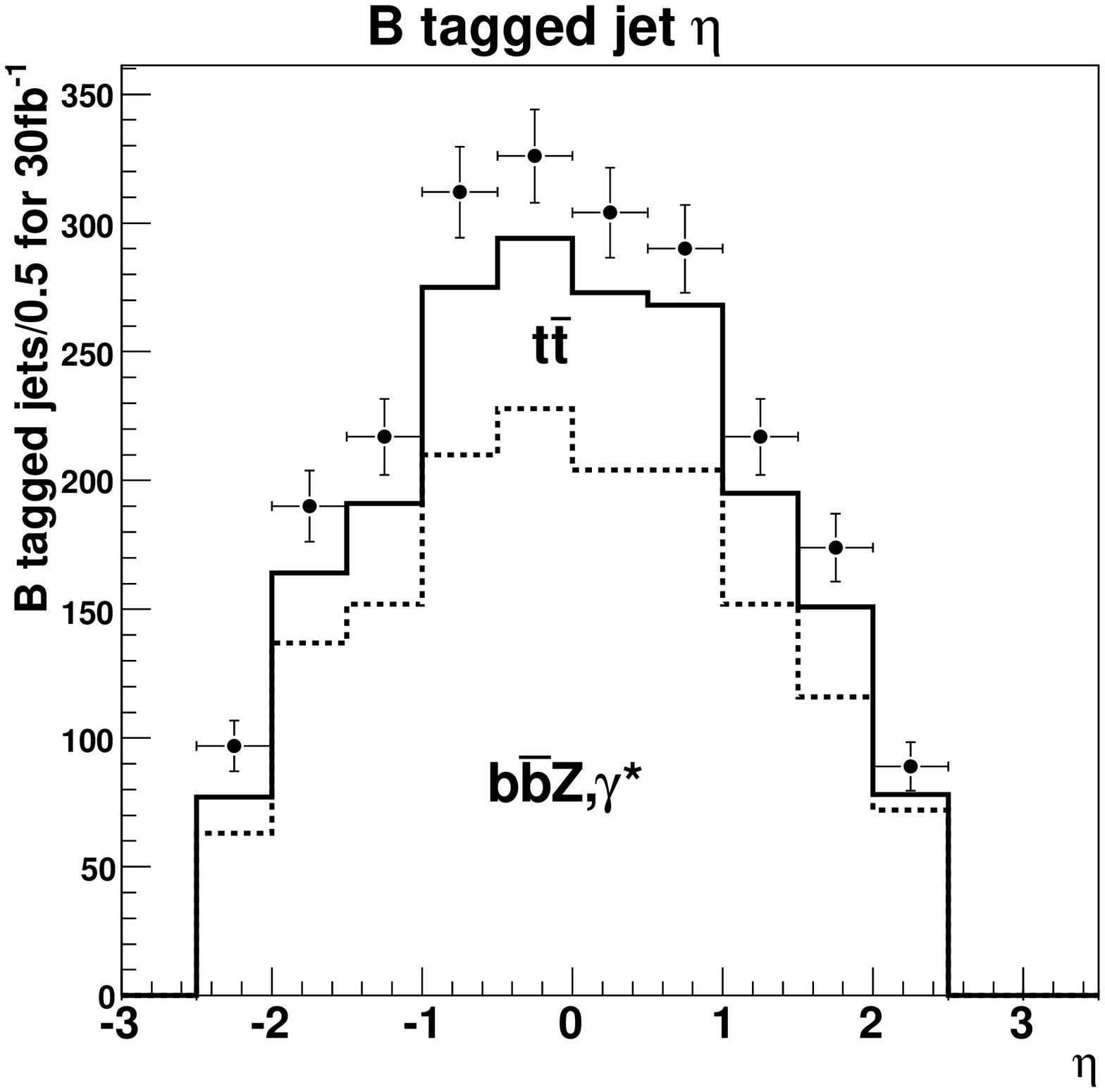}
      \caption{Reconstructed and b tagged jet $\eta$ (points) and the
fraction of $\rm b\bar{\rm b}Z/\gamma^*$ (dashed histogram) and
$\rm t\bar{\rm t}$ events (solid histogram).}
  \label{fig:bjet_eta}
  \end{minipage}
  &
  \begin{minipage}{5cm}
    \centering
  \includegraphics[height=70mm,width=50mm]{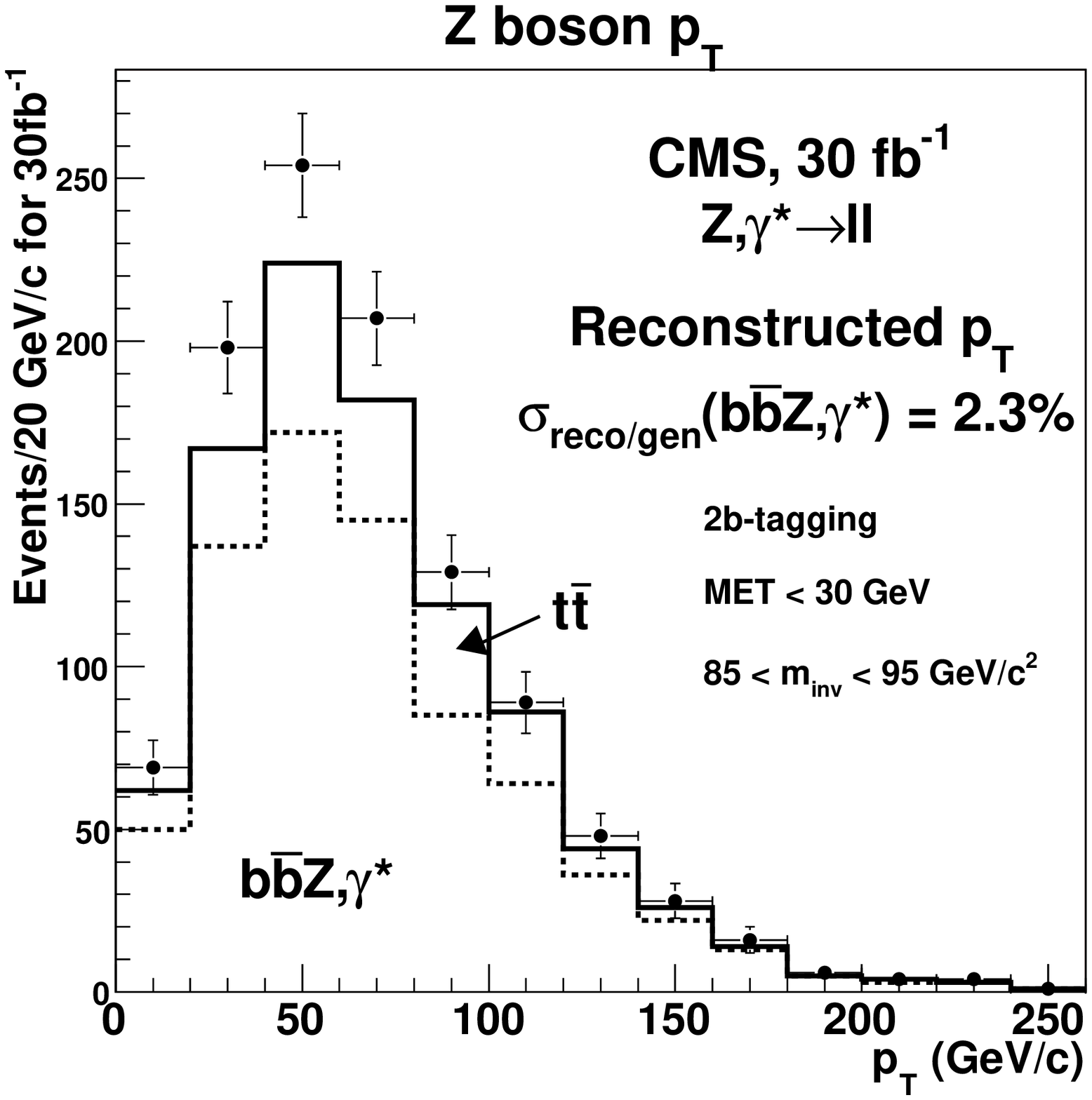}
      \caption{Z boson p$_{\rm T}$ reconstructed from the two leptons
(points) and the
fraction of $\rm b\bar{\rm b}Z/\gamma^*$ (dashed histogram) and
$\rm t\bar{\rm t}$ events (solid histogram).
%The resolution $\rm p_{\rm T}^{reco}/p_{\rm T}^{gen}$ is 2.3 \%.
}
  \label{fig:ZPt}
  \end{minipage}
  \end{tabular}
\end{figure}

Due to a large cross section and a small width, the invariant Z mass peak can be
reconstructed and measured from data with high statistics. 
%The inclusive $\rm Z/\gamma^*$ cross section can therefore be measured from data.
Using the signal selection cuts to extract the $\rm b\bar{\rm b}Z/\gamma^*$
events from the background, the fraction of the $\rm b\bar{\rm b}Z/\gamma^*$
cross section from the inclusive $\rm Z/\gamma^*$ cross section can be measured.
%within the accuracy of the signal selection cuts. 
The number of 
$\rm t\bar{\rm t}$ (+tW) events
can be estimated from the fit shown in Fig.~\ref{fig:invmass}. 
%The statistical error is of the order of 1\%. 
The fraction of $\rm Z/\gamma^*$ events with 
associated light quark and gluon jets can be estimated using the known
b tagging efficiency and mis-tagging rate, 
which can be estimated with
good statistics from the measured $\rm t\bar{\rm t}$ events~\cite{pTDR1}.

The b tagged jet E$_{\rm T}$ and $\eta$ distributions are shown in 
Figs.~\ref{fig:bjet_et} and \ref{fig:bjet_eta}. The measured distributions consist
of 70\% signal events (dashed histograms). 
The contribution from $\rm Z,\gamma^*$ events with soft initial and final
state radiation jets, misidentified as b jets, comes in mostly at low values of jet 
E$_{\rm T}$. This is shown as a gap between the points and the solid histogram
in the figures. Due to the hard b tagging cut, used the purity of the measured b 
jets in $\rm b\bar{\rm b}Z/\gamma^*$ events is very high, close to one, and 
dropping to 0.96 at low values of jet $\rm E_{\rm T}$.

The Z boson p$_{\rm T}$ reconstructed from the two leptons is shown in 
Fig.~\ref{fig:ZPt}. Again the shapes of the background distributions are needed,
and they can be measured from data. The contribution of other backgrounds
is taken into account as described above.
Since the Z boson p$_{\rm T}$ is reconstructed from the two well measured
leptons, the difference between reconstructed and generated Z boson p$_{\rm T}$
is small, about 2.3 \%. 
%Here the signal final
%state consist of 83\% of muons, and only 17\% of electrons, so the accuracy of 
%the measurement is dominated by the muon momentum measurement.

\subsubsection{Systematic uncertainties}

The uncertainty of the signal selection efficiency is related to 
the uncertainty of the lepton identification,
the absolute 
calorimeter scale and the b tagging efficiency. An error in the calorimeter scale 
introduces an error in the jet energy measurement. Here a 1\% error on calorimeter
scale leads to a 3.4\% error on the signal selection efficiency.
The uncertainty of the b tagging efficiency can be estimated from 
$\rm t\bar{\rm t}$ events as in Ref.~\cite{pTDR1}. A value of 5\% can be used
as a conservative estimate.
A lepton identification uncertainty of 2\% is used for both, electrons and muons.

The uncertainty of the $\rm t\bar{\rm t}$ background can be evaluated from the 
signal+background fit shown in Fig.~\ref{fig:invmass}. The error of the fit 
gives the uncertainty. 
The number of $\rm t\bar{\rm t}$ events for 30 fb$^{-1}$ integrated from the 
fit is 8922 and the error $\rm\Delta N_{\rm t\bar{\rm t}}$ = 147.3, which corresponds to a
$\rm t\bar{\rm t}$ background uncertainty 
$\rm\Delta N_{\rm t\bar{\rm t}}/N_{\rm t\bar{\rm t}}$ = 1.7\%.
%The number of $\rm b\bar{\rm b}Z/\gamma^*$ and $\rm jjZ/\gamma^*$ events 
%evaluated
%using the known b tagging efficiency is 1637.1 and 168.9 events, respectively.
The uncertainty of the number of $\rm jjZ/\gamma^*$ events 
N$_{\rm jjZ/\gamma^*}$
= $\rm\epsilon_{\rm mistag}(jet_1)\times\epsilon_{\rm mistag}(jet_2)\times\rm N_{\rm jjZ/\gamma^*}^{\rm no btag}$ is 
$\rm \Delta N_{\rm jjZ/\gamma^*}/N_{\rm jjZ/\gamma^*}$ 
= $\sigma_{\rm fit} \oplus 2\sigma_{\rm mistag}$,
where $\sigma_{\rm fit}$ is the uncertainty of the Z peak fit when no b 
tagging is used.
%$\sigma_{\rm fit}\times\rm N_{\rm jjZ/\gamma^*}^{\rm no btag}$ = 
%$\rm \Delta N_{\rm jjZ/\gamma^*}^{\rm no btag}$.
Assuming a 5\% mistagging uncertainty
and a 1.7\% error from the Z peak fit without b tagging, the uncertainty of 
the $\rm jjZ/\gamma^*$ background is
$\rm \Delta N_{\rm jjZ/\gamma^*}/N_{\rm jjZ/\gamma^*}$= 10.1\%.
% corresponding
%to $\rm \Delta N_{\rm jjZ/\gamma^*}=$ 17.2 events.

The total systematic uncertainty of the above measurements, including the
luminosity uncertainty of 5\%, yields a 14.2\% uncertainty on the cross section 
measurement.

\subsection{Conclusions}
It is shown that the $\rm b\bar{\rm b}Z/\gamma^*$ events can be used as a
benchmark for the MSSM Higgs production $\rm gg/qq\rightarrow b\bar{\rm b}H$.
The $\rm b\bar{\rm b}Z/\gamma^*$ cross section can be measured and compared
with the highest order theoretical calculation available. The associated
jet E$_{\rm T}$ and $\eta$ distributions as well as Z p$_{\rm T}$ distribution
can be measured and compared with the expected theoretical distributions.
Understanding the $\rm b\bar{\rm b}Z/\gamma^*$ events helps us to understand 
and better trust
the theoretical predictions for $\rm b\bar{\rm b}H$ events, 
if a heavy neutral MSSM Higgs boson is found in the 
$\rm H\rightarrow\tau\tau$ decay channel.

\subsection*{Acknowledgements}
The author would like to thank the ARDA/ASAP team for a magnificent grid tool
they have created, and A. Nikitenko and R. Kinnunen for reading the manuscript.

%%%%%%%%%%%%%%%%%%%%%%%%%%%%%%%%%%%%%%%%%%%%%%%%%%%%%%%%%%%%%%%%%%%%%%%%%%%%%
\section[Data-driven background determination in the channel
$H\rightarrow WW\rightarrow l\nu l\nu$ with no hard jets]
{DATA-DRIVEN BACKGROUND DETERMINATION IN THE CHANNEL
$H\rightarrow WW\rightarrow l\nu l\nu$ WITH NO HARD JETS~\protect
\footnote{Contributed by: B.~Mellado, W.~Quayle, S.L.~Wu}}
\subsection{Introduction}
The search for the Higgs boson called for by the Standard Model is arguably one of the most 
important topics in high-energy particle physics today. %~\cite{Englert:1964et,Higgs:1964ia,Higgs:1964pj,Higgs:1966ev,Guralnik:1964eu,Kibble:1967sv}
For a very broad range of masses the dominant decay mode of the Standard Model Higgs 
boson is the decay $H\rightarrow WW$~\cite{Dittmar:1996ss}.  
In this work we study the theoretical uncertainties
involved in a data-driven background determination strategy.  
In Section~\ref{Monte Carlo and Analysis Method}, we describe our Monte Carlo samples, 
event selection, and method for in-situ background determination.
We then discuss the most important systematic errors in 
Sections~\ref{Theoretical Uncertainties in the WW Background} and~\ref{Theoretical Uncertainties in the Top Background}

\subsection{Monte Carlo and Analysis Method}\label{Monte Carlo and Analysis Method}
We consider the following signal and background processes:

\begin{itemize}
\item Higgs production. 
We model the gluon-initiated process with the generator provided in MC@NLO and normalize
the cross-section for the signal to the values given 
in~\cite{ATL-COM-PHYS-2004-062}.  The small contribution from Weak Boson Fusion (VBF) is modelled
with Pythia~\cite{Sjostrand:1993yb,Sjostrand:2000wi}.
\item QCD $WW$ production is modelled with the generator provided in MC@NLO version 3.1~\cite{Frixione:2002ik,Frixione:2003ei}.
A non-negligible number of $WW$ events come from $gg\rightarrow WW$ diagrams that are not included in MC@NLO;
we model this contribution using the generator documented in~\cite{Binoth:2005ua}.
\item $t\overline{t}$ production. The (dominant) doubly-resonant contribution is modelled with MC@NLO. 
To estimate the impact of the singly-resonant and non-resonant $WWbb$ contributions to the background, we perform a
comparison between leading-order calculations of $pp\rightarrow WWbb$ and $pp\rightarrow t\overline{t}\rightarrow WWbb$
using MadEvent~\cite{Maltoni:2002qb,Stelzer:1994ta}.
\item QCD $Z/\gamma$ production, with $Z\rightarrow ee/\mu\mu/\tau\tau$.  We model this background with MC@NLO.
\end{itemize}

\noindent
Although we do not expect detector effects to be important in this study, 
it is convenient to simulate a detector using the last FORTRAN-based release of ATLFAST, and we apply the jet energy corrections
in ATLFAST-B~\cite{ATLFAST}\footnote{We also apply a small correction to the energy of jets for which HERWIG was used
for the parton showering and hadronization; the correction is given by $(1-5\times10^{-5} P_{T}^{jet}+0.042)$ where the
jet $P_{T}$ is measured in GeV.}. 
\begin{table}[t]
\begin{center} 
\caption{Cut flows (in fb) for $M_{H}=160$ GeV in the $e\mu$ channel.}
  \vspace*{1mm}
\begin{tabular}{ c | c c | c c c c c}
\hline
\hline
Cut             & $gg\rightarrow H$     & VBF   & $t\overline{t}$       & EW WW         & $gg\rightarrow WW$	& $qq\rightarrow WW$        & $Z/\gamma^{*}$\\
\hline
\hline
Trigger and $Z$ rej. & 185    & 25.1   & 7586   & 11.4 & 48.5   & 792    & 151   \\
Hard Jet Veto        & 90.0   & 1.48   & 51.6   & 0.16 & 21.2   & 451    & 31.4  \\
B Veto          & 89.6   & 1.46   & 37.6   & 0.16 & 21.1   & 449    & 30.8  \\
$P_{T}^{Higgs}$ & 53.2   & 1.23   & 33.0   & 0.09 & 13.1   & 177    & 23.6  \\
$M_{ll}$        & 42.9   & 1.10   & 7.85   & 0.02 & 6.31   & 65.2   & 22.0  \\
$\Delta\phi_{ll}$ & 33.1 & 0.93   & 5.23   & 0.02 & 5.14   & 42.8   & 0.07  \\
$M_{T}$         & 31.2   & 0.86   & 3.64   & 0.01 & 3.61   & 36.8   & 0.06  \\
\hline
\hline
\end{tabular}
\label{cutflow}
\end{center}
\end{table}
\begin{table}[tbp]
\begin{center}
\caption{Cross-sections (in fb) in the two control samples discussed in Section~\ref{Monte Carlo and Analysis Method}
for $M_{H}=160$ GeV for all lepton flavors.}
  \vspace*{1mm}
\begin{tabular}{ c | c c | c c c c c c }
\hline
\hline
Sample  & $gg\rightarrow H$     & VBF   & $t\overline{t}$       & EW WW         & $gg\rightarrow WW$ & $qq\rightarrow WW$       & $Z/\gamma^{*}$\\
\hline
\hline
Primary          & 1.86   & 0.03        & 33.4   & 0.08 & 6.19   & 121.0  & 7.96\\
b-tagged         & 0.18 & 0.007        & 17.02  & 0.0001      & 0.08 & 1.51  & 1.29\\
\hline
\hline
\end{tabular}
\label{controlSamples}
\end{center}
\end{table}
Our event selection consists of the following cuts:

\begin{itemize}
\item Trigger and topology cuts.  We require that the event has exactly two leptons with transverse 
momentum greater than 15 GeV in the region with $|\eta|<2.5$, and we apply a lepton identification 
efficiency of 90\% for each lepton.  The dilepton invariant mass is required to be less than 300~GeV. 
\item $Z$ rejection.  The event is rejected if the leptons have an invariant mass between 82 and 98 GeV.
We require a large missing transverse
momentum $P_{T}^{miss}>30$~GeV, which is raised to 40 GeV if the two leptons have the same flavor.
(This cut is already included in the first line of the table.)
To reduce the nontrivial background from the decay $Z\rightarrow\tau\tau\rightarrow ll +P_{T}^{miss}$,
we calculate, using the collinear approximation,
$x_{\tau}^{1}$ and $x_{\tau}^{2}$, the energy fractions carried by the visible decay products of the
$\tau$ leptons, and $M_{\tau\tau}$, the invariant mass of the two $\tau$ leptons.  We reject the event
if $x_{\tau}^{1}>0$, $x_{\tau}^{2}>0$, and $|M_{\tau\tau}-M_{Z}|<25$~GeV.
\item Jet veto.  We reject the event if there are any jets with $P_{T}>30$~GeV anywhere in the detector, or
if it contains any b-tagged jets with $P_{T}>20$~GeV and $|\eta|<2.5$.  We assume a b-tagging
efficiency of 60\% with rejections of 10 and 100 against jets from $c$ quarks and light jets, respectively.
\item Transverse momentum of the Higgs candidate, defined as the vector sum of the transverse momenta of the leptons
and the missing $P_{T}$. We require that $P_{T}^{Higgs}>11.1$~GeV.
\end{itemize}

\begin{figure}[tb]
\begin{center}
\epsfig{file=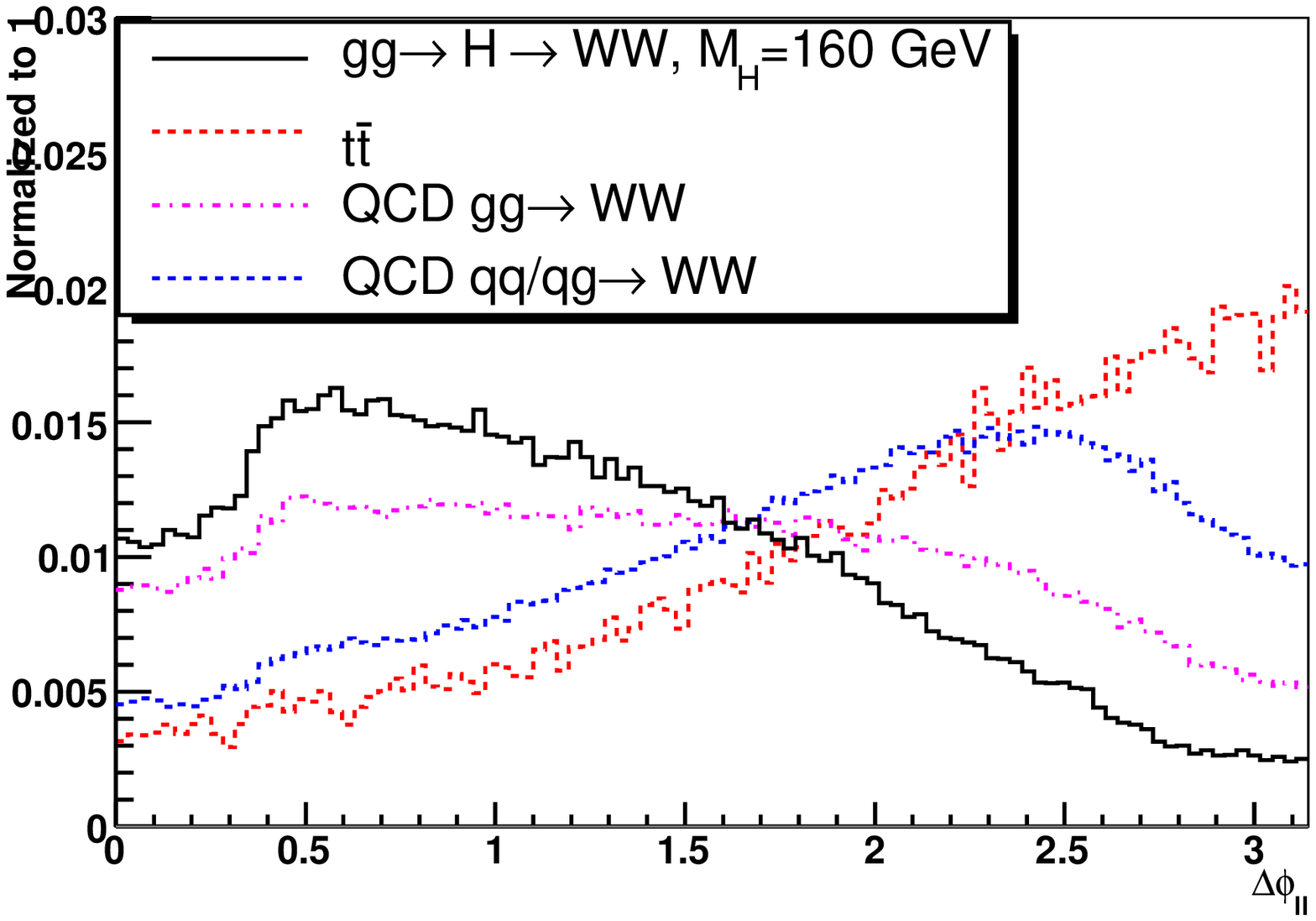, width=12cm}
\caption{The distribution of the azimuthal angle (in the transverse plane) between the leptons after cuts.}
\label{dPhill_control}
\end{center}
\end{figure}

\noindent
In the signal-like region, we apply three more cuts:  we require that the dilepton mass has $6.3<M_{ll}<64.1$~GeV, that
the azimuthal opening angle between the leptons satisfies $\Delta\phi_{ll}<1.5$ radians, and that the transverse mass 
obeys $50<M_{T}<M_{H}+10$~GeV (where $M_{H}$ is the Higgs mass hypothesis).  
The distribution of the dilepton opening angle in the transverse plane, shown in Fig.~\ref{dPhill_control},
owes its discriminating power to the difference in the helicity states of the $W$ pairs in signal and background.
For brevity, we have omitted plots of the other variables.
The cross-sections after successive cuts for a representative Higgs mass of 160~GeV in the $e\mu$ channel
are shown in Table~\ref{cutflow}.
We also consider two control samples:

\begin{itemize}
\item The primary control sample is defined the same way as the signal-like region, but with 
different cuts on the dilepton opening angle in the transverse plane and the dilepton invariant mass. 
We require $\Delta\phi_{ll}>1.5$ radians and $80<M_{ll}<300$~GeV; we remove the cut on the transverse
mass.
\item The b-tagged control sample cuts are the same as in the primary 
control sample, except that instead of applying a b-jet veto, we require that there is a b-tagged
jet with $P_{T}$ between 20 GeV and 30~GeV; we also remove the lower bound on 
the dilepton invariant mass.
\end{itemize}

\noindent
Table~\ref{controlSamples} shows the cross-sections in these two control samples.
In order to make meaningful estimates of systematic errors, it is helpful to 
define the following three quantities:

\begin{itemize}
\item $\alpha_{WW}$:  The ratio of the QCD $WW$ cross-section in the signal-like region over the QCD $WW$ cross-section
in the primary  control sample.
\item $\alpha_{tt}$:  The ratio of the $t\overline{t}$ cross-section in the signal-like region over the $t\overline{t}$
cross-section in the b-tagged control sample.
\item $\alpha_{tt}^{WW}$:  The ratio of the $t\overline{t}$ cross-section in the primary control sample over the
$t\overline{t}$ cross-section in the b-tagged control sample.
\end{itemize}

\noindent
With these ratios taken from Monte Carlo, we estimate the number of $t\overline{t}$ events in the
signal-like region as $N_{tt}^{signal-like}=\alpha_{tt} N_{b-tagged}$ and the number 
of $WW$ background events in the signal-like region as 
\begin{eqnarray}
N_{WW}^{signal-like}=\alpha_{WW} N_{WW}^{primary}=\alpha_{WW} (N_{total}^{primary}-\alpha_{tt}^{WW} N_{b-tagged}-small~backgrounds)\nonumber
\end{eqnarray}
where $N_{tt}^{b-tagged}$ is the number of events in the b-tagged control sample,
$N_{total}^{primary}$ is the total number of events in the primary control sample, 
and the ``$small~backgrounds$'' consist mostly of Drell-Yan events.
Our task is now to estimate the uncertainty in $\alpha_{WW}$, $\alpha_{tt}$, and $\alpha_{tt}^{WW}$.

\subsection{Theoretical Uncertainties in the WW Background}\label{Theoretical Uncertainties in the WW Background}
We begin with the theoretical uncertainties in the extrapolation coefficient $\alpha_{WW}$.
Here, the theoretical error is dominated by the uncertainty in the normalization of
the $gg\rightarrow WW$ contribution;  recent studies
have shown that this contribution can be in
excess of 30\% for the cuts used in those studies~\cite{Binoth:2005ua,Duhrssen:2005bz}.

We compute the the theoretical error as
the sum in quadrature of the uncertainty due to the fit error in the parton density function parameterization
and the uncertainty due to the choice of $Q^{2}$ scale.  To estimate the parton density function (PDF) uncertainty, 
we have used the CTEQ6 PDF set and its error sets; using equation (3) in~\cite{Pumplin:2002vw}, we find that the uncertainty in $\alpha_{WW}$
is 2.8\%.
To assess the uncertainty due to the choice of $Q^{2}$ scale, we have varied the renormalization
and factorization scales by factors of 8.\footnote{This is an unusually large scale variation to choose;
typically, a scale uncertainty will be quoted based on a scale variation of 2 or at most 4.  Our motivation
for this choice is the fact that we expect the K-factor for $gg\rightarrow WW$ to be large, since the
K-factor for $gg\rightarrow\gamma\gamma$ has been calculated and it is slightly less than 2~\cite{Bern:2002jx}.}
We examine four choices of scale variations:
Scale 1 has $Q_{ren}\rightarrow 8 Q_{ren}$, $Q_{fac}\rightarrow Q_{fac}/8$;
Scale 2 has $Q_{ren}\rightarrow Q_{ren}/8$, $Q_{fac}\rightarrow 8 Q_{fac}$;
Scale 3 has $Q_{ren}\rightarrow 8 Q_{ren}$, $Q_{fac}\rightarrow 8 Q_{fac}$; and
Scale 4 has $Q_{ren}\rightarrow Q_{ren}/8$, $Q_{fac}\rightarrow Q_{fac}/8$.
Table~\ref{scale_uncert_24Sep2005} shows the cross-sections before and after cuts in the signal-like region and primary 
control sample for the $gg\rightarrow WW$ and $qq\rightarrow WW$
contributions, with the central-value $Q^{2}$ scales and the four modified scale choices.
The largest variation in $\alpha_{WW}$ we observe is 4.1\%, and we take this to be the theoretical error 
due to the choice of $Q^{2}$ scale.
The total theoretical uncertainty we calculate on the prediction of $\alpha_{WW}$ is therefore 5\%. 
\begin{table}[tbp]
\begin{center}
\caption{Cross-sections before and after cuts 
for the signal-like region and the primary control sample, with the
corresponding extrapolation coefficients, using the nominal assumptions and
the 4 altered scale choices. For historical reasons, the upper bound on the
dilepton invariant mass ($M_{ll}<300$~GeV) is not applied to the control
sample in the values reported in this table.}
  \vspace*{1mm}
\begin{tabular}{ c | c c | c c | c c | c}
\hline
\hline
              & No cuts & & Sig. Reg.  & & Cont. Samp. & & \\
\hline
Scale Choice  & $gg\rightarrow WW$    & $qq\rightarrow WW$ &  $gg$    & $qq$    & $gg$            & $qq$            & $\alpha_{WW}$\\
\hline
Central       & 487.77 & 11302.44 & 6.45                  & 63.20                 & 6.38               & 130.10              & 0.5103\\
scale1        & 239.93 & 12862.82 & 2.92                  & 69.25                 & 3.33               & 143.83              & 0.4904\\
scale2        & 1058.97 & 9076.86 & 14.5                  & 49.03                 & 13.46              & 107.44              & 0.5255\\
scale3        & 278.17 & 11189.52 & 3.81                  & 65.02                 & 3.54               & 131.92              & 0.5081\\
scale4        & 913.38 & 11702.80 & 11.1                  & 61.81                 & 12.66              & 133.51              & 0.4988\\
\hline
\hline
\end{tabular}
\label{scale_uncert_24Sep2005}
\end{center}
\end{table}

\subsection{Theoretical Uncertainties in the Top Background}\label{Theoretical Uncertainties in the Top Background}
We now turn our attention to the uncertainties in 
$\alpha_{tt}$ and $\alpha_{tt}^{WW}$.
Here, the most important question to ask is how to handle single top production.
A procedure for generating both $pp\rightarrow t\overline{t}$ and $pp\rightarrow Wt$
without double-counting at leading order was presented in~\cite{Belyaev:2000me}, and a
calculation including off-shell effects and spin correlations in the $WWbb$ system at
tree level was presented in~\cite{Kauer:2001sp}.
Unfortunately, we know of no event generator available at the time
of this writing which also takes into account the one-loop radiative
corrections to $WWbb$ production, so we will perform our uncertainty estimate at tree-level.

In addition to the $t\overline{t}$ Monte Carlo sample (from MC@NLO) that we have used in the other
sections of this note, we have generated two separate $WWbb$ Monte
Carlo samples using MadGraph.  One includes only doubly-resonant top quark pair production,
and the other includes the full $WWbb$ final state.
For these events, we have allowed the b-quarks to be generated with $P_{T}$ as low
as 1 GeV, and with pseudorapidity as high as 100.  
One would expect a disproportionately large contribution from the region where one b-quark is soft or forward, 
and we therefore feel it is likely that
the single-top contribution is overestimated in our non-resonant $WWbb$ Monte Carlo.
This is exactly what we want if we are to prove that our analysis is robust. 
We have applied the cuts for the signal-like region and both of the control regions
to these two Monte Carlo samples to assess the importance of single-top production in this analysis.  
\begin{table}[tbp]
\begin{center}  
\caption{Cross-sections (in fb) and extrapolation coefficients for the $t\bar{t}$ background for various masses, using MadGraph to model the WWbb background.}
  \vspace*{1mm}
\begin{tabular}{ c | c c c | c c }
\hline                                                                                                                                                    
\hline
Process               & Signal-like   & Cont. Samp.  & b-tagged     & $\alpha_{tt}$ & $\alpha_{tt}^{WW}$\\                            
\hline
$WWbb$                & 13.34         & 109.41        & 47.13         & 0.2829        & 2.3211\\
$tt\rightarrow WWbb$  & 9.80          & 80.77         & 37.72         & 0.2599        & 2.1413\\                                       
\hline                                                                                                                                                    
\hline
\end{tabular}
\label{singletop_uncert_30Aug2005}
\end{center}
\end{table}

Table~\ref{singletop_uncert_30Aug2005} shows the $WWbb$ background cross-sections in the signal-like 
region, the primary control sample, and the b-tagged control sample obtained with the leading-order doubly-resonant $t\overline{t}$ and
inclusive $WWbb$ samples. We note that although the difference in the absolute cross-section given by the 
two samples is approximately 30\%, the corresponding differences in the predictions
of $\alpha_{tt}$ and $\alpha_{tt}^{WW}$ are only about 9\%.  
It is worth noting that this figure is only a general guideline, since the exact values of $\alpha_{tt}$ and $\alpha_{tt}^{WW}$
are strongly dependent on the particulars of the b-tagging algorithm used.  Our intent in this section is only to give a rough 
idea of what the theoretical uncertainty on the extrapolation from a b-tagged sample to a b-vetoed sample should be.  
In practice, this uncertainty should be addressed in detail using full detector simulation
by any experimenter performing a $H\rightarrow WW$ search like the one outlined here.

\subsection{Summary}\label{Summary}
We have proposed a method to estimate the normalization of the dominant backgrounds in the $H\rightarrow WW\rightarrow l\nu l\nu$
channel using two control samples in the data, one b-tagged, and the other b-vetoed; in our approach, the systematic errors
must be given in terms of the ratios $\alpha_{WW}$, $\alpha_{tt}$, and $\alpha_{tt}^{WW}$.
We have computed the theoretical uncertainty on $\alpha_{WW}$; the result is 5\%.
We have shown that, for a b-tagging algorithm operating only on jets with $P_{T}>20$~GeV and $|\eta|<2.5$,
such that $\epsilon_{b}=60$\% and the rejections against light quarks and c-quarks are 100 and 10 
respectively, the effect of singly-resonant and non-resonant $WWbb$ diagrams is less than 10\% on $\alpha_{tt}$ and $\alpha_{tt}^{WW}$.
A study using these uncertainties and this background extraction technique is in progress for the ATLAS experiment; the 
preliminary result is that a Higgs discovery at $M_{H}=160$~GeV would require less than 2 fb$^{-1}$ of integrated luminosity~\cite{ATL-COM-PHYS-2005-074}.
However, final calculations of the uncertainties on these last two extrapolation coefficients, as well as final results on the overall 
sensitivity of the search we have presented here, must be computed within the context of the LHC experiments.

\subsection*{Acknowledgement}\label{Acknowledgement}
The authors are grateful to N. Kauer, and S. Frixione.  This work was supported 
in part by the United States Department of Energy through Grant No.  DE-FG0295-ER40896.

%%%%%%%%%%%%%%%%%%%%%%%%%%%%%%%%%%%%%%%%%%%%%%%%%%%%%%%%%%%%%%%%%%%%%%%%%%%%%
\section[Electroweak corrections to the Higgs decays
   $H\to ZZ/WW\to 4$ leptons]
{ELECTROWEAK CORRECTIONS TO THE HIGGS DECAYS $H\to ZZ/WW\to 4$
LEPTONS~\protect\footnote{Contributed by: A.~Bredenstein, A.~Denner,
S.~Dittmaier, M.M.~Weber}}
\subsection{Introduction}

The primary task of the LHC will be the detection and the
investigation of the Higgs boson. If it is heavier than
$140\,$GeV, it decays dominantly into gauge-boson pairs,
i.e.\ into 4 fermions. These decays offer the largest discovery
potential for a Higgs boson with a mass
$\mathrel{\raisebox{-.3em}{$\stackrel{\displaystyle
      >}{\sim}$}}130\,$GeV \cite{Asai:2004ws,Abdullin:2005yn}, and the decay
$\rm{H}\to\rm{Z}\rm{Z}\to4\ell$ will allow for the most accurate
measurement of the Higgs-boson mass above $130\,$GeV
\cite{Zivkovic:2004sv}.  At an $\rm{e}^+\rm{e}^-$ linear collider,
these decays will enable precision measurements of the corresponding
branching ratios and couplings at the per-cent level.

A kinematical reconstruction of the Higgs boson and of the virtual W
and Z bosons requires the study of distributions.
Thereby, it is important to include radiative corrections, in
particular real photon radiation. In addition, the verification of the
spin and the CP properties of the Higgs boson relies on the study of
angular and invariant-mass distributions
\cite{Barger:1993wt,Choi:2002jk}. As a consequence a Monte Carlo
generator for $\rm{H}\to\rm{Z}\rm{Z}/\rm{W}\rm{W}\to4\rm{f}$ including
electroweak corrections is needed.

In the literature the electroweak ${\cal O}(\alpha)$ corrections are
only known for decays into on-shell gauge bosons
$\rm{H}\to\rm{Z}\rm{Z}/\rm{W}\rm{W}$
\cite{Fleischer:1980ub,Kniehl:1990mq,Kniehl:1991xe,Bardin:1991dp}.  In
this case, also some leading higher-order corrections have been
calculated.  However, below the gauge-boson-pair thresholds only the
leading order is known, and in the threshold region the on-shell
approximation becomes unreliable. Only recently electroweak
corrections to $\rm{H}\to\rm{Z}\rm{Z}/\rm{W}\rm{W}\to4\rm{f}$ have
been considered. Progress on a calculation of the electromagnetic
corrections to $\rm{H}\to\rm{Z}\rm{Z}\to4\rm{f}$ has been reported at
the RADCOR05 conference by Carloni Calame \cite{carlonicalame_talk}.
At this conference we have also presented first results of our
calculation of the complete ${\cal O}(\alpha)$ corrections to the
general $\rm{H}\to4\rm{f}$ processes \cite{axel_talk}. 
In this note we
sketch the calculation and provide some numerical
results. 
More results and details of the calculation can be found in
Ref.~\cite{Bredenstein_paper}. The electroweak corrections
have been implemented into a Monte Carlo generator called {\sl
PROPHECY4f}.

\subsection{Calculational details}

We have calculated the complete ${\cal O}(\alpha)$ virtual and real
photonic corrections to the processes $\rm{H}\to4\rm{f}$. This
includes both the corrections to the decays
$\rm{H}\to\rm{Z}\rm{Z}\to4\rm{f}$ and
$\rm{H}\to\rm{W}\rm{W}\to4\rm{f}$ and their interference.
The calculation of the one-loop diagrams has been performed in the
conventional 't~Hooft--Feynman gauge and  in the 
background-field formalism using the conventions of
Refs.~\cite{Denner:1993kt} and \cite{Denner:1994xt}, respectively.
The masses of the external fermions have been neglected whenever
possible, i.e.\ everywhere but in the mass-singular logarithms.

For the implementation of the finite width of the gauge bosons we use
the ``complex-mass scheme'', which was introduced in
Ref.~\cite{Denner:1999gp} for lowest-order calculations and
generalized to the one-loop level in Ref.~\cite{Denner:2005fg}.  In
this approach the W- and Z-boson masses are consistently considered as
complex quantities, defined as the locations of the propagator poles 
in the complex plane.
To this end, bare real masses are split into complex
renormalized masses and complex counterterms. Since the bare
Lagrangian is not changed, double counting does not occur.
Perturbative calculations can be performed as usual, only parameters
and counterterms, in particular the electroweak mixing angle defined
from the ratio of the W- and Z-boson masses, become complex. Since we
only perform an analytic continuation of the parameters, all relations
that follow from gauge invariance, such as Ward identities, remain
valid. As a consequence the amplitudes are 
gauge independent, and unitarity cancellations are respected.
Moreover, the on-shell renormalization scheme can straightforwardly
be transferred to the complex-mass scheme \cite{Denner:2005fg}.

The amplitudes have been generated with {\sl FeynArts}, using the two
independent versions 1 and 3, as described in
Refs.~\cite{Kublbeck:1990xc} and \cite{Hahn:2000kx}, respectively.
The algebraic evaluation has been performed in two completely
independent ways. One calculation is based on an in-house program
implemented in {\sl Mathematica}, the other has been completed with the
help of {\sl FormCalc} \cite{Hahn:1998yk}.
The amplitudes are expressed in terms of standard matrix elements and
coefficients of tensor integrals \cite{Denner:1993kt}. The reduction
to standard matrix elements is performed as described in the appendix
of Ref.~\cite{Denner:2003iy}.

The tensor coefficients are evaluated as in the calculation of the
corrections to ${\rm e}^+{\rm e}^-\to4\,$fermions \cite{Denner:2005fg}. 
They are recursively reduced to master
integrals at the numerical level.  
The scalar master integrals are evaluated for complex
masses using the methods and results of
Refs.~\cite{'tHooft:1978xw,Beenakker:1988jr,Denner:1991qq}.  UV
divergences are regulated dimensionally and IR divergences with an
infinitesimal photon mass.
Tensor and scalar 5-point functions
are directly expressed in terms of 4-point integrals
\cite{Denner:2002ii}. Tensor 4-point and 3-point integrals are
reduced to scalar integrals with the Passarino--Veltman
algorithm \cite{Passarino:1978jh} as long as no small Gram
determinant appears in the reduction. If small Gram determinants
occur, two alternative schemes are applied \cite{Denner:2005nn}.  
One method makes use of expansions of the tensor coefficients about
the limit of vanishing Gram determinants and possibly other
kinematical determinants. In this way, again all tensor coefficients
can be expressed in terms of the standard scalar
functions. In
the second, alternative method we evaluate a specific tensor
coefficient, the integrand of which is logarithmic in Feynman
parametrization, by numerical integration. Then the remaining
coefficients as well as the standard scalar integral 
are algebraically derived from this coefficient.
The results of the two different codes, based on the different methods
described above are in good numerical agreement. 

Since corrections due to the self-interaction of the Higgs boson
become important for large Higgs masses, we have included the dominant
two-loop corrections to the decay ${\rm H}\to VV$ proportional to
$G_\mu^2 M_{\rm H}^4$ in the large-Higgs-mass limit which were calculated in
Refs.~\cite{Ghinculov:1995bz,Frink:1996sv}.

The matrix elements for the real photonic corrections are evaluated
using the Weyl--van der Waerden spinor technique as formulated in
Ref.~\cite{Dittmaier:1998nn} and have been successfully checked
against the result obtained with {\sl Madgraph}
\cite{Stelzer:1994ta}.  The soft and collinear singularities are
treated both in the dipole subtraction method following
Refs.~\cite{Dittmaier:1999mb,Roth:1999kk} and in the phase-space
slicing method following closely Refs.~\cite{Bohm:1993qx,Dittmaier:1993da,Baur:1998kt}.
For the calculation of non-collinear-safe observables we use the
extension of the subtraction method introduced in Ref.~\cite{Bredenstein:2005zk}.
Final-state radiation beyond ${\cal O}(\alpha)$ is
included at the leading-logarithmic level using the structure
functions given in Ref.~\cite{Beenakker:1996kt} (see also references
therein).

The phase-space integration is performed with Monte Carlo techniques.
One code employs a multi-channel Monte Carlo generator similar to the
one implemented in {\sl RacoonWW} \cite{Denner:1999gp,Roth:1999kk} and
{\sl Lusifer} \cite{Dittmaier:2002ap}, the second one uses the
adaptive multi-dimensional integration program {\sl VEGAS}
\cite{Lepage:1977sw}.

\subsection{Numerical results}

We use the $G_\mu$ scheme, i.e.\ we define the electromagnetic
coupling by $\alpha_{G_\mu}={\sqrt{2}G_\mu M_{\rm W}^2 s_{\rm
    w}^2}/{\pi}$.  Our lowest-order results include the ${\cal
  O}(\alpha)${}-corrected width of the gauge bosons. For the results
presented here, we define distributions in the rest frame of the Higgs
boson and apply no cuts. We show results without photon recombination
and results where the photon has been recombined with the nearest
charged fermion if the invariant mass of the photon--fermion pair is
below 5\,{\rm GeV}.
More details about the setup as well as all input parameters are provided 
in Ref.~\cite{Bredenstein_paper}.

\begin{figure}
\setlength{\unitlength}{1cm}
\centerline{
\begin{picture}(7.7,8)
\put(-1.7,-14.5){\includegraphics{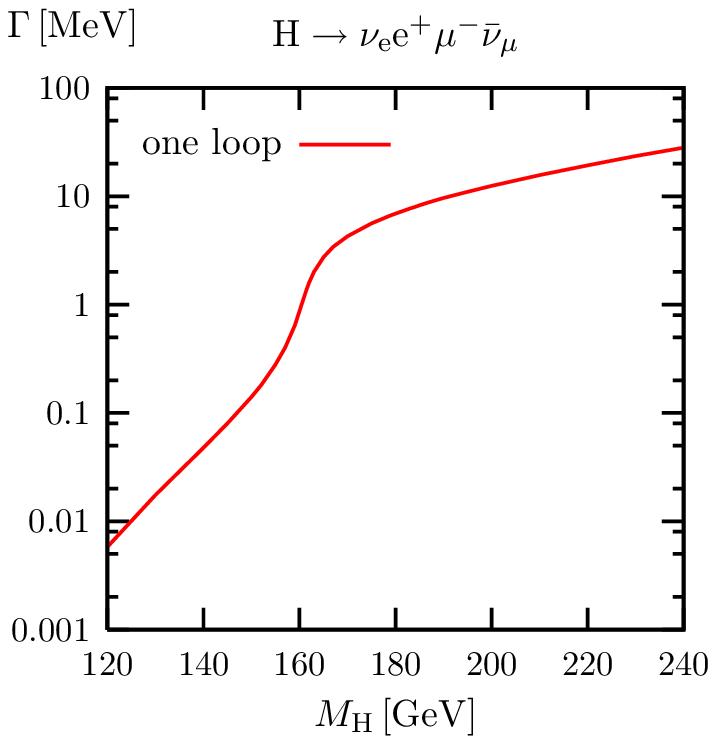}}
\end{picture}
\begin{picture}(7.5,8)
\put(-1.7,-14.5){\includegraphics{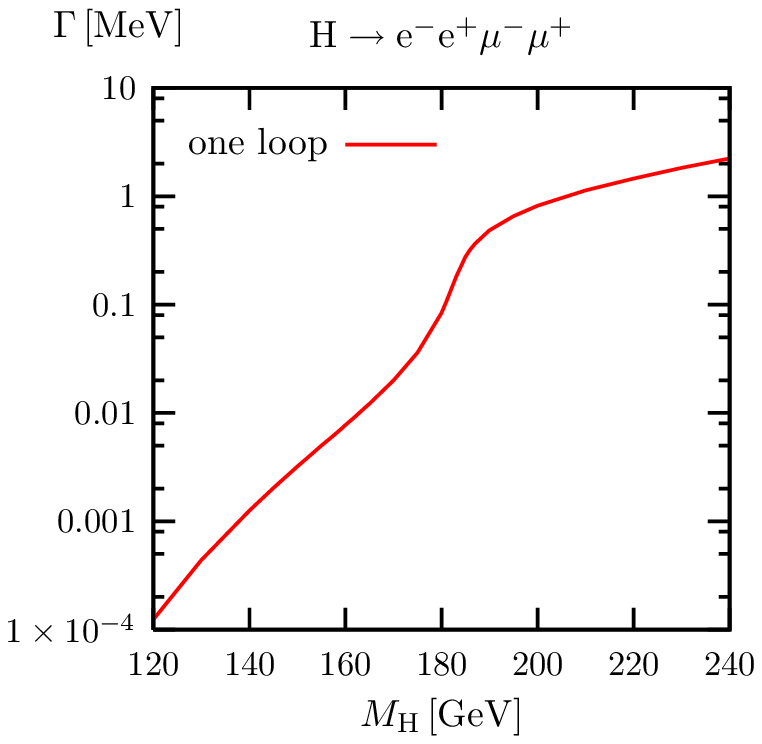}}
\end{picture} }
\centerline{
\begin{picture}(7.7,8)
\put(-1.7,-14.5){\includegraphics{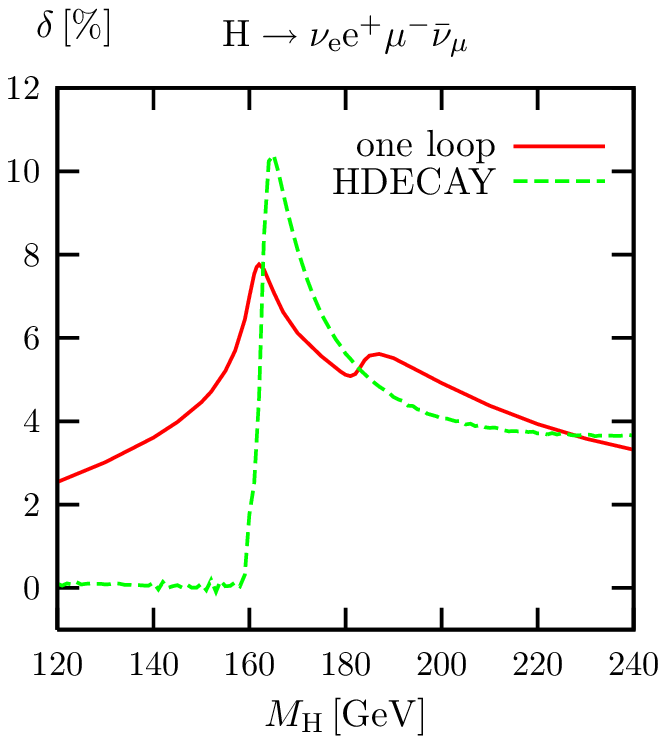}}
\end{picture}
\begin{picture}(7.5,8)
\put(-1.7,-14.5){\includegraphics{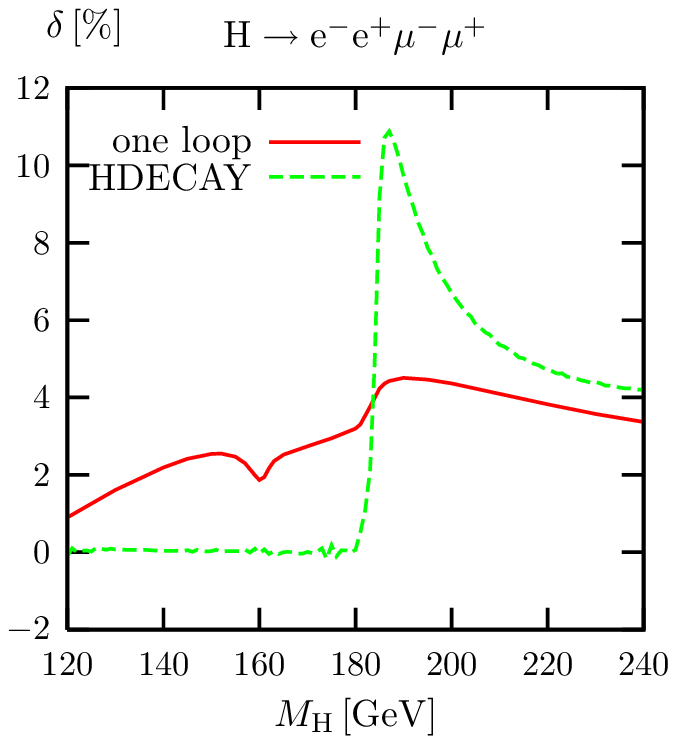}}
\end{picture} }
\caption{Partial decay widths for ${\rm H}\to \nu_{\rm e}{\rm
    e}^+\mu^-\bar{\nu}_{\mu}$ (l.h.s.) and ${\rm H}\to{\rm e}^-{\rm
    e}^+\mu^-\mu^+$ (r.h.s.) as a function of the Higgs-boson mass.
  The upper plots show the absolute predictions including ${\cal
    O}(\alpha)${} corrections, and the lower plots show the relative
  ${\cal O}(\alpha)${} corrections and the predictions of the program
  {\rm HDECAY}{} relative to the complete lowest-order prediction.}
\label{fig:sqrts}
\end{figure}

In the two upper plots of Fig.~\ref{fig:sqrts} we show the partial
decay widths for ${\rm H} \to \nu_{\rm e}{\rm e}^+\mu^-\bar{\nu}_{\mu}$
and ${\rm H} \to {\rm e}^-{\rm e}^+\mu^-\mu^+$ as a function of the
Higgs-boson mass. The lower plots show the corrections relative to the
lowest-order results.  
For ${\rm H} \to \nu_{\rm e}{\rm e}^+\mu^-\bar{\nu}_{\mu}$, the
corrections are at the level of 2--8\% in the considered Higgs-mass region.
Around $160\,{\rm GeV}$, the corrections are
dominated by the Coulomb singularity and  at about $180\,{\rm GeV}$ the ZZ
threshold is visible.
% and at about $360\,{\rm GeV}$ the ${\rm t}\bar{\rm t}$ threshold. 
Note that corrections behave smoothly as a function of the Higgs-boson
mass across the thresholds owing to the use of the complex-mass
scheme.  For the final state ${\rm e}^-{\rm e}^+\mu^-\mu^+$, the
corrections are between 1\% and 5\%.  The effects of the W-pair and
Z-pair
%and top-pair
thresholds are clearly visible.

The lower plots of Fig.~\ref{fig:sqrts} show also a comparison with 
{\rm HDECAY}{} \cite{Djouadi:1997yw}. To this end, we have defined 
\begin{equation}
\Gamma^{\mathrm{HDECAY}} = \Gamma_{{\rm H}{V}{V}}^{\mathrm{HDECAY}}
\frac{\Gamma_{{V} f_1f_2,0}}{\Gamma_{{V},1}}
\frac{\Gamma_{{V} f_3f_4,0}}{\Gamma_{{V},1}}
\end{equation}
and have divided this by the lowest-order width for ${\rm H}\to f_1
\bar{f}_2 f_3 \bar{f}_4$.  {\rm HDECAY}{} agrees well with the
lowest-order ${\rm H}\to 4f$ width below threshold, because there
$\Gamma_{{\rm H}{V}{V}}^{\mathrm{HDECAY}}$ consistently takes into
account the off-shell effects of the gauge bosons.  Above threshold
the gauge bosons are treated as stable, and leading radiative
corrections due to the Higgs-boson self-coupling are incorporated. In
a small window between the two regions {\rm HDECAY}{} interpolates
between the two results.  The large difference between {\rm HDECAY}{}
and our lowest-order prediction above threshold is due to the
difference of the on-shell and off-shell phase space and has nothing
to do with the Coulomb singularity. In particular, for ${\rm H} \to
{\rm e}^-{\rm e}^+\mu^-\mu^+$ there is no Coulomb
singularity, but the phase-space effect with respect to {\rm HDECAY}{}
is present, and the corresponding off-shell effects amount to more
than 5\%.

\begin{figure}
\setlength{\unitlength}{1cm}
\centerline{
\begin{picture}(7.7,8)
\put(-1.7,-14.5){\includegraphics{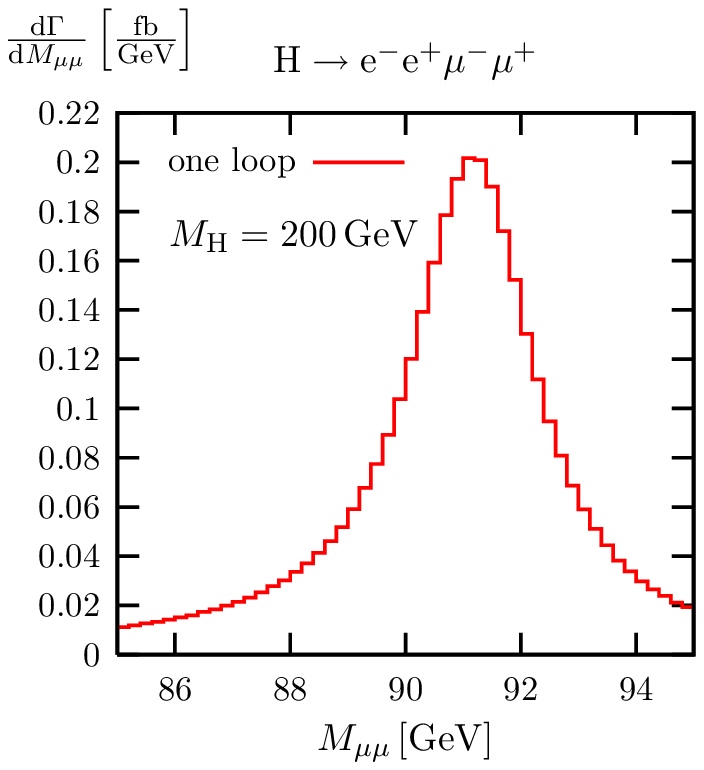}}
\end{picture}
\begin{picture}(7.5,8)
\put(-1.7,-14.5){\includegraphics{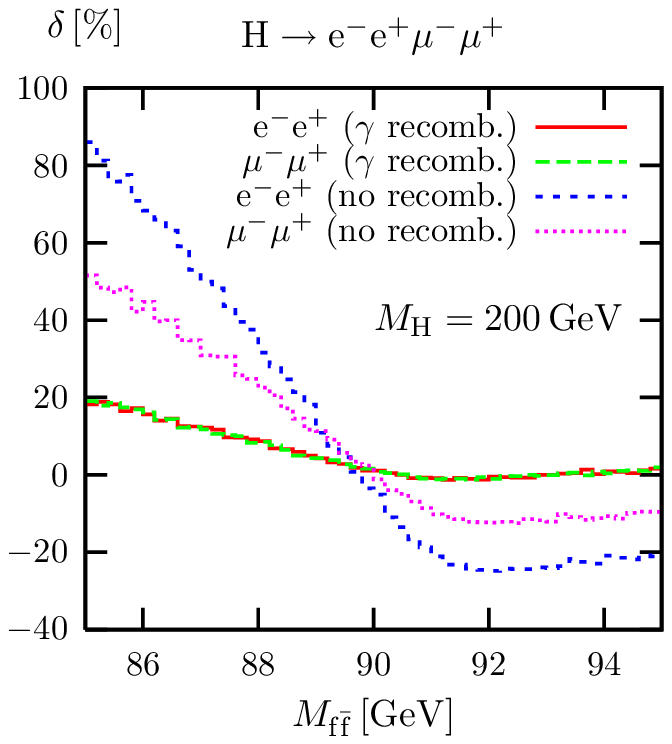}}
\end{picture} }
\caption{Invariant-mass distribution of $\mu^-\mu^+$ (l.h.s.) 
  and relative corrections for the invariant-mass distribution of ${\rm
    e}^-{\rm e}^+$ and $\mu^-\mu^+$ (r.h.s.) in the reaction ${\rm
    H}\to{\rm e}^-{\rm e}^+\mu^-\mu^+$ for 
$M_{\rm H}=200\,{\rm GeV}$.}
\label{fig:zinv}
\end{figure}

In Fig.~\ref{fig:zinv} we study the invariant-mass distribution of the
two fermions resulting from the decay of the Z bosons in the decay
${\rm H}\to{\rm e}^-{\rm e}^+\mu^-{\mu}^+$. The plot on the l.h.s.\ 
shows the distribution for $\mu^-{\mu}^+$ including ${\cal
  O}(\alpha)${} corrections.  The plot on the r.h.s.\ compares the
relative corrections for ${\rm e}^-{\rm e}^+$ and $\mu^-{\mu}^+$. If
we do not recombine photons with collinear fermions, we get very
large corrections for invariant masses below the W-boson mass
\cite{Bredenstein:2005zk,Beenakker:1998gr}.
This is because we define the invariant
mass from the fermion pair excluding the photon. The corrections
depend logarithmically on the fermion masses and are thus different
for electrons and muons. If we recombine the photons with the
fermions,
the corrections are much smaller and independent on the fermion masses.

\begin{figure}
\setlength{\unitlength}{1cm}
\centerline{
\begin{picture}(7.7,8)
\put(-1.7,-14.5){\includegraphics{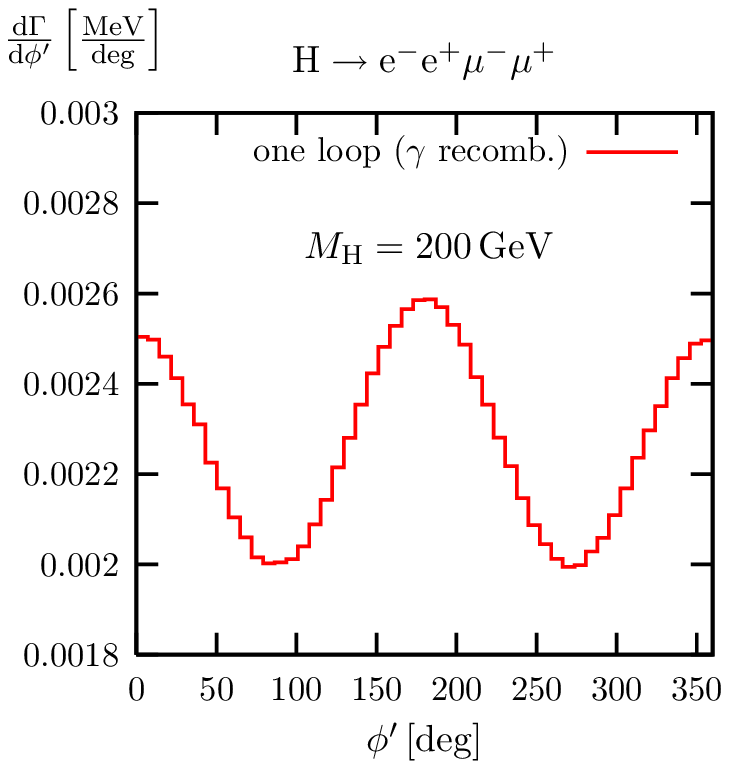}}
\end{picture}
\begin{picture}(7.5,8)
\put(-1.7,-14.5){\includegraphics{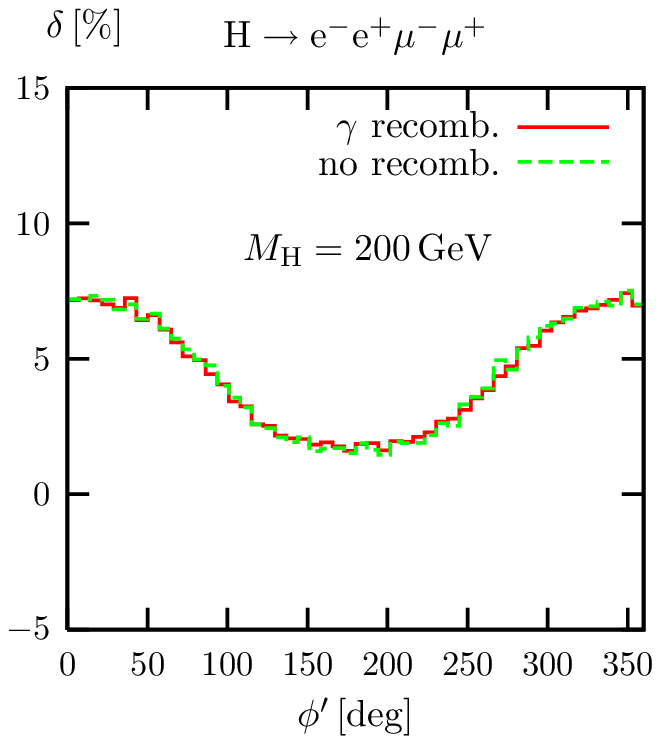}}
\end{picture} }
\caption{Distribution in the angle between the decay planes of the two
  Z bosons in the reaction ${\rm H}\to{\rm e}^-{\rm e}^+\mu^-\mu^+$
  and relative corrections with and without applying photon
  recombination for
%$M_{\rm H}=170\,{\rm GeV}$ and 
$M_{\rm H}=200\,{\rm GeV}$.}
\label{fig:phi}
\end{figure}

The investigation of angular correlations between the fermionic decay
products is an essential tool for investigating the spin and parity
properties of the Higgs boson. In Ref.~\cite{Choi:2002jk} it was
demonstrated how the parity of the Higgs boson can be determined by
looking at the angle between the decay planes of the two Z bosons in
the decay ${\rm H}\to{\rm Z}{\rm Z}$. Including the Z-boson decay,
this angle can be defined as
\begin{eqnarray}
\cos{\phi'} &=& \frac{({\bf k}_+\times{\bf k}_1)({\bf k}_+\times{\bf k}_3)}
                {|{\bf k}_+\times{\bf k}_1||{\bf k}_+\times{\bf k}_3|},
\nonumber\\
\mathop{\mathrm{sgn}}\nolimits(\sin{\phi'}) &=& \mathop{\mathrm{sgn}}\nolimits\{{\bf k}_+\cdot[({\bf k}_+\times{\bf k}_1)\times
                           ({\bf k}_+\times{\bf k}_3)]\},
\label{eq:phi}
\end{eqnarray}
where ${\bf k}_+={\bf k}_1+{\bf k}_2$.  The l.h.s.{} of
Fig.~\ref{fig:phi} shows the decay width for ${\rm H}\to{\rm e}^-{\rm
  e}^+\mu^-\mu^+$ as a function of $\phi'$ revealing a $\cos{2\phi'}$
term. As was noticed in Ref.~\cite{Choi:2002jk} this term is
characteristic for a scalar. For a pseudo scalar a term proportional
to ($-\cos{2\phi'}$) would be present instead.  Photon recombination
has no significant effect for the distribution in $\phi'$,
because adding a soft or collinear photon to a fermion momentum does
not change its direction significantly.

\subsection{Conclusions}

We have presented results from a calculation of the complete
electroweak ${\cal O}(\alpha)$ radiative corrections to the 
Higgs-boson decays into 4 leptons, ${\rm H}\to 4{\ell}$, in the
electroweak Standard Model. We find corrections to the partial widths
in the range of 1--8\%.
If predictions with an accuracy of better than 5\% are
needed, off-shell effects and radiative corrections have to be taken
into account.

%%%%%%%%%%%%%%%%%%%%%%%%%%%%%%%%%%%%%%%%%%%%%%%%%%%%%%%%%%%%%%%%%%%%%%%%%%%%%
\section[Boson boson scattering at the LHC with PHASE]
{BOSON BOSON SCATTERING AT THE LHC WITH PHASE~\protect
\footnote{Contributed by: E.~Accomando, A.~Ballestrero, A.~Belhouari,
         S.~Bolognesi, E.~Maina, C.~Mariotti}}
\subsection{Introduction}
The large energies available at the LHC
will make it possible to access many-particle final states with much higher 
statistics than before. Among these final states, six-fermion signals are of 
particular interest for  
Higgs boson discovery and for analyzing vector boson scattering. At the 
LHC, the SM Higgs production is driven by gluon-gluon fusion. The fusion of 
$\ABBBW$ and $\ABBBZ$ gauge bosons represents 
the second most important contribution to the Higgs production cross section 
\footnote{A detailed review and an extensive
bibliography can be found in 
Ref.~\cite{Djouadi:2005gi}.}.
The Higgs decay channel into $\ABBBWW$, giving rise to two forward-backward jets 
plus four leptons or two leptons and two jets from the $\ABBBW$'s, is 
particularly clean and has been found to be quite promising 
in the low-intermediate mass range (115$ < M_H < $200~GeV). 
If the Higgs boson is not present, the complementary approach to the question of 
electroweak symmetry breaking is to study vector boson scattering. In the 
absence of the Higgs boson, general arguments based on unitarity imply that massive 
gauge bosons 
become strongly interacting at the TeV scale. Processes, mediated by massive 
vector boson scattering $\ABBBVV \rightarrow \ABBBVV$ ($\ABBBV = \ABBBW$,$\ABBBZ$), are 
the most sensitive to the symmetry breaking mechanism.
By analogy with low energy QCD, or adopting one of the many schemes for turning
perturbative scattering amplitudes into amplitudes which satisfy by construction
the unitarity constraints, one is led to expect the presence of resonances in 
\ABBBWWL scattering. Unfortunately, the mass, spin and even number of these
resonances are not uniquely determined~\cite{Butterworth:2002tt}.
 
Six fermion processes are also related to the production of three 
vector bosons and give access to $t \bar{t}$ and single-top 
production, enabling measurements of top mass, $\ABBBW tb$  coupling, decay
branching ratios, rare decays and all other features related to the top quark.
Finally, we should mention that multi-particle final states of this kind 
constitute a direct background to most searches for new physics. 

Three are the key features of \ABBBPhase~\cite{Accomando:2005cc}.
The first one consists in the use of a 
modular helicity formalism for computing matrix elements. Scattering 
amplitudes get contributions from thousands of diagrams and
the computation efficiency has a primary role. The helicity 
method~\cite{Ballestrero:1994jn,Ballestrero:1999md}
we use is suited to compute in a fast and compact way parts of diagrams of
increasing size, and recombine them later in order to obtain the final set. In this
manner, parts common to various diagrams are evaluated just once for all
possible helicity configurations, optimizing the computation 
procedure. The second main feature concerns the integration. We have devised a 
new integration method to address the crucial point of reaching good stability 
and efficiency in event generation. Our integration strategy combines the 
commonly used multichannel approach~\cite{Berends:1994xn} with the adaptivity of 
{\tt VEGAS}. As the number of particles increases, the 
multichannel technique becomes rather cumbersome, given the thousands of 
resonant structures which can appear in the amplitude at the same time.
Conversely, the {\tt VEGAS} adaptivity is not powerful enough to deal with 
all possible peaks of the amplitude. 

\begin{figure}[th]
%\vspace*{-1cm}
\begin{center}
\mbox{
{\epsfig{file=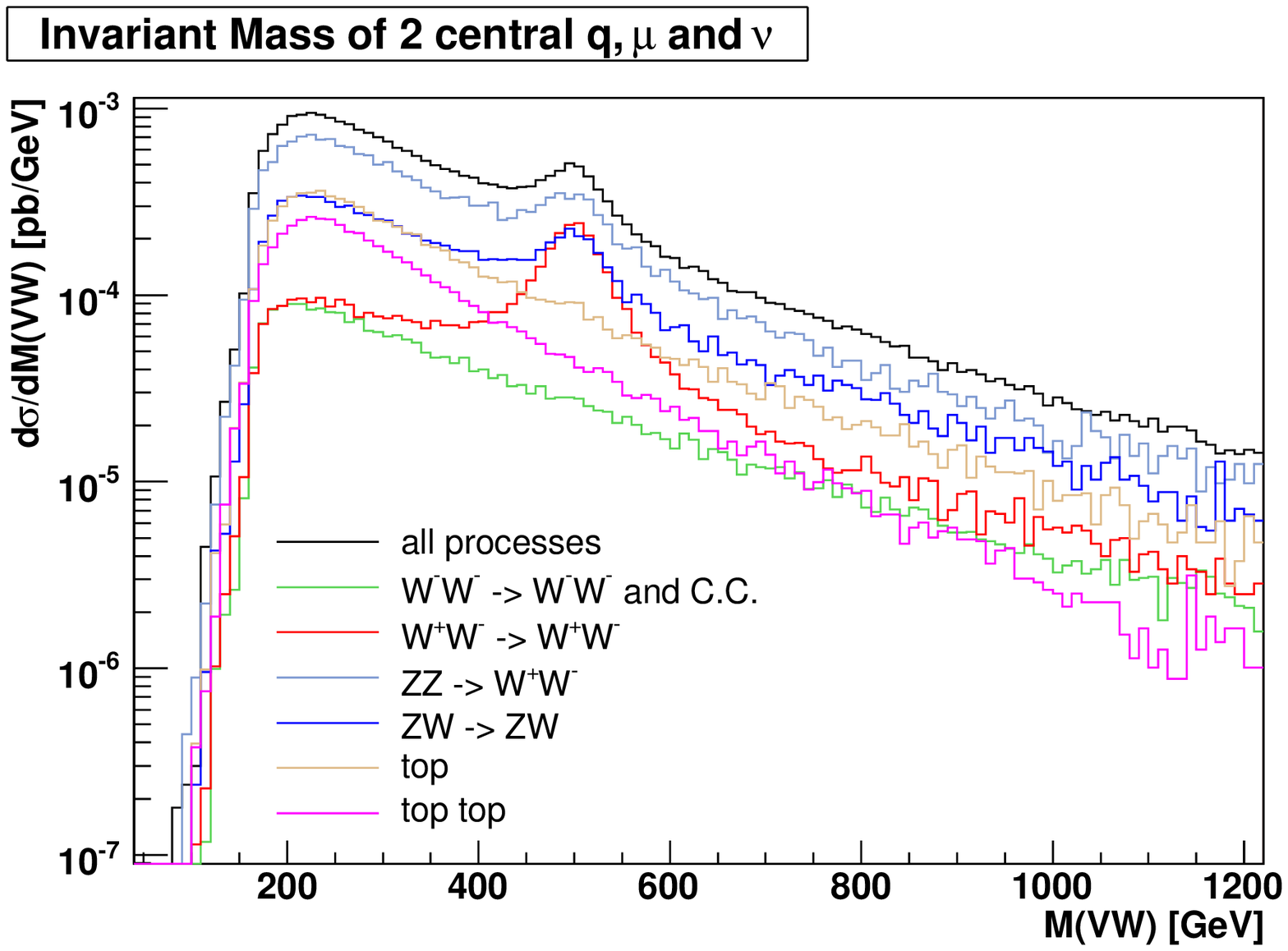,width=7cm}} 
{\epsfig{file=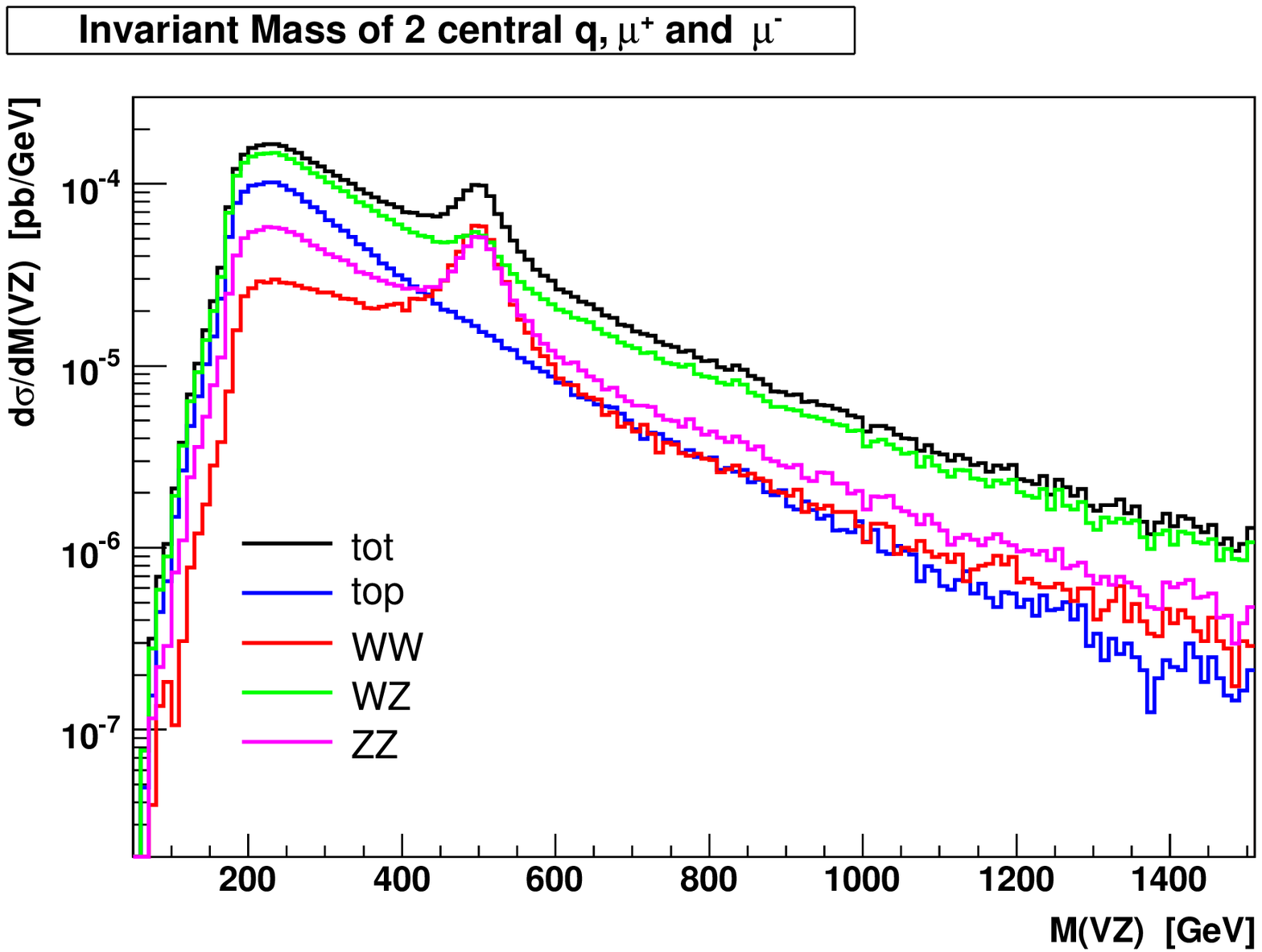,width=7cm}} 
} 
\caption{ Distribution of the invariant mass of the two candidate vector bosons
 for $qq \to 4q +\mu\nu$ and $qq \rightarrow 4q\mu^+ \mu^-$.} 
\label{sub-proc}
\end{center}
\end{figure}

We have merged the two strategies in a single procedure. The outcome is that 
\ABBBPhase adapts to different kinematic cuts and peaks with good 
efficiency, using only a few channels per process.
As third main feature, \ABBBPhase 
employs the {\it one-shot} method developed for 
{\tt WPHACT}~\cite{Accomando:2002sz,Accomando:1996es}.
In this running 
mode, all processes are simultaneously generated in the 
correct relative proportion for any set of experimental cuts, and directly 
interfaced to hadronization and detector simulation programs giving a fully 
comprehensive physical description.
Preliminary results have been presented in~\cite{Accomando:2004my,Accomando:2005za}.

A number of samples of events, representative of all possible processes,
have been produced with \ABBBPhase. In order to comply with the
acceptance and trigger requirements of the CMS experiment,
the cuts in Table~\ref{standard-cuts} have
been applied. We have used the CTEQ5L PDF set with the scale
$Q^2 = M_W^2 + \sum_{i=1}^6 p_{Ti}^2/6$.

\begin{table}[th]
\begin{center}
\caption{Standard acceptance cuts applied in all results. Here lepton refers to
$l^\pm$ only.} 
\vspace*{1mm}
\begin{tabular}{|c|c|c|}
\hline
E(lepton)$ > 20$ GeV &  \ABBBpt(lepton)$ > 10$ GeV & $\vert\eta$(lepton)$\vert < 3$ \\
\hline
E(quarks)$ > 20$ GeV & \ABBBpt(quarks)$ > 10$ GeV & $\vert\eta$(quark)$\vert < 6.5$ \\
\hline
      M($qq$)$ > 20$ GeV & M($l^+l^-$)$ > 20$ GeV &\\
\hline
\end{tabular}
\label{standard-cuts}
\end{center}
\end{table}
 
\subsubsection{Physical sub-processes}
\label{PhysSub}

Many subprocesses 
(i.e. \ABBBWW $\rightarrow$ \ABBBWW, \ABBBZW $\rightarrow$ \ABBBZW, \ABBBZZ $\rightarrow$ \ABBBWW,
\ABBBZZ $\rightarrow$ \ABBBZZ, \ABBBtoptop)
will in general contribute to a specific six fermion reaction. 
It is  impossible to separate and compute individually the cross 
section due to a single subprocess, since there are large interference effects
between the different contributions.
We can however select all complete 2$ \rightarrow $6 processes which include a
specific set of sub-diagrams.
For instance, \ABBBZW$ \rightarrow $\ABBBZW with on shell bosons is described by 5 Feynman diagrams.
These diagrams, with all external vector bosons connected to a fermion
line, constitute the  \ABBBZW$ \rightarrow $\ABBBZW set of 2$ \rightarrow $6 diagrams.
Several sets can contribute to a single
process and therefore the same process can appear in different groups.
Figure~\ref{sub-proc} shows the invariant mass distribution
of the two most central quarks (when ordered in 
$\eta$), the lepton and the neutrino for the reactions 
$PP \rightarrow qq \rightarrow 4q l \nu$ (LHS)
and
$PP \rightarrow qq \rightarrow 4q l^+ l^-$ (RHS).
The distributions for the
different subprocesses as well as the one for the total are presented
for M(H) = 500~GeV.

It should be pointed out that the total cross section in Fig.~\ref{sub-proc}
is smaller than the sum of the cross sections for the various groups.
Notice that the Higgs peak is present in the \ABBBZW$ \rightarrow $\ABBBZW curve.
This is due to
processes that in addition to the \ABBBZW$ \rightarrow $\ABBBZW set of diagrams include also
diagrams describing Higgs production, e.g. 
$u \overline{u} \rightarrow u \overline{u} u \overline{d} \mu^-\overline{\nu}$.
The groups comprising single top and \ABBBtoptop diagrams have a large
cross section. An invariant mass analysis reveals that they 
are indeed dominated by top production. Simple invariant mass vetoes reduce
drastically the EW top background and produce a much sharper Higgs peak.

\begin{figure}[th]
%\vspace*{-1cm}
\begin{center}
\mbox{
{\epsfig{file=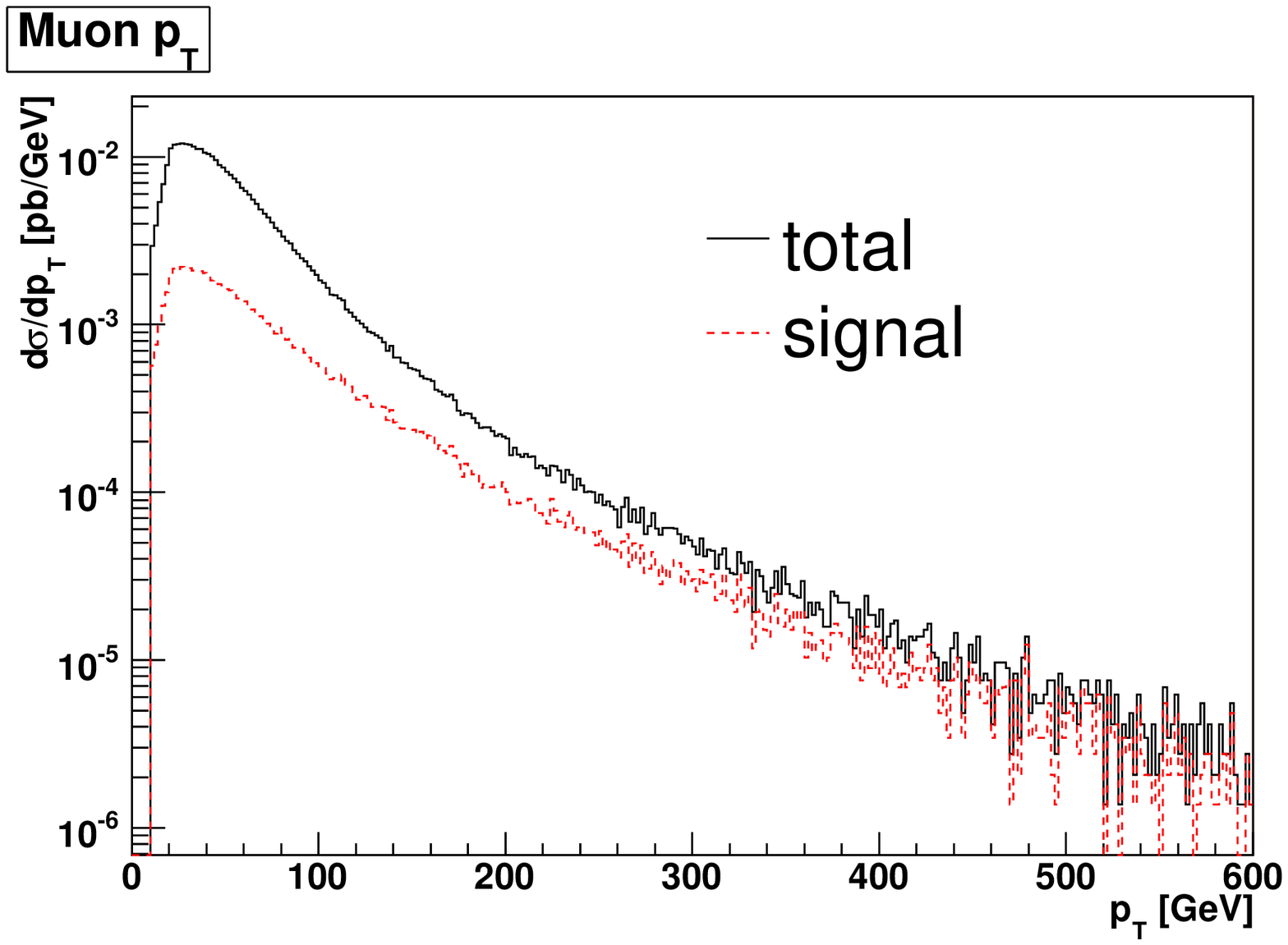,width=7cm}} 
{\epsfig{file=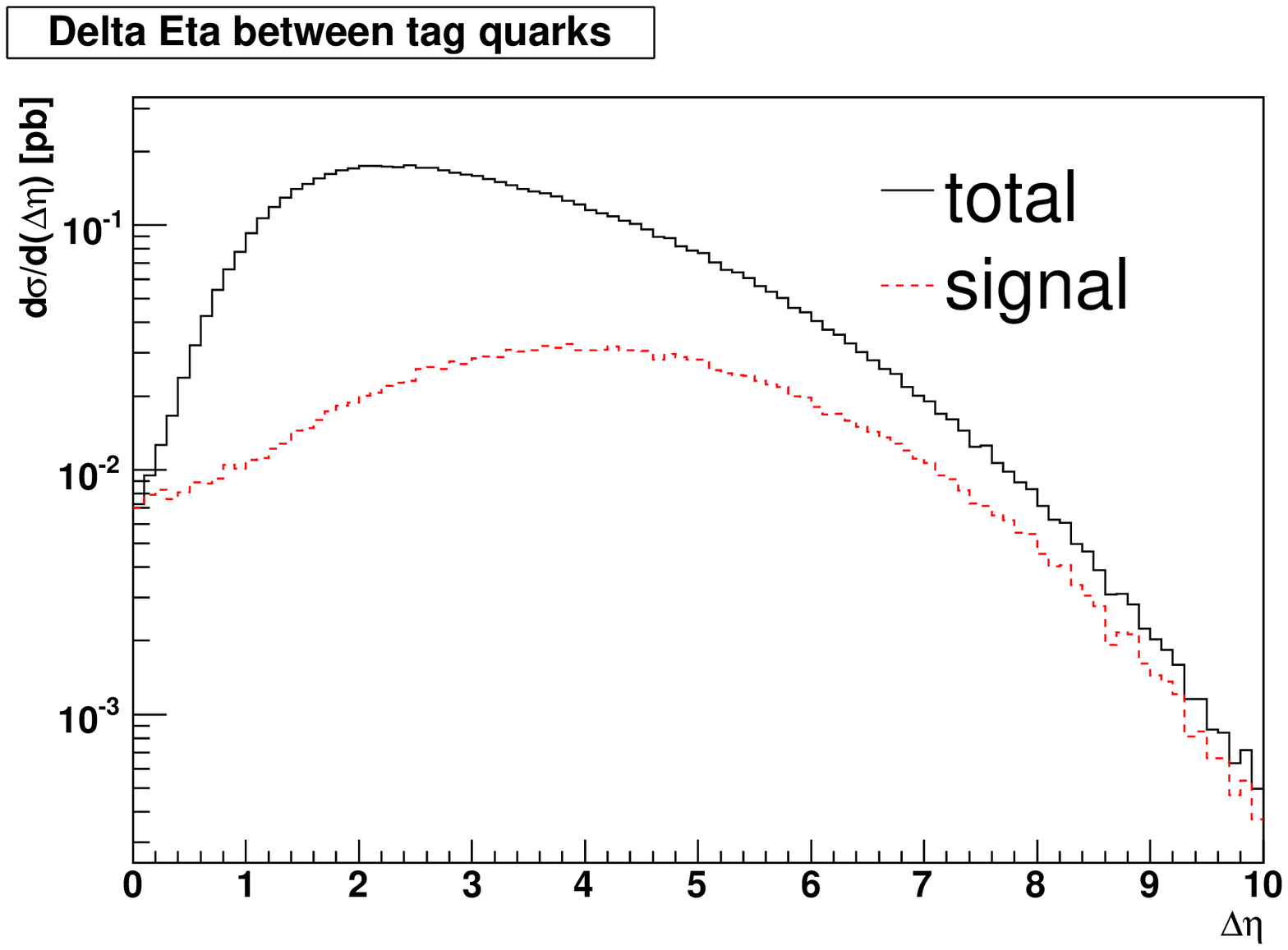,width=7cm}} 
} 
\caption{ Distributions of 
the transverse momentum of the muon,
and of $\Delta\eta$ between the two tagging quarks
for all the events (black) and for the
signal events (red).
} 
\label{m6ferm}
\end{center}
\end{figure}

\subsection{The $\ABBBVV$-fusion signal}
In the absence of firm predictions in the strong scattering
regime, trying to gauge the possibilities of discovering signals of
new physics at the LHC requires the definition of a model of
\ABBBVVL scattering beyond the  SM. Some of these models predict
the formation of spectacular resonances which will be easily detected,
while in other cases only rather small effects are expected.
The simplest approach is to consider the SM in the presence of a very heavy
Higgs boson.
While this entails the violation of perturbative unitarity, the linear rise of
the cross section with $\hat{s}$ - the invariant mass squared in the hard
scattering - will be swamped by the decrease of the parton luminosities at large
momentum fractions and, as a consequence, will be particularly challenging to
detect.
At the LHC for $M_H > $10~TeV, all Born diagrams with Higgs propagators become
completely negligible in the unitary gauge and all expectations reduce to those
in the $M_H \rightarrow \infty$ limit.
We have compared this minimalistic definition of physics beyond the Standard
Model with the predictions of the SM for Higgs masses within the reach of the
LHC.

In addition to the diagrams which are related to the process we would like
to measure, \ABBBVV  fusion, there will be diagrams in which a pair of \ABBBV's  are
produced  without undergoing \ABBBVV scattering.
Furthermore, diagrams related to Higgs production via Higgsstrahlung will also
be present, as well as diagrams which can be interpreted as \ABBBtoptop EW
production or as single top production. Finally, diagrams describing three vector
boson production, which include triple gauge coupling and quartic gauge coupling,
will contribute as well, since they produce the same kind of six fermion final
states.
Depending on the flavour of the quarks the various subprocesses will contribute
and interfere to a different degree. All processes will be experimentally
indistinguishable, apart from heavy quark tagging, and will have to be studied
simultaneously.

In the following, we will concentrate on the $4ql\nu$ final states.
In order to isolate the \ABBBVV fusion process from all the
other six fermion final states and investigate EWSB, different kinematic 
cuts have been applied to the simulated events, after vetoing top quarks.
The invariant mass of the muon and the neutrino has to reconstruct the mass of
a \ABBBW, and  is required to be in the range $M_W \pm 10$~GeV. In \ABBBVV fusion an
additional \ABBBW or a \ABBBZ decaying hadronically is expected to be present.
Therefore, events are required to contain two quarks with the correct flavours
to be produced in the \ABBBW or \ABBBZ decay, with an invariant
mass of $\pm$ 10~GeV around the central value of the appropriate EW boson.
If more than one combination of two quarks satisfies these requirements, the one
closest to the corresponding central mass value is selected. In the following, this combination 
will be assumed to originate from the decay of an EW vector
boson.

In order to reject events which can be identified with the
production of three vector bosons, the flavour content and the invariant mass 
of the two remaining quarks is compared with a \ABBBW and a \ABBBZ in a second step.
If compatible within 
10~GeV with either vector boson, the event is rejected. This happens in about 2\% of the
cases. The events, satisfying all these constraints, will constitute the 
``signal'' sample.

\begin{figure}[th]
%\vspace*{-1cm}
\begin{center}
\mbox{
{\epsfig{file=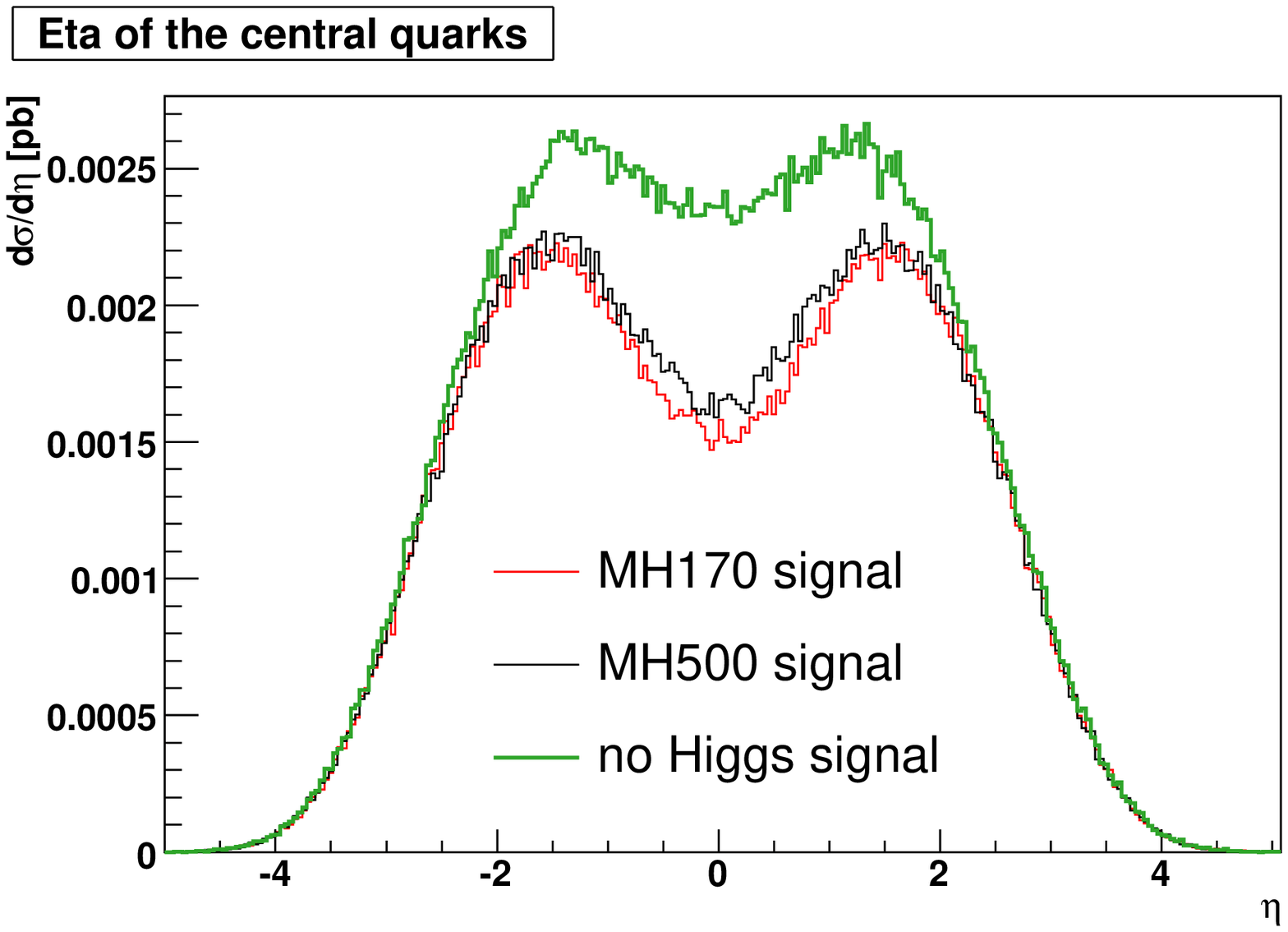,width=7cm}} 
{\epsfig{file=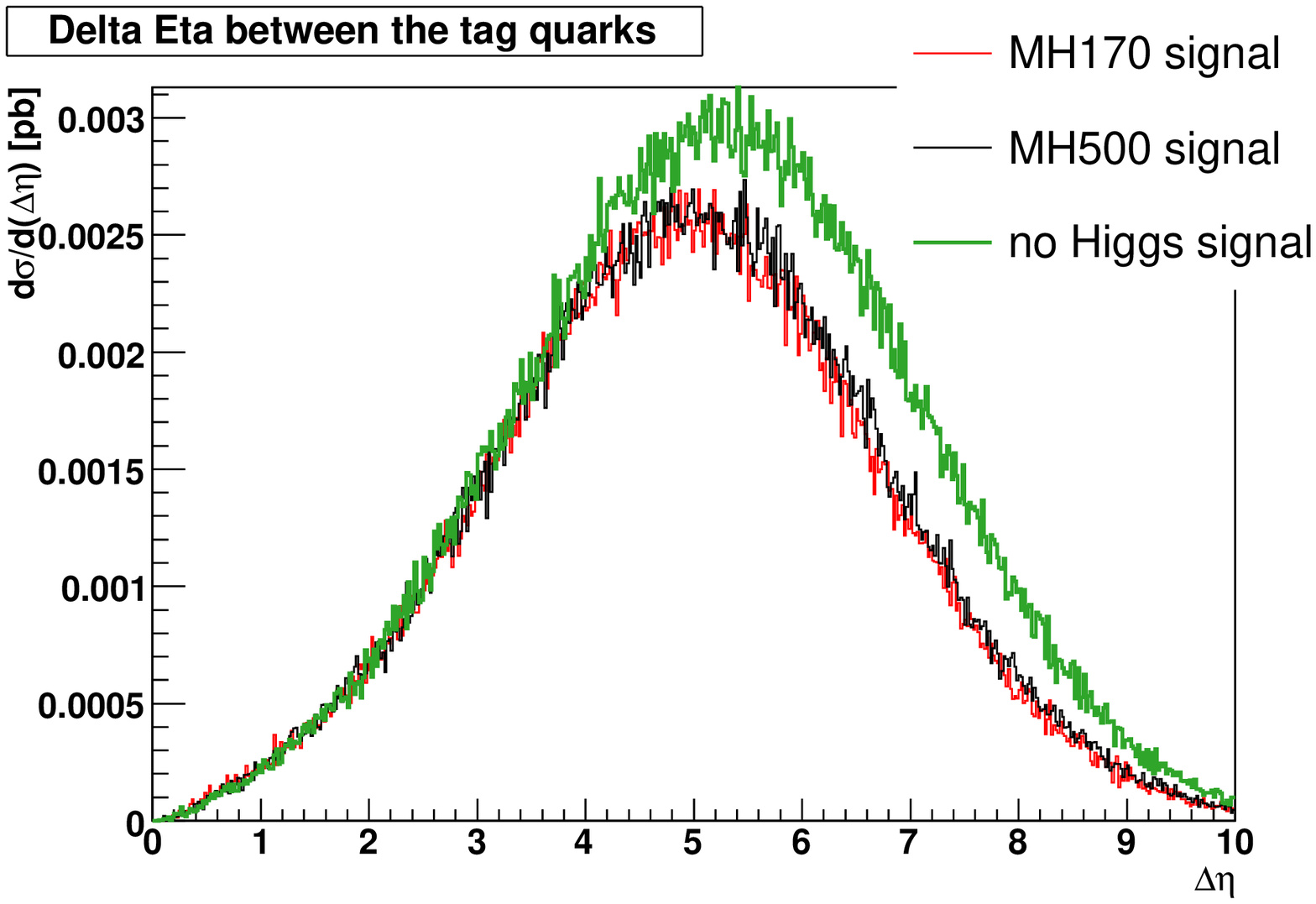,width=7cm}} 
}
\mbox{
{\epsfig{file=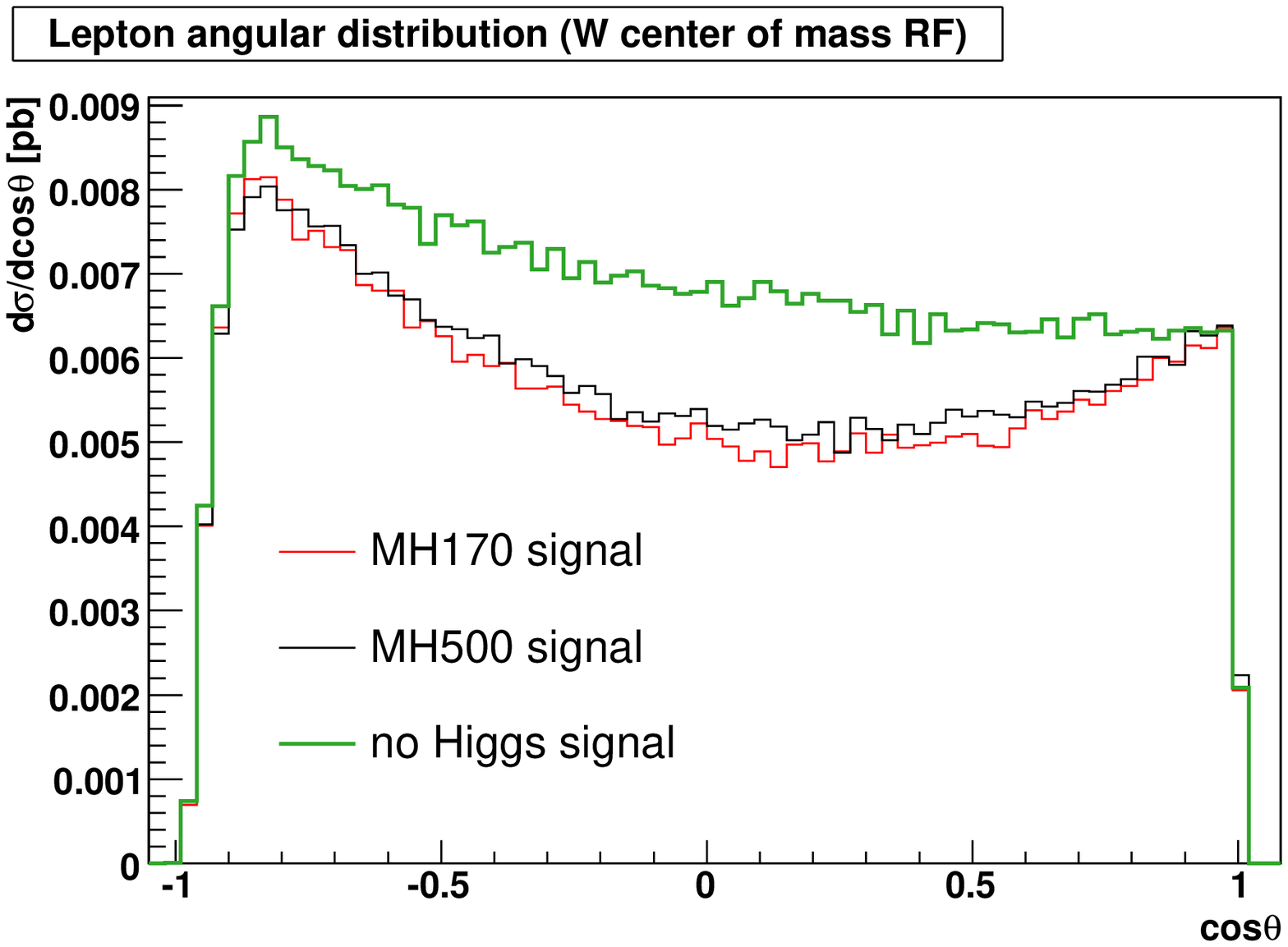,width=7cm}}
{\epsfig{file=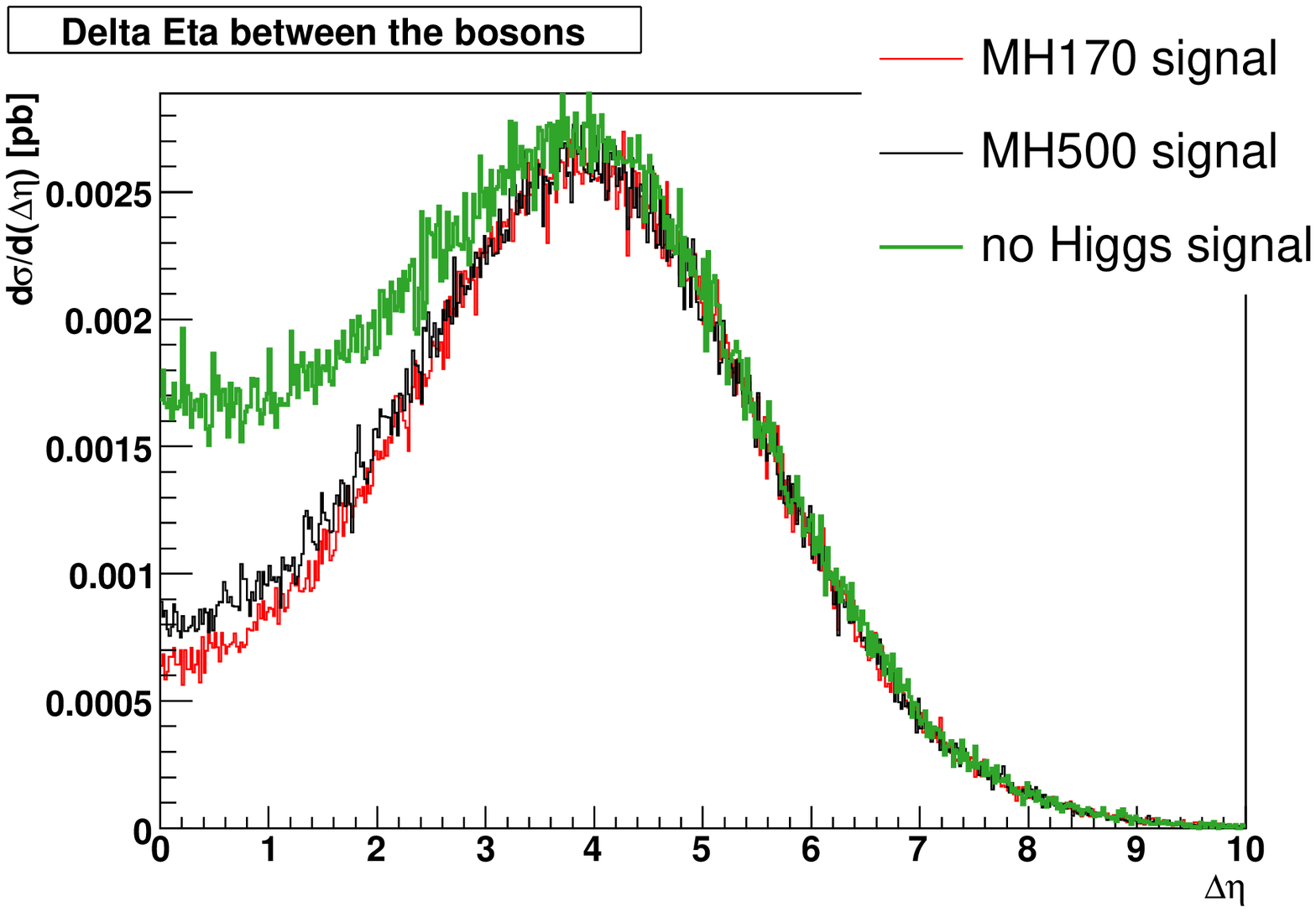,width=7cm}}}
\caption{
The $\eta$ of the two central jets,
the $\Delta\eta$ of the two tagging quarks,
the $\Delta\eta$ of the two vector bosons,
the cosine of the angle between the muon and the W boson
in the W boson rest frame. In green (full line) 
for the no-Higgs case, in black (dashed) for M(H) = 500~GeV and in red
for M(H) = 170~GeV.
All events satisfy $M(VW) > 800$~GeV.
}
\label{dist:code}
\end{center}
\end{figure}

These requirements are not fully realistic: no flavour information will be
available for light quarks and b's will be tagged only in the central part of
the detector. Our aim however, is to define a ``signal'' in the same spirit as
{\tt CC03} was adopted as such at LEP 2, which could be used to compare the results
from the different collaborations. The signal is not necessarily directly
observable but it should be possible to relate it via Monte Carlo to measurable
quantities. If such a definition is to be useful, it must correspond as closely as
possible to the process which needs to be studied and the Monte Carlo
corrections must be small. 
At this stage, we want to isolate the \ABBBVV fusion signal from all other production
channels, while following reasonably close the experimental practice and taking
into account the full set of diagrams required by gauge invariance.
It becomes then  possible to analyze the differences between the \ABBBVV fusion
signal sample and its intrinsic background.
This also provides some preliminary experience at the generator level which 
could guide more realistic and complete studies.

In Fig.~\ref{m6ferm}, we show the distribution of  
the transverse momentum of the muon,
and of the difference in pseudorapidity between the two tagging quarks
for all the events and for the signal.
The ``no-Higgs'' case is chosen as an example, but there are no significant
differences with the case of a finite mass Higgs boson.
The muons in the ``signal'' tend to have a larger $p_T$ than the sum of all subprocesses,
 and the $\Delta \eta$ between tagging quarks tends to be larger.
       
\subsection{Higgs production in \ABBBPhase}

Higgs production in \ABBBVV fusion, followed by the Higgs decay to \ABBBWW or \ABBBZZ, is 
an important channel over the full range of Higgs masses which will be explored
at the LHC. It is the channel with the highest statistical significance for an
intermediate mass Higgs boson~\cite{Asai:2004ws}.
\ABBBPhase is capable of simulating Higgs production
in \ABBBVV fusion together with all its EW irreducible backgrounds
for any Higgs mass and is particularly useful in the intermediate
mass range, where only one of the vector bosons can be approximately
treated in a production  times decay approach.

\subsection{The high mass region}
\label{sec:highmass}

An interesting physics possibility is to investigate, whether an elementary Higgs
boson exists by measuring the \ABBBVW cross section at large M(\ABBBVW). 
For this purpose, kinematic distributions for M(H) = 170~GeV, M(H) = 500~GeV and
the no-Higgs case have been compared for M(\ABBBVW)$ > $800~GeV since the cross
section at large M(\ABBBVW) for M(H) = 170~GeV and M(H) = 500~GeV is essentially due
to transversely polarized vector bosons, while the cross section for the
no-Higgs case is due to a mixture of the two polarizations.
Figure~\ref{dist:code} shows that the distributions are quite insensitive to the value
of the Higgs mass, provided it is much smaller than the invariant mass of the
\ABBBVW system. This raises the stimulating possibility of defining Standard Model
predictions for high invariant mass production of \ABBBVV pairs. These predictions
will obviously suffer from the usual PDF and scale uncertainties, which could 
in principle be controlled by comparing with the cross section of some
appropriate ``standard candle'' process.
  
We have tried several sets of cuts and we believe that using a Neural Network (NN)
is the most effective way of increasing the separation between the no-Higgs case
and the presence of a relatively light Higgs. 
Two samples of events, satisfying the cuts in Table~\ref{standard-cuts}, 
for M(H) = 500~GeV and
the no-Higgs case, respectively, have been employed to train a NN.
The following set of weakly correlated variables have been used  in the training:
the difference in pseudorapidity between the two bosons and between
the two tagging quarks, the transverse momentum of the tagging quarks and the cosine of
the angle between the muon and the \ABBBW boson in the \ABBBW center of mass
system. Some kinematic variables are shown in Fig.~\ref{dist:code} for
the no-Higgs and the M(H) = 500~GeV case.

Applying a cut on the output variable, the NN can enhance
the separation between the heavy and light Higgs case.
In Table~\ref{nn-tab}, the integrated cross section and the number of events
for M(\ABBBVW)$ > $ 800~GeV are shown for different values 
of the cut. 

\begin{table}[bth]
\begin{center}
\caption{Integrated cross section for the number of
  expected events in a year at high luminosity in the two cases and their
  corresponding ratios.}
\vspace*{1mm}
\begin{tabular}{|l|c|c|c|c|c|} 
\cline{2-6}
\multicolumn{1}{c|}{} & $\sigma_{no-Higgs}$ & $\mathcal{L} = 100fb^{-1}$ & 
            $\sigma_{mH = 500 GeV}$ & $\mathcal{L} = 100fb^{-1}$ & ratio  \\
\hline
all events  & 13.6 fb & 1360 $\pm$ 37 & 11.6 fb & 1160 $\pm$ 34 & 1.2\\
\hline
NN $ > $0.54 & 3.17 fb & 317 $\pm$ 18  & 1.95 fb & 195 $\pm$ 14 & 1.6\\
\hline
NN $ > $0.58 & 2.28 fb & 228 $\pm$ 15  & 1.13 fb & 113 $\pm$ 11 & 2.0\\
\hline
\end{tabular}
\label{nn-tab}
\end{center}
\end{table}

\subsection*{Acknowledgements}
This work is supported by MIUR contract 26-01-2001 N.13 and under
contract 2004021808\_009

%%%%%%%%%%%%%%%%%%%%%%%%%%%%%PART%%%%%%%%%%%%%%%%%%%%%%%%%%%%%
\part[NLO AND NNLO QCD COMPUTATIONS]{NLO AND NNLO QCD COMPUTATIONS}
%%%%%%%%%%%%%%%%%%%%%%%%%%%%%PART%%%%%%%%%%%%%%%%%%%%%%%%%%%%%

%%%%%%%%%%%%%%%%%%%%%%%%%%%%%%%%%%%%%%%%%%%%%%%%%%%%%%%%%%%%%%%%%%%%%%%%%%%%%
\section[NLO predictions for many-particle production]
{NLO PREDICTIONS FOR MANY-PARTICLE PRODUCTION~\protect
\footnote{Contributed by: T.~Binoth, F.~Boudjema, A.~Denner, S.~Dittmaier,
    J.~Fujimoto, J.-Ph.~Guillet, G.~Heinrich, J.~Huston, T.~Ishikawa,
    T.~Kaneko, K.~Kato, Y.~Kurihara, E.~Pilon}}
\subsection{Introduction}

At the LHC, most of the interesting processes
%background processes to interesting signatures
will involve multi-particle final states,
either through the decay of resonances or
from direct multi-particle production.
However, only a few theoretical predictions
beyond leading order are available up to now for
processes with more than two particles in the final state.
On the other hand, it is well known that leading order predictions
have serious deficiencies like limited predictive power due to
large scale dependence, large sensitivity
to kinematic cuts and poor jet modelling.
Parton shower Monte Carlos on the other hand
cannot predict the overall normalisation of a process and
do not perform well in describing large angle radiation.
As a consequence of the latter, shapes of certain
distributions may not be well described.
This is especially worrisome if backgrounds  need
to be extrapolated to signal regions using theoretical
predictions.

A lot of activity has been going on recently to
improve this situation. The following sections should serve to
summarize the current situation and to assess the needs and
prospects for the near future.

\subsection{Status}

There are already a few QCD NLO  $2\to 3$ processes in hadronic
collisions that have been calculated so far, see
Table~\ref{multileg-table1}.
\begin{table}[htb]
%\begin{minipage}{\textwidth}
\begin{center}
\caption{List of existing predictions for $pp\to 3$
particles.\newline $^a$ Although this is a $2 \to 3$ process,
because of its colour singlet nature the NLO QCD corrections are
simple and do not involve more than vertex corrections. Strictly
speaking, genuine corrections involving $5$-point loop functions
do appear if one considers the same final state initiated through
$Z$-strahlung. These contributions are small compared to VBF and
have, so far, not been considered.
 \newline $^b$ This refers to the {\em effective} NLO QCD
correction where the Higgs is produced at ``leading order" through
the effective coupling to two gluons. Compared to VBF this process
is characterised by extra central jet activity.}
  \vspace*{1mm}
\begin{tabular}{|lll|}
\hline
\hline
\multicolumn{3}{|c|}{pp$\to$ 3 particles}\\
\hline
\hline
process\,  $(V=Z,W^\pm)$&references&comments\\
\hline
&&\\
$pp\to
3$\,jets&\cite{Bern:1993mq,Kunszt:1994tq,Bern:1994fz,Kilgore:1996sq,Nagy:2001fj}&
public code available\,\cite{Nagy:2001fj}\\
$pp\to V+$ 2\,jets &\cite{Campbell:2003hd,Campbell:2005zv}&public code available\\
$pp\to Z\, b\,\bar{b}$  &\cite{Ellis:1998fv,Campbell:2000bg}&massless $b$ quarks\\
$W^\pm+g^*(\to b\bar{b})$  &\cite{Campbell:2001ik}&massless
$b$ quarks\\
$pp\to H+2$
jets&\cite{Han:1992hr,Figy:2003nv,Berger:2004pc,Figy:2004pt}&
$H$ through VBF, Vector Boson Fusion$^a$ \\
$pp\to H+$ 2\,jets &\cite{Ellis:2005qe}& ``Background" to VBF$^b$, under construction \\
$pp\to \gamma\gamma\, $jet&\cite{DelDuca:2003uz,Binoth:2003xk}&\\
$pp\to t\bar{t}H, b\bar{b}H$&
\cite{Reina:2001bc,Beenakker:2001rj,Beenakker:2002nc,Dawson:2002tg,Dawson:2003kb,Dittmaier:2003ej,Dawson:2004sh,Dawson:2005vi}&\\
$pp\to t\bar{t}\,h^0$&\cite{Peng:2005ti}&SUSY QCD corrections\\
$pp\to t\bar{b}H^-$&\cite{Peng:2006wv}&SUSY QCD corrections\\
$pp\to t\bar{t}\,$jet&\cite{Brandenburg:2004fw}&under
construction\\
&&\\
\hline
\end{tabular}
\label{multileg-table1}
\end{center}
%\end{minipage}
\end{table}

Very recently there has also been  major progress in electroweak
NLO corrections to $2 \to 3$ processes, see
Table~\ref{multileg-table2}. These involve a considerable number
of diagrams with many different mass scales. The past year has
even witnessed the first complete calculation of  NLO electroweak
corrections to $2 \to 4$ processes, Table~\ref{multileg-table3}.
%\vspace*{3mm}

\begin{table}[htb]
\begin{center}
\caption{List of existing calculations for $e^+ e^- \to
\gamma^*\to 4$ jets and $e^+ e^-,\gamma\gamma\to 3$ particles}
  \vspace*{1mm}
\begin{tabular}{|lll|}
\hline \hline
\multicolumn{3}{|c|}{$e^+ e^-$ and $\gamma \gamma$ $\to$ 3 particles.}\\
\hline \hline
process\,  & type of corrections &references\\
\hline
&\\
$e^+ e^- \to 4$\,jets& QCD &
\cite{Glover:1996eh,Campbell:1997tv,Bern:1997sc,Dixon:1997th,Nagy:1998bb,Campbell:1998nn,Weinzierl:1999yf}\\
$e^+ e^- \to \nu \bar{\nu} H$ & EW
&\cite{Belanger:2002me,Belanger:2002ik,Denner:2003yg,Denner:2003iy} \\
$e^+ e^- \to e^+ e^- H$& EW & \cite{Boudjema:2004eb}\\
$e^+ e^- \to t \bar{t} H$ & EW and QCD& \cite{You:2003zq,Belanger:2003nm,Denner:2003ri,Denner:2003zp}\\
$e^+ e^- \to Z H H$&EW & \cite{Zhang:2003jy,Belanger:2003ya}\\
$e^+ e^- \to \nu \bar{\nu} \gamma$
&EW & \cite{Boudjema:2004ba} \\
$\gamma \gamma \to t \bar{t} H$ &EW and QCD &\cite{Chen:2003yd} \\
& &\\
\hline
\end{tabular}
\label{multileg-table2}
\end{center}
\end{table}

%\begin{minipage}{0.62\textwidth}
\begin{table}[htb]
\begin{center}
\caption{$2\to 4$ cross sections presently available.}
  \vspace*{1mm}
\begin{tabular}{|lll|}
\hline \hline
\multicolumn{3}{|c|}{2$\to$ 4}\\
\hline \hline
process&references&comments\\
\hline
&&\\
$e^+e^-\to 4$ fermions&\cite{Denner:2005es,Denner:2005fg}&cross sections\\
&\cite{Boudjema:2004id}&status report\\
$e^+e^-\to H H \nu\bar{\nu}$&\cite{Boudjema:2005rk}& cross
sections \\
&&\\
\hline
\end{tabular}
\label{multileg-table3}
\end{center}
\end{table}
%\end{minipage}

\begin{table}[hbt]
\begin{center}
\caption{Other $2\to 4\, (5)$ calculations.}
  \vspace*{1mm}
\begin{tabular}{|lll|}
\hline \hline
\multicolumn{3}{|c|}{2$\to$ 4\,(5): special models, specific helicity amplitudes, special kinematics.}\\
\hline \hline
process&references&comments\\
\hline
&&\\
N-photon helicity amplitudes&\cite{Mahlon:1993fe}&only specific helicity configurations \\
6- and 7 - gluon amplitudes&\cite{Bern:1994ju,Bern:2004ky}&for non-Susy Yang-Mills only specific \\
&& helicity configurations\\
6- gluon amplitude&\cite{Ellis:2006ss}                   &Result for one phase space point\\
&& (only virtual corrections) \\
6-scalar amplitudes in the Yukawa model&\cite{Binoth:2001vm}&\\
2-photon 4-scalar amplitudes &\cite{Binoth:2002qh}&only specific helicity configurations\\
in the Yukawa model&&\\
&&\\
\hline
\end{tabular}
\label{multileg-table4}
\end{center}
\end{table}

Besides, a few calculation of NLO corrections to $2 \to 4,5$ in
toy models or for specific helicity configurations have been
performed successfully, see Table~\ref{multileg-table4}. One
important development  has been the computation of the {\em
virtual} corrections to the $6$ gluon
amplitude\cite{Ellis:2006ss}.  Results at the amplitude level have
been given for a specific point in phase space. This is an
important step towards the computation of the cross section at
NLO. At the same time, based on the newly developed twistor
techniques\cite{Witten:2003nn,Cachazo:2004kj,Cachazo:2004zb,Brandhuber:2004yw,Cachazo:2004by,Bern:2004ky,Britto:2004nc,
Cachazo:2005ga,Dixon:2005cf}, the major part of the $6$-gluon
amplitude at one-loop has been derived\cite{Britto:2006sj}. We
therefore expect that the NLO correction to the $4$-jet cross
section at the LHC are within sight.

The present panorama of the multi-leg one-loop corrections
together with the recent emergence of new and improved loop
techniques as well as novel approaches such as the twistor string
inspired makes it now possible to tackle $2 \to 3$ and $2 \to 4$
processes that are of importance for the LHC.

%\clearpage

\subsection{A realistic NLO wishlist for multi-particle final states}

A somewhat whimsical experimenter's {\it wishlist} was first
presented at the Run 2 Monte Carlo workshop at Fermilab in
2001~\cite{runiimc}. Since then the list has gathered a great deal
of notoriety and has appeared in numerous LHC-related theory
talks. This list contained a great number of multi-particle
processes with many particles in the final state. Although it was
well motivated from a data analysis point of view, many of the
processes are far beyond present calculational methods and tools.
For example, it is unlikely that $WWW+b\overline{b}+3$\,jets will
be calculated at NLO soon, no matter the level of physics
motivation, but there are a number of high priority calculations,
primarily of backgrounds to new physics, that are accessible with
the present technology. However, the manpower available before the
LHC turns on is limited. Thus, it is necessary to prioritize the
calculations, both in terms of the importance of the calculation
and the effort expected to bring it to completion.

A prioritized list, determined at the Les Houches workshop, is
shown in Table \ref{wishlist}, along with a brief discussion of
the physics motivation. Later in this section, there will be a
discussion of some of the specific theoretical difficulties to
expect. Note that the list contains only $2\rightarrow 3$ and
$2\rightarrow  4$ processes. It will be much more of a challenge
to tackle higher multiplicities before addressing these cross
sections.

First, a few general statements: usually, signatures for new
physics will involve high $p_T$ leptons and jets (especially $b$
jets) and missing transverse momentum. Thus, backgrounds to new
physics will tend to involve (multiple) vector boson production
(with jets) and $t\overline{t}$ pair production (with jets). The
best manner in which to understand the normalization of a cross
section is to measure it; however the rates for some of the
complex final states listed here may be limited and (at least in
the early days) must be known from NLO theory. NLO is the first
order at which both the normalization and shape can be calculated
with any degree of confidence.

\begin{table}[htb]
\begin{center}
\caption{The LHC ``priority" wishlist for which a NLO computation
seems now feasible.  \label{wishlist}}
  \vspace*{1mm}
\begin{tabular}{|l|l|}
\hline
&\\
process&relevant for\\
%\multicolumn{2}{|c|}{}\\
($V\in\{Z,W,\gamma\}$)&\\
\hline
&\\
%\item $2\to 3$
1. $pp\to V\,V$\,jet &  $t\bar{t}H$, new physics\\
%2. $pp\to H+2$\,jets& $H$ production by vector boson fusion (VBF)\\
%\item $2\to 4$
2. $pp\to t\bar{t}\,b\bar{b}$ &  $t\bar{t}H$\\
3. $pp\to t\bar{t}+2$\,jets  &  $t\bar{t}H$\\
4. $pp\to V\,V\,b\bar{b}$ &  VBF$\to H\to VV$,
$t\bar{t}H$, new physics\\
5. $pp\to V\,V+2$\,jets &  VBF$\to H\to VV$\\
6. $pp\to V+3$\,jets &  various new physics signatures\\
7. $pp\to V\,V\,V$ &  SUSY trilepton\\
&\\
\hline
\end{tabular}
\end{center}
\end{table}

\begin{itemize}

\item $pp \rightarrow $ VV + jet: One of the most promising channels for Higgs
production in the low mass range is through the $H\rightarrow
WW^*$ channel, with the W's decaying semi-leptonically. It is
useful to look both in the $H\rightarrow  WW$ exclusive channel,
along with the $H\rightarrow  WW$+jet channel. The calculation of
$pp\rightarrow  WW$+jet will be especially important in
understanding the background to the latter.

%\item $pp \rightarrow  H$+2 jets: A measurement of vector boson fusion (VBF)
%production of the Higgs boson will
%allow the determination of the Higgs coupling to vector bosons. One of the key
%signatures for this process is the presence of forward-backward tagging jets.
%Thus, QCD production of $H$ + 2 jets must be understood, especially as the rates for
%the two are comparable in the kinematic regions of interest.

\item $pp \rightarrow  t\overline{t} b\overline{b}$ and $pp \rightarrow
t\overline{t}$ + 2 jets: Both of these processes serve as background to
$t\overline{t}H$, where the Higgs decays into a $b\overline{b}$ pair. The rate
for $t\overline{t}jj$ is much greater than that for
$t\overline{t}b\overline{b}$ and thus, even if 3 $b$-tags are required, there
may be a significant chance for the heavy flavor mistag of a $t\overline{t}jj$
event to contribute to the background.

\item $pp \rightarrow  VV b\overline{b}$: Such a signature serves as
non-resonant background to $t\overline{t}$ production as well as to possible
new physics.

\item $pp \rightarrow $ VV + 2 jets: The process serves as a background to VBF
production of a Higgs boson.

\item $pp \rightarrow $ V + 3 jets: The process serves as background for
$t\overline{t}$ production where one of the jets may not be reconstructed, as
well as for various new physics signatures involving leptons, jets and missing
transverse momentum.

\item $pp \rightarrow VVV$: The process serves as a background for various new
physics subprocesses such as SUSY tri-lepton production.

\end{itemize}
Work on (at least) the processes 1. to 3. of Table \ref{wishlist}
is already in progress by several groups, and clearly all of them
aim at a setup which allows for an  application to other
processes.

%An overview over the theoretical
%tools which have been developed recently, indicating also
%for which types of processes they are most suitable,
%will be given in section \ref{sec:methods}

\vspace*{3mm}

From an experimentalist's point-of-view, the NLO calculations discussed thus
far may be used to understand changes in normalization and/or shape that occur
for a given process when going from LO to NLO~\cite{Campbell:2004sp}.
Direct comparisons to the data
require either a determination of parton-to-hadron corrections for the theory
or hadron-to-parton corrections for the data~\cite{Flanagan:2005xv}.
(Of course, one is just the
inverse of the other.) Both types of correction take into account the effects
of the underlying event and of fragmentation. For multi-parton final states,
it is also necessary to model the effects of jet algorithms, when two or more
partons may be combined into one jet~\cite{Ellis:2001aa}.

\subsection{Review of theoretical approaches}
\label{sec:methods}

In this section, first a brief overview of the
existing methods to tackle one-loop multi-leg amplitudes
will be given. More detailed descriptions of the individual methods
are given in section \ref{sec:detail}.

The majority of the one-loop cross sections available up to now
has been calculated by following the approach pionneered by Ellis Ross 
and Terrano~\cite{Ellis:1980wv,Ellis:1985er} to calculate real and virtual 
corrections and treat the soft/collinear singularities, applying the 
Passarino-Veltman algorithm\,\cite{Passarino:1978jh}, or
some variation of it, for the tensor reduction and the evaluation of
the one-loop integrals.
This ``traditional" approach consists of the following steps
for the calculation of a $2\to N$ particle process at partonic level:
\begin{enumerate}
\item diagram generation
\item calculation of the real radiation corrections
      (requires $2\to N+1$ amplitudes at tree level and
      subtraction of poles due to soft/collinear massless particles)
\item calculation of the one-loop amplitude
      (involves $(N+2)$--point integrals)
\item combination of real and virtual contributions, integration
      over the phase space
%\item ideally: combination with parton showering
\end{enumerate}
The issue of subtraction of long distance singularities in step 2. above has 
been solved once and for all by the general methods, of phase space slicing 
\cite{Giele:1991vf,Giele:1993dj,Keller:1998tf,Harris:2001sx}, 
and of subtraction 
\cite{Frixione:1995ms,Catani:1996vz,Dittmaier:1999mb,Phaf:2001gc,Catani:2002hc}, 
so that, for multi-particle processes, step 3. is now the bottleneck. 
We will therefore concentrate on the calculation of one-loop amplitudes in 
the following.

A loop integral for an $N$-point function consists of products of
denominators representing the propagators circulating in the loop
and a numerator consisting of a tensor structure that generally
depends on the loop momentum.  When the numerator of an integral
is independent of the loop momentum, it is called a scalar
integral. The traditional method for the calculation of one-loop
tensor integrals consists, through recursion, of  an {\em
algebraic} reduction of the tensor integrals to a set of scalar
"basis integrals". As the basis integrals are known in analytic
form, the virtual amplitude is expressed by analytic functions
which depend on the invariants of the given process, thus having
maximal analytic control when proceeding to the phase space
integration. All tensor integrals can be expressed in terms of
scalar integrals and form factors carrying the Lorentz structure
by solving a system of equations where the unknowns are the tensor
coefficients. The determinant of this system of equations, the
Gram determinant, can vanish or get extremely small for some
particular configurations of phase space. This can lead to
numerical instability as the Gram determinant appears with inverse
powers if the "traditional" reduction method is used. $N$-point
functions for $N\geq 5$, including scalar functions, can also be
reduced to a system of four-and lower-point functions, and in this
process  inverse determinants are generated as well. To deal with
this problem, a few methods have been worked out recently. One can
for example use a Taylor expansion in the Gram determinant
\cite{Giele:2004ub,Denner:2005nn,Ellis:2005zh,Boudjema:2005hb} or
resort to a numerical evaluation of some of the
integrals\,\cite{Binoth:2005ff,Denner:2005nn,Ellis:2005zh}, at
least for these critical points. By stopping the reduction before
only scalar functions are reached and integrating the endpoints of
such a reduction numerically, the occurrence of inverse Gram
determinants can be avoided
completely\,\cite{Binoth:2005ff,Denner:2005nn}.

Another way of tackling the problem is
to revert to a numerical evaluation of all the loop integrals,
which is generally called "semi-numerical" or "numerical" approach,
depending on the extent of algebraic reductions carried out before
evaluating certain integrals numerically.
The borderline between "algebraic" and "semi-numerical"
cannot be drawn in a clear way.
%As there are many variants of the "algebraic"
%approach, only the more recent ones will be listed below.
%
Below we first list the most recent variants of the "algebraic/partly numerical"
approach.
\begin{itemize}
\item Denner, Dittmaier\\ {\it massive and massless, \\
applied to calculate the first cross section for a 6-point process}
\cite{Denner:2005nn,Denner:2005fg,Denner:2005es,Denner:2002ii,Dittmaier:2003bc}
\item Ellis R.K., Giele, Glover, Zanderighi\\ {\it massless propagators only}
\cite{Giele:2004iy,Ellis:2005qe,Ellis:2005zh,Ellis:2006ss}
\item Binoth, Guillet, Heinrich, Pilon, Schubert\\ {\it massless and massive}
\cite{Binoth:1999sp,Binoth:2005ff}
\item GRACE group \\
{\it applications so far massive, massless under development}
\cite{Boudjema:2004id,Kurihara:2005at}
\item Del Aguila, Pittau \cite{delAguila:2004nf};\\
Van Hameren, Vollinga, Weinzierl
\cite{vanHameren:2005ed}\\
{\it based on spinor helicity}
\item Duplancic, Nizic\\ {\it massless propagators only}
\cite{Duplancic:2003tv}
\item Fleischer, Jegerlehner, Tarasov\\ {\it massive only}
\cite{Tarasov:1998nx,Fleischer:1999hq}
\item Bern, Dixon, Kosower\\ {\it massless propagators only}
\cite{Bern:1992em,Bern:1993kr}\\
%{\it This group now pursues a different approach
%based on the construction of loop integrals from unitarity
%cuts, factorisation properties and on-shell recursion relations, see below.}
%\item Davydychev
\end{itemize}
Besides this mostly algebraic approach, there are semi-numerical
methods,  which do split into real and virtual corrections,
but largely rely on the numerical evaluation of
loop integrals, either by doing already
the  integration over the loop momenta numerically,
or by evaluating the Feynman parameter representation
of the integrals numerically.
This requires the elaboration of a scheme
to remove the poles from the integrals before the numerical
integration. The following groups have worked in this
direction recently: {\it (historical order)}
\begin{itemize}
\item   Fujimoto, Shimizu, Kato, Oyanagi \cite{Fujimoto:1991bm}
\item   Ferroglia, Passera, Passarino, Uccirati \cite{Ferroglia:2002mz}
\item   Binoth, Heinrich, Kauer \cite{Binoth:2002xh}
\item   Nagy, Soper \cite{Nagy:2003qn,Nagy:2004ea}
\item Kurihara, Kaneko \cite{Kurihara:2005ja}
%\item   Giele, R.K.~Ellis, Zanderighi \cite{Ellis:2005zh}
\item   Anastasiou, Daleo \cite{Anastasiou:2005cb}
\end{itemize}
Further,
there is  an approach \cite{Soper:1998ye,Soper:1999xk,Soper:2001hu}
which avoids the splitting into real and virtual parts by
first performing the sum over cuts for a given graph and then
integrating over all momenta, including the loop momenta,
numerically.
In this way  unitarity is exploited to cancel soft and collinear
divergences before they show up as explicit poles.
However, this method has only been applied to the process
$e^+e^-\to 3$\,jets at NLO so far.

Very recently, a novel approach to the calculation of one-loop
amplitudes has emerged, which is often referred to as
"twistor-space-inspired" methods
\cite{Witten:2003nn,Cachazo:2004kj,Cachazo:2004zb,Brandhuber:2004yw,Cachazo:2004by,Bern:2004ky,Britto:2004nc,
Cachazo:2005ga,Dixon:2005cf}. %\cite{}.
Using these methods, compact expressions for very complex
tree-level amplitudes could be achieved, and their extension to
loop level has seen a very rapid development
\cite{Bidder:2004tx,Britto:2006sj,Brandhuber:2005jw,Bern:2005cq,Bern:2005ji,Britto:2005ha,Bern:2005hs,
Bern:2004bt,Bedford:2004nh,Bern:2004ky}. In particular, the
unitarity-based method of \cite{Bern:1994cg,Bern:1994zx} has seen
a successful revival due to the use of on-shell recursion
relations \cite{Forde:2005ue,Forde:2005hh,Bern:2005hs}. A very
recent breakthrough is the derivation of the major part of the
$6$-gluon amplitude at NLO (the rational parts are still missing).
However, this approach being very new, it is difficult to judge
whether it can be applied to processes where several different
mass scales are involved.

\subsubsection{More detailed descriptions of recent methods}
\label{sec:detail}

\paragraph{The DD approach}
In the following we describe the salient features of the methods
described in Refs.~\cite{Denner:2002ii,Denner:2005nn} that have
been successfully applied in the calculation of a complete one-loop
correction to a $2\to4$ scattering reaction, viz.\ the
electroweak corrections to the charged-current processes
$e^+e^-\to4f$ \cite{Denner:2005es,Denner:2005fg}.
The described methods, thus, have proven their reliability in
practice.

Particular attention is paid to the issue of numerical stability. For
1- and 2-point integrals of arbitrary tensor rank, general numerically
stable results are used. For 3- and 4-point tensor integrals,
serious numerical instabilities are known to arise in the frequently
used Passarino--Veltman reduction \cite{Passarino:1978jh}
if Gram determinants built of external momenta become small.
While Passarino--Veltman reduction is applied if Gram determinants
are not too small, for the remaining problematic cases dedicated reduction
techniques have been developed. One of the techniques replaces the
standard scalar integral by a specific tensor coefficient that can be
safely evaluated numerically and reduces the remaining tensor
coefficients as well as the standard scalar integral to the new set of
basis integrals. In this scheme no dangerous inverse Gram determinants
occur, but inverse modified Cayley determinants instead. In a second
class of techniques, the basis set of standard scalar integrals is kept,
and the tensor coefficients are iteratively deduced up to terms that are
systematically suppressed by small Gram determinants or by other small
kinematical determinants in specific kinematical configurations; the
numerical accuracy can be systematically improved upon
including higher tensor ranks. For 5- and 6-point tensor integrals,
reductions to 4- and 5-point integrals, respectively, are employed that
do not involve inverse Gram determinants either.

Finally, we summarize some information that is relevant for
the practical use of the methods.
\begin{enumerate}
\item
The methods are valid for massive and massless cases. More precisely,
the formulas given in Refs.~\cite{Denner:2002ii,Denner:2005nn} are valid
without modifications if IR divergences are regularized with mass
parameters or dimensionally.%
\footnote{For the method of Ref.~\cite{Denner:2002ii}, this has been
shown in Ref.~\cite{Dittmaier:2003bc}.}
The IR, i.e.\ soft or collinear, singularities naturally appear in
the standard scalar 2-point, 3-point and 4-point functions.
Finite masses can be either real or complex.
\item
The input and output structure of the methods is the same as for
conventional Passarino--Veltman reduction, where momenta and masses
are used as input and the numerical values of all tensor coefficients
(and the scalar integrals) are delivered as output. This means that
no specific algebraic manipulations are needed in applications, so that
the whole method can be (and in fact is) organized as a numerical
library for scalar integrals and tensor coefficients.
\item
Up to now, the methods are explicitly worked out for $N$-point
integrals with $N\le6$, which is sufficient for $2\to4$ particle
reactions. The extension to higher-point functions is straightforward.
\item
All relevant formulas are published and ready for direct implementation
without further manipulations.
Only the scalar 3- and 4-point functions are needed from elsewhere.
\end{enumerate}

\paragraph{The BGHPS approach}
The method described in\,\cite{Binoth:2005ff} to compute multi-particle
processes relevant for the LHC
at one-loop level has the following main features:
\begin{enumerate}
\item validity for an arbitrary number $N$ of external legs
\item validity for both, massive and massless particles
\item algebraic isolation of IR divergences
\item flexibility in applying reduction algorithms
      algebraically/numerically, depending on phase space regions
\item numerically stable representations of reduction building blocks
\end{enumerate}
In our approach, point 3. above is achieved
by means of an iterative algebraic reduction which decomposes
any $N$-point one-loop
scalar/tensor integral into  an infrared-finite part and an
infrared-divergent part.
No regulator masses for soft and collinear divergences are needed
in our formalism as we regulate all divergences by
working in  $n=4-2\epsilon$ dimensions.
Our reduction endpoints (``basis integrals") are chosen such that
all IR divergences are contained in 3-point and 2-point integrals,
which have simple analytical representations, allowing for
a straightforward isolation of the $1/\epsilon$ poles.
The most complicated building blocks of our reduced amplitude,
the 4-point functions in $n+2$ and $n+4$ dimensions, are always free from
IR divergences.

We thus express all loop diagrams algebraically as linear combinations of
spinors, (contracted) Lorentz tensors and form factors.
The form factors are represented on a basis of 1-,2-,3- and 4-point
functions. The special feature of our set of basis integrals
consists in the fact that it is carefully designed not to
introduce dangerous denominators
which are present in many standard approaches.
As proven in \cite{Binoth:2005ff}, this can only be achieved if
some of the basis integrals are not purely scalar integrals, but do have
Feynman parameters in the numerator. More in detail, our basis integrals are,
apart from trivial 1- and
2-point functions: 3-point functions $I_3$ in $d=n$ and $d=n+2$ dimensions and
4-point functions $I_4$ in $d=n+2$ and $d=n+4$ dimensions, where
$I_3^{n}$ and $I_4^{6}$ can have up to three Feynman parameters in
the numerator, and $I_3^{n+2}$ and $I_4^{n+4}$ can have up to one
Feynman parameter in the numerator. The short-hand notation for this basis set
is thus
\begin{equation}
\label{basis_int}
\{ I_3^{n}(1,j_1,j_1j_2,j_1j_2j_3),I_3^{n+2}(1,j_1) ,
I_4^{6}(1,j_1,j_1j_2,j_1j_2j_3),I_4^{n+4}(1,j_1) \}\;.
\end{equation}
We have shown that  {\it any} $N$-point one-loop amplitude
can be expressed in terms of
this basis, such that no inverse Gram determinants are introduced at all,
and a proliferation  of further higher dimensional integrals is avoided.
The evaluation of any $N$-point amplitude represented in this way
is therefore reduced to
the evaluation of the basis elements. This point refers to
item 4. of the above list. In our approach, the evaluation of the
integrals in (\ref{basis_int})
can be done optionally by further algebraic reduction,
which offers the
possibility to algebraically simplify the expressions further.
This proved useful
in the amplitude computations for $gg\to \gamma\gamma g$
\cite{Binoth:2003xk}
and $gg\to V^*V^* \to l\nu l^\prime \nu^\prime$ \cite{Binoth:2005ua,Drollinger}, both
relevant to
Higgs phenomenology at the LHC. On the other hand, we
provide switches to numerical representations of our building blocks.
The latter are completely free of algebraic objects which might induce
numerical problems.
By combining these two possibilities, using always the one which is more
appropriate in the corresponding phase space regions,
one arrives at algebraic amplitude representations --
which allow for a fast evaluation -- in the bulk of the phase space
and robust numerical representations in critical phase space regions.
For the latter we propose a method based on contour deformation
of multi-dimensional parameter integrals to numerically evaluate our basis
integrals.
The evaluation of processes given in the ``wishlist'' using these methods
is presently  under construction.

\paragraph{The GRACE approach}
The GRACE system is a highly automatised tool for the computation
of total 
%integrated 
cross sections 
%and event generation
both at tree and loop level, starting
from the Feynman diagram generation, complex Dirac and tensor
algebra, loop integration and integration over phase space and
event generation. 
Recently a series of $2 \rightarrow 3$ and a couple of $2 \rightarrow 4$ 
(see Tables~\ref{multileg-table2},\ref{multileg-table3}) processes in
the electroweak sector that involve a very large number of Feynman
diagrams have been computed demonstrating the power of the system.
To carry this success to NLO QCD multi-leg processes, some new
techniques have recently been developed in particular to deal with
massless particles.

For the electro-weak processes with massive particles circulating
in the loop, all tensor reductions of two, three and four-point
functions are performed by solving a system of equations obtained
by taking derivatives with respect to the Feynman
parameters~\cite{Belanger:2003STD}. All higher orders parametric
integrals corresponding to the tensor integrals can then be
recursively derived from the scalar integral. The two-point
integrals are implemented using simple analytical formulas. The
scalar 3-point function and all but the infra-red divergent
4-point scalar functions are evaluated through a call to the FF
package~\cite{vanOldenborgh:1990yc}. For the infrared four-point
function, GRACE supplies its own optimized routines through some
rather simple analytical results~\cite{Fujimoto:1990tb} that lead
to an efficient complete cancellation of infrared divergences
between these loop functions and the infrared factors from the
real soft bremsstrahlung part. Although this tensor reduction can
potentially lead to instabilities due to inverse Gram
determinants, this has not been an issue for the host of multi-leg
processes that we studied.

To extend the system, a fully numerical method to calculate loop
integrals, a numerical contour-integration method, has recently
been developed~\cite{Kurihara:2005ja}. Loop integrals can be
interpreted as a contour integral in a complex plane for an
integrand with multi-poles in the plane. Stable and efficient
numerical integrations along an appropriate contour can be
performed for scalar and tensor integrals appearing in the loop
calculations of the standard model.

For the massless case as would be needed for NLO-QCD processes, a
set of one-loop vertex and box tensor integrals with massless
internal particles has been obtained directly without any
reduction method to scalar-integrals~\cite{Kurihara:2005at}.
Results with one or two massive external lines for the vertex
integral and up to one massive external line for the box integral
have been developed. The dimensionally regularised functions allow
to extract the infra-red and collinear poles. The results are
expressed through very compact formulas for an easy numerical
implementation. The  tensor integrals for the box with two or more
off-shell external legs are under development.

%% 2 %%%%
%Two-loop integral is developed through the numerical method that
%is made by GRACE group for the one-loop
%integral~\cite{Fujimoto:1993fz}. Two methods are proposed to
%control the singularity of the integrand. In the first method, the
%singularity is tamed by symmetrization of the integrand. The
%second method is a kind of hybrid method in which both analytic
%manipulation and numerical integration are used. It is shown that
%methods can evaluate the scalar two-loop functions.

%%% 3 %%%%

%%% 4 %%%

A method to construct event-generators based on next-to-leading
order QCD matrix-elements and leading-logarithmic parton showers
is developed by a GRACE group~\cite{Kurihara:2002ne}. Matrix
elements of loop diagrams as well as those of a tree level can be
generated using an automatic system. A soft/collinear singularity
is treated using a leading-log subtraction method.

The PDF and PS include leading-log(LL) terms of the initial-state
parton emission. If one combines the matrix element with the PDF
or PS very naively, one cannot avoid double-counting of these
LL-terms. Our proposal to solve this problem is to subtract the
LL-terms from the exact matrix elements as
\begin{eqnarray}
\sigma_{LLsub}&=&\frac{1}{(2p^0_1)(2p^0_2)v_{rel}}
\int_{\Omega_{vis}} d\Phi^{(4)}_{N+1} \biggl[ \biggl|{\cal
M}^{(4)}_{N+1}\biggr|^2-
\biggl|{\cal M}^{(4)}_{N}(s x)\biggr|^2 f_{LL}(x,s) \biggr], \nonumber \\
f_{LL}(x,s)&=&16 \pi^2\biggl(\frac{\alpha_s}{2 \pi}\biggr)
\frac{{\tt P}^{
(1)}(x)}{k_\bot^2}\biggl(\frac{1-x}{x}\biggr). \nonumber
\end{eqnarray}
The second term of the integrand is the LL-approximation of the
real-emission matrix-elements under the collinear condition.
Higher soft/collinear correction by the parton shower method is
combined with the NLO matrix-element without any double-counting
in this method.

%\subsubsection{Public programs}

%The only NLO programs for $2\to 3$ processes in hadronic collisions
%which are publicly available so far are
%\begin{tabular}{lll}
%{\tt NLOJET++}& $pp\to 3$\,jets& Z.~Nagy \cite{Nagy:2001fj}\\
%{\tt MCFM}&
%$ pp\to V + 2\,{\rm jets}$, & J.~Campbell et al.
%\cite{Campbell:2000bg,Campbell:2002tg,Campbell:2003hd}\\
%&$pp\to Z+b\bar{b}$&\\
%\end{tabular}

\subsubsection{Combination with parton showers}

To date, experimentalists have been more comfortable with predictions at the
hadron level produced by interfaces to parton shower and hadronization
programs. The Les Houches Accord (2001) provides an interface between matrix
element and parton shower/hadronization programs but cannot be
used directly for the
NLO processes discussed thus far. There is the danger of double-counting of
some of the higher order corrections since these can also be produced by the
parton shower as well as by the matrix element.  In addition,
 the more complex matrix element calculations contain many contributions
with large negative
weights, which are not conducive to a Monte Carlo framework.

In a parton shower interface, a  specific subtraction
scheme must be implemented to  preserve the NLO cross section. As each parton
shower Monte Carlo may produce a different real radiation component,
the subtraction
scheme must necessarily depend on the Monte Carlo program to which the matrix
element program is matched.
The presence of
interference effects with NLO calculations requires that a
relatively small fraction ($\sim$ 10\%) of events have negative weights (of value -1).

Several groups have worked on the subject to consistently
combine partonic NLO calculations with parton showers.
\begin{itemize}
\item  Collins, Zu \cite{Chen:2001nf,Collins:2004vq}
\item  Frixione, Nason, Webber (MC@NLO)
\cite{Frixione:2002ik,Frixione:2003ei}
%\cite{Frixione:2005gz}
\item  Nason \cite{Nason:2004rx}
\item  Kurihara, Fujimoto, Ishikawa, Kato, Kawabata,
Munehisa, Tanaka \cite{Kurihara:2002ne}
\item  Kr\"amer, Soper \cite{Kramer:2003jk,Soper:2003ya,Kramer:2005hw}
\item  Nagy, Soper \cite{Nagy:2005aa,Nagy:2006kb}
\end{itemize}

MC@NLO is the only publically available program that combines NLO calculations
with parton showering and hadronization. The Herwig Monte Carlo is used for
the latter. The processes included to date are: $(W,Z,\gamma*,H,b\overline{b},
t\overline{t},HW,HZ,WW,WZ,ZZ$). Recently, single top hadroproduction has
been added to MC@NLO~\cite{Frixione:2005vw}. This is the first implementation of a process that has
both initial- and final-state singularities. This allows a more general
category of additional processes to be added in the future.
Work is proceeding on inclusion of inclusive
jet production and WW fusion to Higgs. Adding spin correlations to a process
increases the level of difficulty but is important for processes such as
single top production.
%Priorities before the beginning of the LHC include...

%This subject will be treated extensively in other parts of
%the proceedings, therefore no further details will be given here.

%%%%%%%%%%%%%%%%%%%%%%%%%%%%%%%%%%%%%%%%%%%%%%%%%%%%%%%%%%%%%%%%%%%%%%%%%%%%%
\section[One loop gluon initiated corrections in DIPHOX]
{ONE LOOP GLUON INITIATED CORRECTIONS IN DIPHOX~\protect
\footnote{Contributed by: F.~Mahmoudi}}
\subsection{Introduction}
Direct photon pair production is an important background for
the low mass Higgs. An analysis of the diphoton background at LHC has been
performed in \cite{Bern:2002}. A full next to leading order study of direct
photon pair production in hadronic colliders is also incorporated in DIPHOX
\cite{Binoth:2000}, except for $gg \rightarrow \gamma\gamma + X$ at NLO.\\
Because LHC is a gluon factory, gluon fusion is a very important background
for Higgs search at LHC. In this work, we focus on the implementation of the
real contribution of gluon initiated processes of photoproduction, {\it i.e.}
one loop $gg \rightarrow \gamma\gamma + g$ diagrams, into DIPHOX\footnote{Note
that in order to have the complete NLO corrections, one has to add the two
loop virtual corrections as well.}. We consider the result of the direct
calculation of this amplitude \cite{Binoth:2004}. After reviewing the
analytical framework, we will investigate some numerical results from DIPHOX.
\subsection{Analytical framework}
Two topologies are associated to $gg \rightarrow \gamma\gamma + g$: 
%(Fig. \ref{topo}).
%
\begin{figure}[!h]
\begin{center}
\includegraphics[width=0.25\textwidth]{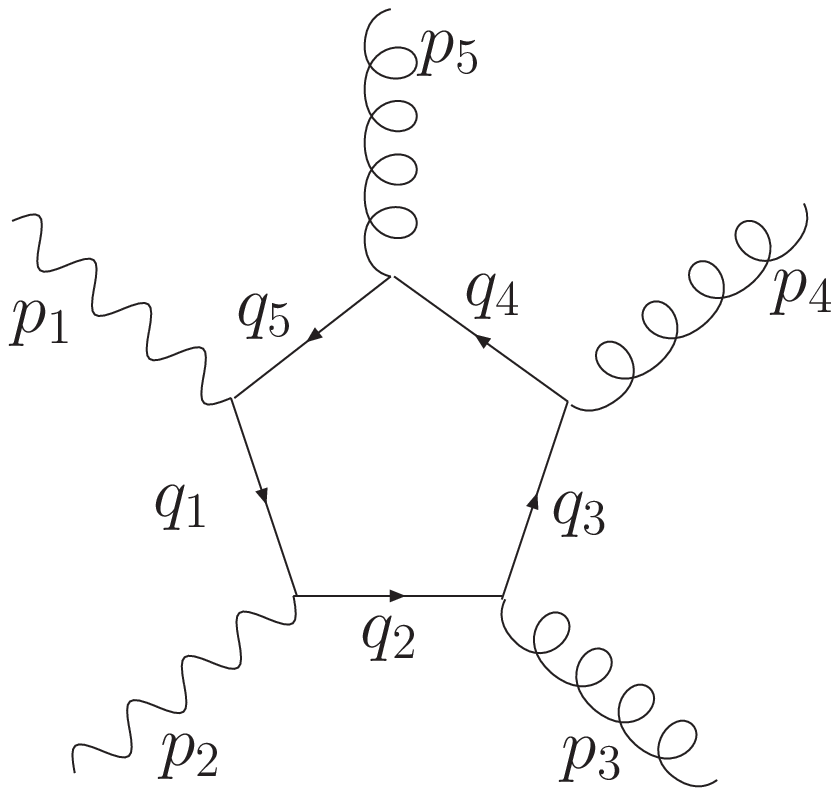}
~~~~~~~~~~~\includegraphics[width=0.25\textwidth]{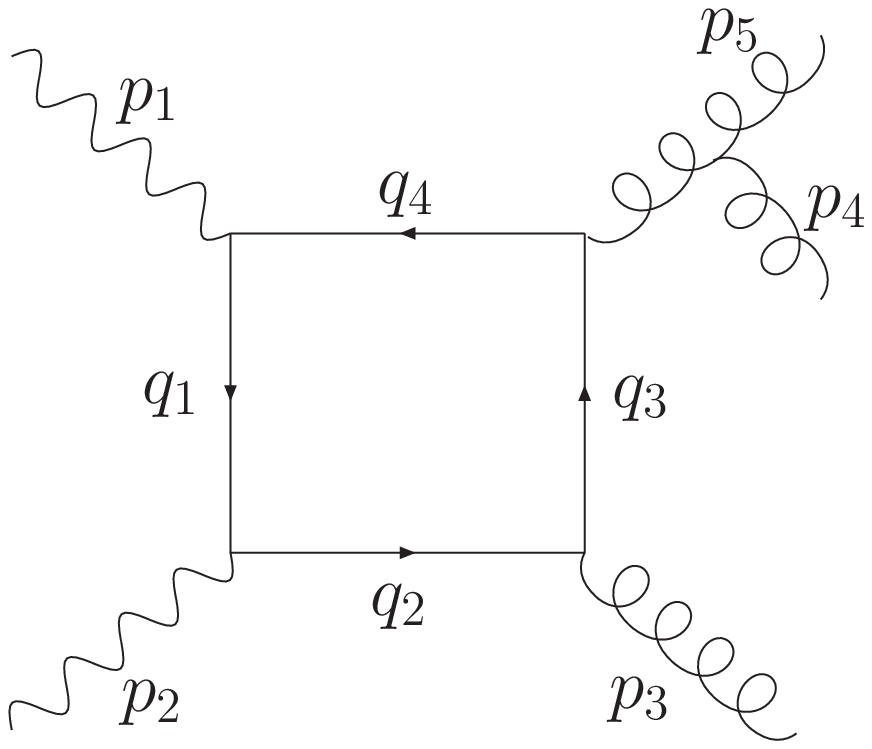}
 \caption{The two topologies of $gg \rightarrow \gamma\gamma + g$.}
\label{topo}
\end{center}
\end{figure}
\\ 
The calculation of the matrix element squared has been performed thanks to
an analytic FORM\cite{Vermaseren:2001}/MAPLE\cite{Maple:2005} code, and a
numerical Fortran code.\\ The analytical amplitudes can be written in function
of the field strength tensor ${\cal F}$, where ${\cal F}$ is defined
as~\cite{Binoth:2004}: \\
\begin{equation}
{\cal F}^{\mu\nu}_j=p^\mu_j\epsilon^\nu_j-p^\nu_j\epsilon^\mu_j \;\;.
\end{equation}
The representation in function of the field strength tensor is not unique, and
it is possible to find a common structure for all helicity amplitudes:
\begin{eqnarray}
{\cal A}^{+++++} &=& \frac{ \mbox{Tr}({\cal F}_1^+{\cal F}_2^+)\mbox{Tr}({\cal F}_3^+{\cal F}_4^+{\cal F}_5^+) }{2\,s_{34}s_{45}s_{35} }\\
{\cal A}^{-++++} &=& \frac{\mbox{Tr}({\cal F}_2^+{\cal F}_3^+) \mbox{Tr}({\cal F}_4^+{\cal F}_5^+)}{s_{23}^2 \, s_{45}^2} \Bigl(C_1^{-++++}\,(p_2 \cdot {\cal F}_1^- \cdot p_4) + C_2^{-++++}\,(p_2 \cdot {\cal F}_1^- \cdot p_5) \Bigr)~~~~~~~~\\
{\cal A}^{++++-} &=& \frac{\mbox{Tr}({\cal F}_1^+{\cal F}_2^+) \mbox{Tr}({\cal F}_3^+{\cal F}_4^+)}{s_{12}^2 \, s_{34}^2} \Bigl(C_1^{++++-}\,(p_1 \cdot {\cal F}_5^- \cdot p_3) + C_2^{++++-}\,(p_1 \cdot {\cal F}_5^- \cdot p_4) \Bigr)~~~~~~~~\\
{\cal A}^{--+++} &=& \frac{\mbox{Tr}({\cal F}_1^-{\cal F}_2^-) \mbox{Tr}({\cal F}_3^+{\cal F}_4^+)}{s_{12}^2 \, s_{34}^2} \Bigl(C_1^{--+++}\,(p_3 \cdot {\cal F}_5^+ \cdot p_2) + C_2^{--+++}\,(p_3 \cdot {\cal F}_5^+ \cdot p_1) \Bigr)~~~~~~~~\\
{\cal A}^{+++--} &=& \frac{\mbox{Tr}({\cal F}_1^+{\cal F}_2^+) \mbox{Tr}({\cal F}_4^-{\cal F}_5^-)}{s_{12}^2 \, s_{45}^2} \Bigl(C_1^{+++--}\,(p_4 \cdot {\cal F}_3^+ \cdot p_2) + C_2^{+++--}\,(p_4 \cdot {\cal F}_3^+ \cdot p_1) \Bigr)~~~~~~~~\\
{\cal A}^{-+++-} &=& \frac{\mbox{Tr}({\cal F}_2^+{\cal F}_3^+) \mbox{Tr}({\cal F}_1^-{\cal F}_5^-)}{s_{23}^2 \, s_{15}^2} \Bigl(C_1^{-+++-}\,(p_2 \cdot {\cal F}_4^+ \cdot p_5) + C_2^{-+++-}\,(p_2 \cdot {\cal F}_4^+ \cdot p_3) \Bigr)~~~~~~~~
\end{eqnarray}\\
One can obtain the coefficients $C_1$ et $C_2$ using the $C$'s of
\cite{Binoth:2004}.  $C_1$ and $C_2$ contain logarithmic terms (from two point
functions), 6 dimensional integrals (four point functions) and constant
terms. These amplitudes share a basic structure in $\mbox{Tr}({\cal
F}_i^\pm{\cal F}_j^\pm)$, $p_i \cdot {\cal F}_k^\pm \cdot p_j$, with real and
complex coefficients. It is possible to write the field strength tensor terms
in function of spinor products \cite{Mahmoudi:2004}.  One can show that:
\begin{equation}
\mbox{Tr}({\cal F}_i^+{\cal F}_j^+) = -\bigl(\langle \, p_i \, p_j \rangle^* \bigr)^2 \;\;,
\end{equation}  
\begin{equation}
\mbox{Tr}({\cal F}_i^-{\cal F}_j^-) = -\bigl(\langle \, p_i \, p_j \rangle \bigr)^2\;\;,
\end{equation} 
and
\begin{equation}
\mbox{Tr}({\cal F}_i^+{\cal F}_j^+{\cal F}_k^+) = \frac{1}{\sqrt2} \, \langle \, p_i \, p_j \rangle^* \, \langle \, p_j \, p_k \rangle^* \, \langle \, p_k \, p_i \rangle^* \;\;.
\end{equation}  
And for the scalar products:
\begin{eqnarray}
p_i \cdot {\cal F}_k^+ \cdot p_j &=& \frac{1}{2\sqrt2} \, \langle \, p_i \, p_j \rangle \, \langle \, p_i \, p_k \rangle^* \, \langle \, p_k \, p_j \rangle^* \;\;,\\
p_i \cdot {\cal F}_k^- \cdot p_j &=& \frac{1}{2\sqrt2} \, \langle \, p_i \, p_j \rangle^* \, \langle \, p_i \, p_k \rangle \, \langle \, p_k \, p_j \rangle \;\;.
\end{eqnarray}
The six dimensional box integrals can be calculated analytically
\cite{Binoth:2001}, and written as:
\begin{equation}
\frac{1}{a+b-c} \, I^{6}_4(a,b,c) = Li_2\Bigl( 1 - \frac{c}{a} \Bigr) + Li_2\Bigl( 1 - \frac{c}{b} \Bigr)
- Li_2\Bigl(- \frac{b}{a} \Bigr) - Li_2\Bigl(- \frac{a}{b} \Bigr) \;\;.
\end{equation}
Thanks to these relations, it is possible to calculate the total squared amplitude. This result has been implemented into DIPHOX.\\
\\
Another interesting study concerns the collinear limits of the amplitudes,
which on one side, allows the study of the cancellations of singularities, and
on the other side, can be used as a cross-check of the results. For example,
in the case of $3\parallel 5$, one can define $p_3 = z \, P_{35}$ and $p_5 = (1-z)
\, P_{35}$ and show that \cite{Mahmoudi:2004}:
\begin{eqnarray}
{\cal A}^{++++-}_{p_3\parallel p_5} &=& \frac{1}{2\sqrt2} \,\frac{1}{\langle \, p_3 \, p_5 \rangle^*} \,  \frac{z^2}{\sqrt{z(1-z)}} \frac{\langle \, p_1 \, p_2 \rangle^*}{\langle \, p_1 \, p_2 \rangle} \, \frac{\langle \, P_{35} \, p_4 \rangle^*}{\langle \, P_{35} \, p_4 \rangle} \nonumber\\
&& + \frac{1}{2\sqrt2} \, \frac{1}{\langle \, p_3 \, p_5 \rangle} \,\frac{(1-z)^2}{\sqrt{z(1-z)}} \, \frac{\langle \, P_{35} \, p_4 \rangle}{\langle \, p_1 \, p_2 \rangle}\, \frac{\langle \, P_{35} \, p_2 \rangle}{\langle \, p_1 \, p_4 \rangle}\, \frac{\langle \, p_2 \, p_4  \rangle^*}{\langle \, p_2 \, p_4 \rangle} \;\;.
\end{eqnarray}\\
%One can notice that in this limit, the amplitude is proportional to $ 1/ \langle \, p_3 \, p_5 \rangle$.  
Thus, the amplitude appears as the superposition of the amplitude of 2 gluons
$\rightarrow$ 2 photons with a positive helicity for particle 3, plus the
amplitude with a negative helicity for particle 3. The coefficients
$z^2/\sqrt{z(1-z)}$ and $(1-z)^2/\sqrt{z(1-z)}$ are related to the well-known
splitting functions \cite{Berends:1988,Bern:1994}, the first one corresponding
to the case where a gluon of positive helicity produces a collinear gluon of
positive helicity and of momentum fraction $z$, whereas the second one
corresponds to the case where a gluon of positive helicity produces a
collinear gluon of negative helicity.  Hence, one can write the amplitude
under the form:
\begin{equation}
{\cal A}_{p_3\parallel p_5}^{\lambda_1 \lambda_2 \lambda_3 \lambda_4 \lambda_5}(p_1,p_2,p_3,p_4,p_5) = \sum_{\lambda=\pm} S_{\lambda}(p_3^{\lambda_3},p_5^{\lambda_5}) \, {\cal A}^{\lambda_1 \lambda_2 (-\lambda) \lambda_4}(p_1,p_2,p_3+p_5,p_4)\;\;,
\end{equation} 
where $\lambda$ refers to the helicity and $S$ are the splitting
functions. Reproducing this well-known result constitutes another test of the
amplitudes.
\subsection{Numerical results}
\begin{figure}
\begin{center}
\includegraphics[width=0.5\textwidth]{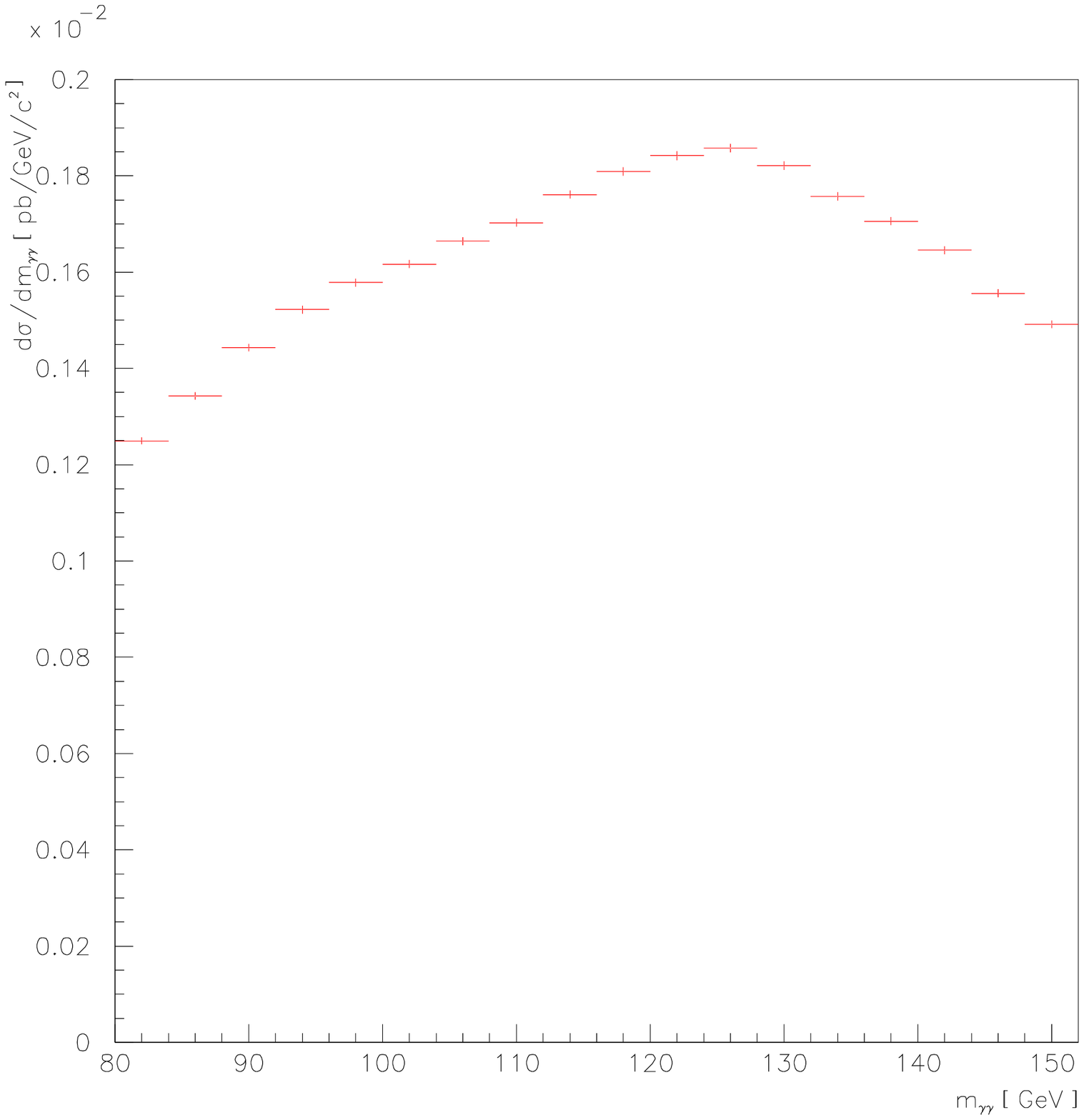}
 \caption{Invariant mass distribution of exclusive production of two photons and one jet.}
\label{mgg}
\end{center}
\end{figure}

\begin{figure}
\begin{center}
\includegraphics[width=0.5\textwidth]{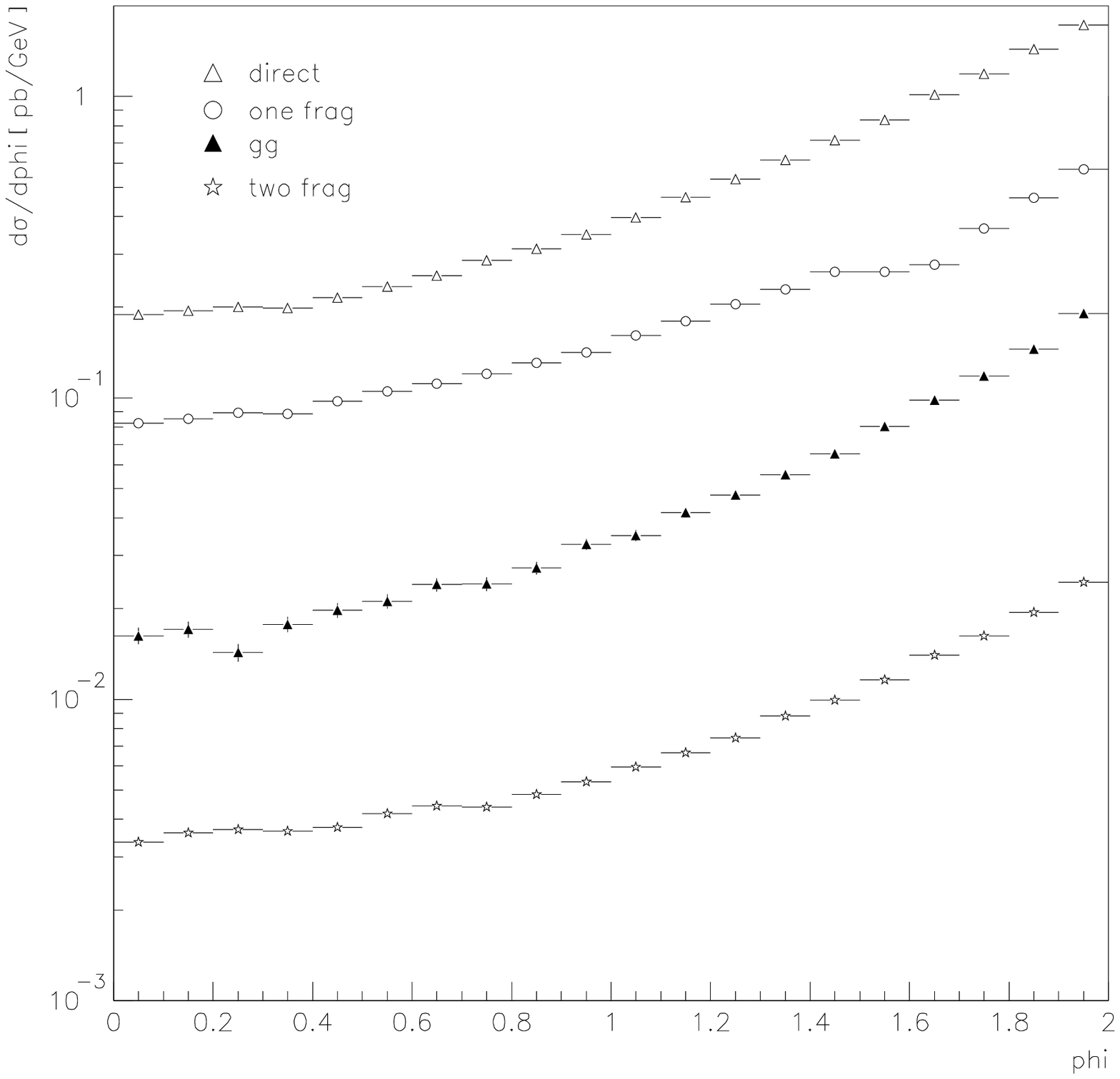}
\includegraphics[width=0.5\textwidth]{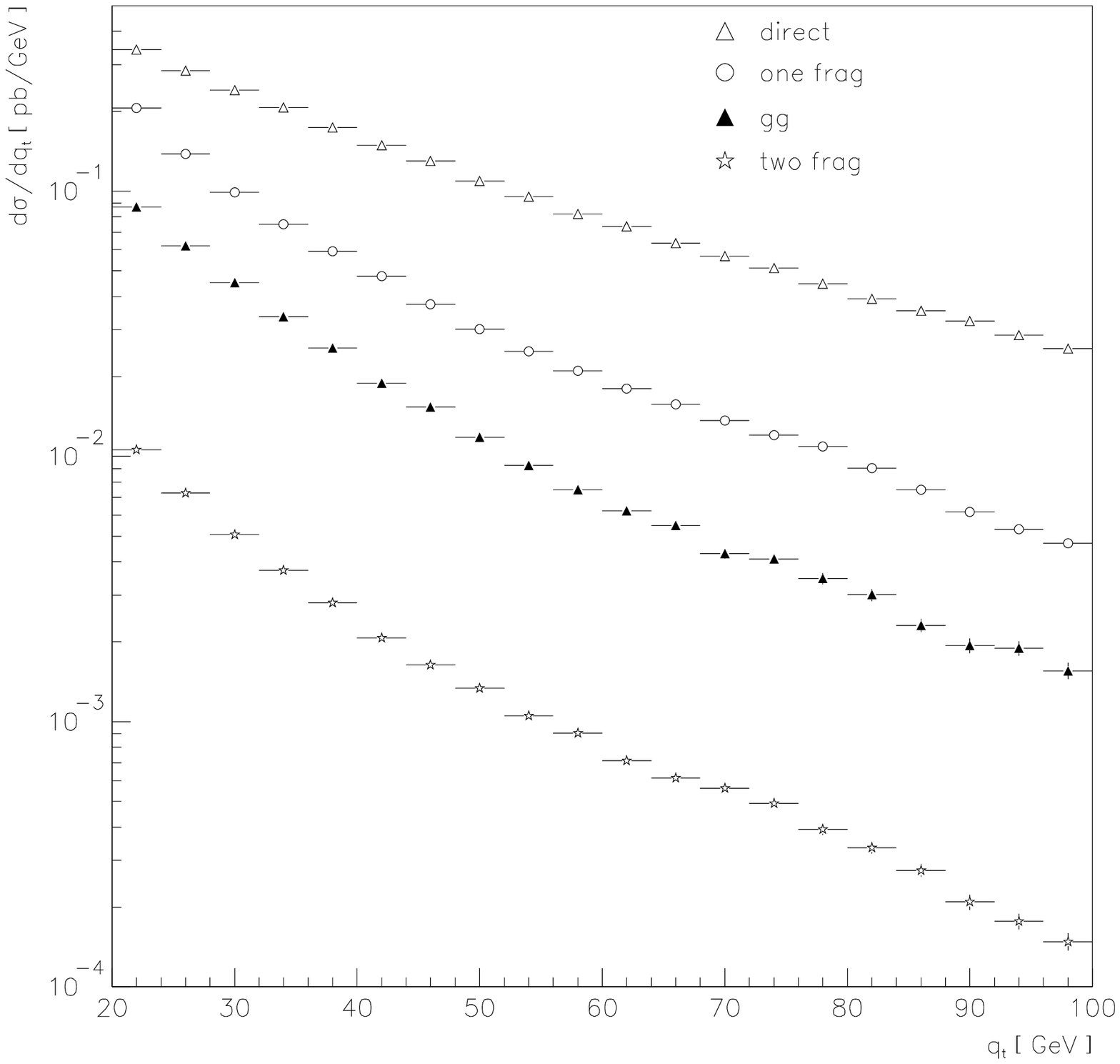}
 \caption{Gluon initiated cross section versus transverse momentum of the photon pair (on the left) and versus the azimuthal angle between the two photons (on the right), compared to the cross sections of direct, one and two fragmentation photoproduction processes, for $\mu^2=M^2=\frac{1}{4} m^2_{\gamma\gamma}$.}
\label{qt}
\end{center}
\end{figure}

We use for the spinor products definitions given in
\cite{Xu:1987}. Considering a quadrivector $k^\mu$ one can use the following
notations:
\begin{equation}
k_\pm = k_0 \pm k_z \;\;,
\end{equation}
and
\begin{equation}
k_\bot = k_x + i \, k_y =| k_\bot | e^{i \varphi_k} = \sqrt{k_+ k_-} e^{i \varphi_k} \;\;.
\end{equation}
With an adequate choice of phase, we can find:
\begin{equation}
| k_+\rangle=
\left[ \begin{array}{c}
\sqrt{k_+} \\  \sqrt{k_-} e^{i \varphi_k} \\ 0 \\ 0 \end{array} \right] \;\;,
\end{equation}
and
\begin{equation}
| k_-\rangle=\left[ \begin{array}{c} 0 \\ 0 \\ \sqrt{k_-} e^{-i \varphi_k} \\ -\sqrt{k_+} \end{array} \right] \;\;.
\end{equation}
Consequently, one has:
\begin{equation}
\langle k_1  k_2\rangle = \langle k_{1-}  | k_{2+}\rangle =\sqrt{k_{1-}k_{2+}} e^{i \varphi_1} - \sqrt{k_{1+}k_{2-}} e^{i \varphi_2} \;\;.
\end{equation}
Using these relations, one can evaluate the numerical value of each amplitude,
and so the matrix element squared. To avoid the numerical difficulties
concerning collinear singularities, we used a multiprecision package
\cite{Bailey:1993} which enables an arbitrary level of numeric precision.\\ \\
To implement this matrix element squared into DIPHOX, we follow the same
procedure as presented in \cite{Binoth:2000}. Let us consider the physical
case:
\begin{equation}
g(p_1)+g(p_2) \longrightarrow \gamma(p_3) + \gamma(p_4) + g(p_5) \;\;.
\end{equation}
In this case, the only possible collinear singularities come from $p_1
\parallel p_5$ or if $p_2 \parallel p_5$. To isolate the part containing
potential collinear or infrared singularities, we consider in the matrix
element squared the coefficient of the eikonal factor:
\begin{equation}
E_{12}  = \frac{p_1.p_2}{p_1.p_5 \; p_2.p_5} \;\;,
\end{equation}
which we call $H_{12}(p_5)$. Therefore the matrix element squared can be written as:
\begin{equation}
|M|^2 =  H_{12}(p_5) \, E_{12} \;\;.
\end{equation}
$H_{12}$ can then be implemented into DIPHOX to obtain the cross section. \\
\\ 
We consider the LHC energy and use the kinematic cuts from the ATLAS and
CMS proposals \cite{CMS:1994,ATLAS:1994}, specifically for the transverse
momentum of the most energetic photon (photon 1): $p_{T\,(\gamma_1)} > 40$
GeV, and the rapidity: $|\eta| < 2.5$. The isolation criterion of photons
requires that the amount of hadronic transverse energy deposed inside a cone
of a certain radius and oriented towards the photon is smaller than
$E_{T\,max}$, fixed by the experiment.  We used the CTEQ 6 set of parton
distribution functions \cite{Pumplin:2002}.\\ \\ In Fig. \ref{mgg} is plotted
the invariant mass distribution of the photon pair for the exclusive
production of two photons and one jet. We require each pair of particles to be
separated by at least 0.3 in the rapidity--azimuthal angle space:
\begin{equation}
\Delta R_{sep} = \sqrt{\Delta\phi^2 + \Delta\eta^2} > 0.3 \;\;,
\end{equation} 
where $\phi$ is the azimuthal angle between the photons. The factorization and renormalization scales are given by:
\begin{equation}
\mu^2=M^2=m^2_{\gamma\gamma} + p_{T\,(\mbox{jet})}^2 \;\;,
\end{equation}
where $m_{\gamma\gamma}$ corresponds to the invariant mass of the photon pair.
The results are in agreement with those of \cite{DeFlorian:1999}.\\
\\
In Fig. \ref{qt} is represented the cross-section of inclusive production of
two photons in function of the transverse momentum of the photon pair, and in
function of the azimuthal angle between the photons. Using the kinematic cuts
of ATLAS and CMS, one imposes for the transverse momentum of photon 1:
\begin{equation}
p_{T\,(\gamma_1)} > 40\mbox{ GeV  ,}
\end{equation}
and for the transverse momentum of photon 2:
\begin{equation}
p_{T\,(\gamma_2)} > 25\mbox{ GeV  .}
\end{equation}
Here we choose the renormalization and factorization scales as:
\begin{equation}
\mu^2=M^2=\frac{1}{4} m^2_{\gamma\gamma}\;\;.
\end{equation}
The results depend on the scale choice. For the isolation criterion of the
photons, we now impose:
\begin{equation}
E_{T\,max} > 15 \mbox{ GeV} \;\;,
\end{equation} 
and, for the radius of the isolation cone:
\begin{equation}
R = \sqrt{\Delta\phi^2 + \Delta\eta^2} > 0.4 \;\;,
\end{equation}
with an acollinearity cut between photons of:
\begin{equation}
m_{\gamma\gamma} > 80 \mbox{ GeV} \;\;.
\end{equation}
For the transverse momentum distribution, we require a transverse momentum cut $q_T > 20$ GeV for the photon pair.\\
\\
As a comparison, the contributions of \textit{direct}, \textit{one
fragmentation} and \textit{two fragmentations} photoproduction processes are
also presented on the figure. One can notice that, for this choice of scales,
the contribution of $g g \longrightarrow \gamma \gamma + jet$ is smaller than
the direct and one fragmentation contributions, but bigger than the two
fragmentation contribution.

\subsection{Conclusions}
The one loop $3 g 2 \gamma$ corrections ($\alpha_S^3$ order) have been
incorporated into DIPHOX using the results of the direct calculation of the
matrix element squared. The results are in agreement with those extracted from
the $5 g$ amplitude. A comparison between transverse momentum and azimuthal
distributions of the photon pair of direct, one fragmentation, two
fragmentation and gluon initiated processes has been performed. The next step
would consist in the implementation of the virtual correction of two loop
gluon initiated process into DIPHOX to obtain the complete NLO accuracy.

\subsection*{Acknowledgements}
I would like to thank Jean-Philippe Guillet for his help regarding DIPHOX, and
for his remarks and suggestions. I would like to thank also Thomas Binoth for
useful discussions. I acknowledge the support of the McCain Fellowship at
Mount Allison University. Feynman diagrams in this paper are drawn with
Jaxodraw \cite{Binosi:2004}.

%%%%%%%%%%%%%%%%%%%%%%%%%%%%%%%%%%%%%%%%%%%%%%%%%%%%%%%%%%%%%%%%%%%%%%%%%%%%%
\section[The architecture of NNLO cross sections]
{THE ARCHITECTURE OF NNLO CROSS SECTIONS~\protect
\footnote{Contributed by: V.~Del~Duca, A.~Gehrmann-De Ridder, T.~Gehrmann,
    E.W.N.~Glover, G.~Heinrich, G.~Somogyi, Z.~Tr\'ocs\'anyi}}
\subsection{Introduction}

QCD is an important component of the Standard Model, and will play a 
fundamental role at the LHC. Within the known Standard Model, it will be
important to have an evaluation as precise as possible of the strong
coupling constant $\alpha_S$, of the parton distributions, of the
electroweak parameters, and of the LHC parton luminosity. Beyond that,
a precise determination of Higgs and New-Physics production, and
particularly of their backgrounds, will be essential, in order to
interpret New-Physics signals.

At high $Q^2$ any production rate can be expressed as a series expansion
in $\alpha_S$. Because QCD is asymptotically free, the simplest approximation
is to evaluate any series expansion at leading order in $\alpha_S$.
However, for most processes a leading-order evaluation yields unreliable 
predictions. The next simplest approximation is a NLO evaluation, which
usually allows for a satisfying assessment of the production rates.
In the past 25 years, many efforts have been made to compute production
rates at NLO and to devise process-independent 
methods~\cite{Kunszt:1992tn,Giele:1991vf,Giele:1993dj,Frixione:1995ms,
Catani:1996vz,Nagy:1996bz,Frixione:1997np,Kosower:1997zr,Campbell:1998nn,Kosower:2003bh} to compute rates
at NLO accuracy (see the NLO section of these proceedings).
In some cases, though, the NLO corrections may not be accurate enough.
Specimen cases are: the extraction of $\alpha_S$ from the data, where 
in order to avoid that the main source of uncertainty be due to the 
NLO evaluation of some production rates, like the event shapes of jet
production in $e^+e^-$ collisions, only observables evaluated at 
next-to-next-to-leading order (NNLO)
accuracy are considered~\cite{Bethke:2004uy}; 
open $b$-quark production at the Tevatron,
where the NLO uncertainty bands are too large to test the 
theory~\cite{Cacciari:2003uh} {\it vs.} the data~\cite{Acosta:2004yw};
Higgs production from gluon fusion in hadron collisions, where it is known
that the NLO corrections are large~\cite{Graudenz:1992pv,Spira:1995rr},
while the NNLO 
corrections~\cite{Harlander:2002wh,Anastasiou:2002yz,Ravindran:2003um}, 
which have been evaluated in the large-$m_t$ limit,
display a modest increase, of the order of less than 20\%, with respect to 
the NLO evaluation; Drell-Yan productions of $W$ and $Z$ vector bosons at LHC,
which can be used as ``standard candles'' to measure the parton luminosity
at the LHC~\cite{Dittmar:1997md,Khoze:2000db,Giele:2001ms,Frixione:2004us}.

In the last few years the NNLO corrections have been computed to the
total cross section~\cite{Harlander:2002wh,Hamberg:1990np} and to the
rapidity distribution~\cite{Anastasiou:2003yy,Anastasiou:2003ds} of
Drell-Yan production, to the total cross section for the production of
a scalar~\cite{Harlander:2002wh,Anastasiou:2002yz,Ravindran:2003um} and
a pseudoscalar~\cite{Harlander:2002vv,Anastasiou:2002wq} Higgs from
gluon fusion as well as to a fully differential cross
section~\cite{Anastasiou:2004xq,Anastasiou:2005qj}, and to jet
production in $e^+e^-$
collisions~\cite{Anastasiou:2004qd,Gehrmann-DeRidder:2004tv,
Gehrmann-DeRidder:2005cm}. The methods which have been used are
disparate: analytic integration, which is the first method to have been
used~\cite{Hamberg:1990np}, cancels the divergences analytically, and
is flexible enough to include a limited class of acceptance cuts by
modelling cuts as propagators~\cite{Anastasiou:2002yz,Anastasiou:2003yy,
Anastasiou:2003ds,Anastasiou:2002wq}; sector
decomposition~\cite{Anastasiou:2004xq,Anastasiou:2004qd,Roth:1996pd,
Binoth:2000ps,Heinrich:2002rc,Anastasiou:2003gr,Gehrmann-DeRidder:2003bm,
Binoth:2004jv}, 
which is flexible enough to include any acceptance cuts and for which
the cancellation of the divergences is performed numerically;
subtraction~\cite{Gehrmann-DeRidder:2005cm,Gehrmann-DeRidder:2003bm,
Kosower:2002su,Weinzierl:2003fx,Weinzierl:2003ra,Kilgore:2004ty,
Frixione:2004is, Gehrmann-DeRidder:2005hi,Gehrmann-DeRidder:2005aw,
Somogyi:2005xz}, for which the cancellation of the divergences is
organised in a process-independent way by exploiting the universal
structure of the infrared divergences of a gauge theory, in particular
the universal structure of the three-parton tree-level splitting
functions~\cite{Berends:1988zn,Gehrmann-DeRidder:1997gf,
Campbell:1997hg,Catani:1998nv,Catani:1999ss,DelDuca:1999ha} and the
two-parton one-loop splitting functions~\cite{Bern:1994zx,Bern:1998sc,
Kosower:1999xi,Kosower:1999rx,Bern:1999ry}. 

The standard approach of subtraction to NNLO relies on defining approximate
cross sections which match the singular behaviour of the QCD cross
sections in all the relevant unresolved limits. In general, the definition
of the approximate cross sections must rely on the single
and double unresolved limits of the QCD squared matrix elements. 
Although, as outlined above, the infrared limits of the QCD matrix elements 
have been extensively studied,
the formulae presented in the literature do not lend themselves directly
for devising the approximate cross sections for two reasons.  The first
problem is that the various single and double soft and/or collinear
limits overlap in a very complicated way and the infrared-factorisation
formulae have to be written in such forms that these overlaps can be
disentangled, so that the double subtraction is avoided.  The second problem
is that even if the factorisation formulae are written such that the double
subtraction does not occur, the expressions cannot straightforwardly
be used as subtraction formulae, because the momenta of the partons in
the factorised matrix elements are unambiguously defined only
in the strict soft and collinear limits. In order to define the
approximate cross sections, one also has to factorise the phase space of
the unresolved partons such that the singular factors can be integrated
and the remaining expressions can be combined with the virtual correction,
leading to cross sections which are finite and integrable in four
dimensions.

In the sector decomposition approach, the singularities are isolated by an 
algebraic procedure acting on the integration variables, which is iterated 
in an automated way. The pole coefficients are integrated numerically. 
This method avoids the manual setup of a subtractionn scheme, but leads to 
rather large expressions as the number of original functions is increased in 
each iteration of the sector decomposition. 

In the following, variants of both the ``standard subtraction" and  the 
sector decomposition approach will be described briefly. 

\subsection{The approach of Del Duca, Somogyi, Tr\'ocs\'anyi~\protect
\footnote{Authors: V.~Del~Duca, G.~Somogyi, Z.~Tr\'ocs\'anyi}}
In the context of the standard subtraction, a subtraction scheme 
was presented in~\cite{Somogyi:2005xz} for processes without
coloured partons in the initial state. Namely, subtraction terms were
explicitly constructed, which regularise the kinematical singularities
of the squared matrix element in all singly- and doubly-unresolved
parts of the phase space, in such a way that the subtraction terms
avoid all possible double and triple subtractions. Thus, the
regularised squared matrix element is integrable over all the phase
space regions where at most two partons become unresolved.  In
particular, new factorisation formulae were presented in the iterated
singly-unresolved limits for the colour-correlated and the
spin-correlated squared matrix elements. It was pointed out that soft
factorisation formulae do not exist for the simultaneously spin- and
colour-correlated squared matrix elements, which indicates that within
the scheme envisaged in~\cite{Somogyi:2005xz} the azimuthally
correlated singly-collinear subtraction terms must not contain colour
correlations. This can be achieved naturally for those processes where
the colour charges in the colour-correlated squared matrix elements can
be factorized completely, which occurs only for processes with no more
than three coloured hard partons~\cite{Ellis:1980wv}, that is for
$e^+e^-\to 2, 3$~jets in this context. For processes with more coloured
hard partons a subtraction scheme that avoids such correlations was
outlined in Ref.~\cite{Somogyi:2005xz}.  However, that paper did not
consider the second problem mentioned above, namely the phase space of
the subtraction terms.  In the following, we outline a possible
solution to that.  In order to avoid a lengthy introduction to the
notation, we use the same notation as Ref.~\cite{Somogyi:2005xz}.

\subsubsection{Subtraction scheme at NNLO}

The NNLO correction to any $m$-jet production rate is
a sum of three contributions, the doubly-real, the one-loop
singly-unresolved real-virtual and the two-loop doubly-virtual terms,
\NNLObeq
\NNLOtsig{NNLO} =
\int_{m+2}\!\NNLOdsig{RR}_{m+2} J_{m+2}
+ \int_{m+1}\!\NNLOdsig{RV}_{m+1} J_{m+1}
+ \int_m\!\NNLOdsig{VV}_m J_m\:.
\label{eq:sigmaNNLO}
\NNLOeeq
The three contributions in \NNLOeqn{eq:sigmaNNLO} are separately divergent, 
but their sum is finite for infrared-safe observables.  As explained in 
\NNLORef{Somogyi:2005xz}, we rewrite \NNLOeqn{eq:sigmaNNLO} as
\NNLObeq
\NNLOtsig{NNLO}=
\int_{m+2}\!\NNLOdsig{NNLO}_{m+2}
+ \int_{m+1}\!\NNLOdsig{NNLO}_{m+1}
+ \int_m\!\NNLOdsig{NNLO}_m\:,
\label{eq:sigmaNNLOfin}
\NNLOeeq
where the integrands
\NNLObeq
\NNLOdsig{NNLO}_{m+2} =
\left[\NNLOdsig{RR}_{m+2} J_{m+2} - \NNLOdsiga{RR}2_{m+2} J_m 
- \NNLOdsiga{RR}1_{m+2} J_{m+1} + \NNLOdsiga{RR}{12}_{m+2} J_m \right]_{\NNLOeps =0}\:,
\label{eq:sigmaNNLOm+2}
\NNLOeeq
\NNLObeq
\NNLOdsig{NNLO}_{m+1} =
\left[\NNLOdsig{RV}_{m+1} J_{m+1} - \NNLOdsiga{RV}{1}_{m+1} J_m
+ \int_1\!\left(\NNLOdsiga{RR}1_{m+2} J_{m+1}
  - \NNLOdsiga{RR}{12}_{m+2} J_m \right) \right]_{\NNLOeps =0}\:,
\label{eq:sigmaNNLO3}
\NNLOeeq
and
\NNLObeq
\NNLOdsig{NNLO}_m =
\left[\NNLOdsig{VV}_m + \int_2\!\NNLOdsiga{RR}2_{m+2}
 + \int_1\!\NNLOdsiga{RV}{1}_{m+1}\right]_{\NNLOeps = 0} J_m
\:,
\label{eq:sigmaNNLO2}
\NNLOeeq
are sparately finite, thus integrable in four dimensions by
construction.  Above $\NNLOdsiga{RR}2_{m+2}$ and $\NNLOdsiga{RR}1_{m+2}$  are
the counterterms regularising the doubly- and singly-unresolved limits
of $\NNLOdsig{RR}_{m+2}$ respectively while the overlap of these
is accounted for by $\NNLOdsiga{RR}{12}_{m+2}$. The singly-unresolved
limits of $\NNLOdsig{RV}_{m+1}$ are regularised  by the counterterm 
$\NNLOdsiga{RV}1_{m+1}$. In this contribution we will deal exclusively with
the subtractions needed to regularise the doubly-real emission.

\subsubsection{Subtraction terms for doubly-real emission}

\paragraph{The general setup}

The cross section $\NNLOdsig{RR}_{m+2}$, is the integral of the tree-level
squared matrix element for $m+2$ parton production over the $m+2$
parton phase space
\NNLObeq
\NNLOdsig{RR}_{m+2} = \NNLOPS{m+2}{}\NNLOM{m+2}{(0)}\,.
\label{eq:tsigRRm+2}
\NNLOeeq
We disentangled the overlap structure of the singularities of
$\NNLOM{m+2}{(0)}$ into the pieces $\NNLObA{2}\NNLOM{m+2}{(0)}$,
$\NNLObA{1}\NNLOM{m+2}{(0)}$ and $\NNLObA{12}\NNLOM{m+2}{(0)}$ in \NNLORef{Somogyi:2005xz}.
These expressions are, as they stand, only defined in the strict soft
and/or collinear limits. To define true counterterms, they need to be
extended over the full phase-space. This extension requires a
phase-space factorisation that maintains momentum conservation exactly,
but such that in addition it respects the structure of delicate
cancallations among the various subtraction terms.  Then the
counterterms may symbolically be written as
\NNLObeeq
\NNLOdsiga{RR}2_{m+2} &=& \NNLOPS{m}{}\:[\NNLOrd p^{(2)}]{\NNLObom{\cal A}}_2 \NNLOM{m+2}{(0)}\,,
\label{eq:dsigRRA2}
\\
\NNLOdsiga{RR}1_{m+2} &=& \NNLOPS{m+1}{}\:[\NNLOrd p^{(1)}]{\NNLObom{\cal A}}_1 \NNLOM{m+2}{(0)}\,,
\label{eq:dsigRRA1}
\NNLOeeeq
and
\NNLObeq
\NNLOdsiga{RR}{12}_{m+2} = \NNLOPS{m}{}\:[\NNLOrd p^{(1)}]\:[\NNLOrd p^{(1)}]{\NNLObom{\cal A}}_{12} 
\NNLOM{m+2}{(0)}\,,
\label{eq:dsigRRA12}
\NNLOeeq
where in \NNLOeqnss{eq:dsigRRA2}{eq:dsigRRA12} we used a calligraphic notation
to indicate the extension of the terms
$\NNLObA{2}\NNLOM{m+2}{(0)}$, $\NNLObA{1}\NNLOM{m+2}{(0)}$ and $\NNLObA{12}\NNLOM{m+2}{(0)}$ over
the whole phase space.

\paragraph{Singly-singular counterterms}

The singly-singular counterterm ${\NNLObom{\cal A}}_{1} \NNLOM{m+2}{(0)}$ reads
\NNLObeq
{\NNLObom{\cal A}}_{1}\NNLOM{m+2}{(0)} =
\sum_{r} \left[\sum_{i\ne r} \frac{1}{2} \NNLOcC{ir}
+ \left(\NNLOcS_{r} - \sum_{i\ne r} \NNLOcC{ir}\NNLOcS_{r}\right) \right]\,.
\label{eq:A1}
\NNLOeeq
Here the singly-collinear term is 
\NNLObeq
\NNLOcC{ir} = 
8\pi\NNLOas\mu^{2\NNLOeps}\frac{1}{s_{ir}}
\NNLObra{m+1}{(0)}{}
\NNLOhP_{f_i f_r}^{(0)}(\NNLOtzz{i}{r},\NNLOtzz{r}{i},\NNLOti{k}_{\perp,ir};\NNLOeps)
\NNLOket{m+1}{(0)}{}\,,
\label{eq:Cir}
\NNLOeeq
where $\NNLOhP_{f_i f_r}^{(0)}(z_i,z_r,\NNLOkT{};\NNLOeps)$ is the Altarelli-Parisi
splitting function. We define the momentum fractions $\NNLOtzz{i}{r}$ and
$\NNLOtzz{r}{i}$ as
\NNLObeq
\NNLOtzz{i}{r} = \frac{s_{iQ}}{s_{iQ} + s_{rQ}}
\qquad\mbox{and}\qquad
\NNLOtzz{r}{i} = \frac{s_{rQ}}{s_{iQ} + s_{rQ}}\,,
\label{eq:ztirztri}
\NNLOeeq
i.e., the energy fractions of the daughter momenta of the splitting
with respect to the energy of the parent parton.  The transverse
momentum $\NNLOkTt{ir}$ is given by
\NNLObeq
\NNLOkTt{ir}^{\mu} = 
  \left(\NNLOtzz{i}{r}-\frac{s_{ir}}{\alpha_{ir}(s_{iQ}+s_{rQ})}\right) p_r^{\mu}
- \left(\NNLOtzz{r}{i}-\frac{s_{ir}}{\alpha_{ir}(s_{iQ}+s_{rQ})}\right) p_i^{\mu}
+(\NNLOtzz{r}{i}-\NNLOtzz{i}{r})\NNLOti{p}_{ir}^{\mu}\,,
\label{eq:kTtir}
\NNLOeeq
where we used the abbreviations $s_{iQ}=2p_i\cdot Q$, $s_{rQ}=2p_r\cdot Q$.

The $m+1$ momenta entering the matrix elements on the right hand side of
\NNLOeqn{eq:Cir} are defined as follows
\NNLObeeq
\NNLOti{p}_{ir}^{\mu} = \frac{1}{1-\alpha_{ir}}(p_i^{\mu} + p_r^{\mu} - \alpha_{ir} Q^{\mu})\,,
\qquad
\NNLOti{p}_n^{\mu} = \frac{1}{1-\alpha_{ir}} p_n^{\mu}\,,
\qquad n\ne i,r\,,
\label{eq:PS_single_coll}
\NNLOeeeq
where
\NNLObeq
\alpha_{ir} =
\frac{(p_i+p_r)\cdot Q-\sqrt{[(p_i+p_r)\cdot Q]^2 - s_{ir}\: s}}{s}
\label{eq:alphair}
\NNLOeeq
and $Q^\mu$ is the total four-momentum of the incoming electron and
positron and $s = Q^2$. Clearly, the total four-momentum is conserved,
\NNLObeq
\NNLOti{p}_{ir}^{\mu} + \sum_n \NNLOti{p}_n^{\mu}
= p_i^\mu + p_r^\mu + \sum_n p_n^\mu\,.
\NNLOeeq

The singly-soft term is
\NNLObeq
\NNLOcS_r = 
-8\pi\NNLOas\mu^{2\NNLOeps}\sum_{i}\sum_{k\ne i} \frac12 \NNLOcS_{ik}(r)
\NNLObra{m+1}{(0)}{} {\NNLObom T}_{i}{\NNLObom T}_{k} \NNLOket{m+1}{(0)}{}\,,
\label{eq:Sr}
\NNLOeeq
if $r$ is a gluon, and $\NNLOcS_r = 0$ if $r$ is a quark or antiquark.
The $m+1$ momenta entering the matrix element on the right hand side of
\NNLOeqn{eq:Sr} are defined in by first rescaling all the hard momenta by a
factor $1/\lambda_r$ and then transforming all of the rescaled
momenta by a Lorentz transformation $\Lambda^{\mu}_{\nu}$,
%
% Switched from lambda -> 1/lambda
%
\NNLObeq
\NNLOti{p}_n^{\mu} = \Lambda^{\mu}_{\nu}[Q,(Q-p_r)/\lambda_r] (p_n^{\nu}/\lambda_r)\,,
\qquad n\ne r\,,
\label{eq:PS_single_soft}
\NNLOeeq
where
\NNLObeq
\lambda_r = \sqrt{1-\frac{s_{rQ}}{s}}\,,
\qquad
\Lambda^{\mu}_{\nu}[K,\NNLOwti{K}] = g^{\mu}_{\nu}
- \frac{2(K+\NNLOwti{K})^{\mu}(K+\NNLOwti{K})_{\nu}}{(K+\NNLOwti{K})^{2}} 
+ \frac{2K^{\mu}\NNLOwti{K}_{\nu}}{K^2}\,.
\label{eq:LambdaKKt}
\NNLOeeq
The matrix $\Lambda^{\mu}_{\nu}[K,\NNLOwti{K}]$ generates a (proper) Lorentz
transformation, provided $K^2 = \NNLOwti{K}^2 \ne 0$. Since $p_r^\mu$ is
massless ($p_r^2 = 0$), the total four-momentum is again conserved.
 
The eikonal factor in \NNLOeqn{eq:Sr} is
\NNLObeq
\NNLOcS_{ik}(r) = \frac{2 s_{ik}}{s_{ir} s_{rk}}\,,
\label{eq:Sikr}
\NNLOeeq
and the sum in \NNLOeqn{eq:Sr} runs over the external partons of the $m+1$ parton 
matrix element on the right hand side.

The soft-collinear subtraction is given by
\NNLObeq
\NNLOcC{ir}\NNLOcS_{r} = 
8\pi\NNLOas\mu^{2\NNLOeps} \frac{1}{s_{ir}}\frac{2\NNLOtzz{i}{r}}{\NNLOtzz{r}{i}}\,\NNLObT_i^2\,
\NNLOM{m+1}{(0)}\,,
\label{eq:CirSr}
\NNLOeeq
if $r$ is a gluon, and $\NNLOcC{ir}\NNLOcS_{r} = 0$ if $r$ is a quark or antiquark. 
The momentum fractions are given by \NNLOeqn{eq:ztirztri}. As pointed out
in \NNLORef{Somogyi:2005xz}, the correct variables in the squared matrix
element in the soft-collinear limit are those that appear in the soft 
limit. Thus the $m+1$ momenta entering the matrix elements on the right hand
side are again given by \NNLOeqn{eq:PS_single_soft}.

The momentum mappings introduced in \NNLOeqns{eq:PS_single_coll}{eq:PS_single_soft}
both lead to exact phase-space factorisation in the form
\NNLObeq
\NNLOPS{m+2}=\NNLOPS{m+1} \: [\NNLOrd p^{(1)}]\,,
\label{eq:PSfact}
\NNLOeeq
where the $m+1$ momenta in the first factor on the right hand side are exactly
those defined in \NNLOeqn{eq:PS_single_coll} or \NNLOeqn{eq:PS_single_soft}. 
%
% Added explicit forms of [dp]
%
The explicit expressions for $[\NNLOrd p^{(1)}]$ read
\NNLObeeq
~[\NNLOrd p^{(1)}] \NNLOaand=
\frac{(1-\alpha_{ir})^{m(d-2)-1}s_{\NNLOwti{ir}Q}}
{\sqrt{(s_{r\NNLOwti{ir}}+s_{\NNLOwti{ir}Q}-s_{rQ})^2+4s_{r\NNLOwti{ir}}(s-s_{\NNLOwti{ir}Q})}}
\,\Theta(1-\alpha_{ir})
\,\frac{\NNLOrd^d p_r}{(2\pi)^{d-1}}\delta_{+}(p_r^2)\,,
\label{eq:dp_coll}
\\ 
~[\NNLOrd p^{(1)}] \NNLOaand=
\lambda_{r}^{m(d-2)-2}\,\Theta(\lambda_{r})
\,\frac{\NNLOrd^d p_r}{(2\pi)^{d-1}}\delta_{+}(p_r^2)\,,
\label{eq:dp_soft}
\NNLOeeeq
for the collinear and soft phase-space factorisations
(\NNLOeqns{eq:PS_single_coll}{eq:PS_single_soft}) respectively. In
\NNLOeqn{eq:dp_coll} $\alpha_{ir}$ is understood to be expressed in terms
of the variable $\NNLOti{p}_{ir}$,
\NNLObeq
\alpha_{ir}=\frac{\sqrt{(s_{r\NNLOwti{ir}}+s_{\NNLOwti{ir}Q}-s_{rQ})^2+4s_{r\NNLOwti{ir}}(s-s_{\NNLOwti{ir}Q})}
-(s_{r\NNLOwti{ir}}+s_{\NNLOwti{ir}Q}-s_{rQ})}{2(s-s_{\NNLOwti{ir}Q})}\,.
\label{eq:alpha_new}
\NNLOeeq

The analytical integration of the counterterms over the factorised
one-parton phase-space $[\NNLOrd p^{(1)}]$ is then possible. Details of these
integrations will be given elsewhere.  

\subsubsection{Doubly-singular and iterated counterterms}

The doubly-singular and interated counterterms are respectively defined by
\NNLObeeq
{\NNLObom{\cal A}}_{2} \NNLOM{m+2}{(0)} \NNLOaand=
\sum_r \sum_{s\ne r} \Bigg\{
  \sum_{i\ne r,s} \Bigg[
  \frac16 \,\NNLOcC{irs}
+\!\!\!\sum_{j\ne i,r,s} \frac18 \,\NNLOcC{ir;js}
%\nn\\&&\qquad\qquad\qquad\quad
+ \frac12 \,\Bigg(\NNLOcCS{ir;s} - \NNLOcC{irs} \NNLOcCS{ir;s}
- \sum_{j\ne i,r,s} \NNLOcC{ir;js} \NNLOcCS{ir;s} \Bigg)\Bigg]
\NNLOnn\\&&\qquad\qquad\quad
+ \frac12 \,\NNLOcS_{rs}
- \sum_{i\ne r,s} \Bigg[
    \NNLOcCS{ir;s} \NNLOcS_{rs}
  + \frac12 \, \NNLOcC{irs}\,\Big(\NNLOcS_{rs}^{\rm N} - \NNLOcS_{rs}^{\rm A}\Big)
%\nn\\&&\qquad\qquad\qquad\qquad\qquad\qquad
  + \sum_{j\ne i,r,s} \frac12 \, \NNLOcC{ir;js} \,\NNLOcS_{rs} \Bigg]
\Bigg\}
\label{eq:A2}
\NNLOeeeq
and
\NNLObeq
{\NNLObom{\cal A}}_{12}\NNLOM{m+2}{(0)}
= \sum_{t}\Bigg[\NNLOcS_{t}{\NNLObom{\cal A}}_{2}\NNLOM{m+2}{(0)} + 
\sum_{k\ne t} \frac12 \,\NNLOcC{kt}{\NNLObom{\cal A}}_{2}\NNLOM{m+2}{(0)} -
\sum_{k\ne t} \NNLOcC{kt}\NNLOcS_{t}{\NNLObom{\cal A}}_{2}\NNLOM{m+2}{(0)}\Bigg]\,,
\label{eq:A12}
\NNLOeeq
where the three terms in \NNLOeqn{eq:A12} each evaluate further into long
expressions. Leaving all details to a further publication, here we only
note that similarly to the singly-unresolved counterterms in
\NNLOeqn{eq:A1}, each term in \NNLOeqns{eq:A2}{eq:A12} represents an extension
of one of the limits discussed in \NNLORef{Somogyi:2005xz}.  The momentum
mappings used for the various terms are either combinations of those
introduced in \NNLOeqns{eq:PS_single_coll}{eq:PS_single_soft} or simple
generalisations thereof. Exact phase-space factorisation is again
possible. Using those factorised phase spaces, it is straightforward to
write \NNLOeqn{eq:sigmaNNLOm+2} explicitly. We have coded
\NNLOeqn{eq:sigmaNNLOm+2} for the case when $\NNLOdsig{RR}_{m+2}$ is the fully
differential cross section for the process $e^+e^- \to q \bar{q} g g g$
($m=3$) and $J_5$ defines the $C$-parameter. We found that the
integral of $\NNLOdsig{NNLO}_{m+2}$ is indeed finite and integrable in four
dimensions using standard Monte Carlo methods.

\subsubsection{Outlook}

The subtraction scheme outlined here uses the known singly- and
doubly-singular limits of the squared matrix elements. These limits
overlap and a way of disentanglement was presented in \NNLORef{Somogyi:2005xz}. 
In this contribution we discussed how to make the next step, namely
we outlined the exact phase space factorisations we propose for the
collinear and soft subtraction terms.  Putting the subtraction terms on
the factorised phase space allows us the integration of the singular
factors such that the remaining expressions can be combined with the
virtual correction. This integration and combination is left for future
work.

\subsection{The antenna subtraction approach~\protect
\footnote{Authors: A.~Gehrmann-De Ridder, T.~Gehrmann, E.W.N.~Glover}}
\subsubsection{Method}
An $m$-jet cross section at NLO is obtained by summing contributions from 
$(m+1)$-parton tree level and $m$-parton one-loop processes:
\begin{displaymath}
{\rm d}\sigma_{NLO}=\int_{{\rm d}\Phi_{m+1}}\left({\rm d}\sigma^{R}_{NLO}
-{\rm d}\sigma^{S}_{NLO}\right) +\left [\int_{{\rm d}\Phi_{m+1}}
{\rm d}\sigma^{S}_{NLO}+\int_{{\rm d}\Phi_{m}}{\rm d}\sigma^{V}_{NLO}\right].
\end{displaymath}
The cross section ${\rm d}\sigma^{R}_{NLO}$ is the $(m+1)$-parton tree-level
cross section, 
while 
${\rm d}\sigma^{V}_{NLO}$ is the one-loop virtual correction to the 
$m$-parton Born cross section ${\rm d}\sigma^{B}$. Both contain infrared 
singularities, which are explicit poles in $1/\epsilon$ in ${\rm d}\sigma^{V}_{NLO}$,
while becoming explicit in ${\rm d}\sigma^{R}_{NLO}$ only after integration 
over the phase space.  In general, this integration  involves the
(often iterative) definition of the jet observable, such that  an analytic
integration is not feasible (and also not appropriate). Instead,   one would
like to have a flexible method that can be easily adapted to  different jet
observables or jet definitions. Therefore, the infrared singularities  
of the real radiation
contributions should be extracted using  infrared subtraction  terms.
One introduces ${\rm d}\sigma^{S}_{NLO}$, which is a counter-term for  
 ${\rm d}\sigma^{R}_{NLO}$, having the same unintegrated
singular behaviour as ${\rm d}\sigma^{R}_{NLO}$ in all appropriate limits.
Their difference is free of divergences 
and can be integrated over the $(m+1)$-parton phase space numerically.
The subtraction term  ${\rm d}\sigma^{S}_{NLO}$ has 
to be integrated analytically over all singular regions of the 
$(m+1)$-parton phase space. 
The resulting cross section added to the virtual contribution 
yields an infrared finite result. 
Several methods for  constructing
 NLO subtraction terms systematically  were proposed in the 
literature~\cite{Campbell:1998nn,Kosower:1997zr,Kosower:2003bh,
Catani:1996vz,
Giele:1991vf,
Kunszt:1992tn,
Frixione:1995ms}. 
For some of these methods, 
extension to NNLO was discussed~\cite{Kosower:2002su,
Weinzierl:2003fx,Kilgore:2004ty,Frixione:2004is,Somogyi:2005xz} 
and partly worked out. 
In this section, focus is  on the  antenna subtraction 
method~\cite{Campbell:1998nn,Kosower:1997zr}, 
which is extended to NNLO~\cite{Gehrmann-DeRidder:2005cm}. 

\begin{figure}[t] 
\begin{center}
\epsfig{file=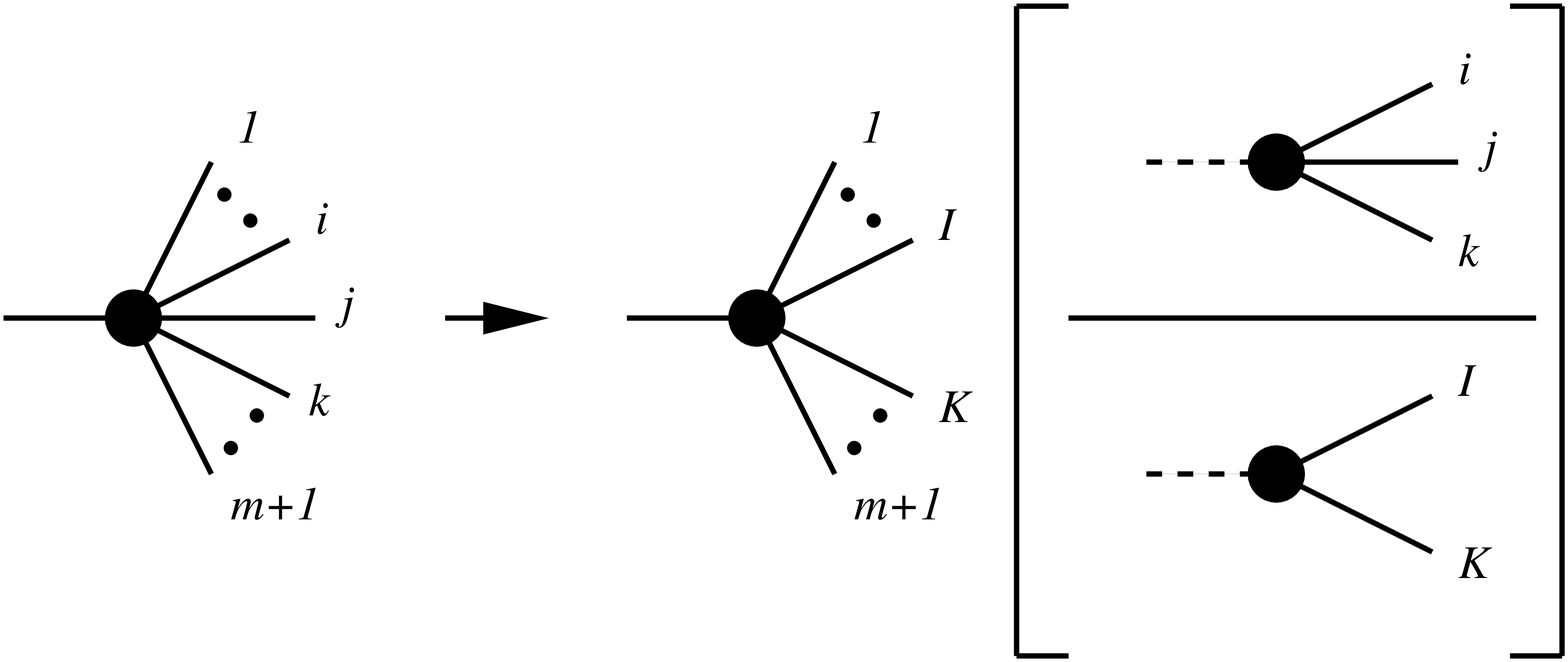,height=3.8cm}
\end{center}
\vspace{-4mm}
\caption{Illustration of NLO antenna factorisation representing the
factorisation of both the squared matrix elements and the 
$(m+1)$-particle phase
space. The term in square brackets
represents both the antenna function and the antenna phase space.
\label{fig:nloant}
}
\vspace{2mm}
\begin{center}
\epsfig{file=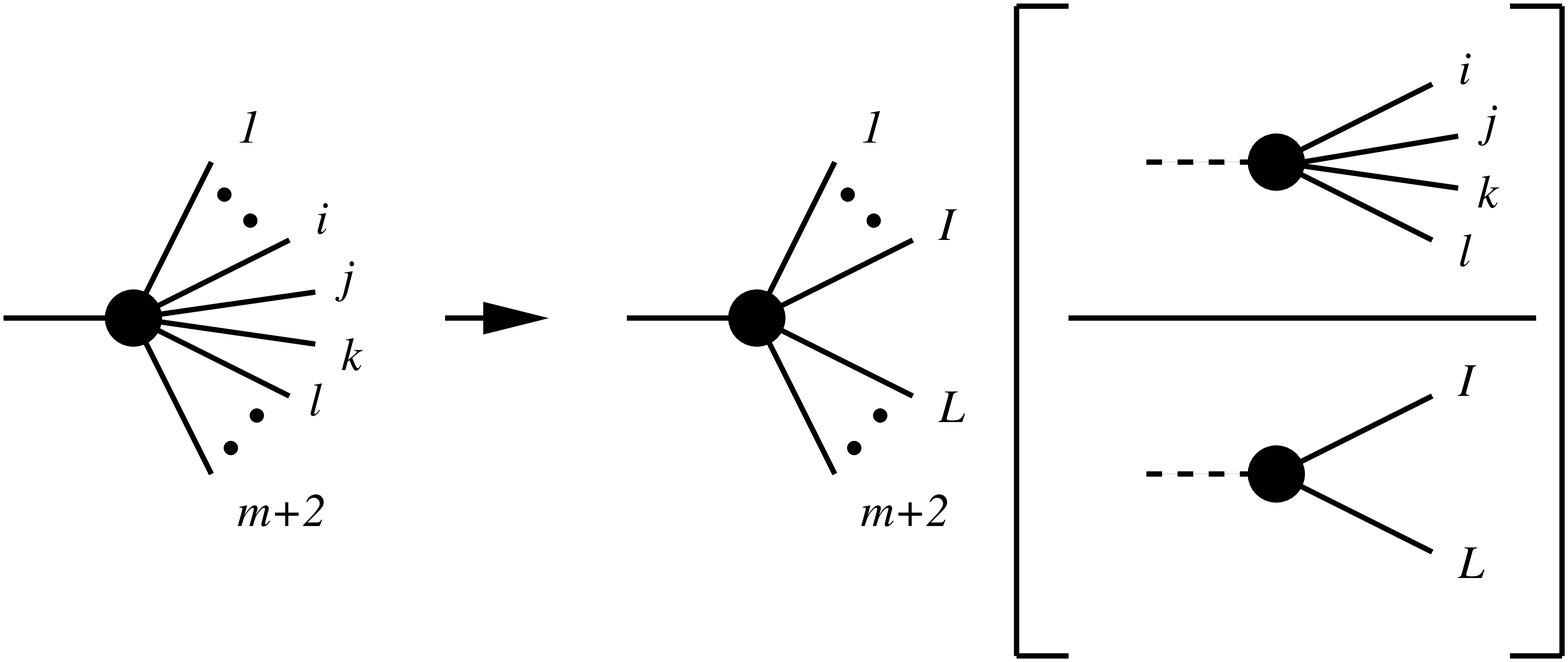,,height=3.8cm}
\end{center}
\vspace{-4mm}
\caption{\label{fig:sub2a} Illustration 
of NNLO antenna factorisation representing the
factorisation of both the squared matrix elements and the $(m+2)$-particle 
phase
space when the unresolved particles are colour connected. 
The term in square brackets
represents both the antenna function and the antenna phase space.}
\end{figure}

The basic idea of the antenna subtraction approach at NLO is to construct 
the subtraction term  
${\rm d}\sigma^{S}_{NLO}$
from antenna functions. Each antenna function encapsulates 
all singular limits due to the 
 emission of one unresolved parton between two colour-connected hard
partons (tree-level three-parton antenna function).
This construction exploits the universal factorisation of 
phase space and squared matrix elements in all unresolved limits,
depicted in Figure~\ref{fig:nloant}.
The individual antenna functions are obtained by normalising 
three-parton tree-level matrix elements to the corresponding two-parton 
tree-level matrix elements. 

At NNLO, the $m$-jet production is induced by final states containing up to
$(m+2)$ partons, including the one-loop virtual corrections to $(m+1)$-parton final 
states. As at NLO, one has to introduce subtraction terms for the 
$(m+1)$- and $(m+2)$-parton contributions. 
Schematically the NNLO $m$-jet cross section reads,
\begin{eqnarray*}
{\rm d}\sigma_{NNLO}&=&\int_{{\rm d}\Phi_{m+2}}\left({\rm d}\sigma^{R}_{NNLO}
-{\rm d}\sigma^{S}_{NNLO}\right) + \int_{{\rm d}\Phi_{m+2}}
{\rm d}\sigma^{S}_{NNLO}\nonumber \\ 
&&+\int_{{\rm d}\Phi_{m+1}}\left({\rm d}\sigma^{V,1}_{NNLO}
-{\rm d}\sigma^{VS,1}_{NNLO}\right)
+\int_{{\rm d}\Phi_{m+1}}{\rm d}\sigma^{VS,1}_{NNLO}  
\nonumber \\&&
+ \int_{{\rm d}\Phi_{m}}{\rm d}\sigma^{V,2}_{NNLO}\;,
\end{eqnarray*}
where $\NNLOd \sigma^{S}_{NNLO}$ denotes the real radiation subtraction term 
coinciding with the $(m+2)$-parton tree level cross section 
 $\NNLOd \sigma^{R}_{NNLO}$ in all singular limits~\cite{Gehrmann-DeRidder:1997gf,
Campbell:1997hg,
Catani:1998nv,
Catani:1999ss,Berends:1988zn,DelDuca:1999ha}. 
Likewise, $\NNLOd \sigma^{VS,1}_{NNLO}$
is the one-loop virtual subtraction term 
coinciding with the one-loop $(m+1)$-parton cross section 
 $\NNLOd \sigma^{V,1}_{NNLO}$ in all singular 
limits
\cite{Bern:1994zx,Kosower:1999xi,Kosower:1999rx,Bern:1998sc,Bern:1999ry}. 
Finally, the two-loop correction 
to the $m$-parton cross section is denoted by ${\rm d}\sigma^{V,2}_{NNLO}$.

Both types of NNLO subtraction terms can be constructed from antenna 
functions. In ${\rm d}\sigma^{S}_{NNLO}$, one has to distinguish four
different types of unresolved configurations:
(a) One unresolved parton but the experimental observable selects only
$m$ jets;
(b) Two colour-connected unresolved partons (colour-connected);
(c) Two unresolved partons that are not colour connected but share a common
radiator (almost colour-unconnected);
(d) Two unresolved partons that are well separated from each other 
in the colour 
chain (colour-unconnected). Among those, configuration (a) is properly 
accounted for by a single tree-level three-parton antenna function 
like used already at NLO. Configuration (b) requires a 
tree-level four-parton antenna function (two unresolved partons emitted 
between a pair of hard partons) 
as shown in Figure~\ref{fig:sub2a}, while (c) and (d) are accounted for by 
products of two tree-level three-parton antenna functions. 

\begin{figure}[t]
\begin{center}
\epsfig{file=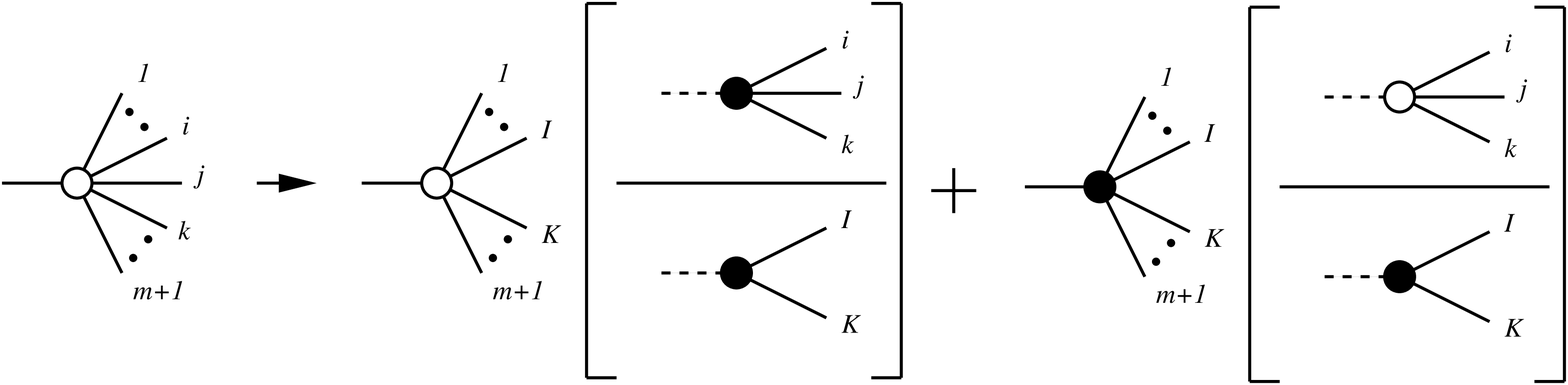,height=3.8cm}
\vspace{-8mm}
\end{center}
\caption{Illustration of NNLO antenna factorisation representing the
factorisation of both the one-loop
``squared" matrix elements (represented by the white blob)
and the $(m+1)$-particle phase
space when the unresolved particles are colour connected. 
\label{fig:subv}}
\end{figure}

In single unresolved limits, the one-loop cross section 
$\NNLOd \sigma^{V,1}_{NNLO}$ is described by the sum of two 
terms
\cite{Bern:1994zx,Kosower:1999xi,Kosower:1999rx,Bern:1998sc,Bern:1999ry}: 
a tree-level splitting function times a one-loop cross section 
and a one-loop splitting function times a tree-level cross section. 
Consequently, the 
one-loop single unresolved subtraction term $\NNLOd \sigma^{VS,1}_{NNLO}$
is constructed from tree-level and one-loop three-parton antenna functions,
as sketched in Figure~\ref{fig:subv}. Several other terms in  
 $\NNLOd \sigma^{VS,1}_{NNLO}$ cancel with the results
from the integration of terms in 
the double real radiation subtraction term  $\NNLOd \sigma^{S}_{NNLO}$
over the phase space appropriate to one of the unresolved partons, thus 
ensuring the cancellation of all explicit infrared poles in the difference 
$\NNLOd \sigma^{V,1}_{NNLO}-\NNLOd \sigma^{VS,1}_{NNLO}$.

Finally, all remaining terms in 
$\NNLOd \sigma^{S}_{NNLO}$ and $\NNLOd \sigma^{VS,1}_{NNLO}$ have to be integrated 
over the four-parton and three-parton antenna phase spaces. After   
integration, the infrared poles are rendered explicit and
cancel with the 
infrared pole terms in the two-loop squared matrix element 
$\NNLOd \sigma^{V,2}_{NNLO}$. 

\subsubsection{Derivation of antenna functions}
The  subtraction terms $\NNLOd \sigma^{S}_{NLO}$,
$\NNLOd \sigma^{S}_{NNLO}$ 
and $\NNLOd \sigma^{VS,1}_{NNLO}$ require three different types of 
antenna functions corresponding to the different pairs of hard partons 
forming the antenna: quark-antiquark, quark-gluon and gluon-gluon antenna 
functions. In the past~\cite{Campbell:1998nn,Kosower:1997zr}, NLO 
antenna functions were 
constructed by imposing definite properties in 
all single unresolved limits (two collinear limits 
and one soft limit for each 
antenna). 
This procedure turns out to be impractical at NNLO, where each antenna 
function must have definite behaviours in a large number of single and 
double unresolved limits. Instead, one can derive these antenna functions in 
a systematic manner from physical matrix elements known to possess the 
correct limits. The quark-antiquark antenna functions can be obtained 
directly from 
the $e^+e^- \to 2j$ real radiation corrections at NLO and 
NNLO~\cite{Gehrmann-DeRidder:2004tv}. 
For quark-gluon and gluon-gluon antenna functions, effective Lagrangians 
are used to obtain tree-level processes yielding a quark-gluon or 
gluon-gluon final state. The antenna functions are then obtained from 
the real radiation corrections to these processes. 
Quark-gluon antenna functions 
were derived~\cite{Gehrmann-DeRidder:2005hi} from the purely QCD 
(i.e.\ non-supersymmetric) NLO and NNLO corrections to the decay of 
a heavy neutralino into a massless gluino plus partons~\cite{Haber:1988px}, 
while 
gluon-gluon antenna functions~\cite{Gehrmann-DeRidder:2005aw} result 
from the QCD corrections 
to Higgs boson decay into partons~\cite{Wilczek:1977zn,Shifman:1978zn}. 

All tree-level three-parton and four-parton antenna functions 
and  three-parton one-loop antenna functions are listed 
in~\cite{Gehrmann-DeRidder:2005cm}, 
where they are  also integrated,  using the 
phase space integration techniques 
described in~\cite{Gehrmann-DeRidder:2003bm}.

\subsubsection{Application to $e^+e^- \to 3$~jets}
To illustrate the application of antenna subtraction  on a 
non-trivial example,  
in~\cite{Gehrmann-DeRidder:2005cm,Gehrmann-DeRidder:2004xe} 
the $1/N^2$-contribution to 
the NNLO corrections to  $e^+e^- \to 3$~jets was 
derived. This colour factor receives
contributions from 
$\gamma^*\to q\bar q ggg$ and $\gamma^*\to q\bar q q\bar qg$
at tree-level~\cite{Hagiwara:1988pp,Berends:1988yn,Falck:1989uz},  
$\gamma^*\to q\bar q gg$ and $\gamma^*\to q\bar q q\bar q$  at
one-loop~\cite{Bern:1996ka,Bern:1997sc,Glover:1996eh,
 Campbell:1997tv,Nagy:1997md}
and $\gamma^*\to q\bar q g$ at two-loops~\cite{Garland:2001tf,Garland:2002ak}. 
The four-parton and 
five-parton final states contain infrared singularities, which need to
be extracted using the antenna subtraction formalism.

In this contribution, all gluons are effectively photon-like, and couple 
only to the quarks, but not to each other. Consequently, only quark-antiquark
antenna functions appear in the construction of the subtraction terms. 

Starting from the program {\tt EERAD2}~\cite{Campbell:1998nn}, which computes 
the four-jet
production at NLO, the NNLO antenna subtraction method 
for the $1/N^2$ colour factor contribution to $e^+e^-\to 3j$
was  implemented. {\tt EERAD2}
already  contains the five-parton and four-parton 
matrix elements relevant here, as well as the NLO-type subtraction terms. 

The implementation contains three channels, classified 
by their partonic multiplicity: 
(a) in the five-parton channel, one
integrates ${\rm d}\sigma_{NNLO}^{R} - {\rm d}\sigma_{NNLO}^{S}$;
(b) in the four-parton channel, one integrates
${\rm d}\sigma_{NNLO}^{V,1} - {\rm d}\sigma_{NNLO}^{VS,1}$;
(c) in the three-parton channel, one integrates
${\rm d}\sigma_{NNLO}^{V,2} +{\rm d}\sigma_{NNLO}^{S}
+ {\rm d}\sigma_{NNLO}^{VS,1}$.
The numerical integration over these channels is carried out by Monte Carlo 
methods. 

By construction, the integrands in the four-parton and 
three-parton channel are free of explicit infrared poles. In the 
five-parton and four-parton channel, the proper implementation of 
the subtraction was tested 
by generating trajectories of phase space points approaching 
a given single or double unresolved limit. 
Along these trajectories, one observes that the 
antenna subtraction terms converge locally towards the physical matrix 
elements, and that the cancellations among individual 
contributions to the subtraction terms take place as expected. 
Moreover, the correctness of the 
subtraction was checked by introducing a 
lower cut (slicing parameter) on the phase space variables, and observing 
that our results are independent of this cut (provided it is 
chosen small enough). This behaviour indicates that the 
subtraction terms ensure that the contribution of potentially singular 
regions of the final state phase space does not contribute to the numerical 
integrals, but is accounted for analytically. 

Finally, is was noted in~\cite{Gehrmann-DeRidder:2005cm} that the 
infrared poles of the two-loop (including one-loop times one-loop) correction
to $\gamma^*\to q\bar qg$ are cancelled in all colour factors by a 
combination of integrated three-parton and four-parton
antenna functions.
This highly non-trivial cancellation 
clearly illustrates that the antenna functions derived 
here 
correctly approximate QCD matrix elements in all infrared singular limits at 
NNLO. They also outline the structure of infrared 
cancellations in $e^+e^-\to 3j$ at NNLO, and indicate the structure of the 
subtraction terms in all colour factors.

\subsubsection{Outlook} 
The antenna subtraction method presented here allows the 
subtraction of infrared 
singularities in the calculation of jet observables at NNLO. It introduces 
subtraction terms for double real radiation at tree level and 
single real radiation at one loop based on 
antenna functions. These antenna 
functions describe the colour-ordered radiation of unresolved 
partons between a 
pair of hard (radiator) partons. All antenna functions at NLO and NNLO 
can be derived systematically from physical matrix elements. 
To demonstrate the application of the new method on a non-trivial example, 
 the NNLO corrections 
to the subleading colour contribution to $e^+e^- \to 3$~jets were implemented. 

An immediate application of the method presented here is 
the calculation of 
the full NNLO corrections to $e^+e^- \to 3$~jets. 
The antenna subtraction method 
can be further generalised to NNLO corrections to jet production in 
 lepton-hadron or hadron-hadron collisions. In these kinematical situations,
the subtraction terms are constructed using the same antenna functions, but 
in different  phase space configurations: instead of the $1\to n$ decay 
kinematics considered here, $2\to n$ scattering kinematics are required, which
can also contain singular configurations due to single or double initial state 
radiation. These require new sets of integrated antenna functions, accounting 
for the different phase space configurations in these cases. 

\subsection{The sector decomposition approach to NNLO cross sections~\protect
\footnote{Author: G.~Heinrich}}

\subsubsection{General aspects}

Sector decomposition is a general method to disentangle and isolate 
overlapping singularities, of both ultraviolet and
infrared nature. 
As the infrared singularities occurring in NNLO 
calculations involving massless particles can be entangled 
in a very complicated way -- in the virtual two-loop integrals 
as well as in the  real radiation parts -- 
sector decomposition is particularly helpful in the context of NNLO 
calculations.   
Originally, it has been conceived by K.~Hepp\,\cite{Hepp:1966eg} for overlapping
ultraviolet singularities. 
Its first phenomenological application can be found in \cite{Roth:1996pd}, 
and in \cite{Binoth:2000ps} it has been developed to an 
automated tool to calculate multi-loop integrals numerically in the 
Euclidean region. 
It has been successfully applied to 
various types of multi-loop integrals 
\cite{Binoth:2003ak,Denner:2004iz,Heinrich:2004iq,Czakon:2004wm,Anastasiou:2005pn}. 
Its application to NNLO phase space integrals, 
first proposed in \cite{Heinrich:2002rc}, 
saw a very rapid development recently\,\cite{Gehrmann-DeRidder:2003bm,
Anastasiou:2003gr,Binoth:2004jv}
and already lead to NNLO results for 
$e^+e^-\to 2$\,jets\,\cite{Anastasiou:2004qd}, 
Higgs production\,\cite{Anastasiou:2005qj} 
and muon decay\,\cite{Anastasiou:2005pn}.
In \cite{Heinrich:2006sw}, first results on its application to 
$e^+e^-\to 3\,$jets were presented. 

The advantages of the sector decomposition approach  
reside in the fact that 
the extraction of the infrared poles is algorithmic, 
being done by an algebraic subroutine,
and that the subtraction terms can be arbitrarily 
complicated as they are integrated only numerically. 
However, the size of the expressions produced by the iterated 
sector decomposition is rather large.
On the other hand, the methods  based on the 
manual construction of an analytic subtraction 
scheme \cite{Gehrmann-DeRidder:2004tv,Gehrmann-DeRidder:2005cm,Somogyi:2005xz,
Kosower:2002su,Weinzierl:2003fx,Weinzierl:2003ra,
Kilgore:2004ty,Frixione:2004is,
Gehrmann-DeRidder:2005hi,
Gehrmann-DeRidder:2005aw,Gehrmann-DeRidder:2004xe}
allow maximal (i.e. analytical) control over the pole terms,  
and  insure a minimal number of subtraction terms. 

\subsubsection{The method}
The universal applicability of sector decomposition goes back 
to the fact that it acts in parameter space by a simple mechanism. 
The parameters can be Feynman parameters in the case of multi-loop 
integrals, or phase space integration parameters, or a combination of both. 
In the following, the working mechanism of sector decomposition
will be outlined only briefly, details can be found 
in \cite{Binoth:2000ps,Binoth:2004jv}.

An overlapping singularity in parameter space is of the type
\begin{eqnarray*}
I&=&
\int_0^1 dx\int_0^1 dy \,x^{-1-\epsilon}\,(x+y)^{-1}\;,
\end{eqnarray*}
where a naive subtraction of the singularity for $x\to 0$ of the form 
\begin{eqnarray}
\int_0^1 dx\int_0^1 dy\, x^{-1-\NNLOeps} f(x,y)&=&-\frac{1}{\NNLOeps}\int_0^1 dy\,\
f(0,y)+\int_0^1 dx\int_0^1 dy\,x^{-\NNLOeps}\,\frac{f(x,y)-f(0,y)}{x}
\label{naive}
\end{eqnarray}
fails. 
To solve this problem, one can split the integration region into sectors 
where the variables $x$ and $y$ are ordered: 
$$I=\int_0^1 dx \int_0^1dy \,x^{-1-\epsilon}\,(x+y)^{-1}
\,[\underbrace{\Theta(x-y)}_{(a)}+\underbrace{\Theta(y-x)}_{(b)}]\;.$$
Then the integration domain  is remapped to the unit cube: 
After the substitutions $y=x\,t$ in sector (a) and 
$x=y\,t$ in sector (b), one has
\begin{eqnarray*}
I&=&\int_0^1 dx\,x^{-1-\epsilon}\int_0^1 dt
\,(1+t)^{-1}+\int_0^1 dy
\,y^{-1-\epsilon}\int_0^1 dt\,t^{-1-\epsilon}\,(1+t)^{-1}\;,
\end{eqnarray*}
where all singularities are factorised. 
For more complicated functions, several iterations 
of this procedure may be necessary, which can be easily implemented 
into an automated subroutine. Once all singularities are factored out, 
subtractions of the type (\ref{naive}) are possible and 
the result can subsequently be expanded in $\epsilon$. Note that  
the subtractions of the pole terms 
naturally lead to plus distributions\,\cite{Anastasiou:2003gr} 
by the identity 
$$x^{-1+\kappa\epsilon}=\frac{1}{\kappa\,
\epsilon}\,\delta(x)+
\sum_{n=0}^{\infty}\frac{(\kappa\epsilon)^n}{n!}
\,\left[\frac{\ln^n(x)}{x}\right]_+\; \mbox{ where } 
\int_0^1 dx \, f(x)\,\left[g(x)/x\right]_+=\int_0^1 dx \, 
\frac{f(x)-f(0)}{x}\,g(x)\;.$$
In this way, a 
Laurent series in $\NNLOeps$ is obtained, where the pole coefficients are 
sums of finite parameter integrals which can be evaluated numerically. 

For the numerical evaluation of loop integrals it has to be assured that 
no integrable singularities 
(e.g. thresholds) are crossed which spoil the numerical convergence. 
For integrals depending only on a single scale, which can be factored out, 
this does not pose a problem at all. 
For integrals with more than one scale, like for example two-loop box 
diagrams, the situation is more difficult, but 
in the case of $e^+e^-$ annihilation to massless final state 
particles, 
evaluation over the whole physical region is possible,  
as the kinematics of these 
processes is such that all Mandelstam variables are always non-negative. 

\subsubsection{Application to $e^+e^-\to 3$\,jets at NNLO}

In order to focus on a concrete example of phenomenological relevance, 
we will discuss the application of sector decomposition to the 
calculation of $e^+e^-\to 3$\,jets at NNLO in the following.

\paragraph{Virtual contributions}

The contributions to the amplitude which involve virtual integrals are composed of 
the two-loop corrections combined with a $1\to 3$ particle phase space, 
and the one-loop corrections combined with a $1\to 4$ particle phase space 
where one particle can become soft and/or collinear. 
In both cases, sector decomposition for loop integrals\,\cite{Binoth:2000ps} 
can serve to extract the poles in $1/\NNLOeps$ from the integrals. 
In what concerns the two-loop integrals, this part could also be 
taken from the literature, as the full two-loop matrix element is known 
analytically\,\cite{Garland:2001tf,Garland:2002ak}. This 
would save a considerable amount of CPU time.
The  two-loop matrix element will only 
depend on the invariants $y_1=s_{12}/q^2, y_2=s_{13}/q^2$ and $y_3=s_{23}/q^2$, 
where $q^2$ is the invariant mass of the $e^+e^-$ system and 
$\sum_{i=1}^3 y_i=1$. The subsequent phase space integration over the $y_i$
is trivial, and the 3-jet measurement function will make sure that 
all events where a singular limit $y_i\to 0$ is approached will be rejected. 

In the case of the one-loop contributions, the most complicated 
objects will be 5-point integrals with one off-shell external leg. 
Sector decomposition will lead to a result in terms of five 
independent scaled Mandelstam invariants $y_i$. This result has to be 
calculated up to order $\NNLOeps^2$, as it will be combined with 
the  $1\to 4$ parton phase space where one parton can become unresolved, 
leading to $1/\NNLOeps^2$ poles. This does not constitute a problem, 
as the expansions to higher order in $\epsilon$, 
as well as  $1\to 4$ parton phase space integrals,  
%over configurations with  unresolved particles, 
are well under control within sector decomposition.  
It is also possible to do parts of the loop integrations analytically to achieve 
a form which is suitable for subsequent 
sector decomposition\,\cite{Heinrich:2004jv,Anastasiou:2005pn}.
However, these contributions have not yet been implemented 
completely into a Monte Carlo program, because 
priority has been given to the most challenging part, which is the 
integration over the $1\to 5$ parton phase space where up to two 
partons can become unresolved.

\begin{figure}[htb]
%\begin{center}
\epsfig{file=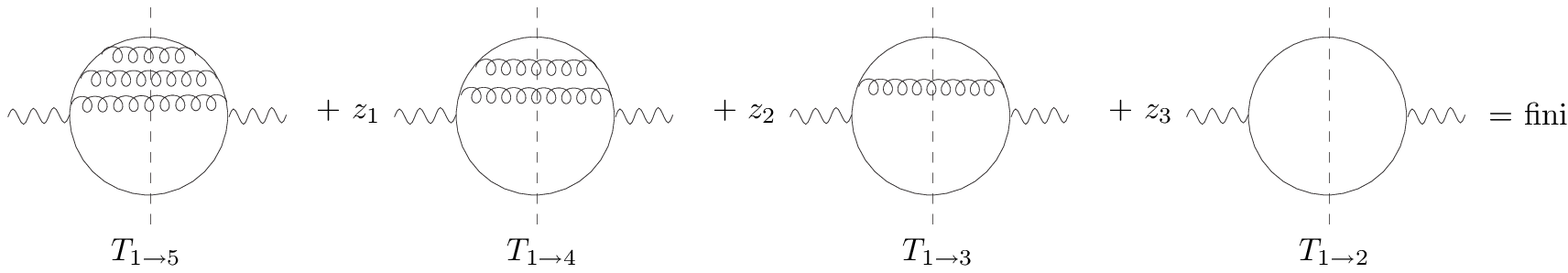,height=2.7cm}
%\end{center}
\caption{Cancellation of IR divergences in the sum over all cuts 
of the renormalised graph\label{NNLOfig2}}
\end{figure}

\paragraph{Real radiation at NNLO}
As mentioned above,  the 
main difficulty in calculating the real radiation part 
of $e^+e^-\to 3$\,jets at NNLO  is  the isolation and 
subtraction of the infrared poles which occur when integrating the 
squared amplitude over the phase space for $\gamma^*\to 5$ partons. 
In \cite{Heinrich:2006sw}, a method has been developed 
to tackle this problem. 
The correctness of the results for the integrals 
over the $1\to 5$ particle phase space can be checked by  exploiting  
the fact that the sum over all cuts of a given 
(UV renormalised) diagram must be infrared finite. 
This is shown in Figure~\ref{NNLOfig2} for a sample diagram:  
Summing over all cuts of this diagram and performing UV renormalisation, 
we obtain the condition 
\begin{eqnarray}
T_{1\to5}+z_1\,T_{1\to 4}+z_2\,T_{1\to 3}+z_3\,T_{1\to 2}&=&\mbox{finite}\;,
\label{cfin}
\end{eqnarray}
where $T_{1\to i}$ denotes the diagram with $i$ cut lines.
The renormalisation constants $z_i$ 
(in Feynman gauge) are given by\,\cite{Binoth:2004jv,Heinrich:2006sw} 
\begin{eqnarray}
z_1=C_F\frac{\NNLOas}{4\pi}\,\frac{1}{\NNLOeps}\, ,\, 
z_2=C_F^2\left(\frac{\NNLOas}{4\pi}\right)^2\,
\left(\frac{1}{2\NNLOeps^2}-\frac{1}{4\NNLOeps}\right)\, ,\,
z_3=C_F^3\left(\frac{\NNLOas}{4\pi}\right)^3\,
\left(\frac{1}{6\NNLOeps^3}-\frac{1}{4\NNLOeps^2}+\frac{1}{6\NNLOeps}\right)\;.
\end{eqnarray}
The important new ingredient 
in  eq.~(\ref{cfin}) is the calculation of $T_{1\to 5}$. 
The sector decompostion method leads to \cite{Heinrich:2006sw}
\begin{eqnarray}
T_{1\to 5}
&=&-C_F^3\left(\frac{\NNLOas}{4\pi}\right)^3\,T_{1\to 2}
\left\{
\frac{0.16662}{\NNLOeps^3}+\frac{1}{\NNLOeps^2}\,
[1.4993-0.4999\,\log{\left(\frac{q^2}{\mu^2}\right)}]\right.\NNLOnn\\
&&\left.+\frac{1}{\NNLOeps}\,[ 5.5959-4.4978\,\log{\left(\frac{q^2}{\mu^2}\right)}+
0.7498\,\log^2{\left(\frac{q^2}{\mu^2}\right)} ]\,+\,\mbox{finite}
\right\}\;,
\label{t5}
\end{eqnarray}
where the numerical accuracy is  1\%. 
The expressions entering eq.~(\ref{cfin}) for $i<5$ 
combine to \cite{Heinrich:2006sw}
\begin{eqnarray}
&&z_1\,T_{1\to 4}+z_2\,T_{1\to 3}+z_3\,T_{1\to 2}=\label{c234}\\
&&C_F^3\left(\frac{\NNLOas}{4\pi}\right)^3\,T_{1\to 2}
\left\{\frac{1}{6\NNLOeps^3}+\frac{1}{2\NNLOeps^2}\,[3-\log{\left(\frac{q^2}{\mu^2}\right)}]
 +\frac{1}{\NNLOeps}\,[5.61-\frac{9}{2}\log{\left(\frac{q^2}{\mu^2}\right)+
\frac{3}{4}\log^2{\left(\frac{q^2}{\mu^2}\right)}}]\,+
\,\mbox{finite}\right\}\,.\NNLOnn
\end{eqnarray}
We can see that  the poles in (\ref{c234}) 
are exactly cancelled by the 5-parton contribution (\ref{t5}) 
within the numerical precision.

\paragraph{Differential cross sections for various observables}

Although the sector decomposition approach is considered to be 
a ``numerical method", as the pole coefficients are only calculated 
numerically, 
the isolation of the poles  is an algebraic 
procedure, leading to a set of finite functions 
for each pole coefficient as well as for the finite part. 
This feature allows the inclusion of any 
(infrared safe) measurement function, at the level of the 
final Monte Carlo program, which means that the 
subtractions and expansions in $\NNLOeps$ do {\it not} have to be 
redone each time a different observable is considered.
However, some optional information about the physical singular limits, 
which does not spoil the above property,  
can be included at the stage of the $\NNLOeps$-expansion, 
thus avoiding the subtraction of certain ``spurious" singularities.

In \cite{Heinrich:2006sw}, it is shown how the four-momenta of 
the final state 
particles in terms of energies and angles can be reconstructed 
from the variables in which the sector decomposition is performed. 
In this way, fully differential information about the final state 
is available,  such that observables  can be calculated 
which cannot be cast into analytic functions, being  
complicated subroutines in the numerical program. 
As an example, the JADE algorithm\,\cite{Bethke:1988zc} to 
define 3--, 4-- and 5--jet events 
has been implemented into a Monte Carlo program built upon 
the output of sector decomposition\,\cite{Heinrich:2006sw}, 
using the 
multi-dimensional integration package BASES\,\cite{Kawabata:1995th}.   
The architecture of the program, being the one of a 
partonic event generator, is such that the JADE 
algorithm can be easily replaced by a different jet algorithm, 
and shape observables can also be defined.

\subsubsection{Outlook}

The method outlined here is a very powerful tool, especially for what 
concerns the double real radiation part of NNLO calculations, as  
it requires neither the manual construction of subtraction terms, 
nor the factorisation of the 
phase space and the analytic integration of the subtraction  terms 
 in the singular limits.  
A disadvantage of the sector secomposition approach is given by the fact that 
it produces very large expressions, as in each decomposition step, 
the number of original functions increases.
Therefore, CPU time is an issue for the treatment of  
processes with a large number of massless particles in the final state.  
However, the method sketched here  relies on a division of the amplitude squared 
into different ``topologies" corresponding to different classes of 
denominator structures, such that the problem is naturally split into smaller 
subparts. If such a ``trivial parallelisation" is not sufficient, 
there is still the possibility to parallelise the evaluation 
of the functions produced by sector decomposition. 
Furthermore, the size of the expressions can be reduced by including 
information about physical limits already at the level of the 
$\NNLOeps$-expansion, 
without loosing any flexibility for what concerns the definition of 
observables at the Monte Carlo level.

For the parts of the full matrix element for $e^+e^-\to 3$\,jets  
at NNLO considered so far, the numerical 
stability is very good. A reason might be that the subtractions within the 
sector decomposition method are local in the sense of plus distributions,  
i.e. the singular limits in each integration variable 
are directly subtracted.

As the method is based on a universal algorithm acting on integration 
variables,  it will surely see a number of interesting applications
in the future, in particular for what concerns the production of 
massive particles.  

%\end{document} 

%\bibliography{nnlo}

%%%%%%%%%%%%%%%%%%%%%%%%%%%%%PART%%%%%%%%%%%%%%%%%%%%%%%%%%%%%
\part[MONTE CARLO ISSUES]{MONTE CARLO ISSUES}
%%%%%%%%%%%%%%%%%%%%%%%%%%%%%PART%%%%%%%%%%%%%%%%%%%%%%%%%%%%%

%%%%%%%%%%%%%%%%%%%%%%%%%%%%%%%%%%%%%%%%%%%%%%%%%%%%%%%%%%%%%%%%%%%%%%%%%%%%%
\section[On reweighting techniques]
{ON REWEIGHTING TECHNIQUES~\protect\footnote{Contributed by: S.~Frixione}}
Fixed-order computations in perturbation theory are characterized
by low-multiplicity, parton-level final states. These are by far and
large unrealistic, and cannot be used in complex simulations such as
those performed by experiments to compute acceptances and to study
detector responses. For such purposes, parton shower Monte Carlos (MCs) 
are used instead. It is well known, however, that MCs lack the
capability of giving reliable predictions for total rates and for
observables sensitive to large-$p_{\SFsss T}$ emissions. To compensate
for this, MC results are typically multiplied by N$^i$LO $K$ factors,
i.e. the ratios of N$^i$LO cross sections over LO ones; this procedure
is called reweighting. Obviously, there are as many $K$ factors as
observables; the standard approach is that of using the $K$ factor
relevant to total rates. It is easy to realize, however, that such a
$K$ factor does not lead to any improvement of the MC results as
far as shapes are concerned. An alternative approach~\cite{Davatz:2004zg}
is that of selecting a given observable $O$, and reweight with the 
``differential'' $K$ factor $K(O)$. This will certainly correct the
shape of $O$ to the N$^i$LO accuracy, as well as the total rate. The
question is what happens to the shapes of other observables (and
even to $O$, in the case in which cuts are applied, which cannot be 
implemented in the fixed-order computation used to obtain $K(O)$).

The purpose of this note is to show that unweighting may actually lead
to worsening, rather than to improving, leading-order Monte Carlo
results. In order to do this, it is sufficient to find an
example in which this happens. Such example can be easily worked out 
in the context of a simple two-dimensional model. Thus, I consider the 
case of two kinematic variables with the following ranges
\begin{equation}
p\,,\;\;\;\;0\le p\le 1\,;\;\;\;\;\;\;
x\,,\;\;\;\;0\le x\le 1\,.
\end{equation}
I assume that the doubly-differential cross section is
\begin{eqnarray}
\frac{d\sigma^{\SFsss T}}{dxdp}&=&\frac{2}{\Gamma(a-1,1)}
\nonumber
\\*&\times&
\frac{e^{-1/p}}{p^a}
\left[\Theta\left(\SFhalf-x\right)\left(\SFhalf+\alpha\frac{1-p}{2}\right)+
\Theta\left(x-\SFhalf\right)\left(\SFhalf-\alpha\frac{1-p}{2}\right)\right]\,,
\label{truedxp}
\end{eqnarray}
where
\begin{equation}
\Gamma(a,x)=\int_x^\infty dt\,t^{a-1}\,e^{-t}
\end{equation}
is the incomplete $\Gamma$ function,
and the superscript T means ``true''. In Eq.~(\ref{truedxp}),
$a$ and $\alpha$ are free parameters; I assume that
\begin{equation}
a\ge 1\,;\;\;\;\;\;\;
-1\le \alpha\le 1\,,
\end{equation}
where the latter condition implies that the cross section is positive
definite. It is a matter of simple algebra to compute the single-inclusive
and total cross sections
\begin{eqnarray}
\frac{d\sigma^{\SFsss T}}{dp}&=&\frac{1}{\Gamma(a-1,1)}\frac{e^{-1/p}}{p^a}\,,
\label{truedp}
\\
\frac{d\sigma^{\SFsss T}}{dx}&=&
\Theta\left(\SFhalf-x\right)\left(1+\alpha-
\alpha\,\frac{\Gamma(a-2,1)}{\Gamma(a-1,1)}\right)
\nonumber \\*&+&
\Theta\left(x-\SFhalf\right)\left(1-\alpha+
\alpha\,\frac{\Gamma(a-2,1)}{\Gamma(a-1,1)}\right)\,,
\label{truedx}
\\
\sigma^{\SFsss T}&=&1\,.
\label{truetot}
\end{eqnarray}
The total rate is equal to one thanks to the prefactor that contains the 
$\Gamma$ function in Eq.~(\ref{truedxp}), which has actually been chosen
for this purpose. Eqs.~(\ref{truedxp})--(\ref{truedx}) suggest that $p$ 
may be seen as a rescaled (transverse) momentum; the larger the parameter $a$, 
the more steeply falling the distribution. The nature of $x$ doesn't need
to be specified here, since what follows actually applies to {\em any}
observable; in order to simplify the discussion, I assume that the cross
section is a constant in the ranges $x<1/2$ and $x>1/2$ (see 
Eq.~(\ref{truedx})); the difference between its values in those ranges
(i.e. the steepness of $d\sigma^{\SFsss T}/dx$) is proportional to $\alpha$. 
Notice that $x$ and $p$ are correlated, and that the slope in $x$ is
flatter the larger $p$.

I now want to apply the reweighting procedure of ref.~\cite{Davatz:2004zg}
to compute $d\sigma/dx$. The correct answer is that of Eq.~(\ref{truedx}),
which in an MC simulation is obtained by filling $x$ bins with the
weights computed with Eq.~(\ref{truedxp}) for all of the phase-space
points $(x,p)$ sampled during the run. However, in order to follow the
procedure of ref.~\cite{Davatz:2004zg}, I must assume that 
the true doubly-differential cross section (i.e., the correct 
MC simulation) is not available. What is available is an MC simulation
which is known to necessitate corrections. In the present simplified
approach, this corresponds to a doubly-differential cross section 
that I write as follows:
\begin{eqnarray}
\frac{d\sigma^{\SFsss U}}{dxdp}&=&\frac{2}{\Gamma(b-1,1)}
\label{wrongdxp}
\\*&\times&
\frac{e^{-1/p}}{p^b}
\left[\Theta\left(\SFhalf-x\right)\left(\SFhalf+\beta\frac{1-p}{2}\right)+
\Theta\left(x-\SFhalf\right)\left(\SFhalf-\beta\frac{1-p}{2}\right)\right]\,,
\nonumber
\end{eqnarray}
where the superscript U stands for ``uncorrected''. The functional form of 
Eq.~(\ref{wrongdxp}) is identical to that of Eq.~(\ref{truedxp}); this
obviously doesn't need to be so, but it simplifies the computations.
The two cross sections are different, however, since in general
$a\ne b$ and $\alpha\ne\beta$. The single-inclusive ``uncorrected''
cross section can be obtained from Eqs.~(\ref{truedp}) and~(\ref{truedx})
with the formal replacements $a\to b$ and $\alpha\to\beta$.
In particular, we have
\begin{eqnarray}
\frac{d\sigma^{\SFsss U}}{dx}&=&
\Theta\left(\SFhalf-x\right)\left(1+\beta-
\beta\,\frac{\Gamma(b-2,1)}{\Gamma(b-1,1)}\right)
\nonumber \\*&+&
\Theta\left(x-\SFhalf\right)\left(1-\beta+
\beta\,\frac{\Gamma(b-2,1)}{\Gamma(b-1,1)}\right)\,.
\label{wrongdx}
\end{eqnarray}
Ref.~\cite{Davatz:2004zg} proceeds by computing the $p$-dependent
correction factor
\begin{equation}
K(p)\equiv\frac{d\sigma^{\SFsss T}}{dp}\Bigg/\frac{d\sigma^{\SFsss U}}{dp}\,=\,
\frac{\Gamma(b-1,1)}{\Gamma(a-1,1)}\,p^{b-a}\,,
\label{Kdef}
\end{equation}
which is then applied event-by-event in the MC simulation. In the
formalism of this note, this is equivalent to defining a ``corrected''
doubly-differential cross section
\begin{eqnarray}
\frac{d\sigma^{\SFsss C}}{dxdp}&\equiv&K(p)\,\frac{d\sigma^{\SFsss U}}{dxdp}
\label{corrdxp}
\\*&=&
\frac{2}{\Gamma(a-1,1)}
\nonumber \\*&\times&
\frac{e^{-1/p}}{p^a}
\left[\Theta\left(\SFhalf-x\right)\left(\SFhalf+\beta\frac{1-p}{2}\right)+
\Theta\left(x-\SFhalf\right)\left(\SFhalf-\beta\frac{1-p}{2}\right)\right]\,,
\nonumber
\end{eqnarray}
from which we obtain the ``corrected'' differential distribution
\begin{eqnarray}
\frac{d\sigma^{\SFsss C}}{dx}&=&
\Theta\left(\SFhalf-x\right)\left(1+\beta-
\beta\,\frac{\Gamma(a-2,1)}{\Gamma(a-1,1)}\right)
\nonumber \\*&+&
\Theta\left(x-\SFhalf\right)\left(1-\beta+
\beta\,\frac{\Gamma(a-2,1)}{\Gamma(a-1,1)}\right)\,.
\label{corrdx}
\end{eqnarray}
We must now understand whether Eq.~(\ref{corrdx}) is a good 
approximation of Eq.~(\ref{truedx}). In order to study this, I introduce
the shorthand notation
\begin{eqnarray}
\frac{d\sigma^{\SFsss A}_<}{dx}&=&\frac{d\sigma^{\SFsss A}}{dx}
\Theta\left(\SFhalf-x\right)\,,
\\
\frac{d\sigma^{\SFsss A}_>}{dx}&=&\frac{d\sigma^{\SFsss A}}{dx}
\Theta\left(x-\SFhalf\right)\,,
\end{eqnarray}
where A = T, U, C and define
\begin{eqnarray}
R^{\SFsss U,C}_<&=&\frac{d\sigma^{\SFsss U,C}_<}{dx}\Bigg/
\frac{d\sigma^{\SFsss T}_<}{dx}\,,
\label{Rmdef}
\\
R^{\SFsss U,C}_>&=&\frac{d\sigma^{\SFsss U,C}_>}{dx}\Bigg/
\frac{d\sigma^{\SFsss T}_>}{dx}\,,
\label{Rpdef}
\\
S^{\SFsss U,C}&=&\left(\frac{d\sigma^{\SFsss U,C}_>}{dx}-
\frac{d\sigma^{\SFsss U,C}_<}{dx}\right)\Bigg/
\left(\frac{d\sigma^{\SFsss T}_>}{dx}-\frac{d\sigma^{\SFsss T}_<}{dx}\right)\,.
\label{Sdef}
\end{eqnarray}
If the ``uncorrected'' cross sections were coincident with the ``true'' ones,
the $R$'s and $S$'s defined in Eqs.~(\ref{Rmdef})--(\ref{Sdef}) would be all 
equal to one. On the other hand, the larger the values of $|R-1|$ and $|S-1|$,
the worse the agreement between the ``true'' cross section and the
``uncorrected'' or ``corrected'' ones. By construction, $R_<$ and $R_>$ are
relevant to the rate for $x<1/2$ and $x>1/2$ respectively, while
$S$ is relevant to the slope.

Using Eqs.~(\ref{truedx}), (\ref{wrongdx}), and~(\ref{corrdx})
we readily get
\begin{eqnarray}
R^{\SFsss U}_<&=&\frac{\Gamma(a-1,1)}{\Gamma(b-1,1)}\,
\frac{(1+\beta)\,\Gamma(b-1,1)-\beta\,\Gamma(b-2,1)}
{(1+\alpha)\,\Gamma(a-1,1)-\alpha\,\Gamma(a-2,1)}\,,
\label{Rwmres}
\\
R^{\SFsss C}_<&=&\frac{(1+\beta)\,\Gamma(a-1,1)-\beta\,\Gamma(a-2,1)}
{(1+\alpha)\,\Gamma(a-1,1)-\alpha\,\Gamma(a-2,1)}\,,
\label{Rcmres}
\\
S^{\SFsss U}&=&\frac{\beta\,\Gamma(a-1,1)}{\alpha\,\Gamma(b-1,1)}\,
\frac{\Gamma(b-1,1)-\Gamma(b-2,1)}
{\Gamma(a-1,1)-\Gamma(a-2,1)}\,,
\label{Swref}
\\
S^{\SFsss C}&=&\frac{\beta}{\alpha}\,,
\label{Scref}
\end{eqnarray}
and the results for $R^{\SFsss U,C}_>$ can be obtained from Eqs.~(\ref{Rwmres})
and~(\ref{Rcmres}) with the formal replacements $\alpha\to -\alpha$,
$\beta\to -\beta$. With the equations above we can explicitly verify
that when $b=a$ and $\beta=\alpha$ (i.e., the ``uncorrected'' cross section
is identical to the ``true'' one), then $R=S=1$.

%%%%%%%%%%%%%%%%%%%%%%%%%%%%%%%%%%%%%%%%%%%%%%%%%%%%%%%%%%%%%%%%%%%
\begin{figure}[htb]
  \begin{center}
      \epsfig{figure=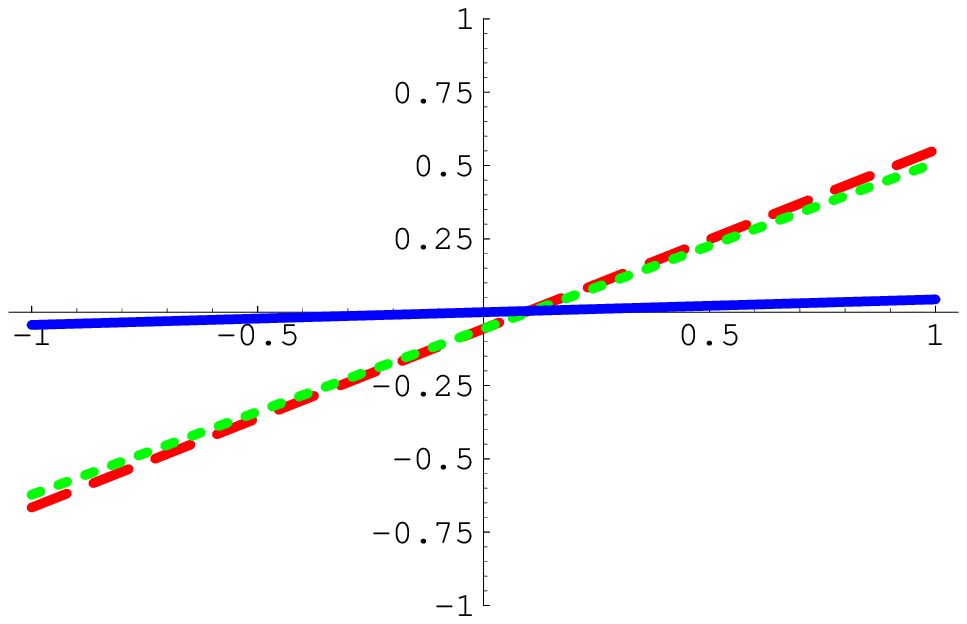,width=0.46\textwidth}
~~~~~~\epsfig{figure=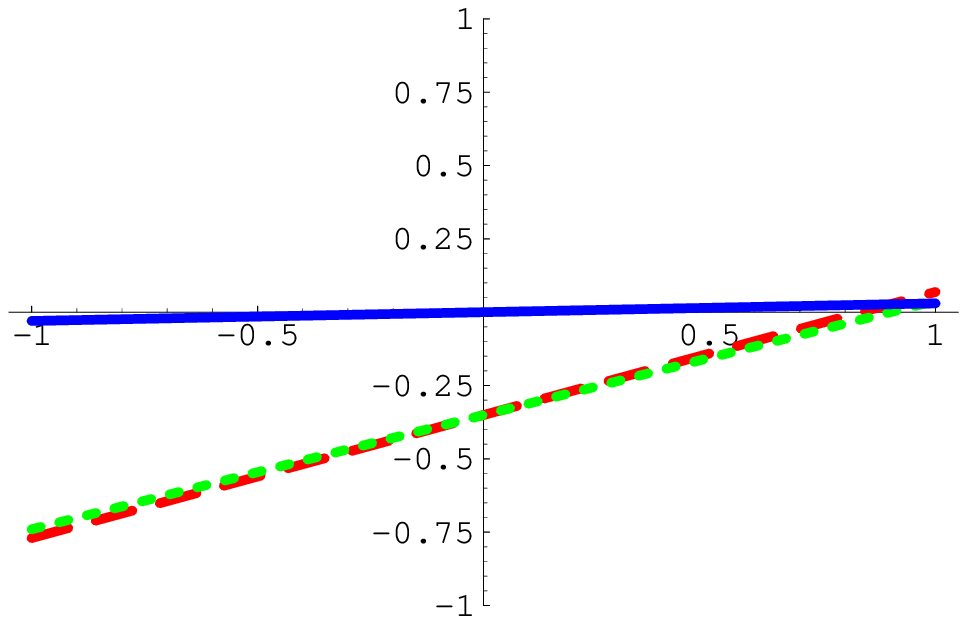,width=0.46\textwidth}
\caption{\label{fig:one} 
I plot here $1-R^{\SFsss U}_<$ (red long-dashed line), $1-R^{\SFsss C}_<$ 
(green short-dashed line), and $R^{\SFsss C}_<-R^{\SFsss U}_<$ 
(blue solid line) as a function of $\beta$, for $a=4$, $b=4.5$, and 
$\alpha=0.1$ (left panel) or $\alpha=0.9$ (right panel).
}
  \end{center}
\end{figure}
%%%%%%%%%%%%%%%%%%%%%%%%%%%%%%%%%%%%%%%%%%%%%%%%%%%%%%%%%%%%%%%%%%%
%%%%%%%%%%%%%%%%%%%%%%%%%%%%%%%%%%%%%%%%%%%%%%%%%%%%%%%%%%%%%%%%%%%
\begin{figure}[htb]
  \begin{center}
      \epsfig{figure=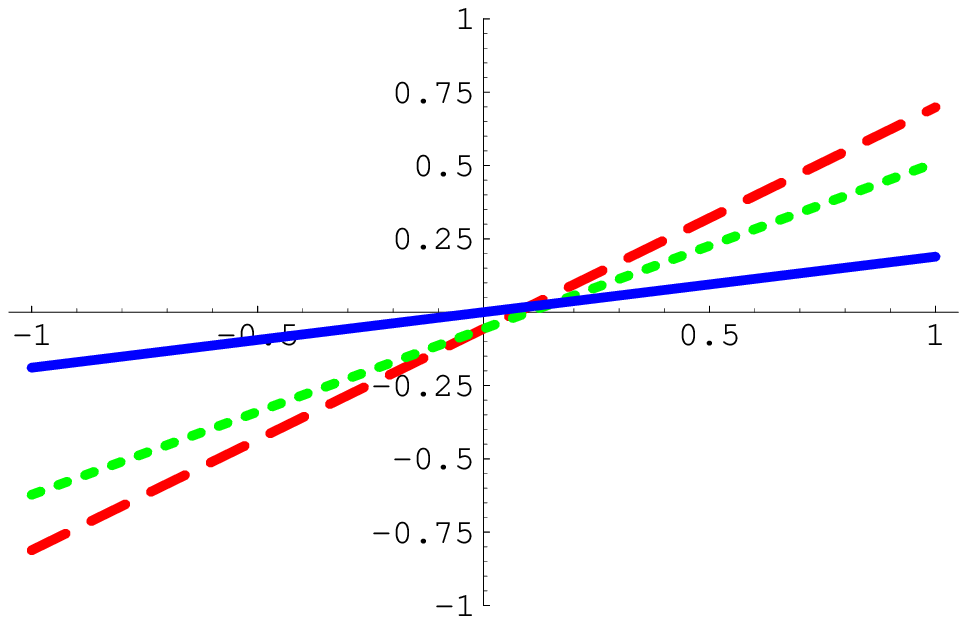,width=0.46\textwidth}
~~~~~~\epsfig{figure=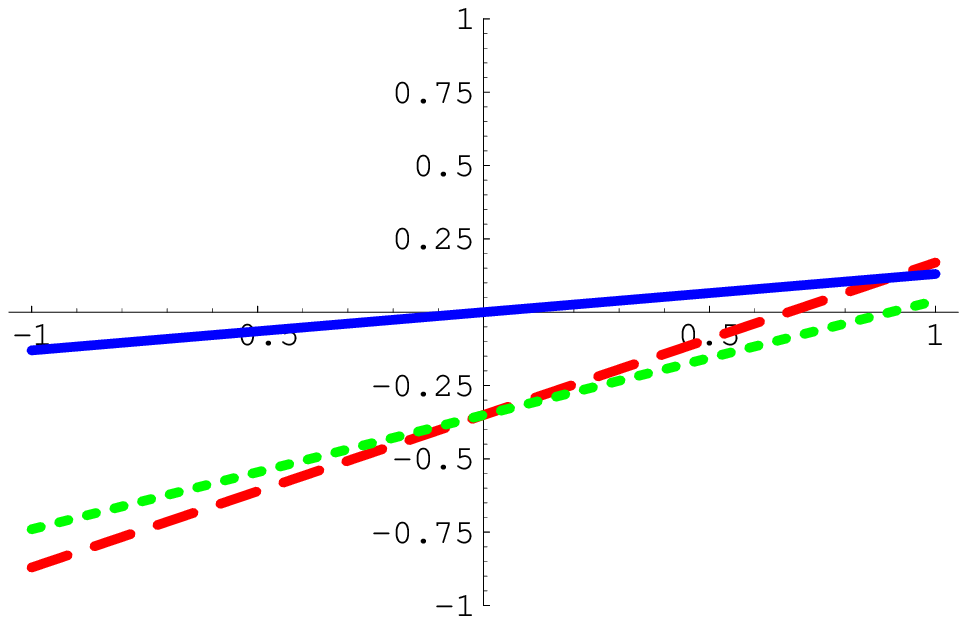,width=0.46\textwidth}
\caption{\label{fig:two}
As in Fig.~\ref{fig:one}, except for the value of $b$; here $b=7$.
}
  \end{center}
\end{figure}
\begin{figure}[htb]
  \begin{center}
      \epsfig{figure=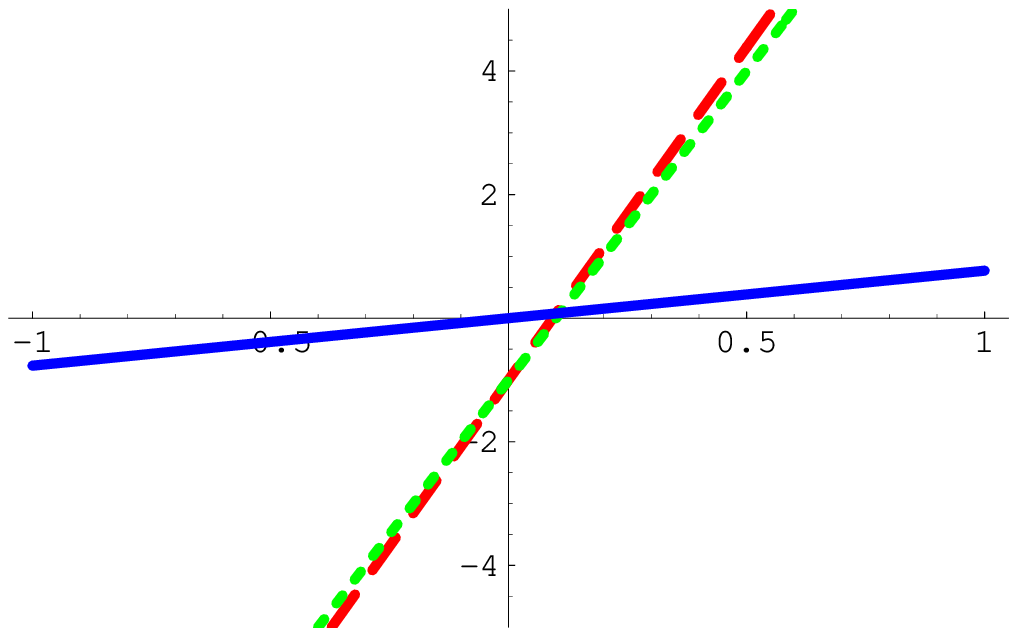,width=0.46\textwidth}
~~~~~~\epsfig{figure=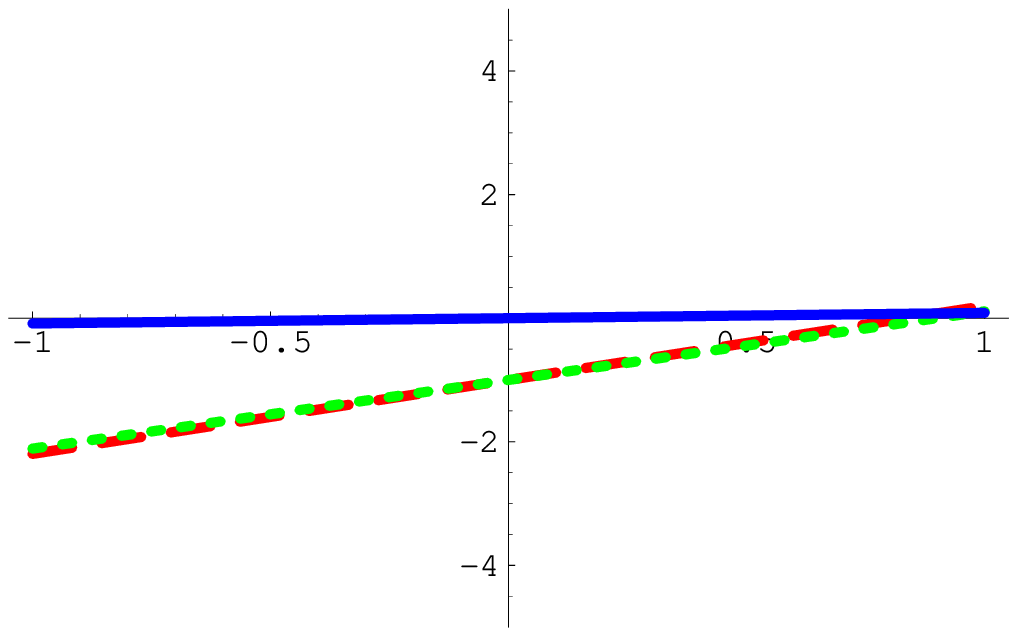,width=0.46\textwidth}
\caption{\label{fig:three} 
As in Fig.~\ref{fig:one}, for $S^{\SFsss U}$ and $S^{\SFsss C}$.
}
  \end{center}
\end{figure}
\begin{figure}[htb]
  \begin{center}
      \epsfig{figure=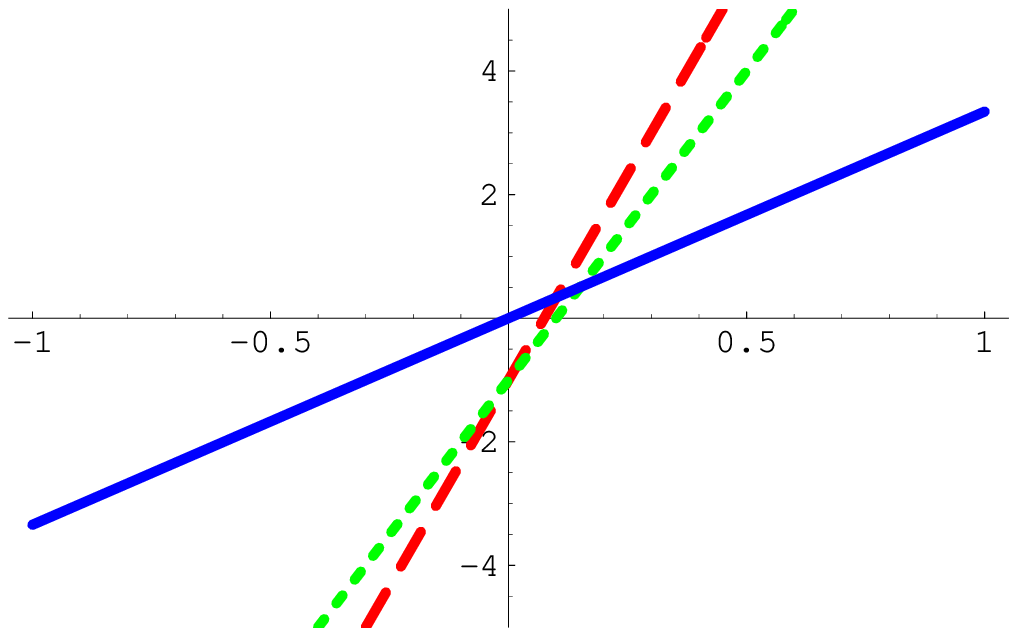,width=0.46\textwidth}
~~~~~~\epsfig{figure=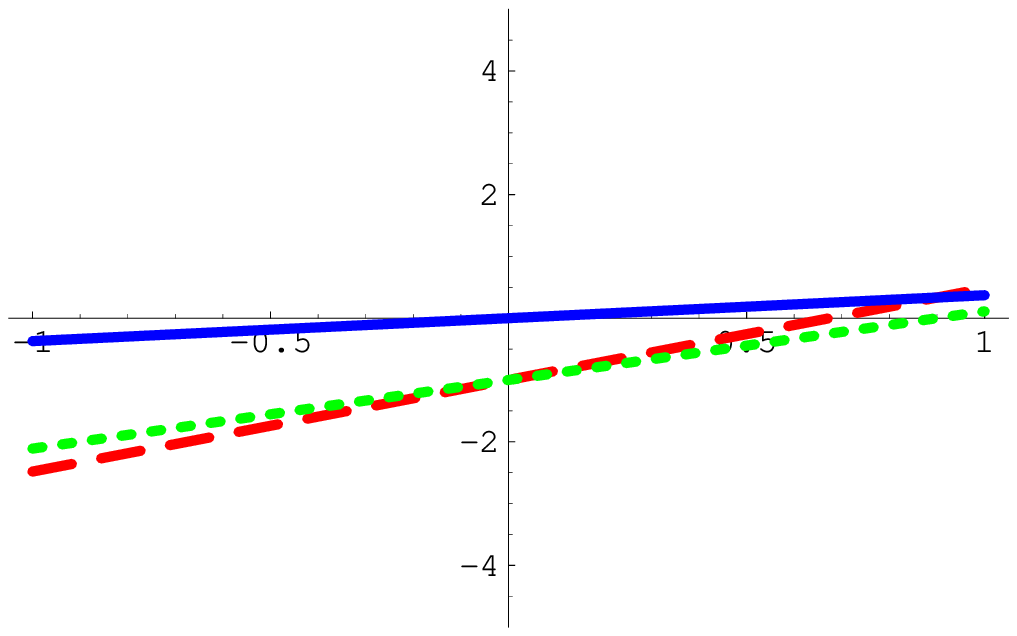,width=0.46\textwidth}
\caption{\label{fig:four}
As in Fig.~\ref{fig:two}, for $S^{\SFsss U}$ and $S^{\SFsss C}$.
}
  \end{center}
\end{figure}

%%%%%%%%%%%%%%%%%%%%%%%%%%%%%%%%%%%%%%%%%%%%%%%%%%%%%%%%%%%%%%%%%%%
I proceed by observing that, if the derivative of the ``true'' $x$ 
distribution has the opposite sign of that of the ``uncorrected'' $x$ 
distribution (say, the former is decreasing while the latter is increasing, 
i.e. $\alpha>0$ and $\beta<0$), then clearly the reweighting cannot correct 
this behaviour, as is most evident from Eq.~(\ref{Scref}). In order
to have an idea of what happens in general, I plot in Figs.~\ref{fig:one}
and~\ref{fig:two} the quantities $1-R^{\SFsss U}_<$, $1-R^{\SFsss C}_<$, and 
$R^{\SFsss U}_<-R^{\SFsss C}_<$, as functions of $\beta$ for given values of 
$\alpha$, $a$, and $b$. In Figs.~\ref{fig:three} and~\ref{fig:four} I plot 
$1-S^{\SFsss U}$, $1-S^{\SFsss C}$, and $S^{\SFsss U}-S^{\SFsss C}$. 
By inspection of the figures, we can see that when
the $p$ ``true'' and ``uncorrected'' distributions are similar ($a=4$,
$b=4.5$), the difference between the ``uncorrected'' and ``corrected'' $x$
distributions is fairly marginal; the ``corrected'' $x$ distribution
may display a disagreement with respect to the ``true'' $x$ distributions that
can be as large as 50\%. The agreement with the ``true'' result
obviously improves when $\beta\simeq\alpha$, but in such a case 
one wouldn't advocate the necessity of a reweighting procedure at all.
In the case in which the $p$ ``uncorrected'' distribution is much steeper than
the ``true'' one ($a=4$, $b=7$), the effect of the reweighting is more
pronounced, but this doesn't imply that the ``corrected'' $x$ distribution
improves the ``uncorrected'' one, since this appears to depend on the value of 
$\alpha$. In any case, the ``corrected'' $x$ distribution agrees better 
with the ``uncorrected'' than with the ``true'' one.

%%%%%%%%%%%%%%%%%%%%%%%%%%%%%%%%%%%%%%%%%%%%%%%%%%%%%%%%%%%%%%%%%%%
%%%%%%%%%%%%%%%%%%%%%%%%%%%%%%%%%%%%%%%%%%%%%%%%%%%%%%%%%%%%%%%%%%%
%%%%%%%%%%%%%%%%%%%%%%%%%%%%%%%%%%%%%%%%%%%%%%%%%%%%%%%%%%%%%%%%%%%
%%%%%%%%%%%%%%%%%%%%%%%%%%%%%%%%%%%%%%%%%%%%%%%%%%%%%%%%%%%%%%%%%%%
In conclusions: it is obvious that the functional form chosen 
here for the cross section is too simplistic to give a proper
description of the complex final state which emerges from a hadronic
collision. It does show, however, that the results of reweighting
may be contrary to expectations, since the corrected cross section
may have a larger disagreement with the physical cross section than
the uncorrected one. As expected, this is more likely to
happen when the reweighting function (Eq.~(\ref{Kdef}) in the context
of the model discussed in this note) is not flat, which is precisely 
when the use of an observable-dependent $K$ factor would be advocated.
For the majority of the parameter choices considered here, reweighting
does improve the ``uncorrected'' result, but the improvement is pretty
marginal. It is impossible to say whether this will also be the 
case for a physical observable. It appears in fact that only the 
comparison with the ``true'' result allows one to assess the accuracy 
with which the reweighting does its job. If the ``true'' result is not 
available, it is impossible to give a sound estimate of the uncertainties 
involved in the procedure. Clearly, the availability of the ``true''
result would render the whole procedure useless; a practical strategy
may then be that of checking that the corrected cross sections obtained
starting from two or more different ``uncorrected'' predictions (say, resulting
from different Monte Carlos) are in better mutual agreement than the 
``uncorrected'' ones; it is clear, however, that such a strategy may 
easily fail.

%%%%%%%%%%%%%%%%%%%%%%%%%%%%%%%%%%%%%%%%%%%%%%%%%%%%%%%%%%%%%%%%%%%%%%%%%%%%%
\section[LCG MCDB --- database of Monte-Carlo simulated events]
{LCG MCDB --- DATABASE OF MONTE-CARLO SIMULATED EVENTS~\protect
\footnote{Contributed by: P.~Bartalini, S.~Belov, L.~Dudko, A.~Gusev,
    A.~Sherstnev}}
\subsection{LCG MCDB Overview}

The LCG MCDB proposal was presented at the Les Houches workshop in 2003~\cite{Dobbs:2004bu,Bartalini:2004nd} 
This paper gives a status report of the LCG MCDB project.

The LCG MCDB project has been created to facilitate communication between
experts of Monte-Carlo (MC) generators and users of the LHC collaborations. 
It provides flexible infrastructure to share generated MC event
samples (MC samples) and the corresponding book-keeping in a convenient way,
with dedicated interfaces to the users and to the authors. 

The LCG MCDB tool is particularly useful for samples that require a frequent interaction
between users and MC experts, or significant CPU resources.
Nowadays, the LCG MCDB project is ready for LHC community and provides many useful interfaces for authors of 
MC samples and for the users. A dedicated web server has been deployed: \verb|http://mcdb.cern.ch|.

The adoption of a central database of MC events is motivated by simulation needs which are specific to the high energy physics domain.
In general, the correct MC simulation  of complex processes requires a rather sophisticated expertise. 
Often, different physics groups in various experimental collaborations approach the same experts and authors of MC generators, respectively, in order to generate MC samples for a particular physics process.
Having these events stored in a public place along with the corresponding documentation, allows for direct cross checks of the performances on reference samples, and prevents possible waste of precious human and computing resources.

The main motivation behind the MCDB project is to make sophisticated MC event samples available for 
various physics groups. For example, the same MC samples of Standard Model (SM) processes 
can be used for the investigations in some SM effects as well as a background for some
studies of new phenomena. Public availability of the event samples helps to speed up the
validation procedure of the events and provides the public stage  for rapid communication
between authors of the samples and their users. 
The previous version~\cite{cms_mcdb_url} of MCDB was launched by the CMS collaboration in 2002.
The main limitations of the CMS MCDB are the AFS based storage supporting
only small size MC samples (basically only parton level events from matrix element tools) and the
lack of search functionalities, mostly based on phonetic keys.

The significant interest shown by the potential users motivated the MCDB migration to the LCG framework,
benefiting from a much more powerful, standardized and exportable 
software tools that are available to all the LHC collaborations.
The LCG MCDB~\cite{lcg_mcdb_url} is now almost ready. In the next sections
we will briefly describe the subsystems and modules of the LCG MCDB, providing instructions for the users.

\subsection{LCG MCDB Description}

The subsystems and software technologies adopted in the LCG MCDB are described in this section. 
The LCG MCDB is based on the following software technologies: WEB, CGI, PERL, SQL, XML, CASTOR and GRID. 
All of the developed  software is available in LCG CVS~\cite{savanna}. The software is organized as a set of
modules with the possibility to export the LCG MCDB software to other sites on the grid. 
We provide a daily backup of the SQL DB and double mirroring of the samples in CASTOR.
 
\begin{figure}
\begin{center}
\includegraphics[width=0.6\textwidth]{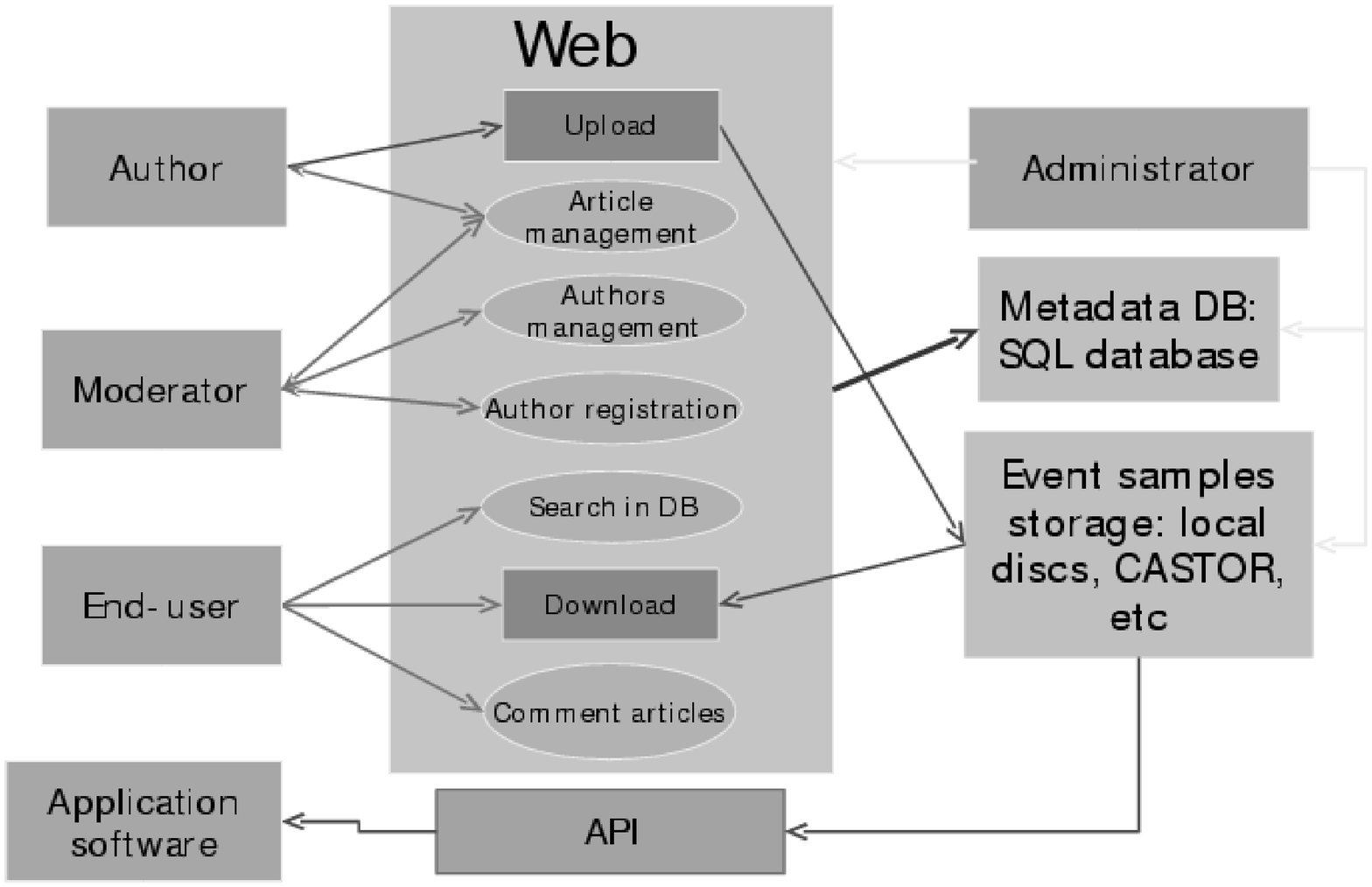}
 \caption{Scheme of the LCG MCDB}
\label{lcg_mcdb_scheme}
\end{center}
\end{figure}

The main concept of the LCG MCDB is the ARTICLE, which is a document describing 
a set of event samples. MCDB articles are divided into CATEGORIES, i.e. a set of articles 
concerning a particular type 
of physics process (e.g. top physics, Higgs physics) or theoretical model 
(e.g. supersymmetry, extra dimensions). There are four different types of permissions 
to access the LCG MCDB.
The USER access is reserved for users who are interested in requesting a new event sample or in downloading or documenting comments to the already published event samples.
The AUTHOR access is reserved to authorized users (MC experts). Only an AUTHOR can upload a new event sample.
The MODERATOR access is reserved to moderators who manage author profiles and monitor other information. 
The ADMINISTRATOR access is reserved to software developers and maintainers who take care of the LCG MCDB itself. 
The scheme of the LCG MCDB is shown in Fig.~\ref{lcg_mcdb_scheme}.

\subsubsection{WEB interface} 
   The main interface of the LCG MCDB is based on WEB technologies. It is split
in two parts:
\begin{itemize}
\item[$\vartriangleright$] The user interface, where any user can apply for
new event samples, search and browse for the already available samples, read the description of the 
events, download the samples, ask a question about the samples and read the previous 
discussion concerning it.

\item[$\vartriangleright$] Author area, where authorized authors can
upload new samples to the database and describe them using a template system. The system
has a lot of pre-entered information. Authors can interact with users on the public
forums attached to each article. With the same interface authors can edit previous
articles or make the articles temporarily inaccessible to users.

\end{itemize}

\subsubsection{SQL DB} 
The LCG MCDB adopts MySQL. 
The SQL technology  provides a possibility to keep information in a very 
structured way. Authors provide documentation of their event 
samples through forms with pre-filled forms from the cache or from the selection menu. 
In this way the description of new events (e.g. MC generator, theoretical model, parameters of generation, kinematic
cuts, etc.) turns out to be much simplified. 

\subsubsection{Storage}
For the native storage of event samples we have selected CASTOR~\cite{Baud:2003ys}, because of the absence of
serious space limitations and considering its popularity in the LHC collaborations.
We provide direct CASTOR paths for all LCG MCDB samples along with the possibility to
get the samples through different interfaces (http, GridFtp etc.). 
A local disk cache system is used to speed up the storage operations.

\subsubsection{Search engine}
Since we use a SQL DB, it is possible to provide the possibility
for a variety of complex search queries, including those specifying relations between DB objects. 
The deployed WEB search interface is realized as 
a dynamic query construction wizard which is based on the JavaScript XML-query constructor.
%than XSLT processing to construct SQL query. 
The development of application programming interfaces to specific external
software (for example a simulation framework of a LHC collaboration) may benefit of similar tools 
in order to simplify the query construction.

\subsubsection{Authorization}
We pay a special attention to the security of the transactions in all LCG MCDB operations. 
There are two possibilities of authorization. The first one is the authorization
with CERN AFS/Kerberos login/password, all of the transactions are encrypted by SSL
technology. The second possibility is to authorize with LCG GRID certificates. Authors can
choose any of these two types of authorizations or both of them. Both of these
authorization methods are standard at CERN and any CERN user can use at least one
of these two methods. 
%An authorized author has access to edit files and articles in LCG MCDB
%only if he/her is the author or coauthor of this article.

\subsubsection{Documentation}
 Most of the LCG MCDB documentation is available from the
 dedicated web server. The information is separated in two parts, corresponding 
 to the technical and the user part. The first part describes the implementation of the LCG MCDB itself.
 The second part is organized as a set of HOW-TOs for users and authors. 
 A separate documentation (available from the CVS repository) is devoted to the developers of the LCG MCDB software.
 
 In most of the cases it is appropriate to refer to the
 set in the HOW-TO for the users and authors, which provides short answers on the
 most common questions with corresponding links to solutions of the problems. 
 A brief start-up manual for not experienced LCG MCDB users is also available
 in the next section of this document. In addition, there are two freely accessible mailing lists
 dedicated to users and developers. Their addresses are available
 in the documentation section at the main web page.

\subsubsection{API to collaboration software}
In the near future, some emphasis will be put on the development of 
application programming interfaces (API) specific to the simulation environments of the LHC collaborations. 
This work will require constant interaction with contact persons in the LHC collaborations. 
The main
idea of this subsystem is to write a set of routines for the collaboration software which
would allow a direct access to the LCG MCDB files during the 
MC production on computer farms.
Access to LCG MCDB samples should not represent a big issues as, at present, collaboration software can already use the
direct CASTOR paths to get the event samples or download the samples from web interfaces.

\subsubsection{HEPML, unified XML format of simulated events}
Another useful feature, which we plan to implement in the LCG MCDB, is a unified
XML format of event samples. At present, each MC generator
supports its own output format. 

Some authors of matrix element tools provide interface programs to pass the events of a 
particular MC generator to the subsequent level of simulation (e.g. showering, hadronization decays) which are based on the Les Houches Accord number one .

However, there is still no agreement on a possible unified format to save intermediate MC information to a file.
%ther'is no unified interface to read all  possible event files and compare physics properties simulated by different MC generators, 
The most appropriate  technology for the unified event format seems to be XML, which provides the possibility to describe
the stored information in a very flexible and standardized way. 
Different MC generators may use the same tag for the description of a physics parameter, or may need to
keep specific information (through the introduction of a dedicated tag).
In other words, the HEPML format should consist of 
many possible XML tags, separated in two different sets describing general and MC specific information respectively. 
Possible internal adaption of this representation to the most popular Monte Carlo generators would
result in a significant improvement of the Monte Carlo documentation and book-keeping.
The LCG MCDB project will in any case support a set of HEPML tags to document event samples internally, and will promote its usage in other environments.
The first practical attempts to introduce a standardized set of MC tags has been performed already,
for example by the CEDAR collaboration~\cite{cedar_url}. 
A dedicated document discussing the details of the requirements and describing the HEPML proposal 
will appear in the near future.

\subsection{How to use the LCG MCDB.}
A user who needs simulated events for a particular process can browse the MCDB categories
and sub-categories (menu at the left side of the main LCG MCDB web page~\cite{lcg_mcdb_url}) and
verify, whether an appropriate sample has been already generated. 
If this is the case, the users may want to read the article describing how the event sample has been prepared
(parameters of the theoretical model, generator name and generation parameters, kinematic cuts, etc.). 
At the bottom of the page a link to the uploaded file(s) is provided, as well as the CASTOR path. 
On top of that, the web page also contains a link to the "Users Comments" interface, where users can
ask questions about the sample and browse the previous discussion on the
article. Users do not need any authorization for the steps described above.

The following procedure has to be followed if one needs to publish a sample in the LCG MCDB (becoming author):

\begin{enumerate}
\item Register as a new author with the link at the right side menu of
the web page~\cite{lcg_mcdb_url}, wait for the confirmation e-mail
\item Login to the LCG MCDB authors area
\item Choose "Create New Article" in the authors menu which will appear
at the right side after the authorization.
\item Fill all necessary fields in the documentation template, 
which will appear (title, generator, theoretical model, cuts, etc.)
\item Upload your event files in the "Event Files" sub-window.
\item Click "Preview/Save" slice and check the box "Publish"
\end{enumerate}

To be authorized in the LCG MCDB, the author needs a valid CERN AFS login or a LCG digital 
certificate. Authors can save unfinished articles in MCDB and resume to correct them at
any moment. Authors can edit their previous articles that are already published on the web or 
make the articles publicly inaccessible for a while.

The LCG MCDB team will appreciate any possible bug report, feedback, comments or suggestions for possible new implementations concerning the LCG MCDB.

%%%%%%%%%%%%%%%%%%%%%%%%%%%%%%%%%%%%%%%%%%%%%%%%%%%%%%%%%%%%%%%%%%%%%%%%%%%%%
\section[Supporting Monte Carlo generators at the LHC]
{SUPPORTING MONTE CARLO GENERATORS AT THE LHC~\protect
\footnote{Contributed by: P.~Bartalini, L.~Dudko, M.~Kirsanov, A.~Sherstnev}}
\subsection{Introduction}

The LCG Simulation project covers a range of activities in the simulation as 
part of the LCG Applications Area, encompassing common development and 
validation work among the LHC experiments on the GEANT4, FLUKA and GARFIELD 
simulation engines as well as on Monte Carlo generators.

The mandate of the LCG Generator project is to collaborate with Monte Carlo (MC) generators authors and with LHC 
experiments in order to prepare, validate and maintain LCG code for both the theoretical and experimental 
communities at the 
LHC, sharing the user support duties, providing assistance for the development of the new object oriented generators and 
guaranteeing the maintenance of the older packages on the LCG supported platforms. 
Contact persons for most of MC generator packages relevant for the LHC and representatives for all the LHC experiments 
have been agreed. Four different work packages (WP) have been defined: 

\begin{itemize}
\item[WP1] Generator services library;
\item[WP2] Event interfaces and particle services; 
\item[WP3] Production, storage and book-keeping of public generator level events; 
\item[WP4] Monte Carlo Validation.
\end{itemize}

\noindent
This paper describes the status and the development guidelines in the four different work packages, concentrating on the 
main activity, i.e. the MC generator services library (GENSER).

\subsection{WP1: the generator services library}

Two different options are available to support Monte Carlo packages in LCG: 
they can be simply stored in the  LCG external area~\cite{spi} or they can be rather
migrated in the dedicated LCG Generator Services module (GENSER)~\cite{chep1}, 
adapting the directory structure according to the LCG policy.
This second solution has been adopted for most of the installed packages. 
However, for each MC package, 
an ad-hoc solution is found taking into account the authors directives and the user requirements. 
Top priority and second priority packages pursued for inclusion in the generator library have been indicated in the 
report of the RTAG 9 working group.

\subsubsection{GENSER}
GENSER is the LCG module for MC generators and generator tools. 
It was the first module in the LCG Simulation CVS repository. 
The sources and the binaries are installed in AFS and the tarballs are made available by the 
Software Process and Infrastructure group (LCG-SPI). This new library has gradually replaced the obsolete CERN 
library for what concerns the MC generators support. In fact GENSER is currently widely adopted as the standard 
Monte Carlo generators library by most of the LHC experiments.

The management of the GENSER releases has been recently improved and is currently coordinated by the central LCG 
librarian from CERN PH/SFT. ~GENSER is fully independent from other large libraries and currently follows a quarterly 
release scheme. Quick bug fixes and special versions can be produced under request. Most of the MC sub-package versions 
produced by the authors are installed. Old versions are maintained as long as they are used.

Configuration and build systems for the librarian and end users 
are based on the SCRAM technology~\cite{Wellisch:2003wb}; future versions of GENSER will support Makefiles as well. 
At the moment GENSER is considered to be at the ``production quality'' stage. 
The current version of GENSER (1.2.1) comprises both shared and static libraries for the platform 
slc3\_ia32\_gcc323.

\subsubsection{External Monte Carlo packages}
The following MC generator packages are stored in the LCG external area, however they are completly supported in GENSER 
with corresponding examples and test suites:
\begin{itemize}
\item   EVTGEN~\cite{Lange:2001uf} version alpha-00-11-07;
\item   SHERPA~\cite{Gleisberg:2003xi} version 1.0.5, 1.0.6;
\item   COMPHEP~\cite{Boos:2004kh} versions 4.2.p1, 4.4.0.
\end{itemize}

\subsubsection{Internal Monte Carlo packages}

The following MC generator packages have been migrated in GENSER, along with the corresponding test and validation 
code:

\begin{itemize}
\item   PYTHIA~\cite{Sjostrand:2000wi,Sjostrand:2004ef} versions 6.205, 6.217, 6.220, 6.221, 6.222, 6.223, 6.224, 6.227,6.304, 6.319, 6.320, 6.321, 6.324;
\item   HERWIG~\cite{Corcella:2000bw} versions 6.500, 6.503, 6.504, 6.505, 6.506, 6.507, 6.508, 6.510;
\item   JIMMY~\cite{Butterworth:1996zw} version 4.1, 4.2;
\item   ISAJET~\cite{Paige:2003mg} versions 7.67, 7.69, 7.71;
\item   HIJING~\cite{Wang:1991ht} versions 1.36, 1.37, 1.383;
\item   MC@NLO~\cite{Frixione:2003ei,Frixione:2002ik} version 2.3.1, 3.1.0;
\item   ALPGEN~\cite{Mangano:2002ea} version 1.3.2, 2.0, 2.01, 2.03, 2.05;
\item   TOPREX~\cite{Slabospitsky:2002ag} version 4.09;
\item   MADGRAPH~\cite{Maltoni:2002qb} version 3.2;
\item   FEYNHIGGS~\cite{Heinemeyer:1998yj} version 2.2.9, 2.2.10;
\item   LHAPDF~\cite{Whalley:2005nh} versions 1.1, 2.0, 3.0, 4.0, 4.1.1, 5.0;
\item   PHOTOS~\cite{Golonka:2005pn} versions 207, 209, 2.14, 2.15;
\item   PHOJET~\cite{Bopp:1998rc} version 1.10;
\item   GLAUBER Xs~\cite{Abdel-Waged:2000pc} version 1.0;
\item   CHARYBDIS~\cite{Harris:2003db} version 1.001;
\item   STAGEN~\cite{Dimopoulos:2001hw} version 1.07 (including TRUENOIR and two GRAVITON codes);
\item   EVTGENLHC~\cite{evtgenlhc} versions 1.2, 1.3.
\end{itemize}

\noindent
In this list, EVTGENLHC represents a special case.
EVTGENLHC is the LHC version of EVTGEN, a Monte Carlo following the spin density matrix formalism that is particularly 
dedicated to the simulation of B decays and specifically designed for B production at the $\Upsilon$(4S) resonance. 
EVTGEN currently comprises one of the largest tuneable and upgradeable collection of decay models.

EVTGENLHC has been set-up and provided in GENSER by the LHCb collaboration. 
It includes an interface to the HEPMC event record that allows for modularization with plug-in to the most popular
general purpose Monte Carlo generators (typically parton shower QCD models).
Mixing description and CP violation implementation have been adapted to the case of incoherent B meson production.

Common work between the LHC collaborations is currently developing as a LCG Generator activity,
with the participation of the orginal EVTGEN authors. This project is 
concentrating on the implementation of particle polarizations 
and on the extension of the decay models to B$_s$, B$_c$ and to B baryons.
LCG Generator also pursues a common initiative 
between experiments at LHC, Tevatron and at the B factories
for the tuning of the EVTGEN decay tables (to be developed in WP4).

%\noindent
%Work has already started for the preparation of the first version of GENSER that will be declared of production quality 
%(1.0.0) to be released by the end of the year 2004. 
%It will extend the support to the Scientific Linux platform~\cite{chep18} and 
%it will contain most of the top priority and second priority packages indicated in the RTAG 9 document. 
%From GENSER 1.0.0 the management of the GENSER releases will be improved, providing appropriate documentation for the 
%development, technical and user support duties and the corresponding task assignment. The full compliance to the LCG 
%standards will be guaranteed. A specific GENSER FAQ will be developed.
%It was broken out from Pythia 7~\cite{chep20} to better factorize the parts which are Pythia-specific from those which are
%general model-independent components of the toolkit and which can be used by any event generator model.

%LCG Generator has set a common milestone with the PHENOGRID~\cite{chep21} initiative for the third quarter of 2005: 
%the first test of ThePEG and EvtGenLHC integration in Herwig++ (and the possible preliminary inclusion of Herwig++ 
%in GENSER).
%The next important milestone deals with the inclusion of the first object oriented general purpose MC generator in GENSER.%SHERPA~\cite{chep22} has been identified as a possible candidate. Work has already started with the authors and should be 
%finalized in the first quarter of 2005.

\subsection{WP2: event interfaces and particle services}
The goal of WP2 is to contribute to the definition of the standards for generator interfaces and formats, collaborating 
in the development of the corresponding application programming interfaces (API).

\subsubsection{ThePEG}
In order to favor the adoption of the new object oriented MC generators in the experiment simulation frameworks, the LCG 
Generator project will share some responsibilities on the development and maintenance of the Toolkit for High Energy 
Physics Event Generation (THEPEG)~\cite{Lonnblad:2004bb}. LCG Generator has set a common milestone with the 
PHENOGRID initiative~\cite{phenogrid} for mid 2005: 
the first test of ThePEG and EvtGenLHC integration in Herwig++.

\subsubsection{HEPML}
HEPML~\cite{Bartalini:2004nd} is a meta-data format where the information is sub-divided in two parts:
\begin{itemize}
\item   The header, that contains the general information concerning the event sample, i.e. author,  creation date, collider description, generator specific data, generation cuts, other physical parameters, parser directives etc.
\item   The event records, i.e. the variable data of events written in some compact format to one string 
(particle momenta, colour chains etc.).
\end{itemize}

\noindent
The header is stored in a text file with XML Syntax. The event records are zip-compressed and attached to the header file.
The HEPML meta-data format provides the basis to the SQL search for public generator level events (WP3).

\subsection{WP3: production, storage and book-keeping of public generator level events}
The goal of WP3 is to produce "certified" public generator event files to be used for benchmarks, comparisons and 
combinations. The format and the structure of the files will be accessible to the simulation frameworks of the LHC 
experiments. Three different activities have been started in this work package: simulation framework, production and book-keeping and storage.

\subsubsection{Simulation framework}
The development of a simple production and validation framework at generator level is
a common software project between LCG and CMS.
A new package has been designed which is relying on HepMC (event interface), 
ROOT and POOL (event storage). 
The beta version of the framework will be available in the end of 2005.

\subsubsection{Production}
A dedicated production centre integrated in the gird middleware will provide the LHC experiments and the other 
end-users with a transparent access to the public event files. This will be essential for those samples requiring an 
huge amount of CPU time and parallelization techniques.

\subsubsection{Book-keeping and storage}
The LCG Monte Carlo Data Base (MCDB)~\cite{Bartalini:2004nd} is a public service for the configuration, 
book-keeping and storage of the generator level event files.  
A prototype is currently in production on a dedicated web server~\cite{mcdbs}. 
Details on MCDB are given in another section of these proceedings.

\subsection{WP4: Monte Carlo validation}
The Monte Carlo validation work package is divided in two different parts:
basic sanity checks and the physics validation.

The activity is currently concentrating on the functional validation of the generator packages 
inserted in GENSER. The basic sanity checks are currently performed in a standalone way. The code is provided by the 
authors, beta testers and librarians and it is stored under the TEST module in the simulation repository. It will be 
subsequently integrated with the simple generator level production framework (developed in WP3).

In the long term the physics validation will be performed with JETWEB~\cite{Butterworth:2002ts}, assuming that it will be interfaced 
to GENSER in a reasonable time scale (i.e. by mid 2006). 

\subsection*{Acknowledgements}

We are indebted with all the other members of the LCG Generator team:
F.~Ambroglini, (INFN-Perugia);
M.~Bargiotti, A.~Pfeiffer and A.~Ribon (CERN);
S.~Belov, S.~Korobov and V.~Uzhinsky (JINR);
J.~Cuevas~Maestro and H.~Naves (Cantabria);
I.~Katchaev and S.~Slabospitsky (IFVE);
S.~Makarychev and I.~Seluzhenkov (ITEP).

We are particularly grateful to all the contributors to the EVTGENLHC
project. In particular we wish to thank D. Lange (Livermore); A. Ryd (Cornell);
P.~Robbe (LHCb, LAL Paris XI); M.~Smizanska and J.~Catmore (ATLAS, Lancaster);
Michela Biglietti (ATLAS, Michigan).

We wish to thank also the LCG-SPI team, as well as our contact persons in the
LHC experiments and in the MC generator projects.

\bibliography{smh05}
\end{document}